\documentclass[a4paper,11pt]{report}
\pdfoutput=1
\usepackage[a4paper,top=2cm, bottom=2cm, left=2.75cm, right=2.5cm]{geometry}
\usepackage{amsthm}
\usepackage{amssymb}
\usepackage{amsmath}
\usepackage{bm}
\usepackage{cite}
\usepackage{graphicx}
\usepackage{tikz}
\usepackage{subfig}
\usepackage{mathrsfs}
\usepackage{ulem}
\usepackage{latexsym,amsfonts,amsmath,amssymb,enumerate,epsfig}
\usepackage{mathdots}

\usepackage{hyperref}

\usepackage[a4paper]{meta-donnees}

\newcommand\bra[1]{{\langle#1|}}
\newcommand\ket[1]{{|#1\rangle}}

\newcommand\bbra[1]{{\langle\!\langle#1|}}
\newcommand\kket[1]{{|#1\rangle\!\rangle}}

\newcommand{\llangle}{\langle\!\langle}
\newcommand{\rrangle}{\rangle\!\rangle}
\newcommand{\steady}{|{\cal S}\rangle}
\newcommand{\finproof}{{\hfill \rule{5pt}{5pt}}}

\newcommand\id{\mathbf{1}}
\newcommand{\eps}{\varepsilon}
\newcommand{\LL}{{\mathbb L}}

\def\cA{{\cal A}}    \def\cB{{\cal B}}    \def\cC{{\cal C}}
\def\cD{{\cal D}}        \def\cF{{\cal F}}
    \def\cH{{\cal H}}    
    \def\cK{{\cal K}}    \def\cL{{\cal L}}
\def\cM{{\cal M}}    \def\cN{{\cal N}}    \def\cO{{\cal O}}
\def\cP{{\cal P}}        \def\cR{{\cal R}}
\def\cS{{\cal S}}    \def\cT{{\cal T}}    \def\cU{{\cal U}}
\def\cV{{\cal V}}        \def\cX{{\cal X}} 
    \def\cZ{{\cal Z}}

\def\fC{{\mathfrak C}}

\def\fP{{\mathfrak P}}

\def\fT{{\mathfrak T}}

\def\fg{{\mathfrak g}}

\def\fm{{\mathfrak m}}

\def\fs{{\mathfrak s}}
\def\ft{{\mathfrak t}}

\newcommand{\CC}{{\mathbb C}}

\newcommand{\II}{{\mathbb I}}

\newcommand{\NN}{{\mathbb N}}

\newcommand{\PP}{{\mathbb P}}

\newcommand{\RR}{{\mathbb R}}

\newcommand{\ZZ}{{\mathbb Z}}

\numberwithin{equation}{section}
\newtheorem{definition}{Definition}[section]
\newtheorem{example}[definition]{Example}
\newtheorem{conjecture}[definition]{Conjecture}
\newtheorem{proposition}[definition]{Proposition}
\newtheorem{theorem}[definition]{Theorem}
\newtheorem{corollary}[definition]{Corollary}
\newtheorem{lemma}[definition]{Lemma}
\newtheorem{remark}[definition]{Remark}

\usepackage{engrec}

\renewcommand{\theequation}{\Roman{chapter}.\Alph{section}.\arabic{equation}}

\begin{document}



\Specialite{Physique th\'{e}orique}
\Arrete{25 mai 2016}
\Auteur{Matthieu Vanicat}
\Directeur{Eric Ragoucy}
\CoDirecteur{Nicolas Cramp\'{e}}    
\Laboratoire{Laboratoire d'Annecy-le-Vieux de Physique Th\'{e}orique }
\EcoleDoctorale{\'{E}cole doctorale de physique de Grenoble}         
\Titre{Approche int\'{e}grabiliste des mod\`{e}les de physique statistique hors d'\'{e}quilibre}
\Depot{30 Juin 2017}       




\Jury{

 
\UGTExaminateur{Eric Bertin}{CNRS/LIPhy} 
\UGTExaminateur{Luigi Cantini}{Universit\'{e} Cergy-Pontoise/LPTM}     
\UGTPresident{Bernard Derrida}{Coll\`{e}ge de France}
\UGTExaminateur{Luc Frappat}{Universit\'{e} Savoie Mont-Blanc/LAPTh}
\UGTRapporteur{Vincent Pasquier}{CEA/IPhT}
\UGTExaminatrice{V\'{e}ronique Terras}{CNRS/LPTMS}     

}

\Rapp{
\UGTRapp{Jan De Gier}{University of Melbourne (Australie)}      
\UGTRapp{Vincent Pasquier}{CEA/IPhT}      

}

\MakeUGthesePDG    

\chapter*{Remerciements}

Je tiens tout d'abord \`{a} remercier Eric Bertin, Luigi Cantini, Bernard Derrida, Luc Frappat, Vincent Pasquier et V\'{e}ronique Terras 
d'avoir accept\'{e} de faire partie du jury et d'avoir donn\'{e} de leur temps pour \'{e}valuer mon travail. Je suis tr\`{e}s reconnaissant \`{a} Bernard Derrida
d'avoir endoss\'{e} le r\^{o}le de pr\'{e}sident du jury. J'adresse aussi un remerciement particulier \`{a} Jan De Gier et Vincent Pasquier d'avoir tr\`{e}s gentiment
accept\'{e} d'\^{e}tre les rapporteurs de ce manuscrit de th\`{e}se, et qui ont, semble-t-il, surv\'{e}cu \`{a} mon anglais parfois un peu approximatif.

Cette th\`{e}se est le fruit de trois ann\'{e}es de travail au cours desquelles j'ai eu la chance d'\'{e}changer et de collaborer avec de nombreuses personnes.
Tout d'abord, bien s\^{u}r, avec mes directeurs de th\`{e}se Eric et Nicolas qui ont \'{e}t\'{e} d'excellents encadrants (j'y reviendrai en d\'{e}tails apr\`{e}s). Ensuite
au sein du LAPTh avec Luc (et sa m\'{e}moire incroyable!) et Caley. Cela a \'{e}t\'{e} un r\'{e}el plaisir de travailler avec vous. Je remercie Vladimir Rittenberg
d'\^{e}tre venu nous rendre visite \`{a} Annecy, ainsi que Tomaz Prosen avec lequel les discussions ont \'{e}t\'{e} tr\`{e}s int\'{e}ressantes. J'ai aussi une pens\'{e}e 
particuli\`{e}re pour Kirone, que j'ai eu la chance de croiser \`{a} de nombreuses occasions, pour sa p\'{e}dagogie, sa comp\'{e}tence et sa gentillesse. 
Je remercie Martin Evans pour m'avoir accueilli \`{a} deux reprises \`{a} \'{E}dimbourg ainsi que pour m'avoir transmis sa soif insatiable de solutions exactes 
en produit de matrices! Je n'oublie pas mes ``grands fr\`{e}res acad\'{e}miques'' Sam et Vincent que je remercie grandement pour l'absence de bizutage,
pour les discussions physiques/math\'{e}matiques et pour les bonnes rigolades. Merci \`{a} Vincent de m'avoir prouv\'{e} que sport et physique th\'{e}orique peuvent 
se marier \`{a} la perfection. Parmi les visiteurs ``venus de loin'' je remercie Michael Wheeler pour son cours sur les polyn\^{o}mes de Macdonald et 
pour le verre de Saint-\'{E}milion. Pour continuer avec la nationalit\'{e} Australienne j'aimerais revenir \`{a} Caley. Apr\`{e}s ma premi\`{e}re d\'{e}ception d'apprendre
que tu ne pratiquais pas le surf, j'ai beaucoup appr\'{e}ci\'{e} de partager mon bureau avec toi durant ces deux derni\`{e}res ann\'{e}es. En plus de cours 
d'anglais pr\'{e}cieux, cela a \'{e}t\'{e} tr\`{e}s b\'{e}n\'{e}fique de collaborer avec toi. Un immense merci \'{e}galement pour la relecture du manuscrit, je sais bien qu'il y 
a de meilleures fa\c{c}ons de passer ses week-ends, surtout \`{a} Annecy.

Cela m'offre une transition pour parler de l'incroyable cadre de travail dont j'ai b\'{e}n\'{e}fici\'{e}. Il y a bien s\^{u}r le lac et les montagnes mais le LAPTh
offre en lui-m\^{e}me un cadre extr\^{e}mement agr\'{e}able par son aspect familial. Les g\^{a}teaux partag\'{e}s autour du caf\'{e} apr\`{e}s le d\'{e}jeuner, agr\'{e}ment\'{e}s par 
des conversations souvent tr\`{e}s dr\^{o}les (bien que parfois \'{e}tranges, merci Philippe!) vont tr\`{e}s certainement me manquer. J'en suis reconnaissant 
\`{a} tous les membres du laboratoire, permanents et non permanents, et notamment \`{a} Fawzi et Luc qui ont successivement dirig\'{e} le LAPTh durant ma p\'{e}riode de th\`{e}se.
Je remercie tout particuli\`{e}rement les th\'{e}sards et post-docs avec qui j'ai partag\'{e} de bons moments: Tom, Luis, 
Vivian et Mathieu (Gauthier) pour le trail, Romain, Vincent (Germain), L\'{e}o, Thibaud, Jordan et M\'{e}ril pour les parties de ping-pong qui donnaient 
parfois le tournis. Yoann pour ses techniques pour aller finir les g\^{a}teaux en douce dans l'apr\`{e}s-midi, Philippe pour son soutien sans faille de 
matheux face \`{a} la horde de th\'{e}sards qui passent leur temps \`{a} rechercher la mati\`{e}re sombre! Une pens\'{e}e aussi pour les th\'{e}sards du labo qui ont fini 
avant moi: Vincent (Bizouard) pour tes propositions de sorties plus improbables les unes que les autres, comme cette randonn\'{e}e de nuit en ski. 
Merci \`{a} Mathieu (Boudaud), avec qui j'ai partag\'{e} mon bureau puis qui est devenu mon colocataire, pour son enthousiasme sans faille et pour les courses
de paret d\'{e}jant\'{e}es au refuge du Danay. Merci aussi \`{a} Dominique, V\'{e}ronique et Virginie pour leur redoutable efficacit\'{e} et gentillesse au service 
administratif du LAPTh. Cela a \'{e}t\'{e} un grand plaisir de discuter hand avec toi Dominique et de venir regarder des matchs dans l'ambiance survolt\'{e}e de 
ton salon!

J'ai aussi eu la chance de passer plusieurs mois au sein du L2C \`{a} Montpellier o\`{u} j'ai toujours \'{e}t\'{e} accueilli tr\`{e}s chaleureusement. P\^{e}le-m\^{e}le, merci 
\`{a} Michele, Lucas, Maxime, Julien, Domi, J\'{e}r\^{o}me, Jean-Charles et tous les autres pour tous les bons moments pass\'{e}s lors des repas/caf\'{e}s partag\'{e}s 
avec vous.

Une grande partie du m\'{e}rite de mon travail de th\`{e}se revient incontestablement \`{a} mes directeurs Eric et Nicolas. Tout d'abord vous m'avez propos\'{e} un 
sujet de th\`{e}se qui m'a \'{e}norm\'{e}ment int\'{e}ress\'{e}. J'ai appris beaucoup de choses, notamment gr\^{a}ce \`{a} vous, et je peux sans h\'{e}siter affirmer que j'ai 
pass\'{e} des ann\'{e}es intellectuellement tr\`{e}s \'{e}panouissantes. Il faut savoir que Eric et Nicolas sont les encadrants les plus disponibles que l'on puisse
imaginer. Durant ces ann\'{e}es de th\`{e}se, la porte de leur bureau a toujours \'{e}t\'{e} grande ouverte, \`{a} n'importe quel moment de la journ\'{e}e, malgr\'{e} leur 
emploi du temps charg\'{e}. Cela a \'{e}t\'{e} un grand plaisir d'apprendre et de travailler \`{a} vos c\^{o}t\'{e}s. Vous avez fait preuve d'une grande gentillesse et je peux
dire, qu'en plus d'avoir \'{e}t\'{e} des professeurs et collaborateurs hors pairs, vous \^{e}tes devenus des amis. Merci \`{a} toi Nicolas de m'avoir gentiment
accueilli chez toi \`{a} Montpellier et fait d\'{e}couvrir quelques vins locaux. Merci \`{a} toi Eric de m'avoir initi\'{e} au ski de randonn\'{e}e et pour tous les 
bons moments pass\'{e}s au bord du lac ou sur les sentiers de randonn\'{e}es.

Il me tient aussi \`{a} c\oe{}ur d'avoir un mot pour mes amis ``picards'': K\'{e}vin, Mickael, Pierre, Audric..., 
mes amis ``parisiens'': Damien, Olivier, Benjamin, Ludovica... 
(merci Damien pour ton talent pour remonter le moral de n'importe qui en racontant de la mani\`{e}re la plus dr\^{o}le 
qui soit tes p\'{e}rip\'{e}ties quotidiennes!) et mes amis ``cachanais'': Maxime, William, Chris, Thibaud, Romain, Pierre, Keurcien, Micka...(m\^{e}me \`{a} 
distance vous parvenez toujours \`{a} me faire rire!).

J'aimerais pour finir adresser un remerciement particulier \`{a} ma famille: mes grands-parents, mes parents ainsi que mon fr\`{e}re pour leur encouragement
et leur soutien sans faille. C'est sans nul doute gr\^{a}ce \`{a} vous que j'ai pu autant m'\'{e}panouir dans mes \'{e}tudes et que je suis parvenu l\`{a}
o\`{u} j'en suis aujourd'hui. Merci enfin \`{a} \'{E}milie, pour tout ce qui a d\'{e}j\`{a} \'{e}t\'{e} mentionn\'{e}, et pour tout le reste.

\chapter*{Preface}

The aim of this thesis is to study out-of-equilibrium statistical physics from the integrable systems point of view.
Integrability is used to obtain exact results on models relevant for non-equilibrium physics.

The theory of equilibrium statistical physics has been very successful to describe the properties of systems at 
thermodynamic equilibrium. Unfortunately, such a theory still eludes us for out-of-equilibrium systems. We lack a theoretical framework and 
fundamental principles (such as the entropy maximization principle for equilibrium physics) describing the behavior of such systems.
We do not know, for instance, how the Boltzmann distribution should be modified.
In the last decades, however, there have been promising attempts at constructing generalizations
of thermodynamic potentials to out-of-equilibrium systems.
The framework of large deviation theory has become very popular and appears particularly efficient to express the out-of-equilibrium properties. It is
seen as a possible unifying formalism to deal with statistical physics systems, both in and out of equilibrium.
A short introduction to out-of-equilibrium statistical physics can be found in chapter one. We define 
the framework of Markov chains and introduce the notion of a non-equilibrium stationary state (NESS) in this context. We point out the relevance of 
the cumulant generating function and the large deviation function for well chosen ``dynamic'' observables to describe these NESS. 
We explain how these quantities can be in principle exactly computed with the help of a 
current-counting deformation of the Markov matrix. We recall also the main tools and properties of the underlying framework, the large 
deviation theory. The macroscopic fluctuation theory (MFT) uses the large deviation theory to state a general 
framework to describe non-equilibrium diffusive systems in the thermodynamic limit. An introduction on MFT can be found in chapter five.

In this context, the role of exactly solvable models (integrable models) is central. They can be used as a benchmark to test the predictions of the 
theories, and they may also help to guess potential fruitful developments. It thus appears important to construct these integrable 
out-of-equilibrium models, and to develop methods to compute their stationary states and dynamical properties analytically. This manuscript 
aims to be part of this process. We build on existing techniques to construct several new examples of integrable out-of-equilibrium models.
An introduction to integrability in the context of Markov processes can be found in chapter two. The key notions, such as 
conserved quantities, $R$ and $K$-matrices, transfer matrices and Bethe ansatz are introduced in the perspective of out-of-equilibrium
statistical physics. We detail the periodic boundary condition case as well as the open boundaries condition case. Some techniques 
for solving the Yang-Baxter and reflection equations are exposed: for example, through the quantum groups framework, and Baxterisation procedures.

To complete the process, it appears highly important to build convenient methods to solve these integrable out-of-equilibrium models and 
extract exact expressions of physical quantities.
In the last few years, the matrix ansatz technique has proven to be very efficient at expressing analytically the stationary state of one dimensional 
interacting particles systems. It can be seen as a bridge making connection between the non-equilibrium stationary states and the theory of 
integrable systems. It will play a key role in this manuscript.
A review of the state of the art of this method can be found in chapter three. 
After introducing it on the totally asymmetric simple exclusion process (TASEP), we give its main properties and we show that it often allows us 
to compute conveniently relevant physical observables, such as the particle currents and densities.
We explain how it can be used in integrable models.
We compute exactly the stationary states of some examples in matrix product form and derive analytical expressions for relevant physical
quantities. These results allow us to test in particular cases some predictions of the MFT.

The outline of the manuscript is as follows. The first chapter is dedicated to non-equilibrium statistical physics. We try to present 
concisely the main relevant concepts. We attempt to give a modest new perspective on the Langevin equation on the ring with a non-conservative force, 
using the rooted trees expansion of the steady state of non-equilibrium Markovian processes.

The second chapter deals with integrability. We review the general theory and provide some new results. 
First we give an interpretation of the transfer matrices of the periodic 
and open TASEP as discrete time Markovian processes. Then we introduce new algebraic structures
to construct solutions to the Yang-Baxter equation and to the reflection equation through Baxterisation procedures. This yields for instance the 
determination of integrable boundary conditions for the multi-species asymmetric simple exclusion process (ASEP). 
We also show how the coordinate Bethe ansatz can be used to find new integrable systems with two species of particles that can react in the bulk.
Finally we introduce new integrable out-of-equilibrium models, given by the resolution of the 
Yang-Baxter and reflection equations: the dissipative symmetric simple exclusion process (DiSSEP),
an open boundaries 2-species TASEP and an open boundaries multi-species symmetric simple exclusion process (multi-species SSEP).

Chapter three presents the matrix ansatz technique, which aims to express analytically non-equilibrium stationary states. 
The connection with the integrable models is explored in detail and suggests a systematic construction of a matrix product steady state in such models.
It relies on two key relations: the Zamolodchikov-Faddeev and Ghoshal-Zamolodchikov relations. 
These general prescriptions are then illustrated by the analytical matrix product 
construction of the stationary state of the three previously introduced integrable stochastic processes:
the DiSSEP, the 2-species TASEP and the multi-species SSEP. It allows us to compute exactly relevant physical quantities in these models, 
such as particle densities and currents.

Chapter four addresses the problem of the exact computation of the cumulants of the current in the open boundaries ASEP from a new perspective.
The current-counting deformation of the Markov matrix is studied through the quantum Knizhnik-Zamolodchikov ($q$KZ) equations. We provide
solutions to these equations in a matrix product form. 
The connection between the solutions of the $q$KZ equations and the Koornwinder polynomials is explored. It yields a matrix product expression for 
certain Koornwinder polynomials. Moreover it suggests an unexpected link between the theory of symmetric polynomials and the 
cumulant generating function of the current in the stationary state. The latter is conjectured to be obtained as a specific limit of 
symmetric Koornwinder polynomials.

The last chapter deals with the hydrodynamic limit. We show how the large system size limit can be performed on the physical observables 
of the three models introduced in chapter three. It yields the mean density and current profiles, but also, depending on the model under 
consideration, the phase diagram and the relaxation rate (gap). We then present a coarse grained description of the diffusive models studied in 
the manuscript, called the macroscopic fluctuation theory. This general approach to non-equilibrium systems in the hydrodynamic limit can 
in principle predict the fluctuations of the particles current and density in the stationary state. We check these predictions for models with 
evaporation and condensation of particles against the exact results obtained in chapter \ref{chap:three} for the cumulant of the current in the 
DiSSEP. We also propose an extension of the theory for multi-species diffusive models based on the analytical results derived for the 
multi-species SSEP.

The thesis is based on the following published papers:

\begin{itemize}
 \item Crampe, N., Ragoucy, E., Vanicat, M., 
 \textsl{Integrable approach to simple exclusion processes with boundaries. Review and progress.}
 J. Stat. Mech. (2014) P11032.
 
 \item Crampe, N., Mallick, K., Ragoucy, E., Vanicat, M., 
 \textsl{Open two-species exclusion processes with integrable boundaries},
 J. Phys. \textbf{A 48} (2015)  175002.
 
 \item Crampe, N., Mallick, K., Ragoucy, E., Vanicat, M.,
 \textsl{Inhomogeneous discrete-time exclusion processes}, 
 J. Phys. \textbf{A 48} (2015) 484002.
 
 \item Crampe, N., Ragoucy, E., Rittenberg, V., Vanicat, M., 
 \textsl{Integrable dissipative exclusion process: Correlation functions and physical properties}, 
 Phys. Rev. \textbf{E 94} (2016) 032102.
 
 \item Crampe, N., Evans, M. R., Mallick, K., Ragoucy, E., Vanicat, M.,
 \textsl{Matrix product solution to a 2-species TASEP with open integrable boundaries},
 J. Phys. \textbf{A 49} (2016) 475001.
 
 \item Crampe, N., Finn, C., Ragoucy, E., Vanicat, M., 
 \textsl{Integrable boundary conditions for multi-species ASEP}, 
 J. Phys. \textbf{A 49} (2016), 375201.
 
 \item Crampe, N., Frappat, L., Ragoucy, E., Vanicat, M.,
 \textsl{3-state Hamiltonians associated to solvable 33-vertex models}, 
 J. Math. Phys. \textbf{57} (2016) 093504.
 
 \item Crampe, N., Frappat, L., Ragoucy, E., Vanicat, M., 
 \textsl{A new braid-like algebra for Baxterisation},
 Commun. Math. Phys. \textbf{349} (2017) 271.
 
 \item Finn, C., Vanicat, M.,
 \textsl{Matrix product construction for Koornwinder polynomials and fluctuations of the current in the open ASEP},
 J. Stat. Mech. (2017) P023102
 
 \item Vanicat, M.,
 \textsl{Exact solution to integrable open multi-species SSEP and macroscopic fluctuation theory}, 
 J. Stat. Phys. \textbf{166} (2017) 1129.
\end{itemize}

\tableofcontents

\chapter{Out-of-equilibrium statistical physics} \label{chap:one}

\section{Equilibrium versus non-equilibrium} \label{sec:equilibrium_vs_nonequilibrium}

\subsection{Physical properties of equilibrium and non equilibrium systems} 

We start by presenting the properties of systems at thermodynamic equilibrium. We recall briefly the fundamental principles of equilibrium 
statistical physics. This will help us to gain some intuition about out-of-equilibrium systems and to motivate the different ideas 
proposed to describe such systems.

\subsubsection{Equilibrium and entropy maximization}

At the macroscopic level, a classical\footnote{The discussion focuses here on classical systems but it can be transposed to quantum systems}
physical system at thermodynamic equilibrium is roughly defined as a system 
at equilibrium for any physical process we can imagine. For instance it should be at thermal, mechanical, electrodynamical and chemical equilibrium. 
In other words this is a state which does not display macroscopic currents of any physical quantities (for example energy, momentum, charges, particles). 
We know that these states can be completely characterized, or described, by a few macroscopic extensive variables such as the energy, the volume, the 
number of particles (or their intensive conjugate variables such as the temperature, the pressure, the chemical potential).
An example of thermodynamic equilibrium is given by a gas of molecules in a closed room at thermal equilibrium with its environment.

In order to have a more precise description of such states, we need to formalize a bit the discussion.
Assume that a system can be in several different configurations $\cC$ and denote by $\fC$ the set of these configurations.
$\fC$ is sometimes called the {\it phase space} or the {\it configuration space} of the system.
For instance, if the system is a gas of molecules contained in a fixed closed room, a configuration $\cC$ could be the knowledge of 
the positions and velocities (plus possibly some internal degrees of freedom) of each of the molecules constituting the gas. Another example
is the human brain, whose configurations can be simply the knowledge of all the connections between the neurons, or can be chosen to 
be the knowledge of the precise chemical content of each neuron and synapse. 

These examples illustrate that a configuration $\cC$ can sometimes be an effective description of the system, which
does not necessarily carry all the information about each of its microscopic degrees of freedom. This is explained by some uncertainties 
we may have on the system (what is the precise composition of a neuron?) or by some assumptions we made to simplify the description 
(some internal freedom of the molecules may have negligible impact on the behavior of the gas). 

This often imposes us to only have a statistical description of the system, trying to model an effective dynamics in this 
simplified configuration space. This motivates the introduction of a probability $\cP(\cC)$ to observe the system in a configuration $\cC$. 

If we assume, for the sake of simplicity, that the set of configurations $\fC$ is finite (we can adapt the discussion below to 
infinite sets), the entropy associated to the probability distribution $\cP$ is defined as\footnote{We give here the definition of the entropy 
used in probability theory. In a more physical context, this definition often involves the 
Boltzmann constant $k$, which is set to $1$ here.}
\begin{equation} \label{eq:entropy}
 \cH(\cP) = -\sum_{\cC \in \fC} \cP(\cC) \ln \left(\cP(\cC)\right).
\end{equation}
The main features of the entropy are the following.
A probability being between $0$ and $1$, it is easy to see that the entropy is always non-negative. The entropy is additive: 
the entropy associated to the joint probability distribution of two independent random variables is the sum of the entropies associated to
the probability distributions of each random variable. In other words, from a physical point of view, if we have two systems that are non-interacting
(far from each other for instance), the entropy of the two systems is equal to the sum of the entropies of each system.

The entropy can be intuitively understood as the amount of disorder (or of lack of information) carried by the system. 
To gain some feeling about this statement, let us consider two extremal cases. If the system is in configuration $\cC$ with probability $1$ (all the 
others configurations having vanishing probabilities), {\it i.e} if we have a full information about the state of the system, the entropy 
is equal to $0$ (it is minimal). Conversely, if we don't know anything about the state of the system, 
{\it i.e} if the probabilities of each configuration are equal, the entropy is given by $\ln \Omega$, where $\Omega$ is the number of configurations.
This value is the maximum of the entropy. Indeed if we want to maximize the entropy under the constraint $\sum_{\cC \in \fC} \cP(\cC) = 1$,
we obtain for all $\cC \in \fC$ the equation
\begin{equation}
 \frac{\partial \cH(\cP)}{\partial \cP(\cC)}-\lambda \frac{\partial}{\partial \cP(\cC)} \left(\sum_{\cC' \in \fC} \cP(\cC') \right) = 0,
\end{equation}
where $\lambda$ is the Lagrange multiplier. It implies that we have for all $\cC$, $\ln \cP(\cC) + 1+\lambda=0$, which proves that 
the probabilities are all equal and their common value is necessarily $1/\Omega$ (due to the sum to $1$ constraint).

This entropy is at the heart of the theory of equilibrium statistical physics. The fundamental law can be stated as follows.
A thermodynamic equilibrium is a state ({\it i.e} a probability distribution) which maximizes the entropy \eqref{eq:entropy} 
under a set of constraints \cite{DiuRGL89}. These constraints are determined by the interactions of the system with its environment 
({\it i.e} the reservoirs) and consist in imposing a fixed value for the average of certain physical observables.

They are roughly established by the following procedure\footnote{Establishing precisely the constraints (for the entropy maximization) 
can be delicate, particularly in systems, 
which possess many conserved quantities when they are isolated 
(typically the integrable systems, see discussion on quenched dynamics in subsection \ref{subsec:framework_out_of_equilibrium}
and also chapter \ref{chap:two}). We consider in such systems Generalized Gibbs Ensembles, see for instance \cite{RigolDYO07,IlievskiDWCEP15}}. 
We have first to identify the physical quantities, which are conserved by the 
dynamics (for instance the energy, the particle number, the charge) when the system is isolated, {\it i.e} disconnected from the reservoir. 
Then, considering the 
system in contact with the reservoirs again, we have to determine which of these physical quantities are exchanged with the reservoir.
The constraints are finally written down by imposing a fixed average value for each of these physical quantities.

For instance, let us consider a gas of particles in a closed room. If the system is completely isolated (no particle or energy exchanges),
there is no constraint and the maximization of the entropy yields $\cP(\cC)=1/\Omega$. This distribution is 
commonly called the {\it microcanonical} distribution.

If the system can exchange energy with the reservoir (heat bath), it imposes a fixed value for the average energy 
$\langle E \rangle := \sum_{\cC \in \fC} \cP(\cC)E(\cC)$, where $E(\cC)$ denotes the energy of the system in configuration $\cC$.
The maximization of the entropy yields the equation 
\begin{equation}
 \frac{\partial \cH(\cP)}{\partial \cP(\cC)}-\lambda \frac{\partial}{\partial \cP(\cC)} \left(\sum_{\cC' \in \fC} \cP(\cC') \right)
 -\beta \frac{\partial}{\partial \cP(\cC)} \left(\sum_{\cC' \in \fC} \cP(\cC')E(\cC') \right)= 0,
\end{equation}
where $\lambda$ and $\beta$ are the Lagrange multipliers. This implies that
\begin{equation}
 \cP(\cC)=\frac{e^{-\beta E(\cC)}}{Z}, \quad \mbox{with} \quad Z(\beta) = \sum_{\cC \in \fC} e^{-\beta E(\cC)}.
\end{equation}
This distribution is commonly called the {\it canonical} distribution (or also the Boltzmann or Gibbs distribution).
$\beta$ is called the inverse temperature. 

If the system can in addition exchange particles with the reservoir, the average number of particles 
$\langle N \rangle := \sum_{\cC \in \fC} \cP(\cC)N(\cC)$ is also fixed ($N(\cC)$ denotes the number of particles in the system in configuration $\cC$). 
A similar computation gives
\begin{equation}
 \cP(\cC)=\frac{e^{-\beta E(\cC)-\mu N(\cC)}}{Z}, \quad \mbox{with} \quad Z(\beta,\mu) = \sum_{\cC \in \fC} e^{-\beta E(\cC)-\mu N(\cC)}.
\end{equation}
This distribution is commonly called the {\it grand-canonical} distribution. $\mu$ is called the chemical potential.
 
$Z$ is called the partition function. It allows us to define the free energy $F= -\ln Z$. 
The free energy is closely related to the following cumulant generating function\footnote{We present here the case of the canonical 
distribution but it can be easily adapted to the grand-canonical distribution}
\begin{eqnarray}
 \ln \langle e^{\nu E(\cC)} \rangle & = & \ln \sum_{\cC \in \fC} e^{\nu E(\cC)} \frac{e^{-\beta E(\cC)}}{Z} \\
 & = & F(\beta)-F(\beta-\nu).
\end{eqnarray}
The free energy thus provides the fluctuations of the energy of the system (average value, variance, and higher order cumulants) 
by taking successive derivative of $F(\beta)$ with respect to the inverse temperature $\beta$.

Note also that the entropy is obtained by the Legendre transform of the free energy (see subsection \ref{subsec:Legendre} for the definition 
and properties of Legendre transformation). We have indeed, from previous computations, that 
\begin{equation}
 \cH(\cP) = \beta \langle E \rangle-F(\beta), \quad \mbox{with} \quad \frac{\partial F}{\partial \beta}(\beta) = \langle E \rangle.
\end{equation}
The intensive variable $\beta$ is conjugated to the extensive variable $\langle E \rangle$. Conversely the free energy can be obtained from 
the Legendre transform of the entropy.

We are now interested in the study of systems which are not at thermodynamic equilibrium. This is the purpose of the next subsection.

\subsubsection{Non equilibrium and macroscopic currents}

An out-of-equilibrium system is basically a system which is not at thermodynamic equilibrium. It can be either relaxing toward 
a stationary state or in a non-equilibrium stationary state (NESS).

At the macroscopic scale, it translates into the presence of non-vanishing currents of physical quantities (such as energy, particles, charges).
A schematic example of such states is given by two particle reservoirs at different densities, connected by a pipe. The 
high density reservoir will pour into the low density one, establishing a non vanishing macroscopic particle current.
In fact, we can find out-of-equilibrium systems everywhere, ranging from the traffic flow on a highway, to the development of the human brain, 
and including propagation of a fire in a forest or dynamics of a group of fish.

Unfortunately, they is no general theory to describe these systems. We do not know what are the relevant variables 
(like the temperature, pressure, chemical potential or other intensive variables in the thermodynamic equilibrium case) to describe these states. 
We would like to proceed from a fundamental law, such as an entropy maximization principle, which could provide us some tools, such as the 
thermodynamic potentials, to describe accurately these systems. We may need to enlarge the configuration space to formulate such 
maximization entropy principle, considering for instance path histories instead of ``static'' configurations, and using some kind 
of action to set the constraints imposed by the environment.

A few promising attempts have been made in the last decades. The discoveries of connections between statistical physics and large 
deviation theory, could provide an efficient framework to develop a theory for out-of-equilibrium systems. It led for instance to the 
formulation of the macroscopic fluctuation theory, which has proven to provide a powerful description of non-equilibrium diffusive systems
in the hydrodynamic limit (see chapter \ref{chap:five}).

But a lot remains to be understood. This motivates the study of particular non-equilibrium systems, focusing on the simplest ones,
to compute relevant physical quantities and try to infer more universal properties. From this perspective exactly solvable models appear 
as very powerful tools, and have been used in a wide range of different contexts and frameworks.

We present below a (non-exhaustive) list of mathematical frameworks commonly used to describe out-of-equilibrium systems.

\subsection{Different frameworks for out-of-equilibrium physics} \label{subsec:framework_out_of_equilibrium}

Many different formalisms had been developed to describe out-of equilibrium systems. We present here only a few of them which are 
widely used in the literature. We try to put, when possible, an emphasis on the role played by integrability and exact solvability 
in these different frameworks.

We stressed at the beginning 
of this chapter, that a system involving a huge number of components and with uncertainties on the exact relevant microscopic content and 
dynamics, is likely to be described by some effective state and some effective dynamics. This motivates the fact of introducing a 
probability distribution (on the configuration set of the system) to describe the statistical properties of the system. The next thing 
to do is to determine how this probability distribution evolves with time, {\it i.e} what is the effective dynamics chosen to 
model the time evolution of the system. In the thermodynamic equilibrium case, the precise choice of this dynamics has basically no impact 
on the long time behavior of the system, which reaches the Boltzmann distribution. But in the non-equilibrium case the 
choice of the framework used to encode the time evolution of the probability distribution is important to model the relaxation toward equilibrium
or the non-equilibrium stationary state. The formalisms presented below correspond to different ways to describe the system dynamics. 

\paragraph*{Langevin equation}

A formalism used to describe out-of-equilibrium systems is the Langevin equation. The idea is to add some random component in the force 
when writing the fundamental law of classical mechanics. The aim is to model some uncertainties in the dynamics at the microscopic scale.
If we want for instance to describe the dynamics of a grain of pollen in a bowl of water,
we do not know precisely the positions and velocities of each of the molecules of water, which are subject to thermal fluctuations. 
This will imply random collisions with the grain of pollen and translate into some random component in the force that acts on the pollen.
This motivates the following description of the dynamics,
where $x$, $v$ and $m$ are respectively the position, the velocity and the mass of the grain of pollen,
$\gamma$ is a friction coefficient, $F$ is a force and 
$\xi(t)$ is a Gaussian white noise which satisfies $\langle \xi(t) \rangle=0$ and $\langle \xi(t)\xi(t') \rangle=\Gamma \delta(t-t')$ 

\begin{equation}
 \begin{cases}
  m \frac{dv}{dt} = -m\gamma v + F(x)+ m\gamma \xi(t) \\
  \frac{dx}{dt} = v.
 \end{cases}
\end{equation}

For a high friction coefficient $\gamma \gg 1$, the acceleration can be neglected and the problem reduces to
\begin{equation} \label{eq:langevin}
 \frac{dx}{dt}= \frac{F(x)}{m\gamma}+\xi(t)
\end{equation}

It can be shown that the probability distribution of the process $\cP(x,t)$ ({\it i.e} the probability to find the particle in position $x$ at time $t$)
satisfies the Fokker-Planck equation \cite{Risken84}
\begin{equation} \label{eq:langevin_FokkerPlanck}
 \frac{\partial}{\partial t}\cP(x,t)=-\frac{\partial}{\partial x}\left(\frac{F(x)}{m\gamma}\cP(x,t)\right)
 +\frac{\Gamma}{2}\frac{\partial^2}{\partial x^2}\cP(x,t)
\end{equation}
The derivation of \eqref{eq:langevin_FokkerPlanck} from equation \eqref{eq:langevin} is not straightforward and relies essentially on the use of 
It\^{o} calculus. We do not provide the details of the computations here. 
Some consistency checks, on the time evolution of the average value of $x$ for instance, can be nevertheless easily performed. 

For the sake of simplicity we can study this Fokker-Planck equation on a ring of perimeter $1$, {\it i.e} imposing periodic boundary condition.
The probability distribution has to be consistent with the periodic geometry of the ring, {\it i.e} $\cP(0,t)=\cP(1,t)$.

We are looking for the stationary state of the problem, {\it i.e} a time-independent distribution $\cS(x)$ satisfying the Fokker-Planck equation.
It is straightforward to deduce that it has to solve the following equation 
\begin{equation} \label{eq:langevin_stationary_FokkerPlanck}
 \frac{\Gamma}{2}\frac{d}{d x}\cS(x)= \frac{F(x)}{m\gamma}\cS(x)+C,
\end{equation}
where $C$ is a constant term.

In the case where the force is conservative, {\it i.e} $\int_0^1 F(u)du=0$, we can define a potential 
$V(x)=-\int_0^x F(u)du$ such that $F(x)=-dV/dx$. Note that this potential fulfills the periodicity condition $V(0)=V(1)$ (this would not 
have been the case for a non-conservative force).
In this case the stationary distribution $\cS(x)$ is a thermal equilibrium and is given, as expected, by the Boltzmann distribution
\begin{equation}
 \cS(x)= \frac{1}{Z}\exp\left(-\frac{V(x)}{kT}\right),
\end{equation}
with temperature given by $2kT=m\gamma\Gamma$.

In the case where the force is not conservative, $\cS(x)$ is not the Boltzmann distribution anymore but a non-equilibrium stationary distribution 
given (up to a normalization) by 
\begin{equation} \label{eq:Langevin_stationary_state}
 \cS(x)= \int_0^x du \exp\left(\int_u^x \frac{2F(y)}{\Gamma m \gamma}dy \right) 
 + \int_x^1 du \exp\left(\int_u^1 \frac{2F(y)}{\Gamma m \gamma}dy + \int_0^x \frac{2F(y)}{\Gamma m \gamma}dy \right).
\end{equation}
It is indeed straightforward to check that it satisfies \eqref{eq:langevin_stationary_FokkerPlanck} with the constant 
$C=\frac{\Gamma}{2}(1-\exp(\int_0^1 \frac{2F(y)}{\Gamma m \gamma}dy))$ and that it fulfills the periodicity condition $\cS(0)=\cS(1)$.
A more intuitive construction of this solution, from a discrete periodic lattice, will be given later in this chapter.  

\paragraph*{Markov chains}

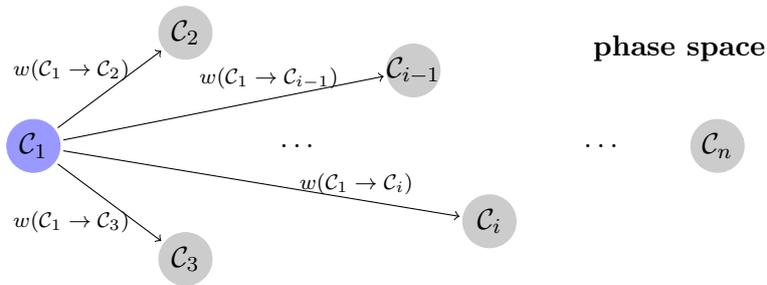
\begin{figure}[htb]
\begin{center}
 \begin{tikzpicture}[scale=1]
 \draw  (0,0) circle (0.35) [fill,circle,blue!40] {}; \node at (0,0) [] {$\cC_1$};
 \draw  (2,1.5) circle (0.35) [fill,circle,gray!40] {}; \node at (2,1.5) [] {$\cC_2$};
 \draw  (2,-1.5) circle (0.35) [fill,circle,gray!40] {}; \node at (2,-1.5) [] {$\cC_3$};
 \draw  (5,1) circle (0.35) [fill,circle,gray!40] {}; \node at (5,1) [] {$\cC_{i-1}$};
 \draw  (6,-1) circle (0.35) [fill,circle,gray!40] {}; \node at (6,-1) [] {$\cC_{i}$};
 \draw  (9,0) circle (0.35) [fill,circle,gray!40] {}; \node at (9,0) [] {$\cC_{n}$};
 \node at (3.5,0) [] {$\dots$};\node at (7.5,0) [] {$\dots$};
 \draw[->] (0.32,0.24) -- (2-0.32,1.5-0.24) ;
 \draw[->] (0.32,-0.24) -- (2-0.32,-1.5+0.24) ;
 \draw[->] (0.392,0.078) -- (5-0.392,1-0.078) ;
 \draw[->] (0.394,-0.065) -- (6-0.394,-1+0.065) ;
 \node at (0.5,1) [] {\scriptsize{$w(\cC_1 \rightarrow \cC_2)$}};
 \node at (0.5,-1) [] {\scriptsize{$w(\cC_1 \rightarrow \cC_3)$}};
 \node at (3.1,0.9) [] {\scriptsize{$w(\cC_1 \rightarrow \cC_{i-1})$}};
 \node at (4.25,-0.5) [] {\scriptsize{$w(\cC_1 \rightarrow \cC_i)$}};
 \node at (8.5,1.3) [] {\textbf{phase space}};
 \end{tikzpicture}
 \caption{Graphical representation of a Markov chain \label{fig:Markov_chain}}
 \end{center}
\end{figure}

Another framework that has become widely used to deal with out-of-equilibrium dynamics is Markov chain theory \cite{VanKampenR83}. 
It models the system's effective dynamics
using the following idea. The system has a given probability $w(\cC \rightarrow \cC')$ to jump from a given configuration $\cC$ 
to another configuration $\cC'$. The main assumption is that the dynamics is supposed to have no memory: the jump probability depends 
only on the starting configuration $\cC$
and on the target configuration $\cC'$ but not on the whole history of the system before arriving in configuration $\cC$. The probability rate 
can also depend on the time when the jump occurs.
Note that the stochastic dynamics of the system has a convenient graphical interpretation: the different configurations of the system are 
interpreted as the vertices of a graph and the jump probabilities as the oriented edges of the graph, see for instance figures \ref{fig:Markov_chain}
and \ref{fig:3states}.

This description is quite general and can be used to model a wide range of systems and dynamics in very different fields, going from 
social science, economy, population dynamics to statistical physics, applied mathematics and including chemistry, biophysics, epidemiology and
genetics.

The advantage is that, the whole information about the stochastic dynamics of the system can be stored in a single matrix, called the Markov matrix,
which contains all the jump probabilities. This allows to recast the time evolution of the probability distribution associated to 
the system as a linear algebra problem, called the master equation, and use the linear algebra machinery to study it.

Note that the 'no memory' assumption can be partially relaxed (while still using the Markovian formalism) by enlarging the configurations set of the system. 
If we want for instance the dynamics to depend on the two last configurations of the system (instead of only the last one), we just need to 
define a Markov chain on 'super configurations', which are defined as a couple of usual configurations of the system. This can be generalized to 
more complicated dynamics with memory.

This Markov chain framework will be the one used in this manuscript. We will see that integrability can play a key-role to study Markovian processes,
because it allows us in some particular cases to compute analytically the stationary state of the model or even to diagonalize exactly the 
Markov matrix. We will focus on one dimensional interacting particle systems, whose Markov matrices are often closely related to quantum spin chain 
Hamiltonians, which allows us to use the quantum integrability machinery. In particular we will see that the non-equilibrium stationary 
states of such models can be often expressed in a matrix product form, with the help of the $R$-matrix, which is the key object in 
quantum integrability. This will be discussed in detail in chapter \ref{chap:two} and chapter \ref{chap:three}.

\paragraph*{Quantum systems: Lindblad equation and quenched dynamics}

In the last few years the study of out-of-equilibrium quantum systems has become a very active field of research. We present very briefly 
two approaches frequently used to model such systems. 

The first one is the quenched dynamics. The idea can be roughly summarized as follows. We would like to study the thermalization of the system, 
{\it i.e} the relaxation of the system toward thermodynamic equilibrium, by pushing it far from equilibrium. This can be 
achieved by choosing adequately the initial condition or by changing quickly a parameter in the Hamilitonian governing the dynamics. 
The whole system is decomposed into two subsystems: one could be thought of as playing the role of the environment ({\it i.e} the reservoir) 
and the other one as playing the role of a non-isolated system in interaction with the reservoir.
The integrable spin chains have a central role in this context. The reason is twofold. Firstly, the quantum Hamiltonian of such models can be
diagonalized exactly using Bethe ansatz, which gives theoretically access to the 
full dynamics of the system and allows us in principle to study analytically the relaxation and equilibrium properties of the system.
Secondly, an important feature of integrable models is the fact that they possess a lot of conserved quantities (also called conserved charges). 
They thus provide toy models to determine what are the relevant charges required in the Boltzmann-like distribution. This led to the 
development of the notion of {\it Generalized Gibbs Ensembles} \cite{RigolDYO07,IlievskiDWCEP15} and to the discovery of the relevance of 
quasi-local charges \cite{ProsenI13}. 

The second method is the Lindblad equation \cite{BreuerP02}. This is an equation for the density matrix of a statistical quantum system.
This roughly corresponds to the master equation \eqref{eq:Markov_master_equation}, which we will encounter in the context of Markov chains.
It can for instance describe a quantum system with stochastic interactions with reservoirs. When the system is in contact with two reservoirs 
at different temperatures (or at different chemical potentials), the density matrix will converge toward a non-equilibrium stationary state
(which is not given by a Boltzmann statistics in general). 
Once again, the integrable quantum spin chains play a privileged role in this context 
\cite{Znidaric10,Znidaric11,Prosen11bis,Prosen11,KarevskiPS13,PopkovKS13,Ilievski14,Ilievski16}. 
The key object of integrability, the $R$-matrix, 
provides an efficient framework, through the RTT algebra (see chapter \ref{chap:two} for details) to construct the stationary density 
matrix in a matrix product form.

We now come back to the case of Markov chains, which will be intensively used in this manuscript. We define more precisely this object and present 
the main tools and properties associated with it.

\section{Markov chains and stationary states}

\subsection{Markov chains} \label{subsec:Markov_chains}

\subsubsection{Discrete time}

We introduced heuristically the concept of Markov chain in the previous section. We now present a more formal mathematical definition.
\begin{definition}
A time-homogeneous discrete time Markov process on a finite state space is a sequence of random variables $(S_n)_{n\geq 0}$
that take values on a finite set $\fC$ and which satisfy the following properties
\begin{equation} \label{eq:Markov_chains_def1}
 \forall n \in \NN, \ \forall \cC_1,\dots,\cC_n \in \fC, \quad \PP(S_n=\cC_n | S_0=\cC_0,\dots,S_{n-1}=\cC_{n-1})=\PP(S_n=\cC_n | S_{n-1}=\cC_{n-1}),
\end{equation}
and 
\begin{equation}
 \forall n \in \NN, \ \forall \cC,\cC' \in \fC, \quad \PP(S_{n+1}=\cC | S_n=\cC')=\PP(S_n=\cC | S_{n-1}=\cC').
\end{equation}
\end{definition}
To be more explicit, $S_n$ stands for the state of the system at time $n$. The set $\fC$ represents the configuration space of the system.
The quantity $\PP(S_n=\cC)$ denotes the probability for the system to be in the configuration $\cC$ at time $n$. To shorten the notation we introduce
\begin{equation}
 P_n(\cC):=\PP(S_n=\cC).
\end{equation}
The conditional probability $\PP(S_n=\cC_n | S_0=\cC_0,\dots,S_{n-1}=\cC_{n-1})$ stands for the probability for the system to be in 
configuration $\cC_n$ at time $n$, knowing that it was in configurations $\cC_0,\cC_1,\dots,\cC_{n-1}$ at time $0,1,\dots,n-1$ respectively
({\it i.e} knowing the whole history of the system).
The property of no memory of the Markov chain is translated by the fact that the latter conditional probability is equal to 
$\PP(S_n=\cC_n | S_{n-1}=\cC_{n-1})$ \eqref{eq:Markov_chains_def1}, which is
the probability for the system to be in configuration $\cC_n$ at time $n$, knowing only that it was in configuration $\cC_{n-1}$ at time $n-1$.

The second property in the definition reflects the fact that the stochastic dynamics is time independent. 
The probability for the system to be in configuration $\cC$ at time $n$, knowing that it was in configuration $\cC'$ at time $n-1$ does not depend
on the time $n$. This allows us to introduce the notation
\begin{equation}
w(\cC' \rightarrow \cC):=\PP(S_n=\cC | S_{n-1}=\cC')
\end{equation}
which stands for the transition probability between configurations $\cC'$ and $\cC$.

\begin{proposition}
 The probability distribution $P_n$ satisfies the master equation
 \begin{equation} \label{eq:Markov_master_equation}
  \forall n \in \NN, \ \forall \cC \in \fC, \quad P_{n+1}(\cC)= \sum_{\cC' \in \fC} w(\cC' \rightarrow \cC)P_n(\cC').
 \end{equation}
\end{proposition}

\proof We have the equality
\begin{eqnarray*}
 P_{n+1}(\cC) & = & \sum_{\cC' \in \fC} \PP(S_{n+1}=\cC,S_n=\cC') \\
 & = &  \sum_{\cC' \in \fC} w(\cC' \rightarrow \cC)P_n(\cC'),
\end{eqnarray*}
where $\PP(S_{n+1}=\cC,S_n=\cC')$ denotes the probability that the system is in configuration $\cC'$ at time $n$ and in configuration $\cC$ at time 
$n+1$.
\finproof

The master equation is the key set of equations (there is one equation per configuration $\cC$) which governs the time evolution 
of the probability distribution. Solving these equations provides the probability distribution of the system at any time. 
This set of equations is linear in the probabilities $P_n(\cC)$ for $\cC \in \fC$, which suggests to use the linear algebra machinery to recast and 
study this problem. This leads to the following definition. 
\begin{definition} \label{def:Markov_chains_vector_space}
 Let $\cV$ be the finite dimensional vector space spanned by the basis $\{\ket{\cC},\cC \in \fC\}$.
 For all $n\in \NN$, we define $\ket{P_n} \in \cV$ by
 \begin{equation}
  \ket{P_n} = \sum_{\cC \in \fC} P_n(\cC)\ket{\cC}.
 \end{equation}
 We also define the Markov matrix
 \begin{equation}
  W= \sum_{\cC,\cC' \in \fC} w(\cC' \rightarrow \cC)\ket{\cC} \bra{\cC'},
 \end{equation}
 where $\bra{\cC'}$ is a vector of the dual space of $\cV$ satisfying the scalar product relation $\langle \cC'|\cC \rangle = \delta_{\cC,\cC'}$.
\end{definition}

We are now equipped to rewrite the time evolution of the probability distribution.
\begin{proposition}
 For all $n\in \NN$, we have the relation
 \begin{equation} \label{eq:Markov_master_equation_matrix}
  \ket{P_{n+1}}=W\ket{P_n}.
 \end{equation}
 This implies that 
 \begin{equation}
  \ket{P_n}=W^n \ket{P_0}.
 \end{equation}
\end{proposition}
 
 \proof
  This is a direct reformulation of the master equation \eqref{eq:Markov_master_equation}.
 \finproof
 
 \begin{remark}
  The transition probabilities $w(\cC \rightarrow \cC')$ satisfy 
  \begin{equation}
   \sum_{\cC' \in \fC} w(\cC \rightarrow \cC') = 1, \forall \cC \in \fC.
  \end{equation}
  This can be reformulated in vector form by introducing the row vector
  \begin{equation}
   \bra{\Sigma} = \sum_{\cC \in \fC}\bra{\cC}.
  \end{equation}
  We have the identity $\bra{\Sigma}W=\bra{\Sigma}$, which means that the entries of each column of the Markov matrix $W$ sum to $1$.
 \end{remark}

\subsubsection{Continuous time}

We would like to derive a continuous time version of the Markov process presented above. One way to address the problem is to say that each time 
step of the previous process corresponds to increase an infinitesimal amount of time, which yields the 
following relabelling of the probability vector $\ket{\cP_{ndt}}:=\ket{P_n}$. We also need to rescale the transition probabilities by  
introducing for $\cC \neq \cC'$ the transition rate $m(\cC \rightarrow \cC')$ such that $w(\cC \rightarrow \cC')=m(\cC \rightarrow \cC')dt$. 
The master equation \eqref{eq:Markov_master_equation_matrix} can be then recast into
\begin{equation} \label{eq:Markov_master_intermediate}
 \ket{\cP_{t+dt}}=\ket{\cP_{t}}+dt\, M\ket{\cP_{t}}, 
\end{equation}
where the (continuous time) Markov matrix $M$ is defined by
\begin{equation} \label{eq:Markov_matrix_continuous_time}
  M= \sum_{\cC,\cC' \in \fC} m(\cC' \rightarrow \cC)\ket{\cC} \bra{\cC'},
 \end{equation}
with 
\begin{equation}
 m(\cC \rightarrow \cC)= -\sum_{\genfrac{}{}{0pt}{}{\cC' \in \fC}{\cC' \neq \cC}}m(\cC \rightarrow \cC').
\end{equation}
This yields the following proposition
\begin{proposition}
 The time evolution of probability vector $\ket{\cP_{t}}$ obeys the master equation
 \begin{equation} \label{eq:Markov_master_equation_continuous}
  \frac{d\ket{\cP_{t}}}{dt} = M\ket{\cP_{t}}.
 \end{equation}
\end{proposition}

\proof
 This equation is obtained by rearranging \eqref{eq:Markov_master_intermediate} as 
 \begin{equation}
  \frac{\ket{\cP_{t+dt}}-\ket{\cP_{t}}}{dt} =  M\ket{\cP_{t}}
 \end{equation}
 and taking the limit $dt \rightarrow 0$.
\finproof

\begin{remark}
 It is easy to check that the sum of the entries of each column of the Markov matrix $M$ is vanishing $\bra{\Sigma}M=0$.
\end{remark}

From a more physical perspective, a continuous time Markov process can be interpreted as follows. During an infinitesimal time $dt$, 
the system in configuration $\cC$ has a probability $m(\cC \rightarrow \cC') \times dt$ to jump to another configuration $\cC'$. 

\subsubsection{Stationary state}

The stationary state is particularly relevant from a physical point of view. This is the probability vector 
$\steady = \sum_{\cC \in \fC} \cS(\cC) \ket{\cC}$ which 
contains the probabilities to observe the system in a given configuration after a very long time. This vector should be stable under 
the dynamics ({\it i.e} the time evolution) of the process. In the discrete time case it thus satisfies $W\steady=\steady$ whereas
in the continuous time case it satisfies $M\steady=0$. In what follows we will present the results in the continuous time framework 
(because this framework will be used intensively in this manuscript and the majority of the models studied will be defined in a 
continuous time setting) but they can be easily transposed to the discrete time framework.

\begin{definition}
 A Markovian process is said to be irreducible if for all configurations $\cC$ and $\cC'$ there exists a sequence of configurations
 $\cC_1,\cC_2,\dots,\cC_i$ which satisfies
 \begin{equation}
 m(\cC \rightarrow \cC_1) m(\cC_1 \rightarrow \cC_2)\dots m(\cC_i \rightarrow \cC') \neq 0.
 \end{equation}
\end{definition}

In other words, the irreducibility condition can be intuitively understood as the fact that there is a non vanishing probability to 
go from any configuration $\cC$ to any other $\cC'$ in a finite number of steps. We are now equipped to state the Perron-Frobenius theorem.

\begin{theorem}
 If $M$ is the Markov matrix associated to a continuous time\footnote{Note that the theorem can be adapted to discrete time Markovian processes.}
 irreducible Markovian process, then $M$ possesses a unique eigenvector with eigenvalue equal to $0$ (the stationary state) and all others 
 eigenvalues of $M$ have negative real parts.
\end{theorem}

The Perron-Frobenius theorem implies in particular that the convergence toward the stationary state is exponentially fast.

Nevertheless we will see that when the size of the system, {\it i.e} the number of configurations in the phase space, tends to infinity,
the behavior of the non-vanishing eigenvalue with largest real part is crucial to determine the relaxation properties of large systems. 

If this largest real part converges to a non-vanishing value, as the size of the system goes to infinity, the system is said to be 
{\it fully gapped}, and the relaxation towards the stationary state in the thermodynamic limit is exponentially fast (the system 
is sometimes said to be {\it massive}).

Conversely, if the largest real part converges to zero, the system is said to be {\it gap-less}. It is then interesting to determine the rate 
of convergence toward zero of the real part, in function of the system size (it provides the relaxation rates of the system in the thermodynamic limit).

Now that the existence of the stationary state is well established for irreducible Markov chains we address the question of 
finding an exact expression for this steady state. To do so, let us go back to the master equation. In the continuous time framework, we would like 
to compute the vector $\steady$ such that $M\steady=0$. This can be written in components as
\begin{eqnarray}
 & & \sum_{\cC' \in \fC} m(\cC' \rightarrow \cC)\cS(\cC')=0, \quad \forall \cC \in \fC \\
 & \Leftrightarrow & \sum_{\genfrac{}{}{0pt}{}{\cC' \in \fC}{\cC' \neq \cC}} \Big( m(\cC' \rightarrow \cC)\cS(\cC')-m(\cC \rightarrow \cC')\cS(\cC) \Big)=0, 
 \quad \forall \cC \in \fC. \label{eq:Markov_stationary_components}
\end{eqnarray}
The last equality is obtained using the fact that 
\begin{equation}
m(\cC \rightarrow \cC)=-\sum_{\genfrac{}{}{0pt}{}{\cC' \in \fC}{\cC' \neq \cC}} m(\cC \rightarrow \cC').
\end{equation}

The equation \eqref{eq:Markov_stationary_components} expresses the fact that the incoming probability flux exactly compensates
the outgoing probability flux (this could be viewed as the analogue of Kirchoff's law). Indeed, if we define the probability current
from configuration $\cC$ to configuration $\cC'$ as $j_{\cC \rightarrow \cC'}=m(\cC \rightarrow \cC')\cS(\cC)-m(\cC' \rightarrow \cC)\cS(\cC')$, then equation 
\eqref{eq:Markov_stationary_components} can be rewritten
\begin{equation}
\sum_{\genfrac{}{}{0pt}{}{\cC' \in \fC}{\cC' \neq \cC}} j_{\cC \rightarrow \cC'}=0, \quad \forall \cC \in \fC.
\end{equation}

\subsection{Thermodynamic equilibrium and detailed balance}

One very particular way for the sums in \eqref{eq:Markov_stationary_components} to vanish is when all the probability currents satisfy 
$j_{\cC \rightarrow \cC'}=0$ independently. This corresponds to the case 
\begin{equation} \label{eq:Markov_detailed_balance}
 m(\cC \rightarrow \cC')\cS(\cC)=m(\cC' \rightarrow \cC)\cS(\cC'), \quad \forall \cC,\cC' \in \fC.
\end{equation}
The latter equation is called detailed balance.
It matches exactly the physical description of a thermodynamic equilibrium given previously in this chapter, where it was stressed 
that this is a state with no macroscopic currents of physical quantities (such as energy, charge, particles) flowing in the system. 
In the stochastic processes context, these physical macroscopic currents are necessarily produced 
at the microscopic level by probability currents between configurations. 

The previous discussion yields the more formal definition of a thermodynamic equilibrium in the context of Markov chains.

\begin{definition}
 The stationary state $\steady$ of a Markov process is a thermodynamic equilibrium if the detailed balance \eqref{eq:Markov_detailed_balance}
 is satisfied.
\end{definition}

\begin{remark}
 The detailed balance condition involves the expression of the steady state distribution and cannot be checked a priori before knowing this 
 distribution. Nevertheless this detailed balance condition can be equivalently recast in a condition depending only on 
 the transition rates $m(\cC \rightarrow \cC')$. We have indeed for any cycle of configurations, if the detailed balance is satisfied, 
 $\cC_1 \rightarrow \cC_2 \rightarrow \dots \rightarrow \cC_n \rightarrow \cC_1$
 \begin{eqnarray}
  \cS(\cC_1) & = & \frac{m(\cC_2 \rightarrow \cC_1)}{m(\cC_1 \rightarrow \cC_2)} \cS(\cC_2) \\
  & = & \frac{m(\cC_2 \rightarrow \cC_1)}{m(\cC_1 \rightarrow \cC_2)}\frac{m(\cC_3 \rightarrow \cC_2)}{m(\cC_2 \rightarrow \cC_3)} \dots \frac{m(\cC_1 \rightarrow \cC_n)}{m(\cC_n \rightarrow \cC_1)}\cS(\cC_1).
 \end{eqnarray}
 Hence we have
 \begin{equation} \label{eq:Markov_chains_cycle}
 m(\cC_1 \rightarrow \cC_2)m(\cC_2 \rightarrow \cC_3) \dots m(\cC_n \rightarrow \cC_1) = m(\cC_1 \rightarrow \cC_n)\dots m(\cC_3 \rightarrow \cC_2) m(\cC_2 \rightarrow \cC_1)
 \end{equation}
 Conversely, if this equality \eqref{eq:Markov_chains_cycle} holds for every cycle, then the detailed balance is satisfied (see next subsection).
 This can be intuitively interpreted as the equality of the probabilities to go along the cycle in one way or in the reverse way.
 This sheds some light on the notion of time reversibility which is developed below. 
\end{remark}

\subsubsection{Time reversibility} \label{subsubsec:time_reversibility}

We are now interested in the implications of the detailed balance condition on the time reversibility of the process. More precisely 
we would like to compare the probability $\cP(\{\cC(t)\})$, to observe a path history $\{\cC(t)\}_{0\leq t \leq T}$ in the stationary state and 
the probability to observe the time reversed path $\{\hat \cC(t)\}_{0\leq t \leq T}$.

\begin{figure}[htb]
\begin{center}
 \begin{tikzpicture}[scale=0.7]
\draw (0,5) -- (1,5) ;
\node at (0.5,5.5) [] {$\cC_1$} ;
\draw (1,5) -- (1,6) ;

\draw (1,6) -- (3,6) ;
\node at (2,6.5) [] {$\cC_2$} ;
\draw (3,6) -- (3,3) ;

\draw (3,3) -- (6,3) ;
\node at (4.5,3.5) [] {$\cC_3$} ;

\draw[dashed] (7,4) -- (9,4) ;
\node at (7,7) [] {path $\{\cC(t)\}_{0\leq t \leq T}$} ;

\draw (10,5) -- (11,5) ;
\node at (10.5,5.5) [] {$\cC_{n-1}$} ;
\draw (11,5) -- (11,4) ;

\draw (11,4) -- (14,4) ;
\node at (12.5,4.5) [] {$\cC_{n}$} ;

\draw[->] (-1,2) -- (15,2) ;
\foreach \i in {0,1,3,6,10,11,14}
{\draw (\i,1.9) -- (\i,2.1) ;}
\node at (0,1.5) [] {$0$} ; \node at (1,1.5) [] {$t_1$} ; \node at (3,1.5) [] {$t_2$} ; \node at (6,1.5) [] {$t_3$} ;
\node at (9.9,1.5) [] {$t_{n-2}$} ; \node at (11.2,1.5) [] {$t_{n-1}$} ; \node at (14,1.5) [] {$T$} ; 

\draw (14,-2) -- (13,-2) ;
\node at (13.5,-1.5) [] {$\cC_1$} ;
\draw (13,-2) -- (13,-1) ;

\draw (13,-1) -- (11,-1) ;
\node at (12,-0.5) [] {$\cC_2$} ;
\draw (11,-1) -- (11,-4) ;

\draw (11,-4) -- (8,-4) ;
\node at (9.5,-3.5) [] {$\cC_3$} ;

\draw[dashed] (7,-3) -- (5,-3) ;
\node at (7,0) [] {path $\{\hat \cC(t)\}_{0\leq t \leq T}$} ;

\draw (4,-2) -- (3,-2) ;
\node at (3.5,-1.5) [] {$\cC_{n-1}$} ;
\draw (3,-2) -- (3,-3) ;

\draw (3,-3) -- (0,-3) ;
\node at (1.5,-2.5) [] {$\cC_{n}$} ;

\draw[->] (-1,-5) -- (15,-5) ;
\foreach \i in {0,1,3,6,10,11,14}
{\draw (14-\i,-5.1) -- (14-\i,-4.9) ;}
\node at (14.1,-5.5) [] {$T$} ; \node at (12.9,-5.5) [] {$T-t_1$} ; \node at (11,-5.5) [] {$T-t_2$} ; \node at (8,-5.5) [] {$T-t_3$} ;
\node at (4.75,-5.5) [] {$T-t_{n-2}$} ; \node at (2.25,-5.5) [] {$T-t_{n-1}$} ; \node at (0,-5.5) [] {$0$} ; 
 \end{tikzpicture}
 \caption{Graphical representation of the trajectories $\{\cC(t)\}_{0\leq t \leq T}$ and $\{\hat \cC(t)\}_{0\leq t \leq T}$ \label{fig:trajectories}}
 \end{center}
\end{figure}
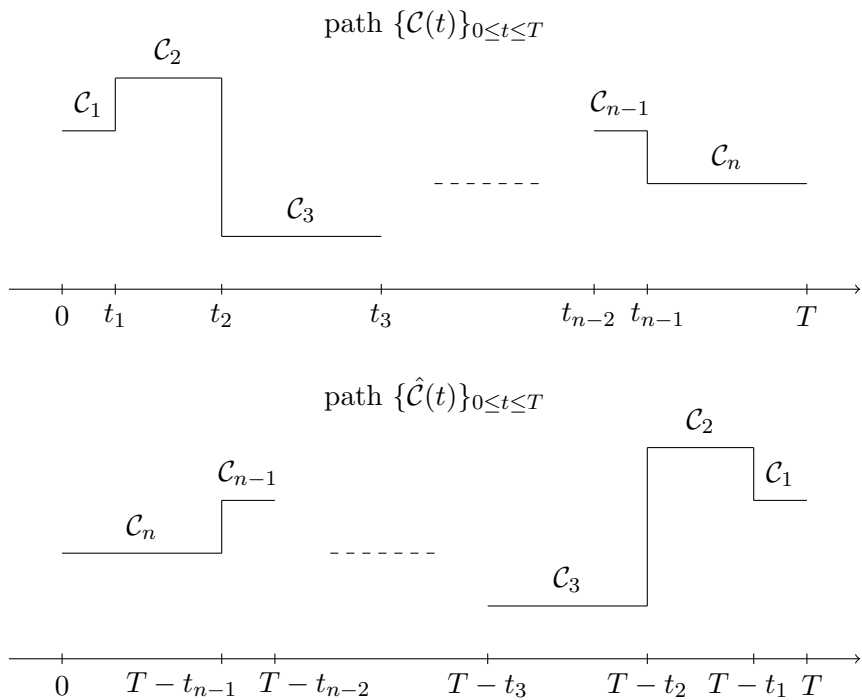

The first thing we need to do is to evaluate $\cP(\{\cC(t)\})$. We are interested in a path
starting at $t=0$ in configuration $\cC_1$, exploring successively the configurations $\cC_1,\cC_2,\dots,\cC_n$, with the 
transition from $\cC_i$ to $\cC_{i+1}$ occurring between times $t_i$ and $t_i+dt_i$ (where $dt_i$ is infinitesimal), and finishing at $t=T$.
For all $1\leq i \leq n$, the system is thus in configuration $\cC_i$ during the 
time interval $[t_{i-1},t_i]$ (where $t_0$ and $t_n$ are defined as $t_0=0$ and $t_n=T$). 
See figure \ref{fig:trajectories} for an illustration of the path history.

A building block of the probability of the whole path is to compute the probability for the system to stay in a 
configuration $\cC$ during a time interval $[t,t']$. The probability for the system to stay in $\cC$ during the 
time interval $[t,t+dt]$ is 
\begin{equation}
 1-dt\sum_{\genfrac{}{}{0pt}{}{\cC' \in \fC}{\cC' \neq \cC}}m(\cC \rightarrow \cC')=1+dt\, m(\cC \rightarrow \cC)
\end{equation}
Hence the probability for the system to remain in $\cC$ during $[t,t']$ can be obtained by multiplying the probabilities that it stays in 
$\cC$ on each infinitesimal intervals $[t+kdt,t+(k+1)dt]$ which yields for $dt\rightarrow 0$
\begin{equation}
 \lim\limits_{dt\rightarrow 0} (1+dt\, m(\cC \rightarrow \cC))^{(t'-t)/dt} = e^{(t'-t)m(\cC \rightarrow \cC)}.
\end{equation}
It is then straightforward to deduce that
\begin{equation} \label{eq:Markov_proba_path}
 \cP(\{\cC(t)\})= \cS(\cC_1) \left(\prod_{i=1}^{n-1} e^{(t_i-t_{i-1})m(\cC_i \rightarrow \cC_i)}m(\cC_i \rightarrow \cC_{i+1})dt_i \right) e^{(t_n-t_{n-1})m(\cC_n \rightarrow \cC_n)}.
\end{equation}
The time reversed path $\{\hat \cC(t)\}_{0\leq t \leq T}$ is defined as $\hat \cC(t)= \cC(T-t)$. Its probability is thus given by
\begin{equation} 
 \cP(\{\hat \cC(t)\})= \cS(\cC_n) \left(\prod_{i=1}^{n-1} e^{(t_i-t_{i-1})m(\cC_i \rightarrow \cC_i)}m(\cC_{i+1} \rightarrow \cC_i)dt_i \right) e^{(t_n-t_{n-1})m(\cC_n \rightarrow \cC_n)}.
\end{equation}
We can then compute
\begin{equation}
 \frac{\cP(\{\cC(t)\})}{\cP(\{\hat \cC(t)\})}= 
 \frac{m(\cC_{n-1} \rightarrow \cC_n)\dots m(\cC_2 \rightarrow \cC_3)m(\cC_1 \rightarrow \cC_2)}{m(\cC_n \rightarrow \cC_{n-1})\dots m(\cC_3 \rightarrow \cC_2)m(\cC_2 \rightarrow \cC_1)} \frac{\cS(\cC_1)}{\cS(\cC_n)}
 =1,
\end{equation}
where the last equality is obtained using the detailed balance condition \eqref{eq:Markov_detailed_balance} (which is thus a key 
property to establish the equality of the path and time reversed path probabilities). Conversely, it is straightforward to 
show that imposing time reversibility in the stationary state implies the detailed balance (we can take for instance a path with only one jump
to prove this fact). 

In conclusion the detailed balance condition is equivalent to the time reversibility of the process in the stationary state.

\subsubsection{Link with the Boltzmann distribution}

The last point that remains to be explored concerning the detailed balance condition is its link with the Boltzmann distribution. 
As stressed previously, when the detailed balance is satisfied, the stationary state describes a physical system at thermodynamic 
equilibrium by definition.
We want to make the connection with the Boltzmann approach presented previously in section \ref{sec:equilibrium_vs_nonequilibrium}.
The system exchanges physical quantities with a single reservoir. For the sake of simplicity, we assume that it only 
exchanges energy with a reservoir at inverse temperature $\beta$ which implies that the stationary distribution should be the Boltzmann distribution
\begin{equation}
 \cS(\cC)= \frac{1}{Z}e^{-\beta E(\cC)}.
\end{equation}
To make connection with the Markov process, we have to relate the relevant physical quantities, basically the energy $E(\cC)$, to  
the transition rates $m(\cC \rightarrow \cC')$ defining the stochastic process. The detailed balance implies that we must have
\begin{equation} \label{eq:detailed_balance_Boltzmann}
 \frac{m(\cC \rightarrow \cC')}{m(\cC' \rightarrow \cC)} = \frac{\cS(\cC')}{\cS(\cC)}= e^{-\beta \Delta E},
\end{equation}
where $\Delta E= E(\cC')-E(\cC)$.

\subsection{Non-equilibrium stationary states}

In this section we are interested in an irreducible Markov process whose stationary state $\steady$ does not satisfy the detailed balance condition, 
{\it i.e} there exist at least two configurations $\cC,\cC' \in \fC$ such that
\begin{equation}
 m(\cC \rightarrow \cC')\cS(\cC) \neq m(\cC' \rightarrow \cC)\cS(\cC').
\end{equation}
In this case the stationary state is called {\it non-equilibrium stationary state}. In opposition to the thermal equilibrium case, there are 
probability currents flowing in the system in the stationary state and the evolution is time irreversible in the stationary state.

The steady state distribution is a priori not of Boltzmann type and cannot be obtained easily. Nevertheless it can be 
expressed exactly using graph theory. We recall now the basic ingredients.

\begin{definition}
A directed graph $G=(\fC,E)$ is a finite set of vertices (or nodes) $\fC$ and a set of directed edges (or arrows) 
$E \subset \fC \times \fC$. By convention we say that the arrow $(\cC',\cC) \in E$ is starting from $\cC$ and ending at $\cC'$
(it will be sometimes denoted as $\cC \rightarrow \cC'$).
\end{definition}

\begin{definition}
 A rooted tree over the set $\fC$ is a directed graph $T=(\fC,E)$ such that
\begin{itemize}
 \item the underlying undirected graph is a tree (i.e. acyclic and connected\footnote{Acyclic means that there is no cycle, 
 {\it i.e.} no sequence of consecutive different edges starting and ending at the same vertex. 
 Connected means that any two distinct vertices are linked by a sequence of consecutive edges.}).
 \item there exists a particular node $r(T)\in \fC$ (called the root of $T$)
 such that all the arrows are oriented toward $r(T)$ (i.e. for any vertex $\cC \in \fC$ there exists a unique
 directed path going from $\cC$ to $r(T)$).
\end{itemize}
For $\cC \in \fC$, let $\mathcal{T}(\cC)$ be the set of rooted trees $T$ over $\fC$ such that $r(T)=\cC$.
\end{definition}

Examples of directed graphs and rooted trees are given in figure \ref{fig:3states} and in figure \ref{fig:arbres} respectively.

\begin{definition}
We consider a Markov matrix $M$ over a finite configurations space $\fC$. The $M$-weight of a given directed graph $G=(\fC,E)$ is
defined by
\begin{equation}
 w(G)= \prod_{(\cC',\cC) \in E} m(\cC \rightarrow \cC'),
\end{equation}
were $m(\cC \rightarrow \cC')$ for $\cC,\cC' \in \fC$ are the entries of Markov matrix (see \eqref{eq:Markov_matrix_continuous_time}).
\end{definition}

\begin{proposition}
The unique stationary measure\footnote{We recall that the stationary measure $\steady=\sum_{\cC \in \fC} \cS(\cC) \ket{\cC}$ 
is defined by the equation $M\steady =0$, where the 
Markov matrix is given in \eqref{eq:Markov_matrix_continuous_time}. Written in components it gives \eqref{eq:Markov_stationary_components}.}
$\steady$ of the irreducible Markov matrix $M$ defined over the finite configurations space $\fC$ is given by
\begin{equation*}
\forall \cC \in \fC, \quad \cS(\cC)= \frac{1}{Z}\sum_{T \in \mathcal{T}(\cC)} w(T),
\end{equation*}
where $Z$ is a normalization constant such that $\sum_{\cC \in \fC}\cS(\cC)=1$.
\end{proposition}

\proof
We fix $\cC' \in \fC$. For $T \in \mathcal{T}(\cC')$ and $\cC \in \fC \backslash \{\cC'\}$, let
$\widetilde{T}$ be the directed graph obtained by adding the arrow $(\cC,\cC')$ to the graph $T$. $\widetilde{T}$ contains
exactly one directed cycle $\cC' \rightarrow \cC \rightarrow \dots \rightarrow \cC'' \rightarrow \cC'$.
Let $\psi(T,\cC)$ be the graph obtained by removing the edge $(\cC',\cC'')$ from $\widetilde{T}$. 

The application
\begin{equation}
\psi \ : \ \mathcal{T}(\cC')\times \fC \backslash \{\cC'\} \rightarrow \bigcup_{\cC \in \fC \backslash \{\cC'\} } \mathcal{T}(\cC)
\end{equation}
is well defined and is bijective.

We indeed observe that $\psi(T,\cC) \in \mathcal{T}(\cC'')$ (where $\cC''$ was introduced previously while defining $\psi(T,\cC)$):
the underlying undirected graph of $\psi(T,\cC)$ is connected and acyclic and the arrows are pointing toward $\cC''$.
Then let us prove that $\psi$ is injective. We show that we can reconstruct $T$ from $\psi(T,\cC)$. The root $\cC''$ of
$\psi(T,\cC)$ is uniquely determined (this is the only node without outgoing arrow). $T$ is obtained by removing 
the arrow $\cC' \rightarrow \cC$ from $\psi(T,\cC)$ and then adding the arrow $\cC'' \rightarrow \cC'$.
Finally the sets $ \mathcal{T}(\cC')\times \fC \backslash \{\cC'\}$ and $\bigcup_{\cC \in \fC \backslash \{\cC'\} } \mathcal{T}(\cC)$
have the same cardinal (because the second set is a disjoint union). 

From the definition of the $M$-weight of a directed graph and of the construction of $\psi(T,\cC)$, we can see that
 $w(T)m(\cC' \rightarrow \cC)=w(\psi(T,\cC))m(\cC'' \rightarrow \cC')$.

We can now complete the proof of the proposition
\begin{eqnarray*}
\sum_{\cC \in \fC \backslash \{\cC'\} } m(\cC' \rightarrow \cC)\cS(\cC') & = & 
 \frac{1}{Z}\sum_{\cC \in \fC \backslash \{\cC'\} } m(\cC' \rightarrow \cC) \left( \sum_{T \in \mathcal{T}(\cC')} w(T) \right) \\
 & = & \frac{1}{Z}\sum_{\cC \in \fC \backslash \{\cC'\} }  \sum_{T \in \mathcal{T}(\cC')} w(\psi(T,\cC))m(r(\psi(T,\cC')) \rightarrow \cC') \\
 & = & \frac{1}{Z}\sum_{\cC'' \in \fC \backslash \{\cC'\} }  \sum_{T \in \mathcal{T}(\cC'')} w(T)m(\cC'' \rightarrow \cC') \\
 & = & \sum_{\cC'' \in \fC \backslash \{\cC'\} } m(\cC'' \rightarrow \cC')\cS(\cC'').
\end{eqnarray*}
The second equality is obtained using $w(T)m(\cC' \rightarrow \cC)=w(\psi(T,\cC))m(r(\psi(T,\cC')) \rightarrow \cC')$. The third equality is obtained using the fact that 
$\psi$ is a bijective map.
\finproof

\begin{example}
In this example we treat the case of a three states Markov chain. This is the simplest situation where we can find
a non reversible Markov matrix (i.e for which the detailed balance is broken). 
The most general Markovian dynamics on three states is illustrated in figure \ref{fig:3states}.
\begin{figure}[htb]
\begin{center}
 \begin{tikzpicture}[scale=1]
\draw  (-5/2,0) circle (1/2) [circle] {}; \node at (-5/2,0) [] {$\cC_3$};
\draw  (5/2,0) circle (1/2) [circle] {}; \node at (5/2,0) [] {$\cC_2$};
\draw  (0,1.732*5/2) circle (1/2) [circle] {}; \node at (0,1.732*5/2) [] {$\cC_1$};
\draw[->] (-1.75,-0.25) -- (1.75,-0.25);
\draw[->] (1.75,0.25) -- (-1.75,0.25);
\draw[->] (-5/2+0.75/2+1.732/2*0.25,0.75*1.732/2-0.25/2) -- (-0.75/2+1.732/2*0.25,1.732*5/2-0.75*1.732/2-0.25/2);
\draw[->] (-0.75/2-1.732/2*0.25,1.732*5/2-0.75*1.732/2+0.25/2) -- (-5/2+0.75/2-1.732/2*0.25,0.75*1.732/2+0.25/2);
\draw[->] (0.75/2-1.732/2*0.25,1.732*5/2-0.75*1.732/2-0.25/2) -- (5/2-0.75/2-1.732/2*0.25,0.75*1.732/2-0.25/2);
\draw[->] (5/2-0.75/2+1.732/2*0.25,0.75*1.732/2+0.25/2) -- (0.75/2+1.732/2*0.25,1.732*5/2-0.75*1.732/2+0.25/2);
\node at (0,-0.55) [] {$m(\cC_3 \rightarrow \cC_2)$};
\node at (0,0.55) [] {$m(\cC_2 \rightarrow \cC_3)$};
\node at (-5/2,1.732*5/4) [] {$m(\cC_1 \rightarrow \cC_3)$}; \node at (-5/2+2.2,1.732*5/4-0.35) [] {$m(\cC_3 \rightarrow \cC_1)$};
\node at (5/2,1.732*5/4) [] {$m(\cC_2 \rightarrow \cC_1)$};  \node at (5/2-1.8,1.732*5/4-1.05) [] {$m(\cC_1 \rightarrow \cC_2)$};
 \end{tikzpicture}
 \end{center}
 \caption{Graphical representation of a three states Markov chain. \label{fig:3states}}
\end{figure}
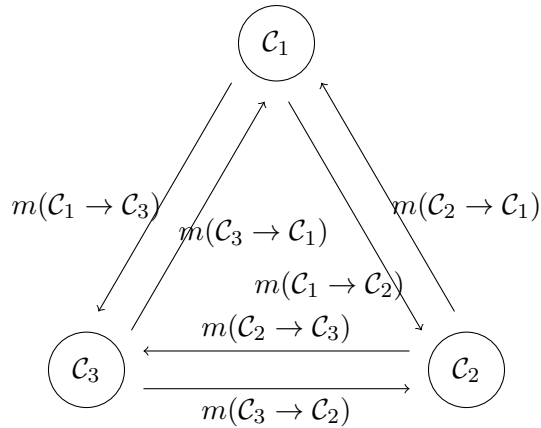
We compute in this case all the quantities introduced previously.
The set of rooted trees $\mathcal{T}(\cC_1)$ contains $3$ elements $T_1$, $T_2$ and $T_3$
which are shown in figure \ref{fig:arbres}.
\begin{figure}[htb]
\begin{center}
 \begin{tikzpicture}[scale=1]
\draw  (0,0) circle (1/2) [circle] {}; \node at (0,0) [] {$\cC_1$};
\draw  (-1.5,-2) circle (1/2) [circle] {}; \node at (-1.5,-2) [] {$\cC_2$};
\draw  (1.5,-2) circle (1/2) [circle] {}; \node at (1.5,-2) [] {$\cC_3$};
\draw[->]  (-1.5+3/4*0.4,-2+0.4) -- (-0.75,-1);\draw  (-0.75,-1) -- (-3/4*0.4,-0.4);
\draw[->]  (1.5-3/4*0.4,-2+0.4) -- (0.75,-1);\draw  (0.75,-1) -- (3/4*0.4,-0.4);
\node at (0,-3) [] {$T_1$};

\draw  (4,0) circle (1/2) [circle] {}; \node at (4,0) [] {$\cC_1$};
\draw  (4,-2) circle (1/2) [circle] {}; \node at (4,-2) [] {$\cC_2$};
\draw  (4,-4) circle (1/2) [circle] {}; \node at (4,-4) [] {$\cC_3$};
\draw[->]  (4,-3.5) -- (4,-3); \draw (4,-3) -- (4,-2.5);
\draw[->]  (4,-1.5) -- (4,-1); \draw (4,-1) -- (4,-0.5);
\node at (4,-5) [] {$T_2$};

\draw  (8,0) circle (1/2) [circle] {}; \node at (8,0) [] {$\cC_1$};
\draw  (8,-2) circle (1/2) [circle] {}; \node at (8,-2) [] {$\cC_3$};
\draw  (8,-4) circle (1/2) [circle] {}; \node at (8,-4) [] {$\cC_2$};
\draw[->]  (8,-3.5) -- (8,-3); \draw (8,-3) -- (8,-2.5);
\draw[->]  (8,-1.5) -- (8,-1); \draw (8,-1) -- (8,-0.5);
\node at (8,-5) [] {$T_3$};
 \end{tikzpicture}
 \end{center}
 \caption{Elements of $\mathcal{T}(\cC_1)$. \label{fig:arbres}}
\end{figure}
The $M$-weight of these trees are respectively
\begin{eqnarray*}
 w(T_1) & = & m(\cC_2 \rightarrow \cC_1)m(\cC_3 \rightarrow \cC_1) \\
 w(T_2) & = & m(\cC_3 \rightarrow \cC_2)m(\cC_2 \rightarrow \cC_1) \\
 w(T_3) & = & m(\cC_2 \rightarrow \cC_3)m(\cC_3 \rightarrow \cC_1).
\end{eqnarray*}
From that we can compute
\begin{equation*}
 \sum_{T \in \mathcal{T}(\cC_1)} w(T) = m(\cC_2 \rightarrow \cC_1)m(\cC_3 \rightarrow \cC_1)+m(\cC_2 \rightarrow \cC_1)m(\cC_3 \rightarrow \cC_2)+m(\cC_3 \rightarrow \cC_1)m(\cC_2 \rightarrow \cC_3).
\end{equation*}
The sets $\mathcal{T}(\cC_2)$ and $\mathcal{T}(\cC_3)$ can be easily deduced from $\mathcal{T}(\cC_1)$ by permutations of
$\cC_1,\cC_2,\cC_3$ on the rooted trees. This leads to
\begin{equation*}
 \sum_{T \in \mathcal{T}(\cC_2)} w(T) = m(\cC_1 \rightarrow \cC_2)m(\cC_3 \rightarrow \cC_2)+m(\cC_1 \rightarrow \cC_2)m(\cC_3 \rightarrow \cC_1)+m(\cC_3 \rightarrow \cC_2)m(\cC_1 \rightarrow \cC_3).
\end{equation*}
and 
\begin{equation*}
 \sum_{T \in \mathcal{T}(\cC_3)} w(T) = m(\cC_2 \rightarrow \cC_3)m(\cC_1 \rightarrow \cC_3)+m(\cC_2 \rightarrow \cC_3)m(\cC_1 \rightarrow \cC_2)+m(\cC_1 \rightarrow \cC_3)m(\cC_2 \rightarrow \cC_1).
\end{equation*}
The normalization factor is thus given by
\begin{eqnarray*}
 Z & = & m(\cC_2 \rightarrow \cC_1)m(\cC_3 \rightarrow \cC_1)+m(\cC_2 \rightarrow \cC_1)m(\cC_3 \rightarrow \cC_2)+m(\cC_3 \rightarrow \cC_1)m(\cC_2 \rightarrow \cC_3) \\
 & & + m(\cC_1 \rightarrow \cC_2)m(\cC_3 \rightarrow \cC_2)+m(\cC_1 \rightarrow \cC_2)m(\cC_3 \rightarrow \cC_1)+m(\cC_3 \rightarrow \cC_2)m(\cC_1 \rightarrow \cC_3) \\
 & & + m(\cC_2 \rightarrow \cC_3)m(\cC_1 \rightarrow \cC_3)+m(\cC_2 \rightarrow \cC_3)m(\cC_1 \rightarrow \cC_2)+m(\cC_1 \rightarrow \cC_3)m(\cC_2 \rightarrow \cC_1).
\end{eqnarray*}
In order to give some intuition of the proof given previously, let us show here that the formula of the proposition gives
the stationary measure in this particular case
\begin{eqnarray*}
 &  & \cS(\cC_1)(m(\cC_1 \rightarrow \cC_2)+m(\cC_1 \rightarrow \cC_3)) \\
 & = & (m(\cC_2 \rightarrow \cC_1)m(\cC_3 \rightarrow \cC_1)+m(\cC_2 \rightarrow \cC_1)m(\cC_3 \rightarrow \cC_2)+m(\cC_3 \rightarrow \cC_1)m(\cC_2 \rightarrow \cC_3)) \\
 & & \times \frac{1}{Z}(m(\cC_1 \rightarrow \cC_2)+m(\cC_1 \rightarrow \cC_3)) \\
 & = & \frac{1}{Z} [(m(\cC_1 \rightarrow \cC_2)m(\cC_3 \rightarrow \cC_2)+m(\cC_1 \rightarrow \cC_2)m(\cC_3 \rightarrow \cC_1)+m(\cC_3 \rightarrow \cC_2)m(\cC_1 \rightarrow \cC_3))m(\cC_2 \rightarrow \cC_1) \\
 & & + (m(\cC_2 \rightarrow \cC_3)m(\cC_1 \rightarrow \cC_3)+m(\cC_2 \rightarrow \cC_3)m(\cC_1 \rightarrow \cC_2)+m(\cC_1 \rightarrow \cC_3)m(\cC_2 \rightarrow \cC_1))m(\cC_3 \rightarrow \cC_1)] \\
 & = & m(\cC_2 \rightarrow \cC_1)\cS(\cC_2)+m(\cC_3 \rightarrow \cC_1)\cS(\cC_3).
\end{eqnarray*}
The computations for $\cS(\cC_2)$ and $\cS(\cC_3)$ work exactly the same way.
\end{example}

\begin{example} \label{ex:random_walker}
 To illustrate the stationary distribution formula on a system containing an arbitrary number of configurations, we present here 
 a 'inhomogeneous random walker' on a ring. The configuration space of this model can be described by $L$ integers $1,2,\dots,L$ denoting 
 the position of the walker on a discrete periodic lattice. We will have by convention $L+1\equiv 1$ and $0\equiv L$. 
 The stochastic dynamics is defined by $m(i \rightarrow i+1)=p_i$ and $m(i \rightarrow i-1)=q_i$ for all $1\leq i\leq L$. All the other transition probabilities are 
 vanishing. It intuitively means that at each time step $dt$, the walker at position $i$ has a probability $p_i\times dt$ to move forward and 
 a probability $q_i\times dt$ to move backward.
 
 If $p_1 p_2 \dots p_L \neq q_1 q_2 \dots q_L$ then the system does not reach a thermodynamic equilibrium (this can be intuitively understood
 as the fact that the random walker is subject to a non-conservative force). The stationary distribution\footnote{The stationary distribution is
 unique thanks to the Perron-Frobenius theorem (the Markov chain is obviously irreducible).} is thus obtained by summing 
 the $M$-weight of rooted trees. There are very few rooted trees with non vanishing $M$-weight. They are simply obtained by 
 cutting one bond of the periodic lattice. This yields the expression
 \begin{equation}
  \cS(i)=\left(\prod_{k=i+1}^L q_k \right)\sum_{j=1}^{i} \left(\prod_{k=j}^{i-1}p_k\right) \times \left(\prod_{k=1}^{j-1}q_k\right)
  +\left(\prod_{k=1}^{i-1} p_k \right)\sum_{j=i+1}^{L} \left(\prod_{k=j}^{L}p_k\right) \times \left(\prod_{k=j-1}^{i+1}q_k\right).
 \end{equation}
 This can be interpreted as the discrete version of the stationary distribution \eqref{eq:Langevin_stationary_state} given in the Langevin 
 equation context.
\end{example}

From a physical point of view, this formula given in terms of rooted trees does not appear completely satisfactory for several reasons.
The first one is that it is not expressed in terms of simple physical quantities (such as the energy in the Boltzmann distribution case).
In other words, the link between the probability rates defining the stochastic model and the relevant physical observables of the system is 
not complete (even if some remarkable progress had been made with the generalized detailed balance, 
see subsection \ref{subsec:generalized_detailed_balance} below).

The second reason is that it is not given together with a maximum entropy principle. Similarly to the thermodynamic equilibrium case,
we would like to know if this stationary distribution can be obtained by maximizing an entropy under a set of physical constraints.
This would be certainly defined on an enlarged phase space containing typically the path history instead of the static configurations.
This is probably intimately linked to the first issue mentioned and may appear as a key to build a general theory to describe 
non-equilibrium stationary states.

Finally the exact formula of the stationary distribution in terms of rooted trees is often very difficult to use for physical computations.
The number of rooted trees is typically increasing exponentially with the number of configurations of the physical system. There are 
indeed $n^{n-2}$ trees that can be constructed from $n$ vertices. Apart from very particular Markov processes for which the directed graph 
underlying the Markov matrix is very simple (in the sense that it contains a very small number of edges, as in example \ref{ex:random_walker}), 
this formula is intractable for computing physical properties in a large system. 

One of the main goals of this manuscript is to investigate these non-equilibrium stationary states in some specific cases, where 
they can be computed exactly in a much simpler form than the rooted trees expansion. The idea is to shed some light on the 
structure of this steady state in simple out-of-equilibrium models.
In chapter \ref{chap:three} we will see a technique, 
called matrix ansatz, which allows us to reduce the exponential complexity of the computation to a polynomial complexity, in some 
particular cases. We hope also to get some intuition from this method about how to relate the non-equilibrium stationary distribution
to relevant physical quantities.
We will give several new examples of non-equilibrium models for which the matrix ansatz can be efficiently used to compute
the stationary state.
 
\section{Toward a description of NESS?}

In this section, we review very briefly several tools that have been developed to describe out-of-equilibrium systems.  

\subsection{Generalized detailed balance and fluctuation theorem} \label{subsec:generalized_detailed_balance}

Even if the detailed balance does not hold for a general Markovian process, it is possible to formulate a generalization of it that is 
always satisfied. For the sake of simplicity, we will consider the particular case where the system is in contact with two reservoirs at 
different inverse temperatures $\beta_1$ and $\beta_2$ and can exchange energy with these reservoirs. The discussion below can be adapted 
to the case of more than two reservoirs and of different physical quantities (such as charges, momentum, particles) exchanged with the reservoirs.
 The generalized detailed balance relation reads
\begin{equation} \label{eq:generalized_balance}
 \frac{m(\cC \rightarrow \cC')}{m(\cC' \rightarrow \cC)} = e^{-\beta_1\Delta E_1-\beta_2\Delta E_2},
\end{equation}
where $\Delta E_1$ (respectively $\Delta E_2$) stands for the energy exchanged with reservoir $1$ (respectively $2$) during the transition from
configuration $\cC$ to configuration $\cC'$. We observe that formula \eqref{eq:generalized_balance} is a generalization of the detailed 
balance relation \eqref{eq:detailed_balance_Boltzmann}, which is easily recovered when $\beta_1=\beta_2$.
Relation \eqref{eq:generalized_balance} can be physically justified by considering the system and the two reservoirs as an isolated system,
whose dynamics should reach the microcanonical distribution. Writing the detailed balance relation for this whole system, and assuming that 
the energy exchanged is small in comparison to the total energy of the reservoirs, yields the generalized detailed balance condition
(see for instance \cite{Derrida07,Mallick15} for details).

This generalized detailed balance relation can be interpreted as the root of the fluctuation theorem, see subsection \ref{subsec:Gallavotti_Cohen}.

\subsection{Dynamic observables and deformed Markov matrix}

In a system at thermodynamic equilibrium, we saw that the stationary distribution is given by the Boltzmann distribution
\begin{equation}
 \cS(\cC) \sim e^{-\beta E(\cC)}
\end{equation}
for a system in contact with a reservoir at inverse temperature $\beta$. It allows us to define the free energy, which is simply the cumulant 
generating function of the energy observable $E$. This free energy is a very efficient tool to describe the properties of the system. 
We would like to build the same kind of tool to describe out-of-equilibrium systems and more particularly non-equilibrium stationary states.
By analogy with equilibrium systems, the generalized balance condition \eqref{eq:generalized_balance} suggests to study the fluctuations of 
the energy exchanged with the reservoirs and to construct the generating function of this observable.

This can be formalized as follows.
We are interested in studying an observable $\cO$, whose value $\cO(\cC \rightarrow \cC')$ depends on the transition 
$\cC \rightarrow \cC'$ under consideration. It can be for instance the energy exchanged with a particular reservoir during the transition 
or the number of particles injected at a particular place during the transition. We would like to compute the fluctuations of the observable 
$\cO$ in the stationary state. To give a precise mathematical meaning to this idea, we first need to give a sense to the value of the 
observable in the stationary state.
One way to achieve that is to begin by defining the value of the observable on a whole path history.
For the path history $\{\cC(t)\}_{0\leq t \leq T}$ defined in 
subsection \ref{subsubsec:time_reversibility}, which explores successively the configurations $\cC_1,\dots,\cC_n$ (with transitions occurring at 
times $t_1,\dots,t_{n-1}$), we define 
\begin{equation}
 \cO_T(\{\cC(t)\}) = \cO(\cC_1 \rightarrow \cC_2)+\dots+\cO(\cC_{n-1} \rightarrow \cC_n).
\end{equation}
The value $\cO_T$ depends on the time evolution, or path history, of the system 
(which is governed by the master equation \eqref{eq:Markov_master_equation_continuous})
on the interval $[0,T]$. It is a random variable. We denote by $\cP_T(O)$ the probability that $\cO_T=O$. Our goal is 
to study this probability distribution in the large time $T$ limit. To determine the time evolution of this probability distribution, we have to 
consider the more precise quantity $\cP_T(O,\cC)$, which denotes the joint probability for the system to be in configuration $\cC$ at time $T$
and for the observable $\cO_T$ to be equal to $O$. The value of $\cP_T(O)$ will be simply recovered through the relation
$\cP_T(O)=\sum_{\cC \in \fC} \cP_T(O,\cC)$.
\begin{proposition}
The time evolution of $\cP_T(O,\cC)$ is given by 
\begin{equation} \label{eq:Markov_joint_master_equation}
 \frac{d\cP_T(O,\cC)}{dT} = \sum_{\genfrac{}{}{0pt}{}{\cC' \in \fC}{\cC' \neq \cC}} m(\cC' \rightarrow \cC) \cP_T(O-\cO(\cC' \rightarrow \cC),\cC')
 -\sum_{\genfrac{}{}{0pt}{}{\cC' \in \fC}{\cC' \neq \cC}} m(\cC \rightarrow \cC') \cP_T(O,\cC).
\end{equation}
\end{proposition}

\proof
 This equation is derived similarly to what was done for \eqref{eq:Markov_master_equation}.
\finproof

This relation can be recast in a more elegant way by introducing the following generating function. 
\begin{definition}
 We define
\begin{equation}
 \hat\cP_T(\mu,\cC) = \sum_{O}e^{\mu O} \cP_T(O,\cC). 
\end{equation}
\end{definition}

\begin{proposition}
The generating function $\hat\cP_T(\mu,\cC)$ fulfills the deformed master equation
\begin{equation}
 \frac{d\hat\cP_T(\mu,\cC)}{dT} = \sum_{\genfrac{}{}{0pt}{}{\cC' \in \fC}{\cC' \neq \cC}} m(\cC' \rightarrow \cC)e^{\mu \cO(\cC' \rightarrow \cC)} \cP_T(\mu,\cC')
 -\sum_{\genfrac{}{}{0pt}{}{\cC' \in \fC}{\cC' \neq \cC}} m(\cC \rightarrow \cC') \cP_T(\mu,\cC).
\end{equation}
\end{proposition}

\proof
 This is established by summing \eqref{eq:Markov_joint_master_equation} over $O$ with the factor $e^{\mu O}$.
\finproof

 It can be rewritten in matrix form by introducing the following vector and deformed Markov matrix.
 \begin{definition}
  We define the deformed probability vector
  \begin{equation}
   \ket{\hat\cP_T(\mu)} = \sum_{\cC \in \fC} \hat\cP_T(\mu,\cC) \ket{\cC},
  \end{equation}
  and the deformed Markov matrix
  \begin{equation}
   M^{\mu} = \sum_{\cC,\cC' \in \fC} m^{\mu}(\cC \rightarrow \cC') \ket{\cC'}\bra{\cC},
  \end{equation}
  with $m^{\mu}(\cC \rightarrow \cC')=m(\cC \rightarrow \cC')e^{\mu \cO(\cC \rightarrow \cC')}$ for $\cC' \neq \cC$ 
  and $m^{\mu}(\cC \rightarrow \cC)=-\sum_{\genfrac{}{}{0pt}{}{\cC' \in \fC}{\cC' \neq \cC}} m(\cC \rightarrow \cC')$.
 \end{definition}

A straightforward computation then yields
\begin{equation}
 \frac{d \ket{\hat\cP_T(\mu)}}{dT} = M^{\mu}\ket{\hat\cP_T(\mu)},
\end{equation}
which provides the following formal expression 
\begin{equation}
 \ket{\hat\cP_T(\mu)} = e^{T M^{\mu}}\ket{\hat\cP_0(\mu)}.
\end{equation}
 
We denote by $E_1(\mu), E_2(\mu),\dots$ the eigenvalues of $M^{\mu}$ ordered by decreasing real part, $\ket{\Psi_1},\ket{\Psi_2},\dots$ the 
corresponding right eigenvectors, and $\bra{\Phi_1},\bra{\Phi_2},\dots$ the corresponding left eigenvectors.

The last equation then becomes
\begin{equation} \label{eq:Markov_deformed_master_equation_sol_intermediate}
 \ket{\hat\cP_T(\mu)}=\langle\Phi_1\ket{\hat\cP_0(\mu)}e^{T E_1(\mu)}\ket{\Psi_1}+\langle\Phi_2\ket{\hat\cP_0(\mu)}e^{T E_2(\mu)}\ket{\Psi_2}+\dots 
\end{equation}
We now would like to deduce, from these computations, the behavior of the probability distribution $\cP_T(O)$ in the long time $T \rightarrow \infty$ 
limit. This can be studied through the generating function of the cumulant defined as
\begin{equation}
 E^T(\mu) := \ln \left(\sum_O e^{\mu O}\cP_T(O) \right) = \ln \left( \langle \Sigma \ket{\hat\cP_T(\mu)} \right),
\end{equation}
where we recall that $\bra{\Sigma} = \sum_{\cC \in \fC} \bra{\cC}$. We observe, from \eqref{eq:Markov_deformed_master_equation_sol_intermediate},
that $E^T(\mu)$ behaves asymptotically as a linear function of $T$. Its long time limit is thus captured by the quantity
\begin{equation} \label{eq:dynamic_observable_cumulant_generating_function}
 E(\mu) := \lim\limits_{T \rightarrow \infty} \frac{E^T(\mu)}{T} = E_1(\mu),
\end{equation}
which is exactly given by the largest eigenvalue of $M^{\mu}$.

This cumulant generating function in the stationary state turns out to be an efficient tool to characterize the macroscopic behavior
of a physical system in the large size limit (thermodynamic limit, see chapter \ref{chap:five}), if the observable $\cO$ is correctly chosen.
In particular the singularities of this function (or of the associated large deviation function, see subsection \ref{subsec:large_deviation})
could determine the phase transitions of the model \cite{GorissenLMV12,Lazarescu15,Lazarescu17}.

In this manuscript, we will be interested in interacting particles systems evolving on a one dimensional lattice. A physical quantity of prime 
interest in such models is the (algebraic) number of particles which cross a particular bond on the lattice during a transition. This 
observable is indeed representative of the non-equilibrium aspects of these interacting particles systems. It provides a quantitative way to 
characterize the particles current and its fluctuation in the stationary state, through the computation of the cumulant generating function.
In chapter \ref{chap:three} and chapter \ref{chap:four} we will address the problem of computing exactly this generating function for 
particular models.

This formalism allows also to state an elegant symmetry on the fluctuations of the entropy production, which we present in the subsection below.

\subsection{Gallavotti-Cohen symmetry and fluctuation theorem} \label{subsec:Gallavotti_Cohen}

Having in mind the generalized balance condition \eqref{eq:generalized_balance}, we now focus on the observable defined by
\begin{equation}
 \cO(\cC \rightarrow \cC') = \ln \frac{m(\cC \rightarrow \cC')}{m(\cC' \rightarrow \cC)}.
\end{equation}
This can be interpreted as the entropy production during the transition $\cC \rightarrow \cC'$ \cite{GallavottiC95}. The associated 
deformed Markov matrix $M^{\mu}$ can be easily computed, its entries are given, for $\cC' \neq \cC$, by
\begin{equation}
 m^{\mu}(\cC \rightarrow \cC') = m(\cC \rightarrow \cC')^{1+\mu} m(\cC' \rightarrow \cC)^{-\mu}.
\end{equation}
We observe that we have the equality $m^{\mu}(\cC \rightarrow \cC') = m^{-1-\mu}(\cC' \rightarrow \cC)$  which translates into
\begin{equation}
 M^{\mu} = \left(M^{-1-\mu}\right)^t,
\end{equation}
where '$\cdot^t$' denotes the matrix transposition. We thus deduce that the spectrum of $M^{\mu}$ and of $M^{-1-\mu}$ are identical.
This proves the following equality satisfied by the largest eigenvalue $E(\mu)$ of $M^{\mu}$
\begin{equation}
 E(\mu) = E(-1-\mu).
\end{equation}
This is known as the Gallavotti-Cohen symmetry \cite{LebowitzS99,Kurchan98,GallavottiC95}. 
We proved here the symmetry involving the entropy production (which holds for any Markov chain) but similar relations can also be derived 
for other physical observables, depending on the model under consideration.
For instance, we will encounter such a symmetry in chapter \ref{chap:four} while studying 
the cumulant generating function of the particle current for the asymmetric simple exclusion process.

The symmetry on the generating function translates into a symmetry on the large deviation function of the entropy production 
(see subsection \ref{subsec:large_deviation} for definition and details), which can be defined as
\begin{equation}
 G(s)= \inf\limits_{\mu} [\mu s-E(\mu)].
\end{equation}
Using the symmetry on the generating function $E$, we obtain
\begin{eqnarray}
 \inf\limits_{\mu} [\mu s-E(\mu)] & = & \inf\limits_{\mu} [(-1-\mu)\cdot (-s)-s-E(-1-\mu)] \\
 & = & \inf\limits_{-1-\mu} [(-1-\mu)\cdot (-s)-E(-1-\mu)]-s.
\end{eqnarray}
This yields the symmetry for the large deviation function
\begin{equation}
 G(s)=G(-s)-s.
\end{equation}
This last equality is known as the fluctuation theorem. This relation was first observed in \cite{EvansCM93} and then proven in \cite{EvansS94}.
It tells us that the probability $\cP_T(s)$ to observe during time interval $[0,T]$ an entropy production equal to $Ts$ satisfies
\begin{equation}
 \frac{\cP_T(-s)}{\cP_T(s)} \underset{T \rightarrow \infty}{\sim} e^{-Ts}
\end{equation}

This fluctuation theorem (and its generalization to other physical observables, depending on the model under consideration) has 
very useful implications. It yields near equilibrium the Einstein fluctuation-dissipation relations, the Onsager
reciprocity relations and the Kubo linear response theory, see for instance \cite{Kubo66}.

\subsection{Large deviation functions} \label{subsec:large_deviation}

In the last decades, the large deviation theory has proven to be a very efficient framework to deal with equilibrium but also non-equilibrium systems.
We present briefly below the main tools (from a statistical physicist
point of view) associated to this theory. The reader is invited to read the very useful review \cite{Touchette09} for details.

\subsubsection{Large deviation principle}

The large deviation theory can be intuitively introduced as a framework to evaluate the probabilities of rare events.
From a mathematical point of view, it could be thought as a refinement of the law of large numbers. From a physical point of view, 
the large deviation functions are seen as possible generalizations of thermodynamic potentials in the out-of-equilibrium context.

\begin{definition}
We consider $(S_n)$ a sequence of random variables taking real values and an interval $A\subset \RR$. 
The probability $\PP(S_n \in A)$ is said to follow a large deviation principle if there exists a real number $G_A$, called \textup{the rate}, such that
\begin{equation}
 -\lim\limits_{n \rightarrow \infty} \frac{1}{n} \ln \PP(S_n \in A) = G_A.
\end{equation}
\end{definition}

The large deviation principle roughly means that the probability $\PP(S_n \in A)$ vanishes exponentially fast with $n$:
\begin{equation}
 \PP(S_n \in A) \sim e^{-nG_A}
\end{equation}

From this perspective it can be interpreted as a refinement of the law of large numbers, which states that the properly normalized sum of
independent identically distributed random variables converges, with probability $1$, to the common expectation value of the variables.
Here, we observe that if $G_A$ vanishes, the probability measure concentrates, with probability $1$, in the set $A$. In particular 
(in the case of the normalized sum of independent identically distributed random variables) if
$A$ is chosen to be any interval containing the expectation value of the variable, the large deviation principle provides the result of 
the law of large number together with the rate of convergence of the probability measure (which explains the denomination 'refinement').
This phenomenon is illustrated in the following example.

\begin{figure}[htbp]
\begin{center}
\includegraphics[width=80mm,height=80mm]{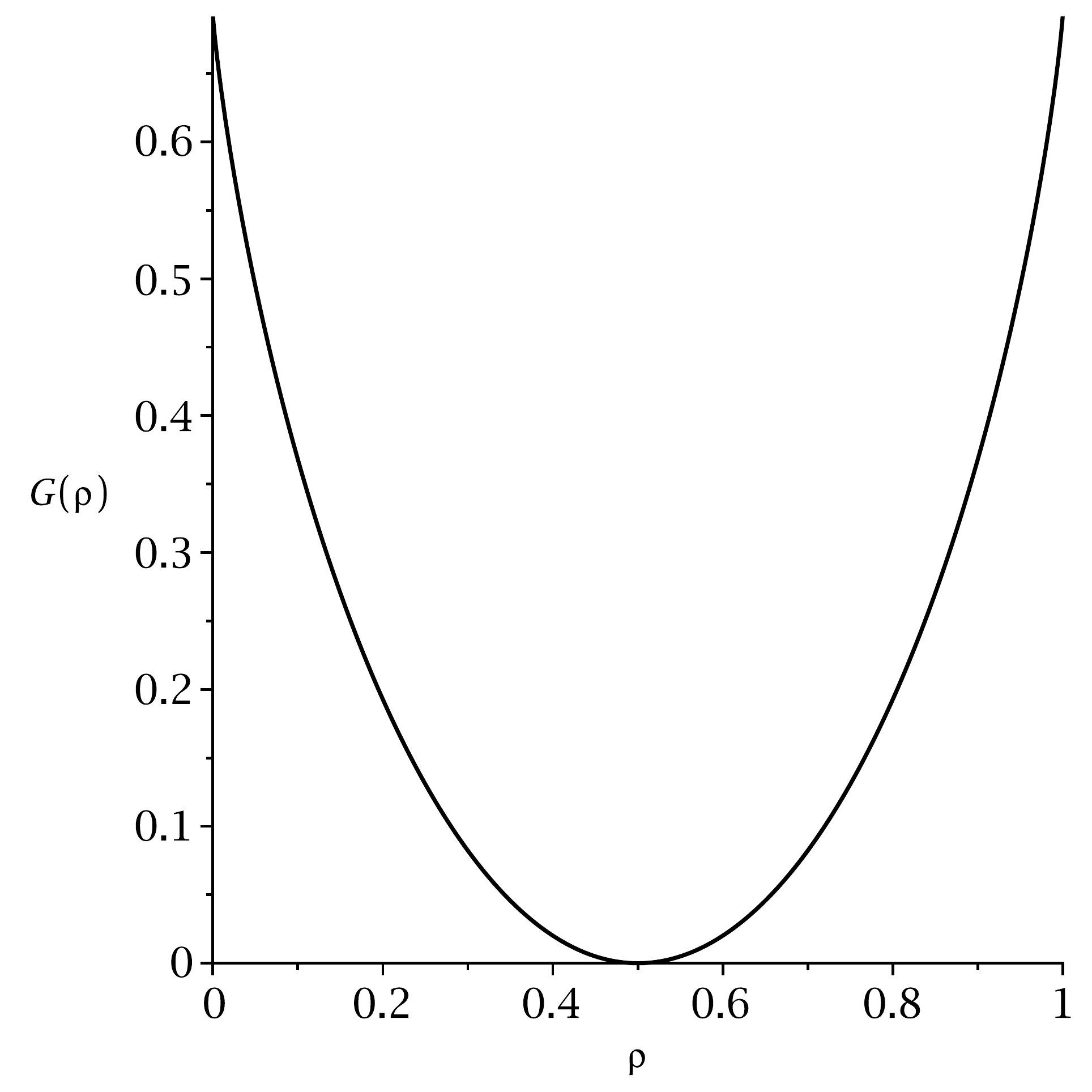}
\end{center}
\caption{Plot of the large deviation function $G(\rho)$. \label{fig:example_large_deviation}}
\end{figure}

\begin{example}
 We present here one of the simplest example of large deviation principle. We consider a sequence of binary independent random variables $(\epsilon_k)$,
 that are equal to $1$ with probability $1/2$ and equal to $0$ with probability $1/2$. We define 
 \begin{equation}
  S_n = \frac{1}{n}\sum_{k=1}^n \epsilon_k.
 \end{equation}
It is straightforward to establish that $S_n$ follows the binomial distribution
\begin{equation}
 \PP(S_n = k/n) = \frac{1}{2^n}\begin{pmatrix}
                                n \\ k
                               \end{pmatrix}.
\end{equation}
We fix $\rho \in [0,1]$ and we set $k=\lfloor \rho n \rfloor$. It means intuitively that for $n$ large enough, $k/n \simeq \rho$.  
We can use the Stirling's formula to estimate the asymptotic behavior of $\PP(S_n = k/n)$, when $n$ is large. This yields the relation
\begin{equation}
  -\lim\limits_{n \rightarrow \infty} \frac{1}{n} \ln \PP(S_n = k/n) = \ln 2 +\rho \ln \rho +(1-\rho)\ln (1-\rho):= G(\rho).
\end{equation}
$G(\rho)$ is the rate function, also called large deviation function. It is a non-negative convex function of $\rho$, which is minimum for $\rho=1/2$ 
(it vanishes for this value), see figure \ref{fig:example_large_deviation}.
It allows us to recover the result of the law of large numbers: the probability measure of $S_n$ concentrates, as $n \rightarrow \infty$,
toward the expectation value of the variables $\epsilon_k$, which is equal to $1/2$.
\end{example}

\subsubsection{Legendre transformation and G\"{a}rtner-Ellis theorem} \label{subsec:Legendre}

We are now going to see that the large deviation function $G$ is closely related to a quantity, which we already encountered  
in this manuscript: the cumulant generating function. This connection appears through the Legendre transformation, that we now define.  
\begin{definition}
If $f$ is a convex function, its Legendre transform $g$ is defined by
\begin{equation}
 g(p) = \sup\limits_{x} (px-f(x))
\end{equation}
\end{definition}
If $f$ is everywhere differentiable the previous definition is equivalent to
\begin{equation}
 g(p) = p x^{*}-f(x^{*}), \quad \mbox{with }x^{*} \mbox{ such that} \quad  \frac{df}{dx}(x^{*}) = p.
\end{equation}
The inverse transform formula is given by 
\begin{equation}
 f(x) = \sup\limits_{p} (px-g(p)),
\end{equation}
or if $g$ is everywhere differentiable
\begin{equation}
 f(x) = x p^{*}-g(p^{*}), \quad \mbox{with }p^{*} \mbox{ such that} \quad  \frac{dg}{dp}(p^{*}) = x.
\end{equation}
It means that the Legendre transform is self-inverse (or involutive).

We now define a quantity which will play an important role in this manuscript, the cumulant generating function.
\begin{definition}
 We consider $(S_n)$ a sequence of real random variables and we define the \textup{cumulant generating function} of $(S_n)$ by
 \begin{equation} \label{eq:cumulant_generating_function}
  E(\mu)= \lim\limits_{n \rightarrow \infty} \frac{1}{n} \ln \left(\sum_{x \in \cX_n} e^{n \mu x} \PP(S_n=x)\right),
 \end{equation}
where $\cX_n$ is the discrete\footnote{For the sake 
 of simplicity, we consider here random variables $S_n$ taking discrete values $x \in \cX_n$ but generalization to 
 continuous random variables is straightforward. We consider also a stochastic process $(S_n)$ parametrized by a discrete variable $n$ but
 the generalization to stochastic processes $(S_t)$ parametrized by a continuous time $t$ is obvious.} configuration set of the random variable $S_n$.
 \end{definition}
It can be shown that the cumulant generating function is always convex (see for instance \cite{Touchette09}).

\begin{remark}
This function is called cumulant generating function because we can obtain the expectation, variance, and higher order cumulants by taking 
successive derivatives with 
respect to the parameter $\mu$. For instance if $\epsilon_k$ is a sequence of independent identically distributed random variables with 
expectation $\langle \epsilon \rangle$ and variance $\mbox{var}(\epsilon)$, and $S_n$ is given by
\begin{equation}
 S_n = \frac{1}{n} \sum_{k=1}^n \epsilon_k,
\end{equation}
then it is straightforward to see that
\begin{equation}
 E'(0) = \langle \epsilon \rangle,
\end{equation}
and 
\begin{equation}
 E''(0) = \mbox{var}(\epsilon).
\end{equation}
\end{remark}

We are now equipped to state the G\"{a}rtner-Ellis theorem, which relates the cumulant generating function of a sequence of random variable to its 
large deviation function.

\begin{theorem}
 Suppose that the cumulant generating function $E(\mu)$ associated to the sequence of random variable $(S_n)$ is differentiable, then 
 $(S_n)$ satisfies a large deviation principle 
 \begin{equation}
 -\lim\limits_{n \rightarrow \infty} \frac{1}{n} \ln \PP(S_n = x) = G(x),
\end{equation}
 where the large deviation function $G$ is given by the Legendre transform of $E$
 \begin{equation}
  G(x) = \sup\limits_{\mu} (\mu x-E(\mu)),
 \end{equation}
 or equivalently
 \begin{equation}
  G(x) = x \mu^{*}-E(\mu^{*}), \quad \mbox{with }\mu^{*} \mbox{ such that} \quad  \frac{dE}{d\mu}(\mu^{*}) = x.
 \end{equation}
\end{theorem}

We do not give a rigorous proof of the theorem but we rather propose an argument to convince the reader of its validity. If $(S_n)$ satisfies
a large deviation principle with large deviation function $G$, then we can evaluate
\begin{equation}
 \sum_{x \in \cX_n} e^{n \mu x} \PP(S_n=x) \sim \sum_{x \in \cX_n} e^{n \mu x} e^{-n G(x)} = \sum_{x \in \cX_n} e^{n (\mu x-G(x))}.
\end{equation}
The behavior of the last sum is obtained through a saddle-point analysis, which yields the following expression
for the cumulant generating function \eqref{eq:cumulant_generating_function}
\begin{equation}
 E(\mu) = \sup\limits_{x} (\mu x-G(x)).
\end{equation}
The Legendre inverse transform formula gives the desired result.

The G\"{a}rtner-Ellis theorem provides a tool to compute large deviation functions through cumulant generating functions. It is commonly used 
in out-of-equilibrium statistical physics, especially to compute large deviation of particles current in the stationary state in interacting 
particles systems.

\subsubsection{Application to out-of-equilibrium statistical physics} 

We present briefly two relevant utilizations of the large deviation theory in the context of out-of-equilibrium statistical physics.
This description is far from being exhaustive. We chose to focus on the applications of the theory which will be used later in 
this manuscript.
The reader is invited to consult the review \cite{Touchette09} for a complete description of the role of large deviation theory in statistical 
physics.

In out-of-equilibrium models,
we saw previously that large deviation theory was the correct tool to deal with the fluctuations of dynamic observables in the stationary
state. It was indeed stressed that the fluctuations of the dynamic observable were encoded in the quantity
\eqref{eq:dynamic_observable_cumulant_generating_function}, which is exactly defined as the cumulant generating function introduced in the 
large deviation context \eqref{eq:cumulant_generating_function}, where the time $T$ plays the role of the large deviation parameter.
The Legendre transformation then provides the large deviation function associated with this observable.
By analogy with equilibrium models, the cumulant generating function may be related to a generalization of the free energy and 
the large deviation function may be connected to an entropy.
These large deviation functions (or cumulant generating functions) are in general difficult to compute exactly: hence it is important
to construct models for which such computations are possible. This motivates the study of {\it integrable systems}, i.e models for which
analytical results can be derived. The next chapter is devoted to the presentation of integrability in the context of Markov chains.
In chapter \ref{chap:three} we will compute the first cumulants of the particle current of the dissipative symmetric simple exclusion process.
We will explore in chapter \ref{chap:four} the connection between the cumulant generating function of the current in the asymmetric simple 
exclusion process and the theory of symmetric polynomials.

A second application of the large deviation theory in out-of equilibrium physics, which will be of particular interest for us, is the 
macroscopic fluctuation theory. This is a theory which aims to provide a coarse grained description of diffusive systems.
The probability to observe a path history, a time evolution, of properly defined macroscopic variables satisfies a large deviation principle.
The large deviation parameter is the size of the system.
The rate function can be heuristically interpreted as an action functional. The macroscopic fluctuation theory provides also predictions about 
fluctuations of the particle current and density in the stationary state. 
This theory will be presented in details in chapter \ref{chap:five}.
The integrable models also play a central role in this context because they appear as benchmarks to test the predictions of the theory. 
We will indeed use in chapter \ref{chap:five} the exact results obtained in chapter \ref{chap:three} on 
the dissipative symmetric simple exclusion process and respectively on the open boundaries multi-species symmetric simple exclusion process 
to check the predictions for dissipative models and respectively to extend the theory to multi-species models.

\chapter{Integrability} \label{chap:two}

First we introduce classical integrability as motivation, then specific models that will be used as examples throughout,
before finally giving the machinery of quantum integrability.

\section{Introduction, motivations and formalism}

\subsection{Conserved quantities}

\subsubsection{Introduction and motivations}

The idea behind integrability goes back to the study of classical mechanical systems.
It was observed that the conservation of the energy in certain simple systems often permits to solve exactly the equations of motion 
(basically Newton's law). This idea was precisely formulated by Liouville \cite{BabelonBT03}, using the Hamiltonian formalism of the classical dynamics.
In this framework a system is described through coordinates of position $q_i$ and momentum $p_i$ ($1\leq i \leq n$) defining the phase space 
(of dimension $2n$). The dynamics is encoded with the Hamiltonian $H(p_i,q_i)$ by the equations
\begin{equation} \label{eq:classical_motion}
 \frac{d q_i}{dt}=\frac{\partial H}{\partial p_i}, \quad  \frac{d p_i}{dt}=-\frac{\partial H}{\partial q_i}.
\end{equation}
The time evolution of a function $F(p_i,q_i)$ of the phase space is obtained by the equation
\begin{equation}
 \frac{d F}{dt} = \{ H,F \},
\end{equation}
where the Poisson bracket is defined as
\begin{equation}
 \{F,G\} = \sum_{i=1}^n \frac{\partial F}{\partial p_i}\frac{\partial G}{\partial q_i}-\frac{\partial G}{\partial p_i}\frac{\partial F}{\partial q_i}
\end{equation}
In this framework, a conserved quantity is thus a function $F$ satisfying $\{H,F\}=0$.

Liouville proved the following theorem.
\begin{theorem}
 If the system possesses $n$ independent\footnote{The independence has here to be understood as the linear independence of the 
 differential forms $d F_i$ (at a generic point)}
 conserved quantities $F_i$ ($1\leq i \leq n$), i.e such that $\{H,F_i\}=0$, in involution
 \begin{equation}
  \{F_i,F_j\}=0,
 \end{equation}
 then the solution of the equations of motion \eqref{eq:classical_motion} can be computed through a ``quadrature''\footnote{Note that the situation
 is much less clear when the phase space is infinite dimensional, in classical field theory for instance.}.
\end{theorem}

\begin{remark}
 The Liouville theorem is proven by performing a change of variables $(p_i,q_i) \rightarrow (F_i,\psi_i)$. The new variables
 are called the action-angle variables and satisfy the equations
 \begin{equation}
  \frac{d F_i}{dt}=0, \quad \frac{d \psi_i}{dt}=\Omega_i,
 \end{equation}
where $\Omega_i$ are time-independent constants. These equations are easily solved. The coordinates $\psi_i$ are constructed through the relation
\begin{equation}
 \psi_i = \frac{\partial S}{\partial F_i},
\end{equation}
 where $S(F,q)$ is defined by the quadrature
 \begin{equation}
  S(F,q)= \int_{q_0}^q \sum_{i=1}^n p_i(F,q)dq_i.
 \end{equation}
\end{remark}

Since this discovery, people have been interested in finding a systematic way to construct Hamiltonians, $H$, 
together with the conserved quantities, $F_i$. 
This has led to the introduction of different key concepts, like the Lax-pair and the classical r-matrix. The reader may refer to \cite{BabelonBT03}
for an introduction to classical integrable systems and a lot of details.

There have been several attempts to adapt and exploit these fundamental concepts in quantum mechanics and statistical physics. 
This has led to remarkable developments ranging from the discovery of quantum $R$-matrices and 
of the Yang-Baxter equation \cite{Baxter82}, to the quantum inverse scattering method \cite{KorepinBI97,FaddeevT07},
and including the quantum groups \cite{ChariP95,Jimbo85,Drinfeld87}.

These works pointed out the relevance of the conserved quantities in the exact solvability of the quantum Hamiltonians.
Unfortunately, we are still lacking a quantum analogue for the Liouville theorem.

In this manuscript, we will make an intensive use of the framework and techniques developed to study the Hamiltonians of quantum 
spin chains. We will be indeed interested in Markovian processes, defined on a one dimensional lattice, which can be studied within 
the quantum spin chain framework (the Markov matrix will often be a similarity transform of quantum spin chain Hamiltonian).

\subsubsection{Markovian case} 

In the context of Markov matrices, a conserved quantity can be interpreted as an observable whose average value is constant in time.
To fix the ideas, let us denote by $\cO$ an observable, {\it i.e} a real valued function $\cO(\cC) \in \mathbb{R}$ of the configurations $\cC$
of the system. The average value of the observable at time $t$ is defined by
\begin{equation}
 \langle \cO \rangle_t = \sum_{\cC} \cO(\cC) P_t(\cC),
\end{equation}
where we recall that $P_t(\cC)$ denotes the probability for the system to be in configuration $\cC$ at time $t$. It will be convenient 
to reformulate that in a matrix form. We introduce a row vector $\bra{\Sigma}$, which stands for the sum of the basis row vectors,
and a diagonal matrix $O$, which encompasses the value of the observable $\cO$
\begin{equation}
 \bra{\Sigma} = \sum_{\cC}\bra{\cC} \quad \mbox{and} \quad O=\sum_{\cC} \cO(\cC) \ket{\cC}\bra{\cC}.
\end{equation}
The average value of $\cO$ at time $t$ can be rewritten
\begin{equation}
 \langle \cO \rangle_t = \bra{\Sigma} O \ket{P_t} = \bra{\Sigma} O \exp (tM) \ket{P_0}, 
\end{equation}
where we recall that $M$ is the Markov matrix encoding the dynamics of the model \eqref{eq:Markov_matrix_continuous_time} 
and $\ket{P_t}$ is the probability distribution of the model at time $t$, see \eqref{eq:Markov_master_equation_continuous}.
$\ket{P_0}$ is the initial probability distribution at time $t=0$.
It is then straightforward to see that, if the matrices $O$ and $M$ commutes, $OM=MO$, then
\begin{equation}
 \langle \cO \rangle_t = \bra{\Sigma} \exp (tM) O \ket{P_0} = \bra{\Sigma} O \ket{P_0} = \langle \cO \rangle_0,
\end{equation}
where we used the Markovian property of $M$, $\bra{\Sigma}M=0$, which implies $\bra{\Sigma}\exp(tM)=\bra{\Sigma}$. In other words the 
average value of $\cO$ is conserved in time.

\begin{remark}
 The observable $\cO$ that we studied is a 'static' observable, i.e it depends on the current configuration $\cC$ of the system. We can adapt 
 the discussion above to 'dynamic' observables, i.e depending on a transition from a configuration $\cC$ to a configuration $\cC'$. We denote 
 by $\cO(\cC \rightarrow \cC')$ the value of the observable associated with this transition. The average value of the observable is now
 defined as 
 \begin{equation}
  \langle \cO \rangle_t = \sum_{\cC, \cC'} \cO(\cC \rightarrow \cC') m(\cC \rightarrow \cC') P_t(\cC),
 \end{equation}
 where we recall that $m(\cC \rightarrow \cC')$ denotes the probability rate from configuration $\cC$ to configuration $\cC'$. 
 It can be rewritten by introducing the matrix
 \begin{equation}
  O=\sum_{\cC} \cO(\cC \rightarrow \cC')m(\cC \rightarrow \cC') \ket{\cC'}\bra{\cC}
 \end{equation}
 in the form $\bra{\Sigma} O \exp (tM) \ket{P_0}$. We thus have the same result as for 'static' observables if $OM=MO$: the 
 average is constant in time.
\end{remark}

\begin{remark}
 Note that when $O$ commutes with $M$, they can be diagonalized in the same basis (if they are diagonalizable). In other words they have the 
 same eigenvectors.
\end{remark}

By analogy with classical Newton's dynamics, where it was shown that the presence of conserved quantities is a key property to integrate exactly
the equation of motion, we would like to construct Markovian models with 'a lot' of conserved observables. In other words we would like 
to construct a Markov matrix together with several independent matrices that all commute with each others. In the case of 
classical Newton's dynamics the number of independent conserved quantities required to exactly solve the equations of motion is precisely 
known, it depends on the dimension of the phase space. The situation is less clear in the case of Markovian processes. 
We are lacking a general theorem (as Liouville theorem) that ensures the exact solvability of the model, {\it i.e} the exact diagonalization
of its Markov matrix, if there are enough conserved quantities.

Nevertheless we will present a framework, taken from quantum integrability, which allows us to produce Markov matrices together with
a family of commuting operators, and which provides tools to diagonalize the Markov matrix in some specific cases.

\subsection{Exclusion processes framework}

\subsubsection{The configuration space}

The first thing to do, if we want to construct integrable Markov processes, is to define the configuration space of the process, 
see chapter \ref{chap:one}. 
The construction of a Markov matrix commuting with a lot of different operators is a very difficult task and requires to deal with
very particular configurations sets. This is the reason we will focus now on what is called {\it exclusion processes} which provide 
a fruitful framework to construct integrable Markov processes. 

These models are defined on a one dimensional lattice\footnote{Note that exclusion processes can be also defined on higher dimensional lattices but 
the one dimensional case reveals to be particularly fruitful in the integrability point of view.}
composed of $L$ sites that are denoted by the variable $i=1,2,\dots,L$. Particles of $N$ different species, labeled  
$1,2,\dots,N$, can evolve on this lattice. The term ``evolve'' is deliberately vague because we focus here on the configuration space of the model, 
we will present the precise stochastic dynamical rules in the next subsection \ref{subsec:Markov_matrix}. 
The particles are subject to a hard core constraint (exclusion principle), {\it i.e} each site is occupied by at most one particle.
Note that this constraint could be relaxed to allow several particles to be on the same site, see for instance the generalized exclusion
processes \cite{SchmittmannZ95,ChouMZ11,Schutz01}.
A site on the lattice can be in $N+1$ different states, depending on its content.
More precisely for each site $i$ we define a local configuration variable $\tau_i \in \{0,1,\dots,N\}$, where 
$\tau_i=0$ if the site $i$ is empty and $\tau_i=s\geq 1$ if the site $i$ is occupied by a particle of species $s$.
 The configuration of the system on the whole lattice are thus in one to one correspondence with the $L$-uplets 
 $(\tau_1,\dots, \tau_L) \in \left\{0,\dots,N\right\}^L$. There are $(N+1)^L$ configurations. 
 
Following the lines of definition \ref{def:Markov_chains_vector_space}, the goal now is to construct a vector space, with a well chosen vector basis,
in order to express the configuration probabilities and the master equation in a concise way.
To each configuration $\cC=(\tau_1,\dots,\tau_L)$ one can associate a basis vector $\ket{\tau_1,\dots,\tau_L}$. 
For such a purpose, we associate to a local configuration variable $\tau_i$ a 
basis vector $\ket{\tau_i}$ of $\CC^{N+1}$.
\begin{definition}
For $\tau=0,1,\dots,N$, the vector $\ket{\tau}$ is defined by 
\begin{equation}
 \ket{\tau}=(\underbrace{0,\dots,0}_{\tau},1,\underbrace{0,\dots,0}_{N-\tau})^t,
\end{equation}
where $.^t$ denotes the usual transposition.
The vectors $\ket{\tau}$, for $\tau=0,\dots,N$ constitute a canonical basis of $\CC^{N+1}$.
\end{definition}
These vectors $\ket{\tau_i}$ represent the building blocks of the vectors $\ket{\tau_1,\dots,\tau_L}$. The construction is 
inspired again by quantum spin chains, where the Hilbert space associated to the whole chain is obtained by tensor products of the 
Hilbert space associated to a single spin.
\begin{definition}
 For $0 \leq \tau_1,\dots,\tau_L \leq N$, the vector $\ket{\tau_1,\dots,\tau_L}$ is defined by
 \begin{equation}
  \ket{\tau_1,\dots,\tau_L}=\ket{\tau_1} \otimes \ket{\tau_2} \otimes \dots \otimes \ket{\tau_L}.
 \end{equation}
The vectors $\ket{\tau_1} \otimes \dots \otimes \ket{\tau_L}$, for $\tau_i=0,\dots,N$ constitute a canonical basis of
$\left(\CC^{N+1}\right)^{\otimes L}$.
\end{definition}
In order for the reader to be familiar with the tensor product notation and with
the conventions used in this manuscript, we give the following explicit examples.

\begin{example} \label{ex:tensor_product}
 \begin{equation}
  \begin{pmatrix}
   a_1 \\ a_2 \\ a_3
  \end{pmatrix} \otimes
  \begin{pmatrix}
   b_1 \\ b_2 \\ b_3
  \end{pmatrix}=
  \begin{pmatrix}
   a_1 b_1 \\ a_1 b_2 \\ a_1 b_3 \\ a_2 b_1 \\ a_2 b_2 \\ a_2 b_3 \\ a_3 b_1 \\ a_3 b_2 \\ a_3 b_3
  \end{pmatrix}
 \end{equation}
 and
 \begin{equation}
  \begin{pmatrix}
   a_1 \\ a_2
  \end{pmatrix} \otimes
  \begin{pmatrix}
   b_1 \\ b_2
  \end{pmatrix} \otimes
  \begin{pmatrix}
   c_1 \\ c_2 
  \end{pmatrix} =
  \begin{pmatrix}
   a_1 b_1 c_1 \\ a_1 b_1 c_2 \\ a_1 b_2 c_1 \\ a_1 b_2 c_2 \\ a_2 b_1 c_1 \\ a_2 b_1 c_2 \\ a_2 b_2 c_1 \\ a_2 b_2 c_2
  \end{pmatrix}
 \end{equation}
\end{example}
In words, the tensor product of two vectors of sizes $n$ and $m$ can be seen as a single vector of size $n \times m$ obtained by multiplying 
each entry of the left vector by the entire right vector. This procedure is used iteratively to construct tensor products of several vectors
as a single vector.  

We now show how to use this machinery to express in a very compact form the probabilities of the different configurations.
We recall that the probability for the system to be in configuration $\cC=(\tau_1,\dots,\tau_L)$ at time $t$ is denoted by 
$P_t(\tau_1,\dots,\tau_L)$.

\begin{definition}
We define a vector $\ket{P_t}$ that contains all the configuration probabilities
\begin{equation} \label{eq:probability_vector_tensor_decomposition}
 \ket{P_t} = \begin{pmatrix}
                P_t(0,\dots,0,0) \\
                P_t(0,\dots,0,1) \\
                \vdots \\
                P_t(N,\dots,N,N)
               \end{pmatrix}
             = \sum_{0\leq \tau_1,\dots,\tau_L \leq N} P_t(\tau_1,\dots,\tau_L)  \, \ket{\tau_1} \otimes \cdots \otimes \ket{\tau_L}.
\end{equation}
\end{definition}
The probabilities of the configurations are stored as the coefficient of a linear expansion.
The coefficient in front of a vector $\ket{\tau_1} \otimes \cdots \otimes \ket{\tau_L}$ is the probability of the corresponding configuration.
This reads formally like the expansion of the wave function of a quantum spin chain system on the Hilbert space basis (the only 
difference being that the probability of a configuration in the latter case is obtained by absolute value of the coefficient {\it squared}).

\begin{example}
The simplest example we can give of such construction is the case with a single species of particles, {\it i.e} for $N=1$. The local configuration
variables $\tau_i$ can only take two values in such models: $\tau_i=0,1$. We have explicitly
\begin{equation}
 \ket{0}=\begin{pmatrix}
          1 \\ 0
         \end{pmatrix}, \quad \mbox{and} \quad 
 \ket{1}=\begin{pmatrix}
          0 \\ 1
         \end{pmatrix},
\end{equation}
which span the vector space $\CC^2$. If we consider the particular case of the basis associated to two sites, we have then 
\begin{equation}
 \begin{pmatrix}
  1 \\ 0
 \end{pmatrix} \otimes
 \begin{pmatrix}
  1 \\ 0
 \end{pmatrix} =
 \begin{pmatrix}
  1 \\ 0 \\ 0 \\ 0
 \end{pmatrix}, \quad 
 \begin{pmatrix}
  1 \\ 0
 \end{pmatrix} \otimes
 \begin{pmatrix}
  0 \\ 1
 \end{pmatrix} =
 \begin{pmatrix}
  0 \\ 1 \\ 0 \\ 0
 \end{pmatrix}, \quad 
  \begin{pmatrix}
  0 \\ 1
 \end{pmatrix} \otimes
 \begin{pmatrix}
  1 \\ 0
 \end{pmatrix} =
 \begin{pmatrix}
  0 \\ 0 \\ 1 \\ 0
 \end{pmatrix}, \quad 
  \begin{pmatrix}
  0 \\ 1
 \end{pmatrix} \otimes
 \begin{pmatrix}
  0 \\ 1
 \end{pmatrix} =
 \begin{pmatrix}
  0 \\ 0 \\ 0 \\ 1
 \end{pmatrix}
\end{equation}
corresponding to empty lattice, one particle on the second site, one particle on the first site and the full lattice respectively.
\end{example}

Let us stress that, even though the exclusion condition can appear as a strong restriction
on the physical systems that can be well described within this framework, the models that can be constructed in this category display 
a rich physical phenomenology and capture the essential aspects of non-equilibrium systems \cite{Schutz01,KrapivskyRB10,ChouMZ11}. 
The framework of exclusion processes thus enjoy a two-fold interest: on physical side it provides relevant non-equilibrium models and 
on mathematical side if offers a good laboratory to investigate integrability and exact solvability.

\subsubsection{Markov matrix} \label{subsec:Markov_matrix}

The configurations set of the system being settled, we are now interested in the stochastic dynamics of the model. All the discussion of 
this subsection will only concern continuous time Markov processes. Nevertheless the machinery that will be developed for the construction
of integrable continuous time processes will also provide at the end examples of integrable discrete time processes, 
see \eqref{eq:TASEP_discrete_time_Markov} and \eqref{eq:TASEP_master_equation_discrete_time_open} for instance. Once again,
it is a hard problem to build a Markov matrix together with a set of commuting operators (which is a strong hint of exact solvability).
We have thus to restrict ourself to particular class of stochastic dynamics that has been revealed to be of great interest from the exact solvability 
point of view. We will be interested in stochastic dynamics allowing only for local configuration changes on the lattice. The particles can 
only jump to their direct neighbor sites, exchange or react only with particles on adjacent sites, and can also be created or annihilated 
locally on a site. The probability rates of such changes depend only on the local configurations on the direct neighbor sites. 
More precisely we are interested in dynamics encoded by a Markov matrix that can be formally decomposed as a sum of local operators, 
sometimes called local jump operators, acting non-trivially on two adjacent sites of the lattice (and trivially on the other sites):
\begin{equation} \label{eq:Markov_matrix_sum_decomposition}
 M=\sum_{i=1}^{L-1}m_{i,i+1},
\end{equation}
with $m_{i,i+1}$ a local jump operator acting on sites $i$ and $i+1$
\begin{equation}
 m_{i,i+1}= \underbrace{\id \otimes \dots \otimes \id}_{i-1} \otimes m \otimes \underbrace{\id \otimes \dots \otimes \id}_{L-i-1},
\end{equation}
where $m$ is a Markov matrix of size $(N+1)^2 \times (N+1)^2$ acting on two adjacent sites, 
{\it i.e} on the vector space $\CC^{N+1} \otimes \CC^{N+1}$ and $\id$ is the $(N+1)\times(N+1)$ identity matrix.
The matrix element $\bra{\upsilon} \otimes \bra{\upsilon'} \, m \, \ket{\tau} \otimes \ket{\tau'}$, with $(\upsilon,\upsilon') \neq (\tau,\tau')$,
is equal to the probability rate that the system jumps from configuration $(\tau_1,\dots,\tau_{i-1},\tau,\tau',\tau_{i+2},\dots,\tau_L)$ to 
configuration $(\tau_1,\dots,\tau_{i-1},\upsilon,\upsilon',\tau_{i+2},\dots,\tau_L)$. Note that this rate depends only on the local configurations
$(\tau,\tau')$ and $(\upsilon,\upsilon')$ and not on the states of the other sites.

This class of model can appear very restrictive at first sight but they revealed to be rich enough to encode the essential feature of non equilibrium
systems \cite{Schutz01,KrapivskyRB10,ChouMZ11}. To fix the ideas and show the physical relevance of this framework, we now present examples of such dynamics.
The models that we introduce will serve as recurrent examples to illustrate the different concepts exposed throughout this manuscript. 
Some of them are widely known in the literature. 

\begin{example}
 The Asymmetric Simple Exclusion Process (ASEP) \cite{Spitzer70,Liggett85} has become over the last decade a paradigmatic model 
 in out-of-equilibrium statistical physics \cite{Derrida98,ChouMZ11,BlytheE07}.
 It was first introduced in the context of biology and gave rise since then to a lot of interest in different fields.
 First of all it displays a very rich physical phenomenology with shock waves \cite{Ferrari91,FerrariKS91} and 
 boundary induced phase transitions \cite{Krug91,DerridaEHP93,SchutzD93}.
 Moreover, from the mathematical point of view, it gave rise to a lot of work in representation theory and combinatorics with the 
 connection to orthogonal polynomials \cite{UchiyamaSW04,CantiniDGW15}, but also in integrability and probability theory with the connection to the 
 Kardar-Parisi-Zhang equation \cite{BorodinFPS07,TracyW09}. 
 
 The model describes particles of a single species that can diffuse on the lattice. During time $dt$, a particle has a probability $p \times dt$ 
 (respectively $q \times dt$) to jump to its right (respectively left) neighbor site, provided that it is empty (this is the exclusion constraint,
 there is at most one particle per site). The asymmetry in the hopping rates mimics a driving force that tends to push the particle in one direction
 rather than in the other. Note that the ASEP is sometimes defined with a right hopping rate $p$ normalized to $1$ (it corresponds to perform 
 a rescaling of the time).
 
 In the vector basis $\ket{0} \otimes \ket{0}$, $\ket{0} \otimes \ket{1}$, $\ket{1} \otimes \ket{0}$, $\ket{1} \otimes \ket{1}$ (ordered this way),
 the local jump operator $m$ is written
 \begin{equation} \label{eq:ASEP_m}
  m = \begin{pmatrix}
       0 & 0 & 0 & 0 \\ 0 & -q & p & 0 \\ 0 & q & -p & 0 \\ 0 & 0 & 0 & 0
      \end{pmatrix}.
 \end{equation}
 The ASEP admits two limits that are of particular interest. They are presented in the following examples.
\end{example}

\begin{example}
 The limit $p=1$ and $q=0$ is called the Totally Asymmetric Simple Exclusion Process (TASEP). This model has been widely studied in the literature 
 because it displays roughly the same physical behavior as the ASEP but the computations often simplify
 drastically. We will observe it in the next section for the expression of the stationary state of the model.
 
 In the vector basis $\ket{0} \otimes \ket{0}$, $\ket{0} \otimes \ket{1}$, $\ket{1} \otimes \ket{0}$, $\ket{1} \otimes \ket{1}$ (ordered this way),
 the local jump operator $m$ is written
 \begin{equation} \label{eq:TASEP_m}
  m = \begin{pmatrix}
       0 & 0 & 0 & 0 \\ 0 & 0 & 1 & 0 \\ 0 & 0 & -1 & 0 \\ 0 & 0 & 0 & 0
      \end{pmatrix}.
 \end{equation}
\end{example}

\begin{example}
 The limit $p=1$ and $q=1$ is called the Symmetric Simple Exclusion Process (SSEP). There is no more driving force in the bulk, the particles have 
 the same probability rates to jump on the left or on the right. The model thus describes particles diffusing on the lattice with an hard-core 
 constraint. The physics is simpler than in the ASEP case, in the sense that the bulk dynamics tends to converge toward a thermodynamic equilibrium, 
 because there is no driving force. The system can only be driven out-of-equilibrium by interactions with particles reservoirs. We will 
 also observe that the computations of physical quantities are much simpler than for the ASEP.
 The local jump operator $m$ is written
  \begin{equation} \label{eq:SSEP_m}
  m = \begin{pmatrix}
       0 & 0 & 0 & 0 \\ 0 & -1 & 1 & 0 \\ 0 & 1 & -1 & 0 \\ 0 & 0 & 0 & 0
      \end{pmatrix}.
 \end{equation}
\end{example}

We observed in the previous examples that the framework is well suited to describe the (driven) diffusion of particle with hard-core repulsion.
But it is not limited to this kind of dynamics and can for instance deal with creation/annihilation of particles or reaction between particles.
The following example is a simple illustration of this fact.

\begin{example}
The Dissipative Symmetric Simple Exclusion Process (DiSSEP) describes particles diffusing in a symmetric way with exclusion constraint (similarly
to the SSEP). In addition to that, particle pairs are allowed to condensate or evaporate from the lattice with the same probability rate $\lambda^2$.
The local jump operator $m$ is written
  \begin{equation} \label{eq:DiSSEP_m}
  m = \begin{pmatrix}
       -\lambda^2 & 0 & 0 & \lambda^2 \\ 0 & -1 & 1 & 0 \\ 0 & 1 & -1 & 0 \\ \lambda^2 & 0 & 0 & -\lambda^2
      \end{pmatrix}.
 \end{equation}
\end{example}

We can also easily define models with several species of particles in interaction. The following example may be thought as one of 
the simplest case we can imagine.

\begin{example}
The $2$-species TASEP is a generalization of the TASEP obtained by adding a second species of particles. During an infinitesimal time $dt$,
a particle of first or second species (labeled respectively $1$ and $2$) can jump to the right with probability $dt$ provided that the 
neighbor site is empty. A particle of species $2$ can also overtake a particle of species $1$ with probability $dt$.
In the vector basis $\ket{0} \otimes \ket{0}$, $\ket{0} \otimes \ket{1}$, $\ket{0} \otimes \ket{2}$, 
$\ket{1} \otimes \ket{0}$, $\ket{1} \otimes \ket{1}$, $\ket{1} \otimes \ket{2}$, 
$\ket{2} \otimes \ket{0}$, $\ket{2} \otimes \ket{1}$, $\ket{2} \otimes \ket{2}$ (ordered this way), the local jump operator $m$ is written
 \begin{equation} \label{eq:2TASEP_m}
  m = \begin{pmatrix}
       \cdot & \cdot & \cdot & \cdot & \cdot & \cdot & \cdot & \cdot & \cdot \\
       \cdot & \cdot & \cdot & 1 & \cdot & \cdot & \cdot & \cdot & \cdot \\
       \cdot & \cdot & \cdot & \cdot & \cdot & \cdot & 1 & \cdot & \cdot \\
       \cdot & \cdot & \cdot & -1 & \cdot & \cdot & \cdot & \cdot & \cdot \\
       \cdot & \cdot & \cdot & \cdot & \cdot & \cdot & \cdot & \cdot & \cdot \\
       \cdot & \cdot & \cdot & \cdot & \cdot & \cdot & \cdot & 1 & \cdot \\
       \cdot & \cdot & \cdot & \cdot & \cdot & \cdot & -1 & \cdot & \cdot \\
       \cdot & \cdot & \cdot & \cdot & \cdot & \cdot & \cdot & -1 & \cdot \\
       \cdot & \cdot & \cdot & \cdot & \cdot & \cdot & \cdot & \cdot & \cdot 
      \end{pmatrix}.
 \end{equation}
\end{example}

In all the models presented, we only dealt with the stochastic dynamics in the bulk of the system, never mentioning what is
happening on the extremities of the lattice. The boundary conditions are of particular interest in out-of-equilibrium statistical physics because 
they can influence the macroscopic behavior of the whole system, even if the interactions at the boundaries are very short range. 
In contrast to equilibrium statistical physics, where the boundary effects are negligible in systems with short range interactions, the system can 
change phase due to boundary effects. We will focus, in what follows, on two kinds of boundary conditions: the periodic boundary condition and 
the open boundaries condition.

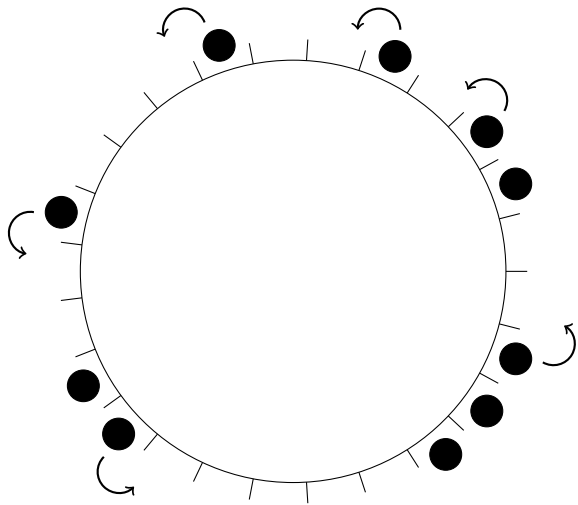
\begin{figure}[htb]
\begin{center}
 \begin{tikzpicture}[scale=0.7]
  \draw (0,0) circle (4);
  \draw (3.874332644, 0.9947595488) -- (4.261765909, 1.094235504);
  \draw   (3.505226720, 1.927014696)-- (3.855749392, 2.119716166);
  \draw   (2.915874509, 2.738188424)-- (3.207461960, 3.012007266);
  \draw   (2.143307180, 3.377311702)-- (2.357637898, 3.715042872);
  \draw   (1.236067975, 3.804226066)-- (1.359674773, 4.184648673);
  \draw  (0.2511620770, 3.992106914)-- (0.2762782847, 4.391317605);
  \draw (-0.7495252568, 3.929149003)-- (-0.8244777825, 4.322063904);
  \draw  (-1.703117167, 3.619308209)-- (-1.873428884, 3.981239030);
  \draw  (-2.549695959, 3.082052971)-- (-2.804665555, 3.390258268);
  \draw  (-3.236067977, 2.351141010)-- (-3.559674775, 2.586255111);
  \draw  (-3.719105944, 1.472498211)-- (-4.091016538, 1.619748032);
  \draw (-3.968458805, 0.5013329344)-- (-4.365304686, 0.5514662278);
  \draw(-3.968458805, -0.5013329344)-- (-4.365304686, -0.5514662278);
  \draw (-3.719105944, -1.472498211)-- (-4.091016538, -1.619748032);
  \draw (-3.236067977, -2.351141010)-- (-3.559674775, -2.586255111);
  \draw (-2.549695959, -3.082052971)-- (-2.804665555, -3.390258268);
  \draw (-1.703117167, -3.619308209)-- (-1.873428884, -3.981239030);
  \draw(-0.7495252568, -3.929149003)-- (-0.8244777825, -4.322063904);
  \draw (0.2511620770, -3.992106914)-- (0.2762782847, -4.391317605);
  \draw  (1.236067975, -3.804226066)-- (1.359674773, -4.184648673);
  \draw  (2.143307180, -3.377311702)-- (2.357637898, -3.715042872);
  \draw  (2.915874509, -2.738188424)-- (3.207461960, -3.012007266);
  \draw  (3.505226720, -1.927014696)-- (3.855749392, -2.119716166);
  \draw (3.874332644, -0.9947595488)-- (4.261765909, -1.094235504);
  \draw                   (4., 0.)-- (4.4, 0.);
   \draw               (4.183994187, 1.656560487) circle (0.3) [fill,circle] {};
   \draw               (3.640576474, 2.645033636) circle (0.3) [fill,circle] {};
   \draw               (1.916006809, 4.071721738) circle (0.3) [fill,circle] {};
   \draw               (-1.390576474, 4.279754323) circle (0.3) [fill,circle] {};
   \draw               (-4.358624226, 1.119104489) circle (0.3) [fill,circle] {};
   \draw               (-3.943380058, -2.167891537) circle (0.3) [fill,circle] {};
   \draw               (-3.280358822, -3.080461978) circle (0.3) [fill,circle] {};
   \draw               (2.868407953, -3.467309593) circle (0.3) [fill,circle] {};
   \draw               (3.640576473, -2.645033637) circle (0.3) [fill,circle] {};
   \draw               (4.183994185, -1.656560490) circle (0.3) [fill,circle] {};
    \draw[->,thick]              (3.969951993, 3.039651488) arc (-30:150:0.4);
    \draw[->,thick]               (2.014532176, 4.576205864) arc (0:180:0.4);
    \draw[->,thick]              (-1.664097726, 4.714952678) arc (20:200:0.4);
    \draw[->,thick]              (-4.872634364, 1.121353802) arc (80:260:0.4);
    \draw[->,thick]              (-3.557678385, -3.513249850) arc (135:315:0.4);
    \draw[->,thick]              (4.693669292, -1.723214606) arc (240:420:0.4);
 \end{tikzpicture}
 \end{center}
 \caption{Dynamical rules of the periodic TASEP.}
 \label{fig:TASEP_periodic}
\end{figure}

The periodic boundary condition describes a lattice with a ring shape. The last site $L$ and the first site $1$ are neighbors:
the first site $1$ plays the role of a site $L+1$ (see figure \ref{fig:TASEP_periodic}). The Markov matrix is formally written, in this case,
\begin{equation} \label{eq:periodic_Markov_matrix_sum_decomposition}
 M=\sum_{i=1}^{L}m_{i,i+1}.
\end{equation}
In this equation, it is understood that $L+1 \equiv 1$.

The open boundaries condition describes a lattice that is coupled with particle reservoirs at its extremities: at the first site, $1$, and 
at the last site, $L$. Particles of different species can be injected, extracted or exchanged at these two extremities with a probability
rate that depends only on the content of the first (respectively last) site, and not on the local configurations on the other sites. In this case
the Markov matrix is expressed as
\begin{equation} \label{eq:open_Markov_matrix_sum_decomposition}
 M=B_1+\sum_{i=1}^{L-1}m_{i,i+1} +\overline{B}_L,
\end{equation}
with 
\begin{equation}
 B_1 = B \otimes \underbrace{\id \otimes \dots \otimes \id}_{L-1}, \quad \mbox{and} \quad 
 \overline{B}_L = \underbrace{\id \otimes \dots \otimes \id}_{L-1} \otimes \overline{B},
\end{equation}
where $B$ and $\overline{B}$ are Markov matrices of size $(N+1) \times (N+1)$ acting on a single site, 
{\it i.e} on the vector space $\CC^{N+1}$. 
The matrix element $\bra{\upsilon} \, B \, \ket{\tau}$ (respectively $\bra{\upsilon} \, \overline{B} \, \ket{\tau}$), with $\upsilon \neq \tau$,
is equal to the probability rate that the system jumps from configuration $(\tau,\tau_2,\dots,\tau_L)$ to 
configuration $(\upsilon,\tau_2,\dots,\tau_L)$ (respectively from $(\tau_1,\dots,\tau_{L-1},\tau)$ to $(\tau_1,\dots,\tau_{L-1},\upsilon)$).
Note that this rate depends only on the local configurations $\tau$ and $\upsilon$ and not on the states of the other sites.

\begin{example}
In the ASEP, we can define the following dynamics at the boundaries. During an infinitesimal time $dt$, a particle located on site $1$ can be 
absorbed by the left reservoir with probability $\gamma \times dt$, and conversely the left reservoir can inject a particle on the first site, if 
it is empty, with probability $\alpha \times dt$. Similarly on the site $L$, the right reservoir can inject a particle (provided that the site is 
empty) with probability $\delta \times dt$ and remove a particle with probability $\beta \times dt$. In the basis $\ket{0}$, $\ket{1}$ (ordered this 
way) the boundary local jump operators $B$ and $\overline{B}$ thus are written 
 \begin{equation} \label{eq:ASEP_B_Bb}
  B= \begin{pmatrix}
      -\alpha & \gamma \\
      \alpha & -\gamma
     \end{pmatrix} \quad \mbox{and} \quad 
  \overline{B}= \begin{pmatrix}
      -\delta & \beta \\
      \delta & -\beta
     \end{pmatrix}   
 \end{equation}
 Note that these boundary local jump operators are also suitable to encode the dynamics with the reservoirs in the case of the SSEP and of the DiSSEP.
\end{example}

\begin{example}
In the TASEP case, the dynamics of the system at the boundaries should not be in opposition with the dynamics in the bulk, where particles are only
allowed to jump to the right. In other words, the left reservoir can only inject particles on the lattice and the right reservoir can only 
remove particle from the lattice. This translates into the fact that the boundary rates $\gamma$ and $\delta$ introduced for the ASEP are now 
vanishing. The boundary local jump operators thus are written
 \begin{equation} \label{eq:TASEP_B_Bb}
  B= \begin{pmatrix}
      -\alpha & 0 \\
      \alpha & 0
     \end{pmatrix} \quad \mbox{and} \quad 
  \overline{B}= \begin{pmatrix}
      0 & \beta \\
      0 & -\beta
     \end{pmatrix}   
 \end{equation}
\end{example}

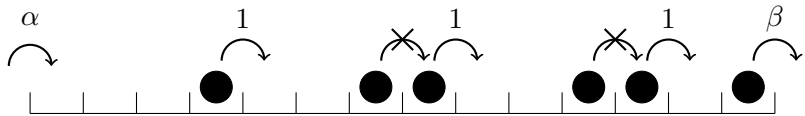
\begin{figure}[htb]
\begin{center}
 \begin{tikzpicture}[scale=0.7]
\draw (-2,0) -- (12,0) ;
\foreach \i in {-2,-1,...,12}
{\draw (\i,0) -- (\i,0.4) ;}
\draw[->,thick] (-2.4,0.9) arc (180:0:0.4) ; \node at (-2.,1.8) [] {$\alpha$};
\draw  (1.5,0.5) circle (0.3) [fill,circle] {};
\draw  (4.5,0.5) circle (0.3) [fill,circle] {};
\draw  (5.5,0.5) circle (0.3) [fill,circle] {};
\draw  (11.5,0.5) circle (0.3) [fill,circle] {};
\draw  (8.5,0.5) circle (0.3) [fill,circle] {};
\draw  (9.5,0.5) circle (0.3) [fill,circle] {};
\draw[->,thick] (1.6,1) arc (180:0:0.4); \node at (2.,1.8) [] {$1$};
\draw[->,thick] (5.6,1) arc (180:0:0.4); \node at (6.,1.8) [] {$1$};
\draw[->,thick] (9.6,1) arc (180:0:0.4); \node at (10.,1.8) [] {$1$};
\draw[->,thick] (4.6,1) arc (180:0:0.4); \draw[thick] (4.8,1.2) -- (5.2,1.6) ;\draw[thick] (4.8,1.6) -- (5.2,1.2) ;
\draw[->,thick] (8.6,1) arc (180:0:0.4); \draw[thick] (8.8,1.2) -- (9.2,1.6) ;\draw[thick] (8.8,1.6) -- (9.2,1.2) ;
\draw[->,thick] (11.6,1) arc (180:0:0.4) ; \node at (12.,1.8) [] {$\beta$};
 \end{tikzpicture}
 \end{center}
 \caption{Dynamical rules of the open TASEP.}
 \label{fig:TASEP_open}
\end{figure}

We know give an example of boundary jump operators for a multi-species ({\it i.e} $N>1$) model.
\begin{example}
For the $2$-species TASEP, we do not want the dynamics on the boundaries to be opposite with respect to the dynamics in the bulk
(similarly to the single species TASEP). In other words,
the left reservoir can only inject particles of both species on the first site (if it is empty) and exchange a particle of species $1$ with 
a particle of species $2$, in agreement with the overtake rule in the bulk. Similarly the right reservoir can only remove particles from 
the last site or exchange particle of species $2$ with particles of species $1$. An illustration of very specific boundary local jump operators 
is given (in the basis $\ket{0}$, $\ket{1}$, $\ket{2}$) by 
 \begin{equation} \label{eq:2TASEP_B_Bb_example}
  B= \begin{pmatrix}
      -1 & \cdot & \cdot \\
      1-\alpha & -\alpha & \cdot \\
      \alpha & \alpha & \cdot 
     \end{pmatrix} \quad \mbox{and} \quad 
  \overline{B}= \begin{pmatrix}
      \cdot & \beta & \beta \\
      \cdot & -\beta & \cdot \\
      \cdot & \cdot & -\beta 
     \end{pmatrix}   
 \end{equation}
\end{example}

We have presented so far the physical framework that we will be interested in throughout this manuscript.
We also introduced the main mathematical tools, which we use to define precisely the Markov matrices
encoding the stochastic dynamics of the physical systems. 

We will now be interested in exactly solvable models. We saw at the beginning of this chapter that the existence of ``many'' conserved 
quantities is a strong hint of exact solvability.
We see below a systematic way to produce Markov matrices belonging to a set of commuting operators. This construction provides also 
(at least in specific cases) tools to diagonalize exactly the Markov matrix.
       
\section{Integrability for periodic boundary conditions}

We start to present the construction of integrable Markov matrices with the case of periodic boundary conditions. 
The method relies on the construction of a set of commuting operators containing the Markov matrix. This set is conveniently 
generated by a {\it transfer matrix}, which depends on a variable called spectral parameter. The Markov matrix is recovered 
from the transfer matrix by taking the derivative with respect to the spectral parameter. The essential feature of this transfer 
matrix is that it commutes for different values of the spectral parameter. Similarly to the Markov matrix, which is defined in terms of local
jump operators, the transfer matrix will be defined with the help of operators, called $R$ matrices, which also act locally. 
The $R$ matrix can be thought as a spectral parameter upgrading of the local jump operators $m$.
Moreover the commutation property of the transfer matrix is a direct consequence of a local property satisfied by the $R$-matrices:
the Yang-Baxter equation.

\subsection{R-matrix and transfer matrix}

\subsubsection{R-matrix and Yang-Baxter equation}

We introduce in this paragraph the key object of integrability: the $R$-matrix. It is the building block of the transfer matrix, which 
generates the Markov matrix together with a set of commuting operators. It is also directly related to the bulk local jump operators $m$ and 
thus appears as the guarantee of the integrability of the local stochastic rules of the model.

\begin{definition}
A matrix $R(z,z')$ of size $(N+1)^2 \times (N+1)^2$, i.e acting on $\CC^{N+1} \otimes \CC^{N+1}$, satisfies the Yang-Baxter equation if 
\begin{equation} \label{eq:Yang_Baxter_2spectralparameters}
 R_{1,2}(z_1,z_2) \, R_{1,3}(z_1,z_3) \, R_{2,3}(z_2,z_3) =
 R_{2,3}(z_2,z_3) \, R_{1,3}(z_1,z_3) \, R_{1,2}(z_1,z_2).
\end{equation}
\end{definition}
The Yang-Baxter equation states an equality between products of operators acting on the vector space $\CC^{N+1} \otimes \CC^{N+1} \otimes \CC^{N+1}$.
The subscript indices indicate on which copies of $\CC^{N+1}$ the operators are acting non trivially. For instance 
\begin{equation}
 R_{1,2}(z,z')=R(z,z') \otimes \id, \qquad R_{2,3}(z,z')=\id \otimes R(z,z'), \qquad \dots 
\end{equation}

\begin{figure}[htb]
\begin{center}
 \begin{tikzpicture}[scale=0.7]
 \draw  (-2,0) -- (2,0) ;
 \draw[->]  (-2,0) -- (-1,0) ;
 \draw  (0,-2) -- (0,2) ;
 \draw[->]  (0,-2) -- (0,-1) ;
 \node at (-2.5,0) [] {$i$};
 \node at (0,-2.5) [] {$j$};
 \node at (-1,0.5) [] {$z_i$};
 \node at (0.5,-1) [] {$z_j$};
 \end{tikzpicture}
 \end{center}
 \caption{Graphical representation of the matrix $R_{ij}(z_i,z_j)$.}
 \label{fig:Rmatrix}
\end{figure}
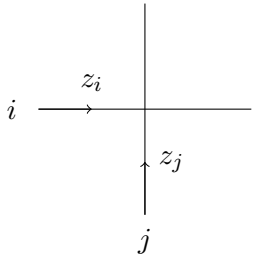

We would like to give a pictorial representation of the Yang-Baxter equation.
The action of the $R$-matrix $R_{ij}(z_i,z_j)$ can be represented graphically in figure \ref{fig:Rmatrix}.
The $R$-matrix is drawn as a vertex. This vertex is defined by two lines labeled $i$ and $j$, which correspond to the tensor space component
numbers $i$ and $j$ respectively. Each line is oriented by an arrow and carry a spectral parameter. The incoming half line (according 
to the arrow direction) labeled with $i$ stands for a vector $\ket{\tau}$ of the $i$-th tensor space component, and can thus 
be in $N+1$ different states. Similarly for the incoming half line labeled with $j$, which stands for a vector $\ket{\tau'}$.
The out-going half lines stand for the vectors $\bra{\upsilon}$, respectively $\bra{\upsilon'}$, which belong to $i$-th, respectively $j$-th 
tensor space components. When the vectors $\ket{\tau}$, $\ket{\tau'}$ and $\bra{\upsilon}$, $\bra{\upsilon'}$ are specified, 
the vertex represents the matrix element $\bra{\upsilon}_i \bra{\upsilon'}_j R_{ij}(z_i,z_j)\ket{\tau}_i\ket{\tau'}_j$.

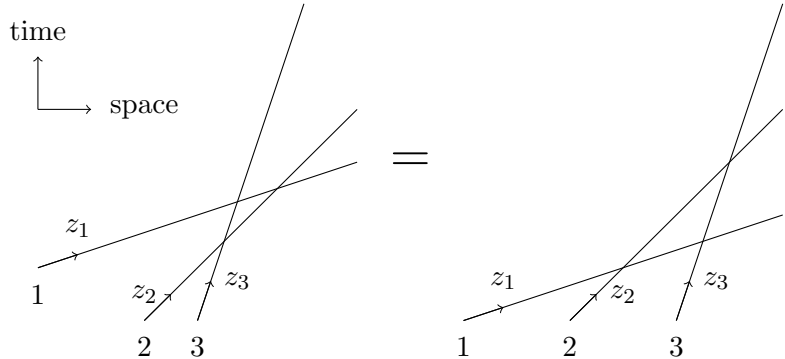
\begin{figure}[htb]
\begin{center}
 \begin{tikzpicture}[scale=0.7]
 \draw (-3+8,-3) -- (-1+8,3) ; \draw[->] (-3+8,-3) -- (-2.75+8,-2.25) ; \node at (-2.25+8,-2.25) [] {$z_3$}; \node at (-3+8,-3.5) [] {$3$};
 \draw (-5+8,-3) -- (-1+8,1) ; \draw[->] (-5+8,-3) -- (-4.5+8,-2.5) ; \node at (-4+8,-2.5) [] {$z_2$}; \node at (-5+8,-3.5) [] {$2$};
 \draw (-7+8,-3) -- (-1+8,-1) ; \draw[->] (-7+8,-3) -- (-6.25+8,-2.75) ; \node at (-6.25+8,-2.25) [] {$z_1$}; \node at (-7+8,-3.5) [] {$1$};
 \node at (0,0) [] {\huge{$=$}};
 \draw (4-8,-3) -- (6-8,3) ; \draw[->] (4-8,-3) -- (4.25-8,-2.25) ; \node at (4.75-8,-2.25) [] {$z_3$}; \node at (4-8,-3.5) [] {$3$};
 \draw (3-8,-3) -- (7-8,1) ; \draw[->] (3-8,-3) -- (3.5-8,-2.5) ; \node at (3-8,-2.5) [] {$z_2$}; \node at (3-8,-3.5) [] {$2$};
 \draw (1-8,-2) -- (7-8,0) ; \draw[->] (1-8,-2) -- (1.75-8,-1.75) ; \node at (1.75-8,-1.25) [] {$z_1$}; \node at (1-8,-2.5) [] {$1$};
 \draw[->] (-7,1) -- (-7,2); \node at (-7,2.5) [] {time};
 \draw[->] (-7,1) -- (-6,1); \node at (-5,1) [] {space};
 \end{tikzpicture}
 \end{center}
 \caption{Graphical representation of the Yang-Baxter equation.}
 \label{fig:Yang_Baxter}
\end{figure}

With this graphical interpretation, we will be able to compute efficiently a matrix element of a product of $R$ matrices acting in different
components of the tensor product. It also provides a meaningful illustration of the Yang-Baxter equation, see figure \ref{fig:Yang_Baxter}.
This pictorial representation of the $R$-matrix had been widely used in the context of two-dimensional vertex models in equilibrium
statistical physics \cite{Baxter82}.

The Yang-Baxter equation has a nice intuitive interpretation (see figure \ref{fig:Yang_Baxter}) coming from integrable quantum field theory. In this context
the $R$-matrix $R(z_1,z_2)$ is the diffusion matrix between two particles with rapidities $z_1$ and $z_2$. 
The integrability is the fact that the diffusion matrix of 3 particles factorizes in this R-matrix ({\it i.e} the scattering of three particles 
can be decomposed into two-particle scatterings). The Yang-Baxter equation is the consistency relation of this factorization, which ensures 
the independence of the result with respect to the order of the two-particles scatterings.

From a more algebraic side, the $R$-matrix can be thought as a convenient object to encode the commutation relations of an algebra 
(see chapter \ref{chap:three} for instance with the Zamolodchikov-Faddeev relation). The Yang-Baxter relation appears in this case as a 
consistency relation ensuring the associativity of the algebra.

We would like now to relate the $R$-matrix to the bulk local jump operator $m$. To do so, we need one more definition.

\begin{definition}
 For $R(z,z')$ satisfying the Yang-Baxter equation, we introduce the braided $R$-matrix 
 \begin{equation}
  \check R(z,z') = P.R(z,z'),
 \end{equation}
 where $P$ is the permutation operator in $\CC^{N+1} \otimes \CC^{N+1}$ that is $P \ket{\tau} \otimes \ket{\tau'} = \ket{\tau'} \otimes \ket{\tau}$,
 for all $\tau,\tau'=0,\dots,N$.
  
 Note that the braided matrix $\check R(z,z')$ satisfies the braided Yang-Baxter equation
 \begin{equation} \label{eq:braided_ybe}
  \check R_{12}(z_1,z_2) \check R_{23}(z_1,z_3)\check R_{12}(z_2,z_3)=\check R_{23}(z_2,z_3)\check R_{12}(z_1,z_3)\check R_{23}(z_1,z_2)\;. 
 \end{equation}
\end{definition}

We are now equipped to state one of the key points of this subsection. It provides a precise definition of integrability for a bulk local 
jump operator $m$ and relates it to the $R$-matrix.

\begin{definition}
 A bulk local jump operator $m$ is said to be integrable if there exists an $R$-matrix $R(z,z')$ satisfying the Yang-Baxter equation 
 \eqref{eq:Yang_Baxter_2spectralparameters}, a constant $\theta$ and a complex number $z'$ such that
 \begin{equation} \label{eq:derivative_bulk_local_jump}
  m = \theta \left. \frac{\partial}{\partial z} \check R(z,z') \right|_{z=z'}.
 \end{equation}
\end{definition}
In other words, the integrable local jump operators $m$ are obtained by taking the derivative of a braided $R$-matrix with respect to a spectral parameter.
Conversely, we could wonder whether it is possible, starting from a local jump operator $m$, to upgrade it to a spectral parameter dependent $R$-matrix.
This will be partially answered in subsection \ref{subsubsec:baxterisation}.

We would like to stress that taking the derivative of a braided $R$-matrix does not provide always a local Markovian matrix. The 
sum of the entries of each column of the local Markovian matrix should indeed vanish. It is straightforward to check that if the 
sum of the entries of each column of the $R$-matrix is equal to $1$ then the derivative enjoys the sum to $0$ property (but we still 
have to check that the off-diagonal entries of the derivative are non-negative).
This motivates the following definition
\begin{definition}
 A matrix $R(z,z')$ acting on $\CC^{N+1} \otimes \CC^{N+1}$ satisfies the Markovian property if
 \begin{equation}
  \bra{\sigma} \otimes \bra{\sigma} R(z,z') = \bra{\sigma} \otimes \bra{\sigma}
 \end{equation}
 where $\bra{\sigma}$ defined as $\bra{\sigma}=\sum_{\upsilon=0}^N \bra{\upsilon}$ stands for the sum over all the local configurations on one site.
\end{definition}
Note that such a $R$-matrix satisfies the requirement of a discrete time Markovian process, provided that its entries are non-negative.
We will see below that it can indeed be used as the building block of discrete time Markov matrices defined on the whole lattice. 

We now list a set of properties that will be always satisfied by the $R$-matrices we will consider. When defining the transfer matrix below, 
these properties will be essential to ensure its commutation relation and its link with the Markov matrix of the model.
\begin{definition}
 A matrix $R(z,z')$ acting on $\CC^{N+1} \otimes \CC^{N+1}$ satisfies the regularity property if
 \begin{equation} \label{eq:regularity_2spectralparameters}
  R(z,z)=P,
 \end{equation}
 where $P$ is the permutation operator.
\end{definition}

The permutation operator $P$ allows us to define the matrix $R_{21}(z,z'):=P.R_{12}(z,z').P$. We will often observe in the following that 
this matrix is closely related to the inverse of the $R$-matrix. The following definition specifies this connection.

\begin{definition}
  A matrix $R(z,z')$ acting on $\CC^{N+1} \otimes \CC^{N+1}$ satisfies the unitarity property if
 \begin{equation} \label{eq:unitarity_2spectralparameters}
  R_{12}(z,z').R_{21}(z',z)=\id.
 \end{equation}
 Note that for a braided $R$-matrix, it reads $\check R(z,z'). \check R(z',z)=\id$.
\end{definition}

In most of the known examples, the dependence of the $R$-matrix on the two spectral parameters $z$ and $z'$ is simpler than expected, in the 
sense that it depends only on the ratio $z/z'$ or on the difference $z-z'$. This motivates the following definitions.

\begin{definition}
A matrix $R(z,z')$ is said to be \textup{multiplicative} in the spectral parameters if
\begin{equation}
 R(z,z')=R\left(\frac{z}{z'}\right).
\end{equation}
It is straightforward to simplify the properties introduced above in this case:
\begin{itemize}
\item The Yang-Baxter equation becomes
\begin{equation} \label{eq:Yang_Baxter}
 R_{1,2}\left(\frac{z_1}{z_2}\right) \, R_{1,3}\left(\frac{z_1}{z_3}\right) \, R_{2,3}\left(\frac{z_2}{z_3}\right) =
 R_{2,3}\left(\frac{z_2}{z_3}\right) \, R_{1,3}\left(\frac{z_1}{z_3}\right) \, R_{1,2}\left(\frac{z_1}{z_2}\right).
\end{equation}
\item The regularity property becomes $R(1)=P$.
\item The unitarity property becomes $R_{12}(z).R_{21}(1/z)= \id$
\item The Markovian property becomes $\bra{\sigma} \otimes \bra{\sigma} R(z) = \bra{\sigma} \otimes \bra{\sigma}$.
\end{itemize}
\end{definition}

\begin{definition}
A matrix $R(z,z')$ is said to be \textup{additive} in the spectral parameters if
\begin{equation}
 R(z,z')=R(z-z').
\end{equation}
It is straightforward to simplify the properties introduced above in this case:
\begin{itemize}
\item The Yang-Baxter equation becomes
\begin{equation} \label{eq:Yang_Baxter_additif}
 R_{1,2}(z_1-z_2) \, R_{1,3}(z_1-z_3) \, R_{2,3}(z_2-z_3) =
 R_{2,3}(z_2-z_3) \, R_{1,3}(z_1-z_3) \, R_{1,2}(z_1-z_2).
\end{equation}
\item The regularity property becomes $R(0)=P$
\item The unitarity property becomes $R_{12}(z).R_{21}(-z)= \id$
\item The Markovian property becomes $\bra{\sigma} \otimes \bra{\sigma} R(z) = \bra{\sigma} \otimes \bra{\sigma}$.
\end{itemize}
\end{definition}

We now provide examples of such $R$-matrices, more particularly those related to the stochastic models already introduced in this manuscript.

\begin{example}
 The $R$-matrix related to the ASEP is multiplicative in the spectral parameters and is given by
 \begin{equation} \label{eq:ASEP_R}
  R(z) = \begin{pmatrix}
          1 & 0 & 0 & 0 \\ 
          0 & \frac{(1-z)q}{p-qz} & \frac{z(p-q)}{p-qz} & 0 \\
          0 & \frac{p-q}{p-qz} & \frac{(1-z)p}{p-qz} & 0 \\
          0 & 0 & 0 & 1
         \end{pmatrix}
 \end{equation}
 This matrix satisfies the Yang-Baxter equation \eqref{eq:Yang_Baxter}, the regularity, unitarity and Markovian properties. 
 The link with the bulk local jump operator \eqref{eq:ASEP_m} is given by
 $(q-p)\check R'(1)=m$, i.e it corresponds to a value $\theta = q-p$.
\end{example}

\begin{example}
 For the TASEP, the $R$ matrix is obtained by taking the limit $q=0$ and $p=1$ on the matrix \eqref{eq:ASEP_R}. It yields to the simple expression 
  \begin{equation} \label{eq:TASEP_R}
  R(z) = \begin{pmatrix}
          1 & 0 & 0 & 0 \\ 
          0 & 0 & z & 0 \\
          0 & 1 & 1-z & 0 \\
          0 & 0 & 0 & 1
         \end{pmatrix}
 \end{equation}
 This matrix satisfies the Yang-Baxter equation \eqref{eq:Yang_Baxter}, the regularity, unitarity and Markovian properties. 
 The link with the bulk local jump operator \eqref{eq:TASEP_m} is given by
 $-\check R'(1)=m$, i.e $\theta=-1$. 
 
 The $R$-matrix is graphically represented, as explained previously at the beginning of this subsection, 
 in figure \ref{fig:TASEP_matrix_elements} 
 (in the figure the missing vertices correspond to vanishing matrix elements of the $R$-matrix of the TASEP).
In this particular case of single species models, {\it i.e} for $N=1$, the graphical interpretation can be specified as follows.
A dashed incoming line (according to the arrow direction) denotes the vector $|0\rangle$ (or equivalently an empty site),
whereas a continuous thick line denotes the vector $|1\rangle$ (or equivalently an occupied site).
In a similar way, the out-going lines (after the crossing point) represent the states $\langle0|$ and $\langle1|$ on which we are contracting 
the matrix $R$.
Note that, in models where the number of particles is conserved by the dynamics, the number of incoming continuous thick lines is equal to
the number of out-going continuous thick lines.
 
 \begin{figure}[htb]
\begin{center}
 \begin{tikzpicture}[scale=0.7]
\node at (-4,2) [thick] {$\langle00|R_{ij}(\frac{z_1}{z_2})|00\rangle$};
\node at (0,2) [thick] {$\langle10|R_{ij}(\frac{z_1}{z_2})|10\rangle$};
\node at (4,2) [thick] {$\langle01|R_{ij}(\frac{z_1}{z_2})|10\rangle$};
\node at (8,2) [thick] {$\langle10|R_{ij}(\frac{z_1}{z_2})|01\rangle$};
\node at (12,2) [thick] {$\langle11|R_{ij}(\frac{z_1}{z_2})|11\rangle$};
\foreach \i in {-4,0,8}
{\draw[dashed] (\i,0) -- (\i,1) ;}
\foreach \i in {-4,4}
{\draw[dashed] (\i,0) -- (\i+1,0) ;}
\foreach \i in {4,12}
{\draw[ultra thick] (\i,0) -- (\i,1) ;}
\foreach \i in {0,8,12}
{\draw[ultra thick] (\i,0) -- (\i+1,0) ;}
\foreach \i in {-4,0,4}
{\draw[->,dashed] (\i,-1) -- (\i,-0.5) ; \draw[dashed] (\i,-0.5) -- (\i,0) ;}
\foreach \i in {-4,8}
{\draw[->,dashed] (\i-1,0) -- (\i-0.5,0) ; \draw[dashed] (\i-0.5,0) -- (\i,0) ;}
\foreach \i in {8,12}
{\draw[->, ultra thick] (\i,-1) -- (\i,-0.5) ; \draw[ultra thick] (\i,-0.5) -- (\i,0) ;}
\foreach \i in {0,4,12}
{\draw[->, ultra thick] (\i-1,0) -- (\i-0.5,0) ; \draw[ultra thick] (\i-0.5,0) -- (\i,0) ;}
\foreach \i in {-4,0,4,8,12}
{\node at (\i-1.25,0) [] {\footnotesize{$i$}};\node at (\i,-1.25) [] {\footnotesize{$j$}};
\node at (\i-0.5,0.3) [] {\footnotesize{$z_1$}};\node at (\i +0.3,-0.5) [] {\footnotesize{$z_2$}};}
\node at (-4,-2.5) [thick] {\Large{$1$}};
\node at (0,-2.5) [thick] {\Large{$1-\frac{z_1}{z_2}$}};
\node at (4,-2.5) [thick] {\Large{$\frac{z_1}{z_2}$}};
\node at (8,-2.5) [thick] {\Large{$1$}};
\node at (12,-2.5) [thick] {\Large{$1$}};
 \end{tikzpicture}
 \end{center}
\caption{Non-vanishing vertices associated with the $R$-matrix of the TASEP. \label{fig:TASEP_matrix_elements}}
\end{figure}
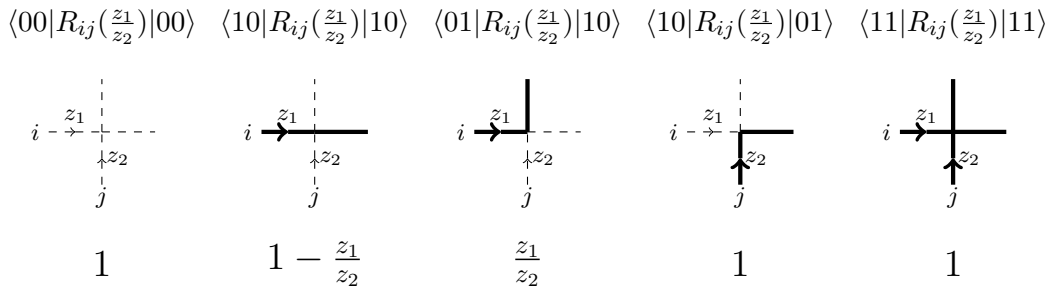
\end{example}

\begin{example}
 For the SSEP, the $R$-matrix is additive in the spectral parameters and is given by
  \begin{equation} \label{eq:SSEP_R}
  R(z) = \begin{pmatrix}
          1 & 0 & 0 & 0 \\ 
          0 & \frac{z}{z+1} & \frac{1}{z+1} & 0 \\
          0 & \frac{1}{z+1} & \frac{z}{z+1} & 0 \\
          0 & 0 & 0 & 1
         \end{pmatrix}
 \end{equation}
  This matrix satisfies the Yang-Baxter equation \eqref{eq:Yang_Baxter_additif}, the regularity, unitarity and Markovian properties. 
 The link with the bulk local jump operator \eqref{eq:SSEP_m} is given by
 $\check R'(0)=m$, i.e $\theta=1$. Note that the $R$-matrix can be written concisely as 
 \begin{equation}
 R(z)= \frac{z+P}{z+1},
 \end{equation}
 where $P$ is the permutation operator. This writing leads to a straightforward generalization to the multi-species case (see chapter \ref{chap:three}).
 The $R$-matrix of the SSEP can also be obtained from that of the ASEP by taking the limit $p=q=1$. This limit has to be 
 carefully taken (if not we obtain the permutation matrix) by introducing the following scaling
 \begin{equation}
  R^{SSEP}(z) = \lim\limits_{h \rightarrow 0} R^{ASEP}(e^{hz})|_{q=pe^{h}}.
 \end{equation}
 This scaling transforms a multiplicative dependence in the spectral parameter into an additive one, as expected.
\end{example}

All the previous examples were related to single species models. We provide here an example of an $R$-matrix related to a multi-species model.

\begin{example}
 The $R$-matrix of the $2$-species TASEP is multiplicative in the spectral parameter and is given by
  \begin{equation} \label{eq:2TASEP_R}
  R(z) = \begin{pmatrix}
       1 & \cdot & \cdot & \cdot & \cdot & \cdot & \cdot & \cdot & \cdot \\
       \cdot & \cdot & \cdot & z & \cdot & \cdot & \cdot & \cdot & \cdot \\
       \cdot & \cdot & \cdot & \cdot & \cdot & \cdot & z & \cdot & \cdot \\
       \cdot & 1 & \cdot & 1-z & \cdot & \cdot & \cdot & \cdot & \cdot \\
       \cdot & \cdot & \cdot & \cdot & 1 & \cdot & \cdot & \cdot & \cdot \\
       \cdot & \cdot & \cdot & \cdot & \cdot & \cdot & \cdot & z & \cdot \\
       \cdot & \cdot & 1 & \cdot & \cdot & \cdot & 1-z & \cdot & \cdot \\
       \cdot & \cdot & \cdot & \cdot & \cdot & 1 & \cdot & 1-z & \cdot \\
       \cdot & \cdot & \cdot & \cdot & \cdot & \cdot & \cdot & \cdot & 1 
      \end{pmatrix}
 \end{equation}
 This matrix satisfies the Yang-Baxter equation \eqref{eq:Yang_Baxter}, the regularity, unitarity and Markovian properties. 
 The link with the bulk local jump operator \eqref{eq:2TASEP_m} is given by
 $-\check R'(1)=m$, i.e $\theta=-1$.
\end{example}

\subsubsection{Transfer matrix}

We defined previously the integrability of a local jump operator $m$ as being the derivative of some $R$-matrix. We will justify this definition in this subsection.
We will indeed see that in this case it is possible to construct a transfer matrix, which generates
a set of commuting operators including the Markov matrix.
The $R$-matrix is the key building block of this transfer matrix as explained in the following definition.

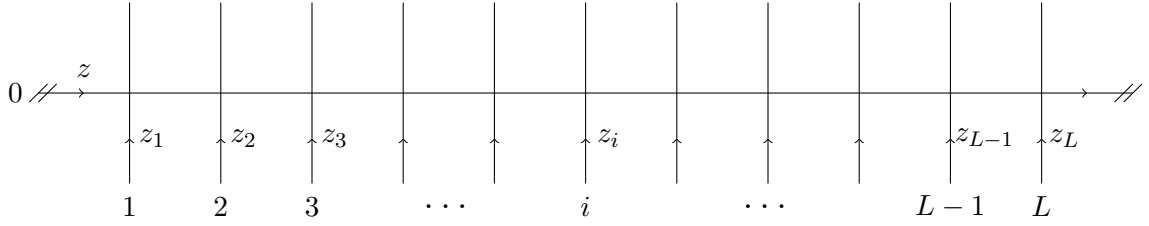
\begin{figure}[htb]
\begin{center}
 \begin{tikzpicture}[scale=0.6] 
 \draw[] (-4,-2) -- (16,-2) ;
 \draw[->] (-6,-2) -- (-5,-2) ; \draw[] (-5,-2) -- (-4,-2) ; 
 \draw[->] (16,-2) -- (17,-2) ; \draw[] (17,-2) -- (18,-2) ;
 \draw (17.8,-2.2) -- (18.2,-1.8); \draw (17.6,-2.2) -- (18.0,-1.8);
 \draw (-6.2,-2.2) -- (-5.8,-1.8); \draw (-6,-2.2) -- (-5.6,-1.8); 
 \foreach \i in {-4,-2,...,16}
{\draw[->] (\i,-4) -- (\i,-3) ; \draw[] (\i,-3) -- (\i,0) ;}
\node at (-6.5,-2) [thick] {$0$};
\node at (-4,-4.5) [thick] {$1$};
\node at (-2,-4.5) [thick] {$2$};
\node at (0,-4.5) [thick] {$3$};
\node at (6,-4.5) [thick] {$i$};
\node at (3,-4.5) [thick] {\Large{$\dots$}};
\node at (10,-4.5) [thick] {\Large{$\dots$}};
\node at (14,-4.5) [thick] {$L-1$};
\node at (16,-4.5) [thick] {$L$};
\node at (-5,-1.5) [thick] {$z$};
\node at (-3.5,-3) [thick] {$z_1$};
\node at (-2+0.5,-3) [thick] {$z_2$};
\node at (0.5,-3) [thick] {$z_3$};
\node at (6.5,-3) [thick] {$z_i$};
\node at (14.75,-3) [thick] {$z_{L-1}$};
\node at (16.5,-3) [thick] {$z_L$};
 \end{tikzpicture}
 \end{center}
\caption{Graphical representation for $t(z|\mathbf{z})$. Due to the trace in the definition of the transfer matrix, 
the two double slash should be considered as linked.} \label{fig:transfer_matrix_periodic}
\end{figure}

\begin{definition}
 The inhomogeneous transfer matrix is an operator acting on the whole lattice $(\CC^{N+1})^{\otimes L}$ and is given by
 \begin{equation} \label{eq:inhomogeneous_transfer_matrix_2spectralparameters}
 t(z|z_1,z_2,\dots,z_L)= \mbox{tr}_{0} \left(R_{0,L}(z,z_L) \, R_{0,L-1}(z,z_{L-1}) \, \dots \, R_{0,1}(z,z_1)\right).
\end{equation}
The parameters $z_1,\dots,z_L$ are called the inhomogeneity parameters, each of them being associated to a particular site of the lattice 
(see figure \ref{fig:transfer_matrix_periodic}). We introduce $\mathbf{z}=(z_1,\dots,z_L)$ to shorten the notation $t(z|\mathbf{z})=t(z|z_1,\dots,z_L)$.
\end{definition}

We have introduced in the definition of the transfer matrix an auxiliary space $\CC^{N+1}$, labeled by $0$, which gets traced out. 
See figure \ref{fig:transfer_matrix_periodic} for a graphical illustration.

Note that when the $R$-matrix is multiplicative in the spectral parameter, the inhomogeneous transfer matrix is given by 
\begin{equation} \label{eq:inhomogeneous_transfer_matrix}
 t(z|\mathbf{z})= \mbox{tr}_{0} \left(R_{0,L}\left(\frac{z}{z_L}\right) \, R_{0,L-1}\left(\frac{z}{z_{L-1}}\right) \, \dots \, 
 R_{0,1}\left(\frac{z}{z_1}\right)\right).
\end{equation}

We now state the main property of this inhomogeneous transfer matrix.
\begin{proposition}
If the matrix $R(z,z')$ satisfies the Yang-Baxter equation and the unitarity property, the inhomogeneous transfer matrix satisfies the 
commutation relation
\begin{equation}
 [t(z|\mathbf{z}),t(z'|\mathbf{z})]=0
\end{equation}
\end{proposition}
\proof
We compute explicitly
\begin{eqnarray*}
 t(z|\mathbf{z})t(z'|\mathbf{z}) & = & \mbox{tr}_{0} \left(R_{0,L}\left(\frac{z}{z_L}\right)\dots R_{0,1}\left(\frac{z}{z_1}\right)\right)
 \mbox{tr}_{0'} \left(R_{0',L}\left(\frac{z'}{z_L}\right)\dots R_{0',1}\left(\frac{z'}{z_1}\right)\right) \\
 & = & \mbox{tr}_{0,0'} \left(R_{0,L}\left(\frac{z}{z_L}\right)R_{0',L}\left(\frac{z'}{z_L}\right)
 \dots R_{0,1}\left(\frac{z}{z_1}\right)R_{0',1}\left(\frac{z'}{z_1}\right)\right) \\
 & = & \mbox{tr}_{0,0'} \left(R_{0,0'}\left(\frac{z}{z'}\right)^{-1}R_{0,0'}\left(\frac{z}{z'}\right) \right. \times \\
 & & \hspace{2cm}  \left. R_{0,L}\left(\frac{z}{z_L}\right)R_{0',L}\left(\frac{z'}{z_L}\right)
 \dots R_{0,1}\left(\frac{z}{z_1}\right)R_{0',1}\left(\frac{z'}{z_1}\right)\right)  \\
 & = & \mbox{tr}_{0,0'} \left(R_{0,0'}\left(\frac{z}{z'}\right)^{-1} \times \right. \\
 & & \hspace{1cm} \left. R_{0',L}\left(\frac{z'}{z_L}\right)R_{0,L}\left(\frac{z}{z_L}\right)
 \dots R_{0',1}\left(\frac{z'}{z_1}\right)R_{0,1}\left(\frac{z}{z_1}\right)R_{0,0'}\left(\frac{z}{z'}\right)\right) \\
 & = & \mbox{tr}_{0,0'} \left(R_{0',L}\left(\frac{z'}{z_L}\right)R_{0,L}\left(\frac{z}{z_L}\right)
 \dots R_{0',1}\left(\frac{z'}{z_1}\right)R_{0,1}\left(\frac{z}{z_1}\right)\right) \\
 & = & t(z'|\mathbf{z})t(z|\mathbf{z}).
\end{eqnarray*}
\finproof

This property tells us that, expanding the transfer matrix as a polynomial (up to an overall normalization) in the spectral parameter provides operators which 
commute with each other. The next step will be to see that the Markov matrix is one of these operators. This is the purpose of 
the following proposition.

\begin{proposition}
The Markov matrix is related to the transfer matrix in the following way 
 \begin{equation}
  \theta \left.\frac{d\ln t(z)}{dz}\right|_{z=1} = \theta t'(1)t(1)^{-1}= \sum_{k=1}^L m_{k,k+1} =M,
 \end{equation}
 where $t(z)$ is the homogeneous transfer matrix defined as $t(z)=t(z|1,1,\dots,1)$ (this is the inhomogeneous transfer matrix where 
 all the inhomogeneity parameters are set to $1$).
\end{proposition}
\proof
A straightforward computation yields the equality 
\begin{equation}
 \left.\frac{d\ln t(z)}{dz}\right|_{z=1} = t(1)^{-1}t'(1).
\end{equation}
The first step is thus to evaluate
\begin{eqnarray*}
 t(1) & = & \mbox{tr}_{0}(P_{0,L}\dots P_{0,2} P_{0,1}) \\
 & = & \mbox{tr}_{0}(P_{0,1} P_{1,L}\dots P_{1,2}) \\
 & = & \mbox{tr}_{0}(P_{0,1}) P_{1,L}\dots P_{1,2} \\
 & = & P_{1,L}\dots P_{1,2}.
\end{eqnarray*}
so that $t(1)^{-1}=P_{1,2}\dots P_{1,L}$. This is also possible to show that $t(1)=P_{L,L-1}\dots P_{L,1}$, playing with $P_{0,L}$ instead of $P_{0,1}$. 
This yields the other expression $t(1)^{-1}=P_{L,1}\dots P_{L,L-1}$.

We need also to calculate
\begin{eqnarray*}
 t'(1) & = & \sum_{k=2}^L \mbox{tr}_{0}(P_{0,L}\dots P_{0,k+1}R_{0,k}'(1)P_{0,k-1} \dots P_{0,1})+\mbox{tr}_{0}(P_{0,L}\dots P_{0,2}R_{0,1}'(1)) \\
 & = & \sum_{k=2}^L P_{1,L}\dots P_{1,k+1}R_{1,k}'(1)P_{1,k-1}\dots P_{1,2} + P_{L,L-1}\dots P_{L,2} R_{L,1}'(1).
\end{eqnarray*}
Using the first expression of $t(1)^{-1}$ that we derived for the sum over $k$ and the second expression of $t(1)^{-1}$ for the last term in the 
previous equation, we obtain 
\begin{eqnarray}
 \theta t(1)^{-1}t'(1) & = & \theta \sum_{k=2}^L P_{k-1,k}R_{k-1,k}'(1) + \theta P_{L,1}R_{L,1}'(1) = M.
\end{eqnarray}
\finproof

\paragraph*{Refinement with twist}

The transfer matrix defined on a periodic lattice can be modified to 'twist' the periodic boundary conditions. 
This can be intuitively understood as the fact that the site $L+1$ is not identified with site $1$ anymore 
(usual periodic condition $\tau_{L+1}=\tau_1$) but imposing instead for instance $\tau_{L+1} \equiv \tau_1 +1$ modulo $N$, or a more 
complicated relation.

This twist is achieved by adding an operator $\cT(z)$, that may or may not depend on a spectral parameter $z$, in the expression of the transfer matrix.
This twist operator $\cT(z)$ has to be carefully chosen in order not to break the integrability of the model, {\it i.e} the commutation property of 
the transfer matrix.
This yields the definitions below.

\begin{definition}
 An integrable twist operator $\cT(z)$ is an invertible $(N+1)\times (N+1)$ matrix that satisfies the relation 
 \begin{equation} \label{eq:twist_operator_exchange_relation}
  \check R(z_1,z_2) \cT(z_1) \otimes \cT(z_2) = \cT(z_2) \otimes \cT(z_1) \check R(z_1,z_2).
 \end{equation}
 Note that this relation is a particular (one dimensional or scalar) representation of the so called RTT relation defined later. 
\end{definition}

\begin{definition}
 A twisted transfer matrix is defined as
 \begin{equation}
  t_{\cT}(z|\mathbf{z})= \mbox{tr}_{0} \left(R_{0,L}\left(\frac{z}{z_L}\right) \, R_{0,L-1}\left(\frac{z}{z_{L-1}}\right) \, \dots \, 
 R_{0,1}\left(\frac{z}{z_1}\right)\cT_{0}(z)\right).
 \end{equation}
where $\cT(z)$ is an integrable twist operator.
\end{definition}

Note that taking $\cT(z)=1$ fulfills the relation \eqref{eq:twist_operator_exchange_relation} and allows us to recover the 
usual expression of the non-twisted transfer matrix.

\begin{proposition}
 A twisted transfer matrix defines a family of commuting operators
 \begin{equation}
 [t_{\cT}(z|\mathbf{z}),t_{\cT}(z'|\mathbf{z})]=0. 
 \end{equation}
\end{proposition}

\proof
 This is done in a exactly similar way as for the non-twisted transfer matrix, using in addition the exchange relation  
 \eqref{eq:twist_operator_exchange_relation}
\finproof

From the out-of-equilibrium physics point of view, the interest of such a twist deformation of the transfer matrix is the 
possibility (often restricted by integrability) to modify the local stochastic dynamics of the model on the particular sites $1$ and $L$.
This is formalized by the following property.

\begin{proposition}
 We have the relation
 \begin{equation}
  \theta \left.\frac{d\ln t_{\cT}(z)}{dz}\right|_{z=1} = \theta t_{\cT}'(1)t_{\cT}(1)^{-1}
  = \sum_{k=1}^{L-1} m_{k,k+1} +\cT_L(1)^{-1} m_{L,1} \cT_L(1) +\theta \cT_L(1)^{-1} \cT_L'(1).
 \end{equation}
\end{proposition}

\proof
 The derivation of this formula is done in a similar way as for the untwisted case.
\finproof

%

\paragraph*{Transfer matrix as discrete time Markov matrix} 

We stressed in the previous paragraphs that the transfer matrix can be used to define (through its derivative) continuous time 
Markov matrices.
We are now going to see that the transfer matrix (or a closely related operator) can be used itself in some particular cases to define 
a discrete time Markov matrix.
Rather than generic considerations, we will focus on a particular example related to the transfer matrix of the TASEP.

From the R-matrix associated to the TASEP\eqref{eq:TASEP_R}, 
we build the inhomogeneous periodic transfer matrix \eqref{eq:inhomogeneous_transfer_matrix}.
The homogeneous case (\textit{i.e.} $z_i=1$) was studied (from the discrete time Markovian process point of view) in \cite{GolinelliM07}. 
We now  introduce the following operator
\begin{eqnarray}
 M(z|\mathbf{z}) &=& t(z|\mathbf{z}) t(z_1|\mathbf{z})^{-1}   \label{eq:TASEP_discrete_time_Markov} \\
&=&  tr_0 \left[ R_{0,L}\left(\frac{z}{z_L}\right)R_{0,L-1}\left(\frac{z}{z_{L-1}}\right)
\dots R_{0,1}\left(\frac{z}{z_1}\right) \right]
  R_{2,1}(\frac{z_2}{z_1})\dots R_{L,1}(\frac{z_L}{z_1}). \nonumber
\end{eqnarray}
Note that we have normalized $t(z|\mathbf{z}) $ using $t(z_1|\mathbf{z})$, but a different choice $t(z_j|\mathbf{z})$, $j=2,3,...,L$ leads to a 
similar rotated matrix $M(z|\mathbf{z})$.  
Obviously $M(z|\mathbf{z})$  commutes with $t(z'|\mathbf{z})$ and
 has the same eigenvectors. We choose below to use the ``normalized matrix'' $M(z|\mathbf{z})$ instead of $t(z|\mathbf{z})$ 
because it allows one to construct easily local jump operators.
A direct computation yields indeed that $-M'(1|1,\dots,1)$, where the derivative is taken with respect to the spectral parameter $z$, 
is the Markov matrix of the continuous time TASEP. 

\begin{figure}[htb]
\begin{center}
 \begin{tikzpicture}[scale=0.6]
 \draw[] (-4,0) -- (16,0) ; 
 \draw[] (-2,-2) -- (16,-2) ;
 \draw[->] (-6,0) -- (-5,0) ; \draw[] (-5,0) -- (-4,0) ; 
 \draw[->] (16,0) -- (17,0) ; \draw[] (17,0) -- (18,0) ;
 \draw (17.8,-0.2) -- (18.2,0.2); \draw (17.6,-0.2) -- (18.0,0.2);
 \draw (-6.2,-0.2) -- (-5.8,0.2); \draw (-6,-0.2) -- (-5.6,0.2);
 \draw[] (-4,0) -- (-4,2) ; \draw[->] (18,-2) -- (17,-2) ;\draw[] (17,-2) -- (16,-2) ;
 \draw (-4,0) arc (180:270:2cm);
 \foreach \i in {-2,0,...,16}
{\draw[] (\i,0) -- (\i,2) ; \draw[->] (\i,-4) -- (\i,-3) ; \draw[] (\i,-3) -- (\i,0) ;}
\node at (-6.5,0) [thick] {$0$};
\node at (17.7,-2.5) [thick] {$1$};
\node at (-2,-4.5) [thick] {$2$};
\node at (0,-4.5) [thick] {$3$};
\node at (6,-4.5) [thick] {$i$};
\node at (3,-4.5) [thick] {\Large{$\dots$}};
\node at (10,-4.5) [thick] {\Large{$\dots$}};
\node at (14,-4.5) [thick] {$L-1$};
\node at (16,-4.5) [thick] {$L$};
\node at (-5,0.5) [thick] {$z$};
\node at (17,-1.5) [thick] {$z_1$};
\node at (-2+0.5,-3) [thick] {$z_2$};
\node at (0.5,-3) [thick] {$z_3$};
\node at (6.5,-3) [thick] {$z_i$};
\node at (14.75,-3) [thick] {$z_{L-1}$};
\node at (16.5,-3) [thick] {$z_L$};
 \end{tikzpicture}
 \end{center}
\caption{Graphical representation for $M(z|\mathbf{z})$. Due to the trace in the definition of the transfer matrix, 
the two double slash should be considered as linked.} \label{fig:TASEP_discrete_time_Markov}
\end{figure}
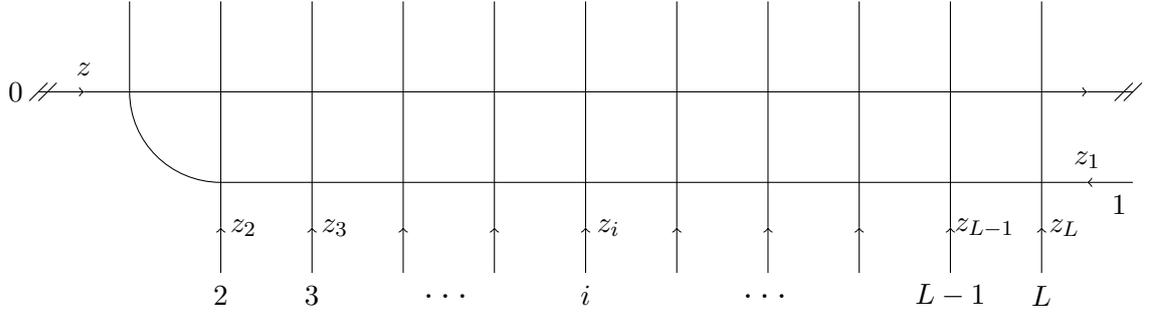

From expression \eqref{eq:TASEP_discrete_time_Markov} and fig. \ref{fig:TASEP_matrix_elements} we can deduce a graphical
representation for $M(z|\mathbf{z})$. The starting point is the lattice illustrated in fig. \ref{fig:TASEP_discrete_time_Markov} 
that one has to fill according to
the matrix element one wants to compute. Instead of explaining it in full generality, we take below a concrete example.

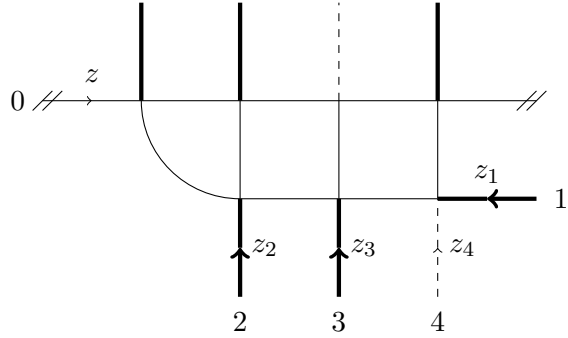
\begin{figure}[htb]
\begin{center}
 \begin{tikzpicture}[scale=0.65]
 \draw[->] (0,0) -- (1,0) ; 
 \draw[] (1,0) -- (10,0) ;
 \draw[ultra thick] (2,0) -- (2,2) ;
 \draw[->,ultra thick] (4,-4) -- (4,-3) ; \draw[ultra thick] (4,-3) -- (4,-2) ;\draw[] (4,-2) -- (4,0) ;\draw[ultra thick] (4,0) -- (4,2) ;
 \draw[->,ultra thick] (6,-4) -- (6,-3) ; \draw[ultra thick] (6,-3) -- (6,-2) ;
 \draw[] (8,-2) -- (8,0) ; \draw[ultra thick] (8,0) -- (8,2) ; 
 \draw[] (4,-2) -- (6,-2) ;
 \draw[->,ultra thick] (10,-2) -- (9,-2) ; \draw[ultra thick] (9,-2) -- (8,-2) ;
 \draw[] (2,0) arc (180:270:2cm);
 \draw[] (6,-2) -- (6,0); \draw[dashed] (6,0) -- (6,2);
 \draw[->,dashed] (8,-4) -- (8,-3);\draw[dashed] (8,-3) -- (8,-2);
 \draw[] (6,-2) -- (8,-2);
 \draw (9.8,-0.2) --(10.2,0.2);\draw (9.6,-0.2) --(10,0.2);
 \draw (-0.2,-0.2) --(0.2,0.2);\draw (0,-0.2) --(0.4,0.2);
 \node at (4.5,-3) [thick] {$z_2$};
 \node at (6.5,-3) [thick] {$z_3$};
 \node at (8.5,-3) [thick] {$z_4$};
 \node at (4,-4.5) [thick] {$2$};
 \node at (6,-4.5) [thick] {$3$};
 \node at (8,-4.5) [thick] {$4$};
 \node at (-0.5,0) [thick] {$0$};
 \node at (1,0.5) [thick] {$z$};
 \node at (10.5,-2) [thick] {$1$};
 \node at (9,-1.5) [thick] {$z_1$};
 \end{tikzpicture}
 \end{center}
 \caption{Starting point for the computation of  $\langle 1101 |M(z|\mathbf{z}) | 1110 \rangle$.\label{fig:TASEP_discrete_time_Markov_ex}}
\end{figure}

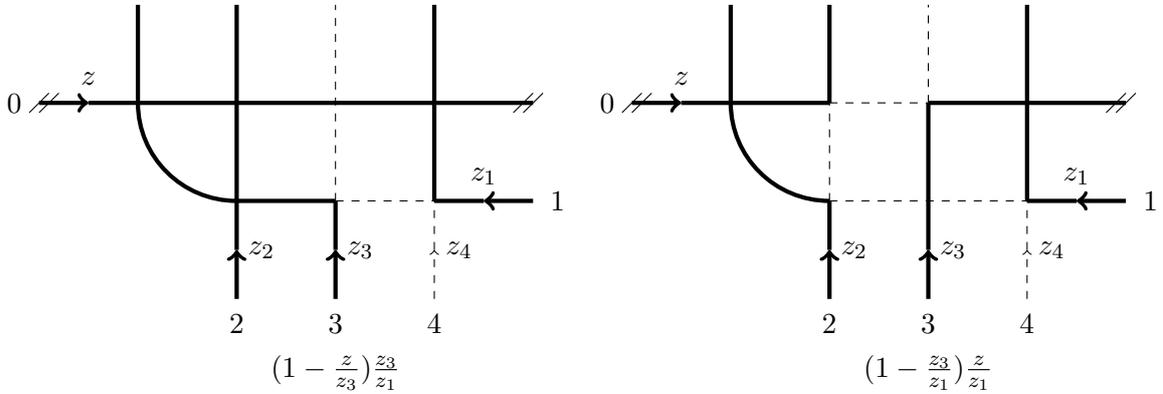
\begin{figure}[htb]
\begin{center}
 \begin{tikzpicture}[scale=0.65]
 \draw[->,ultra thick] (0,0) -- (1,0) ; 
 \draw[ultra thick] (1,0) -- (10,0) ;
 \draw[ultra thick] (2,0) -- (2,2) ;
 \draw[->,ultra thick] (4,-4) -- (4,-3) ; \draw[ultra thick] (4,-3) -- (4,2) ;
 \draw[->,ultra thick] (6,-4) -- (6,-3) ; \draw[ultra thick] (6,-3) -- (6,-2) ;
 \draw[ultra thick] (8,-2) -- (8,2) ; 
 \draw[ultra thick] (4,-2) -- (6,-2) ;
 \draw[->,ultra thick] (10,-2) -- (9,-2) ; \draw[ultra thick] (9,-2) -- (8,-2) ;
 \draw[ultra thick] (2,0) arc (180:270:2cm);
 \draw[dashed] (6,-2) -- (6,2);
 \draw[->,dashed] (8,-4) -- (8,-3);\draw[dashed] (8,-3) -- (8,-2);
 \draw[dashed] (6,-2) -- (8,-2);
 \draw (9.8,-0.2) --(10.2,0.2);\draw (9.6,-0.2) --(10,0.2);
 \draw (-0.2,-0.2) --(0.2,0.2);\draw (0,-0.2) --(0.4,0.2);
 \node at (4.5,-3) [thick] {$z_2$};
 \node at (6.5,-3) [thick] {$z_3$};
 \node at (8.5,-3) [thick] {$z_4$};
 \node at (4,-4.5) [thick] {$2$};
 \node at (6,-4.5) [thick] {$3$};
 \node at (8,-4.5) [thick] {$4$};
 \node at (-0.5,0) [thick] {$0$};
 \node at (1,0.5) [thick] {$z$};
 \node at (10.5,-2) [thick] {$1$};
 \node at (9,-1.5) [thick] {$z_1$};
 \node at (6,-5.5) [thick] {$(1-\frac{z}{z_3})\frac{z_3}{z_1}$};
 
 \draw[->,ultra thick] (12,0) -- (13,0) ; 
 \draw[ultra thick] (13,0) -- (16,0) ;
 \draw[ultra thick] (14,0) -- (14,2) ;
 \draw[->,ultra thick] (16,-4) -- (16,-3) ; \draw[ultra thick] (16,-3) -- (16,-2) ;
 \draw[->,ultra thick] (18,-4) -- (18,-3) ; \draw[ultra thick] (18,-3) -- (18,0) ;
 \draw[ultra thick] (20,-2) -- (20,2) ; 
 \draw[ultra thick] (18,0) -- (22,0) ;
 \draw[ultra thick] (16,0) -- (16,2) ;
 \draw[->,ultra thick] (22,-2) -- (21,-2) ; \draw[ultra thick] (21,-2) -- (20,-2) ;
 \draw[ultra thick] (14,0) arc (180:270:2cm);
 \draw[dashed] (18,0) -- (18,2);
 \draw[->,dashed] (20,-4) -- (20,-3);\draw[dashed] (20,-3) -- (20,-2);
 \draw[dashed] (16,-2) -- (20,-2);
 \draw[dashed] (16,-2) -- (16,0);
 \draw[dashed] (16,0) -- (18,0);
 \draw (21.8,-0.2) --(22.2,0.2);\draw (21.6,-0.2) --(22,0.2);
 \draw (11.8,-0.2) --(12.2,0.2);\draw (12,-0.2) --(12.4,0.2);
 \node at (16.5,-3) [thick] {$z_2$};
 \node at (18.5,-3) [thick] {$z_3$};
 \node at (20.5,-3) [thick] {$z_4$};
 \node at (16,-4.5) [thick] {$2$};
 \node at (18,-4.5) [thick] {$3$};
 \node at (20,-4.5) [thick] {$4$};
 \node at (11.5,0) [thick] {$0$};
 \node at (13,0.5) [thick] {$z$};
 \node at (22.5,-2) [thick] {$1$};
 \node at (21,-1.5) [thick] {$z_1$};
  \node at (18,-5.5) [thick] {$(1-\frac{z_3}{z_1})\frac{z}{z_1}$};
 \end{tikzpicture}
 \caption{The two different drawings involved in the computation of the transition rate
 $\langle 1101 |M(z|\mathbf{z}) | 1110 \rangle$ with their respective weights.
 \label{fig:TASEP_discrete_time_Markov_weights}}
 \end{center}
\end{figure}

As an example we take $L=4$ and use the  graphical interpretation to compute the transition rate between the initial configuration $(1,1,1,0)$ and the 
final configuration $(1,1,0,1)$. The initial (resp. final) configuration fixes the form of the incoming (resp. out-going) external lines
(dashed or thick) as in fig. \ref{fig:TASEP_discrete_time_Markov_ex}.
 Then, we look for drawings of the form given in fig. \ref{fig:TASEP_discrete_time_Markov_ex}
where the remaining thin lines have to be replaced by thick or dashed lines in such a way that  the weights 
(as given in fig.\ref{fig:TASEP_matrix_elements})
of all the vertices do not vanish. The total weight of a given possible drawing is then the product of all these weights.
It is easy to see that there are only two possible drawings, given in fig. \ref{fig:TASEP_discrete_time_Markov_weights} 
together with their corresponding weights.
Finally the weight of $\langle 1101 |M(z|\mathbf{z}) | 1110 \rangle$ is the sum of the weights of the possible drawings.\\

Using this graphical interpretation, we are able to compute all the possible rates between any two  configurations. We remark in particular that 
the number of particles is conserved (as mentioned previously each non-vanishing vertex preserves the number of particles). 
Therefore, we restrict ourselves to a given sector with a fixed number of particles. We can also show that all the rates starting from a 
given configuration (with at least one particle) sum to one which proves that $M(z|\mathbf{z})$ can be used as a discrete time Markov matrix. 
We have to impose also
\begin{equation}\label{eq:TASEP_discrete_time_Markov_constraints}
0\leq \frac{z_i}{z_1}\leq1 \quad\mbox{and}\quad 0\leq \frac{z}{z_i}\leq1 \,,\quad i=1,2,...L,
\end{equation}
so that the probabilities are positive and less than 1. The sector with no particles
is special: its  dimension is $1$ and the matrix $M(z|\mathbf{z})$ is reduced to the scalar 
 $1+\prod_{i=1}^{L}(1-z/z_i)$. Therefore, it cannot be considered as a Markov matrix in the empty sector.
From now, we consider only the cases with at least one particle.

The Markov process given by $M(z|\mathbf{z})$ can be  interpreted as a discrete time process with sequential update.
The configuration at the time $t+1$ is obtained from the one at time $t$ by the following dynamics:
\begin{itemize}
 \item \textup{Particle update:} starting from  right to  left (i.e. from the site $L$ to the site $L-1$ 
and so on), a particle at the site $i$ jumps to the right on the neighboring site
 with a probability $1-z_i/z_1$ provided this site is empty. The particle does not jump
 with the probability $z_i/z_1$. We remind that we are on a periodic lattice, so that
 the  site on the right of the site $L$ is the site $1$. Note that a particle located on site 1 does not move.

 \item \textup{Hole update:}  once the particle update is done, one performs the hole update. 
 Contrary to the particle update, we do not necessarily start  and finish at the sites 1 or  $L$. 
 Let $r$ be the site number of a particle (we recall that we restrict ourselves to the case with at least one particle).
 Starting from the site $r$, we go from  left to 
  right up to the site $r-1$, using periodicity and knowing that the site on the 
  right of the site $L$ is the site 1.   A hole at the site $i$ jumps to 
 the left on the neighboring site with the probability
  $1-z/z_{i-1}$ provided the site is occupied. The hole may stay at site $i$ with probability $z/z_{i-1}$. 
 By convention, we set $z_0=z_L$.
\end{itemize}

As mentioned previously, we would like to emphasize that, due to the inhomogeneities in the transfer matrix, the rates depend on the site 
where the particle or the hole is situated.
 All the probabilities are positive and less than 1 thanks to \eqref{eq:TASEP_discrete_time_Markov_constraints}.
Let us also mention that for the homogeneous case, the update simplifies.
 Indeed, the first step becomes trivial: the particles do not move.

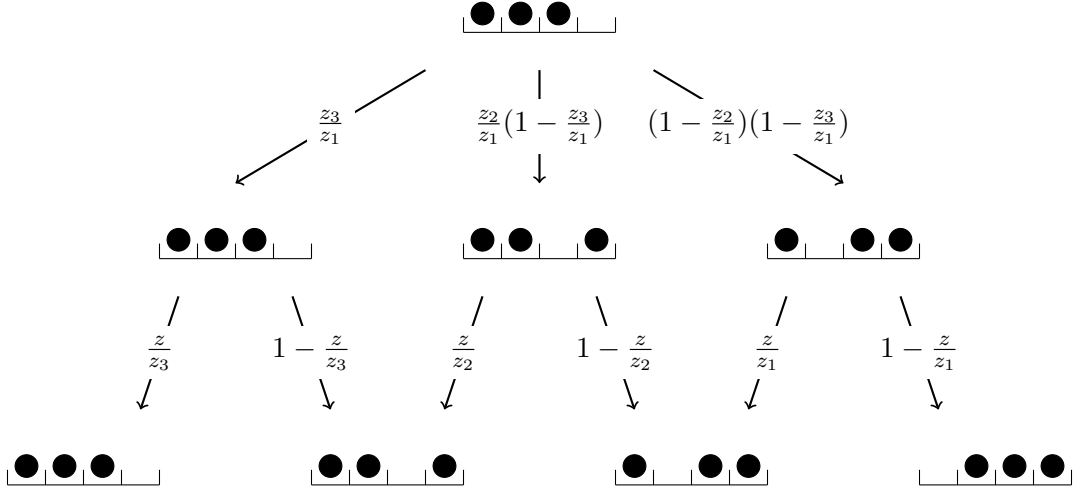
\begin{figure}[htb]
\begin{center}
 \begin{tikzpicture}[scale=1]
  \draw (5,6) -- (7,6) ;
\foreach \i in {5,5.5,...,7}
{\draw (\i,6) -- (\i,6.2) ;}
\draw  (5.25,6.25) circle (0.15) [fill,circle] {};
\draw  (5.75,6.25) circle (0.15) [fill,circle] {};
\draw  (6.25,6.25) circle (0.15) [fill,circle] {};

\draw[->,thick] (4.5,5.5)--  node [fill=white] {$\frac{z_3}{z_1}$} (2,4);
\draw[->,thick] (6,5.5)--  node [fill=white] {$\frac{z_2}{z_1}(1-\frac{z_3}{z_1})$}(6,4);
\draw[->,thick] (7.5,5.5)--  node [fill=white] {$(1-\frac{z_2}{z_1})(1-\frac{z_3}{z_1})$}(10,4);

 \draw (1,3) -- (3,3) ;
\foreach \i in {1,1.5,...,3}
{\draw (\i,3) -- (\i,3.2) ;}
\draw  (1.25,3.25) circle (0.15) [fill,circle] {};
\draw  (1.75,3.25) circle (0.15) [fill,circle] {};
\draw  (2.25,3.25) circle (0.15) [fill,circle] {};
 \draw (5,3) -- (7,3) ;
\foreach \i in {5,5.5,...,7}
{\draw (\i,3) -- (\i,3.2) ;}
\draw  (5.25,3.25) circle (0.15) [fill,circle] {};
\draw  (5.75,3.25) circle (0.15) [fill,circle] {};
\draw  (6.75,3.25) circle (0.15) [fill,circle] {};
 \draw (9,3) -- (11,3) ;
\foreach \i in {9,9.5,...,11}
{\draw (\i,3) -- (\i,3.2) ;}
\draw  (9.25,3.25) circle (0.15) [fill,circle] {};
\draw  (10.25,3.25) circle (0.15) [fill,circle] {};
\draw  (10.75,3.25) circle (0.15) [fill,circle] {};

\draw[->,thick] (1.25,2.5)--  node [fill=white] {$\frac{z}{z_3}$} (0.75,1);
\draw[->,thick] (2.75,2.5)--  node [fill=white] {$1-\frac{z}{z_3}$} (3.25,1);
\draw[->,thick] (5.25,2.5)--  node [fill=white] {$\frac{z}{z_2}$} (4.75,1);
\draw[->,thick] (6.75,2.5)--  node [fill=white] {$1-\frac{z}{z_2}$} (7.25,1);
\draw[->,thick] (9.25,2.5)--  node [fill=white] {$\frac{z}{z_1}$} (8.75,1);
\draw[->,thick] (10.75,2.5)--  node [fill=white] {$1-\frac{z}{z_1}$} (11.25,1);

\draw (-1,0) -- (1,0) ;
\foreach \i in {-1,-0.5,...,1}
{\draw (\i,0) -- (\i,0.2) ;}
\draw  (-0.75,0.25) circle (0.15) [fill,circle] {};
\draw  (-0.25,0.25) circle (0.15) [fill,circle] {};
\draw  (0.25,0.25) circle (0.15) [fill,circle] {};
\draw (3,0) -- (5,0) ;
\foreach \i in {3,3.5,...,5}
{\draw (\i,0) -- (\i,0.2) ;}
\draw  (3.25,0.25) circle (0.15) [fill,circle] {};
\draw  (3.75,0.25) circle (0.15) [fill,circle] {};
\draw  (4.75,0.25) circle (0.15) [fill,circle] {};
\draw (7,0) -- (9,0) ;
\foreach \i in {7,7.5,...,9}
{\draw (\i,0) -- (\i,0.2) ;}
\draw  (7.25,0.25) circle (0.15) [fill,circle] {};
\draw  (8.25,0.25) circle (0.15) [fill,circle] {};
\draw  (8.75,0.25) circle (0.15) [fill,circle] {};
\draw (11,0) -- (13,0) ;
\foreach \i in {11,11.5,...,13}
{\draw (\i,0) -- (\i,0.2) ;}
\draw  (11.75,0.25) circle (0.15) [fill,circle] {};
\draw  (12.25,0.25) circle (0.15) [fill,circle] {};
\draw  (12.75,0.25) circle (0.15) [fill,circle] {};
 \end{tikzpicture}
 \caption{An example of sequential update corresponding to the Markov matrix $M(z|\mathbf{z})$. The first line is
 the configuration at time $t$ and the third line shows the possible configurations at  time $t+1$.
 The second line corresponds to the intermediate configurations after the update of the particles and the hole update still to be done.
 The label of the arrows provides the rate of the corresponding change of configurations.
 \label{fig:TASEP_sequential_update}}
 \end{center}
\end{figure}

 These rules are illustrated in  figure \ref{fig:TASEP_sequential_update} for a  chain with 4 sites in the configuration $(1,1,1,0)$
at time $t$. 
We deduce from this figure the different possible rates between the configurations
 which correspond  to the entries of $M(z|\mathbf{z})$.
One gets
\begin{equation}
 M(z|\mathbf{z})\big( (1,1,0,1),(1,1,1,0)\big)=\langle 1101 |M(z|\mathbf{z}) | 1110 \rangle
=\left(1-\frac{z}{z_3}\right)\frac{z_3}{z_1}+\left(1-\frac{z_3}{z_1}\right)\frac{z}{z_1}
\end{equation}
in accordance with the calculation done previously (see figure \ref{fig:TASEP_discrete_time_Markov_weights}).

\textup{Justification of the sequential up-date}: The sequential update described above can be easily identified when considering
the Markov matrix $M(z|\mathbf{z})$ at the special point $z=z_j$. Indeed, we write 
$M(z_j|\mathbf{z}) = \Big( t(z_j|\mathbf{z})\fP\Big)\Big(\fP^{-1} t(z_1|\mathbf{z})^{-1}\Big)$ where $\fP$ is the cyclic permutation. 
Then, from the explicit expressions
\begin{eqnarray}
t(z_j|\mathbf{z}) &=& R_{j,j-1}(\frac{z_j}{z_{j-1}})\cdots R_{j,1}(\frac{z_j}{z_{1}})\,R_{j,L}(\frac{z_j}{z_{L}})\cdots R_{j,j+1}(\frac{z_j}{z_{j+1}})
\\
t(z_1|\mathbf{z})^{-1} &=& R_{2,1}(\frac{z_2}{z_{1}})\, R_{3,1}(\frac{z_{3}}{z_{1}})\cdots R_{L,1}(\frac{z_L}{z_{1}})
\end{eqnarray}
it is easy to see that $\fP^{-1} t(z_1|\mathbf{z})^{-1}$ corresponds to the particle update, while $t(z_j|\mathbf{z})\fP$ 
corresponds to the hole update at $z=z_j$.  

Let us remark that such a simple sequential update is specific to the totally asymmetric exclusion process.
For the partially asymmetric case, it would be much more involved.

Some properties of the stationary state of the Markov chain defined by $M(z|\mathbf{z})$ will be investigated in chapter \ref{chap:three}.

\subsection{How to find R-matrices?} \label{subsec:Find_Rmatrix}

We stressed, all along this section, that the $R$-matrix and the Yang-Baxter equation are the cornerstone of integrability. The $R$-matrix 
is the building block\footnote{In full generality the building block of the transfer matrix is a Lax operator satisfying the RTT relation 
\eqref{eq:RTT_relation} and which can be chosen more generally than just equal to $R$.}
of the transfer matrix. The Yang-Baxter equation ensures the commutation property of the transfer matrix. In order to 
produce integrable Markovian processes, it thus appears essential to determine $R$-matrix solutions to the Yang-Baxter equation.
This equation is very hard to solve because it involves cubic relations in the entries of the $R$-matrix. It attracted a lot of attention over the 
last decades and several solutions have been constructed in specific cases \cite{KulishS82,Jimbo86bis,BellonMV91,MartinsP11,FonsecaFR15}. 
But we are still lacking a general classification of the solutions of this equation.

However different methods have been developed, relying often on algebraic structures hidden behind the Yang-Baxter equation,
and providing solutions in some particular situations. We present three of the most commonly used below.

\subsubsection{Direct resolution of the Yang-Baxter equation}

We present a method developed by Baxter to compute the $R$-matrix of the six-vertex and eight-vertex models. The main idea could be summarized 
as follows. We would like to isolate some constraints satisfied by the entries of the $R$-matrix, which are implied by the Yang-Baxter equation.
In other words, the $(N+1)^2 \times (N+1)^2$ entries of the $R$-matrix are lying on some manifold. The spectral parameter parametrizes the manifold:
a given spectral parameter corresponds to one 
point on the manifold and the manifold is spanned when the spectral parameter varies. 

Nothing is better than an example to see how it works. We reproduce briefly here the solution of Baxter for the eight-vertex model \cite{Baxter82} 
(in the Markovian case). 

Let us define 
\begin{equation}
 R = \begin{pmatrix}
      a & 0 & 0 & d \\
      0 & b & c & 0 \\
      0 & c & b & 0 \\
      d & 0 & 0 & a 
     \end{pmatrix}
\end{equation}
We also introduce the matrix $R'$ (respectively the matrix $R''$) which is equal to the matrix $R$ but with entries $a$, $b$, $c$, $d$ replaced
by $a'$, $b'$, $c'$, $d'$ (respectively by $a''$, $b''$, $c''$, $d''$). We would like to solve the Yang-Baxter equation
\begin{equation} \label{eq:Yang_Baxter_manifold}
 R_{12} R_{13}' R_{23}'' = R_{23}'' R_{13}' R_{12}.
\end{equation}
The matrix $R$ intuitively corresponds to $R(z_1/z_2)$ , the matrix $R'$ to $R(z_1/z_3)$ and the matrix $R''$ to $R(z_2/z_3)$ 
(equation \eqref{eq:Yang_Baxter_manifold} thus appears as some kind of generalization of \eqref{eq:Yang_Baxter}).
Writing equation \eqref{eq:Yang_Baxter_manifold} in components, we can show that it is equivalent to the following set of equations
\begin{eqnarray}
 ac'a''+da'd'' & = & bc'b''+ca'c'' \label{eq:Yang_Baxter_manifold_eq1} \\ 
 ab'c''+dd'b'' & = & ba'c''+cc'b'' \label{eq:Yang_Baxter_manifold_eq2} \\
 cb'a''+bd'd'' & = & ca'b''+bc'c'' \label{eq:Yang_Baxter_manifold_eq3} \\
 ad'b''+db'c'' & = & bd'a''+cb'd'' \label{eq:Yang_Baxter_manifold_eq4} \\
 aa'd''+dc'a'' & = & bb'd''+cd'a'' \label{eq:Yang_Baxter_manifold_eq5} \\
 da'a''+ac'd'' & = & db'b''+ad'c'' \label{eq:Yang_Baxter_manifold_eq6}
\end{eqnarray}

Following the lines of \cite{Baxter82}, we observe that these equations are linear in $a''$, $b''$, $c''$, $d''$. If we want these equations 
to have non vanishing solutions, the determinant of any linear system composed of four of these equations should be equal to zero. For instance
the determinant of the linear system defined by equations \eqref{eq:Yang_Baxter_manifold_eq1}, \eqref{eq:Yang_Baxter_manifold_eq3}, 
\eqref{eq:Yang_Baxter_manifold_eq4} and \eqref{eq:Yang_Baxter_manifold_eq6} is given by
\begin{equation} \label{eq:Yang_Baxter_manifold_determinant}
 (cda'b'-abc'd')[(a^2-b^2)(c'^2-d'^2)+(c^2-d^2)(a'^2-b'^2)]
\end{equation}
To decide which of these two factors should be equal to zero, we consider the case where $a=a'$, $b=b'$, $c=c'$ and $d=d'$. It corresponds 
intuitively to the case $z_2=z_3$ in the equation \eqref{eq:Yang_Baxter}. In this case, we expect from the regularity condition that
$R''=R(z_2/z_3)=R(1)=P$ so that equation \eqref{eq:Yang_Baxter_manifold} becomes trivial. Indeed we observe in this case that the condition 
\eqref{eq:Yang_Baxter_manifold_determinant} is satisfied because the first factor vanishes. For continuity reasons, this factor should thus 
also vanish in the general case $a\neq a'$, $b\neq b'$, $c\neq c'$ and $d\neq d'$. This yields the condition
\begin{equation} \label{eq:Yang_Baxter_manifold_constraint1}
 \frac{ab}{cd} = \frac{a'b'}{c'd'}
\end{equation}
When this condition is satisfied, we can solve for $a''$, $b''$, $c''$, $d''$. Substituting back in the remaining equations 
\eqref{eq:Yang_Baxter_manifold_eq2} and \eqref{eq:Yang_Baxter_manifold_eq5} yields to the unique constraint
\begin{equation} \label{eq:Yang_Baxter_manifold_constraint2}
 \frac{a^2+b^2-c^2-d^2}{ab} = \frac{a'^2+b'^2-c'^2-d'^2}{a'b'}
\end{equation}
Following Baxter's work \cite{Baxter82}, we introduce the following quantities
\begin{equation} \label{eq:Yang_Baxter_manifold_Delta}
 \Delta = \frac{a^2+b^2-c^2-d^2}{2(ab+cd)}
\end{equation}
and
\begin{equation} \label{eq:Yang_Baxter_manifold_Gamma}
 \Gamma = \frac{ab-cd}{ab+cd}
\end{equation}
We can also define $\Delta'$ and $\Gamma'$ (respectively $\Delta''$ and $\Gamma''$) where the parameters $a$, $b$, $c$, $d$ replaced by  
$a'$, $b'$, $c'$, $d'$ (respectively by $a''$, $b''$, $c''$, $d''$).
We can show that the constraints \eqref{eq:Yang_Baxter_manifold_constraint1} and \eqref{eq:Yang_Baxter_manifold_constraint2} are equivalent to
\begin{equation}
 \Delta = \Delta' \quad \mbox{and} \quad \Gamma = \Gamma'.
\end{equation}
When these constraints are satisfied, we recall that we can solve for $a''$, $b''$, $c''$, $d''$. It is then straightforward to check that
\begin{equation}
 \Delta = \Delta' = \Delta'' \quad \mbox{and} \quad \Gamma = \Gamma' = \Gamma''.
\end{equation}

The equations \eqref{eq:Yang_Baxter_manifold_Delta} and \eqref{eq:Yang_Baxter_manifold_Gamma} thus define the manifold on which are lying the 
entries of the $R$-matrix. The remaining part of the work is to find a parametrization of the manifold with the help of spectral parameters.
The hard part is to determine a parametrization which is additive or multiplicative in the spectral parameter (see the computations below).
Baxter succeeded to find one for this eight-vertex model with the help of elliptic functions. 

To simplify the computation and also to be consistent with the purpose of this manuscript, we will be interested in the Markovian case,
{\it i.e} when $d=1-a$ and $c=1-b$ (and same conditions for the prime and double prime variables), so that the entries of each 
column of the $R$-matrix sum to $1$. In this particular case, the constraints 
\eqref{eq:Yang_Baxter_manifold_Delta} and \eqref{eq:Yang_Baxter_manifold_Gamma} reduce to
\begin{equation}
 \Delta = \Gamma = \frac{a+b-1}{2ab-a-b+1} = \frac{a'+b'-1}{2a'b'-a'-b'+1} = \frac{a''+b''-1}{2a''b''-a''-b''+1}.
\end{equation}
This can be easily solved in $b$, $b'$ and $b''$
\begin{equation}
 b= \frac{(\Delta+1)(a-1)}{\Delta(2a-1)-1},
\end{equation}
and similarly $b'$ and $b''$.
The equation \eqref{eq:Yang_Baxter_manifold} is then solved by expressing $a''$ in function of $a$ and $a'$
\begin{equation}
 a''=\frac{\Delta(2aa'-a-a'+1)-a-a'+1}{\Delta(2aa'-2a'+1)-2a+1}.
\end{equation}
Keeping in mind that we want to find a parametrization with ``multiplicative'' spectral parameter, we would like to find a function $f$ such that
\begin{equation}
 f(a'')=\frac{f(a')}{f(a)}.
\end{equation}
This is achieved by the function
\begin{equation}
 f \ : \ a \rightarrow \frac{a(\lambda-1)+1}{a(\lambda+1)-1},
\end{equation}
where $\lambda$ is defined by the relation
\begin{equation}
 \Delta = \frac{1-\lambda^2}{1+\lambda^2}.
\end{equation}
We are now equipped to introduce the spectral parameters $z_1=f(a)$, $z_2=f(a')$ and $z_3=f(a'')$. This allows us to write 
$R=R(z_1/z_2)$, $R'=R(z_1/z_3)$ and $R''=R(z_2/z_3)$ where $R(z)$ is defined by
 \begin{equation} \label{eq:DiSSEP_R}
  R(z) = \begin{pmatrix}
          \frac{z+1}{\lambda(z-1)+z+1} & 0 & 0 & \frac{\lambda(z-1)}{\lambda(z-1)+z+1} \\
          0 & \frac{z-1}{\lambda(z+1)+z-1} & \frac{\lambda(z+1)}{\lambda(z+1)+z-1} & 0 \\
          0 & \frac{\lambda(z+1)}{\lambda(z+1)+z-1} & \frac{z-1}{\lambda(z+1)+z-1} & 0 \\
          \frac{\lambda(z-1)}{\lambda(z-1)+z+1} & 0 & 0 & \frac{z+1}{\lambda(z-1)+z+1}
         \end{pmatrix}
 \end{equation}
 
 It satisfies the Yang-Baxter equation \eqref{eq:Yang_Baxter}, the unitarity, regularity and Markovian properties. It is the $R$-matrix 
 associated to the DiSSEP. We have indeed 
 \begin{equation}
  2\lambda \check R'(1) = m
 \end{equation}
where $m$ is the bulk local jump operator of the DiSSEP introduced in \eqref{eq:DiSSEP_m}.

\subsubsection{Quantum groups}

Quantum groups are the theoretical algebraic framework behind the $R$-matrices. We will briefly define this algebraic structure and then 
argue on how it can be used to generate solutions to the Yang-Baxter equation.

We first briefly introduce the notion of Hopf algebra, starting from the basics to fix the notations. The reader is invited to refer to \cite{ChariP95} 
for details

\begin{definition}
 An \textup{algebra} $\cA$ over the field $\CC$ is a $\CC$-vector space equipped with two linear maps $\mu \ : \ \cA \otimes \cA \rightarrow \cA$,
 called the \textup{multiplication}, and $\iota \ : \ \CC \rightarrow \cA$, called the \textup{unit}, 
 such that the following consistency relations hold\footnote{ '$\cdot$' in the first equation denotes the external 
 composition law of the $\CC$-vectorial space $\cA$.}
 \begin{eqnarray}
  && \mu(a \otimes \iota(z)) = \mu(\iota(z) \otimes a) = z\cdot a, \quad \forall a \in \cA, \ \forall z \in \CC \\
  && \mu(a_1 \otimes \mu(a_2 \otimes a_3))=\mu(\mu(a_1 \otimes a_2) \otimes a_3), \quad \forall a_1,a_2,a_3 \in \cA.
 \end{eqnarray}
The last property is usually called \textup{associativity} of the algebra.

An algebra is said to be \textup{commutative} if $\mu(a_1 \otimes a_2)= \mu(\sigma(a_1 \otimes a_2))$, for all $a_1,a_2 \in \cA$,
where $\sigma$ is the permutation operator, i.e $\sigma(a_1 \otimes a_2)=a_2 \otimes a_1$. 

If $\cA$ and $\cB$ are two algebras, an homomorphism of algebras is a linear map $\psi \ : \ \cA \rightarrow \cB$ satisfying
$\psi \circ \mu^{\cA} = \mu^{\cB} \circ (\psi \otimes \psi)$ and $\iota^{\cB} \circ \psi = \iota^{\cA}$ 
where $\mu^{\cA}$ and $\iota^{\cA}$ (respectively $\mu^{\cB}$ and $\iota^{\cB}$) denote the multiplication and the unit in $\cA$
(respectively in $\cB$).
\end{definition}

This definition can appear a bit trivial but we put it here to stress the symmetry with the following definition.

\begin{definition}
 A \textup{coalgebra} $\cA$ over the field $\CC$ is a $\CC$-vector space equipped with two linear maps $\Delta \ : \ \cA \rightarrow \cA \otimes \cA$,
 called the \textup{comultiplication}, and $\epsilon \ : \ \cA \rightarrow \CC$, called the \textup{counit},
 such that the following consistency relations hold\footnote{To be fully rigorous $z \otimes a$ and $a \otimes z$ are 
 identified with $z \cdot a$, for all $a \in \cA$ and $z\in \CC$ in the first relation}
 \begin{eqnarray}
  && (\II \otimes \epsilon) \, (\Delta(a))= (\epsilon \otimes \II) \, (\Delta(a))=a, \quad \forall a \in \cA \\
  && (\II \otimes \Delta) \, (\Delta(a))= (\Delta \otimes \II) \, (\Delta(a)), \quad \forall a \in \cA.
 \end{eqnarray}
 The last property is usually called \textup{coassociativity} of the coalgebra.
 
 A coalgebra is said to be \textup{cocommutative} if $\Delta(a)= \sigma(\Delta(a))$, for all $a \in \cA$.
 
 If $\cA$ and $\cB$ are two coalgebras, an homomorphism of coalgebras is a linear map $\psi \ : \ \cA \rightarrow \cB$ satisfying
$(\psi \otimes \psi) \circ \Delta^{\cA} = \Delta^{\cB} \circ \psi$ and $\epsilon^{\cB} \circ \psi = \epsilon^{\cA}$ 
where $\Delta^{\cA}$ and $\epsilon^{\cA}$ (respectively $\Delta^{\cB}$ and $\epsilon^{\cB}$) denote the comultiplication and the counit
in $\cA$ (respectively in $\cB$).
\end{definition}

\begin{definition}
 A \textup{Hopf algebra} $\cA$ over the field $\CC$ is a $\CC$-vector space satisfying the following properties
 \begin{itemize}
  \item $\cA$ is simultaneously an algebra and a coalgebra (over the field $\CC$);
  \item the multiplication $\mu \ : \ \cA \otimes \cA \rightarrow \cA$ and the unit $\iota \ : \ \CC \rightarrow \cA$ are both
  homomorphisms of coalgebras;
  \item the comultiplication $\Delta \ : \ \cA \rightarrow \cA \otimes \cA$ and the counit $\epsilon \ : \ \cA \rightarrow \CC$ are both
  homomorphisms of algebras;
  \item there exists a linear map $S \ : \ \cA \rightarrow \cA$, called the antipode, such that the following relations hold
  \begin{equation}
  \mu \circ (S \otimes \II) \, (\Delta(a)) =  \mu \circ (\II \otimes S) \, (\Delta(a)) = \iota \circ \epsilon \, (a), \quad \forall a \in \cA.
  \end{equation}
 \end{itemize}
\end{definition}

\begin{example}
 If $\fg$ is a Lie algebra, we can define a Hopf structure on its universal enveloping algebra $U(\fg)$ (which is defined as the quotient of 
 of $\bigoplus_{n\geq 0} \fg^{\otimes n}$ by the ideal generated by the relations $x\otimes y-y \otimes x -[x,y]$, see for instance 
 \cite{ChariP95} for details). The universal enveloping algebra is intuitively understood as the algebra generated by all the polynomial relations 
 in the Lie algebra generators. It is thus sufficient to give the definition of the structure maps on the Lie algebra $\fg$ solely (the definition
 of these maps on elements of $U(\fg)$ can be deduced immediately using the homomorphism property)
 \begin{equation}
  \Delta(g)=g \otimes 1 + 1 \otimes g, \quad S(g)=-g, \quad \epsilon(g)=0, \quad \forall g \in \fg.
 \end{equation}
 Note that this defines a cocommutative Hopf algebra.
\end{example}

\begin{definition}
 We call \textup{almost cocommutative} a Hopf algebra $\cA$ for which there exists an invertible element $\cR \in \cA \otimes \cA$ satisfying
 \begin{equation}
  \sigma \circ \Delta (a) = \cR \Delta(a) \cR^{-1}, \quad \forall a \in \cA.
 \end{equation}
 If $\cR$ satisfies in addition\footnote{The relations hold in $\cA \otimes \cA \otimes \cA$, the subscript on $\cR \in \cA \otimes \cA$ indicate
 on which tensor spaces it belongs, for instance $\cR_{12}=\cR \otimes 1$, $\cR_{23}=1 \otimes \cR$...}
 \begin{eqnarray}
  && (\Delta \otimes \II) \, (\cR) = \cR_{13}\cR_{23}, \\
  && (\II \otimes \Delta) \, (\cR) = \cR_{13}\cR_{12},
 \end{eqnarray}
 then $\cA$ is said to be \textup{quasitriangular} and $\cR$ is called the \textup{universal R-matrix}. 
\end{definition}

\begin{proposition}
 If $\cA$ is a quasitriangular Hopf algebra with universal R-matrix $\cR$ then
 \begin{eqnarray}
  && \cR_{12}\cR_{13}\cR_{23} = \cR_{23}\cR_{13}\cR_{12} \label{eq:Hopf_ybe}\\
  && (\epsilon \otimes \II) \, (\cR)=(\II \otimes \epsilon) \, (\cR) = 1 \\
  && (S \otimes \II) \, (\cR) = (\II \otimes S^{-1}) \, (\cR) = \cR^{-1} \\
  && (S \otimes S) \, (\cR) = \cR
 \end{eqnarray}
\end{proposition}

The first relation \eqref{eq:Hopf_ybe} of the proposition is at the root of the construction of solutions to the Yang-Baxter equation. 
We sketch here the general idea in order to motivate the discussion below. We recall that a representation $\pi \ : \cA \rightarrow End(V)$ of
an algebra $\cA$ over a vector space $V$ is an homomorphism of algebras between $\cA$ and the algebra of matrices acting on $V$, denoted $End(V)$.
The irreducible representations\footnote{An irreducible representation $\pi \ : \cA \rightarrow End(V)$ is a representation which has no non-trivial 
stable subspace, {\it i.e} subspace $W\neq \{0\}$ strictly included in $V$ such that for all $a\in \cA$, $\pi (a) (W) \subseteq W$.}
of the quasitriangular Hopf algebras usually depend (at least for the well-known examples of deformation of affine Lie algebras presented 
briefly below) on a spectral parameter $z$. If we denote by $\pi_z$ this irreducible representation, we can define 
$R(z,z'):= \pi_z \otimes \pi_{z'} \, (\cR)$. Applying the representation $\pi_{z_1} \otimes \pi_{z_2} \otimes \pi_{z_3}$ to the equality 
\eqref{eq:Hopf_ybe} yields the Yang-Baxter relation for the matrix $R(z,z')$
\begin{equation}
 R_{12}(z_1,z_2)R_{13}(z_1,z_3)R_{23}(z_2,z_3)=R_{23}(z_2,z_3)R_{13}(z_1,z_3)R_{12}(z_1,z_2).
\end{equation}

It thus appears very important, from the integrable systems point of view, to construct quasitriangular Hopf algebras and study their representation 
theory. Jimbo \cite{Jimbo85} and Drinfeld \cite{Drinfeld87} made a remarkable breakthrough in finding a whole class of examples of non-commutative
and non-cocommutative 
quasitriangular Hopf algebras (only very few examples were known before their work). Their idea can be roughly summarized as follows. They 
discovered how to deform the commutation relations of an affine Lie algebra $\hat\fg$, which is co-commutative Hopf algebra, 
to obtain a non co-commutative Hopf algebra.

We now present very briefly one of the main deformations of an affine Lie algebra $\hat\fg$: $U_q(\hat\fg)$.
Rather than general considerations we will deal with a case of particular interest: the affine Lie algebra $\hat{sl}_{N+1}$.

We recall that $\hat{sl}_{N+1}$ is generated by $e_i$, $f_i$ and $h_i$, for $i \in \ZZ /(N+1) \ZZ$, which are subject to the relations
\begin{equation}
 [h_i,h_j]=0, \quad [h_i,e_j]=C_{i,j}e_j, \quad  [h_i,f_j]=-C_{i,j}f_j, \quad [e_i,f_j]=\delta_{i,j} h_i,
\end{equation}
and the Serre relations. $C_{i,j}=2\delta_{i,j}-\delta_{i,j-1}-\delta_{i,j+1}$ are the entries of the Cartan matrix and 
$\delta_{i,j}$ is equal to $1$ if $i-j\in (N+1)\ZZ$ and $0$ otherwise.

The algebra\footnote{We use here a deformation parameter denoted by $t^{1/2}$ instead the usual notation $q$ to ease with forthcoming computations}
$U_{t^{1/2}}(\hat{sl}_{N+1})$ is obtained by deforming these commutation relations (with a deformation parameter $t^{1/2}$). 
More precisely, it is generated by $e_i$, $f_i$ and $k_i^{\pm 1}$ for $i \in \ZZ /(N+1) \ZZ$, which are subject to the relations
\begin{eqnarray*}
 & & k_i k_i^{-1}=k_i^{-1}k_i=1, \quad [k_i,k_j]=0, \\
 & & k_i e_j=t^{C_{i,j}/2}e_j k_i, \quad  k_i f_j=t^{-C_{i,j}/2}f_j k_i, 
 \quad [e_i,f_j]=\delta_{i,j} \frac{k_i-k_i^{-1}}{t^{1/2}-t^{-1/2}},
\end{eqnarray*}
and the Serre relations. Note that to recover the non-deformed algebra, we have to set $k_i=t^{h_i/2}$ and take the limit $t \rightarrow 1$.

This algebra can be endowed with a Hopf structure by defining the coproduct
\begin{equation}
 \Delta k_i^{\pm 1} =  k_i^{\pm 1} \otimes  k_i^{\pm 1}, \quad \Delta e_i = 1 \otimes e_i + e_i \otimes k_i, \quad 
 \Delta f_i = f_i \otimes 1 + k_i^{-1} \otimes f_i.
\end{equation}
This defines a non cocommutative algebra.

The algebra $U_{t^{1/2}}(\hat{sl}_{N+1})$ admits finite dimensional irreducible representations depending on a spectral parameter $z$ and labeled 
by an integer $l$ (see for instance \cite{KunibaMMO16,KunibaOW17} for a review in context of non-equilibrium statistical mechanics). 
We first need to define the associated representation space $V_l$. For such a purpose we introduce
\begin{equation}
 I_l = \{ \boldsymbol{\rho} = (\rho_0,\dots,\rho_N) \in \NN^{N+1} \ | \ \rho_0+\dots+\rho_N=l \}. 
\end{equation}
For each $\boldsymbol{\rho} \in I_l$ we associate a vector $\ket{\boldsymbol{\rho}}$ and we define
\begin{equation}
 V_l = \bigoplus_{\boldsymbol{\rho} \in I_l} \CC \ket{\boldsymbol{\rho}}.
\end{equation}

The irreducible representation $\pi^{(l)}_z \ : \ U_q(\hat{sl}_{N+1}) \rightarrow End(V_l)$ is given by
\begin{eqnarray}
& &\pi^{(l)}_z(e_i)\ket{\boldsymbol{\rho}}=z^{\delta_{i,0}}\frac{t^{\rho_i/2}-t^{-\rho_i/2}}{t^{1/2}-t^{-1/2}}\ket{\boldsymbol{\rho}-\boldsymbol{i}} \\
& &\pi^{(l)}_z(f_i)\ket{\boldsymbol{\rho}}=z^{-\delta_{i,0}}\frac{t^{\rho_{i+1}/2}-t^{-\rho_{i+1}/2}}{t^{1/2}-t^{-1/2}}\ket{\boldsymbol{\rho}+\boldsymbol{i}} \\
& & \pi^{(l)}_z(k_i)\ket{\boldsymbol{\rho}}=z^{\rho_{i+1}-\rho_i}\ket{\boldsymbol{\rho}},
\end{eqnarray}
where $\boldsymbol{i} \in \ZZ^{N+1}$ has vanishing entries except the $i$-th which is equal to $1$ and the $i+1$-th which is equal to $-1$ (we 
recall that the indices have to be understood modulo $N+1$). We set by convention $\ket{\boldsymbol{\rho}\pm\boldsymbol{i}}=0$ if 
$\boldsymbol{\rho}\pm\boldsymbol{i}$ has some negative entry. 

It allows us to define the matrix $R^{(l),(l')}(z,z'):= \pi^{(l)}_z \otimes \pi^{(l')}_{z'} \, (\cR)$. It was shown that it depends in fact 
only on the ratio of the spectral parameters $z$ and $z'$ : $R^{(l),(l')}(z,z')=R^{(l),(l')}(z/z')$. By taking irreducible representations of 
different dimension on each of the tensor space components in \eqref{eq:Hopf_ybe}, we obtain a whole class
of generalized (they involve different $R$-matrices) Yang-Baxter equations
\begin{equation}
 R_{12}^{(l),(m)}(z_1/z_2)R_{13}^{(l),(n)}(z_1/z_3)R_{23}^{(m),(n)}(z_2/z_3)
 =R_{23}^{(m),(n)}(z_2/z_3)R_{13}^{(l),(n)}(z_1/z_3)R_{12}^{(l),(m)}(z_1/z_2).
\end{equation}

\begin{example}
We give example of such matrices in the particular case $N=1$. To be consistent with the purpose of this manuscript, the matrices are given 
after a similarity transformation, discovered in \cite{KunibaMMO16}, which provides them the Markovian property. 
For $l=l'=1$, we have
\begin{equation}
 R^{(1),(1)}(z) = \begin{pmatrix}
         1 & 0 & 0 & 0 \\
         0 & \frac{1-z}{t-z} & \frac{z(t-1)}{t-z} & 0 \\
         0 & \frac{t-1}{t-z} & \frac{(1-z)t}{t-z} & 0 \\
         0 & 0 & 0 & 1
        \end{pmatrix}.
\end{equation}
If we set $t=p/q$, we recover the $R$-matrix associated to the ASEP \eqref{eq:ASEP_R}. For $l=1$ and $l'=2$, we have
\begin{equation}
R^{(1),(2)}(z)= \begin{pmatrix}
 1 & 0 & 0 & 0 & 0 & 0 \\
 0 & \frac{t^{1/2}-z}{t^{3/2}-z} & 0 & \frac{z(t^2-1)}{t^{3/2}-z} & 0 & 0 \\
 0 & 0 & \frac{t^{-1/2}-z}{t^{3/2}-z} & 0 & \frac{z(t-1)}{t^{3/2}-z} & 0 \\
 0 & \frac{t^{1/2}(t-1)}{t^{3/2}-z} & 0 & \frac{(t^{-1/2}-z)t^2}{t^{3/2}-z} & 0 & 0 \\
 0 & 0 & \frac{t^{-1/2}(t^2-1)}{t^{3/2}-z} & 0 & \frac{(t^{1/2}-z)t}{t^{3/2}-z} & 0 \\
 0 & 0 & 0 & 0 & 0 & 1 
 \end{pmatrix}.
\end{equation}
For $l=2$ and $l'=1$, we have
\begin{equation}
R^{(2),(1)}(z)= \begin{pmatrix}
  1 & 0 & 0 & 0 & 0 & 0 \\
  0 & \frac{t^{-1/2}-z}{t^{3/2}-z} & \frac{z(t-1)}{t^{3/2}-1} & 0 & 0 & 0 \\
  0 & \frac{t^{-1/2}(t^2-1)}{t^{3/2}-1} & \frac{t(t^{1/2}-z}{t^{3/2}-z} & 0 & 0 & 0 \\
  0 & 0 & 0 & \frac{t^{1/2}-z}{t^{3/2}-z} & \frac{z(t^2-1)}{t^{3/2}-z} & 0 \\
  0 & 0 & 0 & \frac{t^{1/2}(t-1)}{t^{3/2}-z} & \frac{(t^{-1/2}-z)t^2}{t^{3/2}-z} & 0 \\
  0 & 0 & 0 & 0 & 0 & 1
 \end{pmatrix}.
\end{equation}
Finally for $l=l'=2$, we have $R^{(2),(2)}(z)=$
\begin{equation}
 \begin{pmatrix}
  1 & \cdot & \cdot & \cdot & \cdot & \cdot & \cdot & \cdot & \cdot \\
  \cdot & \frac{1-z}{t^2-z} & \cdot & \frac{z(t^2-1)}{t^2-z} & \cdot & \cdot & \cdot & \cdot & \cdot \\
  \cdot & \cdot & \frac{(1-z)(1-tz)}{t(t-z)(t^2-z)} & \cdot & \frac{z(1-z)(t-1)}{(t-z)(t^2-z)} & \cdot & \frac{z^2(t-1)(t^2-1)}{(t-z)(t^2-z)} & \cdot & \cdot \\
  \cdot & \frac{t^2-1}{t^2-z} & \cdot & \frac{(1-z)t^2}{t^2-z} & \cdot & \cdot & \cdot & \cdot & \cdot \\
  \cdot & \cdot & \frac{(1-z)(t+1)(t^2-1)}{t(t-z)(t^2-z)} & \cdot & \frac{t^3 z+(1-2z)t^2+z(z-2)t+z}{(t-z)(t^2-z)} & \cdot & \frac{z(1-z)(t+1)(t^2-1)t}{(t-z)(t^2-z)} & \cdot & \cdot \\
  \cdot & \cdot & \cdot & \cdot & \cdot & \frac{1-z}{t^2-z} & \cdot & \frac{z(t^2-1)}{t^2-z} & \cdot \\
  \cdot & \cdot & \frac{(t-1)(t^2-1)}{(t-z)(t^2-z)} & \cdot & \frac{(1-z)(t-1)t^2}{(t-z)(t^2-z)} & \cdot & \frac{(1-z)(1-tz)t^3}{(t-z)(t^2-z)} & \cdot & \cdot \\
  \cdot & \cdot & \cdot & \cdot & \cdot & \frac{t^2-1}{t^2-z} & \cdot & \frac{(1-z)t^2}{t^2-z} & \cdot \\
  \cdot & \cdot & \cdot & \cdot & \cdot & \cdot & \cdot & \cdot & 1 \\
 \end{pmatrix}.
\end{equation}
\end{example}

\begin{remark}
 Applying the representation $\pi_{z_1}^{(l)} \otimes \pi_{z_2}^{(l)} \otimes \II$, where $\pi_{z}^{(l)}$ is the  
 fundamental representation (the irreducible representation of smallest dimension) of the Hopf algebra, to the relation \eqref{eq:Hopf_ybe},
 yields the so-called RTT relation
 \begin{equation} \label{eq:RTT_relation}
  R_{12}^{(l),(l)}(z_1,z_2) T_1(z_1) T_2(z_2) = T_2(z_2) T_1(z_1) R_{12}^{(l),(l)}(z_1,z_2),
 \end{equation}
 where $T_1(z_1)= \pi_{z_1}^{(l)} \otimes \pi_{z_2}^{(l)} \otimes \II \, (\cR_{13})$ 
 and $T_2(z_2)= \pi_{z_1}^{(l)} \otimes \pi_{z_2}^{(l)} \otimes \II \, (\cR_{23})$. The matrix $T(z)$ can be seen as a $(l+1) \times (l+1)$ matrix
 with entries living in the Hopf algebra $\cA$. The RTT relation can be interpreted as a way to encode the commutation relations of the
 elements of $\cA$. If the generators of $\cA$ are correctly stored in the matrix $T(z)$, the RTT relation is then equivalent to the defining relations
 of the algebra. This provides a very elegant presentation of the algebra, sometimes called the FRT presentation \cite{FaddeevRT90}.
 This gives also a convenient way to define the coproduct of the algebra
 \begin{equation}
  \Delta(T(z)) = T(z) \dot \otimes T(z),
 \end{equation}
 where '$\dot \otimes$' denotes the usual matrix product between the matrices $T(z)$ and $T(z)$ but taking tensor product of the algebraic entries.
 For instance
\begin{equation}
 \begin{pmatrix}
  a & b \\ c & d 
 \end{pmatrix} \dot \otimes
 \begin{pmatrix}
  e & f \\ g & h 
 \end{pmatrix} = 
 \begin{pmatrix}
  a \otimes e + b \otimes g & a \otimes f + b \otimes h \\ c \otimes e + d \otimes g & c \otimes f + d \otimes h
 \end{pmatrix}
\end{equation}
The counit is given by
\begin{equation}
 \epsilon (T(z)) = \II,
\end{equation}
and the antipode is given by
\begin{equation}
 S(T(z))=T(z)^{-1}.
\end{equation}
This presentation is often used to define the commutation relations of another deformation of a (affine) Lie algebra $\fg$: the Yangian $Y(\fg)$.
The reader can for instance refer to \cite{Molev07} for details.
\end{remark}

\subsubsection{Baxterization} \label{subsubsec:baxterisation}

A particularly interesting technique to construct solution to the Yang-Baxter equation was proposed by V.F.R. Jones \cite{Jones90} 
in a framework of knot theory, which is known as Baxterisation. It allows one to obtain solutions of the YBE with spectral parameter 
from representations of the braid group or quotients thereof. 
Important cases are the ones of the Hecke algebra, the Temperley--Lieb algebra or the Birman--Murakami--Wenzl algebra \cite{Jimbo86,Isaev04}.
Since then, many authors tried to generalize or produce other suitable formulae that may lead to solutions of the Yang-Baxter equation, 
see e.g. \cite{ChengGX91,ZhangGB91,Li93,BoukraaM01,ArnaudonCDM03,KulishMN10,FonsecaFR15}.

\paragraph*{Hecke algebra}

In this paragraph we introduce the Hecke algebra and its Baxterisation. The Hecke algebra has found a lot of applications
in integrable systems (see below), in combinatorics \cite{Cherednik95} and in knot theory \cite{Jones85}.

\begin{definition} 
 For any integer $n\geq1$, and complex $\omega \in \CC$, $\cH_n(\omega)$ is the unital associative algebra over $\CC$ 
 with generators $\sigma_{1},\dots,\sigma_{n-1}$ and subject to the relations
 \begin{eqnarray}
&& \sigma_{i}\sigma_{i+1}\sigma_{i}=\sigma_{i+1}\sigma_{i}\sigma_{i+1} \,,\qquad i=1,...,n-2 \label{eq:Hecke_braid}\\
&&  [\sigma_{i}\,,\,\sigma_{j}] = 0\,,\quad |i-j|>1 \label{eq:Hecke_commutation} \\
&& \sigma_{i}\sigma_{i}^{-1}=\sigma_{i}^{-1}\sigma_{i}=1 \label{eq:Hecke_inverse} \\
&& \sigma_{i}^2=\omega \sigma_{i}+1. \label{eq:Hecke_polynomial}
 \end{eqnarray}
\end{definition}

\begin{remark}
The Hecke algebra can be seen as the quotient of the group algebra build on the Artin's braid group $\mathfrak{B}_n$ by the 
relation \eqref{eq:Hecke_polynomial}.
The Artin's braid group, which is at the heart of the construction of knot invariants, is indeed defined by the relations
\eqref{eq:Hecke_braid}, \eqref{eq:Hecke_commutation} and \eqref{eq:Hecke_inverse}.
\end{remark}

\begin{remark}
 The Hecke algebra appears as a deformation of the group algebra build from the permutation group $\mathfrak{S}_n$. For $\omega=0$,
 the defining relations of the Hecke algebra coincide indeed with those of the permutation group (where $\sigma_{i}$ stands for 
 the permutation of $i$ and $i+1$). We have in particular $\sigma_{i}^2=1$.
\end{remark}

One of the most important feature of the Hecke algebra (at least in the context of integrable systems) lies in the fact that 
it produces solutions to the (spectral parameter dependent) braided Yang-Baxter equation. The construction is exposed in 
the following theorem

 \begin{theorem}\label{th:Hecke_main}
   If $\sigma_i$ satisfy the relations of $\mathcal{H}_n(\omega)$ and $t$ is such that $\omega=t^{-1/2}-t^{1/2}$, then 
    \begin{eqnarray}
 \check{R}_i(z) = \frac{(z-1)\sigma_i+t^{-1/2}-t^{1/2}}{zt^{-1/2}-t^{1/2}}, \label{eq:Hecke_Rmatrix}
 \end{eqnarray} 
satisfies the braided Yang--Baxter equation
   \begin{equation}
   \check R_{i}(z_1) \check R_{i+1}(z_1 z_2)\check R_{i}(z_2)=\check R_{i+1}(z_2)\check R_{i}(z_1 z_2)\check R_{i+1}(z_1)\;. \label{eq:Hecke_bybe}
   \end{equation}
Moreover the following properties hold: 
\begin{eqnarray}
&&\mbox{ -- unitarity }\
  \check  R_{i}(z)\check R_{i}(1/z)=1\;,\\[1.2ex]
&&\mbox{ -- regularity }
   \check R_{i}(1)=1\;,\\[1.2ex]
&&\mbox{ -- locality }\quad
   \check R_i(z)\check R_j(z')=\check R_j(z')\check R_i(z) \quad\text{for}\quad |i-j|>1\;.\label{eq:Hecke_loca}
\end{eqnarray}
\end{theorem}

\proof
The regularity and locality properties are obvious.\\
The unitarity  and the braided Yang--Baxter equation are established through a direct computation, using the relations \eqref{eq:Hecke_braid} 
and \eqref{eq:Hecke_polynomial}.
\finproof

\begin{example}
We can show that the bulk local jump operator $m$ of the ASEP \eqref{eq:ASEP_m} provides an explicit representation of the Hecke algebra 
$\cH_n(\omega)$ with $\omega=\sqrt{\frac{q}{p}}-\sqrt{\frac{p}{q}}$ in the tensor space $End(\CC^2)^{\otimes n}$
\begin{eqnarray}
 \cH_n(\omega) &\rightarrow& End(\CC^2)^{\otimes n}\nonumber\\
 \sigma_i &\mapsto& \II ^{\otimes i-1} \otimes S \otimes \II ^{\otimes n-i-1}
\end{eqnarray}
where $S$ is a $4\times 4$ matrix (acting on $\CC^2 \otimes \CC^2$) given by
\begin{equation}\label{eq:eiNASEP}
  S = \frac{1}{\sqrt{pq}}(m+q),
\end{equation}
 Then the Baxterised $R$-matrix 
 \begin{equation}
  \check R(z)=\frac{(z-1)S+t^{-1/2}-t^{1/2}}{zt^{-1/2}-t^{1/2}}
 \end{equation}
 corresponding to \eqref{eq:Hecke_Rmatrix} with $t^{1/2}=\sqrt{\frac{p}{q}}$ coincides with the expression of the $R$-matrix
 of the ASEP given in \eqref{eq:ASEP_R} (up to the multiplication by the permutation operator to obtain the non-braided $R$-matrix).
 
 This representation remains valid for the homogeneous $N$-species ASEP which is defined by the bulk local jump operator $m$ acting 
 on $\CC^{N+1} \otimes \CC^{N+1}$
 \begin{eqnarray}
  m & = & \sum_{0 \leq i<j \leq N} \Big[ \Big( q \, \ket{j}\bra{i} \otimes \ket{i}\bra{j} - q\, \ket{i}\bra{i} \otimes \ket{j}\bra{j} \Big) 
  \label{eq:mASEP_m} \\
  & & \hspace{2cm} + \Big( p \, \ket{i}\bra{j} \otimes \ket{j}\bra{i} - p\, \ket{j}\bra{j} \otimes \ket{i}\bra{i} \Big) \Big] 
 \end{eqnarray}
 It provides through the Baxterisation procedure the expression of the $R$-matrix of the $N$-species ASEP 
 \begin{eqnarray}
  \check R(z) & = & 
  \sum_{0 \leq i<j \leq N} \Big[ \Big( \frac{(1-z)q}{p-qz} \, \ket{j}\bra{i} \otimes \ket{i}\bra{j} 
  + \frac{p-q}{p-qz}\, \ket{i}\bra{i} \otimes \ket{j}\bra{j} \Big) 
  \label{eq:mASEP_R} \\
  & & \hspace{2cm} + \Big( \frac{(1-z)p}{p-qz} \, \ket{i} \bra{j} \otimes \ket{j}\bra{i} 
  + \frac{z(p-q)}{p-qz}\, \ket{j}\bra{j} \otimes \ket{i}\bra{i} \Big) \Big]. 
 \end{eqnarray}
\end{example} 

To summarize, the idea of the Baxterisation is to get a solution of the Yang-Baxter equation 
(\textit{i.e.} an $R$-matrix depending on a spectral parameter) from a representation of the Hecke algebra.
This idea has been intensively used and generalized to try to classify the solutions of the Yang-Baxter equation
\cite{ChengGX91,ZhangGB91,Li93,BoukraaM01,ArnaudonCDM03,CrampeFRV17}. We will see below some of these generalizations.

This procedure strongly motivates the study of the representations in tensor space\footnote{Note that representation in non-tensor spaces play
also a very important role, e.g. the Temperley-Lieb algebra (which is a quotient of the Hecke algebra) produces integral loop models.}

of the Hecke algebra because they
produce integrable models.

We present here the classification of such representations in the particular case $N=1$, {\it i.e}
\begin{eqnarray}
 \cH_n(\omega) &\rightarrow& End(\CC^2)^{\otimes n}\nonumber\\
 \sigma_i &\mapsto& \II ^{\otimes i-1} \otimes S \otimes \II ^{\otimes n-i-1}\label{eq:Hecke_rep}
\end{eqnarray}
where $S$ is acting on $\CC^2 \otimes \CC^2$. The matrix $S$ has thus to be invertible and to satisfy the relations
\begin{eqnarray}
 && S_{12}S_{23}S_{12}=S_{23}S_{12}S_{23} \label{eq:Hecke_braid_rep} \\
 && S^2=\omega S+1. \label{eq:Hecke_polynomial_rep}
\end{eqnarray}

The classification of the constant braided Yang--Baxter equation \eqref{eq:Hecke_braid_rep} for $N=1$ was done in \cite{Hietarinta93} 
(to be precise the unbraided relation was classified there but we just need to multiply the solutions on the left by the permutation matrix $P$
to obtain the desired classification).
We just need to pick up the $4\times 4$ matrices satisfying also the relation \eqref{eq:Hecke_polynomial_rep}.

The classification of \cite{Hietarinta93} provides 23 solutions, up to transformations
\begin{equation}\label{eq:Hecke_transfo_rep}
S \mapsto \lambda\, g \otimes g\ S\ g^{-1}\otimes g^{-1} \quad \mbox{;} \quad 
S \mapsto S^{t_1t_2} \quad \mbox{and} \quad S \mapsto P\,S\,P,
\end{equation}
where $\lambda$ is a complex non-zero parameter, $g$ is any invertible $2\times2$ matrix and $P$ is the permutation matrix.
Among these solutions, only 9 are also solution to \eqref{eq:Hecke_polynomial_rep} and this supplementary relation imposes (in general)
constraints on their parameters.
Using the notations of \cite{Hietarinta93}, these solutions are (up to the transformations  \eqref{eq:Hecke_transfo_rep}):
\begin{equation}\label{eq:Hecke_constraints_hieta}
\begin{tabular}{| c | c | c |}
\hline
 \rule{0mm}{2.5ex} Matrix & Constraints on the parameters & Value of $\omega$ \\[.7ex] \hline 
 \rule{0mm}{2.4ex} $R_{H3,1}$ & $pq=1$, $\quad k^2=1$, $\quad s^2=1$ & $0$ \\[.4ex]  \hline 
 \rule{0mm}{2.4ex} $R_{H2,1}$ & $k^2pq=1$ & $k^2-pq$ \\[.4ex] \hline 
 \rule{0mm}{2.4ex} $R_{H2,2}$ & $k^2pq=1$ & $k^2-pq$ \\[.4ex] \hline 
 \rule{0mm}{2.4ex} $R_{H2,3}$ & $k^2=1$, $\quad p+q=0$, $\quad 2ks+p^2+q^2=0$ & $0$ \\[.4ex] \hline
 \rule{0mm}{2.4ex} $R_{H1,1}$ & $4p^2q^2=1$ & $2(p^2-q^2)$ \\[.4ex] \hline
 \rule{0mm}{2.4ex} $R_{H1,2}$ & $pq=1$ & $p-q$ \\[.4ex] \hline 
 \rule{0mm}{2.4ex} $R_{H1,3}$ & $k^4=1$ & $0$ \\[.4ex] \hline 
 \rule{0mm}{2.4ex} $R_{H1,4}$ & $pq=1$, $\quad k^2=1$ & $0$ \\[.4ex] \hline 
 \rule{0mm}{2.4ex} $R_{H0,3}$ & no constraint & $0$ \\[.4ex] 
 \hline 
\end{tabular}
\end{equation}

 After imposing these constraints, the previous matrices are not independent anymore: they can be all 
obtained  from the 7 matrices $R_{H2,1}$, $R_{H2,2}$, $R_{H1,1}$, $R_{H1,2}$, $R_{H1,3}$, $R_{H1,4}$, $R_{H0,3}$ 
(subjected to the constraints given in \eqref{eq:Hecke_constraints_hieta}). 

They are explicitly given by
\begin{equation}
\label{eq:Hecke_solu_dim4}
\begin{aligned}
&R_{H1,1}= \begin{pmatrix} \sinh(\theta)+\eps &0 & 0& \sinh(\theta) \\
 0 & \sinh(\theta) & \cosh(\theta) & 0 \\ 0 & \cosh(\theta) & \sinh(\theta) & 0 \\ \sinh(\theta) & 0 & 0 & \sinh(\theta)-\eps
 \end{pmatrix} \quad \mbox{;} \quad
 \\[1ex]
&R_{H1,2}=\begin{pmatrix} a &0 & 0& b \\
 0 & 0 & a^{-1} & 0 \\ 0 & a & a-a^{-1} & 0 \\ 0 & 0 & 0 & -a^{-1} 
 \end{pmatrix} \quad \mbox{;} \quad 
R_{H1,3}= \eps\begin{pmatrix} 1 &b & -b& ab \\
 0 & 0 & 1 & -a \\ 0 & 1 & 0 & a \\ 0 & 0 & 0 & 1 
 \end{pmatrix} \quad \mbox{;} \quad 
 \\[1ex]
&R_{H2,1/H2,2}= \begin{pmatrix} a &0 & 0& 0 \\
 0 & 0 & b^{-1} & 0 \\ 0 & b & a-a^{-1} & 0 \\ 0 & 0 & 0 & \eps\,a^\eps 
 \end{pmatrix} \quad \mbox{;} \quad 
 R_{H1,4}= \begin{pmatrix} 0 &0 & 0& a \\
 0 & 0 & \eps & 0 \\ 0 & \eps & 0 & 0 \\ a^{-1} & 0 & 0 & 0
 \end{pmatrix}
\\[1ex]
&
R_{H0,3}=\II_4,
\end{aligned}
\end{equation}
where $a$, $b$ and $\theta$ are free complex parameters and $\eps=\pm1$ that are expressed in terms of $p$, $q$ and $k$ through the change
of variables:
\begin{eqnarray}
R_{H1,1} &:& \exp\theta = 2p^2 \quad \mbox{and} \quad q=\frac{\eps}{2p}
\\
R_{H1,2} &:& a=p\,,\quad q=\frac{1}{a} \quad \mbox{and} \quad k=b
\\
R_{H1,3} &:& \eps=k^2\,,\quad a=\eps kq \quad \mbox{and} \quad b=\eps kp
\\
R_{H2,1/H2,2} &:& a=k^2 \quad \mbox{and} \quad kp=b
\\
R_{H1,4} &:& a=p.
\end{eqnarray}

\paragraph*{${\cal S}_n$ and ${\cal T}_n$ algebras}

Here, we introduce new braid-like algebras ${\cal S}_n$ and ${\cal T}_n$. We produce, as the main result of this paragraph
a new Baxterisation formula that leads to R-matrices depending genuinely on two spectral parameters. 
The obtained R-matrices satisfy the usual properties of unitarity, regularity and locality. 
Moreover, we show that the matrix representations of ${\cal S}_n$ are determined only by the defining relations of ${\cal S}_3$. 
A classification thereof in terms of $4 \times 4$ matrices is given together with the expressions of the corresponding R-matrices. 
We also get the corresponding Hamiltonians.
In the general case, a particular $m \times m$ representation is exhibited, that appears to be linked to some generalizations of the multi-species 
Totally Asymmetric Simple Exclusion Process (TASEP). The results presented here are extracted from the work \cite{CrampeFRV17}.

\begin{definition}\label{def:Sn_algebra}
For any integer $n\geq1$, $\mathcal{S}_n$ is the unital associative algebra over $\CC$ with generators $\sigma_{1},\dots,\sigma_{n-1}$ and subject to
the relations
 \begin{eqnarray}
&&  [\sigma_{i+1}\,\sigma_{i}\,,\,\sigma_{i}+\sigma_{i+1}]=0\,,\qquad i=1,...,n-2 \label{eq:Sn_relation1}\\
&&  [\sigma_{i}\,,\,\sigma_{j}] = 0\,,\quad |i-j|>1 \label{eq:Sn_relation2}
 \end{eqnarray}
 where $[\,.\,,\,.\,]$ stands for the commutator. 
\end{definition}
Let us stress that the definition of the algebra $\cS_n$ does not need the existence of the inverse generators $\sigma_i^{-1}$:
there are interesting realizations of this algebra where represented generators are non-invertible
(see e.g. proposition \ref{pro:Sn_representation_m} below).
For $n=1$, one has $\cS_1 \simeq \CC$.

Relation \eqref{eq:Sn_relation1} can be written equivalently as
\begin{equation}
 \sigma_{i}\,\sigma_{i+1}\, \sigma_{i}-\sigma_{i+1}\,\sigma_{i}^2=\sigma_{i+1}\,\sigma_{i}\, \sigma_{i+1}-\sigma_{i+1}^2\,\sigma_{i}\;,
\end{equation}
which can be seen as a modification\footnote{Unfortunately, we cannot implement a free parameter in this relation and 
keep the baxterisation procedure. Then, $\cS_n$ cannot be viewed as a deformation of the braid group.} of the defining relations
of the braid group $\cB_n$ (however without the inverse generators). This justifies the terminology we used for the algebra $\cS_n$ as 
a braid-like algebra.
Let us also mention that $\cS_n$, like $\cB_n$, is infinite dimensional.

\begin{proposition}\label{pro:Sn_iso2}
Let $\gamma\in\CC$ be such that the generators $1+\gamma \sigma_i$ are invertible $\forall i$
with its inverse understood as the the following formal series:
\begin{equation}
 (1+\gamma \sigma_i)^{-1}=\sum_{k=0}^\infty \left(-{\gamma}\ \sigma_i\right)^k\;.
 \label{eq:Sn_inverse}
\end{equation} 
Then, the M\"obius map 
\begin{equation}\label{eq:Sn_iso2}
 \fm_{\alpha,\beta,\gamma} : \ 
\begin{array}{ccl}
\mathcal{S}_n&\ \rightarrow\ &\mathcal{S}_n\\
 \sigma_{i} &\mapsto &(\alpha\ + \beta \sigma_i)\ (1+\gamma \sigma_i)^{-1}
\end{array}
\end{equation}
with $\alpha, \beta \in \CC$ is an algebra homomorphism.
\end{proposition}

 \proof 
 Remark that $\gamma=0$ ensures the existence of at least one parameter $\gamma$ such that the condition on the invertibility 
 of $1 + \gamma \sigma_i$ is fulfilled.  We divide the proof into two parts, depending on whether $\gamma$ is null or not. 

When $\gamma=0$, one gets $\fm_{\alpha,\beta,0}=\fs_{\alpha,\beta}$,
where $\fs_{\alpha,\beta}(\sigma)=\alpha+\beta\sigma$. It is straightforward to show that 
$[ \fs_{\alpha,\beta}(\sigma_{i+1}) \fs_{\alpha,\beta}(\sigma_i)\,,\, \fs_{\alpha,\beta}(\sigma_i)+ \fs_{\alpha,\beta}(\sigma_{i+1})]=0$. 
Thus $\fs_{\alpha,\beta}$ is an homomorphism of $S_n$.
 
For $\gamma\neq 0$, the M\"obius transformation is the composition of two maps:
\begin{equation}
\fm_{\alpha,\beta,\gamma}=\fs_{\beta/\gamma,\alpha-\beta/\gamma}\   \circ\ \fs_{1,\gamma}^{-1}
\end{equation}
where  $\fs_{1,\gamma}^{-1}(\sigma)=(1+\gamma\,\sigma)^{-1}$, as given by the expansion \eqref{eq:Sn_inverse}. 
To prove that $\fs_{1,\gamma}^{-1}$ is also an homomorphism, we start from \eqref{eq:Sn_relation1} written 
for $\fs_{1,\gamma}(\sigma)$ and multiply this relation on the right by 
$\fs_{1,\gamma}(\sigma_i)^{-1}\,\fs_{1,\gamma}(\sigma_{i+1})^{-1}\,\fs_{1,\gamma}(\sigma_i)^{-1}$ and on the left by 
$\fs_{1,\gamma}(\sigma_{i+1})^{-1}\,\fs_{1,\gamma}(\sigma_{i})^{-1}\,\fs_{1,\gamma}(\sigma_{i+1})^{-1}$.
This shows that $\fs_{1,\gamma}(\sigma_i)^{-1}$ also verifies \eqref{eq:Sn_relation1}. Note that 
$\fs_{1,\gamma}(\sigma)$ is an invertible element by hypothesis.
Then $\fm_{\alpha,\beta,\gamma}$ is 
an homomorphism since it is a composition of homomorphisms.
\finproof

We also introduce another algebra $\mathcal{T}_n$, defined as follows
\begin{definition}
For any integer $n\geq1$, $\mathcal{T}_n$ is the unital associative algebra over $\CC$ with generators $\tau_{1},\dots,\tau_{n-1}$ and subject to
the relations
 \begin{eqnarray}
&&  [\tau_{i}\, \tau_{i+1}\,,\,\tau_{i}+\tau_{i+1}]=0\,,\qquad i=1,...,n-2 \label{eq:Tn_relation1}\\
&&  [\tau_{i}\,,\,\tau_{j}] = 0\,,\quad |i-j|>1. \label{eq:Tn_relation2}
 \end{eqnarray}
\end{definition}

This algebra is closely related to $\cS_n$ as stated in the following proposition:

\begin{proposition}\label{pro:Sn_iso}
\begin{equation}
\mbox{The map }\quad
\phi :\quad \begin{cases} \mathcal{S}_n&\rightarrow \mathcal{T}_n\\ \sigma_{i} &\mapsto \tau_{n-i}\end{cases}
\quad\mbox{ is an algebra isomorphism.}
\end{equation}
\end{proposition}

 \proof 
 The isomorphism is proved by direct computations.
\finproof

The following theorem contains the main result of this paragraph and justifies the introduction of the algebra $\cS_n$.
 \begin{theorem}\label{th:Sn_main}
   If $\sigma_i$ satisfy the relations of $\mathcal{S}_n$, then 
    \begin{eqnarray}
 \check R_i^\sigma(x,y)= \Sigma_i(y)  \Sigma_i(x)^{-1}\quad\text{where}\quad \Sigma_i(x)=1-x \sigma_i \label{eq:Sn_Rmatrix}
 \end{eqnarray} 
 satisfy the braided Yang--Baxter equation
   \begin{equation}
   \check R_{i}(x,y) \check R_{i+1}(x,z)\check R_{i}(y,z)=\check R_{i+1}(y,z)\check R_{i}(x,z)\check R_{i+1}(x,y)\;, \label{eq:Sn_bybe}
   \end{equation}
 and also the locality, unitarity and regularity properties.
\end{theorem}

\proof
The unitarity, regularity and locality properties are obvious.\\
To prove the braided Yang--Baxter equation \eqref{eq:Sn_bybe}, let us remark that, 
after multiplication  on the right by $\Sigma_i(y)$ and on the left by $\Sigma_{i+1}(y)$, it is equivalent to
\begin{eqnarray}
 A_i(y)A_i(x)^{-1}A_i(z)=A_i(z)A_i(x)^{-1}A_i(y)\label{eq:Sn_A}
\end{eqnarray}
where $A_i(x)=\Sigma_{i+1}(x)\Sigma_i(x)$.
Relation \eqref{eq:Sn_A} is equivalent to
\begin{equation}
 [\ A_i(x)\ ,\ A_i(y)\ ]=0\;. \label{eq:Sn_Abis}
\end{equation}
Indeed, setting $z=0$ in \eqref{eq:Sn_A} leads to \eqref{eq:Sn_Abis}, which obviously implies \eqref{eq:Sn_A}.
Let us notice now that
\begin{equation}
 A_i(x)=1-x(\sigma_i+\sigma_{i+1})+x^2 \sigma_{i+1}\sigma_i
\end{equation}
Therefore, the defining relation \eqref{eq:Sn_relation1} of $\cS_n$ implies relation \eqref{eq:Sn_Abis} and then the braided Yang--Baxter equation.
\finproof

We want to emphasize that the Baxterisation introduced here depends 
\textsl{separately} on the two spectral parameters. This is a new feature in comparison to the previous 
Baxterisations \cite{Jones90,Jimbo86,Isaev04}.

Theorem \ref{th:Sn_main} gives a sufficient condition to obtain the braided Yang--Baxter equation. The following proposition proves that it is also a
necessary condition.

\begin{proposition}\label{pro:Sn_1}
If $\check R_i^\sigma(x,y)$ given by \eqref{eq:Sn_Rmatrix} satisfies the braided Yang--Baxter equation \eqref{eq:Sn_bybe}
and the locality property, then the generators
$\sigma_i$ satisfy $\cS_n$.
\end{proposition}

\proof 
We have already seen that the braided Yang--Baxter equation implies \eqref{eq:Sn_Abis}. The different coefficients of 
\eqref{eq:Sn_Abis} w.r.t. $x$ and $y$ imply \eqref{eq:Sn_relation1}. The locality implies \eqref{eq:Sn_relation2} which concludes the proof.
\finproof

Until now, we used the algebra $\cS_n$ to get a solution of the braided Yang--Baxter equation, but we can use similarly the algebra $\cT_n$:
\begin{theorem}\label{th:Tn_main}
The generators $\tau_i$ satisfy the algebra $\cT_n$ if and only if
\begin{eqnarray}
\check R_i^\tau(x,y)= \fT_i(x)  \fT_i(y)^{-1}\quad\text{where}\quad \fT_i(x)=1-x \tau_i \label{eq:Tn_Rmatrix}
\end{eqnarray} 
are unitary, regular and local solutions of the braided Yang--Baxter equation.
\end{theorem}

\proof
Direct consequence of theorem \ref{th:Sn_main} and proposition \ref{pro:Sn_1}, using the isomorphism of proposition \ref{pro:Sn_iso}.
\finproof

Let us stress that there is a flip between the spectral parameters in definitions \eqref{eq:Sn_Rmatrix} and \eqref{eq:Tn_Rmatrix}. 

We are now interested in some matrix representations of $\cS_n$ and in particular the ones useful in the context of integrable systems,
{\it i.e} representations of $\cS_n$ in $End(\CC^m)^{\otimes n}$. 

More precisely, we look for representations of the following type
\begin{eqnarray}
 \cS_n &\rightarrow& End(\CC^m)^{\otimes n}\nonumber\\
 \sigma_i &\mapsto& \II ^{\otimes i-1} \otimes S \otimes \II ^{\otimes n-i-1}\label{eq:Sn_rep}
\end{eqnarray}
where $\II$ is the identity of $End(\CC^m)$ and $S\in End(\CC^m)\otimes End(\CC^m)$.
We use the following notation $S_{j,j+1}=\II^{\otimes j-1} \otimes S \otimes \II^{\otimes n-j-1}$:
the indices in $S_{j,j+1}$ label the copies of $End(\CC^m)$ in which the operator $S$ acts non-trivially.
To look for such matrix representations of $\cS_n$ ($n\geq 3$), it is necessary  and sufficient to find $S$ satisfying the single relation of $\cS_3$:
\begin{equation} \label{eq:Sn}
  [S_{23}\,S_{12}\,,\,S_{12}+S_{23}]=0\;.
\end{equation}
We give below a classification of the solutions of this equation for $m=2$ and some solutions for any $m$. 

Before that, let us remark that  proposition \ref{pro:Sn_iso2} implies that if $S$ is a solution of \eqref{eq:Sn}, then so is $\fm_{\alpha,\beta,\gamma}(S)$.
We have loosely used the same notation for the M\"obius map acting on the algebra and the ones acting on the representation.

Using the Baxterisation introduced in theorem \ref{th:Sn_main} and the realization of $\cS_n$ given by \eqref{eq:Sn_rep}, the matrix
\begin{equation}
 \check R(x,y)=(\II-yS)(\II-xS)^{-1}\;\label{eq:Sn_Rc2}
\end{equation}
is a solution of the braided Yang--Baxter equation
  \begin{equation}
   \check R_{i,i+1}(x,y) \check R_{i+1,i+2}(x,z)\check R_{i,i+1}(y,z)=\check R_{i+1,i+2}(y,z)\check R_{i,i+1}(x,z)\check R_{i+1,i+2}(x,y)\;, 
   \label{eq:Sn_bybe2}
   \end{equation}
 where the indices stand for the spaces on which the matrix $\check R(x,y)$ acts non-trivially.
 
 Similarly, one gets matrix representations for $\cT_n$. Indeed, we look for representations of the following type
 \begin{eqnarray}
 \cT_n &\rightarrow& End(\CC^m)^{\otimes n}\nonumber\\
 \tau_i &\mapsto& \II^{\otimes i-1} \otimes T \otimes \II^{\otimes n-i-1}\label{eq:Tn_rep}
\end{eqnarray}
In this case, one has to solve the equation 
 \begin{equation} \label{eq:Tn_1}
  [T_{12}\,T_{23}\,,\,T_{12}+T_{23}]=0\;
\end{equation}
the associated $R$-matrix being now 
\begin{equation}
 \check R(x,y)=(\II-xT)(\II-yT)^{-1}\;. \label{eq:Tn_Rc}
\end{equation}

Relation \eqref{eq:Sn} is difficult to solve in general: there are $(m^3)^2$ cubic relations in terms of the $(m^2)^2$ entries of the matrix $S$.
However, for $m=2$, using symmetry transformations and a direct resolution with a formal mathematical software, we are able to compute all the solutions 
which are presented in the following theorem. Note that we do not impose
a priori that $S$ is invertible and indeed some particular cases of the solutions are not, see for instance remark \ref{rmk:Sn_TASEP}.
It could be interesting to study if the finite dimensional representations found by 'brute force' computations below 
arise as natural quotients of the algebra $\cS_n$.
 
\begin{theorem}\label{th:Sn_classification}
The whole set of representations of $\cS_n$ of type \eqref{eq:Sn_rep}, for $m=2$, is obtained by applying the following transformations 
\begin{itemize}
\item $S\mapsto S^{-1}$, when $S$ is invertible, 
\item $S\mapsto S_{21}^{t_1t_2}$, where $(.)^{t_1t_2}$ is the transposition in the space $End(\CC^m)\otimes End(\CC^m)$,
\item $S\mapsto Q_1Q_2\,S\,Q_2^{-1}Q_1^{-1}$, where $Q$ is any invertible element of $End(\CC^m)$.
\end{itemize}
or M\"obius transformation to the seven matrices below:
\begin{itemize}
\item two $4$-parameter matrices
\begin{equation}
 S^{(1)}=\begin{pmatrix}
                0 & \cdot & \cdot & \cdot \\
                b & c & \cdot & \cdot \\
                d & \cdot & 0 & \cdot \\
                \cdot & a & \cdot & 0 
               \end{pmatrix}, \quad
 S^{(2)}=\begin{pmatrix}
                0 & b-c & b+c & d \\
                \cdot & 0 & \cdot & a+b \\
                \cdot & \cdot & 0 & b-a \\
                \cdot & \cdot & \cdot & 0
               \end{pmatrix}
\end{equation}
\item four $3$-parameter matrices
\begin{equation}
 S^{(3)}=\begin{pmatrix}
          0 & \cdot & \cdot & \cdot \\
          b & \frac{ab}{c} & \cdot & \cdot \\
          c & \cdot & a & \cdot \\
          \cdot & \cdot & \cdot & 0
         \end{pmatrix}, \quad 
 S^{(4)}=\begin{pmatrix}
          0 & \cdot & \cdot & \cdot \\
          \cdot & b & \cdot & \cdot \\
          \cdot & c & 0 & \cdot \\
          \cdot & \cdot & \cdot & a
         \end{pmatrix}, 
         \end{equation}
  \begin{equation}
 S^{(5)}=\begin{pmatrix}
          a & \cdot & \cdot & \cdot \\
          \cdot & b & \cdot & \cdot \\
          \cdot & \cdot & c & \cdot \\
          \cdot & \cdot & \cdot & 0
         \end{pmatrix}, \quad 
S^{(6)}=\begin{pmatrix}
          0 & \cdot & \cdot & \cdot \\
          b & c & \cdot & a \\
          \cdot & \cdot & 0 & \cdot \\
          \cdot & -b & b & 0 
         \end{pmatrix}, \quad 
  \end{equation}
\item one $2$-parameter matrix
\begin{equation}
 S^{(7)}=\begin{pmatrix}
          -a & \cdot & \cdot & \cdot \\
          b & 0 & a & \cdot \\
          \cdot & \cdot & -a & \cdot \\
          \cdot & \cdot & \cdot & 0 
         \end{pmatrix},
\end{equation}
\end{itemize}
where $a,b,c,d$ are free complex parameters.
\end{theorem}

\proof
The proof consists in finding all the $4\times 4$ matrices $S$ solution of equation \eqref{eq:Sn}.
We use a technique introduced in \cite{Hietarinta93}  to classify the constant $4\times 4$ 
solutions of the Yang--Baxter equation.
Let $s_{ij}$ (for $i,j=1,2,3,4$) be the entries of $S$.
Using the transformations exposed in the theorem \ref{th:Sn_classification}, we can always look for solution of the equation with $s_{41}=0$. Indeed if $S$ is a 
solution with $s_{41} \neq0$, then $S$ is related through a transformation to a solution $S^{new}$ with $s^{new}_{41}=0$.
More precisely if $s_{14}=0$ we set
$S^{new}=(S_{21})^{t}$ and we have $s^{new}_{41}=0$ because the transformation exchanges $s_{41}$ and $s_{14}$. If $s_{14}s_{41}\neq 0$, we set
$S^{new}=Q\otimes Q\,S\,Q^{-1}\otimes Q^{-1}$, with 
\begin{equation}
 Q=\begin{pmatrix}
    1 & 0 \\
    B & 1
   \end{pmatrix}.
\end{equation}
We have $s^{new}_{41}=s_{14}B^4+(s_{24}+s_{34}-s_{12}-s_{13})B^3+(s_{44}-s_{22}-s_{23}-s_{32}-s_{33}+s_{11})B^2+(s_{21}+s_{31}-s_{42}-s_{43})B+s_{41}$.
Since we are in the case $s_{14} \neq 0$, it is always possible to find $B$ such that $s^{new}_{41}=0$.
Therefore, without loss of generality, we can now set $s_{41}=0$.

At this stage, we use a computer software to solve \eqref{eq:Sn} with $s_{41}=0$. Then, using the transformations given above in theorem
\ref{th:Sn_classification}, we select the solutions 
which are not related by transformations and get the seven solutions presented in the theorem. 
\finproof

\begin{remark}\label{rmk:Sn_TASEP} 
The solution $S^{(4)}$ provides the TASEP Markovian matrix as a subcase:
\begin{equation}
S^{(4)}\Big|_{a=0;\,c=1;\,b=-1}=\begin{pmatrix}
          0 & \cdot & \cdot & \cdot \\
          \cdot & -1 & \cdot & \cdot \\
          \cdot & 1 & 0 & \cdot \\
          \cdot & \cdot & \cdot & 0
         \end{pmatrix}.
\end{equation}
\end{remark}

As explained previously in this chapter, each $S$ is an integrable Hamiltonian. Then, from any $4\times4$ integrable Hamiltonian $h$ 
obtained in this classification, we can construct new ones by action of the group of transformations generated by
\begin{eqnarray}\label{eq:Sn_transfo_h}
&&h \mapsto Q \otimes Q\, h\, (Q\otimes Q)^{-1} \mbox{ ; } h \mapsto \ h^{t_1t_2} \mbox{ ; } h \mapsto \ P\,h\,P
 \mbox{ ; } h \mapsto (\alpha + \beta h)(1+\gamma h)^{-1}.\qquad
\end{eqnarray}
These transformations are indeed symmetry transformations for the relation \eqref{eq:Sn} or map a solution of \eqref{eq:Sn} to a solution of 
\eqref{eq:Tn_1}.

Using theorem \ref{th:Sn_main} and the previous classification theorem \ref{th:Sn_classification}, we get a set of solutions 
to the braided Yang--Baxter 
equation. The braided $R$-matrices are easily computed using \eqref{eq:Sn_Rmatrix}.
One has to keep in mind that one can construct other $R$-matrices from these ones by using the symmetry transformations 
described previously.

As explained previously, the resolution of equation \eqref{eq:Sn} is complicated and a complete classification for any $m$ seems impossible.
However, it is still possible to find some solutions. In the following proposition, we present particular non-trivial matrix representations for any $m$:
\begin{proposition}\label{pro:Sn_representation_m}
Let $E_{ij}$ be the canonical basis of $End(\CC^m)$ and let $\rho_i$ and $\mu_{i,j}$, $1\leq i<j\leq m$ be some complex numbers.
We define
\begin{eqnarray}
 {S}&=&\sum_{1\leq i<j \leq m} \rho_i\ E_{ii}\otimes E_{jj} +\mu_{i,j}\ E_{ji}\otimes E_{ij}\label{eq:Sn_inhTASEP}
\end{eqnarray}
Then, the map $\sigma_i\mapsto {S}_{i,i+1}$ is an homomorphism of algebra $\mathcal{S}_n\rightarrow End\big((\CC^m)^{\otimes n}\big)$. 
\end{proposition}

\proof 
We prove by direct computations that $S$ given by \eqref{eq:Sn_inhTASEP} verifies \eqref{eq:Sn}.
\finproof

Remark that $S$ is non-invertible.
Obviously, the use of the transformations given in theorem \ref{th:Sn_classification} provides new representations isomorphic to \eqref{eq:Sn_inhTASEP}.

There is a similar type of representation for $\cT_n$. 
\begin{proposition}\label{pro:Tn_representation_m}
Let $E_{ij}$ be the canonical basis of $End(\CC^m)$ and let $\zeta_j$ and $\nu_{i,j}$, $1\leq i<j\leq m$ be some complex numbers.
We define
\begin{equation}
T=\sum_{1\leq i<j \leq m} \zeta_j\ E_{ii}\otimes E_{jj} +\nu_{i,j}\ E_{ji}\otimes E_{ij}\;,\label{eq:Tn_inhTASEP}
\end{equation}
Then, the map $\tau_i\mapsto T_{i,i+1}$ is an homomorphism of algebra 
$\mathcal{T}_n\rightarrow End\big((\CC^m)^{\otimes n}\big)$. 
\end{proposition}

\begin{remark}
The representation given above, when restricted to the Markovian condition $\zeta_j=-\nu_{ij}$, $\forall i,j$, 
corresponds to the Markov matrix studied in \cite{AritaM13}. Hence, the R-matrix $\check R^\zeta(x,z)$ 
constructed here provides an R-matrix for this model. Note that the $T$-matrix \eqref{eq:Tn_inhTASEP} does not obey the Hecke algebra
except when all $\zeta_j$ are equal. To deal with genuine inhomogeneous hopping rates, the Hecke algebra is not sufficient: 
one needs the $\cT_n$ algebra introduced in this paper.
\end{remark}

From $S$ (resp. $T$) given by \eqref{eq:Sn_inhTASEP} (resp. \eqref{eq:Tn_inhTASEP}), one gets an $R$-matrix using Baxterisation \eqref{eq:Sn_Rc2} 
(resp. \eqref{eq:Tn_Rc}). Surprisingly enough, from these two $R$-matrices, we can obtain another one, as stated in the following theorem.

\begin{theorem}\label{th:Sn_prod}
Let us define the matrix
\begin{equation}
 \check R(x,z)=\check R^\rho(x,z)\check R^\zeta(x,z)\; ,
\end{equation} 
where  $\check R^\rho(x,y)=(\II-yS)(\II-xS)^{-1}$ and  $\check R^\zeta(x,y)=(\II-xT)(\II-yT)^{-1}$
are respectively the Baxterisations of $S$ given by \eqref{eq:Sn_inhTASEP} and of $T$ given by \eqref{eq:Tn_inhTASEP}.

If the relations $\rho_i\nu_{i,j}=\mu_{i,j}\zeta_j$ for $1\leq i<j\leq m$ hold, $\check R(x,z)$
satisfies the braided Yang--Baxter equation and is unitary.
\end{theorem}

\proof 
It is proven by direct computations.
\finproof

Let us remark that the relations $\rho_i\nu_{i,j}=\mu_{i,j}\zeta_j$ are equivalent to $[T,S]=0$.

The $R$-matrix introduced in the theorem \ref{th:Sn_prod} is a generalization of the matrix introduced in \cite{Cantini16} 
to study the multi-species totally 
asymmetric exclusion process with different hopping rates. More explicitly, setting $\mu_{i,j}=-\rho_i$ and  $\nu_{i,j}=-\zeta_j$, 
the corresponding $R$-matrix provided by the theorem \ref{th:Sn_prod}
gives the integrable local jump operator of the multi-species totally 
asymmetric exclusion process with different hopping rates:
\begin{equation}
\left.\frac{\partial \check R(x,z)}{\partial x}\right|_{x=z=0} = m
= \sum_{1\leq i<j \leq m}    (\rho_{i}-\zeta_j)\ E_{ii}\otimes E_{jj} -(\rho_{i}-\zeta_j)\ E_{ji}\otimes E_{ij}\;.
\end{equation}
It describes a Markovian model with $m$ species of particles, where the local exchange rules are given by 
\begin{equation}
 i j  \xrightarrow{\ \rho_{i}-\zeta_j \ }  j i, \quad \mbox{ if } i<j.
\end{equation}
The $R$-matrix was already presented in \cite{Cantini16}, but without the factorization provided by the theorem \ref{th:Sn_prod}.
Using the property $[T,S]=0$, it can be rewritten as
\begin{equation}
\check R(x,y)=\Sigma(x,y)\,\Sigma(y,x)^{-1} \mbox{ with } \Sigma(x,y)=(\II-yS)(\II-xT).
\end{equation}

To conclude this paragraph, we can mention that numerous  questions are still open. 
The Hecke algebra has been a very useful tool in different contexts: {\it e.g.} it is the centralizer of the quantum group 
$\cU_q(gl_N)$ \cite{Jimbo86} or it permits to construct link invariants \cite{Jones87}. We believe that the algebra $\cS_n$ 
we have introduced here should have similar fields of applications.
We think that the classification of its irreducible representations should be also interesting.
The defining relations of the algebra $\cS_n$ look also very similar to the ones of the braid group: 
the connections between these two algebras should also be explored.

In theorem \ref{th:Sn_prod}, we show that the product
of two $R$-matrices based on $\cS_n$ and $\cT_n$ provides a new $R$-matrix if a simple condition on the parameters holds. 
It would be interesting to understand if this feature is 
associated to the special representation used in the theorem or if it is still true at the algebraic level.

Finally, the list of $R$-matrices provided in this paragraph may be 
used to introduce also new models in the context of quantum mechanics (spin chains) or 2D-statistical models (loop or vertex models). 
The knowledge of their associated $R$-matrix may allow one to solve them using, for example, 
the algebraic Bethe ansatz \cite{SklyaninTF79} or the matrix ansatz \cite{DerridaEHP93,BlytheE07} (see chapter \ref{chap:three}).   

\paragraph*{${\cal M}_n$ algebra}

\begin{definition}\label{def:Mn_algebra}
For any integer $n\geq1$, $\mathcal{M}_n$ is the unital associative algebra over $\CC$ with generators 
$\sigma_{1},\dots,\sigma_{n-1}$ and subject to
the relations
 \begin{eqnarray}
&&  \sigma_{i}\sigma_{i+1}\sigma_{i}-\sigma_{i+1}\sigma_{i}\sigma_{i+1}=\lambda(\sigma_{i+1}^2-\sigma_{i}^2)+\mu(\sigma_{i+1}-\sigma_{i}) \,,
\qquad i=1,...,n-2 \qquad \label{eq:Mn_relation1}\\
&&   \sigma_{i}\sigma_{i+1}^2=\sigma_{i}^2\sigma_{i+1}, \qquad \sigma_{i+1}\sigma_{i}^2=\sigma_{i+1}^2\sigma_{i} \label{eq:Mn_relation2} \\
&&  [\sigma_{i}\,,\,\sigma_{j}] = 0\,,\quad |i-j|>1. \label{eq:Mn_relation3}
 \end{eqnarray}
\end{definition}

The following theorem contains the main result of this paragraph and justifies the introduction of the algebra $\cM_n$.
 \begin{theorem}\label{th:Mn_main}
   If $\sigma_i$ satisfy the relations of $\mathcal{M}_n$, then 
    \begin{eqnarray}
 \check R_i(x,y)= \Sigma_i(x,y)\Sigma_i(y,x)^{-1}\quad\text{where}\quad \Sigma_i(x,y)=1-\mu xy-(x+\lambda xy)\sigma_i \label{eq:Mn_Rmatrix}
 \end{eqnarray} 
 satisfy the braided Yang--Baxter equation
   \begin{equation}
   \check R_{i}(x,y) \check R_{i+1}(x,z)\check R_{i}(y,z)=\check R_{i+1}(y,z)\check R_{i}(x,z)\check R_{i+1}(x,y)\;. \label{eq:Mn_bybe}
   \end{equation}
and also the locality, unitarity and regularity properties.
\end{theorem}

Note that the inverse $\Sigma_i(y,x)^{-1}$ in the definition of $\check R$ has to be understood as the formal series
\begin{equation}
 \Sigma_i(y,x)^{-1}=\sum_{k=0}^{+\infty} (\mu xy+(y+\lambda xy)\sigma_i)^k.
\end{equation}

\proof
The first step is to establish the relation
\begin{equation} \label{eq:Mn_eq1}
 \sigma_{i}^n\sigma_{i+1}\sigma_{i}-\sigma_{i+1}\sigma_{i}\sigma_{i+1}^n=\lambda(\sigma_{i+1}^{n+1}-\sigma_{i}^{n+1})+\mu(\sigma_{i+1}^n-\sigma_{i}^n),
\end{equation}
for any integer $n\geq 1$
This can be done by induction: the $n=1$ case is nothing else but the algebraic relation \eqref{eq:Mn_relation1}. Then if we assume that 
the relation holds for a given $n\geq 1$, we have that
\begin{eqnarray*}
& & \sigma_{i}^{n+1}\sigma_{i+1}\sigma_{i}= \sigma_{i}\sigma_{i+1}\sigma_{i}\sigma_{i+1}^n
 +\lambda(\sigma_{i}\sigma_{i+1}^{n+1}-\sigma_{i}^{n+2})+\mu(\sigma_{i}\sigma_{i+1}^n-\sigma_{i}^{n+1}) \\
 &=&  \Big[\sigma_{i+1}\sigma_{i}\sigma_{i+1}+\lambda(\sigma_{i+1}^2-\sigma_{i}^2)+\mu(\sigma_{i+1}-\sigma_{i})\Big]\sigma_{i+1}^n
 +\lambda(\sigma_{i}\sigma_{i+1}^{n+1}-\sigma_{i}^{n+2})+\mu(\sigma_{i}\sigma_{i+1}^n-\sigma_{i}^{n+1}) \\
 &=&\sigma_{i+1}\sigma_{i}\sigma_{i+1}^{n+1}
 +\lambda(\sigma_{i+1}^{n+2}-\sigma_{i}^{n+2})+\mu(\sigma_{i+1}^{n+1}-\sigma_{i}^{n+1}),
\end{eqnarray*}
where the last equality is obtained by pushing all the powers to the left using the relations
\begin{equation} \label{eq:Mn_eq2}
 \sigma_{i}^p\sigma_{i+1}^q=\sigma_{i}^{p+q-1}\sigma_{i+1}, \quad  \sigma_{i+1}^p\sigma_{i}^q=\sigma_{i+1}^{p+q-1}\sigma_{i}, 
 \quad \mbox{for } p,q\geq 1.
\end{equation}
These last equations are directly derived from \eqref{eq:Mn_relation2}.

Equations \eqref{eq:Mn_eq1} and \eqref{eq:Mn_eq2} immediately imply that
\begin{equation*}
 \sigma_{i}^p\sigma_{i+1}^q\sigma_{i}^r-\sigma_{i+1}^r\sigma_{i}^q\sigma_{i+1}^p=
 \lambda(\sigma_{i+1}^{p+q+r-1}-\sigma_{i}^{p+q+r-1})+\mu(\sigma_{i+1}^{p+q+r-2}-\sigma_{i}^{p+q+r-2}), \quad \mbox{for } p,q,r\geq 1
\end{equation*}
This is equivalent to the relation
\begin{eqnarray} \label{eq:Mn_eqbis}
& & \frac{\sigma_{i}}{1-X\sigma_{i}} \cdot \frac{\sigma_{i+1}}{1-Y\sigma_{i+1}} \cdot \frac{\sigma_{i}}{1-Z\sigma_{i}}
 -\frac{\sigma_{i+1}}{1-Z\sigma_{i+1}} \cdot \frac{\sigma_{i}}{1-Y\sigma_{i}} \cdot \frac{\sigma_{i+1}}{1-X\sigma_{i+1}} \\
&=& \frac{\lambda \sigma_{i+1}^2+\mu \sigma_{i+1}}{(1-X\sigma_{i+1})(1-Y\sigma_{i+1})(1-Z\sigma_{i+1})}
 -\frac{\lambda \sigma_{i}^2+\mu \sigma_{i}}{(1-X\sigma_{i})(1-Y\sigma_{i})(1-Z\sigma_{i})}.
\end{eqnarray}
and the relation \eqref{eq:Mn_eq2} can be equivalently rewritten as
\begin{equation}
 \frac{\sigma_{i}}{1-X\sigma_{i}}\cdot \frac{\sigma_{i+1}}{1-Y\sigma_{i+1}}=\frac{\sigma_{i}}{1-Y\sigma_{i}} \cdot \frac{\sigma_{i+1}}{1-X\sigma_{i+1}}.
\end{equation}
 We are now ready to prove the Yang-Baxter equation. It will be convenient to write 
 \begin{equation}
  \check R_i(x,y)= 1+ \frac{(y-x)\sigma_i}{\Sigma_i(y,x)}.
 \end{equation}
 We can then expand
 \begin{eqnarray}
  & & \check R_{i}(x,y) \check R_{i+1}(x,z)\check R_{i}(y,z)-\check R_{i+1}(y,z)\check R_{i}(x,z)\check R_{i+1}(x,y) \\
  &=& \hspace{-3mm}(y-x)(z-x)(z-y)
  \left[\frac{\sigma_{i}}{\Sigma_{i}(y,x)} \cdot \frac{\sigma_{i+1}}{\Sigma_{i+1}(z,x)} \cdot \frac{\sigma_{i}}{\Sigma_{i}(z,y)}
  -\frac{\sigma_{i+1}}{\Sigma_{i+1}(z,y)} \cdot \frac{\sigma_{i}}{\Sigma_{i}(z,x)} \cdot \frac{\sigma_{i+1}}{\Sigma_{i+1}(y,x)}\right] \nonumber \\
  +& & \hspace{-10mm}(y-x)(z-y)\left[ \frac{\sigma_{i}}{\Sigma_{i}(y,x)} \cdot \frac{\sigma_{i}}{\Sigma_{i}(z,y)}
  -\frac{\sigma_{i+1}}{\Sigma_{i+1}(z,y)} \cdot \frac{\sigma_{i+1}}{\Sigma_{i+1}(y,x)} \right] 
  +(z-x)\left[ \frac{\sigma_{i+1}}{\Sigma_{i+1}(z,x)} - \frac{\sigma_{i}}{\Sigma_{i}(z,x)} \right] \nonumber \\
  +& & \hspace{-10mm}(y-x) \left[ \frac{\sigma_{i}}{\Sigma_{i}(y,x)} - \frac{\sigma_{i+1}}{\Sigma_{i+1}(y,x)} \right]
  +(z-y) \left[ \frac{\sigma_{i}}{\Sigma_{i}(z,y)} - \frac{\sigma_{i+1}}{\Sigma_{i+1}(z,y)} \right]. \nonumber 
 \end{eqnarray}
 Using relation \eqref{eq:Mn_eqbis}, we are left to prove that
 \begin{eqnarray}
  & &  (y-x)(z-x)(z-y)\frac{\lambda \sigma_{i}^2+\mu \sigma_{i}}{\Sigma_{i}(y,x)\Sigma_{i}(z,x)\Sigma_{i}(z,y)}  \\
 & = & (y-x)(z-y) \frac{\sigma_{i}^2}{\Sigma_{i}(y,x)\Sigma_{i}(z,y)}+(y-x)\frac{\sigma_{i}}{\Sigma_{i}(y,x)}+(z-y)\frac{\sigma_{i}}{\Sigma_{i}(z,y)}
 -(z-x)\frac{\sigma_{i}}{\Sigma_{i}(z,x)}, \nonumber 
 \end{eqnarray}
 as well as the same equation for $\sigma_{i+1}$.
 This is done by a direct computation (all the quantities involved commute with each other).
\finproof

\begin{remark}
 Note that contrary to Hecke and BMW algebras, the defining relations for the ${\cal M}_n$, ${\cal S}_n$, ${\cal T}_n$ algebras do not imply 
 necessarily the existence of a minimal polynomial for $\sigma_i$.
 Moreover the Baxterisation of the latter algebras depends on two spectral parameters.
\end{remark}

\subsection{Diagonalization of the transfer matrix}

We presented in the previous sections the construction of integrable Markov matrix. The integrability was defined as the fact that the Markov 
matrix enters a set of commuting operators, generated by a transfer matrix. We stressed that this strong property translates into 
a lot of conserved quantities and is thus a hint for an exact solvability of the model. This also means that the Markov matrix and the transfer matrix
share common eigenvectors (these eigenvectors are independent of the spectral parameter). This justifies the following definition.

\begin{definition}
 We denote by $\{\ket{\Psi_i}\}_{1\leq i \leq 2^L}$ (respectively $\{E_i(z)\}_{1\leq i \leq 2^L}$) the eigenvectors (respectively the eigenvalues) 
 of the transfer matrix $t(z)$
 \begin{equation}
  t(z)\ket{\Psi_i}= E_i(z) \ket{\Psi_i}.
 \end{equation}
 We thus have for the Markov matrix
 \begin{equation}
  M\ket{\Psi_i}= \theta E_i'(1) \ket{\Psi_i}. 
 \end{equation}
\end{definition}

\begin{remark}
 We are dealing here with models defined on a periodic lattice. In practice we will encounter a lot of systems for which the stochastic dynamics 
 conserves the number of particles of each species on the periodic lattice (in simple models the particles are often only allowed to jump 
 from site to site but not to appear or disappear). We define
 \begin{equation} \label{eq:diago_particle_number}
  \cN(\tau) = \sum_{i=1}^L n_i(\tau),
 \end{equation}
 where $n_i(\tau)$ is a diagonal $(N+1)\times(N+1)$ matrix acting non-trivially on site $i$ (and trivially on other sites) as
 $diag(\underbrace{0,\dots,0}_{\tau},1,\underbrace{0,\dots,0}_{N-\tau})$. The operator $\cN(\tau)$ is constructed to count the number of 
 particles of species $\tau$. Then we have for all $\tau,\tau'$
 \begin{equation}
  [t(z),\cN(\tau)]=0, \quad \mbox{and} \quad [\cN(\tau),\cN(\tau')]=0.
 \end{equation}
 The commutation properties stated just above 
 proves that for such systems the action the transfer matrix is block diagonal (one block corresponds to a fixed number of particles of 
 each species that we call sometimes sector) and that the diagonalization can be performed independently on each sector.
\end{remark}

We would like to determine the complete set of eigenvectors of the Markov matrix.
Several methods can be used to diagonalize the Markov matrix (they have been historically developed to deal with quantum Hamiltonians).
We present below two of them: the coordinate Bethe ansatz and the algebraic Bethe ansatz and illustrate them on examples. We briefly 
introduce in a second time other diagonalization techniques.

Note that we focus here on the diagonalization of the homogeneous transfer matrix (or equivalently of the Markov matrix) but some of the methods 
presented below can be also transposed to the case of inhomogeneity parameters.

\subsubsection{Coordinate Bethe ansatz}

The problem of finding the eigenvectors of a Markov matrix is very similar (at least for the models studied here) to 
the case of quantum Hamiltonians. Stochastic models defined on a one dimensional lattice are indeed often related through a similarity transformation
to one dimensional quantum spin chains. 

The analytical computation of the eigenvectors and eigenvalues of a quantum Hamiltonian is a rather difficult question, 
that is even usually impossible to achieve.
However, in \cite{Bethe31}, H. Bethe succeeded in computing exactly the spectrum of the one-dimensional 
Heisenberg quantum spin chain \cite{Heisenberg28}, introducing a method that is now called after him: the coordinate Bethe ansatz (CBA). 
This technique can be thought as a generalized Fourier transform and opened basically a new field of research. Since then the method had been 
widely used in different contexts (quantum spin chains, quantum systems, Markovian processes). 
We point out the pioneering works \cite{YangY66,Yang67,Sutherland68,Gaudin71bis,Gaudin1971bisbis,Sutherland75,Gaudin83}.

We begin to present the coordinate Bethe ansatz in systems involving a single species of particle, {\it i.e} $N=1$ in our notation
(we recall that in this particular case the occupation variables $\tau_i$ take only two values $0$ for holes and $1$ for particles). 
A more involved case with two-species of particles will be studied below.
We are interested in models where the stochastic dynamics encoded in the Markov matrix conserves the number of particles. The discussion
on the block diagonal decomposition of the Markov matrix suggests the following definition.

\begin{definition}
For $1\leq x\leq L$ we introduce the vector
\begin{equation}
 \ket{\{x\}}=\underbrace{\ket{0} \otimes \dots \otimes \ket{0}}_{x-1} \otimes \ket{1} \otimes \underbrace{\ket{0} \otimes \dots \otimes \ket{0}}_{L-x}
\end{equation}
and more generally for $1\leq x_1<x_2<\dots<x_r\leq L$ we introduce the vector
\begin{equation}
 \ket{\{x_1,x_2,\dots,x_r\}}= 
 \underbrace{\ket{0} \otimes \dots \otimes \ket{0}}_{x_1-1} \otimes \ket{1} \otimes \underbrace{\ket{0} \otimes \dots \otimes \ket{0}}_{x_2-x_1-1}
 \otimes \ket{1} \otimes \dots \otimes \ket{1} \otimes \underbrace{\ket{0} \otimes \dots \otimes \ket{0}}_{L-x_r}.
\end{equation}
\end{definition}

\begin{proposition}
The vectors $\ket{\{x_1,x_2,\dots,x_r\}}$ define a basis of the sector with $r$ particles (i.e of the eigenspace associated to the eigenvalue $r$
of the operator $\cN(1)$ introduced in \eqref{eq:diago_particle_number}).
For each vector $\ket{\Psi}$ living in this sector (i.e satisfying $\cN(1)\ket{\Psi}=r\ket{\Psi}$)
we have the expansion 
\begin{equation}
 \ket{\Psi}=\sum_{1\leq x_1<x_2<\dots<x_r\leq L} a(x_1,x_2,\dots,x_r) \ket{\{x_1,x_2,\dots,x_r\}}.
\end{equation}
\end{proposition}

The coordinate Bethe ansatz consists in assuming the following Fourier-like decomposition of the coefficient $a(x_1,x_2,\dots,x_r)$
\begin{equation} \label{eq:CBA_ansatz}
 a(x_1,x_2,\dots,x_r) = \sum_{\sigma \in \mathfrak{S}_r} A_{\sigma} \  u_{\sigma(1)}^{x_1}u_{\sigma(2)}^{x_2} \dots u_{\sigma(r)}^{x_r}
\end{equation}
$\mathfrak{S}_r$ denotes the permutation group of the elements $\{1,2,\dots,r\}$.
The parameters $u_1$, $u_2$,..., $u_r$ are called the Bethe roots and are expected to satisfy a set of polynomial algebraic relations called
the Bethe equations (see examples below). The coefficients $A_\sigma$ are expected to be determined by exchange or scattering relations
$A_{\sigma t_i} = \check S(u_{\sigma(i)},u_{\sigma(i+1)}) A_\sigma$, where $t_i$ is the permutation $i\leftrightarrow i+1$ (this last relation 
is sufficient to express any coefficient $A_\sigma$ in function of the coefficient $A_{Id}$ since the permutations
$t_1,t_2,\dots,t_{r-1}$ generate the group $\mathfrak{S}_r$).

\paragraph*{The example of the ASEP}

The ASEP is a stochastic process whose Markov matrix has been defined in \eqref{eq:Markov_matrix_sum_decomposition}, with the local jump operator $m$
given in \eqref{eq:ASEP_m}. We argued previously that it raised a lot of interest in both 
physical and mathematical community. On the physical side it is one of the simplest out-of-equilibrium model that can be defined on a ring.
On a mathematical side it is integrable and give rise to exact computations. 
We present the exact diagonalization of this model using CBA. 

\begin{example}
As a warm-up we begin to expose the method in the sector with $1$ particle. The eigenvalue equation $M\ket{\Psi}=E\ket{\Psi}$, where
$M$ is the Markov matrix encoding the dynamics of the ASEP and $\ket{\Psi}= \sum_{x=1}^L a(x) \ket{\{x\}}$, can be written in components as
\begin{equation}
 E a(x) = p a(x-1)+q a(x+1)-(p+q)a(x), \quad \mbox{for} \quad x=1,\dots,L 
\end{equation}
where we have assumed the periodicity constraints $a(L+1)=a(1)$ and $a(0)=a(L)$ for this equation to hold also for $x=1$ and $x=L$. 
Plugging the Bethe ansatz $a(x)=Au^x$ in this last equation yields
\begin{equation}
 E \ Au^x = pAu^{x-1}+qAu^{x+1}-(p+q)Au^x.
\end{equation}
Dividing by $Au^x$ provides the expression of the eigenvalue
\begin{equation}
 E=\frac{p}{u}+qu-(p+q).
\end{equation}
The periodicity imposes $a(x+L)=a(x)$ which translates into the Bethe equation $u^L=1$. The Bethe equation admits $L$ different solutions
(Bethe roots) which provide for each one an eigenvector of the Markov matrix in the sector with $1$ particle. This sector being of dimension $L$, 
we thus have diagonalized completely the Markov matrix in this subspace.
\end{example}

In order to get used to the Bethe ansatz and to understand the general structure we present now the case of the $2$ particles sector.
This is the simplest situation where the coefficients $A_{\sigma}$ play an important role.

\begin{example}
For $1\leq x_1<x_2 \leq L$ and $x_1$, $x_2$ not nearest neighbors on the ring, {\it i.e}  $x_2 \neq x_1+1$ and $(x_1,x_2) \neq (1,L)$ 
(this particular case is studied below), the eigenvalue equation reads
\begin{equation} \label{eq:ASEP_CBA_eigen_eq_2_particles}
 E \ a(x_1,x_2) = p a(x_1-1,x_2)+p a(x_1,x_2-1)+q a(x_1+1,x_2)+q a(x_1,x_2+1)-2(p+q)a(x_1,x_2).
\end{equation}
where we have assumed the periodicity constraints $a(x_1,L+1)=a(1,x_1)$ and $a(0,x_2)=a(x_2,L)$.
The ansatz $a(x_1,x_2)=A_{12} \ u_1^{x_1}u_2^{x_2}+A_{21} \ u_2^{x_1}u_1^{x_2}$ (note that we made a slight abuse of notation by writing
for clarity and convenience $A_{12}$ instead of $A_{Id}$ and $A_{21}$ instead of $A_{t_1}$) fulfills the eigenvalue equation provided that the 
eigenvalue is equal to
\begin{equation} \label{eq:ASEP_CBA_eigenvalue_2_particles}
 E=p\left(\frac{1}{u_1}+\frac{1}{u_2}\right)+q(u_1+u_2)-2(p+q).
\end{equation}
When $x_1$ and $x_2$ are close together, {\it i.e} when $x_2=x_1+1$, the eigenvalue equation takes a slightly different form
\begin{equation}
 E \ a(x_1,x_2) = p a(x_1-1,x_2)+q a(x_1,x_2+1)-(p+q)a(x_1,x_2).
\end{equation}
Subtracting this equation to \eqref{eq:ASEP_CBA_eigen_eq_2_particles} (which holds for every $x_1,x_2$ thanks to the ansatz and the eigenvalue 
being fixed to \eqref{eq:ASEP_CBA_eigenvalue_2_particles}) yields the 'boundary' condition
\begin{equation}
 p a(x_1,x_1)+q a(x_1+1,x_1+1)-(p+q)a(x_1,x_1+1)=0.
\end{equation}
This latter equation is solved if the coefficients $A_{12}$ and $A_{21}$ satisfy
\begin{equation}
 A_{21}=-\frac{p+qu_1u_2-(p+q)u_2}{p+qu_1u_2-(p+q)u_1}A_{12}.
\end{equation}
The periodicity constraints $a(x_1,L+1)=a(1,x_1)$ and $a(0,x_2)=a(x_2,L)$ imply the Bethe equations
\begin{equation}
 u_1^L=\frac{A_{12}}{A_{21}}=-\frac{p+qu_1u_2-(p+q)u_1}{p+qu_1u_2-(p+q)u_2}
\end{equation}
and
\begin{equation}
 u_2^L=\frac{A_{21}}{A_{12}}=-\frac{p+qu_1u_2-(p+q)u_2}{p+qu_1u_2-(p+q)u_1}.
\end{equation}
The problem of diagonalizing the Markov matrix in the $2$ particles sector is thus reduced to the resolution of these two polynomial equations
of degree $L+2$. This is still a hard task but this reduction is a huge step forward because the direct diagonalization of the Markov matrix in this 
sector requires to find the roots of the characteristic polynomial of degree $L(L-1)/2$ (which grows much faster than $L$ for $L$ large). 
\end{example}

Although the $3$ particles case could appear as a straightforward generalization of the $2$ particles case, it is in fact a crucial step to check 
toward the validity of the ansatz for a general sector with a given number of particles. It is indeed necessary to verify that the 
'three-body interaction' (see details below) does not bring new constraints and factorizes into two-body interactions.  

\begin{example}
For $1\leq x_1<x_2<x_3\leq L$ far from each other, the eigenvalue equation can be written in components as
\begin{eqnarray*}
 E a(x_1,x_2,x_3) & = & p\left[a(x_1-1,x_2,x_3)+a(x_1,x_2-1,x_3)+a(x_1,x_2,x_3-1)\right] \\
 & & + q\left[a(x_1+1,x_2,x_3)+a(x_1,x_2+1,x_3)+a(x_1,x_2,x_3+1)\right] \\
 & & -3(p+q)a(x_1,x_2,x_3) =0
\end{eqnarray*}
The ansatz for the expression of $a(x_1,x_2,x_3)$ given by the general formula \eqref{eq:CBA_ansatz} contains $3!=6$ terms and 
imposes the following expression for the eigenvalue
\begin{equation}
 E= p\left(\frac{1}{u_1}+\frac{1}{u_2}+\frac{1}{u_3}\right)+q(u_1+u_2+u_3)-3(p+q).
\end{equation}
The two-body interaction, given by the case $x_2=x_1+1$ and $x_3$ far from $x_1$ and $x_2$, yields the 'boundary' condition
\begin{equation} \label{eq:ASEP_CBA_twobody_12}
 pa(x_1,x_1,x_3)+qa(x_1+1,x_1+1,x_3)-(p+q)a(x_1,x_1+1,x_3)=0.
\end{equation}
This equation is solved if
\begin{eqnarray*}
& &\frac{A_{213}}{A_{123}}=-\frac{p+qu_1u_2-(p+q)u_2}{p+qu_1u_2-(p+q)u_1}, \quad 
\frac{A_{132}}{A_{312}}=-\frac{p+qu_1u_3-(p+q)u_1}{p+qu_1u_3-(p+q)u_3}, \\
& &\frac{A_{321}}{A_{231}}=-\frac{p+qu_2u_3-(p+q)u_3}{p+qu_2u_3-(p+q)u_2}.
\end{eqnarray*}
Once again we used the notation $A_{\sigma(1)\sigma(2)\sigma(3)}$ instead of $A_{\sigma}$ for $\sigma \in \mathfrak{S}_3$.
Similarly, the two-body interaction, given by the case $x_3=x_2+1$ and $x_1$ far from $x_2$ and $x_3$, yields the 'boundary' condition
\begin{equation} \label{eq:ASEP_CBA_twobody_23}
 pa(x_1,x_2,x_2)+qa(x_1,x_2+1,x_2+1)-(p+q)a(x_1,x_2,x_2+1)=0.
\end{equation}
This equation is again solved if
\begin{eqnarray*}
& &\frac{A_{132}}{A_{123}}=-\frac{p+qu_2u_3-(p+q)u_3}{p+qu_2u_3-(p+q)u_2}, \quad 
\frac{A_{321}}{A_{312}}=-\frac{p+qu_1u_2-(p+q)u_2}{p+qu_1u_2-(p+q)u_1}, \\
& &\frac{A_{213}}{A_{231}}=-\frac{p+qu_1u_3-(p+q)u_1}{p+qu_1u_3-(p+q)u_3}.
\end{eqnarray*}

The three-body interaction given by the case $x_2=x_1+1$ and $x_3=x_1+2$ yields the boundary equation
\begin{eqnarray*}
& & q\left[a(x_1+1,x_1+1,x_1+2)+a(x_1,x_1+2,x_1+2)\right] \\
& & +p\left[a(x_1,x_1,x_1+2)+a(x_1,x_1+1,x_1+1)\right]-2(p+q)a(x_1,x_1+1,x_1+2)=0
\end{eqnarray*}
This appears to be exactly the sum of the equations \eqref{eq:ASEP_CBA_twobody_12} and \eqref{eq:ASEP_CBA_twobody_23} given by 
the two-body interactions and hence it does not bring any new constraint and it is automatically fulfilled.

The periodicity constraints $a(0,x_2,x_3)=a(x_2,x_3,L)$ and $a(x_1,x_2,L+1)=a(1,x_1,x_2)$ impose the Bethe equations
\begin{equation}
 u_1^L=\frac{A_{123}}{A_{231}}=\frac{A_{123}}{A_{213}}\cdot \frac{A_{213}}{A_{231}}= \frac{p+qu_1u_2-(p+q)u_1}{p+qu_1u_2-(p+q)u_2} \cdot 
 \frac{p+qu_1u_3-(p+q)u_1}{p+qu_1u_3-(p+q)u_3}
\end{equation}
and similarly
\begin{equation}
  u_2^L= \frac{p+qu_1u_2-(p+q)u_2}{p+qu_1u_2-(p+q)u_1} \cdot \frac{p+qu_2u_3-(p+q)u_2}{p+qu_2u_3-(p+q)u_3}
\end{equation}
and
\begin{equation}
  u_3^L= \frac{p+qu_1u_3-(p+q)u_3}{p+qu_1u_3-(p+q)u_1} \cdot \frac{p+qu_2u_3-(p+q)u_3}{p+qu_2u_3-(p+q)u_2}.
\end{equation}
\end{example}

We can now move to the general case.

\begin{proposition}
In the $r$ particles sector, the eigenvectors of the Markov matrix $M$ are given by
\begin{equation}
\ket{\Psi}=\sum_{1\leq x_1<x_2<\dots<x_r\leq L} \sum_{\sigma \in \mathfrak{S}_r}
A_{\sigma} \  u_{\sigma(1)}^{x_1}u_{\sigma(2)}^{x_2} \dots u_{\sigma(r)}^{x_r} \ket{\{x_1,x_2,\dots,x_r\}},
\end{equation}
where the Bethe roots $u_1,\dots,u_r$ satisfy the Bethe equations
\begin{equation} \label{eq:CBA_Bethe_equations}
 u_i^L=(-1)^{r-1}\prod_{\genfrac{}{}{0pt}{}{j=1}{j \neq i}}^r \frac{p+qu_iu_j-(p+q)u_i}{p+qu_iu_j-(p+q)u_j}
\end{equation}
and the coefficients $A_{\sigma}$ are determined (up to an overall normalization) by the relations
\begin{equation}
 A_{\sigma \circ t_i}=-\frac{p+qu_{\sigma(i)}u_{\sigma(i+1)}-(p+q)u_{\sigma(i+1)}}{p+qu_{\sigma(i)}u_{\sigma(i+1)}-(p+q)u_{\sigma(i)}}A_{\sigma}.
\end{equation}
The associated eigenvalue is given by
\begin{equation} \label{eq:CBA_eigenvalue}
 E= \sum_{i=1}^r \left(\frac{p}{u_i}+qu_i-(p+q)\right).
\end{equation}
\end{proposition}

\proof
This can be proved by showing that the $r$-body interaction reduces to the sum of two-body interactions (similarly to what was done previously for the 
three-body interaction).
It has been done in details for the XXZ spin chain \cite{YangY66} which is known to be similar, up to a gauge transformation, to the Markov matrix
of the ASEP.
\finproof

We stress that the diagonalization problem has been reduced to the resolution a set of algebraic equation of degree $L+2(r-1)$ 
(that has to be put in contrast with finding the roots of the characteristic polynomial of the Markov matrix
which is of degree $\left(\genfrac{}{}{0pt}{}{L}{r}\right)$ in this sector). This is still a very hard task that we cannot handle in general. 
However, several methods have been developed to study the Bethe equations in the thermodynamic limit \cite{YangY66bis,Takahashi71,Gaudin71,Baxter82}
and allow exact computations in this limit. Moreover this set of equations can be efficiently studied with numerical methods.

\paragraph*{The CBA as a tool to construct integrable models}

The goal of this paragraph is to show how the CBA can be used to construct exactly solvable models. It presents the main results of 
\cite{CrampeFRV16bis}

Since the work of Bethe \cite{Bethe31}, the classification of solvable one-dimensional systems has been the heart of a lot of researches.
General methods are now known but involve huge computations 
which, in general, do not permit to provide a classification of solvable models. See subsection \ref{subsec:Find_Rmatrix} for details.

In this section, we are interested in finding Markov matrix $M$ such that the following eigenvalue problem
\begin{equation}\label{eq:H33_eigen}
 M\Psi =E \Psi
\end{equation}
can be solved exactly. The Markov matrices under consideration correspond to
 one-dimensional lattice gases with nearest neighbor interactions \cite{KatzLS84,KrapivskyRB10,SchmittmannZ95}, and are written as follows
\begin{equation}
 M=\sum_{\ell=1}^L m_{\ell,\ell+1}\;
\end{equation}
where we assume periodic boundary conditions (by convention $L+1\equiv 1$) and the indices $\ell,\ell+1$ indicate on which sites the 
local operator $m$ acts. This type of problem also appears in the context of two-dimensional equilibrium statistical models 
or one-dimensional spin chains. The archetypes of such models are respectively the 6-vertex model \cite{Baxter82} 
or the Heisenberg spin chain \cite{Heisenberg28}.
Here, we focus on the case where $m$ is a $9\times 9$ matrix which means that each site may take three different values.

The cases when $M$ commutes with some charges of the following type
\begin{equation}\label{eq:H33_charge}
 \cN=\sum_{\ell=1}^L  n_{\ell}
\end{equation}
are of particular interest. 
In the context of out-of-equilibrium models, the number of conserved charges 
corresponds to the number of conserved species of particles (one of the value is for the empty site).

When $M$ commutes with two different charges, the number of non vanishing elements of $m$ is reduced to $15$ (in the context of statistical mechanics, 
it is called the 15-vertex model). In out-of-equilibrium statistical physics, it corresponds to models where there are two classes of 
conserved particles (such as two species ASEP).

When $M$ commutes with only one charge, they are several possibilities depending on the degeneracy of the eigenvalues of $n$. 
One usually takes $n$ with three different eigenvalues: the local jump operator $m$ has then $19$ non vanishing entries. 
Such solvable models have been classified and studied previously in \cite{IdzumiTA94, PimentaM11, Martins13, CrampeFR13, FonsecaFR15}.  
They correspond to out-of-equilibrium models with only one species of particle, but where two particles can occupy the same site.

We concentrate here on Markov matrices $M$ commuting with one charge that has one eigenvalue degenerated twice: 
$m$ has then $33$ non vanishing entries. 
Usually, the denomination $33$-vertex model is dedicated to integrable models whose $R$-matrix has 
$33$ non-vanishing entries. However, it appears that for the models solvable by CBA, when an $R$-matrix can be constructed, 
its non-vanishing entries coincide with the ones of $m$, see construction in \cite{FonsecaFR15,CrampeFRV17}. 
Although this property has not been proven in full generality, but only checked case by case, we will call our Markov matrices, $33$-vertex 
Markov matrices.
These models may be interpreted, in the context of out-of-equilibrium model, as diffusing particles possessing two internal degrees of freedom.  
Let us emphasize that such models  have been also introduced 
to study mRNA translation in \cite{KlumppCL08, CiandriniSR10}. 
Therefore, we hope that the solvable models introduced here may be helpful in this context or to describe other phenomena.

Let us remark that the three cases described above exhaust all the non trivial cases when $M$ possesses conserved charge(s). 
To fix the notations, we use the canonical basis 
\begin{equation}\label{eq:H33_basis}
 |0\rangle=\begin{pmatrix}1\\0\\0\\ \end{pmatrix}\quad,\qquad |1\rangle=\begin{pmatrix} 0\\1\\0\\ \end{pmatrix}\quad\text{and}\qquad 
|2\rangle=\begin{pmatrix} 0\\0\\1\\ \end{pmatrix}\;.
\end{equation}
The vector $|0\rangle$ will correspond to the empty site whereas $|1\rangle$ and $|2\rangle$ correspond to a 
particle in different internal states. In this context, the most general Markov matrices which preserve the number of particles are the ones
which commute with the charge \eqref{eq:H33_charge} with
\begin{equation} \label{def:H33_q}
n=\begin{pmatrix} 0 &0 &0\\ 0 &1 &0\\ 0 &0 &1\\ \end{pmatrix}\;.
\end{equation}
Therefore the local Hamiltonian $m$ takes the following form
\begin{equation}
m = 
\begin{pmatrix}
m_{11} & 0 & 0 & 0 & 0 & 0 & 0 & 0 & 0 \\
0 & m_{22} & m_{23} & m_{24} & 0 & 0 & m_{27} & 0 & 0 \\
0 & m_{32} & m_{33} & m_{34} & 0 & 0 & m_{37} & 0 & 0 \\
0 & m_{42} & m_{43} & m_{44} & 0 & 0 & m_{47} & 0 & 0 \\
0 & 0 & 0 & 0 & m_{55} & m_{56} & 0 & m_{58} & m_{59} \\
0 & 0 & 0 & 0 & m_{65} & m_{66} & 0 & m_{68} & m_{69} \\
0 & m_{72} & m_{73} & m_{74} & 0 & 0 & m_{77} & 0 & 0 \\
0 & 0 & 0 & 0 & m_{85} & m_{86} & 0 & m_{88} & m_{89} \\
0 & 0 & 0 & 0 & m_{95} & m_{96} & 0 & m_{98} & m_{99} \\
\end{pmatrix}.
\label{eq:H33_m12}
\end{equation}
Note that the $15$-vertex model is a sub-case of the problem studied here in opposition to the $19$-vertex model that exhibits different 
non vanishing entries. 

The goal of this paragraph consists in classifying all such models which are solvable by Coordinate Bethe Ansatz.
 The main result is exposed right below.
The proof is then detailed in several parts. In a first time we perform the first step of the CBA (i.e. the nesting), which 
leads to a reduced problem dealing with $4\times4$ $R$-matrices possessing a specific form. In a second time we classify these $R$-matrices, 
solutions of a braided Yang--Baxter equation with spectral parameters.

{\it Main result :}
We present the final result, after gathering the different constraints that should satisfy the parameters (the computations
are detailed below). We obtain the following classification of 33-vertex Markov matrices.
The jump operator \eqref{eq:H33_m12} is solvable by CBA if and only if its entries obey the following constraints:
\begin{enumerate}
\item[$(i)$] We must have:
\begin{equation}\begin{aligned}
& m_{23}+m_{47}=0\quad&;\qquad m_{32}+m_{74}=0\quad&;\qquad m_{27}=m_{34}=m_{43}=m_{72}=0\\
& m_{24}=m_{37}\quad&;\quad \qquad m_{42}=m_{73}\qquad&;\qquad m_{22}+m_{44}=m_{33}+m_{77},
\end{aligned}
\label{eq:H33_1st.constraint}
\end{equation}
\item[$(ii)$] The matrix
\begin{equation}\label{eq:H33_Tmat}
T= \begin{pmatrix}t_{55} & t_{56} & t_{58} & t_{59} \\
 t_{65} & t_{66} & t_{68} & t_{69} \\
 t_{85} & t_{86} & t_{88} & t_{89} \\
 t_{95} & t_{96} & t_{98} & t_{99} \end{pmatrix}
\end{equation}
where
\begin{equation}\label{eq:H33_m.tilde}
\begin{aligned} 
t_{55} = m_{55} +m_{11} - m_{22} - m_{44} \ ;\quad
t_{58} = m_{58} +m_{23} \ ;\quad t_{85} = m_{85} + m_{32}\ ;\quad t_{59} = m_{59}\\
t_{66} = m_{66} +m_{11} - m_{33} -  m_{44} \ ;\quad
t_{69} = m_{69} +m_{23} \ ;\quad t_{96} = m_{96} + m_{32}\ ;\quad t_{68} = m_{68}\\
t_{88} = m_{88} +m_{11} - m_{22} - m_{77}\ ;\quad
t_{56} = m_{56} -m_{23} \ ;\quad t_{65} = m_{65} - m_{32}\ ;\quad t_{86} = m_{86}\\
t_{99} = m_{99} +m_{11} - m_{33} - m_{77}\ ;\quad
t_{89} = m_{89} -m_{23} \ ;\quad t_{98} = m_{98} - m_{32} \ ;\quad t_{95} = m_{95}
\end{aligned}
\end{equation}
must be a representation of one of these three algebras:
\begin{enumerate}
\item {\it Hecke algebras}
\begin{equation}\label{eq:H33_rel_hecke}
T_{12}T_{23}T_{12}-m_{24}m_{42}\,T_{12}=T_{23}T_{12}T_{23}-m_{24}m_{42}\,T_{23} \quad \mbox{and} \quad
T^2=\mu\,T
\end{equation}
\item {\it $\cT_n$ algebras}
\begin{equation} \label{eq:H33_rel_Tn}
T_{12}T_{23}T_{12}+T_{12}\, (T_{23})^2=T_{23}T_{12}T_{23}+(T_{12})^2\, T_{23}
\end{equation}
\item {\it $\cS_n$ algebras}
\begin{equation} \label{eq:H33_rel_Sn}
T_{12}T_{23}T_{12}+ (T_{23})^2\, T_{12}=T_{23}T_{12}T_{23}+T_{23}\,(T_{12})^2
\end{equation}
\end{enumerate}
Fortunately, the $4\times4$ solutions to relations \eqref{eq:H33_rel_hecke}, \eqref{eq:H33_rel_Tn} or \eqref{eq:H33_rel_Sn} have been classified: 
all the possible expressions of $T$ can be then recovered from these classifications. 
For Hecke algebras, the classification of $4\times 4$ matrices is given in \eqref{eq:Hecke_solu_dim4} and for 
$\cS_n$ and $\cT_n$ algebras the classification is given in theorem \ref{th:Sn_classification}.
\end{enumerate}

We present now the proofs of this result. Constraint $(i)$ is found while performing the first step of the CBA whereas the constraint $(ii)$ 
is obtained by solving the Yang-Baxter equation for the reduced problem.

{\it Coordinate Bethe ansatz :}
As mentioned above, we now focus on jump operators \eqref{eq:H33_m12} that commute with $\cN$, defined by \eqref{eq:H33_charge} and \eqref{def:H33_q}. 
The basis \eqref{eq:H33_basis} allows us to introduce the following elementary states
\begin{equation}\label{eq:H33_elemstate}
 \vert x_1,i_1;x_2,i_2;\dots;x_r,i_r \rangle= |0\rangle^{\otimes x_1-1} \otimes |i_1\rangle\otimes |0\rangle^{\otimes x_2-x_1-1}\otimes |i_2\rangle
 \dots \otimes |i_r\rangle \otimes |0\rangle^{\otimes L-x_r}
\end{equation}
where $i_k=1,2$ and $x_k=1,2,\dots,L$. 
In words, the state $\vert x_1,i_1;x_2,i_2;\dots;x_r,i_r \rangle$ stands for the configuration where the particles are in positions 
$x_1, x_2,\dots,x_r$ 
with internal states $i_1,i_2,\dots i_r$ respectively\footnote{We remind that $|0\rangle$ stands for an empty site.}. 

Notice that $\vert x_1,i_1;x_2,i_2;\dots;x_r,i_r \rangle$ is an eigenvector of $\cN$ with eigenvalue $r$ whatever the values of $x_k$'s and $i_k$'s are. 
In fact, the set of states $\vert x_1,i_1;x_2,i_2;\dots;x_r,i_r \rangle$ spans the vector space with $r$ particles.
Therefore, an Markov matrix eigenstate $\Psi_r$  in a given sector with $r$ particles can be written as a linear combination 
of the elementary states \eqref{eq:H33_elemstate} with coefficients $a(x_1,\dots,x_r)$, which are complex-valued functions to be determined:
\begin{equation}\label{eq:H33_psiM}
\Psi_r = \sum_{1 \leq x_1 < \dots < x_r \le L}\ \sum_{i_1,i_2,\dots,i_r=1,2}\  a_{i_1,i_2,\dots,i_r}(x_1,x_2,\dots,x_r) 
\vert x_1,i_1;x_2,i_2;\dots;x_r,i_r \rangle.
\end{equation}
The coordinate Bethe ansatz \cite{Bethe31} consists in assuming a plane wave decomposition for the functions $a_{i_1,i_2,\dots,i_r}(x_1,x_2,\dots,x_r)$:
\begin{equation}
a_{i_1,i_2,\dots,i_r}(x_1,x_2,\dots,x_r) = \sum_{\sigma \in {\mathfrak S}_r} A_\sigma^{(i_1,i_2,\dots,i_r)} \prod_{k=1}^r u_{\sigma(k)}^{x_k} ,
\label{eq:H33_planewave}
\end{equation}
where ${\mathfrak S}_r$ is the permutation group of $r$ elements.
The unknowns $A_\sigma^{(i_1,i_2,\dots,i_r)}$ are functions on the symmetric group algebra depending 
on the parameters $u_1, u_2, \dots, u_r$ called rapidities 
and which are solutions of the Bethe equations determined below.
To simplify the following computations, we encompass the $2^r$ different unknowns $A_\sigma^{(i_1,i_2,\dots,i_r)}$ 
for a given $\sigma$ in the following
vector
\begin{equation}
 \boldsymbol{A_\sigma}=\sum_{i_1,i_2,\dots,i_r=1,2} A_\sigma^{(i_1,i_2,\dots,i_r)}  
 |\overline{i}_1\rangle\otimes  |\overline{i}_2\rangle \otimes \dots \otimes  |\overline{i}_r\rangle
\end{equation}
where
\begin{equation}\label{eq:H33_basis_bis}
\qquad |\overline{1}\rangle=\begin{pmatrix} 1\\0\\ \end{pmatrix}\quad\text{and}\qquad 
|\overline{2}\rangle=\begin{pmatrix} 0\\1\\ \end{pmatrix}\;.
\end{equation}

As usual, we project the eigenvalue problem \eqref{eq:H33_eigen} on the different elementary states, the eigenvector having the form \eqref{eq:H33_psiM} 
with \eqref{eq:H33_planewave}. We do not detail the calculations, since they are similar to the nested coordinate Bethe ansatz developed 
in \cite{Yang67, Sutherland68, Sutherland75} based on the ideas of \cite{Bethe31}. The computations are divided into two main steps:
\begin{itemize}
 \item we reduce the original eigenvalue problem with $L$ sites allowing three different states, to an eigenvalue problem for a system with 
 a smaller number of sites, that allow only two different states
 (this system is called the reduced problem);
 \item we determine (and classify) when the  reduced problem is integrable. 
\end{itemize}
In the following, we  sketch these two steps and give the  main results.

{\it First step :}
Performing the CBA on the position of the particles (and not looking at their internal states), we get a first set of constraints 
on the parameters of the local Hamiltonian. This corresponds to the constraints \eqref{eq:H33_1st.constraint} given previously.
This first step allows us to determine the energy of the state $\Psi_r$:
\begin{equation}
E_r = L \, m_{11} + \sum_{k=1}^r \epsilon(u_k) \qquad \text{with} \qquad 
\epsilon(u) = m_{22}+m_{44}-2m_{11} + \frac{m_{24}}{u} + m_{42}\, u
\label{eq:H33_energie}
\end{equation}
provided the coefficients $\boldsymbol{A_\sigma}$ 
are related by
\begin{align}
& \boldsymbol{A_{\sigma \ft_j}}= \check S_{j,j+1}(u_{\sigma(j)},u_{\sigma(j+1)})\,\boldsymbol{A_{\sigma}} \label{eq:H33_AAS}\\
& \check S(z_1,z_2) = -\frac{z_2}{z_1}\,\Lambda(z_1,z_2)\,\Lambda(z_2,z_1)^{-1}\,,\label{eq:H33_Smatrix}
\end{align}
where $\ft_j \in {\mathfrak S}_r$ denotes the transposition $(j,j+1)$ and
\begin{align}\label{def:H33_Lambda}
&\Lambda(z_1,z_2) = T-\left(m_{42}\, z_1+\frac{m_{24}}{z_2}\right)\,\II_4 .
\end{align}
$T$  is the $4\times4$ constant matrix \eqref{eq:H33_Tmat} whose entries depend on the entries of $m$, as stated in \eqref{eq:H33_m.tilde}.  

Let us describe more precisely the meaning of the indices of $\check S_{j,j+1}$ in \eqref{eq:H33_AAS}: they indicate in which spaces the $4\times4$ 
matrix $\check S$ acts non trivially in the tensor product $(\CC^2)^{\otimes r}$ spanned by 
$\{|\overline{i}_1\rangle\otimes  |\overline{i}_2\rangle \otimes \dots \otimes  |\overline{i}_r\rangle\ |\ i_k=1,2\}$.
 Explicitly, we get
 \begin{equation}
  \check S_{j,j+1}(z_1,z_2)=\id_2^{\otimes j-1}\otimes \check S(z_1,z_2)\otimes \id_2^{\otimes L-j-1}\;.
 \end{equation}

Due to the defining relations of the permutation group ${\mathfrak S}_r$ 
$$\ft_j^2=id,\qquad [\ft_j,\ft_k]=0,\ |k-j|>1,\qquad \ft_j\ft_{j+1}\ft_j=\ft_{j+1}\ft_j\ft_{j+1},$$
relations \eqref{eq:H33_AAS} gives constraint on $\check S$:
\begin{eqnarray}
 &&\check S_{j,j+1}(z_j,z_{j+1})\check S_{j,j+1}(z_{j+1},z_j)=1\quad,\qquad \big[ \check S_{j,j+1}(z_j,z_{j+1}), \check S_{k,k+1}(z_k,z_{k+1}) \big] = 0 \qquad \nonumber \\
 &&\check S_{12}(z_1,z_2) \check S_{23}(z_1,z_3) \check S_{12}(z_2,z_3)=\check S_{23}(z_2,z_3) \check S_{12}(z_1,z_3)\check S_{23}(z_1,z_2).\label{eq:H33_YBE}
\end{eqnarray}
The first two relations are trivially satisfied by \eqref{eq:H33_Smatrix}. The third one \eqref{eq:H33_YBE}, called braided Yang--Baxter equation, holds only
if supplementary constraints on the entries of $T$ are satisfied. We postpone the study of these constraints and
suppose from now on that they are indeed satisfied.
  
Because of the periodicity of the model, the rapidities $u_j$ are quantified and must obey the first set of Bethe equations 
\begin{equation}
u_{j}^{L}\, \boldsymbol{A_{id}}= 
 S_{j+1,j}(u_{j+1},u_j)\cdots S_{r,j}(u_r,u_j)\, S_{1,j}(u_{1},u_j)\cdots S_{j-1,j}(u_{j-1},u_j)\,\boldsymbol{A_{id}}\,, 
\quad j=1,...,r.
\label{eq:H33_BAE}
\end{equation}
where
\begin{equation}\label{eq:H33_S-Scheck}
 S(x,y)= P\check S(x,y) \quad\text{and}\qquad 
 P= \begin{pmatrix}1 &0 & 0& 0 \\
 0 & 0 & 1 & 0 \\ 0 & 1 & 0 & 0 \\ 0 & 0 & 0 & 1 \end{pmatrix}.
\end{equation}
This set of eigenvalue problems is the reduced problem. The matrix $S$ is a $4\times 4$ matrix.

{\it Second step (nesting) :}
The matrix $ S(x,y)$ is obviously regular, $S(x,x)=-P$, 
so that the set of eigenvalue problems \eqref{eq:H33_BAE} can be recasted using a transfer matrix
\begin{equation}\label{eq:H33_tran}
t(z;\{u_1,...u_r\})=tr_0\Big(S_{10}(u_1,z)\cdots  S_{M0}(u_r,z)\Big)\;.
 \end{equation}
 This is the transfer matrix introduced in \eqref{eq:inhomogeneous_transfer_matrix_2spectralparameters} where the Bethe roots play the role 
 of inhomogeneity parameters. We recall that
because of the Yang--Baxter equation \eqref{eq:H33_YBE}, the transfer matrix commute for different values of $z$:
\begin{equation}
 \big[ t(z;\{u_1,...u_r\}),t(z';\{u_1,...u_r\}) \big] = 0\;.
\end{equation}

Therefore, the Bethe equations \eqref{eq:H33_BAE} reduced to  
\begin{equation}
u_{j}^{L}\, \boldsymbol{A_{id}}=-t(u_j;\{u_1,...u_r\})\boldsymbol{A_{id}}
\end{equation}
are compatible since we can diagonalize simultaneously all the $t(u_j;\{u_1,...u_r\})$.
Let us remark that the transfer matrix \eqref{eq:H33_tran} may be used to define a new integrable system.

To finish the computations, we should solve the reduced problem using, for example, the Bethe ansatz once again.
However, in the cases treated here, the reduced problem has no conserved charge and the resolution becomes much harder.    
How to apply the Bethe ansatz in these cases is still an open question. However, let us mention that new methods appeared 
recently in order to solve similar problems where there is no conserved 
charge due to the boundary conditions (generalization of the CBA \cite{CrampeRS10, CrampeR12}, Onsager approach \cite{Baseilhac06}, 
separation of variables \cite{Niccoli12}, inhomogeneous Bethe equation \cite{CaoYSW13}, 
modified algebraic Bethe ansatz \cite{BelliardCR13, BelliardC13, AvanBGP15}). 
A generalization of these methods may be possible to deal with the eigenvalue problem \eqref{eq:H33_BAE}.  
In the case of Markovian processes, the matrix ansatz developed in \cite{DerridaEHP93} with its link with integrability 
\cite{SasamotoW97,CrampeRV14} may be also helpful for the resolution of this eigenvalue problem.

{\it Braided Yang--Baxter equation :}

This part is devoted to the classification of the matrices $T$ such that the braided Yang--Baxter equation \eqref{eq:H33_YBE} holds.
We split the problem into three subcases: $i$) $m_{24}m_{42}\neq 0$, then $ii$) $m_{42}=0, m_{24}\neq 0$ and $iii$) $m_{24}=0, m_{42}\neq 0$. 
The case $m_{24}=m_{42}= 0$ is
excluded since it corresponds to an energy \eqref{eq:H33_energie} which does not depend on the rapidities (there is no diffusion of particles).

$\bullet$ {\it Case $\boldsymbol{m_{24}\,m_{42}\neq 0}$}  
By taking different expansions w.r.t. $z_1$, $z_2$ and $z_3$ 
 in \eqref{eq:H33_YBE} and after algebraic manipulations,
we find that the braided Yang--Baxter equation \eqref{eq:H33_YBE} holds if and only if $T$ satisfies:
\begin{equation}
\label{eq:H33_heckem1}
\begin{split}
 & T_{12}T_{23}T_{12}-m_{24}m_{42}T_{12}=T_{23}T_{12}T_{23}-m_{24}m_{42}T_{23}\,, \\
 & (T_{12})^2\ T_{23}=T_{12}\ (T_{23})^2\,, \\
 & (T_{23})^2\ T_{12}=T_{23}\ (T_{12})^2\;.
\end{split}
\end{equation}
This is a particular case of the algebra $\cM_n$ introduced in \eqref{eq:Mn_relation1}.
Note that these relations come directly from the form \eqref{eq:H33_Smatrix} of the $S$-matrix, and do not depend on the size of $T$. 
In general, classifying the solutions of equations \eqref{eq:H33_heckem1} is a difficult task. However, in the case of $4\times4$ 
matrices treated here, 
it is  possible to use formal mathematical software to deal with them. Firstly, we prove that all the solutions of \eqref{eq:H33_heckem1}
verify
\begin{equation}\label{eq:H33_mu}
 T^2= \mu T\;.
\end{equation}
Secondly, we perform the following transformation
\begin{equation}\label{eq:H33_T-Ttilde}
 T=\tau \widetilde T + \rho
\end{equation}
with $\rho$ solution of $\rho^2-\mu\rho+m_{24}m_{42}=0$ and $\tau=\pm\sqrt{\rho^2+m_{24}m_{42}}$.
The matrix $\widetilde T$ satisfies the relations of the Hecke algebra
\begin{eqnarray}
&& \widetilde T_{12}\widetilde T_{23}\widetilde T_{12}=\widetilde T_{23}\widetilde T_{12}\widetilde T_{23}\quad,\\
&& \widetilde T-\widetilde T^{-1}=\frac{\mu-2\rho}{\tau}\equiv \widetilde \mu\;.\label{eq:H33_hecke}
\end{eqnarray}
The classification of the $4\times 4$ matrices satisfying these relations was given in \eqref{eq:Hecke_solu_dim4}.

The value of $\mu$ in \eqref{eq:H33_mu} is reconstructed from these data. Indeed, simple algebraic manipulations show that 
\begin{equation}\label{eq:H33_mutau}
\mu^2 = \frac{m_{24}\,m_{42}} {x^2(1-x^2)}\,,\quad  \rho=\mu\,x^2\,,\quad \tau =x\,\mu 
 \quad \mbox{where $x$ is a solution of} \quad  2x^2+\widetilde \mu\, x-1 = 0.
\end{equation}

Finally, we remark that for any previous solutions of the Hecke algebra, the matrix $\check S$ verifies the Yang--Baxter equation \eqref{eq:H33_YBE}. 
Performing the change of variable
\begin{equation}
z_j=\frac{\mu(1-x_j)+\delta(1+x_j)}{2m_{42}(1-x_j)}, \qquad \mbox{with} \quad \delta=\sqrt{\mu^2-4m_{24}m_{42}},
\end{equation}
one can verify that $\check S(z_1,z_2)$ depends only on the ratio $\dfrac{x_1}{x_2}$ (up to a normalisation factor).
In fact, in terms of the variables $x_j$,  equation \eqref{eq:H33_Smatrix} is equivalent to the usual Baxterization of the Hecke algebra 
recalled in \eqref{eq:Hecke_Rmatrix}.

$\bullet$ {\it Case $\boldsymbol{m_{24}=0}$ and  $\boldsymbol{m_{42}\neq 0}$.}
Now, the different expansions w.r.t. $z_1$, $z_2$ and $z_3$ 
 in \eqref{eq:H33_YBE}, lead to the sole relation:
\begin{eqnarray}\label{eq:H33_m24=0}
T_{12}T_{23}T_{12}+T_{12}\ (T_{23})^2=T_{23}T_{12}T_{23}+(T_{12})^2\ T_{23}
\end{eqnarray}

One recognizes in \eqref{eq:H33_m24=0} the algebra $\cT_3$ defined in \eqref{eq:Tn_relation1}. The $4\times 4$ solutions have been classified 
in theorem \ref{th:Sn_classification}. 

$\bullet$ {\it Case $\boldsymbol{m_{42}=0}$ and  $\boldsymbol{m_{24}\neq 0}$}

This case can be deduced from the case 
$m_{24}=0, m_{42}\neq 0$ in the following way. From any solution $\check S(z_1,z_2)$ to the braided Yang--Baxter equation, one can construct a new one 
\begin{equation}\label{eq:H33_transf}
\check \Sigma(z_1,z_2)=\check S^{t_1t_2}(1/z_2,1/z_1)\;,
\end{equation}
where $(.)^{t_j}$ denotes the transposition in space $j$. Then, starting from 
\begin{eqnarray}
\check S(z_1,z_2) &=& -\frac{z_2}{z_1}\,\Lambda(z_1,z_2)\,\Lambda(z_2,z_1)^{-1}\quad \mbox{with} \quad
\Lambda(z_1,z_2) = T-\frac{m_{24}}{z_2}\,\II_4 
\end{eqnarray}
we get 
\begin{eqnarray}
\check \Sigma(z_1,z_2) &=& - \frac{z_2}{z_1}\,\bar\Lambda(z_1,z_2)\,\bar\Lambda(z_2,z_1)^{-1}\quad \mbox{with} \quad 
 \bar\Lambda(z_1,z_2) = T^{t_1t_2}-m_{24}\, z_1\,\II_4 \; .\qquad
\end{eqnarray}
It is clear that, up to a replacement $m_{24} \leftrightarrow m_{42}$,  $\check \Sigma(z_1,z_2)$ corresponds to the expression of 
the S-matrix \eqref{eq:H33_Smatrix} with $m_{24}=0$. 
Therefore, any solution $\check S(z_1,z_2)$ of the Yang--Baxter equation for $m_{42}=0$ is obtained 
from a solution $\check \Sigma(z_1,z_2)$ for $m_{24}=0$ (classified in the previous paragraph).

For the case $m_{42}=0$, the condition on $T$  reads
\begin{eqnarray}\label{eq:H33_m42=0}
 T_{12} T_{23} T_{12}+ (T_{23})^2 \  T_{12} = T_{23} T_{12} T_{23}+ T_{23} \ (T_{12})^2
\end{eqnarray}
which is deduced from relation \eqref{eq:H33_m24=0} by transposition, as expected.
One recognizes now in \eqref{eq:H33_m42=0} the defining relations of the algebra $\cS_3$, see \eqref{eq:Sn_relation1}.

\subsubsection{Algebraic Bethe ansatz}

We now present another method to diagonalize the Markov matrix (or equivalently the transfer matrix) of an integrable process. This method 
is called algebraic Bethe ansatz (ABA) \cite{SklyaninTF79,KulishS79,TakhtadzhanF79} and relies heavily on the algebraic structure associated to the integrable model.
The main idea is to use the so-called RTT relation that we recall now the key features. Once again we treat here the simple case of single species models
({\it i.e} $N=1$), but generalizations to multi-species models have been developed through the nesting procedure \cite{KulishR86,BelliardR08}.

\begin{definition}
 We define a $2 \times 2$ matrix
 \begin{equation} \label{eq:ABA_T}
  T(z) = \begin{pmatrix}
          A(z) & B(z) \\ C(z) & D(z)
         \end{pmatrix},
 \end{equation}
 with entries $A(z),B(z),C(z),D(z)$ depending on a spectral parameter $z$ and belonging to a non-commutative algebra $\mathcal{A}$ defined
 with the help of a $R$-matrix (satisfying the braided Yang-Baxter equation) through the relation
 \begin{equation} \label{eq:ABA_RTT}
  \check R \left(\frac{z_1}{z_2}\right) T(z_1) \otimes T(z_2) = T(z_2) \otimes T(z_1) \check R \left(\frac{z_1}{z_2}\right).
 \end{equation}
\end{definition}
Note that the associativity of the algebra is ensured by the Yang-Baxter equation (see chapter \ref{chap:three} for details about 
a similar statement).

\begin{example}
 For the single species ASEP, the $R$-matrix is defined in \eqref{eq:ASEP_R} and the associated RTT relation is equivalent to 
 \begin{eqnarray*}
  & & A(z_1)A(z_2)=A(z_2)A(z_1), \quad B(z_1)B(z_2)=B(z_2)B(z_1), \\
  & & C(z_1)C(z_2)=C(z_2)C(z_1), \quad D(z_1)D(z_2)=D(z_2)D(z_1), \\
  & & A(z_1)B(z_2)=f\left(\frac{z_2}{z_1}\right)B(z_2)A(z_1)+g\left(\frac{z_1}{z_2}\right)B(z_1)A(z_2), \\
  & & D(z_1)B(z_2)=f\left(\frac{z_1}{z_2}\right)B(z_2)D(z_1)-g\left(\frac{z_1}{z_2}\right)B(z_1)D(z_2), \\
  & & C(z_1)A(z_2)=f\left(\frac{z_1}{z_2}\right)A(z_2)C(z_1)-g\left(\frac{z_1}{z_2}\right)A(z_1)C(z_2), \\
  & & C(z_1)D(z_2)=f\left(\frac{z_2}{z_1}\right)D(z_2)C(z_1)+g\left(\frac{z_1}{z_2}\right)D(z_1)C(z_2), \\
  & & B(z_1)C(z_2)=\frac{p}{q}C(z_2)B(z_1)+\frac{p}{q}\frac{z_1}{z_2}g\left(\frac{z_1}{z_2}\right)\left[A(z_1)D(z_2)-A(z_2)D(z_1)\right], \\
  & & D(z_1)A(z_2)=A(z_2)D(z_1)-\frac{p}{q}g\left(\frac{z_1}{z_2}\right)\left[C(z_2)B(z_1)-\frac{z_1}{z_2}C(z_1)B(z_2)\right],
 \end{eqnarray*}
 with 
 \begin{equation}
  f(z)=\frac{p-qz}{p(1-z)}, \qquad g(z)=\frac{p-q}{p(1-z)}.
 \end{equation}
 \end{example}
 
 We stress that a representation of the RTT algebra on the vector space $\left(\mathbb{C}^2\right)^{\otimes L}$ can always be obtained 
 with the help of the $R$-matrix.
 
 \begin{proposition}
  The matrix defined by
  \begin{equation} \label{eq:ABA_RTT_representation}
   T(z)=R_{0,L}(z)R_{0,L-1}(z)\dots R_{0,1}(z),
  \end{equation}
  provides a representation of the RTT algebra defined in \eqref{eq:ABA_RTT} on the vector space $\left(\mathbb{C}^2\right)^{\otimes L}$.
 \end{proposition}

 \proof 
 This is shown by direct computation using the Yang-Baxter equation and the relation between the braided and non-braided $R$ matrices: 
 $\check R(z)=PR(z)$, where $P$ is the permutation operator. 
 \finproof
 
 Note that $T(z)$ is then seen as a $2\times 2$ matrix in tensor component space labeled by $0$ with entries 
 $A(z),B(z),C(z),D(z)$ (as defined in \eqref{eq:ABA_T}) which
 are operators on the vector space $\left(\mathbb{C}^2\right)^{\otimes L}$.
 The interest of this explicit representation relies mainly on the following fact.
 
 \begin{lemma}
  The transfer matrix for a periodic model $t(z)$ defined in \eqref{eq:inhomogeneous_transfer_matrix} (with the inhomogeneity parameters $z_i=1$) 
  is obtained from the representation \eqref{eq:ABA_RTT_representation} through
  the relation
  \begin{equation}
   t(z)=tr_{0} T(z)=A(z)+D(z).
  \end{equation}
 \end{lemma}

 The idea of the ABA is to express the eigenvectors of the transfer matrix as
 \begin{equation}
  \ket{\Psi}= B(u_1)B(u_2)\dots B(u_r)\ket{\Omega},
 \end{equation}
where $\ket{\Omega}$ is chosen to be an eigenvector of $A(z)$ and $D(z)$ and to satisfy $C(z)\ket{\Omega}=0$ (it is sometimes called the
highest weight vector of the representation). The parameters $u_1,u_2,\dots,u_r$ are called Bethe roots and are solution
to a set of algebraic equations (the Bethe equations).
The details of the construction depend on the model under consideration. To fix the ideas on a concrete example, we present now the case of the 
single species ASEP on a periodic lattice.

\begin{proposition}
 The eigenvectors of the transfer matrix $t(z)$ of the ASEP in the sector with $r$ particles are given by
 \begin{equation}
  \ket{\Psi}= B(u_1)B(u_2)\dots B(u_r)\ket{\Omega},
 \end{equation}
 where $B(z)$ is constructed from the representation \eqref{eq:ABA_RTT_representation} and 
 \begin{equation}
  \ket{\Omega} = \underbrace{\ket{0} \otimes \ket{0} \otimes \dots \otimes \ket{0}}_{L \mbox{ times}}.
 \end{equation}
 The Bethe roots are solutions to the Bethe equations
 \begin{equation}
  \left(\frac{p(1-u_i)}{p-qu_i}\right)^L=(-1)^{r-1} \prod_{\genfrac{}{}{0pt}{}{j=1}{j \neq i}}^r \frac{pu_i-qu_j}{pu_j-qu_i}.
 \end{equation}
  The associated eigenvalue is given by
 \begin{equation}
  E(z)=\prod_{j=1}^r \frac{pz-qu_j}{p(z-u_j)}+\left(\frac{p(1-z)}{p-qz}\right)^L \prod_{j=1}^r \frac{pu_j-qz}{p(u_j-z)}.
 \end{equation}
\end{proposition}

\proof
First note that $C(z)\ket{\Omega}=0$ and 
\begin{equation}
 A(z)\ket{\Omega}=\ket{\Omega}, \qquad D(z)\ket{\Omega}=\left(\frac{p(1-z)}{p-qz}\right)^L \ket{\Omega}.
\end{equation}
The next step is to compute the action of $A(z)$ on $\ket{\Psi}$ using the commutation relation between $A(z)$ and $B(u_i)$ 
\begin{equation}
 A(z)B(u_i)=\frac{pz-qu_i}{p(z-u_i)}B(u_i)A(z)-\frac{u_i(p-q)}{p(z-u_i)}B(z)A(u_i),
\end{equation}
which allows us to push $A(z)$ completely to the right through all the $B(u_i)$. Because of the relation $B(u_i)B(u_j)=B(u_j)B(u_i)$ the 
vector $A(z)\ket{\Psi}$ is completely symmetric with respect to the Bethe roots. This allows us to write
\begin{eqnarray} \label{eq:ASEP_ABA_actionA}
& & A(z)\ket{\Psi} = A(z)B(u_1)\dots B(u_r)\ket{\Omega} \\
& = & \hspace{-3mm}\prod_{j=1}^r \frac{pz-qu_j}{p(z-u_j)} \ket{\Psi}
-\sum_{i=1}^r \frac{u_i(p-q)}{p(z-u_i)}\prod_{\genfrac{}{}{0pt}{}{j=1}{j \neq i}}^r \frac{pu_i-qu_j}{p(u_i-u_j)}
B(u_1)\dots B(u_{i-1})B(z)B(u_{i+1})\dots B(u_r)\ket{\Omega}. \nonumber 
\end{eqnarray}
Note that the result is split between a 'wanted' part (proportional to $\ket{\Psi}$ as expected) and an 'unwanted' part (where a parameter 
$u_i$ has been replaced by $z$ in the expression of $\ket{\Psi}$). All the game be to cancel this last part using the action of $D(z)$ on $\ket{\Psi}$.
This can be computed using the commutation relation between $D(z)$ and $B(u_i)$
\begin{equation}
 D(z)B(u_i)=\frac{pu_i-qz}{p(u_i-z)}B(u_i)D(z)-\frac{u_i(p-q)}{p(u_i-z)}B(z)D(u_i),
\end{equation}
which allows us to push $D(z)$ completely to the right through all the $B(u_i)$. This yields
\begin{eqnarray} \label{eq:ASEP_ABA_actionD}
& & D(z)\ket{\Psi} = D(z)B(u_1)\dots B(u_r)\ket{\Omega}
 =  \left(\frac{p(1-z)}{p-qz}\right)^L \prod_{j=1}^r \frac{pu_j-qz}{p(u_j-z)} \ket{\Psi} \\
& - & \sum_{i=1}^r \frac{u_i(p-q)}{p(u_i-z)}\left(\frac{p(1-u_i)}{p-qu_i}\right)^L
\prod_{\genfrac{}{}{0pt}{}{j=1}{j \neq i}}^r \frac{pu_j-qu_i}{p(u_j-u_i)}
B(u_1)\dots B(u_{i-1})B(z)B(u_{i+1})\dots B(u_r)\ket{\Omega}, \nonumber
\end{eqnarray}
where we observe again a 'wanted' part and an 'unwanted' part. It is then direct to see that $\ket{\Psi}$ is an eigenvector of
$A(z)+D(z)$ if and only if the 'unwanted' parts in \eqref{eq:ASEP_ABA_actionA} and \eqref{eq:ASEP_ABA_actionD} cancel one with each other,
 {\it i.e} if the Bethe equations hold
 \begin{equation}
  \left(\frac{p(1-u_i)}{p-qu_i}\right)^L= (-1)^{r-1}\prod_{\genfrac{}{}{0pt}{}{j=1}{j \neq i}}^r \frac{pu_i-qu_j}{pu_j-qu_i}.
 \end{equation}
 The associated eigenvalue is then given by
 \begin{equation}
  E(z)=\prod_{j=1}^r \frac{pz-qu_j}{p(z-u_j)}+\left(\frac{p(1-z)}{p-qz}\right)^L \prod_{j=1}^r \frac{pu_j-qz}{p(u_j-z)}.
 \end{equation}
\finproof

\begin{remark}
 We recall that the Markov matrix $M$ of the ASEP is obtained from the transfer matrix $t(z)$ through the relation
 \begin{equation}
  M=(q-p)t'(1)t(1)^{-1} = (q-p)\left.\frac{d \ln t(z)}{dz}\right|_{z=1}.
 \end{equation}
 The vector $\ket{\Psi}$ is thus an eigenvector of $M$ with eigenvalue
 \begin{equation}
  E=(q-p)\frac{E'(1)}{E(1)}=(q-p)^2 \sum_{i=1}^r\frac{u_i}{(1-u_i)(p-qu_i)}.
 \end{equation}
\end{remark}

\begin{remark}
 The Bethe roots $u_1,\dots,u_r$ are directly related to the ones introduced in the coordinate Bethe ansatz through the change of variables
 \begin{equation}
   u_i = \frac{p(1-\tilde u_i)}{p-q\tilde u_i}. 
 \end{equation}
 The parameters $\tilde u_1,\dots,\tilde u_r$ satisfy indeed the equations
 \begin{equation}
  \tilde u_i^L= (-1)^{r-1}\prod_{\genfrac{}{}{0pt}{}{j=1}{j \neq i}}^r
  \frac{p+q\tilde u_i \tilde u_j-(p+q)\tilde u_i}{p+q\tilde u_i \tilde u_j-(p+q)\tilde u_j},
 \end{equation}
which are exactly the Bethe equations obtained in the coordinate Bethe ansatz \eqref{eq:CBA_Bethe_equations}.
The eigenvalue of the Markov matrix is expressed in these new variables as
\begin{equation}
 E=\sum_{i=1}^r \left( \frac{p}{\tilde u_i}+q\tilde u_i -(p+q) \right),
\end{equation}
in exact agreement with the expression derived from the coordinate Bethe ansatz \eqref{eq:CBA_eigenvalue}.
The precise link between the two methods (ABA and CBA) has been investigated in \cite{EsslerFGKK05,Ovchinnikov10}.
\end{remark}

Note that the algebraic Bethe ansatz provides also a fruitful framework to compute physical observables, such as correlation functions. We can mention 
the works \cite{KitanineMT99,KitanineMT00,MailletT00,Kitanine01,CastroM07}.

\subsubsection{Other methods}

We present briefly here other methods, which have been successfully developed to diagonalize exactly models with periodic boundary conditions.

While solving the six-vertex and eight-vertex models, Baxter noticed that the eigenvalues of the transfer matrices satisfied some simple 
functional relations \cite{Baxter82}. These functional difference equation gave basically rise to a new angle to study 
the integrable models: the functional Bethe ansatz \cite{Reshetikhin83}. 
It provides another interpretation of the Bethe equations, which appear in 
this context as a necessary condition to ensure the vanishing of unwanted residues.

Baxter interpreted these functional relation at the algebraic level \cite{Baxter82} through the construction of the Q-operator.
The idea is to build an operator having convenient algebraic relations with the transfer matrix, called the t-Q relations.  
In the last few years this Q-operator was understood to as being some particular transfer matrix
with infinite dimensional auxiliary space. It is constructed in a matrix product form (often with q-deformed oscillators) using infinite dimensional
representation of RTT algebras \cite{BazhanovLMS10,BazhanovFLMS11,LazarescuP14}.

The deep algebraic structure carried by the integrable systems was also exploited to diagonalize quantum Hamiltonians through the vertex operators
\cite{DaviesFJMN93} and compute the correlation functions \cite{JimboMMN92,BoosGKS06}. The connection between this approach and the matrix product method 
presented in chapters \ref{chap:three} and \ref{chap:four}, which are both related to the Knizhnik-Zamolodchikov equations, remains to be fully
understood.

Another method was also introduced to diagonalize models with periodic boundary conditions: the separation of variables (SoV) \cite{Sklyanin92}. 
This method aims to give an unifying framework for the resolution of classical and quantum integrable systems.
 
\section{Integrability for open systems}

We present in this section the construction of integrable Markov matrices in the case of open boundary conditions. Similarly to the periodic case,
the method relies on the construction of a set of commuting operators (containing the Markov matrix). This set is also generated by 
a transfer matrix which commutes for different values of the spectral parameter. The Markov matrix is again recovered by taking the derivative of 
this transfer matrix with respect to the spectral parameter. The transfer matrix is build from two key operators, which act locally on the lattice : 
the $R$-matrix, which guarantees the integrability of the bulk dynamics, and the $K$-matrices, which ensure the integrability of the 
boundary dynamics with the reservoirs. We have indeed already seen that the $R$-matrix is directly related to the bulk local jump operator $m$. 
We will point out a similar connection between the $K$-matrices and the boundary local jump operators $B$ and $\overline B$.
The commutation property of the transfer matrix is a direct consequence of two local properties: the Yang-Baxter equation, satisfied by the $R$-matrix,
and the reflection equation \cite{Cherednik84}, satisfied by the $K$-matrices.
 
\subsection{Reflection matrices and transfer matrix}

\subsubsection{K-matrices and reflection equation}

The $R$-matrix and its properties had already been introduced in the previous section to deal with models defined on the periodic lattice.
All along this subsection, the $R$-matrix $R(z)$ stands for a matrix satisfying the Yang-Baxter equation \eqref{eq:Yang_Baxter}, the 
unitarity, regularity and Markovian properties. We will also assume that $R(z)$ is related to a bulk local jump operator through 
the relation \eqref{eq:derivative_bulk_local_jump}.

We present here the key object to deal with the integrability of the boundary dynamics: the $K$-matrix. It will be together with $R(z)$ the building 
block of the transfer matrix.

\begin{definition}
A matrix $K(z)$ of size $(N+1) \times (N+1)$, i.e acting on $\CC^{N+1}$, satisfies the reflection equation if 
 \begin{equation} \label{eq:reflection_equation}
  R_{1,2}\left(\frac{z_1}{z_2}\right) K_1(z_1) R_{2,1}(z_1 z_2) K_2(z_2) =
  K_2(z_2) R_{1,2}(z_1 z_2) K_1(z_1) R_{2,1} \left(\frac{z_1}{z_2}\right).
 \end{equation}
\end{definition}
The reflection equation states an equality between product of operators acting in $\CC^{N+1} \otimes \CC^{N+1}$. We recall that the subscript indices
indicate on which copies $\CC^{N+1}$ of the tensor space the operators are acting non-trivially. For instance
\begin{equation}
 K_1(z) = K(z) \otimes \id, \quad K_2(z) = \id \otimes K(z).
\end{equation}

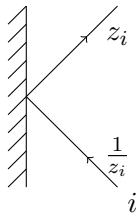
\begin{figure}[htb]
\begin{center}
 \begin{tikzpicture}[scale=0.8]
\draw (0,-1.5) -- (0,1.5) ;
\foreach \j in {-1.2,-0.9,...,1.6}
{\draw (0-0.3,\j-0.3) -- (0,\j);}
\draw[->] (1.5,-1.5) -- (1,-1); \draw (1,-1) -- (0,0);
\node at (1.5,-1) [] {$\frac{1}{z_i}$} ;
\draw[->] (0,0) -- (1,1); \draw (1,1) -- (1.5,1.5);
\node at (1.5,1) [] {$z_i$} ;
\node at (1.75,-1.75) [] {$i$};
 \end{tikzpicture}
 \caption{Graphical representation of the matrix $K_i(z_i)$. \label{fig:Kmatrix}}
 \end{center}
\end{figure}

We would like to give a pictorial representation of the reflection equation. The action of the $K$-matrix $K_i(z_i)$ can be represented graphically 
in figure \ref{fig:Kmatrix}. 
The $K$-matrix is drawn as the reflection of a line labeled $i$, which corresponds to the tensor space component number $i$, on the 
boundary. The line is oriented by an arrow and carries a spectral parameter $z_i$. The incoming half line (according to the arrow direction)
stands for a vector $\ket{\tau}$ of the $i$-th tensor space component, and can thus be in $N+1$ different states. The out-going half line stands for
the vector $\bra{\upsilon}$, which belongs to  the $i$-th tensor space component. When the vectors $\ket{\tau}$ and $\bra{\upsilon}$ 
are specified, the left-reflection diagram represents the matrix element $\bra{\upsilon}K(z_i)\ket{\tau}$.

For instance, in the particular case of single species models, {\it i.e} for $N=1$, this graphical interpretation can be specified as follows.
A dashed line corresponds to the vector $|0\rangle$ (or equivalently to an empty site), whereas a continuous thick line corresponds to $|1\rangle$ 
 (or equivalently to an occupied site). In a similar way, the out-going lines (after the reflection point)
 represent the state of the vector with which we are contracting to the left the matrix $K$ : dashed line for $\bra{0}$ and continuous 
 thick line for $\bra{1}$. Example of such graphical representation is given explicitly for the TASEP in fig. \ref{fig:TASEP_matrix_elements_K}.

\begin{figure}[htb]
\begin{center}
 \begin{tikzpicture}[scale=0.8]
\draw (0,-3) -- (0,3) ;
\foreach \j in {-2.7,-2.4,...,3.1}
{\draw (0-0.3,\j-0.3) -- (0,\j);}

\draw[->] (3,-3) -- (2.5,-2.5); \draw (2.5,-2.5) -- (0,0);
\node at (2,-2.5) [] {$\frac{1}{z_2}$} ;
\draw[->] (0,0) -- (2.5,2.5); \draw (2.5,2.5) -- (3,3);
\node at (2,2.5) [] {$z_2$} ;
\node at (3.25,-3.25) [] {$2$};

\draw[->] (3,-2) -- (2.25,-1.75); \draw (2.25,-1.75) -- (0,-1);
\node at (2.75,-1.5) [] {$\frac{1}{z_1}$} ;
\draw[->] (0,-1) -- (2.25,-0.25); \draw (2.25,-0.25) -- (3,0);
\node at (2.25,0.15) [] {$z_1$} ;
\node at (3.25,-2.25) [] {$1$};

 \draw[->] (4,1.5) -- (4,2.5); \node at (4,2.75) [] {time};
 \draw[->] (4,1.5) -- (5,1.5); \node at (5.75,1.5) [] {space};
\node at (5,0) [] {\huge{$=$}};

\draw (7,-3) -- (7,3) ;
\foreach \j in {-2.7,-2.4,...,3.1}
{\draw (7-0.3,\j-0.3) -- (7,\j);}

\draw[->] (10,-3) -- (9.5,-2.5); \draw (9.5,-2.5) -- (7,0);
\node at (9,-2.5) [] {$\frac{1}{z_2}$} ;
\draw[->] (7,0) -- (9.5,2.5); \draw (9.5,2.5) -- (10,3);
\node at (9,2.5) [] {$z_2$} ;
\node at (10.25,-3.25) [] {$2$};

\draw[->] (10,0) -- (9.25,0.25); \draw (9.25,0.25) -- (7,1);
\node at (9.,-0.15) [] {$\frac{1}{z_1}$} ;
\draw[->] (7,1) -- (9.25,1.75); \draw (9.25,1.75) -- (10,2);
\node at (9.25,1.35) [] {$z_1$} ;
\node at (10.25,-0.25) [] {$1$};
 \end{tikzpicture}
 \caption{Graphical representation of the reflection equation. \label{fig:reflection_equation}}
 \end{center}
\end{figure}
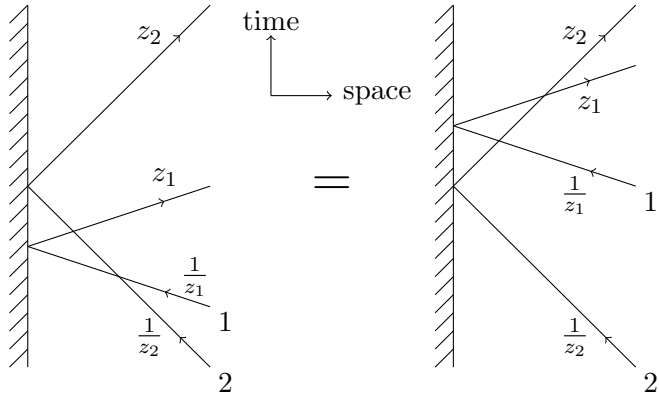

 As for the Yang-Baxter equation, there exists a nice intuitive interpretation for the reflection equation coming from quantum field theory 
(see figure \ref{fig:reflection_equation}): the $K$-matrix $K(z)$ is the scattering matrix of a particle with rapidity $z$ on the boundary.
The integrability is the fact that the simultaneous scattering of 2 particles on the boundary factorizes in this $K$-matrix 
({\it i.e} the scattering of two particles on the boundary can be decomposed into single-particle scatterings on the boundary).
The reflection equation is the consistency relation for this factorization, which ensures the independence of the result with respect to 
the order of the different events (scatterings between the two particles or scatterings on the boundary).

We now present the connection of the $K$-matrix with the boundary local jump operator $B$.

\begin{definition}
 A boundary local jump operator $B$ is said to be integrable if there exists a $K$-matrix $K(z)$ satisfying the reflection equation 
 \eqref{eq:reflection_equation} such that
 \begin{equation} \label{eq:Kmatrix_localjump}
  B = \frac{\theta}{2} K'(1),
 \end{equation}
 where the constant $\theta$ is defined in \eqref{eq:derivative_bulk_local_jump}.
\end{definition}

In other words, the integrable boundary local jump operators $B$ are obtained by taking the derivative of a $K$-matrix with respect to the 
spectral parameter. Conversely, we could wonder how is it possible, starting from a local jump operator $B$, to upgrade it to a spectral parameter
dependent $K$-matrix. This will be partially answered with the Baxterisation procedure presented in subsection \ref{subsubsec:Baxterization_boundary}.
We would like to stress that taking the derivative of a $K$-matrix does note provide always a local Markovian matrix. 
The sum of the entries of each column of the local Markovian matrix should indeed vanish.
It is straightforward to check that if the sum of the entries of each column of the $K$-matrix is equal to 1 then the derivative enjoys 
the sum to 0 property (but we still have to check that the off-diagonal entries of the derivative are non-negative).
This motivates the following definition

\begin{definition}
 A matrix $K(z)$ acting on $\CC^{N+1}$ satisfies the Markovian property if
 \begin{equation}
  \bra{\sigma} K(z) = \bra{\sigma},
 \end{equation}
 where we recall that $\bra{\sigma}=\sum_{\upsilon=0}^{N}\bra{\upsilon}$ achieves the sum over all the local configurations on one site.
\end{definition}

Note that a such $K$-matrix satisfies the requirement of a discrete time Markovian process (provided that its entries are non-negative).
We will see below that it can indeed be used as the building block (together with the $R$-matrix) of discrete time Markov matrices defined on 
the whole lattice with open boundaries.

We now list a set of properties that can be satisfied by a $K$-matrix. The usefulness of these properties will make sense while defining the 
transfer matrix below. They will be essential to connect the transfer matrix to the Markov matrix of the model.

\begin{definition}
 A matrix $K(z)$ acting on $\CC^{N+1}$ satisfies the regularity property if
 \begin{equation}
  K(1) = \id.
 \end{equation}
\end{definition}

\begin{definition}
 A matrix $K(z)$ acting on $\CC^{N+1}$ satisfies the unitarity property if
 \begin{equation} \label{eq:Kmatrix_unitarity}
  K(z).K(1/z) = \id.
 \end{equation}
\end{definition}

\begin{remark}
 We chose to introduce all the definition with $R$-matrix and $K$-matrix which are 'multiplicative' in the spectral parameter. We 
 can write similar definition for matrices which are 'additive' in the spectral parameter. In this case a matrix $K(z)$ 
 satisfies the reflection equation if 
  \begin{equation} \label{eq:reflection_equation_additif}
  R_{1,2}(z_1-z_2) K_1(z_1) R_{2,1}(z_1+z_2) K_2(z_2) =
  K_2(z_2) R_{1,2}(z_1+z_2) K_1(z_1) R_{2,1}(z_1-z_2).
 \end{equation}
 It satisfies the regularity property if $K(0)=\id$ and the unitarity property if $K(z).K(-z)=\id$.
\end{remark}

Up to now, we dealt only with the integrability of the left boundary local jump operator $B$. We can study in a similar way the right 
boundary local jump operator $\overline B$. This motivates the following definitions.

\begin{definition}
A matrix $\overline K(z)$ of size $(N+1) \times (N+1)$, i.e acting on $\CC^{N+1}$, satisfies the reversed reflection equation if 
 \begin{equation} \label{eq:reflection_equation_reversed}
  R_{1,2}\left(\frac{z_1}{z_2}\right)^{-1} \overline K_1(z_1) R_{2,1}(z_1 z_2)^{-1} \overline K_2(z_2) =
  \overline K_2(z_2) R_{1,2}(z_1 z_2)^{-1} \overline K_1(z_1) R_{2,1}\left(\frac{z_1}{z_2}\right)^{-1},
 \end{equation}
 which can be rewritten without inverse on the $R$-matrix using the unitarity property $R_{2,1}(z)^{-1}=R_{1,2}(1/z)$.
\end{definition}

\begin{definition}
 A right boundary local jump operator $\overline B$ is said to be integrable if there exists a $K$-matrix $\overline K(z)$ satisfying the 
 reversed reflection equation \eqref{eq:reflection_equation_reversed} such that
 \begin{equation} \label{eq:Kbmatrix_localjump}
  \overline B = -\frac{\theta}{2} \overline K'(1),
 \end{equation}
 where the constant $\theta$ is defined in \eqref{eq:derivative_bulk_local_jump}.
\end{definition}

The unitarity, regularity and Markovian properties are defined for $\overline K(z)$ in the exact same way as for $K(z)$.

We now provides examples of such $K$-matrices, more particularly those related to the stochastic models already introduced in this manuscript.

\begin{example}
The $K$-matrices related to the ASEP are multiplicative in the spectral parameter. The matrix $K(z)$ is given by
 \begin{equation} \label{eq:ASEP_K}
  K(z) = \begin{pmatrix}
          \frac{z(z(\gamma-\alpha)+\alpha-\gamma+q-p)}{\gamma z^2+z(\alpha-\gamma+q-p)-\alpha} & \frac{(z^2-1)\gamma}{\gamma z^2+z(\alpha-\gamma+q-p)-\alpha} \\
          \frac{(z^2-1)\alpha}{\gamma z^2+z(\alpha-\gamma+q-p)-\alpha} & \frac{\gamma-\alpha+z(\alpha-\gamma+q-p)}{\gamma z^2+z(\alpha-\gamma+q-p)-\alpha}
         \end{pmatrix}.
 \end{equation}
 It satisfies the reflection equation \eqref{eq:reflection_equation} and the unitarity, regularity and Markovian properties.
 It is related to the left boundary local jump operator $B$ defined in \eqref{eq:ASEP_B_Bb} through the relation
 \begin{equation}
  \frac{q-p}{2}K'(1)=B,
 \end{equation}
 which corresponds to a value $\theta=q-p$. The right matrix $\overline K(z)$ is given by
 \begin{equation} \label{eq:ASEP_Kb}
  \overline{K}(z) = \begin{pmatrix}
                   \frac{z(z(\beta-\delta)+\delta-\beta+p-q)}{\beta z^2+z(\delta-\beta+p-q)-\delta} & \frac{(z^2-1)\beta}{\beta z^2+z(\delta-\beta+p-q)-\delta} \\
                   \frac{(z^2-1)\delta}{\beta z^2+z(\delta-\beta+p-q)-\delta} & \frac{\beta-\delta+z(\delta-\beta+p-q)}{\beta z^2+z(\delta-\beta+p-q)-\delta}
                  \end{pmatrix}.
 \end{equation}
 It satisfies the reversed reflection equation \eqref{eq:reflection_equation_reversed} and the unitarity, regularity and Markovian properties.
 It is related to the right boundary local jump operator $\overline B$ defined in \eqref{eq:ASEP_B_Bb} through the relation
 \begin{equation}
  -\frac{q-p}{2}\overline{K}'(1)=\overline{B}.
 \end{equation}
\end{example}

\begin{example}
For the TASEP, it will appear convenient to introduce the parameters $a$ and $b$ related to the injection/extraction rates $\alpha$ and $\beta$ 
through the relations
\begin{equation}
 a=\frac{1-\alpha}{\alpha}, \qquad b=\frac{1-\beta}{\beta}.
\end{equation}
The $K$-matrices can be then expressed as
 \begin{equation} \label{eq:TASEP_K}
  K(z) = \begin{pmatrix}
          \frac{z(z+a)}{za+1} & 0 \\
          \frac{1-z^2}{za+1} & 1
         \end{pmatrix}
 \end{equation}
 It satisfies the reflection equation \eqref{eq:reflection_equation} and the unitarity, regularity and Markovian properties.
 It is connected to the left boundary local jump operator $B$ introduced in \eqref{eq:TASEP_B_Bb} by
 \begin{equation}
  -\frac{1}{2}K'(1)=B,
 \end{equation}
 which corresponds to a value $\theta=-1$. A graphical representation of this $K$-matrix is given in figure \ref{fig:TASEP_matrix_elements_K}.
 
 \begin{figure}[htb]
\begin{center}
 \begin{tikzpicture}[scale=0.7]
\foreach \i in {-6,0,6,12}
{\draw (\i,-1.5) -- (\i,1.5) ;
\foreach \j in {-1.2,-0.9,...,1.6}
{\draw (\i-0.3,\j-0.3) -- (\i,\j);} ;}
\foreach \i in {-6,0}
{\draw[->,dashed] (\i+1.5,-1.5) -- (\i+0.75,-0.75) ; \draw[dashed] (\i+0.75,-0.75) -- (\i,0) ;}
\foreach \i in {6,12}
{\draw[->,ultra thick] (\i+1.5,-1.5) -- (\i+0.75,-0.75) ; \draw[ultra thick] (\i+0.75,-0.75) -- (\i,0) ;}
\foreach \i in {-6,6}
{\draw[->,dashed] (\i,0) -- (\i+0.75,0.75) ; \draw[dashed] (\i+0.75,0.75) -- (\i+1.5,1.5) ;}
\foreach \i in {0,12}
{\draw[->,ultra thick] (\i,0) -- (\i+0.75,0.75) ; \draw[ultra thick] (\i+0.75,0.75) -- (\i+1.5,1.5) ;}
\foreach \i in {-6,0,6,12}
{\node at (\i+1.4,-0.75) [] {\footnotesize{$1/z$}};   }
\foreach \i in {-6,0,6,12}
{\node at (\i+0.3,0.75) [] {\footnotesize{$z$}};   }
\node at (-5.5,-2.5) [] {$\frac{(a+z)z}{za+1}$} ;
\node at (0.5,-2.5) [] {$\frac{1-z^2}{za+1}$} ;
\node at (6.5,-2.5) [] {$0$} ;
\node at (12.5,-2.5) [] {$1$} ;
\node at (-5.5,2.5) [] {$\langle0| K(z)|0\rangle$} ;
\node at (0.5,2.5) [] {$\langle1| K(z)|0\rangle$} ;
\node at (6.5,2.5) [] {$\langle0| K(z)|1\rangle$} ;
\node at (12.5,2.5) [] {$\langle1| K(z)|1\rangle$} ;
 \end{tikzpicture}
 \caption{Graphical representation of the $K$-matrix of the TASEP. \label{fig:TASEP_matrix_elements_K}}
 \end{center}
\end{figure}
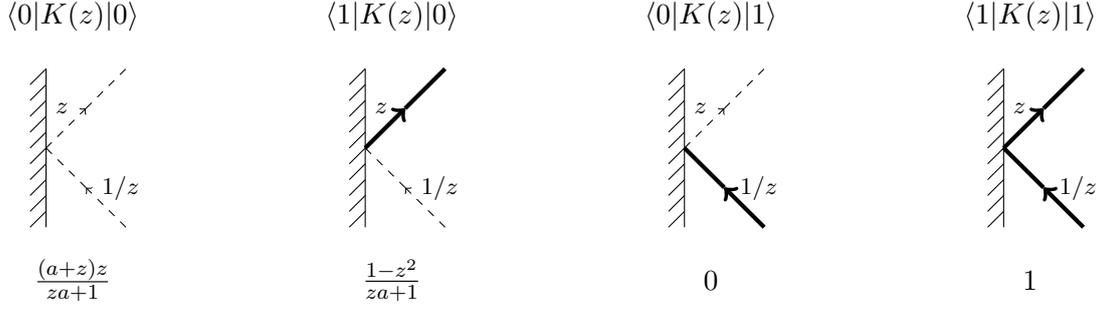
The matrix $\overline K(z)$ is given by
 \begin{equation} \label{eq:TASEP_Kb}
  \overline{K}(z) = \begin{pmatrix}
                  1 & \frac{z^2-1}{z(z+b)} \\
                  0 & \frac{zb+1}{z(z+b)}
                  \end{pmatrix}
 \end{equation}
 It satisfies the reversed reflection equation \eqref{eq:reflection_equation_reversed} and the unitarity, regularity and Markovian properties.
 It is related to the right boundary local jump operator $\overline B$ defined in \eqref{eq:TASEP_B_Bb} through the relation
 \begin{equation}
  \frac{1}{2}\overline{K}'(1)=\overline{B}.
 \end{equation}
 These $K$-matrices can be obtained from the ones of the ASEP by setting $p=1$, $q=0$, $\gamma=0$ and $\delta=0$.
\end{example}

\begin{example}
In the case of the SSEP, the $K$-matrices are additive in the spectral parameter. The left boundary matrix $K(z)$ reads
 \begin{equation} \label{eq:SSEP_K}
  K(z) = \begin{pmatrix}
          \frac{z(\gamma-\alpha)+1}{z(\alpha+\gamma)+1} & \frac{2z\gamma}{z(\alpha+\gamma)+1} \\
          \frac{2z\alpha}{z(\alpha+\gamma)+1} & \frac{z(\alpha-\gamma)+1}{z(\alpha+\gamma)+1}
         \end{pmatrix}.
 \end{equation}
 It satisfies the reflection equation \eqref{eq:reflection_equation_additif} and the unitarity, regularity and Markovian properties.
 It is related to the left boundary local jump operator $B$ defined in \eqref{eq:ASEP_B_Bb} through the relation
 \begin{equation}
 \frac{1}{2} K'(0)=B,
 \end{equation}
 which corresponds to the value $\theta=1$. The right boundary matrix $\overline K(z)$ reads
 \begin{equation} \label{eq:SSEP_Kb}
  \overline{K}(z) = \begin{pmatrix}
                   \frac{z(\beta-\delta)-1}{z(\beta+\delta)-1} & \frac{2z\beta}{z(\beta+\delta)-1} \\
                   \frac{2z\delta}{z(\beta+\delta)-1} & \frac{z(\delta-\beta)-1}{z(\beta+\delta)-1}
                  \end{pmatrix}
 \end{equation}
 It satisfies the reversed reflection equation and the unitarity, regularity and Markovian properties.
 It is related to the right boundary local jump operator $\overline B$ defined in \eqref{eq:ASEP_B_Bb} through the relation
 \begin{equation}
  -\frac{1}{2}\overline{K}'(0)=\overline{B}.
 \end{equation}
 The $K$-matrices of the SSEP are obtained from the ones of the ASEP by the scaling limit 
  \begin{equation}
  K^{SSEP}(z) = \lim\limits_{h \rightarrow 0} K^{ASEP}(e^{hz})|_{q=e^{h},p=1},
 \end{equation}
 and similarly for $\overline K(z)$.
 This scaling transforms a multiplicative dependence in the spectral parameter into an additive one, as expected.
\end{example}

All the previous examples were related to single species models. We provide here an example of R-matrix related to a multi-species model.

\begin{example}
 A $K$-matrix for the $2$-species TASEP is given by
 \begin{equation} \label{eq:2TASEP_K_example}
  K(z)=\begin {pmatrix} 
  {z}^{2}&0&0\\ 
  \noalign{\medskip}-{\frac {a z \left( {z}^{2}-1 \right) }{za+1}}&{\frac {z \left( a+z \right) }{za+1}}&0\\ 
  \noalign{\medskip}-{\frac {{z}^{2}-1}{za+1}}&-{\frac {{z}^{2}-1}{za+1}}&1
  \end{pmatrix} 
\end{equation}
  It satisfies the reflection equation \eqref{eq:reflection_equation} and the unitarity, regularity and Markovian properties.
 It is connected to the left boundary local jump operator $B$ introduced in \eqref{eq:2TASEP_B_Bb_example} by
 \begin{equation}
  -\frac{1}{2}K'(1)=B,
 \end{equation}
 which corresponds to a value $\theta=-1$.
\end{example}

\subsubsection{Transfer matrix}

We defined previously the integrability of the boundary local jump operators $B$ and $\overline B$ as being the derivative
of some $K$-matrices. This definition will make sense in this subsection. We will indeed see that
in this case it is possible to construct a transfer matrix, which generates a set of commuting
operators including the Markov matrix. The $K$-matrices, together with the $R$-matrix, are the key building blocks of this transfer
matrix as explained in the following definition.

\begin{definition}
 The inhomogeneous transfer matrix for a system with open boundaries is an operator acting on the whole lattice $\left(\CC^{N+1}\right)^{\otimes L}$
 and is given by 
 \begin{equation} \label{eq:inhomogeneous_transfer_matrix_open}
  t(z|\mathbf{z})=tr_0 \left(\widetilde{K}_0(z) R_{0,L}\left(\frac{z}{z_L}\right) \dots R_{0,1}\left(\frac{z}{z_1}\right)
  K_0(z) R_{1,0}(z z_1) \dots R_{L,0}(z z_L) \right),
 \end{equation}
 where\footnote{We recall that $\cdot^{t_i}$ denotes the usual matrix transposition in the $i$-th tensor space component}
 \begin{equation} \label{eq:Ktilde_from_Kb}
  \widetilde K_1(z)= tr_0\left(\overline{K}_0\left(\frac{1}{z}\right)\left(\left(R_{0,1}(z^2)^{t_1}\right)^{-1}\right)^{t_1}P_{0,1}\right),
 \end{equation}
  or equivalently 
 \begin{equation} \label{eq:Kb_from_Ktilde}
  \overline{K}_1(z)= tr_0\left( \widetilde{K}_0\left(\frac{1}{z}\right)R_{01}\left(\frac{1}{z^2}\right)P_{01} \right).
 \end{equation}
\end{definition}

 \begin{remark}
  The matrix $\widetilde{K}(z)$ satisfies the dual reflection equation
  \begin{equation} \label{eq:reflection_equation_dual}
   \widetilde{K}_2(z_2)\,\left(R_{21}^{t_1}(z_1z_2)^{-1}\right)^{t_1}\,\widetilde{K}_1(z_1)\,R_{21}\left(\frac{z_2}{z_1}\right)
   =R_{12}\left(\frac{z_2}{z_1}\right)\,\widetilde{K}_1(z_1)\,\left(R_{12}^{t_2}(z_1z_2)^{-1}\right)^{t_2}\,\widetilde{K}_2(z_2).
  \end{equation}
 \end{remark}
 
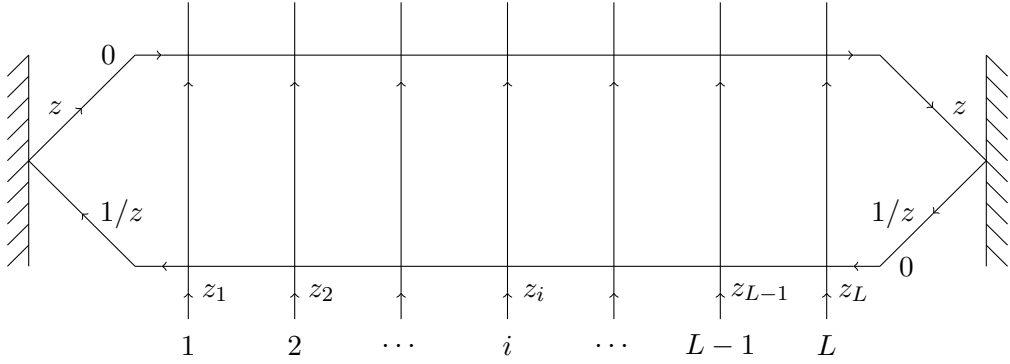
\begin{figure}[htb]
\begin{center}
 \begin{tikzpicture}[scale=0.7]
\draw (-6,-2) -- (-6,2) ;
\draw (12,-2) -- (12,2) ;
\foreach \i in {-1.6,-1.2,...,2.1}
{\draw (-6-0.4,\i-0.4) -- (-6,\i) ; \draw (12+0.4,\i-0.4) -- (12,\i) ;}
\draw[->] (-4,-2) -- (-5,-1) ; \draw[] (-5,-1) -- (-6,0) ; 
\draw[->] (-6,0) -- (-5,1) ; \draw[] (-5,1) -- (-4,2) ; 
\draw[->] (10,2) -- (11,1) ; \draw[] (11,1) -- (12,0) ; 
\draw[->] (12,0) -- (11,-1) ; \draw[] (11,-1) -- (10,-2) ;
\draw[->] (10,-2) -- (9.5,-2) ; \draw[->] (9.5,-2) -- (-3.5,-2) ; \draw[] (-3.5,-2) -- (-4,-2) ;
\draw[->] (-4,2) -- (-3.5,2) ; \draw[->] (-3.5,2) -- (9.5,2) ; \draw[] (9.5,2) -- (10,2) ;
\foreach \i in {-3,-1,...,10}
{\draw[->] (\i,-3) -- (\i,-2.5) ; \draw[->] (\i,-2.5) -- (\i,1.5) ; \draw[] (\i,1.5) -- (\i,3) ; }
\node at (-5.5,1) [] {$z$} ; \node at (11.5,1) [] {$z$} ;
\node at (-4.25,-1) [] {$1/z$} ; \node at (10.25,-1) [] {$1/z$} ;
\node at (-2.5,-2.5) [] {$z_1$} ; \node at (-0.5,-2.5) [] {$z_2$} ; \node at (3.5,-2.5) [] {$z_i$} ;
\node at (7.75,-2.5) [] {$z_{L-1}$} ; \node at (9.5,-2.5) [] {$z_L$} ;
\node at (-3,-3.5) [] {$1$} ; \node at (-1,-3.5) [] {$2$} ; \node at (3,-3.5) [] {$i$} ;
\node at (7,-3.5) [] {$L-1$} ; \node at (9,-3.5) [] {$L$} ;
\node at (1,-3.5) [] {$\dots$} ; \node at (5,-3.5) [] {$\dots$} ;
\node at (-4.5,2) [] {$0$} ; \node at (10.5,-2) [] {$0$} ;
 \end{tikzpicture}
 \caption{Graphical representation of the transfer matrix of the open TASEP.\label{fig:TASEP_open_transfer_matrix}}
 \end{center}
\end{figure}

 A pictorial representation of the transfer matrix with open boundaries is given in figure \ref{fig:TASEP_open_transfer_matrix}.
 
 The main feature of the inhomogeneous transfer matrix is that it generates a set of commuting operators. This is expressed in the 
 following proposition
 
 \begin{proposition}
  The inhomogeneous transfer matrix satisfies the commutation relation
  \begin{equation}
   [t(z|\mathbf{z}),t(z'|\mathbf{z})]=0.
  \end{equation}
 \end{proposition}
 
 \proof
 Let us define
 \begin{equation}
  \cK_0(z) =  R_{0,L}\left(\frac{z}{z_L}\right) \dots R_{0,1}\left(\frac{z}{z_1}\right) K_0(z) R_{1,0}(z z_1) \dots R_{L,0}(z z_L).
 \end{equation}
 It is straightforward to check that this dressed $K$-matrix satisfies the reflection equation
  \begin{equation} \label{eq:reflection_equation_dressed}
  R_{0,0'}\left(\frac{z}{z'}\right) \cK_0(z) R_{0',0}(z z') \cK_{0'}(z') =
  \cK_{0'}(z') R_{0,0'}(z z') \cK_0(z) R_{0',0} \left(\frac{z}{z'}\right).
 \end{equation}
 We then follow the lines of \cite{Sklyanin88} to compute
 \begin{eqnarray*}
  t(z|\mathbf{z})t(z'|\mathbf{z}) & = & tr_0\left(\widetilde K_0(z) \cK_0(z)\right) tr_{0'}\left(\widetilde K_{0'}(z') \cK_{0'}(z')\right) \\
  & = & tr_0\left(\widetilde K_0(z)^{t_0} \cK_0(z)^{t_0}\right) tr_{0'}\left(\widetilde K_{0'}(z') \cK_{0'}(z')\right) \\
  & = & tr_{0,0'}\left(\widetilde K_0(z)^{t_0} \widetilde K_{0'}(z') \cK_0(z)^{t_0} \cK_{0'}(z')\right) \\
  & = & tr_{0,0'}\left(\widetilde K_0(z)^{t_0} \widetilde K_{0'}(z')\times \left(R_{0',0}(z z')^{t_0}\right)^{-1} \left(R_{0',0}(z z')^{t_0}\right)
  \times\cK_0(z)^{t_0} \cK_{0'}(z')\right) \\
  & = & tr_{0,0'}\left(\left(\widetilde K_{0'}(z')\left(\left(R_{0',0}(z z')^{t_0}\right)^{-1})\right)^{t_0}\widetilde K_0(z)\right)^{t_0}
  \left(\cK_0(z)R_{0',0}(z z')\cK_{0'}(z')\right)^{t_0} \right) \\
  & = & tr_{0,0'}\left(\left(\widetilde K_{0'}(z')\left(\left(R_{0',0}(z z')^{t_0}\right)^{-1})\right)^{t_0}
  \widetilde K_0(z)R_{0',0}\left(\frac{z'}{z}\right)\right) \right. \\
  & & \hspace{2cm}\times \left. \left(R_{0,0'}\left(\frac{z}{z'}\right)\cK_0(z)R_{0',0}(z z')\cK_{0'}(z')\right) \right),
 \end{eqnarray*}
 where the last equality is obtained by applying the transposition $\cdot^{t_0}$ in auxiliary space $0$ and then
 by inserting the unitarity relation
 \begin{equation}
  R_{0',0}\left(\frac{z'}{z}\right)R_{0,0'}\left(\frac{z}{z'}\right)=1.
 \end{equation}
 We observe that we are now in position to use the reflection equation \eqref{eq:reflection_equation_dressed} and respectively the
 dual reflection equation \eqref{eq:reflection_equation_dual} to exchange the positions of the matrices $\cK_{0}(z)$ and $\cK_{0'}(z')$
 and respectively of the matrices $\widetilde K_{0}(z)$ and $\widetilde K_{0'}(z')$. Then we repeat the whole sequence of transformations
 in reverse order to finally obtain $t(z'|\mathbf{z})t(z|\mathbf{z})$.
 \finproof

 The previous proposition tells us that, the set of commuting operators can be obtained by expanding the 
 transfer matrix with respect to the spectral parameter.
 
 \begin{remark}
  In the case where the $R$-matrix and the $K$-matrices are additive in the spectral parameter, the inhomogeneous transfer matrix is defined by
  \begin{equation} \label{eq:transfer_matrix_open_additive}
  t(z|\mathbf{z})=tr_0 \left(\widetilde{K}_0(z) R_{0,L}(z-z_L) \dots R_{0,1}(z-z_1)
  K_0(z) R_{1,0}(z+z_1) \dots R_{L,0}(z+z_L) \right),
 \end{equation}
 with 
 \begin{equation}
  \widetilde K_1(z)= tr_0\left(\overline{K}_0\left(-z\right)\left(\left(R_{0,1}(2z)^{t_1}\right)^{-1}\right)^{t_1}P_{0,1}\right),
 \end{equation}
  or equivalently 
 \begin{equation}
  \overline{K}_1(z)= tr_0\left( \widetilde{K}_0\left(-z\right)R_{01}\left(-2z\right)P_{01} \right).
 \end{equation}
 \end{remark}
 
 We are now equipped to state the connection between the transfer matrix and the Markov matrix of the model.
 
 \begin{proposition}
  The Markov matrix is related to the transfer matrix in the simple following way
  \begin{equation} \label{eq:transfer_matrix_to_Markov_matrix_open}
   \frac{\theta}{2}t'(1) = B_1+\sum_{k=1}^{L-1} m_{k,k+1}+\overline{B}_L = M,
  \end{equation}
  where the homogeneous transfer matrix is defined as $t(z)=t(z|1,\dots,1)$.
 \end{proposition}
 
 \proof
 Using the regularity property of the matrix $R$ and $K$, we have
 \begin{eqnarray*}
  t'(1) & = & tr_0(\widetilde{K}_0(1)P_{0L}\dots P_{01}\cdot K_0'(1) \cdot P_{10} \dots P_{L0}) \\
  & + & \sum_{k=1}^{L-1} \left( tr_0(\widetilde{K}_0(1)P_{0L}\dots P_{0,k+1}R_{0k}'(1)P_{0,k-1}\dots P_{01}\cdot 1 \cdot P_{10} \dots P_{L0}) \right. \\
  & &  + \left. tr_0(\widetilde{K}_0(1)P_{0L}\dots P_{01}\cdot 1 \cdot P_{10}\dots P_{k-1,0}R_{k0}'(1)P_{k+1,0} \dots P_{L0}) \right) \\
  & + & tr_0(\widetilde{K}_0(1)R_{0L}'(1)P_{0,L-1}\dots P_{01}\cdot 1 \cdot P_{10}\dots P_{L0}) \\
  & + & tr_0(\widetilde{K}_0(1)P_{0L}\dots P_{01}\cdot 1 \cdot P_{10}\dots P_{L-1,0}R_{L0}'(1)) \\
  & + & tr_0(\widetilde{K}_0'(1)P_{0L}\dots P_{01}\cdot 1 \cdot P_{10}\dots P_{L0}) \\
  & = & tr_0(\widetilde{K}_0(1)) K_1'(1) + 2\sum_{k=1}^{L-1} tr_0(\widetilde{K}_0(1)) \check R_{k,k+1}'(1) \\ 
  & & - \left.\frac{d}{dz}tr_0\left( \widetilde{K}_0\left(\frac{1}{z}\right)R_{0L}\left(\frac{1}{z^2}\right)P_{0L} \right) \right|_{z=1} \\
  & = &  K_1'(1) + 2\sum_{k=1}^{L-1}\check R_{k,k+1}'(1) - \overline{K}_L'(1).
 \end{eqnarray*}
 The last equality is obtained thanks to the regularity property of $\overline{K}$
 \begin{equation}
  tr_0(\widetilde{K}_0(1)) = \overline{K}(1)=1.
 \end{equation}
 The relations \eqref{eq:Kmatrix_localjump}, \eqref{eq:Kbmatrix_localjump} and \eqref{eq:derivative_bulk_local_jump} allow us to conclude 
 the proof.
 \finproof
 
 \begin{example}
  For the single species TASEP, the matrix $\widetilde{K}(z)$ is given by
  \begin{equation} \label{eq:TASEP_Ktilde}
   \widetilde{K}(z) = \begin{pmatrix}
                       \frac{1}{zb+1} & \frac{1}{zb+1} \\
                       0 & \frac{zb}{zb+1}
                      \end{pmatrix}
  \end{equation}
  
 The  graphical representation for the matrix $\widetilde{K}$ is given in fig. \ref{fig:TASEP_matrix_elements_Ktilde}.

\begin{figure}[htb]
\begin{center}
 \begin{tikzpicture}[scale=0.7]
\foreach \i in {-6,0,6,12}
{\draw (\i,-1.5) -- (\i,1.5) ;
\foreach \j in {-1.2,-0.9,...,1.6}
{\draw (\i+0.3,\j-0.3) -- (\i,\j);} ;}
\foreach \i in {-6,6}
{\draw[->,dashed] (\i,0) -- (\i-0.75,-0.75) ; \draw[dashed] (\i-0.75,-0.75) -- (\i-1.5,-1.5) ;}
\foreach \i in {0,12}
{\draw[->,ultra thick] (\i,0) -- (\i-0.75,-0.75) ; \draw[ultra thick] (\i-0.75,-0.75) -- (\i-1.5,-1.5) ;}
\foreach \i in {-6,0}
{\draw[->,dashed] (\i-1.5,1.5) -- (\i-0.75,0.75) ; \draw[dashed] (\i-0.75,0.75) -- (\i,0) ;}
\foreach \i in {6,12}
{\draw[->,ultra thick] (\i-1.5,1.5) -- (\i-0.75,0.75) ; \draw[ultra thick] (\i-0.75,0.75) -- (\i,0) ;}
\foreach \i in {-6,0,6,12}
{\node at (\i-1.4,-0.75) [] {\footnotesize{$1/z$}};   }
\foreach \i in {-6,0,6,12}
{\node at (\i-0.3,0.75) [] {\footnotesize{$z$}};   }
\node at (-6.5,-2.5) [] {$\frac{1}{zb+1}$} ;
\node at (-0.5,-2.5) [] {$0$} ;
\node at (5.5,-2.5) [] {$\frac{1}{zb+1}$} ;
\node at (11.5,-2.5) [] {$\frac{zb}{zb+1}$} ;
\node at (-6.5,2.5) [] {$\langle0|\widetilde K(z)|0\rangle$} ;
\node at (-0.5,2.5) [] {$\langle0|\widetilde K(z)|1\rangle$} ;
\node at (5.5,2.5) [] {$\langle1|\widetilde K(z)|0\rangle$} ;
\node at (11.5,2.5) [] {$\langle1|\widetilde K(z)|1\rangle$} ;
 \end{tikzpicture}
 \caption{Graphical representation of the $\widetilde K$-matrix of the TASEP. \label{fig:TASEP_matrix_elements_Ktilde}}
 \end{center}
\end{figure}
 \end{example}

 \paragraph*{Transfer matrix as discrete time Markov matrix}

We showed in the previous paragraphs how the transfer matrix can be used to define (through its derivative) continuous time 
Markov matrix.
We are now going to see on the particular case of the single species open TASEP that the inhomogeneous transfer matrix can be used itself
to define a discrete time Markov matrix.

The building blocks of the transfer matrix for the open case are the $R$ matrix defined in \eqref{eq:TASEP_R} and
the boundary matrices $K$ and $\widetilde{K}$ defined in \eqref{eq:TASEP_K} and \eqref{eq:TASEP_Ktilde}.
They enter the construction of the inhomogeneous open transfer matrix $t(z|\mathbf{z})$ defined in \eqref{eq:inhomogeneous_transfer_matrix_open}. 

We use the operator $t(z|\mathbf{z})$ to define the following discrete time Markov process
 \begin{equation} \label{eq:TASEP_master_equation_discrete_time_open}
  |P_{t+1}\rangle=t(z|\mathbf{z})|P_{t}\rangle\;.
 \end{equation}
 The parameters must satisfy the following constraints
 \begin{equation}
  0\leq zz_i\leq 1,\quad0\leq \frac{z}{z_i}\leq 1\quad\text{and}\qquad az,bz\geq 0
 \end{equation}
to ensure that the entries of $t(z|\mathbf{z})$ are probabilities.
We can also show that the entries on each column of $t(z|\mathbf{z})$ sum to one, which guarantees
the conservation of the probability $|P_{t}\rangle$.
 
 We now draw diagrams to represent the action of the transfer matrix.
 The general picture is displayed in fig. \ref{fig:TASEP_open_transfer_matrix}. 

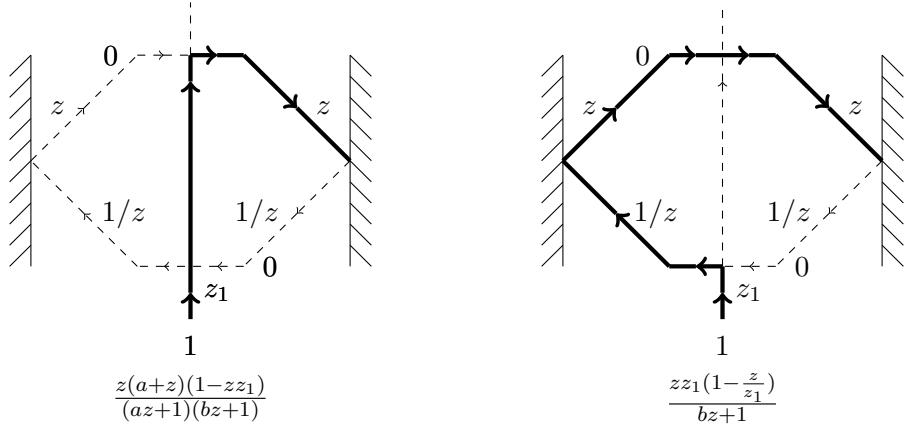
\begin{figure}[htb]
\begin{center}
 \begin{tikzpicture}[scale=0.7]
\draw (0,-2) -- (0,2) ;
\draw (6,-2) -- (6,2) ;
\foreach \i in {-1.6,-1.2,...,2.1}
{\draw (-0.4,\i-0.4) -- (0,\i) ; \draw (6+0.4,\i-0.4) -- (6,\i) ;}
\draw[->,dashed] (2,-2) -- (1,-1) ; \draw[dashed] (1,-1) -- (0,0) ; 
\draw[->,dashed] (0,0) -- (1,1) ; \draw[dashed] (1,1) -- (2,2) ; 
\draw[->,ultra thick] (4,2) -- (5,1) ; \draw[ultra thick] (5,1) -- (6,0) ; 
\draw[->,dashed] (6,0) -- (5,-1) ; \draw[dashed] (5,-1) -- (4,-2) ;
\draw[->,dashed] (4,-2) -- (3.5,-2) ; \draw[->,dashed] (3.5,-2) -- (2.5,-2) ; \draw[dashed] (2.5,-2) -- (2,-2) ;
\draw[->,dashed] (2,2) -- (2.5,2) ; \draw[dashed] (2.5,2) -- (3,2) ;\draw[->,ultra thick] (3,2) -- (3.5,2) ; \draw[ultra thick] (3.5,2) -- (4,2) ;
\foreach \i in {3}
{\draw[->,ultra thick] (\i,-3) -- (\i,-2.5) ; \draw[->,ultra thick] (\i,-2.5) -- (\i,1.5) ; \draw[ultra thick] (\i,1.5) -- (\i,2) ; \draw[dashed] (\i,2) -- (\i,3) ;}
\node at (0.5,1) [] {$z$} ; \node at (5.5,1) [] {$z$} ;
\node at (1.75,-1) [] {$1/z$} ; \node at (4.25,-1) [] {$1/z$} ;
\node at (3.5,-2.5) [] {$z_1$} ; 
\node at (3,-3.5) [] {$1$} ; 
\node at (1.5,2) [] {$0$} ; \node at (4.5,-2) [] {$0$} ;
\node at (3.5,-2.5) [] {$z_1$} ; 
\node at (3,-3.5) [] {$1$} ; 
\node at (1.5,2) [] {$0$} ; \node at (4.5,-2) [] {$0$} ;
\node at (3,-4.5) [] {$\frac{z(a+z)(1-zz_1)}{(az+1)(bz+1)}$} ;

\draw (10,-2) -- (10,2) ;
\draw (16,-2) -- (16,2) ;
\foreach \i in {-1.6,-1.2,...,2.1}
{\draw (10-0.4,\i-0.4) -- (10,\i) ; \draw (16+0.4,\i-0.4) -- (16,\i) ;}
\draw[->,ultra thick] (12,-2) -- (11,-1) ; \draw[ultra thick] (11,-1) -- (10,0) ; 
\draw[->,ultra thick] (10,0) -- (11,1) ; \draw[ultra thick] (11,1) -- (12,2) ; 
\draw[->,ultra thick] (14,2) -- (15,1) ; \draw[ultra thick] (15,1) -- (16,0) ; 
\draw[->,dashed] (16,0) -- (15,-1) ; \draw[dashed] (15,-1) -- (14,-2) ;
\draw[->,dashed] (14,-2) -- (13.5,-2) ;\draw[dashed] (13.5,-2) -- (13,-2) ; \draw[->,ultra thick] (13,-2) -- (12.5,-2) ; \draw[ultra thick] (12.5,-2) -- (12,-2) ;
\draw[->,ultra thick] (12,2) -- (12.5,2) ; \draw[->,ultra thick] (12.5,2) -- (13.5,2) ; \draw[ultra thick] (13.5,2) -- (14,2) ;
\foreach \i in {13}
{\draw[->,ultra thick] (\i,-3) -- (\i,-2.5) ; \draw[ultra thick] (\i,-2.5) -- (\i,-2) ;\draw[->,dashed] (\i,-2) -- (\i,1.5) ; \draw[dashed] (\i,1.5) -- (\i,3) ; }
\node at (10.5,1) [] {$z$} ; \node at (15.5,1) [] {$z$} ;
\node at (11.75,-1) [] {$1/z$} ; \node at (14.25,-1) [] {$1/z$} ;
\node at (13.5,-2.5) [] {$z_1$} ; 
\node at (13,-3.5) [] {$1$} ; 
\node at (11.5,2) [] {$0$} ; \node at (14.5,-2) [] {$0$} ;
\node at (13,-4.5) [] {$\frac{zz_1(1-\frac{z}{z_1})}{bz+1}$} ;
 \end{tikzpicture}
 \caption{Graphical computation of the transition rate
 $\langle 0|t(z|\mathbf{z})|1\rangle$ for $L=1$.
 The two different contributions are represented with
 their respective weights.\label{fig:TASEP_open-L=1}}
 \end{center}
\end{figure}

As an example we can compute graphically for $L=1$ the transition rate $\langle 0|t(z|\mathbf{z})|1\rangle$
between the initial configuration $(1)$ and the final configuration $(0)$
 (see fig.~\ref{fig:TASEP_open-L=1}). The sum of both contributions drawn in fig.~\ref{fig:TASEP_open-L=1} gives
\begin{equation} \label{eq:TASEP_weight_example_open}
\langle 0|t(z|\mathbf{z})|1\rangle  =  \frac{z(a+z)(1-zz_1)}{(az+1)(bz+1)}+\frac{zz_1(1-\frac{z}{z_1})}{bz+1} 
 =  (1-z^2)\frac{z(a+z_1)}{(az+1)(bz+1)}.
\end{equation}

\begin{figure}[htbp]
\begin{center}
 \begin{tikzpicture}[scale=0.85]
\draw (9.8,9.8) -- (10.2,9.8) ;\draw (9.8,9.8) -- (9.8,10) ;\draw (10.2,9.8) -- (10.2,10) ;
\draw  (10,10) circle (0.125) [fill,circle] {};

\draw (2.3,6.3) -- (2.7,6.3) ;\draw (2.3,6.3) -- (2.3,6.5) ;\draw (2.7,6.3) -- (2.7,6.5) ;
\draw[dashed] (1.9,6.3) -- (2.3,6.3) ;\draw[dashed] (1.9,6.3) -- (1.9,6.5) ;
\draw[dashed] (2.7,6.3) -- (3.1,6.3) ;\draw[dashed] (3.1,6.3) -- (3.1,6.5) ;
\draw  (2.5,6.5) circle (0.125) [fill,circle] {};

\draw (6.9,6.3) -- (7.3,6.3) ;\draw (6.9,6.3) -- (6.9,6.5) ;\draw (7.3,6.3) -- (7.3,6.5) ;
\draw[dashed] (6.5,6.3) -- (6.9,6.3) ;\draw[dashed] (6.5,6.3) -- (6.5,6.5) ;
\draw[dashed] (7.3,6.3) -- (7.7,6.3) ;\draw[dashed] (7.7,6.3) -- (7.7,6.5) ;
\draw  (7.1,6.5) circle (0.125) [fill,circle] {};
\draw  (7.5,6.5) circle (0.125) [fill,circle] {};

\draw (12.7,6.3) -- (13.1,6.3) ;\draw (12.7,6.3) -- (12.7,6.5) ;\draw (13.1,6.3) -- (13.1,6.5) ;
\draw[dashed] (12.3,6.3) -- (12.7,6.3) ;\draw[dashed] (12.3,6.3) -- (12.3,6.5) ;
\draw[dashed] (13.1,6.3) -- (13.5,6.3) ;\draw[dashed] (13.5,6.3) -- (13.5,6.5) ;
\draw  (12.5,6.5) circle (0.125) [fill,circle] {};
\draw  (12.9,6.5) circle (0.125) [fill,circle] {};

\draw (17.3,6.3) -- (17.7,6.3) ;\draw (17.3,6.3) -- (17.3,6.5) ;\draw (17.7,6.3) -- (17.7,6.5) ;
\draw[dashed] (16.9,6.3) -- (17.3,6.3) ;\draw[dashed] (16.9,6.3) -- (16.9,6.5) ;
\draw[dashed] (17.7,6.3) -- (18.1,6.3) ;\draw[dashed] (18.1,6.3) -- (18.1,6.5) ;
\draw  (17.1,6.5) circle (0.125) [fill,circle] {};
\draw  (17.5,6.5) circle (0.125) [fill,circle] {};
\draw  (17.9,6.5) circle (0.125) [fill,circle] {};

\draw (1.05,2.8) -- (1.45,2.8) ;\draw (1.05,2.8) -- (1.05,3) ;\draw (1.45,2.8) -- (1.45,3) ;
\draw[dashed] (0.65,2.8) -- (1.05,2.8) ;\draw[dashed] (0.65,2.8) -- (0.65,3) ;
\draw[dashed] (1.45,2.8) -- (1.85,2.8) ;\draw[dashed] (1.85,2.8) -- (1.85,3) ;
\draw  (1.25,3) circle (0.125) [fill,circle] {};

\draw (3.75,2.8) -- (4.15,2.8) ;\draw (3.75,2.8) -- (3.75,3) ;\draw (4.15,2.8) -- (4.15,3) ;
\draw[dashed] (3.35,2.8) -- (3.75,2.8) ;\draw[dashed] (3.35,2.8) -- (3.35,3) ;
\draw[dashed] (4.15,2.8) -- (4.55,2.8) ;\draw[dashed] (4.55,2.8) -- (4.55,3) ;
\draw  (4.35,3) circle (0.125) [fill,circle] {};

\draw (6.9,2.8) -- (7.3,2.8) ;\draw (6.9,2.8) -- (6.9,3) ;\draw (7.3,2.8) -- (7.3,3) ;
\draw[dashed] (6.5,2.8) -- (6.9,2.8) ;\draw[dashed] (6.5,2.8) -- (6.5,3) ;
\draw[dashed] (7.3,2.8) -- (7.7,2.8) ;\draw[dashed] (7.7,2.8) -- (7.7,3) ;
\draw  (7.1,3) circle (0.125) [fill,circle] {};
\draw  (7.5,3) circle (0.125) [fill,circle] {};

\draw (12.7,2.8) -- (13.1,2.8) ;\draw (12.7,2.8) -- (12.7,3) ;\draw (13.1,2.8) -- (13.1,3) ;
\draw[dashed] (12.3,2.8) -- (12.7,2.8) ;\draw[dashed] (12.3,2.8) -- (12.3,3) ;
\draw[dashed] (13.1,2.8) -- (13.5,2.8) ;\draw[dashed] (13.5,2.8) -- (13.5,3) ;
\draw  (12.5,3) circle (0.125) [fill,circle] {};
\draw  (13.3,3) circle (0.125) [fill,circle] {};

\draw (15.45,2.8) -- (15.85,2.8) ;\draw (15.45,2.8) -- (15.45,3) ;\draw (15.85,2.8) -- (15.85,3) ;
\draw[dashed] (15.05,2.8) -- (15.45,2.8) ;\draw[dashed] (15.05,2.8) -- (15.05,3) ;
\draw[dashed] (15.85,2.8) -- (16.25,2.8) ;\draw[dashed] (16.25,2.8) -- (16.25,3) ;
\draw  (15.25,3) circle (0.125) [fill,circle] {};
\draw  (15.65,3) circle (0.125) [fill,circle] {};

\draw (18.55,2.8) -- (18.95,2.8) ;\draw (18.55,2.8) -- (18.55,3) ;\draw (18.95,2.8) -- (18.95,3) ;
\draw[dashed] (18.15,2.8) -- (18.55,2.8) ;\draw[dashed] (18.15,2.8) -- (18.15,3) ;
\draw[dashed] (18.95,2.8) -- (19.35,2.8) ;\draw[dashed] (19.35,2.8) -- (19.35,3) ;
\draw  (18.35,3) circle (0.125) [fill,circle] {};
\draw  (18.75,3) circle (0.125) [fill,circle] {};
\draw  (19.15,3) circle (0.125) [fill,circle] {};

\draw (1.05,-0.7) -- (1.45,-0.7) ;\draw (1.05,-0.7) -- (1.05,-0.5) ;\draw (1.45,-0.7) -- (1.45,-0.5) ;
\draw[dashed] (0.65,-0.7) -- (1.05,-0.7) ;\draw[dashed] (0.65,-0.7) -- (0.65,-0.5) ;
\draw[dashed] (1.45,-0.7) -- (1.85,-0.7) ;\draw[dashed] (1.85,-0.7) -- (1.85,-0.5) ;
\draw  (1.25,-0.5) circle (0.125) [fill,circle] {};

\draw (3.75,-0.7) -- (4.15,-0.7) ;\draw (3.75,-0.7) -- (3.75,-0.5) ;\draw (4.15,-0.7) -- (4.15,-0.5) ;
\draw[dashed] (3.35,-0.7) -- (3.75,-0.7) ;\draw[dashed] (3.35,-0.7) -- (3.35,-0.5) ;
\draw[dashed] (4.15,-0.7) -- (4.55,-0.7) ;\draw[dashed] (4.55,-0.7) -- (4.55,-0.5) ;
\draw  (4.35,-0.5) circle (0.125) [fill,circle] {};

\draw (6.9,-0.7) -- (7.3,-0.7) ;\draw (6.9,-0.7) -- (6.9,-0.5) ;\draw (7.3,-0.7) -- (7.3,-0.5) ;
\draw[dashed] (6.5,-0.7) -- (6.9,-0.7) ;\draw[dashed] (6.5,-0.7) -- (6.5,-0.5) ;
\draw[dashed] (7.3,-0.7) -- (7.7,-0.7) ;\draw[dashed] (7.7,-0.7) -- (7.7,-0.5) ;
\draw  (7.1,-0.5) circle (0.125) [fill,circle] {};
\draw  (7.5,-0.5) circle (0.125) [fill,circle] {};

\draw (12.7,-0.7) -- (13.1,-0.7) ;\draw (12.7,-0.7) -- (12.7,-0.5) ;\draw (13.1,-0.7) -- (13.1,-0.5) ;
\draw[dashed] (12.3,-0.7) -- (12.7,-0.7) ;\draw[dashed] (12.3,-0.7) -- (12.3,-0.5) ;
\draw[dashed] (13.1,-0.7) -- (13.5,-0.7) ;\draw[dashed] (13.5,-0.7) -- (13.5,-0.5) ;
\draw  (12.5,-0.5) circle (0.125) [fill,circle] {};
\draw  (13.3,-0.5) circle (0.125) [fill,circle] {};

\draw (15.45,-0.7) -- (15.85,-0.7) ;\draw (15.45,-0.7) -- (15.45,-0.5) ;\draw (15.85,-0.7) -- (15.85,-0.5) ;
\draw[dashed] (15.05,-0.7) -- (15.45,-0.7) ;\draw[dashed] (15.05,-0.7) -- (15.05,-0.5) ;
\draw[dashed] (15.85,-0.7) -- (16.25,-0.7) ;\draw[dashed] (16.25,-0.7) -- (16.25,-0.5) ;
\draw  (15.25,-0.5) circle (0.125) [fill,circle] {};
\draw  (15.65,-0.5) circle (0.125) [fill,circle] {};

\draw (18.55,-0.7) -- (18.95,-0.7) ;\draw (18.55,-0.7) -- (18.55,-0.5) ;\draw (18.95,-0.7) -- (18.95,-0.5) ;
\draw[dashed] (18.15,-0.7) -- (18.55,-0.7) ;\draw[dashed] (18.15,-0.7) -- (18.15,-0.5) ;
\draw[dashed] (18.95,-0.7) -- (19.35,-0.7) ;\draw[dashed] (19.35,-0.7) -- (19.35,-0.5) ;
\draw  (18.35,-0.5) circle (0.125) [fill,circle] {};
\draw  (18.75,-0.5) circle (0.125) [fill,circle] {};
\draw  (19.15,-0.5) circle (0.125) [fill,circle] {};

\draw (6.9,-4.2) -- (7.3,-4.2) ;\draw (6.9,-4.2) -- (6.9,-4) ;\draw (7.3,-4.2) -- (7.3,-4) ;

\draw (12.7,-4.2) -- (13.1,-4.2) ;\draw (12.7,-4.2) -- (12.7,-4) ;\draw (13.1,-4.2) -- (13.1,-4) ;
\draw  (12.9,-4) circle (0.125) [fill,circle] {};

\node at (5,8.25) [] {$\frac{az}{(az+1)(bz+1)}$} ;
\node at (8.25,8.25) [] {$\frac{abz^2}{(az+1)(bz+1)}$} ;
\node at (11.75,8.25) [] {$\frac{1}{(az+1)(bz+1)}$} ;
\node at (15,8.25) [] {$\frac{bz}{(az+1)(bz+1)}$} ;

\draw[thick] (9.5,9.75) -- (6.5,8.55); \draw[->,thick] (4.5,7.75) -- (2.5,6.95);
\draw[thick] (9.8,9.5) -- (8.8,8.5);  \draw[->,thick] (8,7.7) -- (7.25,6.95);
\draw[thick] (10.2,9.5) -- (11.2,8.5);  \draw[->,thick] (12,7.7) -- (12.75,6.95);
\draw[thick] (10.5,9.75) -- (13.5,8.55); \draw[->,thick] (15.5,7.75) -- (17.5,6.95);

\node at (1.8,4.75) [] {$zz_1$} ;
\node at (3.8,4.75) [] {$1-zz_1$} ;
\node at (7.1,4.75) [] {$1$} ;
\node at (9.4,4.75) [] {$(1-zz_0)(1-zz_1)$} ;
\node at (12.9,4.75) [] {$zz_0(1-zz_1)$} ;
\node at (14.8,4.75) [] {$zz_1$} ;
\node at (18.25,4.75) [] {$1$} ;

\draw[thick] (2.3,6.1) -- (1.8,5.1); \draw[->,thick] (1.5,4.5) -- (1.05,3.6);
\draw[thick] (2.7,6.1) -- (3.2,5.1); \draw[->,thick] (3.5,4.5) -- (3.95,3.6);
\draw[thick] (7.1,6.1) -- (7.1,5.2); \draw[->,thick] (7.1,4.4) -- (7.1,3.6);
\draw[thick] (12.9,6.1) -- (12.9,5.2); \draw[->,thick] (12.9,4.4) -- (12.9,3.6);
\draw[thick] (12.3,6.1) -- (10.8,5.1); \draw[->,thick] (9.9,4.5) -- (8.55,3.6);
\draw[thick] (13.5,6.1) -- (14.5,5.1); \draw[->,thick] (15.1,4.5) -- (16,3.6);
\draw[thick] (17.5,6.1) -- (18,5.1); \draw[->,thick] (18.35,4.4) -- (18.75,3.6);

\node at (1.25,1.25) [] {$\frac{z}{z_1}$} ;
\node at (2.5,1.25) [] {$1-\frac{z}{z_1}$} ;
\node at (3.95,1.25) [] {$1$} ;
\node at (7.1,1.25) [] {$1$} ;
\node at (10,1.25) [] {$1-\frac{z}{z_0}$} ;
\node at (12.9,1.25) [] {$\frac{z}{z_0}$} ;
\node at (14.3,1.25) [] {$1-\frac{z}{z_1}$} ;
\node at (15.65,1.25) [] {$\frac{z}{z_1}$} ;
\node at (18.75,1.25) [] {$1$} ;

\draw[thick] (1.25,2.6) -- (1.25,1.7); \draw[->,thick] (1.25,0.9) -- (1.25,0.1);
\draw[thick] (1.85,2.6) -- (2.35,1.6); \draw[->,thick] (2.65,1) -- (3.1,0.1);
\draw[thick] (3.95,2.6) -- (3.95,1.7); \draw[->,thick] (3.95,0.9) -- (3.95,0.1);
\draw[thick] (7.1,2.6) -- (7.1,1.7); \draw[->,thick] (7.1,0.9) -- (7.1,0.1);
\draw[thick] (12.9,2.6) -- (12.9,1.7); \draw[->,thick] (12.9,0.9) -- (12.9,0.1);
\draw[thick] (15.65,2.6) -- (15.65,1.7); \draw[->,thick] (15.65,0.9) -- (15.65,0.1);
\draw[thick] (18.75,2.6) -- (18.75,1.7); \draw[->,thick] (18.75,0.9) -- (18.75,0.1);
\draw[thick] (12.3,2.6) -- (10.8,1.6); \draw[->,thick] (9.9,1) -- (8.55,0.1);
\draw[thick] (15,2.6) -- (14.5,1.6); \draw[->,thick] (14.2,1) -- (13.75,0.1);

\draw[->,thick] (3.95,-1) -- (6.7,-3.9);
\draw[->,thick] (12.9,-1) -- (7.5,-3.9);
\draw[->,thick] (15.65,-1) -- (13.1,-3.7);
\draw[->,thick] (18.75,-1) -- (13.3,-3.9);
\draw[->,thick] (7.1,-1) -- (12.7,-3.7);
\draw[->,thick] (1.25,-1) -- (12.5,-3.9);
 \end{tikzpicture}
 \caption{An example of sequential update corresponding to the Markov matrix $t(z|\mathbf{z})$. The first line is
 the configuration at time $t$. The second line represents the possible configurations after adding the two supplementary sites 
 corresponding to the boundaries.
 The third line corresponds to the intermediate configurations after the updates of the particles.
 The fourth line represents  the possible configurations after the updates of the holes.
 The label of the arrows provides the rate of the corresponding change of configurations. The last line represents the final
 configurations at time $t+1$ after the summation step.}
 \label{fig:TASEP_seq_open}
 \end{center}
\end{figure}
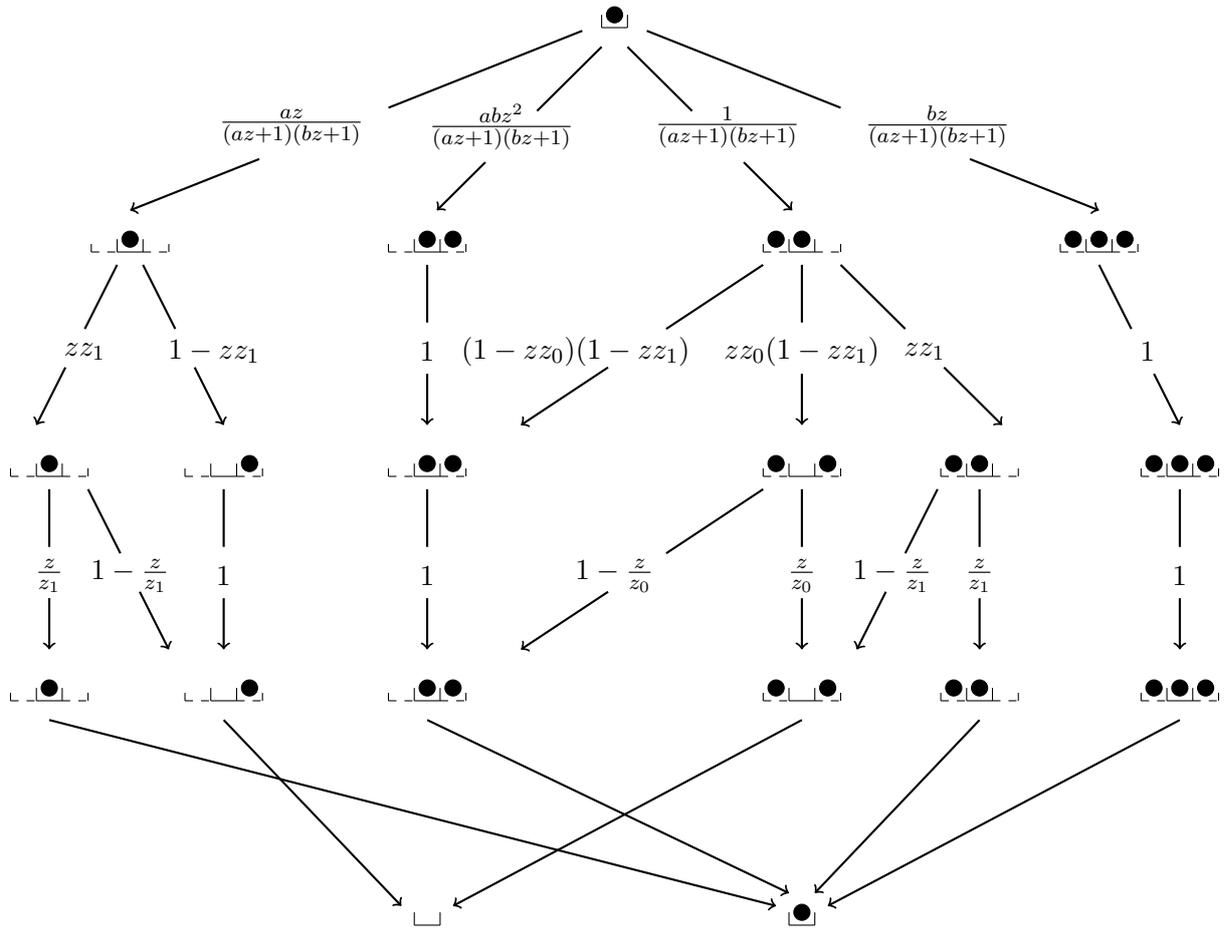

 The transfer matrix $t(z|\mathbf{z})$ defines a discrete time Markov process on the finite size lattice with open boundaries. The corresponding
 stochastic dynamics can be described explicitly using a sequential update: 
starting from a given configuration at time $t$, the configuration at time $t+1$ is obtained  by the  following  stochastic  rules
\begin{itemize}
 \item {\it Initialisation:} 
 \begin{itemize}
 \item The left boundary is replaced by an additional site (the site $0$ with inhomogeneity parameter 1) occupied by a particle
 with probability $\frac{1}{az+1}$ and unoccupied with probability $\frac{az}{az+1}$.
 \item The right boundary is replaced by an additional site (the site $L+1$ with inhomogeneity parameter 1) occupied by a particle
 with probability $\frac{bz}{bz+1}$ and unoccupied with probability $\frac{1}{bz+1}$.
 \end{itemize}
 \item {\it Particle update:} starting from right to left (from site $L$ to site $0$), a particle at site $i$ can jump to the right on the site $i+1$ 
 (provided that the site is empty) with probability $1-zz_i$ and stay at the same place with probability $zz_i$.
 \item {\it Hole update:} once arrived at site $0$, we go the other way starting from left to right (from site $1$ to site $L+1$): a hole at site $i$ 
 can jump to the left on the site $i-1$ (provided that the site is occupied) with probability $1-\frac{z}{z_{i-1}}$ and stay at the same place with 
 probability $\frac{z}{z_{i-1}}$.
 \item {\it Summation:} Then we have to drop the additional sites $0$ and $L+1$ and  to sum the weights corresponding to 
 the same final configuration.
\end{itemize}
An example of such sequential update is given in fig. \ref{fig:TASEP_seq_open} for $L=1$. From this figure we can compute the transition rates

\begin{eqnarray}
 \langle 0|t(z|\mathbf{z})|1\rangle & = & \frac{az}{(az+1)(bz+1)}\left[zz_1\left(1-\frac{z}{z_1}\right)+1-zz_1\right] \nonumber \\
& & +\frac{1}{(az+1)(bz+1)}\left[(1-zz_1)z^2+zz_1\left(1-\frac{z}{z_1}\right)\right] \\
& = & (1-z^2)\frac{z(a+z_1)}{(az+1)(bz+1)},\label{eq:TASEP_0t1}\\
 \langle 1|t(z|\mathbf{z})|1\rangle & = & \frac{z(z+b)}{bz+1} +\frac{(1-z^2)(1-zz_1)}{(az+1)(bz+1)}.
\end{eqnarray}

 The equation  \eqref{eq:TASEP_0t1} is, of course, identical to
  the expression \eqref{eq:TASEP_weight_example_open}
 derived using the graphical representation in fig.~\ref{fig:TASEP_open-L=1}.

\subsection{How to find K-matrices?}

\subsubsection{Direct resolution of the reflection equation}

Written in components, the reflection equation \eqref{eq:reflection_equation} gives equations that are quadratic in the entries of 
the $K$-matrix. In practice it is thus much simpler to solve than the Yang-Baxter equation (which gives cubic equations in the entries 
of the $R$-matrix). However there is so far no general method to systematically solve this reflection equation and it has to be studied case
by case. Several classification have been provided for particular models, see for instance \cite{AbadR95,MintchevRS01}, 
and the reader can find in these references a bench
of methods to help solving this reflection equation. We present below example of solutions of the reflection equation for the DiSSEP and 
for the multi-species SSEP. These $K$-matrices were derived with a technique that is roughly summarized as follows: we performed 
elementary algebraic manipulations on the quadratic equations (given by \eqref{eq:reflection_equation}) to obtain an equality 
between something depending only on the spectral parameter $z_1$ on the left hand side and 
something depending only on the spectral parameter $z_2$ on the right hand side (which proves that the two terms are constant).

\begin{example}
 The $K$-matrices associated to the DiSSEP are multiplicative in the spectral parameter. The left matrix reads 
 \begin{equation} \label{eq:DiSSEP_K}
  K(z) = \begin{pmatrix}
           \frac{(z^2+1)((z^2-1)(\gamma-\alpha)+4z\lambda)}{2z((z^2-1)(\alpha+\gamma)+2\lambda(z^2+1))} & \frac{(z^2-1)((z^2+1)(\gamma-\alpha)+2z(\alpha+\gamma))}{2z((z^2-1)(\alpha+\gamma)+2\lambda(z^2+1))} \\
           \frac{(z^2-1)((z^2+1)(\alpha-\gamma)+2z(\alpha+\gamma))}{2z((z^2-1)(\alpha+\gamma)+2\lambda(z^2+1))} & \frac{(z^2+1)((z^2-1)(\alpha-\gamma)+4z\lambda)}{2z((z^2-1)(\alpha+\gamma)+2\lambda(z^2+1))}
         \end{pmatrix}
 \end{equation}
 which satisfies the reflection equation, the regularity, unitarity and Markovian properties. It generates the left boundary local jump operator $B$
 through the relation
 \begin{equation}
  \lambda K'(1)=B
 \end{equation}
 which corresponds to a value $\theta=2\lambda$. The right matrix reads
 \begin{equation} \label{eq:DiSSEP_Kb}
  \overline{K}(z) = \begin{pmatrix}
                     \frac{(z^2+1)((z^2-1)(\delta-\beta)+4z\lambda)}{2z((1-z^2)(\beta+\delta)+2\lambda(z^2+1))}  & \frac{(z^2-1)((z^2+1)(\delta-\beta)-2z(\beta+\delta))}{2z((1-z^2)(\beta+\delta)+2\lambda(z^2+1))} \\
                     \frac{(z^2-1)((z^2+1)(\beta-\delta)-2z(\beta+\delta))}{2z((1-z^2)(\beta+\delta)+2\lambda(z^2+1))} & \frac{(z^2+1)((z^2-1)(\beta-\delta)+4z\lambda)}{2z((1-z^2)(\beta+\delta)+2\lambda(z^2+1))}
                    \end{pmatrix}
 \end{equation}
 which satisfies the reversed reflection equation, the regularity, unitarity and Markovian properties. 
 It generates the right boundary local jump operator $\overline B$ as follows
 \begin{equation}
  -\lambda \overline{K}'(1)=\overline{B}.
 \end{equation}
\end{example}

\begin{example}
We are interested in the integrable stochastic boundaries associated to the multi-species SSEP. The multi-species SSEP
is a multi-species generalization of the SSEP, whose bulk local jump operator $m$ is given by
$m=P-\II$, where $P$ is the permutation matrix acting on $\CC^{N+1} \otimes \CC^{N+1}$. A more physical description of the 
dynamics will be given in chapter \ref{chap:three}.
The bulk dynamic is integrable : the $R$-matrix is given by
\begin{equation}
 R(z)= \frac{z+P}{z+1}.
\end{equation}
It is additive in the spectral parameter and satisfies the Yang-Baxter equation \eqref{eq:Yang_Baxter_additif}, 
the regularity, unitarity and Markovian properties.
The local jump operator $m$ can be recovered through the relation $m=\check R'(0)$.

The solutions of the reflection equation (associated with this $R$-matrix) have been classified in \cite{MintchevRS01}. 
We present here, without proof, classes of integrable stochastic boundaries $B$ and $\overline B$ among this classification.  
We divide the $N+1$ species (and holes) into $p$ distinct families $F_1,\dots,F_p$ of non-vanishing cardinalities $f_1,\dots,f_p$ at the left boundary 
and into $q$ distinct families $G_1,\dots,G_q$ of non-vanishing cardinalities $g_1,\dots,g_q$
at the right boundary. We hence have two different partitions $\{0,\dots,N\}=\bigsqcup_{k=1}^p F_k=\bigsqcup_{k=1}^q G_k$.
We define $2(N+1)$ non negative numbers $\alpha_0,\dots,\alpha_N$ for the left boundary and $\beta_0,\dots,\beta_N$ for the right boundary with
the constraints
\begin{equation}
 \mbox{for all }1\leq k \leq p, \quad  \sum_{s\in F_k} \alpha_s=1,
\end{equation}
and
\begin{equation}
 \mbox{for all }1\leq k \leq q, \quad  \sum_{s\in G_k} \beta_s=1.
\end{equation}

The left boundary conditions are given by
\begin{equation} \label{eq:mSSEP_left_general_boundary}
 B|s'\rangle =-\frac{1}{a}|s'\rangle+ \sum_{s\in F_k} \frac{\alpha_s}{a}|s\rangle,  \qquad 1\leq s' \leq N.
\end{equation}
$k$ in \eqref{eq:mSSEP_left_general_boundary} is such that $s'\in F_k$.
Remark that in the particular case where the family of $s'$ contains only one species, i.e. $F_k=\{s'\}$,
we get from the constraints that $\alpha_{s'}=1$ and hence $ B|s'\rangle = 0$.

In the same way, the right boundary conditions are given by
\begin{equation}
 \overline B|s'\rangle = -\frac{1}{b}|s'\rangle+\sum_{s\in G_k} \frac{\beta_s}{b}|s\rangle,  \qquad 1\leq s' \leq N,
\end{equation}
with $k$ such that $s'\in G_k$.
Note that when we have a single family on the left and a single family on the right, 
i.e. when $p=q=1$, then the boundary conditions reduce to the one studied
in details in chapter \ref{chap:three} of this manuscript. 
To illustrate these boundary conditions, we give some examples in the case $N=3$ for the left boundary:
\begin{eqnarray}
 B & = & \frac{1}{a}\begin{pmatrix}
  \alpha_0-1 & \alpha_0 & 0 & \alpha_0 \\
  \alpha_1 &\alpha_1-1 & 0 & \alpha_1 \\
  0 & 0 & 0 & 0 \\
  \alpha_3 & \alpha_3 & 0 & \alpha_3-1
 \end{pmatrix}, \quad \mbox{with} \quad \alpha_0+\alpha_1+\alpha_3=1, \label{eq:mSSEP_mat1} \\
 B & = & \frac{1}{a}\begin{pmatrix}
  \alpha_0-1 & \alpha_0 & 0 & 0 \\
  \alpha_1 &\alpha_1-1 & 0 & 0 \\
  0 & 0 & \alpha_2-1 & \alpha_2 \\
  0 & 0 & \alpha_3  & \alpha_3-1
 \end{pmatrix}, \quad \mbox{with} \quad \alpha_0+\alpha_1=1 \mbox{ and } \alpha_2+\alpha_3=1, \\
  B & = & \frac{1}{a}\begin{pmatrix}
  \alpha_0-1 & 0 & \alpha_0 & 0 \\
  0 & \alpha_1-1 & 0 & \alpha_1 \\
  \alpha_2 & 0 & \alpha_2-1 & 0 \\
  0 & \alpha_3 & 0 & \alpha_3-1
 \end{pmatrix}, \quad \mbox{with} \quad \alpha_0+\alpha_2=1 \mbox{ and } \alpha_1+\alpha_3=1, \\
   B & = & \frac{1}{a}\begin{pmatrix}
  0 & 0 & 0 & 0 \\
  0 & \alpha_1-1 & \alpha_1 & 0 \\
  0 & \alpha_2 & \alpha_2-1 & 0 \\
  0 & 0 & 0 & 0
 \end{pmatrix}, \quad \mbox{with} \quad \alpha_1+\alpha_2=1, \label{eq:mSSEP_mat4}  
\end{eqnarray}
Examples of right boundaries in the case $N=3$ are obtained by replacing $\alpha_i$ by $\beta_i$ and $a$ by $b$ in the matrices 
\eqref{eq:mSSEP_mat1}-\eqref{eq:mSSEP_mat4} above.

Coming back to the general multi-species case $N\geq 1$, the $K$-matrices associated to the boundary local jump operators $B$ and $\overline B$ are
given by 
\begin{equation} \label{eq:mSSEP_K_Kb_general}
 K(z)= \II+\frac{2zaB}{z+a} \quad \mbox{and} \quad \overline{K}(z) = \II+\frac{2zb\overline{B}}{z-b}.
\end{equation}
They satisfy the reflection equation, as well as the regularity, unitarity and Markovian properties. 
Note the similarity of these expressions with the one of the $R$-matrix.
 We have the relations
 \begin{equation}
  B = \frac{1}{2}K'(0) \quad \mbox{and} \quad \overline{B} = -\frac{1}{2} \overline{K}'(0).
 \end{equation}
\end{example}

\subsubsection{Baxterization} \label{subsubsec:Baxterization_boundary}

We have seen in the previous section the construction of solutions to the Yang-Baxter equation starting from 
a local jump operator satisfying specific algebraic relations.
This procedure has been extended to the reflection equation \cite{LevyM94} through the boundary Hecke algebra \cite{MartinS93,MartinS94},
that we now introduce.

\paragraph*{Boundary Hecke algebra}

\begin{definition}
 For $n\geq 1$, the boundary Hecke algebra $b\cH_n(\omega,\omega_0)$ is the unital associative algebra over $\CC$ 
 with generators $\sigma_{0},\sigma_1,\dots,\sigma_{n-1}$,
 where $\sigma_1,\dots,\sigma_{n-1}$ satisfy the defining relations \eqref{eq:Hecke_braid}, \eqref{eq:Hecke_commutation},
 \eqref{eq:Hecke_inverse} and \eqref{eq:Hecke_polynomial} of the Hecke algebra $\cH(\omega)$ 
 and subject to the additional relations
 \begin{eqnarray}
  && \sigma_0 \sigma_1 \sigma_0 \sigma_1 = \sigma_1 \sigma_0 \sigma_1 \sigma_0, \label{eq:bHecke_braid} \\
  && \sigma_0 \sigma_i = \sigma_i \sigma_0, \quad i>1, \label{eq:bHecke_commutation} \\
  && \sigma_0 \sigma_0^{-1} = \sigma_0^{-1} \sigma_0 = 1, \label{eq:bHecke_inverse} \\
  && \sigma_0^2=\omega_0 \sigma_0 +1, \label{eq:bHecke_polynomial}
 \end{eqnarray}
 where $\omega_0$ is a complex number not necessarily equal to $\omega$.
\end{definition}

One of the main interest of this algebra, in the context of integrable systems, is the fact that it produces solution to the 
(spectral parameter dependent) braided reflection equation through the following Baxterisation procedure.

\begin{theorem}\label{th:bHecke_main}
   If $\sigma_i$ satisfy the relations of the boundary Hecke algebra, then 
    \begin{eqnarray}
 K_0(z) = \frac{(z^2-1)\sigma_0+z(u_0^{-1/2}-u_0^{1/2})+t_0^{-1/2}-t_0^{1/2}}{z^2t_0^{-1/2}+z(u_0^{-1/2}-u_0^{1/2})-t_0^{1/2}}, \label{eq:bHecke_Kmatrix}
 \end{eqnarray} 
 where $t_0$ is such that $\omega_0=t_0^{-1/2}-t_0^{1/2}$ and $u_0$ is a free complex parameter, satisfies the braided reflection equation
   \begin{equation}
   \check R_1(z_1/z_2) K_0(z_1)\check R_1(z_1z_2)  K_0(z_2)\ = \ K_0(z_2)\check R_1(z_1z_2) K_0(z_1)\check R_1(z_1/z_2)\;. \label{eq:bHecke_bre}
   \end{equation}
Moreover the following properties hold: 
\begin{eqnarray}
&&\mbox{ -- unitarity }\
  K_0(z)K_0(1/z)=1\;,\\[1.2ex]
&&\mbox{ -- regularity }
   K_0(1)=1\;,\\[1.2ex]
&&\mbox{ -- locality }\quad
   K_0(z)\check R_i(z')=\check R_i(z')K_0(z) \quad\text{for}\quad i>1\;.\label{eq:bHecke_loca}
\end{eqnarray}
\end{theorem}

\proof
The regularity and locality properties are obvious.\\
The unitarity  and the braided Yang--Baxter equation are established through a direct computation, using the relations \eqref{eq:bHecke_braid}, 
\eqref{eq:Hecke_polynomial} and \eqref{eq:bHecke_polynomial}.
\finproof

\begin{example}
We can show that the bulk local jump operator $m$ \eqref{eq:ASEP_m} and the boundary local jump operator $B$ \eqref{eq:ASEP_B_Bb} of the ASEP
provide an explicit representation of the boundary Hecke algebra 
$b\cH_n(\omega,\omega_0)$ with $\omega=\sqrt{\frac{q}{p}}-\sqrt{\frac{p}{q}}$ and $\omega_0=\sqrt{\frac{\gamma}{\alpha}}-\sqrt{\frac{\alpha}{\gamma}}$
in the tensor space $End(\CC^2)^{\otimes n}$
\begin{eqnarray}
 b\cH_n(\omega,\omega_0) &\rightarrow& End(\CC^2)^{\otimes n}\nonumber\\
 \sigma_i &\mapsto& \II ^{\otimes i-1} \otimes S \otimes \II ^{\otimes n-i-1}\label{eq:bHecke_rep} \\
 \sigma_0 &\mapsto&  W \otimes \II ^{\otimes n-1}
\end{eqnarray}
where $S$ is a $4\times 4$ matrix (acting on $\CC^2 \otimes \CC^2$) given by
\begin{equation}
  S = \frac{1}{\sqrt{pq}}(m+q),
\end{equation}
and is $W$ a $2\times 2$ matrix (acting on $\CC^2$) given by
\begin{equation}
W= \frac{1}{\sqrt{\alpha\gamma}}(B+\gamma) 
\end{equation}
 Then the Baxterised $K$-matrix 
 \begin{equation}
   K(z)=\frac{(z^2-1)W+z(u_0^{-1/2}-u_0^{1/2})+t_0^{-1/2}-t_0^{1/2}}{z^2t_0^{-1/2}+z(u_0^{-1/2}-u_0^{1/2})-t_0^{1/2}}
 \end{equation}
 corresponding to \eqref{eq:bHecke_Kmatrix}, with $t_0^{1/2}=\sqrt{\frac{\alpha}{\gamma}}$ and $u_0$ such that
 $u_0^{1/2}-u_0^{-1/2}=\frac{p-q+\gamma-\alpha}{\sqrt{\alpha\gamma}} $, coincides with the expression of the $K$-matrix 
 of the ASEP given in \eqref{eq:ASEP_K}.
 \end{example}
 
 \begin{remark}
  The Baxterization procedure presented in theorem \ref{th:bHecke_main} can be extended to a more general boundary Hecke algebra, called 
  cyclotomic algebra, where the relation \eqref{eq:bHecke_polynomial} is generalized to
  \begin{equation}
   \sum_{k=0}^r a_k\, (\sigma_0)^k=0\label{eq:bHecke_cyclo}
  \end{equation}
for some fixed $r=2,3,\dots$ and $a_0$, ... $a_r$ free parameters.
For an introduction to cyclotomic algebras, see e.g. \cite{Ariki99}.
Then, a K-matrix can be constructed as a polynomial in $\sigma_0$ \cite{KulishM06}.
When $r=2$, the cyclotomic Hecke algebra is just the boundary Hecke algebra.
 \end{remark}

\paragraph*{Generalization of the boundary algebra}

We present in this paragraph a generalization of the cyclotomic Hecke algebra, introduced in \cite{IsaevO07}, 
where the relation related to the minimal polynomial is not required. We present also the Baxterization of this algebra given in \cite{IsaevO07}.

\begin{definition}
 We consider a generalization of the boundary Hecke algebra, called $\tilde b\cH_n(\omega)$, which is a unital associative algebra over $\CC$ 
 with generators $\sigma_{0},\sigma_1,\dots,\sigma_{n-1}$,
 where $\sigma_1,\dots,\sigma_{n-1}$ satisfy the defining relations \eqref{eq:Hecke_braid}, \eqref{eq:Hecke_commutation},
 \eqref{eq:Hecke_inverse} and \eqref{eq:Hecke_polynomial} of the Hecke algebra $\cH(\omega)$ 
 and subject to the additional relations
 \begin{eqnarray}
  && \sigma_0 \sigma_1 \sigma_0 \sigma_1 = \sigma_1 \sigma_0 \sigma_1 \sigma_0, \label{eq:tbHecke_braid} \\
  && \sigma_0 \sigma_i = \sigma_i \sigma_0, \quad i>1, \\
  && \sigma_0 \sigma_0^{-1} = \sigma_0^{-1} \sigma_0 = 1. 
 \end{eqnarray}
\end{definition}

The previous algebra is the boundary Hecke algebra where the relation \eqref{eq:bHecke_polynomial} has been taken out. More precisely,
the boundary Hecke algebra 
or the cyclotomic Hecke algebras are quotients of this algebra obtained by imposing a polynomial relation \eqref{eq:bHecke_cyclo}.
We present below a Baxterisation procedure of $\tilde b\cH_n(\omega)$.

\begin{theorem} \label{thm:tbHecke_main}
   If $\sigma_i$ satisfy the relations of $\tilde b\cH_n(\omega)$, then 
 \begin{eqnarray}
   K_0(z)=(1-z \sigma_0)\left(1-\frac{1}{z}\sigma_0\right)^{-1}, \label{eq:tbHecke_Kmatrix}
 \end{eqnarray} 
 satisfies the braided reflection equation
   \begin{equation}
   \check R_1(z_1/z_2) K_0(z_1)\check R_1(z_1z_2)  K_0(z_2)\ = \ K_0(z_2)\check R_1(z_1z_2) K_0(z_1)\check R_1(z_1/z_2)\;. \label{eq:tbHecke_bre}
   \end{equation}
Moreover the following properties hold: 
\begin{eqnarray}
&&\mbox{ -- unitarity }\
  K_0(z)K_0(1/z)=1\;,\\[1.2ex]
&&\mbox{ -- regularity }
   K_0(1)=1\;,\\[1.2ex]
&&\mbox{ -- locality }\quad
   K_0(z)\check R_i(z')=\check R_i(z')K_0(z) \quad\text{for}\quad i>1\;.\label{eq:tbHecke_loca}
\end{eqnarray}
\end{theorem}

\proof
The unitarity, regularity and locality properties are obvious.\\
The braided Yang--Baxter equation will be established below with theorem \ref{thm:bbHecke_main} and proposition \ref{pro:bbHecke_tbHecke}.
\finproof

We present now another generalization of the boundary Hecke algebra, introduced in \cite{CrampeFRV16}, which will be very useful to deal with
the multi-species ASEP. 

\begin{theorem}\label{thm:bbHecke_main}
 Let $\sigma_i$ ($i=1,\dots,L-1$) be the generators of the Hecke algebra satisfying \eqref{eq:Hecke_braid}, \eqref{eq:Hecke_polynomial} 
 and $\check R_i(x)$ the associated braided R-matrices \eqref{eq:Hecke_Rmatrix}.
 Let us also define
 \begin{equation}\label{eq:bbHecke_Kmatrix}
   K_0(z) = (1-(z-1)\sigma_0)\left(1-\left(\frac{1}{z}-1\right)\sigma_0\right)^{-1}
 \end{equation}
with $\sigma_0$ a supplementary generator.
The inverse in \eqref{eq:bbHecke_Kmatrix} is understood as the formal series
\begin{equation}\label{eq:bbHecke_expression_inverse}
 \left(1-\left(\frac{1}{z}-1\right)\sigma_0\right)^{-1}=z\,\Big(1-(1-z)(\sigma_0+1)\Big)^{-1}
 =(y+1)\sum_{n=0}^\infty (-y)^n (\sigma_0+1)^n\;,
\end{equation}
where $y=z-1$.

Then $K_0(x)$ is a solution of the braided reflection equation
\begin{equation}\label{eq:bbHecke_bre}
 \check R_1(z_1/z_2) K_0(z_1)\check R_1(z_1z_2) K_0(z_2)\ = \ K_0(z_2)\check R_1(z_1z_2) K_0(z_1)\check R_1(z_1/z_2)
\end{equation}
if and only if the supplementary generator $\sigma_0$ satisfies
 \begin{equation}\label{eq:bbHecke_braid}
  \sigma_1\ \sigma_0\ \sigma_1\ \sigma_0\ -\  \sigma_0\ \sigma_1\ \sigma_0\ \sigma_1\ =\ \omega(\ \sigma_0^2\ \sigma_1\ \sigma_0\ -\ \sigma_0\ \sigma_1\ \sigma_0^2\ )\;.
 \end{equation}
 Moreover the operator $K_0(z)$ is unitary:
\begin{equation}
 K_0(z)K_0(1/z)=1,
\end{equation}
and satisfies the regularity property
\begin{equation}
 K_0(1)=1.
\end{equation}
We denote by $\overline b\cH_n(\omega)$ the algebra generated by $\sigma_0,\sigma_1,\dots,\sigma_{L-1}$
\end{theorem}

\proof
We multiply both sides of the braided reflection equation \eqref{eq:bbHecke_bre} on the left and on the right by 
\begin{equation}
\frac{z_2}{z_1}\left(1-\left(\frac{1}{z_2}-1\right)\sigma_0\right)=\frac1{z_1}(1+(z_2-1)(\sigma_0+1))
\end{equation}
and use \eqref{eq:Hecke_Rmatrix}, \eqref{eq:bbHecke_Kmatrix} 
to get the following equivalent relation 
\begin{eqnarray}
&&(1+y_2(\sigma_0+1))\ ((z_1-z_2)\sigma_1+\omega z_2)\ \frac{1}{z_1}\check K(z_1)\ ((z_1 z_2-1)\sigma_1+\omega)\ (1-y_2 \sigma_0)\nonumber\\
&=&\!\!\!(1-y_2\sigma_0)\ ((z_1z_2-1)\sigma_1+\omega)\ \frac{1}{z_1}\check K(z_1)\ ((z_1-z_2)\sigma_1+\omega z_2)\ (1+y_2(\sigma_0+1))\qquad\label{eq:bbHecke_bre_eq}
\end{eqnarray}
where $y_i=z_i-1$. Then, we use the expansion \eqref{eq:bbHecke_expression_inverse} of $\frac{1}{z_1}\check K(z_1)$ in terms of $y_1$. The coefficient
of $y_1y_2^3$ in \eqref{eq:bbHecke_bre_eq} provides relation \eqref{eq:bbHecke_braid}, which proves that \eqref{eq:bbHecke_bre} implies \eqref{eq:bbHecke_braid}.\\

To prove the reverse implication, we use the following lemma:
\begin{lemma}\label{lem:bbHecke}
 Relation \eqref{eq:bbHecke_braid} implies, for $k=0,1,2,\dots$,
 \begin{eqnarray}
  \sigma_1\ \sigma_0\ \sigma_1\ \sigma_0^k\ -\  \sigma_0^k\ \sigma_1\ \sigma_0\ \sigma_1 &=& \omega(\ \sigma_0^{k+1}\ \sigma_1\ \sigma_0\ -\ 
  \sigma_0\ \sigma_1\ \sigma_0^{k+1}\ ),\label{eq:bbHecke_lem1}\qquad \quad\\
   \sigma_1\ \sigma_0^k\ \sigma_1\ \sigma_0\ -\  \sigma_0\ \sigma_1\ \sigma_0^k\ \sigma_1 &=& \omega(\ \sigma_0^{k+1}\ \sigma_1\ \sigma_0\ -\ 
   \sigma_0\ \sigma_1\ \sigma_0^{k+1}\ \label{eq:bbHecke_lem2}\\
   &&\hspace{1cm}+\ \sigma_0^{k}\ \sigma_1\ \sigma_0\ -\ \sigma_0\ \sigma_1\ \sigma_0^{k}\ ),\nonumber\\
   \sigma_1\ (\sigma_0+1)^k\ \sigma_1\ \sigma_0 -\ \sigma_0\ \sigma_1\ (\sigma_0+1)^k\ \sigma_1 &=& 
   \omega(\ (\sigma_0+1)^{k+1}\ \sigma_1\ \sigma_0\ \nonumber\\
   &&\hspace{1cm}-\ \sigma_0\ \sigma_1\ (\sigma_0+1)^{k+1}\ )\;,\label{eq:bbHecke_lem3}\\
   \sigma_1\ \sigma_0\,(\sigma_0+1)^k\ \sigma_1\ \sigma_0 -\ \sigma_0\ \sigma_1\ \sigma_0\,(\sigma_0+1)^k\ \sigma_1 &=& 
   \omega(\ \sigma_0\ (\sigma_0+1)^{k+1}\ \sigma_1\ \sigma_0\ \nonumber\\
   &&\hspace{1cm}-\ \sigma_0\ \sigma_1\ \sigma_0\ (\sigma_0+1)^{k+1}\ )\;.\ \label{eq:bbHecke_lem4}
 \end{eqnarray}
\end{lemma}
The first relation of the lemma \eqref{eq:bbHecke_lem1} is proven by recursion using \eqref{eq:bbHecke_braid}. 
Relation \eqref{eq:bbHecke_lem2} is proven also by
recursion with \eqref{eq:bbHecke_lem1}, \eqref{eq:Hecke_braid} and \eqref{eq:Hecke_polynomial}. The third and the fourth are proven
by expanding $(\sigma_0+1)^k$ and using \eqref{eq:bbHecke_lem2}.

The lemma allows us to prove that
\begin{equation}\label{eq:bbHecke_ek}
 \sigma_1 K(z)\sigma_1\sigma_0-\sigma_0\sigma_1 K(z)\sigma_1=\omega\left((\sigma_0+1) K(z)\sigma_1\sigma_0-\sigma_0\sigma_1(\sigma_0+1) K(z)   \right)\;.
\end{equation}

Finally, by expanding \eqref{eq:bbHecke_bre_eq} and by using relation \eqref{eq:bbHecke_ek}, we prove that equation \eqref{eq:bbHecke_braid}
implies \eqref{eq:bbHecke_bre} which concludes the proof of the theorem.
\finproof

The connection between the Baxterisations presented in theorems \ref{thm:tbHecke_main} and \ref{thm:bbHecke_main} is detailed in the following 
proposition.

\begin{proposition} \label{pro:bbHecke_tbHecke}
 Assume that $\sigma_0 \in \overline b\cH_n(\omega)$ is such that $\sigma_0 +1$ is invertible. Then the following map is an algebra morphism
 \begin{eqnarray}
 \overline b\cH_n(\omega) &\rightarrow& \tilde b\cH_n(\omega) \nonumber\\
 \sigma_i &\mapsto& \sigma_i, \quad 1\leq i \leq n-1 \label{eq:bbHecke_tbHecke} \\
 \sigma_0 &\mapsto& \sigma_0 (1+\sigma_0)^{-1}.
\end{eqnarray}
Moreover this map transforms the Baxterised $K$-matrix \eqref{eq:bbHecke_Kmatrix} into \eqref{eq:tbHecke_Kmatrix}, up to a normalization factor.
\end{proposition}
\proof
This can be shown by using relation \eqref{eq:bbHecke_braid} for $\sigma_0$ and lemma \ref{lem:bbHecke}.
\finproof

We now present an application of this algebraic machinery to out-of-equilibrium statistical physics. The algebra $\overline b \cH_n(\omega)$ is
indeed at the root of integrable boundary matrices for the multi-species ASEP. This is detailed in the following paragraph.

\paragraph*{Integrable boundary conditions for the multi-species ASEP}

We wish to give explicit solutions for integrable Markovian boundary matrices $B$ for the multi-species ASEP. 
These solutions are obtained from $K$-matrices obeying the braided reflection equation \eqref{eq:bbHecke_bre}
(with the $R$-matrix associated to the $N$-species ASEP \eqref{eq:mASEP_R} obtained from the Baxterisation \eqref{eq:Hecke_Rmatrix}) 
through the relation
\begin{equation}
 B= \frac{q-p}{2}K'(1).
\end{equation}
We first present the integrable Markovian boundary conditions $B$. We will then argue that they belong (up to a shift and normalization)
to a $\overline b \cH_n(\omega)$ algebra. This will provide through the Baxterisation procedure an explicit expression of the associated $K$-matrices.

The integrable boundary conditions depend on two free real positive parameters (rates) $\alpha$ and $\gamma$, and four positive integers 
$s_1,\ s_2,\ f_1$ and $f_2$, that label two special slow ($s$) and two special fast ($f$) species, with the conditions
\begin{equation}
0\leq s_1\leq s_2<f_2\leq f_1\leq N\mbox{ and }f_1-f_2 = s_2-s_1.\label{eq:mASEP_cont_sf}
\end{equation}
The four special species will be essentially created on the boundary, while the remaining species will
essentially (but not only) decay onto these four types. Any species in between $s_1$ and $s_2$ will be paired
with one species in between $f_2$ and $f_1$,  allowing a transmutation (on the boundary) between the pairs.
Finally, in between $s_2$ and $f_2$, either nothing happens, or the species decay to $s_2$ and $f_2$.

More specifically, integrability is preserved when we have the following rules and rates on the boundary:
\begin{itemize}
\item \textit{Class of very slow species:} for species $\tau$ with $1\leq \tau<s_1$, we have:
\begin{equation}
\tau\ \xrightarrow{\gamma}\ s_1 \mbox{ and } \tau\ \xrightarrow{\alpha}\ f_1.
\end{equation}
\item  \textit{Class of slow species:} for species $\tau$ with $s_1\leq \tau\leq s_2$, we have:
\begin{equation}
\tau\ \xrightarrow{\alpha}\ \overline \tau=s_1+f_1-\tau=s_2+f_2-\tau.
\end{equation}
\item  \textit{Class of intermediate species:} for species $\tau$ with $s_2< \tau<f_2$, we have the two possibilities:
\begin{enumerate}
\item $\tau\ \xrightarrow{0}\ \tau'$, $\forall \tau'$ (no decay, creation or transmutation).
\item $\tau\ \xrightarrow{\widetilde\gamma}\ s_2$ and $\tau\ \xrightarrow{\alpha}\ f_2$.
\end{enumerate}
\item \textit{Class of  fast species:} for species $\tau$ with $f_2\leq \tau\leq f_1$, we have:
\begin{equation}
\tau\ \xrightarrow{\widetilde\gamma}\ \overline \tau=s_1+f_1-\tau.
\end{equation}
\item \textit{Class of very fast species:} for species $\tau$ with $f_1< \tau\leq N$, we have:
\begin{equation}
\tau\ \xrightarrow{\widetilde\gamma} \ s_1 \mbox{ and } \tau\ \xrightarrow{\widetilde\alpha} \ f_1.
\end{equation}
\end{itemize}
We have introduced the following combination of the rates:
\begin{equation}\label{eq:tilde}
    \tilde{\alpha} = \frac{(\alpha + \gamma + q - p)\alpha}{\alpha + \gamma},
    \qquad
    \tilde{\gamma} = \frac{(\alpha + \gamma + q - p)\gamma }{\alpha + \gamma}.
\end{equation}
This implies that $\alpha$, $\gamma$, $p$, $q$ are constrained such that $\tilde\alpha$, $\tilde\gamma$ are
positive.

Note that, depending on the choice of $s_1$, $s_2$, $f_2$ and $f_1$, some classes of species may not occur: for instance if $s_1=0$, there is no very 
slow species. In the same way, if $f_2=s_2+1$, there are no intermediate species.

Due to the second constraint in \eqref{eq:mASEP_cont_sf}, the number of slow species coincides with the number of fast species, in accordance with
the pairing mentioned above.
By counting the number of possibilities for $s_1$, $s_2$, $f_1$ and $f_2$ with the constraints \eqref{eq:mASEP_cont_sf}, we can deduce that, for
multi-species ASEP there exist\footnote{We have included in the counting the two possible choices for the intermediate species when $f_2>s_2+1$.} 
$\begin{pmatrix}  N+2\\  3   \end{pmatrix}$ 
different integrable boundaries, each of them depending on two real parameters.

Note that in any transition, the number of particles for the species in the very slow and very fast classes can only decrease. 
It may stay constant for the slow, fast and intermediate classes. For the four special types
it may increase.\\

To summarize, these rates are gathered in the two following types of boundary matrices, depending on the two possibilities for intermediate species:

\begin{eqnarray}
&&B^0(\alpha,\gamma|s_1, s_2, f_2, f_1)=
\label{eq:mASEP_Bempty}\\
&&\left(
\begin{array}{ccc|cccc|ccc|cccc|ccc}
\mbox{-}\sigma & & & & & & & & & & & & & & & &
\\
 &\ddots & & & & & & & & & & & & & & &
\\
 & & \mbox{-}\sigma & & & & & & & & & & & & & &
\\
\hline
\gamma & \cdots&\gamma & \mbox{-}\alpha & & & & & & & & & &\widetilde\gamma & \widetilde\gamma& \cdots & \widetilde\gamma
\\
 & & & & \mbox{-}\alpha& & & & & & & &\widetilde\gamma & & & & 
\\
 & & & & &\ddots & & & & & &\iddots & & & & & 
\\
 & & & & & & \mbox{-}\alpha & & & & \widetilde\gamma& & & & & & 
\\
\hline
 & & & & & & &0 &\cdots& 0 & & & & & & & 
\\
 & & & & & & &\vdots & & \vdots & & & & & & & 
\\
 & & & & & & &0 &\cdots& 0 & & & & & & & 
\\
\hline
 & & & & & & \alpha& & & & \mbox{-}\widetilde\gamma& & & & & & 
\\
 & & & & &\iddots & & & & & &\ddots & & & & & 
\\
 & & & &\alpha & & & & & & & & \mbox{-}\widetilde\gamma& & & & 
\\
\alpha & \cdots& \alpha& \alpha& & & & & & & & & &\mbox{-}\widetilde\gamma & \widetilde\alpha& \cdots& \widetilde\alpha
\\
\hline
 & & & & & & & & & & & & & & \mbox{-}\widetilde\sigma & &
\\
 & & & & & & & & & & & & & & & \ddots&
\\
 & & & & & & & & & & & & & & & &  \mbox{-}\widetilde\sigma
\end{array}
\right)\nonumber
\end{eqnarray}

\begin{eqnarray}
&&B(\alpha,\gamma|s_1, s_2, f_2, f_1)=
\label{eq:mASEP_Bfull}\\
&&\left(
\begin{array}{ccc|cccc|ccc|cccc|ccc}
\mbox{-}\sigma & & & & & & & & & & & & & & & &
\\
 &\ddots & & & & & & & & & & & & & & &
\\
 & & \mbox{-}\sigma & & & & & & & & & & & & & &
\\
\hline
\gamma & \cdots&\gamma & \mbox{-}\alpha & & & & & & & & & &\widetilde\gamma & \widetilde\gamma& \cdots & \widetilde\gamma
\\
 & & & & \mbox{-}\alpha& & & & & & & &\widetilde\gamma & & & & 
\\
 & & & & &\ddots & & & & & &\iddots & & & & & 
\\
 & & & & & & \mbox{-}\alpha & \widetilde\gamma & \cdots & \widetilde\gamma & \widetilde\gamma& & & & & & 
\\
\hline
 & & & & & & & \mbox{-}\sigma' & & & & & & & & & 
\\
 & & & & & & & & \ddots & & & & & & & & 
\\
 & & & & & & & & & \mbox{-}\sigma' & & & & & & & 
\\
\hline
 & & & & & & \alpha & \alpha & \cdots & \alpha & \mbox{-}\widetilde\gamma& & & & & & 
\\
 & & & & &\iddots & & & & & &\ddots & & & & & 
\\
 & & & &\alpha & & & & & & & & \mbox{-}\widetilde\gamma& & & & 
\\
\alpha & \cdots& \alpha& \alpha& & & & & & & & & &\mbox{-}\widetilde\gamma & \widetilde\alpha& \cdots& \widetilde\alpha
\\
\hline
 & & & & & & & & & & & & & & \mbox{-}\widetilde\sigma & &
\\
 & & & & & & & & & & & & & & & \ddots&
\\
 & & & & & & & & & & & & & & & &  \mbox{-}\widetilde\sigma
\end{array}
\right)\nonumber
\end{eqnarray}
We have introduced $\sigma=\alpha+\gamma$, $\widetilde\sigma=\widetilde\alpha+\widetilde\gamma$ and $\sigma'=\alpha+\widetilde\gamma$.
The empty spaces in the matrices above are filled with zeros, and the lines indicate the positions of the four special types of species.

\begin{remark}
We remark that the rates can be written in a more symmetric way by introducing the combination of parameters\footnote{Note that in our parameters, 
$\widetilde\gamma$ corresponds to $\gamma$ in the one-species ASEP. This choice avoids the presence of a square root in $a$ and $c$ and in the rates
appearing in the boundary matrices.}
\begin{equation}
 a = -\frac{\alpha}{\gamma},\qquad c = \frac{\alpha+\gamma}{\alpha+\gamma+q-p},
\end{equation}
which have previously appeared as the parameters of the Askey-Wilson polynomials in the context of the ASEP stationary state \cite{UchiyamaSW04}. Then 
\begin{equation}\label{eq:mASEP_np1}
 \alpha=\frac{ac(p-q)}{(a-1)(c-1)}\qquad\text{and}\qquad\gamma=-\frac{c(p-q)}{(a-1)(c-1)},
\end{equation}
and the parameters $\tilde{\alpha}$ and $\tilde{\gamma}$ become
\begin{equation}\label{eq:mASEP_np2}
 \tilde{\alpha}=\frac{a(p-q)}{(a-1)(c-1)}\qquad\text{and}\qquad\tilde{\gamma}=-\frac{(p-q)}{(a-1)(c-1)}\;.
\end{equation}
\end{remark}

\begin{remark}
We can produce more integrable solutions using conjugation by any diagonal invertible matrix $V$. Indeed, due to the invariance 
of the R-matrix \eqref{eq:mASEP_R} by the conjugation by $V_1V_2$, $VK(z)V^{-1}$ is solution of the reflection equation if $K(z)$ is also a solution. 
However, the resulting conjugated matrix may not be Markovian. Nonetheless, we remark that conjugation by the diagonal matrix 
$diag(e^{s_1},e^{s_2},\dots,e^{s_N})$ provides a deformed integrable boundary matrix that 
allows one to compute the cumulants of the currents at the boundary for the different species. 
\end{remark}

\begin{remark}
We presented the integrable left boundary matrices of the multi-species ASEP but the integrable right boundary matrices can be directly
deduced from them. They are indeed obtained from a right reflection matrix $\overline K(z)$ which satisfies the reflection equation 
\eqref{eq:reflection_equation_reversed}. Due to the symmetry relation (for the $R$-matrix of the multi-species ASEP)
\begin{equation} \label{eq:mASEP_symmetry_R}
 R_{21}(z)= U_1 U_2 R_{12}(z) U_1 U_2, \mbox{ with } U=\begin{pmatrix}
          &&1\\
          &\iddots&\\
          1&&
         \end{pmatrix},
\end{equation}
 the matrix $\overline K(z)$ can be constructed from a solution $K(z)$ of the reflection equation \eqref{eq:reflection_equation} through the relation
\begin{equation}\label{eq:mASEP_rel_K_Kbar}
\overline K(z)=U K\left(1/z\right)\,U. 
\end{equation}
A right boundary matrix $\overline B$ is obtained with the relation
\begin{equation}
 \overline B=-\frac{q-p}{2}\overline K'(1).
\end{equation}
Explicitly, the right boundary matrices are deduced from relation \eqref{eq:mASEP_rel_K_Kbar} 
and take the form 
\begin{equation}
\overline B(\beta,\delta|s_1', s_2', f_2', f_1') = U\,B(\beta,\delta|s''_1, s''_2, f''_2, f''_1)\,U
\end{equation}
where $U$ is defined in \eqref{eq:mASEP_symmetry_R}. The conjugation by $U$ implies
$f''_j=N-s'_j$ and $s''_j=N-f'_j$, $j=1,2$.

 Let us stress that the parameters entering $\overline B$ are independent from the ones used in the left boundary $B$.
Altogether we will have four real parameters: $\alpha,\gamma$ for the left boundary, and $\beta,\delta$ for the right one. 
In the same way, the labels $s_1', s_2', f_2', f_1'$ of the four special species in the right boundary are independent from the four special
species labels $s_1, s_2, f_2, f_1$ in the left boundary. 

The bijection between right and left boundaries can be seen in the following identity
\begin{equation}
\overline B(\beta,\delta|s_1, s_2, f_2, f_1)\equiv  B(\delta,\beta|s_1, s_2, f_2, f_1)\Big|_{z\leftrightarrow \widetilde z}
\end{equation}
where $z\leftrightarrow \widetilde z$  means that we interchange $\beta$ with $\widetilde \beta$ and $\delta$ with $\widetilde\delta$. 
As in the case of left boundaries, we use the notation
\begin{equation}
\widetilde \beta =  \frac{(\beta + \delta + q - p)\beta }{\beta + \delta}
\mbox{ and }\widetilde \delta =  \frac{(\beta + \delta + q - p)\delta }{\beta + \delta}.
\end{equation}
\end{remark}

\begin{example}
For the case $N=1$, we recover the one-species ASEP. We get only one possible choice for $s_1$, $s_2$, $f_1$ and $f_2$ 
given by $s_1=s_2=0$ and $f_1=f_2=1$. Then, in the language used in this paragraph, the particle 0 (vacancy) is slow and the particle 1 is fast
and the rates at the boundary are given by 
\begin{equation}
0\ \xrightarrow{\alpha} \ 1
\mbox{ and } 1\ \xrightarrow{\widetilde\gamma} \ 0.
\end{equation}
 One recovers that for the one-species ASEP,
the generic boundary is integrable. The boundary matrix has the form
\begin{equation}
B=\begin{pmatrix} -\alpha & \widetilde\gamma \\ \alpha & -\widetilde\gamma \end{pmatrix}.
\end{equation}
One can use Bethe ansatz method to compute the eigenvalues and compute for example the spectral gap \cite{DeGierE06}.

Conjugation by a diagonal matrix provides the non-Markovian boundary matrix used to compute the cumulant of the 
current \cite{GorissenLMV12}:
\begin{equation}
B(s)=\begin{pmatrix} -\alpha & e^s\,\widetilde\gamma \\ e^{-s}\,\alpha & -\widetilde\gamma \end{pmatrix}.
\end{equation}
It still corresponds to an integrable boundary.
\end{example}
 
\begin{example} 
For the case $N=2$, we obtain the two-species ASEP. There are four possibilities summarized in table \ref{table:mASEP_N2}.
We recover the boundaries found in \cite{CrampeMRV15}.
\begin{table}[htb] 
\begin{center}
   \begin{tabular}{|c|c|c|c|c|}
     \hline
     &$s_1=s_2=0$ & $s_1=s_2=1$  &  \multicolumn{2}{c|}{$s_1=s_2=0$}     \\
     &$f_1=f_2=1$ & $f_1=f_2=2$  &  \multicolumn{2}{c|}{$s_1=s_2=2$ }    \\
     \hline
\begin{tabular}{c} Type of  \\   part. \end{tabular}  
&\begin{tabular}{c} part. 0 slow  \\ part. 1 fast \\ part. 2 very fast \end{tabular}  
& \begin{tabular}{c} part. 0 very slow \\ part. 1 slow \\ part. 2 fast \end{tabular}  
& \multicolumn{2}{c|}{\begin{tabular}{c}part. 0 slow \\ part. 1 intermediate \\ part. 2 fast \end{tabular}  }\\
     \hline
    & \rule{0ex}{3.5ex}0\ $\xrightarrow{\alpha}$\ 1    & 0\ $\xrightarrow{\gamma}$\ 1
      && 0\ $\xrightarrow{\alpha}$\ 2 \\
Rates     & 1\ $\xrightarrow{\widetilde\gamma}$\ 0& 0\ $\xrightarrow{\alpha}$\ 2
      &  0\ $\xrightarrow{\alpha}$\ 2& 1\ $\xrightarrow{\widetilde\gamma}$\ 0\\
     & 2\ $\xrightarrow{\widetilde\gamma}$\ 0& 1\ $\xrightarrow{\alpha}$\ 2
      & 2\ $\xrightarrow{\widetilde\gamma}$\ 0& 1\ $\xrightarrow{\alpha}$\ 2\\
     & 2\ $\xrightarrow{\widetilde\alpha}$\ 1& 2\ $\xrightarrow{\widetilde\gamma}$\ 1
      &&2\ $\xrightarrow{\widetilde\gamma}$\ 0\\
      \hline
     \rule{0ex}{3ex}Name in \cite{CrampeMRV15}&$L_1$&$L_2$&$L_4$&$L_3$\\
      \hline
   \end{tabular}
   \caption{\label{table:mASEP_N2} The four integrable boundaries in the case $N=2$. The last row corresponds 
   to the names of these boundaries in \cite{CrampeMRV15}.}
\end{center}
\end{table}
\end{example}

\begin{example}
Some of the boundary matrices can be related to former studies 
of boundary Hecke algebras \cite{DipperJ92,Lambropoulou99}. In our notation, they correspond to 
the matrices $B(\alpha,\gamma|0,s_2,N-s_2,N)$ or $B^0(\alpha,\gamma|0,s_2,N-s_2,N)$. Among them, some have been 
considered: $B^0(\alpha,\gamma|0,1,N-1,N)$ was analyzed in \cite{Mandelshtam15}, 
and for the two-species ASEP ($N=2$) $B^0(\alpha,\gamma|0,0,2,2)$ was studied in \cite{Uchiyama08,CorteelMW15,Cantini15}.
\end{example}

\begin{proposition} \label{pro:mASEP_bbHecke}
For any matrix $B = B(\alpha, \beta | s_1, s_2, f_2, f_1)$ or $B = B^0(\alpha, \beta | s_1, s_2, f_2, f_1)$, the generators
\begin{equation}\label{eq:e0NASEP}
    \sigma_0 = \frac{B + \alpha + \gamma + q - p}{p - q}
    \mbox{ and } \sigma_1 = (m+q)/\sqrt{pq}
\end{equation}
obey relation \eqref{eq:bbHecke_braid}, where $m\equiv m_{12}$ is given in \eqref{eq:mASEP_m} and $B$ acts non trivially in space 1.
\end{proposition}

\proof
The matrices $\sigma_1$ and $\sigma_0$ given in \eqref{eq:eiNASEP} and \eqref{eq:e0NASEP} act
on two site multi-species ASEP configurations.
For a given start state, $\tau \tau'$, we can find a subset of the particle species
$\cS = \{\tau, \tau', \tau'', \ldots \}$ such that for \textit{any} polynomial in $\sigma_1$ and $\sigma_0$
acting on this state, these are the only species involved in the resulting configurations.

For all of the boundary matrices we consider, the subset $\cS$ turns out to be small, and related to the
different classes of particles we introduced above: the non-diagonal part of $\sigma_1$ exchanges particles on
sites 1 and 2, as allowed by bulk matrix $m$; the non-diagonal part of $\sigma_0$ injects and removes particles
at site 1 as allowed by the boundary transitions given previously. The idea of the proof is
then to project the `global' matrices $\sigma_0$, $\sigma_1$ down to the smaller number of species in $\cS$.  If for
every starting state we can show that the resulting projected $\sigma_0$, $\sigma_1$ satisfy \eqref{eq:bbHecke_braid}, then
this implies that the `global' matrices also satisfy \eqref{eq:bbHecke_braid}.

At this point, the proof decomposes into different steps:
\begin{itemize}
    \item
We remark that for any start state $\tau \tau'$, the set $\cS$ falls into one of three categories:
\begin{eqnarray}
&& \mathcal{S} = \{\tau, \tau', s_1,s_2, f_1,f_2\},\\
&& \mathcal{S} = \{\tau, s_1+ f_1-\tau,\tau',s_1+ f_1-\tau' \},\\
&& \mathcal{S} = \{\tau, s_1+ f_1-\tau,\tau', s,f\},\mbox{ with } (s,f)=(s_1, f_1) \mbox{ or } (s_2, f_2)
   \label{eq:mASEP_Smixed}
\end{eqnarray}
Note that these sets can be reduced depending on the class of the species $\tau$ and $\tau'$. For instance,
if $\tau$ and $\tau'$ are of very slow class, then $\mathcal{S} = \{\tau, \tau', s_1,f_1\}$. Note also
that the ordering of the start state does not change $\cS$ so $\tau$, $\tau'$ are interchangeable in
\eqref{eq:mASEP_Smixed}.
    \item
Projecting the boundary matrix, $B$, corresponding to $\sigma_0$ down to the species in $\cS$ results in a boundary
matrix of size $|\cS|$ of type \eqref{eq:mASEP_Bempty} or \eqref{eq:mASEP_Bfull}.  To see this, we perform the projection
by `deleting' species from $B$ by removing the corresponding row and column: 
 we  use  the following operations which preserve the forms \eqref{eq:mASEP_Bempty} or \eqref{eq:mASEP_Bfull}:
    \begin{itemize}
        \item
    Deleting any species in the very slow, intermediate, or very fast class;
        \item
    Deleting a species, $\widetilde\tau$, in the slow or fast class with $\widetilde\tau \ne s_1, f_1$ if we also delete
    the species $s_1 + f_1 - \widetilde\tau$;
        \item
    Deleting species $s_1$ and $f_1$ together, if $s_1 = 1$, $f_1 = N$, and $f_1 - f_2 = s_2 - s_1 > 0$.
        \item
    Deleting species $s_2$ and $f_2$ together, if $f_2 = s_2 + 1$, and $f_1 - f_2 = s_2 - s_1 > 0$.
    \end{itemize}
    These operations are always sufficient to project down to any subsets $\cS$ as defined above.
The projected $\sigma_0$ is then obtained from the projected $B$ through \eqref{eq:e0NASEP}.
    \item
For the local bulk matrix $m$ (giving $\sigma_1$) we can delete any number of species, preserving the form
\eqref{eq:mASEP_m}.
    \item
To complete the proof all we need to do is to verify that all boundary matrices in this family give $\sigma_0$
matrices which satisfy  \eqref{eq:bbHecke_braid} for size $2$ up to $6$ (the maximum $|\cS|$).  We have done this by a direct
computation with a formal mathematical software package.
\end{itemize}
To illustrate the projection on $\cS$, we consider the following boundary matrix
\begin{eqnarray}
&&B=
\left(
\begin{array}{c|cc|c|cc|c}
 \mbox{-}\sigma & & & & & &
\\
\hline
\gamma & \mbox{-}\alpha && & &\widetilde\gamma & \widetilde\gamma
\\
 & &\mbox{-}\alpha & \widetilde\gamma & \widetilde\gamma & &
\\
\hline
 & & & \mbox{-}\sigma' & &&
\\
\hline
 & &\alpha & \alpha & \mbox{-}\widetilde\gamma & &
\\
 \alpha& \alpha& & & &\mbox{-}\widetilde\gamma & \widetilde\alpha
 \\
 \hline
 & & & & & & -\widetilde\sigma
\end{array}
\right)
\end{eqnarray}
and give some examples of start state $(\tau,\tau')$ and the resulting subset $\cS$ and corresponding reduced matrix.
In the case where $(\tau,\tau')=(0,3)$, we obtain $\cS=\{0,3,s_1=1,s_2=2,f_1=5,f_2=4\}$ and the reduced matrix reads
\begin{equation}
\left(
\begin{array}{c|cc|c|cc}
 \mbox{-}\sigma & & & & & 
\\
\hline
\gamma & \mbox{-}\alpha && & &\widetilde\gamma 
\\
 & &\mbox{-}\alpha & \widetilde\gamma & \widetilde\gamma & 
\\
\hline
 & & & \mbox{-}\sigma' & &
\\
\hline
 & &\alpha & \alpha & \mbox{-}\widetilde\gamma & 
\\
 \alpha& \alpha& & & &\mbox{-}\widetilde\gamma 
\end{array}
\right).
\end{equation}
In the case where  $(\tau,\tau')=(1,5)$, we obtain $\cS=\{1,5\}$ and the reduced matrix reads
\begin{equation}
\begin{pmatrix} \mbox{-}\alpha & \widetilde\gamma \\ \alpha & \mbox{-}\widetilde\gamma \end{pmatrix}.
\end{equation}
Finally, in the case where $(\tau,\tau')=(2,3)$, we obtain $\cS=\{2,3,4\}$ and the reduced matrix reads
\begin{equation}
\begin{pmatrix} \mbox{-}\alpha & \widetilde\gamma& \widetilde\gamma \\ 
0 &\mbox{-}\sigma' &0 \\ \alpha & \alpha & \mbox{-}\widetilde\gamma \end{pmatrix}.
\end{equation}
\finproof

From theorem \ref{thm:bbHecke_main}, we can now give an expression for the Baxterised $K$-matrices of the multi-species ASEP.

\begin{proposition}
 For any matrix $B = B(\alpha, \beta | s_1, s_2, f_2, f_1)$ or $B = B^0(\alpha, \beta | s_1, s_2, f_2, f_1)$, the Baxterised $K$-matrix defined by
 \begin{equation}\label{eq:mASEP_K}
    K(z) = \frac{(\alpha + \gamma + q - p)(\frac{1}{z} - 1) + q - p}{(\alpha + \gamma + q - p)(z - 1) + q - p}
           \left(\frac{1 - (z - 1)\sigma_0}{1 - (\frac{1}{z} - 1) \sigma_0}\right),
\end{equation}
where
\begin{equation}
    \sigma_0 = \frac{B + \alpha + \gamma + q - p}{p - q},
\end{equation}
satisfies the reflection equation, the unitarity, regularity and Markovian properties. Moreover, the normalization factor in \eqref{eq:mASEP_K}
ensures that we have the relation
\begin{equation}
 B= \frac{q-p}{2}K'(1).
\end{equation}
\end{proposition}
\proof
It is straightforward to check the unitarity and regularity properties. We deduce immediately from theorem \ref{thm:bbHecke_main} and 
proposition \ref{pro:mASEP_bbHecke} that the $K$-matrix satisfies the reflection equation. The Markovian property is deduced from the fact that
the sum of the entries of each column of $B$ is $0$. The last relation is obtained by direct computation.
\finproof

\begin{remark}
 The $K$-matrix given in the previous proposition can be more precisely evaluated (the resulting expression does not contain an inverse anymore)
 \begin{align}
    & K(z) = 1 + k(z)\,\Big(b_0 + z \,b_0^+ + \frac1z \,b_0^-\Big),\label{eq:mASEP_K_developed}\\
    & \mbox{with}\quad 
k(z) = \frac{\left(z^2 - 1\right) \left(\alpha + \gamma\right)}
      {\left(\gamma z + \alpha\right)\left((\alpha + \gamma)(z - 1) + (q - p)z \right)}.
\end{align}
The matrices $b_0$, $b_0^+$ and $b_0^-$ are such that $B=b_0 + b_0^+ + b_0^-$ and are given by
\begin{equation}
b_0^+=\left(\begin{array}{ccc|c|c|c|c} \mbox{-}\gamma& & & & & & \\ & \ddots& & & & & \\ & &\mbox{-}\gamma & & & & \\ 
\hline \gamma& \cdots& \gamma& & & & \\ 
& & & & & & \\ \hline & & & & & & \\ \hline & & & & & &\\ \hline & & & & & & \end{array}\right)
\mbox{ and }
b_0^-=\left(\begin{array}{c|c|c|c|ccc} & & & & & & \\ \hline & & & & & & \\ \hline & & & & & & \\ 
\hline& & & & & & \\ 
& & & & \widetilde\alpha& \cdots&\widetilde\alpha \\  \hline & & & & \mbox{-}\widetilde\alpha& & \\ & & & & & \ddots & \\  & & & & & & \mbox{-}\widetilde\alpha
\end{array}\right)
\label{eq:bpm}
\end{equation}
where we draw symbolically the lines corresponding to the four special types of particles,
to indicate which part of the matrix we picked up in the boundary matrix to construct $b_0^\pm$. 
Again, the empty spaces are all filled with zeros. The remaining part is $b_0=B-(b_0^+ + b_0^-)$, 
where $B$ is either \eqref{eq:mASEP_Bempty} or \eqref{eq:mASEP_Bfull}.
Note that the decomposition is done in such a way that each matrix $b_0$, $b_0^\pm$ is Markovian.

The expression \eqref{eq:mASEP_K_developed} of the $K$-matrix can be established using the algebraic relations
\begin{equation}\label{poly:e0b}
\begin{aligned}
    & b_0^2 = -\left(\alpha + \tilde{\gamma}\right) b_0 + \tilde{\alpha} b_0^+ + \gamma b_0^-,
      \qquad
      \left(b_0^+\right)^2 = - \gamma b_0^+,
      \qquad
      \left(b_0^-\right)^2 = -\tilde{\alpha} b_0^-,
      \\
    & b_0 b_0^+ = b_0^+ b_0 = -\alpha b_0^+,
      \qquad
      b_0 b_0^- = b_0^- b_0 = -\tilde{\gamma} b_0^-,
      \\
    &
      b_0^+ b_0^- = b_0^- b_0^+ = 0.
\end{aligned}
\end{equation}
The expression \eqref{eq:mASEP_K_developed} makes connection, in the single species case $N=1$,
with the Baxterisation of the boundary Hecke algebra \eqref{eq:bHecke_Kmatrix}.
\end{remark}

\subsection{Diagonalization of the transfer matrix}

The diagonalization of the transfer matrix in the open case is much harder than in the periodic case. The main reason is the fact that the 
particles number may not be conserved by the dynamics on the boundaries (even if the particles number is conserved by the bulk dynamics). 
The situation is thus different from the periodic case where the diagonalization had been achieved at fixed number of particles, separately
in each sector (we recall that the transfer matrix was block diagonal). The most commonly used methods in the periodic case, the coordinate 
Bethe ansatz and the algebraic Bethe ansatz cannot be applied directly here.

However the coordinate Bethe ansatz had been modified successfully in some particular cases, 
such as model with diagonal boundaries \cite{Gaudin71}. More recently it had been applied to
the open single species ASEP \cite{Simon09,CrampeRS10}, providing partial results on the eigenvectors.

The algebraic Bethe ansatz had also been adapted successfully to diagonalize transfer matrices of certain open models, such as the open XXX and 
XXZ spin chain, first for triangular boundaries \cite{BelliardCR13}, and later for general boundaries \cite{BelliardC13,AvanBGP15}. 
It has also been applied to the open single species TASEP \cite{Crampe15}.

We present in the example below how the results of the algebraic Bethe ansatz for the open XXZ spins chain can be used to compute 
the spectrum of the DiSSEP with open boundaries.

\begin{example}
We encountered the DiSSEP several times in this manuscript, through the presentation of its local jump operators $m$ in
\eqref{eq:DiSSEP_m} and $B$, $\overline B$ in \eqref{eq:ASEP_B_Bb}, and then through the associated $R$-matrix \eqref{eq:DiSSEP_R} 
and $K$-matrices \eqref{eq:DiSSEP_K} and \eqref{eq:DiSSEP_Kb}. This proved the integrability of the model.
The reader may refer to subsection \ref{subsec:DiSSEP} for details about the dynamics and properties of the DiSSEP.
The integrability of this model is also revealed through its unexpected connexion with the XXZ spins chain. To be more precise, let us introduce the 
following quantum Hamiltonian $H$
\begin{eqnarray}
  H&=&(\alpha-\gamma)\sigma_1^+ -\frac{\alpha+\gamma}{2}(\sigma^z_1+1)\ +\ (\delta-\beta)\sigma_L^+ -\frac{\delta+\beta}{2}(\sigma^z_L+1)\nonumber\\
  &&-\frac{\lambda^2-1}{2}\ \sum_{k=1}^{L-1}
 \Big( \sigma_k^x\sigma_{k+1}^x+\sigma_k^y\sigma_{k+1}^y-\frac{\lambda^2+1}{\lambda^2-1}(\sigma_k^z\sigma_{k+1}^z-1) \Big)\label{eq:DiSSEP_H_XXZ}
\end{eqnarray}
where $\sigma^{x,y,z,+,-}$ are the Pauli matrices. It corresponds to the open XXZ spins chain with upper triangular boundaries.
This Hamiltonian $H$ is conjugated to the Markov matrix $M$ of the DiSSEP defined as usual by
\begin{equation}
 M=B_1+\sum_{k=1}^{L-1} m_{k,k+1} + \overline{B}_L.
\end{equation}
Namely, one has
\begin{equation}\label{eq:DiSSEP_conjugation_XXZ}
  H=Q_1Q_2\dots Q_L M Q_1^{-1}Q_2^{-1}\dots Q_L^{-1}\quad\text{where}\quad Q=\begin{pmatrix} -1&1\\1&1 \end{pmatrix} \;.
\end{equation}
Let us also mention that the XXZ Model for particular choices of boundaries is conjugated to the Markov matrix of the open ASEP. 
However, for the boundaries 
present in \eqref{eq:DiSSEP_H_XXZ}, the conjugation provides non-Markovian boundaries (the sum to $0$ property is not fulfilled).  

We deduce from \eqref{eq:DiSSEP_conjugation_XXZ} that the spectrum of $M$ is identical to the one of $H$. Moreover, 
the eigenvalues (but not the eigenvectors) of XXZ spin chain with upper triangular boundaries are the same that the ones for diagonal boundaries 
and one can use the results of \cite{Gaudin83,Sklyanin88,AlcarazBBBQ87}. 
Note also that for $\lambda^2=1$, the bulk Hamiltonian becomes diagonal, and the full Hamiltonian triangular, 
allowing to get its spectrum easily without Bethe ansatz, in accordance with the results of subsection \ref{subsec:DiSSEP}.

The eigenvalues of $H$ with diagonal boundaries
can be parametrized in two different ways depending on the choice of the pseudo-vacuum:
\begin{itemize}
 \item For the pseudo-vacuum with all the spins up and in the notations of the present paper,
 the eigenvalues of $H$ are given by
 \begin{equation}\label{eq:DiSSEP_eigenvalue1}
  E=-\alpha-\beta-\gamma-\delta+4(\phi-1)^2 \sum_{i=1}^N \frac{u_i}{(u_i-\phi^2)(u_i-1)}
 \end{equation}
where $N=0,1,\dots,L$ and $u_i$ are the Bethe roots. The Bethe roots must satisfy the following Bethe equations
\begin{eqnarray}\label{eq:DiSSEP_bethe_equations1}
\frac{u_i+a\phi^2}{\phi(au_i+1)}\ \frac{u_i+b\phi^2}{\phi(bu_i+1)} \left(\frac{\phi(u_i-1)}{u_i-\phi^2} \right)^{2L}=
\prod_{\genfrac{}{}{0pt}{}{j=1}{j\neq i}}^N \frac{\phi^2(\phi^2u_i-u_j)(u_iu_j-1)}{(u_i-\phi^2u_j)(u_iu_j-\phi^4)}
\end{eqnarray}
where $i=1,2,\dots,N$ and $a$ and $b$ are defined in \eqref{eq:DiSSEP_boundaries_relations_G}.
\item
For the pseudo-vacuum with all the spins down, the eigenvalues of $H$ are given by
 \begin{equation}\label{eq:DiSSEP_eigenvalue2}
  E=4(\phi-1)^2 \sum_{i=1}^N \frac{v_i}{(v_i-\phi^2)(v_i-1)}
 \end{equation}
where $v_i$ satisfy the following Bethe equations
\begin{eqnarray}\label{eq:DiSSEP_bethe_equations2}
\frac{av_i+\phi^2}{\phi(v_i+a)}\ \frac{bv_i+\phi^2}{\phi(v_i+b)} \left(\frac{\phi(v_i-1)}{v_i-\phi^2} \right)^{2L}=
\prod_{\genfrac{}{}{0pt}{}{j=1}{j\neq i}}^N \frac{\phi^2(\phi^2v_i-v_j)(v_iv_j-1)}{(v_i-\phi^2v_j)(v_iv_j-\phi^4)}\;.
\end{eqnarray}
\end{itemize}

Let us stress again that, although the spectrum of the XXZ spin chain is the same for diagonal or upper boundaries, the eigenvectors are different. 
For the XXZ spin chains with upper triangular boundaries, the eigenvectors associated to the parametrization \eqref{eq:DiSSEP_eigenvalue1} 
and \eqref{eq:DiSSEP_bethe_equations1} of the eigenvalues 
were computed only recently by algebraic Bethe ansatz in \cite{Belliard15} based on the previous results for the XXX spin chain 
\cite{BelliardCR13,BelliardC13}. The computation of the eigenvectors associated to the parametrization \eqref{eq:DiSSEP_eigenvalue2} 
and \eqref{eq:DiSSEP_bethe_equations2} is still an open problem.
\end{example}

In the past few years, other techniques had been used to tackle this diagonalization problem of open systems. 
In the spirit of functional Bethe ansatz, the Off-diagonal Bethe ansatz was developed and successfully applied to several
open models \cite{CaoYSW13,WenYCCY15}.

The Baxter Q-operator idea was adapted, using matrix product expressions obtained through infinite dimensional representations of RTT algebras, 
to the (current counting deformation) of the open ASEP, in a pioneering work \cite{LazarescuP14}. 

New ideas also emerged to deal with these open systems. The separation of variables method, inspired from the 
action-angle variables construction in classical integrable systems, was successfully used to solve several models 
\cite{FrahmSW08,Niccoli12,FaldellaKN14,KitanineMN14}, 
see also \cite{Sklyanin95} for an introduction. This method could provide an unifying framework for the exact resolution of both 
classical and quantum integrable systems, but a lot remains to be understood.

Finally, the q-Onsager algebra provided another point of view on the exact resolution of open models \cite{BaseilhacK07,BaseilhacB13}.

All the diagonalization techniques presented in this subsection have proven to be efficient on particular models but may 
appear difficult to adapt to new models. A somehow hard analysis has to be perform each time. Moreover these methods 
provides the spectrum of the model parametrized with Bethe roots, which have to satisfy the Bethe equations. We saw previously that these Bethe  
equations are non-linear coupled algebraic relations, which appear (apart from very particular cases) impossible to solve exactly. 
These equations can be analyzed for large system sizes, {\it i.e} in the thermodynamic limit (see chapter \ref{chap:five} for details).
They have been studied for periodic models,
introducing for instance the Bethe roots density \cite{YangY66bis,Takahashi71,Gaudin71,Baxter82}, which yields exact results for physical observables.
However, for finite size systems, the exact computations of physical quantities cannot be performed completely, they are always expressed in 
terms of Bethe roots. 

In the following chapter we present an alternative method, which can be applied to integrable Markovian processes: the matrix ansatz.
This method provides the stationary state of the model analytically. Its huge advantage is that it does not require to solve any Bethe equations and
gives access to exact expressions of physical observables for finite size systems. Moreover its range of application seems to be quite large and, 
in some sens, quite model independent: it seems to be possible to use it for any integrable Markovian exclusion process at the price of 
solving two key relations, the Zamolodchikov-Faddeev and the Ghoshal-Zamolodchikov relations. However this method gives only one eigenvector of the 
transfer matrix, the stationary state, and does not provide the complete spectrum and eigenvectors of the model. We do not have access to the full 
dynamics of the model but only to the stationary state. Moreover this method had been applied, so far, only on Markovian models (or on current-counting
deformation of the Markovian models) but not on general integrable model like quantum spin chains.

\chapter{Matrix ansatz for non-equilibrium steady states} \label{chap:three}

\section{Presentation of the method and link with integrability} \label{sec:Matrix_ansatz}

The matrix ansatz (also called matrix product state) has become over the last decades an incredibly powerful tool in a lot of 
different fields. In physics, this method can be roughly described as a way to encode spatial correlations in probability distributions, and 
thus goes far beyond the mean-field approximation. In mathematics such states could be interpreted as generalized characters associated to representation
of various algebras and endorse the role of a keystone at the interplay between combinatorics, representation theory, integrability and
stochastic processes.
The matrix product states were simultaneously and independently introduced in the context of one dimensional Markovian systems on one side and of
one dimensional quantum Hamiltonians on the other side. 

In statistical physics of classical systems, the matrix ansatz was developed to express exactly the steady state of stochastic processes 
describing particles in interaction. It was used for the first time in \cite{DerridaEHP93} and led to a rigorous description of 
phase transitions in an out-of-equilibrium system \cite{DerridaDM92,SchutzD93}.
This chapter will be devoted to this topic, more details and references about 
the relevant literature are given in the next section.

At the same time, in quantum physics, the Density Matrix Renormalization Group (DMRG) algorithm \cite{White92} has proven to be very efficient 
in computing, with a high numerical accuracy, the ground state and low energy states of one dimensional Hamiltonian with short range interactions 
\cite{WhiteH93, SorensenA93}. 
The convergence and validity of this algorithm has been proven to be directly related to the existence of a matrix product expression 
for the eigenvectors under consideration \cite{RommerO97}.
During the same period, on the exact result side, the ground state of several one dimensional Hamiltonians had been analytically 
expressed with a matrix ansatz \cite{FannesNW89, FannesNW92, LangeKZ94}, which permitted exact computation of physical quantities such as 
correlation lengths. More recently, in statistical physics of open quantum systems,
very interesting results have been obtained in the context of the Lindblad equation (that describes a quantum system coupled to reservoirs) where
the density matrix has been expressed analytically in a matrix product form \cite{Prosen11, KarevskiPS13, Prosen14}. These works also pointed 
out the link of the matrix ansatz with integrability.

We can also mention that, in integrable models, the eigenvectors obtained with the algebraic Bethe ansatz technique can also be reformulated using
matrix product states \cite{AlcarazL04, GolinelliM06}. It is also important to point out that recent progress have been made in integrable 
models by computing explicitly Baxter Q operator in a matrix product form \cite{LazarescuP14} and \cite{BazhanovFLMS11}. In these works the Q operator is constructed as
a generalized transfer matrix with infinite dimensional auxiliary space.

In representation theory and combinatorics, recent progress has been made in expressing Macdonald polynomials (which form
a family of multivariate polynomials, symmetric under permutation of variables, and containing 
Schur, Hall-Littlewood and Jack polynomials as specializations) in a matrix product form \cite{CantiniDGW15}. 
This led in particular to explicit formulas and sum rules \cite{DeGierW16, CantiniDGW16}. This matrix product expression gave also 
the framework to introduce a new family of polynomials \cite{GarbaliDGW16}, that encompasses the Macdonald polynomials.
In a similar way Koornwinder polynomials\footnote{these polynomials will be defined in details in chapter \ref{chap:four}} 
(which form a family of multivariate Laurent polynomials, symmetric under permutation and
inversion of variables) have been expressed in some particular cases in a matrix product form \cite{Cantini15, FinnV16}. 
All these works shed light on connections and interplay between symmetric polynomials, representation theory of Hecke algebras, 
quantum Knizhnik-Zamolodchikov equations and integrable stochastic processes. The chapter \ref{chap:four} of this manuscript will be devoted 
to this topic.

Finally, in knot theory the study of the so-called ``quantum invariants'' initiated by the celebrated Jones polynomial \cite{Jones85} appeared as a 
small revolution in the field. These knot invariants are polynomials constructed in a matrix product form using representations of the braid group and 
its quotients (Hecke algebra, Birman-Murakami-Wenzl algebra,...). 

In the present chapter we will focus on the role of the matrix ansatz for exact solvability in classical non-equilibrium steady states. We will 
particularly investigate its link with integrability. The idea of the method will be introduced on the specific example of the TASEP from which 
we will draw the general framework. Some new examples will be given as an illustration.

\subsection{General idea and example}

\subsubsection{General idea}

In the context of exact results in out-of-equilibrium statistical physics, 
the matrix ansatz is a technique introduced to express analytically the stationary distribution of certain exclusion processes.
It was first successfully used in the seminal paper \cite{DerridaEHP93} to compute exactly the stationary state of the TASEP. 
Since this pioneering work, it has been widely used in the context of Markovian dynamics. During the past few years
the computation of the stationary distribution of the multi-species (T)ASEP on a periodic lattice has attracted a lot of interest. 
Initiated in \cite{DerridaEHP93} and \cite{MallickMR99} for respectively two and three species of particles,
the steady state was first expressed in the general case in a matrix product form \cite{EvansFM09, ProlhacEM09, AritaAMP11, AritaAMP12} 
by reformulating a pushing procedure. Another matrix product solution
was found later by studying the interplay between the steady state of the multi-species ASEP and the Macdonald polynomials \cite{CantiniDGW15}.
The connection to three dimensional integrability of the multi-species TASEP also gave rise to another matrix product expression 
\cite{KunibaMO15, KunibaMO16}.
Several works tackled particular multi-species ASEPs with open boundaries, such as \cite{Arita12} for reflexive boundary conditions,
or \cite{Uchiyama08, AyyerLS09, AyyerLS12} for semi-permeable boundaries and finally \cite{CrampeMRV15, CrampeEMRV16} for integrable boundaries 
for 2-species TASEP.
Some generalizations of the multi-species TASEP with inhomogeneous hopping rates have been solved using a matrix ansatz on the periodic lattice 
\cite{AritaM13}, and also for very specific boundary conditions \cite{Karimipour99}. 
We can also point out here the matrix product solution of the ABC model on the ring with equal densities of each species \cite{EvansKKM98}.
In the simpler case of the multi-species SSEP, a matrix product expression of the steady state was recently proposed \cite{Vanicat17} for a particular 
class of integrable boundary conditions.
We can also mention some work related to single species ASEP-like models with discrete time dynamics \cite{DeGierN99, CrampeMRV15inhomogeneous}, 
or deterministic dynamics in the bulk with stochastic boundaries \cite{DeGier01} or with impurity \cite{HinrichsenS97}.
A matrix product expression of the steady state of a cellular automaton coupled
with stochastic boundaries was also recently found \cite{ProsenM16}.
In \cite{KunibaO16} the steady state of an integrable two-species zero range process, introduced in \cite{KunibaMMO16} 
was constructed in a matrix product form, 
generalizing the well known factorized stationary distribution of the single species zero range process \cite{EvansH05}.
The matrix ansatz has also been widely used to deal with reaction-diffusion stochastic models. Several coagulation-decoagulation models 
were solved this way in \cite{HinrichsenSP96} or in \cite{IsaevPR01} where a classification of such ``matrix product'' exactly solvable models 
is provided. Recently a model with pair creation and annihilation of particles was exactly solved using a matrix ansatz in \cite{CrampeRRV16}. 
The reader may refer to the review \cite{BlytheE07} for more details about the use of matrix product states in non-equilibrium steady states.

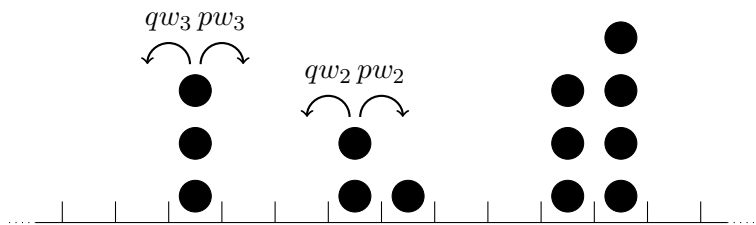
\begin{figure}[htb]
\begin{center}
 \begin{tikzpicture}[scale=0.7]
\draw (-1.5,0) -- (11.5,0) ; \draw[dotted] (-2,0) -- (-1.5,0) ; \draw[dotted] (11.5,0) -- (12,0) ; 
\foreach \i in {-1,0,...,11}
{\draw (\i,0) -- (\i,0.4) ;}
\draw (1.5,0.5) circle (0.3) [fill,circle] {};\draw (1.5,1.5) circle (0.3) [fill,circle] {};\draw (1.5,2.5) circle (0.3) [fill,circle] {};
\draw (4.5,0.5) circle (0.3) [fill,circle] {};\draw (4.5,1.5) circle (0.3) [fill,circle] {};
\draw (5.5,0.5) circle (0.3) [fill,circle] {};
\draw (8.5,0.5) circle (0.3) [fill,circle] {};\draw (8.5,1.5) circle (0.3) [fill,circle] {};\draw (8.5,2.5) circle (0.3) [fill,circle] {};
\draw (9.5,0.5) circle (0.3) [fill,circle] {};\draw (9.5,1.5) circle (0.3) [fill,circle] {};\draw (9.5,2.5) circle (0.3) [fill,circle] {};\draw (9.5,3.5) circle (0.3) [fill,circle] {};
\draw[->,thick] (1.4,3) arc (0:180:0.4); \node at (1.,3.8) [] {$qw_3$};
\draw[->,thick] (1.6,3) arc (180:0:0.4); \node at (2.,3.8) [] {$pw_3$};
\draw[->,thick] (4.4,2) arc (0:180:0.4); \node at (4.,2.8) [] {$qw_2$};
\draw[->,thick] (4.6,2) arc (180:0:0.4); \node at (5.,2.8) [] {$pw_2$};
 \end{tikzpicture}
 \end{center}
 \caption{Dynamical rules of the zero range process.}
 \label{fig:ZRP}
\end{figure}

The matrix product ansatz arises as a natural generalization of the factorized distribution
\begin{equation} \label{eq:factorized_distribution}
 \cS(\tau_1,\tau_2,\dots,\tau_L)=\frac{1}{Z}f(\tau_1)f(\tau_2)\dots f(\tau_L),
\end{equation}
where $f$ is a real function and $Z$ is a normalization such that the sum of the weights equals $1$. 
\begin{remark} \label{rem:ZRP}
 Despite its apparently simple form, the distribution \eqref{eq:factorized_distribution} has found many important applications
 in non-equilibrium steady states. A paradigmatic example is given by the stationary distribution of the (single species) zero range
 process on a ring \cite{Spitzer70}. In this model, there are no constraints on the number of particles lying on each site of the lattice, the particles
 can hop to the left (respectively right) nearest neighbor site with probability rate $q w_n$ (respectively $p w_n$), where 
 $n$ denotes the number of particles sitting on the departure site and $w_n$ is a non-negative number, see figure \ref{fig:ZRP}. The essential feature 
 of such models is that the hopping rates depend only on the departure site and not on the target site (in contrast with the exclusion processes).
 It thus describes a system in which the interactions between the particles have zero range. Several generalization of this model have been 
 proposed: with open boundaries \cite{LevineMS05}, with discrete time dynamics \cite{EvansMZ04,ZiaEM04}, 
 with continuous mass on each site \cite{EvansMZ04,ZiaEM04}, 
 with particle number constrained on each site \cite{SchutzRB96}, or with several species of particles \cite{KunibaO16}.
 The reader may refer to the review \cite{EvansH05} for more details.
\end{remark}

This distribution \eqref{eq:factorized_distribution} is that of independent and identically distributed random variables $\tau_1,\dots,\tau_L$ 
with probability distribution 
given by the (properly normalized) function $f$. This framework cannot describe any correlated variables. The fundamental idea of the 
matrix ansatz is to replace the real valued function $f$ by a matrix valued function. 
More precisely, to each local configuration variable $\tau_i$ is associated a matrix $X_{\tau_i}$. The variable $\tau_i$ can 
take $N+1$ different values $0,1,\dots,N$, hence there are $N+1$ different matrices\footnote{Note that this could be generalized to the case where 
the local configuration $\tau_i$ takes continuous values. We have in this case an infinity of matrices $X_t$, parametrized by a continuous parameter $t$}
$X_0,X_1,\dots,X_N$. 
This leads to the following definition.
\begin{definition} \label{def:Matrix_ansatz}
 A probability distribution $\cS(\tau_1,\tau_2,\dots,\tau_L)$, $0\leq \tau_i \leq N$, is said to have a matrix product form if there exist
 matrices $X_0,X_1,\dots,X_N$ and $U$ and a trace operator $tr$ (which satisfies the cyclic property $tr(AB)=tr(BA)$) with non-negative values such that
  \begin{equation} 
 \cS(\tau_1,\tau_2,\dots,\tau_L)=\frac{1}{Z_L}\mbox{tr}\left( U X_{\tau_1}X_{\tau_2} \dots X_{\tau_L}\right),
 \end{equation}
 where the normalization $Z_L$ ensures that the sum of the probabilities gives $1$. 
 In practice, when the model under consideration is irreducible (it implies in particular that none of the particle species number is conserved by the dynamics,
 which is often the case for models with open boundaries), $Z_L$ is equal to $\mbox{tr}(U C^L)$, with
 \begin{equation}
 C=X_0+X_1+\dots+X_N. 
 \end{equation}
 \end{definition}
 
 \begin{remark}
 The matrix product expression above encompasses the cases of systems with open boundaries or systems on a ring.
 
 For models defined on a lattice with open boundaries, the matrix $U$ is written using
 a row vector $\llangle W|$ and a column vector $|V\rrangle$ in the form $U=|V\rrangle \llangle W|$. 
 The distribution then reads
 \begin{equation} \label{eq:Matrix_ansatz}
 \cS(\tau_1,\tau_2,\dots,\tau_L)=\frac{1}{Z_L}\llangle W| X_{\tau_1}X_{\tau_2} \dots X_{\tau_L} |V\rrangle,
 \end{equation}
 where the normalization $Z_L$ is equal to $\llangle W| C^L |V\rrangle$, with
 \begin{equation}
 C=X_0+X_1+\dots+X_N. 
 \end{equation}
 
 For models defined on a periodic lattice, the matrix $U$ is taken equal to the identity matrix.
 The distribution then reads
   \begin{equation} 
 \cS(\tau_1,\tau_2,\dots,\tau_L)=\frac{1}{Z_L}\mbox{tr}\left(X_{\tau_1}X_{\tau_2} \dots X_{\tau_L}\right).
 \end{equation}
 In this case the normalization has often a more subtle expression if the particle number of the species is conserved by the dynamics: 
 we have to sum the weights of the lattice configurations that have a given particle number of each species. 
\end{remark}
The vectors  $\llangle W|$ and $|V\rrangle$ (or the trace operator $tr$) are needed to contract the matrix product to get a real valued probability.
We use the notation with a double bracket on the vectors $\llangle W|$ and $|V\rrangle$ to distinguish the vector space in which they belong 
(the vector space on which the matrices $X_\tau$ act linearly) from the vector space of physical configurations whose vectors are written
with a single bracket (for instance the vector $\steady$).

The matrix product form of a distribution can capture correlations between the local configuration variables through the non commutative 
structure of the algebra generated by $X_0,\dots,X_N$. The order of the matrices in the product is thus very important. We stress that 
this matrix product framework can thus be used to deal with systems that do not have Boltzmann statistics and that lie beyond the scope of the 
central limit theorem (which describes uncorrelated variables). This is for instance well illustrated by the work \cite{AngelettiBA14} 
in which the statistics of a sum of correlated variables (with matrix product distribution) was studied.

In order to express the stationary distribution of an exclusion process using this matrix product formalism, it is of course crucial to 
carefully choose the matrices $X_{\tau}$ and the vectors $\llangle W|$ and $|V\rrangle$ (for systems with open boundaries).
It turns out that the algebraic relations that these objects must
fulfill are closely related to the Markov matrix of the model. This last point will be discussed in details in the next sections.

The matrix ansatz formulation of the steady state reveals also to be very convenient to express the multi-point correlation functions in 
the stationary state. If we define the variable $\rho_s^{(i)}$ such that $\rho_s^{(i)}=1$ if there is a particle of species $s$ on 
site $i$ and $\rho_s^{(i)}=0$ otherwise, then the $k$-point correlation function is by definition equal to the quantity
$\langle \rho_{s_1}^{(i_1)}\dots \rho_{s_k}^{(i_k)}\rangle$, where the bracket $\langle \cdot \rangle$ stands for the expectation value in
the stationary distribution. The matrix product structure of the steady state allows us to write concisely
\begin{equation}
 \langle \rho_{s_1}^{(i_1)}\dots \rho_{s_k}^{(i_k)}\rangle =
 \frac{1}{Z_L}\llangle W|C^{i_1-1}X_{s_1}C^{i_2-i_1-1}X_{s_2} \dots C^{i_k-i_{k-1}-1}X_{s_k}C^{L-i_k}|V\rrangle. 
\end{equation}
In particular the mean density (one-point correlation function) of particle of species $s$ at site $i$ reads
\begin{equation}
 \langle \rho_s^{(i)} \rangle = \frac{1}{Z_L}\llangle W|C^{i-1}X_s C^{L-i}|V\rrangle.
\end{equation}
Note that in the single species case $N=1$, that we encountered many times with the examples of the ASEP, TASEP and SSEP along this manuscript,
$\tau_i=0,1$ and we have always $\rho_1^{(i)}=\tau_i$. In this single species case, to lighten the notations,
the multi-point correlation function will be simply written $\langle \tau_{i_1} \dots \tau_{i_k} \rangle$.

We now present the example of the TASEP to fix the ideas and get used to the different notions introduced. 

\subsubsection{Historical example of the TASEP}

The stationary distribution of the TASEP with open boundary conditions was analytically expressed using a matrix product form in the
pioneering work \cite{DerridaEHP93}. 
We recall that the Markov matrix governing the stochastic dynamics of this model is given in the form
\begin{equation} \label{eq:TASEP_Markov_matrix}
 M=B_1+\sum_{k=1}^{L-1}m_{k,k+1}+\overline{B}_L
\end{equation}
where $m$ has been introduced in \eqref{eq:TASEP_m} and $B$, $\overline{B}$ in \eqref{eq:TASEP_B_Bb}. We reproduce hereafter some of 
the results obtained in \cite{DerridaEHP93}.

\begin{proposition} \label{prop:TASEP_matrix_ansatz}
 The stationary distribution of the TASEP with open boundary condition \eqref{eq:TASEP_Markov_matrix} is given by the matrix product
 expression
 \begin{equation} \label{eq:TASEP_matrix_ansatz}
  \cS(\tau_1,\tau_2,\dots,\tau_L)=\frac{1}{Z_L}\llangle W| \prod_{i=1}^L \left(\tau_i D +(1-\tau_i)E\right) |V\rrangle,
 \end{equation}
 where the matrices $E$, $D$ and the boundary vectors $\llangle W|$ and $|V\rrangle$ satisfy the algebraic relations
 \begin{equation} \label{eq:TASEP_algebraic_relations}
  DE=D+E, \qquad \llangle W|E=\frac{1}{\alpha}\llangle W|, \qquad D|V\rrangle=\frac{1}{\beta}|V\rrangle,
 \end{equation}
 and the normalization is given by $Z_L=\llangle W|C^L|V\rrangle$, with $C=D+E$.
\end{proposition}
Note that in the case studied here of the single species TASEP, the local variables $\tau_i$ can only take two values, $0$ and $1$. The expression
\eqref{eq:TASEP_matrix_ansatz} is thus completely equivalent to the one in definition \ref{def:Matrix_ansatz} if we set $X_0=E$ and $X_1=D$ 
(we used the name $E$ and $D$ for the matrices $X_0$ and $X_1$ to stick to the notation introduced in \cite{DerridaEHP93}).
The proof of the proposition \ref{prop:TASEP_matrix_ansatz} will be given in subsection \ref{subsec:teslescopic_relations}.
We now recall an explicit representation of $E$, $D$, $\llangle W|$ and $|V\rrangle$ as infinite dimensional matrices and vectors respectively.
The operators $E$ and $D$ can be expressed as acting linearly on a Fock space, with basis denoted by $\{|k\rrangle\}_{k\geq 0}$
\begin{equation} \label{eq:TASEP_representation_DE}
 E=\sum_{k=0}^{+\infty} \Big[|k\rrangle \llangle k| + |k+1\rrangle \llangle k|\Big], \qquad 
 D=\sum_{k=0}^{+\infty} \Big[|k\rrangle \llangle k| + |k\rrangle \llangle k+1|\Big].
\end{equation}
In matrix notation it gives explicitly
\begin{equation} 
 E=\begin{pmatrix}
    1 & 0 & 0 & 0 & \hdots \\
    1 & 1 & 0 & 0 &  \\
    0 & 1 & 1 & 0 &  \\
    0 & 0 & 1 & 1 &  \\
    \vdots & & & & \ddots 
   \end{pmatrix},
   \qquad 
 D=\begin{pmatrix}
    1 & 1 & 0 & 0 & \hdots \\
    0 & 1 & 1 & 0 &  \\
    0 & 0 & 1 & 1 &  \\
    0 & 0 & 0 & 1 &  \\
    \vdots & & & & \ddots 
   \end{pmatrix}. 
\end{equation}
The boundary vectors $\llangle W|$ and $|V\rrangle$ are represented as
\begin{equation} \label{eq:TASEP_representation_vectors}
 \llangle W| = \sum_{k=0}^{+\infty} \left(\frac{1-\alpha}{\alpha}\right)^k \llangle k|, \qquad 
 | V \rrangle = \sum_{k=0}^{+\infty} \left(\frac{1-\beta}{\beta}\right)^k |k\rrangle,
\end{equation}
or more explicitly
\begin{equation}
 \llangle W| = \begin{pmatrix}
                1 & \left(\frac{1-\alpha}{\alpha}\right) & \left(\frac{1-\alpha}{\alpha}\right)^2 & \hdots
               \end{pmatrix}, \qquad
 |V\rrangle = \begin{pmatrix}
                1 \\ \left(\frac{1-\beta}{\beta}\right) \\ \left(\frac{1-\beta}{\beta}\right)^2 \\ \vdots
               \end{pmatrix}.
\end{equation}
It is straightforward to check that the expressions \eqref{eq:TASEP_representation_DE} and \eqref{eq:TASEP_representation_vectors} fulfill
the algebraic relations \eqref{eq:TASEP_algebraic_relations}. It proves that these algebraic relations are consistent. Moreover the 
scalar product $\llangle W|V\rrangle=\alpha\beta/(\alpha+\beta-1)$ is non-vanishing.

The matrix product expression of the stationary state of the open TASEP has proven to be a very powerful tool with which to compute physical quantities and
to study the macroscopic behavior of the system. It permitted, in particular, (see chapter \ref{chap:five}) the rigorous derivation of the phase diagram of the model.
We recall here the calculation of the normalization, of the particle current and of the particle density given in \cite{DerridaEHP93}.
We begin by showing the efficiency of the algebraic relations \eqref{eq:TASEP_algebraic_relations}, to analytically compute the stationary weights,
on a particular example:
\begin{eqnarray} \label{eq:TASEP_particular_weight}
 Z_5 \, \cS(0,1,0,1,1) & = & \llangle W| EDEDD |V\rrangle \\
 & = & \frac{1}{\alpha \beta^2} \llangle W| DE |V\rrangle \\
 & = & \frac{1}{\alpha \beta^2} \llangle W| D+E |V\rrangle \\
 & = & \frac{1}{\alpha \beta^2}\left(\frac{1}{\alpha}+\frac{1}{\beta}\right) \llangle W|V \rrangle.
\end{eqnarray}
The equality between line $1$ and line $2$ is obtained using the action of the matrices $E$ and $D$ on the boundary vectors $\llangle W|$ and 
$|V\rrangle$ respectively. We get the third line because of the algebraic relation $DE=D+E$. The last equality is again obtained because 
of the relations on the boundary vectors. 

\begin{proposition}
 The normalization $Z_L$ is given in a closed form by
 \begin{equation} \label{eq:TASEP_normalization}
  Z_L= \sum_{p=1}^L \frac{p(2L-1-p)!}{L!(L-p)!}
  \frac{\left(\frac{1}{\beta}\right)^{p+1}-\left(\frac{1}{\alpha}\right)^{p+1}}{\frac{1}{\beta}-\frac{1}{\alpha}}\llangle W|V\rrangle.
 \end{equation}
\end{proposition}
We choose to present now the detailed proof of formula \eqref{eq:TASEP_normalization} to illustrate the efficiency of the algebraic relations
to compute physical quantities.
\proof
Following the lines of \cite{DerridaEHP93}, the first step is to prove the relation
\begin{equation} \label{eq:TASEP_Cn}
 C^n=\sum_{p=0}^n B_{n,p} \sum_{q=0}^p E^q D^{p-q}, 
\end{equation}
where $B_{n,p}$ is a combinatorial factor given explicitly by
\begin{equation} \label{eq:TASEP_ballot_number}
 B_{n,p}=\frac{p(2n-1-p)!}{n!(n-p)!}.
\end{equation}
This relation is obtained by direct reordering of the matrices $E$ and $D$ in the product $(E+D)^n$. We push $E$ to the left (respectively
$D$ to the right) because it behaves conveniently on the left vector $\llangle W|$ (respectively on the right vector $|V\rrangle$). 
This can be achieved thanks to the fact that $DE=D+E$. The relation \eqref{eq:TASEP_Cn} is proved by induction.

The case $n=1$ is easily verified. Assume now that the formula \eqref{eq:TASEP_Cn} is established for a given $n\geq 1$.
We can then compute
\begin{equation}
 C^{n+1}=C^n C= \sum_{p=0}^n B_{n,p} \sum_{q=0}^p E^q D^{p-q} (E+D).
\end{equation}
It thus appears that we need to reorder the products of the form $D^k E$. It is straightforward to prove (by induction for instance)
that 
\begin{equation}
 D^k E=D^k+D^{k-1}+\dots+D+E.
\end{equation}
Hence, by elementary manipulations on sums, we obtain
\begin{equation}
 C^{n+1}=\sum_{p=1}^{n+1} \left(\sum_{l=p-1}^n B_{n,l}\right) \sum_{q=0}^p E^q D^{p-q}.
\end{equation}
We thus need to compute 
\begin{equation}
 \sum_{l=p-1}^n B_{n,l}=B_{n+1,p}.
\end{equation}
This last equality can be readily checked using the recursive relation $B_{n+1,p}=B_{n+1,p+1}+B_{n,p-1}$ and the fact that $B_{n,n}=1=B_{n+1,n+1}$.
At the end we obtain
\begin{equation}
 C^{n+1}=\sum_{p=0}^{n+1} B_{n+1,p} \sum_{q=0}^p E^q D^{p-q},
\end{equation}
which ends the induction proof of \eqref{eq:TASEP_Cn}.
We are now in position to contract $C^L$ with the boundary vectors $\llangle W|$ and $|V\rrangle$. Using the fact that
\begin{equation}
 \llangle W|E^q=\frac{1}{\alpha^q}\llangle W|, \quad \mbox{and} \quad D^{p-q}|V\rrangle=\frac{1}{\beta^{p-q}}|V\rrangle,
\end{equation}
and the formula
\begin{equation}
 \sum_{q=0}^p \frac{1}{\alpha^q}\frac{1}{\beta^{p-q}}=\frac{\frac{1}{\alpha^{p+1}}-\frac{1}{\beta^{p+1}}}{\frac{1}{\alpha}-\frac{1}{\beta}},
\end{equation}
we establish the desired result \eqref{eq:TASEP_normalization}.
\finproof

\begin{proposition}
The mean particle current between sites $i$ and $i+1$ is given by
\begin{equation} \label{eq:TASEP_current}
 \langle j \rangle= \frac{Z_{L-1}}{Z_L},
\end{equation}
where $Z_L$ is the normalization computed in \eqref{eq:TASEP_normalization}. Note that the current does not depend on the sites 
where it is measured, as expected from the particle conservation in the bulk.
\end{proposition}
\proof
The mean number of particles that cross the bond between sites $i$ and $i+1$ per unit of time is given by
the sum of probabilities for configurations with a particle on site $i$ and a hole on site $i+1$ ({\it i.e} configurations for 
which it is possible that a particle jumps from site $i$ to $i+1$ during the next transition). This has then to be multiplied by the 'rate of a jump'
which is $1$. This gives
\begin{eqnarray}
\langle j \rangle & = & \sum_{\tau_1,\dots,\tau_{i-1},\tau_{i+2},\dots,\tau_L=0,1} \cS(\tau_1,\dots,\tau_{i-1},1,0,\tau_{i+2},\dots,\tau_L) \\
& = & \frac{\llangle W|(E+D)\dots (E+D) DE (E+D) \dots (E+D)|V\rrangle}{Z_L} \\
& = & \frac{\llangle W| C^{i-1}DE C^{L-i-1}|V\rrangle}{Z_L} \\
& = & \frac{\llangle W| C^{i-1}(D+E) C^{L-i-1}|V\rrangle}{Z_L} \\
& = & \frac{Z_{L-1}}{Z_L}.
\end{eqnarray}
\finproof

\begin{proposition}
 The mean density of particle at site $i$ is expressed analytically as
 \begin{equation} \label{eq:TASEP_density}
  \langle \tau_i \rangle = \sum_{k=1}^{L-i} B_{k,1} \frac{Z_{L-k}}{Z_L} + \frac{Z_{i-1}}{Z_L} \sum_{k=1}^{L-i} B_{L-i,k}\frac{1}{\beta^{k+1}},
 \end{equation}
 where the combinatorial coefficient $B_{n,p}$ is given in \eqref{eq:TASEP_ballot_number}.
\end{proposition}
Note that here and in what follows, the notation $\langle \cdot \rangle$ stands for the expectation with respect to the stationary distribution.
\proof
\begin{eqnarray}
 \langle \tau_i \rangle & = & \sum_{\tau_1,\dots,\tau_L=0,1} \tau_i \cS(\tau_1,\dots,\tau_L) \\
 & = & \sum_{\tau_1,\dots,\tau_{i-1},\tau_{i+1},\dots,\tau_L=0,1} \cS(\tau_1,\dots,\tau_{i-1},1,\tau_{i+1},\dots,\tau_L) \\
 & = & \frac{\llangle W|(E+D)\dots (E+D) D (E+D)\dots (E+D)|V\rrangle}{Z_L} \\
 & = & \frac{\llangle W|C^{i-1} D C^{L-i}|V\rrangle}{Z_L}.
\end{eqnarray}
The proof relies essentially on the identity
\begin{equation}
 D C^n = \sum_{k=0}^{n-1} B_{k+1,1} C^{n-k} + \sum_{k=2}^{n+1} B_{n,k-1} D^k,
\end{equation}
which can be proven by induction on $n$. We thus have
\begin{equation}
 \llangle W|C^{i-1} D C^{L-i}|V\rrangle  =  \sum_{k=0}^{L-i-1}B_{k+1,1}\llangle W|C^{L-1-k}|V\rrangle
 + \sum_{k=2}^{L-i+1}\frac{B_{L-i,k-1}}{\beta^k} \llangle W|C^{i-1}|V\rrangle 
\end{equation}
from which it is easy to establish the desired result \eqref{eq:TASEP_density}.
\finproof

We stress that the results \eqref{eq:TASEP_normalization}, \eqref{eq:TASEP_current} and \eqref{eq:TASEP_density} are exact and valid for any size
$L$ of the lattice. These physical observables were computed efficiently using the matrix product formulation of the steady state. 
We will see in chapter \ref{chap:five} that the analytical expressions obtained can be studied in the thermodynamic limit ({\it i.e} in 
the large lattice size $L$ limit), to derive asymptotic expressions of the observables and compute rigorously the phase diagram of the system. In 
particular we will see that the model displays boundary induced phase transitions.

\subsubsection{Pushing procedure for the open TASEP}

We present here (without any proof) a combinatorial interpretation of the stationary weights of the open single species TASEP, called the pushing 
procedure. This procedure is well known for the multi-species TASEP on the ring \cite{EvansFM09} and has also been successfully applied
to understand the combinatorics of the stationary weights of the single species open TASEP in the particular case $\alpha=\beta=1$.
We start by presenting the method in this particular case and then we generalize to free parameters $\alpha$ and $\beta$.

To shorten the notations, a configuration on the lattice can be seen as a binary string of $0$ and $1$ of length $L$. For instance the binary string
$1010$ corresponds to the configuration $\cC=(1,0,1,0)$. The stationary weight of a given configuration is obtained by enumerating all the 
possible binary strings that we get by pushing the $1$'s to the right through the $0$'s or by doing nothing. For instance starting from $110$ we can obtain
$110$, $101$ and $011$. Hence we have the weight $3$ for the configuration $110$ which is consistent with the matrix product computation
\begin{eqnarray}
 \llangle W|DDE|V\rrangle & = & \llangle W|D(D+E)|V\rrangle = \llangle W|D^2|V\rrangle + \llangle W|DE|V\rrangle \\
 & = & \llangle W|D^2|V\rrangle + \llangle W|D|V\rrangle + \llangle W|E|V\rrangle = 3 \llangle W|V\rrangle,
\end{eqnarray}
using the fact that $\llangle W|E=\llangle W|$ and $D|V\rrangle=|V\rrangle$ because $\alpha=\beta=1$.

We give another example with the binary string $1010$. We can obtain $1010$, $1001$, $0110$, $0101$ and $0011$. This gives us a weight equal 
to $5$ which is consistent with
\begin{eqnarray}
 \llangle W|DEDE|V\rrangle & = & \llangle W|(D+E)(D+E)|V\rrangle = \llangle W|(D^2+DE+ED+E^2)|V\rrangle \\
 & = & \llangle W|D^2|V\rrangle + \llangle W|D|V\rrangle + \llangle W|E|V\rrangle + \llangle W|ED|V\rrangle + \llangle W|E^2|V\rrangle \nonumber \\
 & = & 5 \llangle W|V\rrangle.
\end{eqnarray}

\paragraph*{Pushing procedure for generic $\alpha$ and $\beta$}

We present a generalization of the pushing procedure for the cases where $\alpha$ and $\beta$ are generic. To the best of our knowledge this 
generalization was not known up to now. 

In addition to the rules given above, the $1$'s are now allowed to enter the binary string to the left with a weight
$a=(1-\alpha)/\alpha$ (note that $a=0$ when $\alpha=1$) and leave the binary string to the right with a weight $b=(1-\beta)/\beta$ (again $b=0$ if
$\beta=1$).
Note that now, starting from a given binary string, we can access all the other binary strings of same length (by pushing the $1$'s to the right and making them 
enter or leave the binary string). Moreover, there are several ways to get a binary string starting from the initial one: we choose the one 
with the minimal number of $1$'s leaving or entering the system. This point will be detailed at the end of the paragraph. 

Nothing is better than a concrete example to understand the procedure.
From the binary string $110$, we can obtain $110$, $101$ and $011$ with weight $1$. But we can also obtain $100$, $010$ and $001$ with weight $b$
(by making the right most $1$ leave the binary string). Making the two $1$'s leave the binary string we get $000$ with weight $b^2$.
We can also make a new $1$ enter the system to obtain $111$ with weight $a$. Summing all these contributions we end up with the stationary weight
$3+3b+b^2+a$ for the configuration $110$ which is consistent with the matrix product computation
\begin{eqnarray}
 \llangle W|DDE|V\rrangle & = & \llangle W|D^2|V\rrangle + \llangle W|D|V\rrangle + \llangle W|E|V\rrangle \\
 & = & \big( (1+b)^2+(1+b)+(1+a) \big)\llangle W|V\rrangle \\
 & = & \big( 3+3b+b^2+a \big)\llangle W|V\rrangle,
\end{eqnarray}
using the fact that $\llangle W|E=\frac{1}{\alpha}\llangle W| = (1+a)\llangle W|$ and $D|V\rrangle=\frac{1}{\beta}|V\rrangle = (1+b)|V\rrangle$.

We give another example with the configuration $1010$. From this binary string we can obtain the string $1010$, $1001$, $0110$, $0101$ and
$0011$ with weight $1$. We can also make the right most $1$ leave the binary string to get $1000$, $0100$, $0010$ and $0001$ with weight $b$.
Pushing the two $1$'s out of the string gives $0000$ with weight $b^2$. We can also make a new $1$ appear on the left on the string to obtain
$1110$, $1101$, $1011$ and $0111$ with weight $a$. If two $1$'s are injected on the left we get $1111$ with weight $a^2$. The last binary 
string $1100$ is obtained with weight $ab$ by making a $1$ appear on the left and the right most $1$ leave on the right. Summing all these 
contributions gives the total stationary weight $5+4b+b^2+4a+a^2+ab$ in agreement with
\begin{eqnarray}
 \llangle W|DEDE|V\rrangle & = &
 \llangle W|D^2|V\rrangle + \llangle W|D|V\rrangle + \llangle W|E|V\rrangle + \llangle W|ED|V\rrangle + \llangle W|E^2|V\rrangle \nonumber \\
 & = & \big( (1+b)^2+(1+b)+(1+a)+(1+a)(1+b)+(1+a)^2 \big) \llangle W|V\rrangle \nonumber \\
 & = & \big( 5+4b+b^2+4a+a^2+ab \big) \llangle W|V\rrangle.
\end{eqnarray}
It is important to recall that in this procedure we always select the simplest path to go from the starting binary string to any other 
binary string. Simplest path should be understood here as the path whose weight has the minimal powers of $a$ and $b$ (the minimal sum 
of the powers of $a$ and $b$ to be fully rigorous: a path of weight $a^nb^m$ is preferred against a path of weight $a^{n'}b^{m'}$ if $n+m<n'+m'$). 
For instance, it is indeed easy to see that there are several possibilities to go from 
the binary string $1010$ to the binary string $0110$. We could first choose to make both $1$'s leave the string on the right and then make 
two new $1$'s appear on the left and push them at the correct position, giving a weight $a^2b^2$. But we saw that it is also possible to 
just push one step the left most $1$ to the right, with weight $1=a^0b^0$. Since $0+0<2+2$ we keep the second path 
(which is manifestly the simplest path). 

\subsection{Telescopic relations} \label{subsec:teslescopic_relations}

In this subsection we will discuss the algebraic constraints that should be satisfied by the matrices $X_0,X_1,\dots,X_N$ and the boundary vectors
$\llangle W|$ and $|V\rrangle$ in order that the matrix product state \eqref{eq:Matrix_ansatz} computes correctly the stationary distribution.
An efficient formalism to present these algebraic relations is the one of the tensor product. We already encountered this framework when 
defining the vector basis associated to the physical configurations on the lattice, see chapter \ref{chap:two}. 
We will use it here to write the stationary
distribution in a concise way.
\begin{proposition}
 The matrix product state defined in \eqref{eq:Matrix_ansatz} can be recast in the form
 \begin{equation}
  \steady =\frac{1}{Z_L}\llangle W| \begin{pmatrix}
                                     X_0 \\ X_1 \\ \vdots \\ X_N
                                    \end{pmatrix} \otimes
\begin{pmatrix}
                                     X_0 \\ X_1 \\ \vdots \\ X_N
                                    \end{pmatrix} \otimes \dots \otimes  
\begin{pmatrix}
                                     X_0 \\ X_1 \\ \vdots \\ X_N
                                    \end{pmatrix} |V\rrangle.                                    
 \end{equation}
\end{proposition}
\proof
This is a direct consequence of the definition of the tensor product, see for instance example \ref{ex:tensor_product}. We thus have
\begin{equation}
 \llangle W| \begin{pmatrix}
                                     X_0 \\ X_1 \\ \vdots \\ X_N
                                    \end{pmatrix} \otimes
\begin{pmatrix}
                                     X_0 \\ X_1 \\ \vdots \\ X_N
                                    \end{pmatrix} \otimes \dots \otimes  
\begin{pmatrix}
                                     X_0 \\ X_1 \\ \vdots \\ X_N
                                    \end{pmatrix} |V\rrangle =
\begin{pmatrix}
 \llangle W| X_0 \dots X_0 X_0 |V\rrangle \\
  \llangle W| X_0 \dots X_0 X_1 |V\rrangle \\
  \vdots \\
   \llangle W| X_N \dots X_N X_N |V\rrangle
\end{pmatrix},
\end{equation}
in agreement with the matrix product expression \ref{eq:Matrix_ansatz}.
\finproof

We are now equipped to present the root of the algebraic relations involved in the matrix product formulation.

\subsubsection{Particular example of the TASEP}

We begin with the case of the single species open TASEP

\begin{proposition}
 The algebraic constraints stated in \eqref{eq:TASEP_algebraic_relations} are equivalent to the telescopic relations:
 \begin{itemize}
  \item in the bulk
 \begin{equation} \label{eq:TASEP_telescopic_relation_bulk}
 DE=D+E \quad \Leftrightarrow \quad 
  m\begin{pmatrix}
            E \\ D
           \end{pmatrix} \otimes
\begin{pmatrix}
            E \\ D
           \end{pmatrix} = 
           \begin{pmatrix}
            E \\ D
           \end{pmatrix} \otimes
           \begin{pmatrix}
            -1 \\ 1
           \end{pmatrix}-
           \begin{pmatrix}
            -1 \\ 1
           \end{pmatrix} \otimes
           \begin{pmatrix}
            E \\ D
           \end{pmatrix}
 \end{equation}
 \item at the left boundary
 \begin{equation} \label{eq:TASEP_telescopic_relation_left}
 \llangle W|E=\frac{1}{\alpha}\llangle W| \quad \Leftrightarrow \quad \llangle W| B \begin{pmatrix}
            E \\ D
           \end{pmatrix} = 
           \llangle W| \begin{pmatrix}
            -1 \\ 1
           \end{pmatrix}
 \end{equation}
 \item at the right boundary
  \begin{equation} \label{eq:TASEP_telescopic_relation_right}
 D|V\rrangle =\frac{1}{\beta}|V\rrangle \quad \Leftrightarrow \quad  \overline{B} \begin{pmatrix}
            E \\ D
           \end{pmatrix}|V\rrangle = 
           -\begin{pmatrix}
            -1 \\ 1
           \end{pmatrix}|V\rrangle
 \end{equation}
 \end{itemize}
 where $m$, $B$ and $\overline{B}$ are defined in \eqref{eq:TASEP_m} and \eqref{eq:TASEP_B_Bb}.
\end{proposition}
\proof
We compute explicitly
\begin{equation}
 m \begin{pmatrix}
            E \\ D
           \end{pmatrix} \otimes
\begin{pmatrix}
            E \\ D
           \end{pmatrix}  =  
           \begin{pmatrix}
            0 & 0 & 0 & 0 \\
            0 & 0 & 1 & 0 \\
            0 & 0 & -1 & 0 \\
            0 & 0 & 0 & 0
           \end{pmatrix}
           \begin{pmatrix}
            E^2 \\ ED \\ DE \\ D^2
           \end{pmatrix} = 
           \begin{pmatrix}
            0 \\ DE \\ -DE \\ 0
           \end{pmatrix}
\end{equation}
and
\begin{equation}
           \begin{pmatrix}
            E \\ D
           \end{pmatrix} \otimes
           \begin{pmatrix}
            -1 \\ 1
           \end{pmatrix}-
           \begin{pmatrix}
            -1 \\ 1
           \end{pmatrix} \otimes
           \begin{pmatrix}
            E \\ D
           \end{pmatrix}=
           \begin{pmatrix}
            -E \\ E \\ -D \\ D
           \end{pmatrix}-
           \begin{pmatrix}
            -E \\ -D \\ E \\ D
           \end{pmatrix} =
           \begin{pmatrix}
            0 \\ E+D \\ -E-D \\ 0
           \end{pmatrix}.
\end{equation}
We thus obtain the desired result \eqref{eq:TASEP_telescopic_relation_bulk}. In the same way for the relation on the left boundary, we compute
\begin{equation}
 \llangle W| B \begin{pmatrix}
            E \\ D
           \end{pmatrix} =
 \llangle W| \begin{pmatrix}
              -\alpha & 0 \\ \alpha & 0
             \end{pmatrix}
             \begin{pmatrix}
              E \\ D
             \end{pmatrix}=
 \llangle W| \begin{pmatrix}
              -\alpha E \\ \alpha E
             \end{pmatrix} =
             \begin{pmatrix}
              -\alpha \llangle W|E \\ \alpha \llangle W|E
             \end{pmatrix}
\end{equation}
and 
\begin{equation}
 \llangle W| \begin{pmatrix}
              -1 \\ 1
             \end{pmatrix} =
             \begin{pmatrix}
              -\llangle W| \\ \llangle W|
             \end{pmatrix},
\end{equation}
which yields the property \eqref{eq:TASEP_telescopic_relation_left}. The equation \eqref{eq:TASEP_telescopic_relation_right} is derived in a very 
similar way.
\finproof

Note that these relations are called telescopic because of their ``divergence like'' form. This appellation will also make sense with the proof of the 
following proposition. 
At first sight, it may appear a bit complicated to encode single relations in complicated tensor forms, like \eqref{eq:TASEP_telescopic_relation_bulk}
instead of $DE=D+E$. However, it allows simple generalization to other models, as well as a nice and simple proof of the stationary measure property
(see below).
\begin{proposition} \label{prop:TASEP_steady_state}
 The matrix product state 
 \begin{equation}
  \steady = \frac{1}{Z_L} \llangle W| \begin{pmatrix}
                                       E \\ D 
                                      \end{pmatrix} \otimes
                                      \begin{pmatrix}
                                       E \\ D 
                                      \end{pmatrix} \otimes \dots \otimes
                                      \begin{pmatrix}
                                       E \\ D 
                                      \end{pmatrix} |V\rrangle,
 \end{equation}
 where $E$, $D$, $\llangle W|$ and $|V\rrangle$ satisfy the telescopic relations 
 \eqref{eq:TASEP_telescopic_relation_bulk}, \eqref{eq:TASEP_telescopic_relation_left} and \eqref{eq:TASEP_telescopic_relation_right}, 
 is the stationary state of the model, i.e $M\steady=0$.
\end{proposition}
Note that this proposition is strictly equivalent to the proposition \ref{prop:TASEP_matrix_ansatz}. It has just been reformulated using
the tensor product formalism.
\proof 
\begin{eqnarray*}
 & & \left( B_1+\sum_{k=1}^{L-1} m_{k,k+1} + \overline{B}_L \right) \llangle W| \begin{pmatrix}
                                       E \\ D 
                                      \end{pmatrix} \otimes
                                      \begin{pmatrix}
                                       E \\ D 
                                      \end{pmatrix} \otimes \dots \otimes
                                      \begin{pmatrix}
                                       E \\ D 
                                      \end{pmatrix} |V\rrangle \\ 
 & = & \llangle W|\left[B\begin{pmatrix}
                    E \\ D
                   \end{pmatrix}\right] \otimes
                   \begin{pmatrix}
                    E \\ D
                   \end{pmatrix} \otimes \dots \otimes
                   \begin{pmatrix}
                    E \\ D 
                   \end{pmatrix} |V\rrangle
       +\llangle W|\begin{pmatrix}
                    E \\ D
                   \end{pmatrix} \otimes \dots \otimes
                   \begin{pmatrix}
                    E \\ D
                   \end{pmatrix} \otimes 
                   \left[\overline{B}\begin{pmatrix}
                    E \\ D 
                   \end{pmatrix}\right] |V\rrangle \\
 &+& \hspace{-1mm} \sum_{k=1}^{L-1} \llangle W|\underbrace{\begin{pmatrix}
                    E \\ D
                   \end{pmatrix} \otimes \dots \otimes
                   \begin{pmatrix}
                    E \\ D
                   \end{pmatrix}}_{k-1} \otimes 
                   \left[m\begin{pmatrix}
                    E \\ D 
                   \end{pmatrix} \otimes
                   \begin{pmatrix}
                    E \\ D
                   \end{pmatrix}\right] \otimes
                   \underbrace{\begin{pmatrix}
                    E \\ D 
                   \end{pmatrix} \otimes \dots \otimes
                   \begin{pmatrix}
                    E \\ D 
                   \end{pmatrix}}_{L-k-1}
                   |V\rrangle \\
 & = & \llangle W|\begin{pmatrix}
                    -1 \\ 1
                   \end{pmatrix} \otimes
                   \begin{pmatrix}
                    E \\ D
                   \end{pmatrix} \otimes \dots \otimes
                   \begin{pmatrix}
                    E \\ D 
                   \end{pmatrix} |V\rrangle   
       -\llangle W|\begin{pmatrix}
                    E \\ D
                   \end{pmatrix} \otimes \dots \otimes
                   \begin{pmatrix}
                    E \\ D
                   \end{pmatrix} \otimes 
                   \begin{pmatrix}
                    -1 \\ 1 
                   \end{pmatrix} |V\rrangle \\
 &+& \hspace{-1mm} \sum_{k=1}^{L-1} \llangle W|\underbrace{\begin{pmatrix}
                    E \\ D
                   \end{pmatrix} \otimes \dots \otimes
                   \begin{pmatrix}
                    E \\ D
                   \end{pmatrix}}_{k-1} \otimes 
                   \left[\begin{pmatrix}
                    E \\ D 
                   \end{pmatrix} \otimes
                   \begin{pmatrix}
                    -1 \\ 1 
                   \end{pmatrix}-
                   \begin{pmatrix}
                    -1 \\ 1 
                   \end{pmatrix} \otimes 
                   \begin{pmatrix}
                    E \\ D 
                   \end{pmatrix} \right] \otimes
                   \underbrace{\begin{pmatrix}
                    E \\ D 
                   \end{pmatrix} \otimes \dots \otimes
                   \begin{pmatrix}
                    E \\ D 
                   \end{pmatrix}}_{L-k-1}
                   |V\rrangle \\
 & = & \sum_{k=1}^L \llangle W| \underbrace{\begin{pmatrix}
                                 E \\ D
                                \end{pmatrix} \otimes \dots \otimes 
                                \begin{pmatrix}
                                 E \\ D
                                \end{pmatrix}}_{k-1} \otimes 
                                \begin{pmatrix}
                                 -1 \\ 1
                                \end{pmatrix} \otimes
                                \underbrace{\begin{pmatrix}
                                 E \\ D
                                \end{pmatrix} \otimes \dots \otimes 
                                \begin{pmatrix}
                                 E \\ D
                                \end{pmatrix}}_{L-k} |V\rrangle \\
 & & - \sum_{k=1}^L \llangle W| \underbrace{\begin{pmatrix}
                                 E \\ D
                                \end{pmatrix} \otimes \dots \otimes 
                                \begin{pmatrix}
                                 E \\ D
                                \end{pmatrix}}_{k-1} \otimes 
                                \begin{pmatrix}
                                 -1 \\ 1
                                \end{pmatrix} \otimes
                                \underbrace{\begin{pmatrix}
                                 E \\ D
                                \end{pmatrix} \otimes \dots \otimes 
                                \begin{pmatrix}
                                 E \\ D
                                \end{pmatrix}}_{L-k} |V\rrangle \\
 & = & 0.
\end{eqnarray*}
The action of the Markov matrix on the matrix product state leads to a telescopic sum.
\finproof

\subsubsection{General case}

The previous cancellation scheme is quite general and goes far beyond the single species TASEP case. In fact the algebraic structure 
defined by the telescopic relations (in the bulk and on the boundaries) appears in all known examples of matrix product expression 
of the stationary state (of continuous time Markov processes). To be more precise, we consider a stochastic process described by the Markov matrix
\eqref{eq:open_Markov_matrix_sum_decomposition} where $m$ is a $(N+1)^2 \times (N+1)^2$ matrix, while $B$ and $\overline{B}$ 
are $(N+1) \times (N+1)$ matrices. We first introduce two key vectors.
\begin{definition}
 We introduce two vectors $\mathbf{X}$ and $\mathbf{\overline{X}}$ 
 \begin{equation}
  \mathbf{X}=\begin{pmatrix}
              X_0 \\ X_1 \\ \vdots \\ X_N
             \end{pmatrix}, \qquad 
  \mathbf{\overline{X}}= \begin{pmatrix}
                          \overline{X}_0 \\ \overline{X}_1 \\ \vdots \\ \overline{X}_N
                         \end{pmatrix}.
 \end{equation}
For a given bulk local jump operator $m$, we say that $\mathbf{X}$ and $\mathbf{\overline{X}}$ satisfy the bulk telescopic relation if we have
\begin{equation} \label{eq:telescopic_relation_bulk}
 m \mathbf{X} \otimes \mathbf{X} = \mathbf{X} \otimes \mathbf{\overline{X}} - \mathbf{\overline{X}} \otimes \mathbf{X}.
\end{equation}
For given boundary local jump operators $B$ and $\overline{B}$, we say that 
$\mathbf{X}$ and $\mathbf{\overline{X}}$ satisfy the boundary telescopic relations if there exist two boundary vectors 
$\llangle W|$ and $|V\rrangle$ such that
\begin{equation} \label{eq:telescopic_relation_boundaries}
 \llangle W| B \mathbf{X} = \llangle W| \mathbf{\overline{X}}, \qquad \overline{B}\mathbf{X}|V\rrangle = -\mathbf{\overline{X}}|V\rrangle.
\end{equation}
\end{definition}
We are now equipped to state the generalization of proposition \ref{prop:TASEP_steady_state}.
\begin{proposition} \label{prop:steady_state}
 If the vectors $\mathbf{X}$ and $\mathbf{\overline{X}}$ satisfy 
the telescopic relations \eqref{eq:telescopic_relation_bulk} and \eqref{eq:telescopic_relation_boundaries},
then the matrix product state 
\begin{equation} \label{eq:steady_state}
 \steady = \frac{1}{Z_L}\llangle W| \mathbf{X} \otimes \mathbf{X} \otimes \dots \otimes \mathbf{X} |V\rrangle
\end{equation}
satisfies
\begin{equation}
 M\steady=0,
\end{equation}
where the Markov matrix is given by
 \begin{equation}
  M=B_1+\sum_{k=1}^{L-1}m_{k,k+1} +\overline{B}_L.
 \end{equation}
If $\steady$ is not vanishing, it thus provides the stationary state\footnote{If the Markov matrix is irreducible, we know that the stationary state
is unique. If it is not irreducible, it provides one or several of the stationary states.} associated to the Markov matrix.
\end{proposition}
\proof
 The property is proven using exactly the same cancellation scheme as for \ref{prop:TASEP_steady_state}, replacing the vector $(E,D)^t$
by $\mathbf{X}$ and the vector $(-1,1)^t$ by $\mathbf{\overline{X}}$.
\finproof

\begin{remark}
 In \cite{KrebsS97} it was proven that for any Markov matrix $M$ that can be decomposed into the sum of operators acting locally on the lattice 
 \begin{equation}
  M=B_1+\sum_{k=1}^{L-1}m_{k,k+1} +\overline{B}_L,
 \end{equation}
it is possible to construct explicitly matrices $X_0,\dots,X_N$ and $\overline{X}_0,\dots,\overline{X}_N$ 
and vectors $\llangle W|$ and $|V\rrangle$ such that 
the telescopic relations \eqref{eq:telescopic_relation_bulk} and \eqref{eq:telescopic_relation_boundaries} are fulfilled.
Thanks to the previous property, we know that the stationary state is given in a matrix product form by \eqref{eq:steady_state} (if it is not vanishing). 
In practice, the construction of the vector $\llangle W|$ in \cite{KrebsS97} requires the prior knowledge of the stationary distribution of the model...
Hence the result of \cite{KrebsS97} appears more as an existence theorem (the existence of a matrix product expression of the steady state) than as 
a useful tool for explicit computations of physical quantities. Indeed the algebraic relations given by 
\eqref{eq:telescopic_relation_bulk} and \eqref{eq:telescopic_relation_boundaries} satisfied by the matrices
$X_0,\dots,X_N$ and $\overline{X}_0,\dots,\overline{X}_N$ and the vectors $\llangle W|$ and $|V\rrangle$ are not 
sufficient to fix the value of the stationary weights (we cannot perform in general algebraic computations as we showed in 
\eqref{eq:TASEP_particular_weight} for the TASEP),
the explicit representation is needed (or further algebraic relations, as we will see below).
\end{remark}

\subsubsection{Other examples}

We have already seen in the above subsection the matrix product expression of the steady state of the open single species TASEP. We now
present the case of the ASEP and of the SSEP. We give the matrix product solution within the developed framework of telescopic relations.

\begin{example} \label{ex:ASEP_MA}
 We recall that the stochastic dynamics of the ASEP is encoded in the local jump operators $m$ defined in \eqref{eq:ASEP_m}
 and $B$, $\overline{B}$ defined in \eqref{eq:ASEP_B_Bb}.
 We define the vector
 \begin{equation}
  \mathbf{X}= \begin{pmatrix}
               E \\ D
              \end{pmatrix}
 \end{equation}
 where the algebraic elements $E$ and $D$ satisfy the relation $pDE-qED=(p-q)(E+D)$. Similarly to the TASEP case, representation of such
 operators as infinite matrices exists and will be given in chapter \ref{chap:four}.
 We also introduce the vector
  \begin{equation}
  \overline{\mathbf{X}}= \begin{pmatrix}
               q-p \\ p-q
              \end{pmatrix},
 \end{equation}
 and the boundary vectors $\llangle W|$ and $|V\rrangle$ which satisfy 
 \begin{equation} \label{eq:ASEP_relation_boundaries}
  \llangle W|\left(\alpha E-\gamma D\right)=(p-q)\llangle W|, \quad \mbox{and} \quad \left(\beta D-\delta E\right)|V\rrangle=(p-q)|V\rrangle.
 \end{equation}
 Representation of $\llangle W|$ and $|V\rrangle$ as infinite row and column vectors will also be given in chapter \ref{chap:four}.
 
 Then it is easy to check that the telescopic relations \eqref{eq:telescopic_relation_bulk} and \eqref{eq:telescopic_relation_boundaries} hold
 and hence the steady state is given by \eqref{eq:steady_state}.
\end{example}

\begin{example} \label{ex:SSEP_MA}
 Similar results hold for the SSEP with matrices $E$, $D$ and boundary vectors $\llangle W|$ and $|V\rrangle$ that fulfill
 \begin{equation} \label{eq:SSEP_algebraic_relations}
  DE-ED=E+D, \quad \llangle W|(\alpha E-\gamma D)=\llangle W|, \quad \mbox{and} \quad (\beta D-\delta E)|V\rrangle=|V\rrangle.
 \end{equation}
 We have in this case
 \begin{equation}
  \mathbf{X} = \begin{pmatrix}
                E \\ D
               \end{pmatrix}, \quad \mbox{and} \quad 
  \overline{\mathbf{X}} = \begin{pmatrix}
                -1 \\ 1
               \end{pmatrix}             
 \end{equation}
\end{example}

We will encounter in section \ref{sec:examples_MA} more complex examples where the auxiliary vector $\overline{\mathbf{X}}$ 
(sometimes called hat vector) contains non scalar operators ({\it i.e} which do not commute with the matrices $X_i$), 
in contrast to what we saw up to now in the examples.

\subsection{Thermodynamic equilibrium case}

We now make some remarks about the matrix ansatz for a system that reaches a thermodynamic equilibrium in the long time limit. To fix the 
ideas we present as usual the example of the SSEP and ASEP.

\subsubsection{Some examples}

\begin{example}
 The SSEP reaches in the long time limit a thermodynamic equilibrium if and only if the particle densities of the two reservoirs are equal, i.e
 \begin{equation}
  \frac{\alpha}{\alpha+\gamma} = \frac{\delta}{\beta+\delta}.
 \end{equation}
 This condition can be rewritten more concisely $\alpha \beta = \delta \gamma$. This constraint can be derived using the 
 detailed balance condition (that defines rigorously a thermodynamic equilibrium in the Markov chain context, see chapter \ref{chap:one}).
 We have indeed 
 \begin{eqnarray}
  \cS(0,\dots,0) & = & \frac{\gamma}{\alpha}\cS(1,0,\dots,0) \\
                 & = & \frac{\gamma}{\alpha}\cS(0,1,0,\dots,0) \\
                 & \dots & \\
                 & = & \frac{\gamma}{\alpha}\cS(0,\dots,0,1) \\
                 & = & \frac{\gamma \delta}{\alpha \beta}\cS(0,\dots,0),
 \end{eqnarray}
 which proves that $\frac{\gamma \delta}{\alpha \beta}=1$ is necessary. In this particular case the matrix product formulation of 
 the steady state simplifies drastically:
 \begin{equation}
  \steady = \begin{pmatrix}
             1-r \\r
            \end{pmatrix} \otimes \dots \otimes
            \begin{pmatrix}
             1-r \\r
            \end{pmatrix},
 \end{equation}
where $r=\frac{\alpha}{\alpha+\gamma}=\frac{\delta}{\beta+\delta}$. We observe that the matrices $E$ and $D$ can thus be chosen as scalars in this 
case ($E=1-r$, $D=r$) and the auxiliary vector $\overline{\mathbf{X}}=0$. The telescopic relations read indeed
\begin{equation}
 B\begin{pmatrix}
   1-r \\ r
  \end{pmatrix} = 0, \qquad
 m\begin{pmatrix}
   1-r \\ r
  \end{pmatrix} \otimes
  \begin{pmatrix}
   1-r \\ r
  \end{pmatrix} = 0, \qquad 
 \overline B\begin{pmatrix}
   1-r \\ r
  \end{pmatrix} = 0. 
\end{equation}
\end{example}

\begin{example}
 The ASEP reaches a thermodynamic equilibrium if and only if the parameters $\alpha,\beta,\gamma,\delta$ and $q$ fulfill the constraint
 \begin{equation} \label{eq:ASEP_eq_thermo_constraint}
  \frac{\gamma \delta}{\alpha \beta}\left(\frac{q}{p}\right)^{L-1}=1.
 \end{equation}
 This can be intuitively understood as the reservoir densities compensating exactly the driving force in the bulk. 
Once again this constraint can be determined using the detailed balance condition (that must hold in a thermodynamic equilibrium).
 We have indeed 
 \begin{eqnarray}
  \cS(0,\dots,0) & = & \frac{\gamma}{\alpha}\cS(1,0,\dots,0) \\
                 & = & \frac{\gamma q}{\alpha p}\cS(0,1,0,\dots,0) \\
                 & \dots & \\
                 & = & \frac{\gamma q^{L-1}}{\alpha p^{L-1}}\cS(0,\dots,0,1) \\
                 & = & \frac{\gamma \delta}{\alpha \beta}\left(\frac{q}{p}\right)^{L-1}\cS(0,\dots,0),
 \end{eqnarray}
 which proves that $\frac{\gamma \delta}{\alpha \beta}\left(\frac{q}{p}\right)^{L-1}=1$. In this particular case the stationary state can still be given
 in matrix product form
 \begin{equation} \label{eq:ASEP_steady_thermo}
  \steady = \frac{1}{Z_L}\llangle W| \mathbf{X} \otimes \dots \otimes \mathbf{X} |V\rrangle,
 \end{equation}
with
\begin{equation}
 \mathbf{X} = \begin{pmatrix}
                E \\ D
              \end{pmatrix}
\end{equation}
where the operators $E$, $D$ and the boundary vectors $\llangle W|$ and $|V\rrangle$ satisfy
\begin{equation}
 pDE=qED, \qquad \llangle W|(\alpha E-\gamma D)=0, \qquad (\delta E-\beta D)|V\rrangle = 0.
\end{equation}
The vector $\mathbf{X}$ above satisfies the relations $m \mathbf{X}\otimes\mathbf{X}=0$, $\llangle W|B\mathbf{X}=0$ and
$\overline{B}\mathbf{X}|V\rrangle=0$ which ensure that \eqref{eq:ASEP_steady_thermo} is the stationary state of the model. An explicit representation
of such matrices and vector can be found on the $L+1$ dimensional vector space spanned by the basis $\{\kket{k}\}_{0\leq k\leq L}$
\begin{equation}
 E=\sum_{k=0}^{L-1} \kket{k+1}\bbra{k}, \qquad D=\sum_{k=0}^{L-1}\frac{\delta}{\beta}\left(\frac{q}{p}\right)^k \kket{k+1}\bbra{k},
\end{equation}
and 
\begin{equation}
 \bbra{W}=\bbra{L}, \qquad \kket{V}=\kket{0}.
\end{equation}
In matrix notation it gives explicitly
\begin{equation}
 E= \begin{pmatrix}
     0 & 0 & 0 & \hdots & 0 \\
     1 & 0 & 0 &  & \vdots \\
     0 & 1 & 0 & \ddots & \vdots \\
     \vdots &  & \ddots & \ddots & 0 \\
     0 & \hdots & 0 & 1 & 0
    \end{pmatrix}, \qquad 
 D= \begin{pmatrix}
     0 & 0 & 0 & \hdots & 0 \\
     \frac{\delta}{\beta} & 0 & 0 &  & \vdots \\
     0 & \frac{\delta}{\beta}\frac{q}{p} & 0 & \ddots & \vdots \\
     \vdots &  & \ddots & \ddots & 0 \\
     0 & \hdots & 0 & \frac{\delta}{\beta}\left(\frac{q}{p}\right)^{L-1} & 0
    \end{pmatrix},   
\end{equation}
and
\begin{equation}
 \bbra{W}=\begin{pmatrix}
           0 & \hdots & 0 & 1
          \end{pmatrix}, \qquad
 \kket{V}=\begin{pmatrix}
           1 \\ 0 \\ \vdots \\ 0
          \end{pmatrix}.
\end{equation}
It is easy to see that this representation works only when the constraint \eqref{eq:ASEP_eq_thermo_constraint} is satisfied. Indeed the 
relation $\llangle W|(\alpha E-\gamma D)=0$ is only fulfilled in this case.

Note that the matrix product state \eqref{eq:ASEP_steady_thermo} can be rewritten as an inhomogeneous factorized state
\begin{equation}
 \steady = \begin{pmatrix}
            1 \\ \frac{\delta}{\beta}\left(\frac{q}{p}\right)^{L-1}
           \end{pmatrix} \otimes \dots \otimes
           \begin{pmatrix}
            1 \\ \frac{\delta}{\beta}\frac{q}{p}
           \end{pmatrix} \otimes
           \begin{pmatrix}
            1 \\ \frac{\delta}{\beta}
           \end{pmatrix}
\end{equation}
or equivalently using the constraint \eqref{eq:ASEP_eq_thermo_constraint}
\begin{equation}
 \steady = \begin{pmatrix}
            1 \\ \frac{\alpha}{\gamma}
           \end{pmatrix} \otimes
           \begin{pmatrix}
            1 \\ \frac{\alpha}{\gamma}\frac{p}{q}
           \end{pmatrix} \otimes \dots \otimes
           \begin{pmatrix}
            1 \\ \frac{\alpha}{\gamma}\left(\frac{p}{q}\right)^{L-1}
           \end{pmatrix}.
\end{equation}
\end{example}

The two previous examples both share the feature of having a vanishing auxiliary vector $\overline{\mathbf{X}}$ in the telescopic relations associated
to the matrix product formulation of the steady state. This seems to be always the case in systems that reach a thermodynamic equilibrium and 
whose steady state can be written in matrix product form (although we do not have any general proof of this fact). 
The relation $m \mathbf{X} \otimes \mathbf{X}=0$ (telescopic relation with vanishing auxiliary vector) gave in the previous examples commutation
relations of the type $X_i X_j=\exp(-\Delta E/kT)X_jX_i$ where $\exp(-\Delta E/kT)=\frac{\cS(\dots,i,j,\dots)}{\cS(\dots,j,i,\dots)}$ 
is the Boltzmann factor obtained when exchanging particles of
species $i$ and $j$ lying on adjacent sites on the lattice (see chapter \ref{chap:one} for details). Similar observations hold also for 
the relations on the boundary vectors $\llangle W|$ and $|V\rrangle$.
We will encounter this fact again in the remark below.

\subsubsection{A new look on the open TASEP stationary distribution}

We end up this subsection on matrix product steady states with an intriguing observation. 
The stationary distribution of the single species open TASEP can be obtained as the marginal of the Boltzmann distribution of a more complex 
process at thermodynamic equilibrium.
To be more precise this process is also defined on a one dimensional lattice with $L$ sites. Each site $i$ of the lattice can be in three different 
states: $\tau_i=-,0,+$. The stochastic dynamics of the process is defined as usual locally (on two adjacent sites in the bulk and on the 
single extremal sites at the boundaries) by the following rules
\begin{equation}
 \begin{array}{|c |c| c| }
 \hline \text{Left} & \text{Bulk} & \text{Right} \\
 \hline
 -\, \xrightarrow{\ 1 \ }\, 0&  0+\, \overset{\ 1 \ }{\longleftrightarrow} \,+0&+\, \xrightarrow{\ 1 \ } \,0\\
 0\, \xrightarrow{\ a \ }\, -&0-\, \overset{\ 1 \ }{\longleftrightarrow}\, -0&0\, \xrightarrow{\ b \ }\, +\\ 
  & +-\, \overset{\ 1 \ }{\longleftrightarrow}\, 00 &  \\ \hline
 \end{array}
 \end{equation}
where the parameters $a$ and $b$, that appeared previously in the pushing procedure of the TASEP, are given by
\begin{equation}
 a=\frac{1-\alpha}{\alpha}, \quad \mbox{and} \quad b=\frac{1-\beta}{\beta}.
\end{equation}
In the basis $\ket{--}$, $\ket{-0}$, $\ket{-+}$, $\ket{0-}$, $\ket{00}$, $\ket{0+}$, $\ket{+-}$, $\ket{+0}$, $\ket{++}$ ordered this way,
the bulk local jump operator $m$ associated to the above dynamical rules reads
\begin{equation}
 m= \begin{pmatrix}
     \cdot & \cdot & \cdot & \cdot & \cdot & \cdot & \cdot & \cdot & \cdot  \\
     \cdot & -1 & \cdot & 1 & \cdot & \cdot & \cdot & \cdot & \cdot  \\
     \cdot & \cdot & \cdot & \cdot & \cdot & \cdot & \cdot & \cdot & \cdot  \\
     \cdot & 1 & \cdot & -1 & \cdot & \cdot & \cdot & \cdot & \cdot  \\
     \cdot & \cdot & \cdot & \cdot & -1 & \cdot & 1 & \cdot & \cdot  \\
     \cdot & \cdot & \cdot & \cdot & \cdot & -1 & \cdot & 1 & \cdot  \\
     \cdot & \cdot & \cdot & \cdot & 1 & \cdot & -1 & \cdot & \cdot  \\
     \cdot & \cdot & \cdot & \cdot & \cdot & 1 & \cdot & -1 & \cdot  \\
     \cdot & \cdot & \cdot & \cdot & \cdot & \cdot & \cdot & \cdot & \cdot  
    \end{pmatrix}.
\end{equation}
In the basis $\ket{-}$, $\ket{0}$, $\ket{+}$ ordered like this, the boundary local jump operators read
\begin{equation}
 B = \begin{pmatrix}
      -1 & a & \cdot \\
      1 & -a & \cdot \\
      \cdot & \cdot & \cdot 
     \end{pmatrix}, \quad \mbox{and} \quad 
 \overline{B} = \begin{pmatrix}
                 \cdot & \cdot & \cdot \\
                 \cdot & -b & 1 \\
                 \cdot & b & -1 
                \end{pmatrix}.
\end{equation}
The stationary state of this model can be expressed in a matrix product form. We define $e$ and $d$ two algebraic elements such that
$d e=1$ and two boundary vectors $\llangle W|$ and $|V\rrangle$ such that $\llangle W|e = a \llangle W|$ and 
$d|V\rrangle=b|V\rrangle$. Representation of such algebraic objects are $e=E-1$ and $d=D-1$ with $E$, $D$ given in 
\eqref{eq:TASEP_representation_DE} and $\llangle W|$ and $|V\rrangle$ are the same as the ones given in \eqref{eq:TASEP_representation_vectors}.
We define
\begin{equation}
 \mathbf{X} = \begin{pmatrix}
               X_{-} \\ X_{0} \\ X_{+}
              \end{pmatrix}
            = \begin{pmatrix}
               e \\ 1 \\ d
              \end{pmatrix}.
\end{equation}
We then have
\begin{equation}
 m \mathbf{X} \otimes \mathbf{X} = m \begin{pmatrix}
                                      e^2 \\ e \\ ed \\ e \\ 1 \\ d \\ d e \\ d \\ d^2
                                     \end{pmatrix} = 
 \begin{pmatrix}
  0 \\ -e +e \\ 0 \\ e-e \\ -1+d e \\ -d+d \\ 1-d e \\ d-d \\ 0
 \end{pmatrix} = 0.
\end{equation}
We have also
\begin{equation}
 \llangle W|B \mathbf{X} = \llangle W| \begin{pmatrix}
                                        -e +a \\ e-a \\ 0
                                       \end{pmatrix} = 0, \quad \mbox{and} \quad 
 \overline{B}\mathbf{X}|V\rrangle = \begin{pmatrix}
                                     0 \\ -b+d \\ b-d
                                    \end{pmatrix} |V\rrangle = 0.
\end{equation}
Hence the steady state is given by
\begin{equation}
 \steady = \frac{1}{Z_L}\llangle W| \mathbf{X} \otimes \dots \otimes \mathbf{X} |V\rrangle. 
\end{equation}
This is a thermodynamic equilibrium, the detailed balance is indeed satisfied. To prove this statement we have to check
that the identity $m(\cC \rightarrow \cC') \cS(\cC)=m(\cC' \rightarrow \cC)\cS(\cC')$ holds for every allowed transition 
$\cC \rightarrow \cC'$.
We have for instance
\begin{eqnarray}
 & & \cS(\tau_1,\dots,\tau_{i-1},0,+,\tau_{i+2},\dots,\tau_L)-\cS(\tau_1,\dots,\tau_{i-1},+,0,\tau_{i+2},\dots,\tau_L) \\
 & = & \frac{1}{Z_L}\llangle W|X_{\tau_1}\dots X_{\tau_{i-1}}(1d-d 1)X_{\tau_{i+2}}\dots X_{\tau_L}|V\rrangle \\
 & = & 0,
\end{eqnarray}
 or
\begin{eqnarray}
 & & \cS(\tau_1,\dots,\tau_{i-1},+,-,\tau_{i+2},\dots,\tau_L)-\cS(\tau_1,\dots,\tau_{i-1},0,0,\tau_{i+2},\dots,\tau_L) \\
 & = & \frac{1}{Z_L}\llangle W|X_{\tau_1}\dots X_{\tau_{i-1}}(d e-1\cdot 1)X_{\tau_{i+2}}\dots X_{\tau_L}|V\rrangle \\
 & = & 0, 
\end{eqnarray}
or for transitions involving the boundaries
\begin{eqnarray}
 & & a\cS(0,\tau_2,\dots,\tau_L) - \cS(-,\tau_2,\dots,\tau_L) \\
 & = & \frac{1}{Z_L}\llangle W|(a-e)X_{\tau_2}\dots X_{\tau_L}|V\rrangle \\
 & = & 0.
\end{eqnarray}
Similar computations apply also for other allowed transitions.

From the identities $E=1+e$ and $D=1+d$ it is straightforward to see that the stationary weights of the TASEP are obtained by summing 
several Boltzmann weights of this process. For instance 
\begin{equation}
\llangle W|DE|V\rrangle = \llangle W|1\cdot 1|V\rrangle + \llangle W|1\cdot e|V\rrangle +\llangle W|d\cdot 1|V\rrangle+\llangle W|d\cdot e|V\rrangle.
\end{equation}
Each of the term in the last sum is a Boltzmann weight of the '$(-,0,+)$' process. 

It would be interesting to study if the same kind of approach could be generalized to other non-equilibrium models. This would support the 
fact that we need to enlarge the configuration space of out-of-equilibrium systems to obtain an efficient description of their stationary
distribution. A lot remains to be understood in this direction.

\subsection{Link with integrability}

Although the matrix ansatz is often associated with integrable models because of the numerous examples of matrix product 
stationary states that can be found in this class, it does not seem to exist only in such privileged systems. There are indeed 
a few example of a matrix product formulation of the steady state in models that are not known to be integrable (but are not 
proven to be non-integrable...). Among them can be mentioned the ABC model on the ring with equal densities of each particle species \cite{EvansKKM98}
or a two species TASEP with open boundaries \cite{AyyerLS12}.
However we will argue in this subsection that, for stochastic integrable models, the construction of the steady state in a matrix product form
is made easier by the use of the $R$ and $K$ matrices. These key objects play indeed a central role in encoding the algebraic relations that must
be satisfied by the matrices and boundary vectors involved in the matrix ansatz. 

\subsubsection{Algebraic setup} 

The starting point of the construction is to define a vector which depends on the spectral parameter $z$
and that can be seen as an upgrading of the vector $\mathbf{X}$
\begin{equation}
 \mathbf{A}(z)= \begin{pmatrix}
                 A_0(z) \\ A_1(z) \\ \vdots \\ A_N(z)
                \end{pmatrix}
\end{equation}
The entries $A_i(z)$ of this vector belongs to a non-commutative algebra $\cA$ and have a finite expansion with respect to 
the spectral parameter $z$:
\begin{equation}
 A_i(z)= \sum_{k=-p}^q G_{i,k}z^k, \quad \mbox{with} \quad G_{i,k} \in \cA.
\end{equation}
We will see that the generators $G_{i,k}$ and their commutation relations play a crucial role in the matrix product construction: they are the 
building blocks of the matrices $X_0,\dots,X_N$ involved in the matrix ansatz.
The number of components $N+1$ of the vector is directly related to the number of different possible local configurations on a single site
($N$ different species of particles plus the hole).
The vector $\mathbf{A}(z)$ can be thought as a spectral parameter dependent generalization of the vector 
\begin{equation}
 \mathbf{X}=\begin{pmatrix}
             X_0 \\ X_1 \\ \vdots \\ X_N
            \end{pmatrix}
\end{equation}
involved in the matrix product expression of the steady state. 
The precise relation between these two vectors will be given in subsection \ref{subsec:ZF}.

We now give examples of such vectors $\mathbf{A}(z)$ that are relevant in the study of the TASEP, ASEP and SSEP. These three models will 
be systematically used all along this subsection to illustrate the different notions introduced. They are indeed probably familiar to the 
reader and will help to fix the ideas.

\begin{example}
The specific expression of the vector $\mathbf{A}(z)$ that is relevant to the matrix product formulation of the stationary state in the TASEP and
in the ASEP reads \cite{CrampeRV14}
\begin{equation} \label{eq:TASEP_vector_A}
 \mathbf{A}(z)=\begin{pmatrix}
                z+e \\ \frac{1}{z}+d
               \end{pmatrix}.
\end{equation}
Note that the commutation relations satisfied by the generators $e$ and $d$ are not the same in the case of the TASEP and of the ASEP, 
see subsection \ref{subsec:ZF}.
\end{example}

\begin{example}
 In the case of the SSEP, the vector $\mathbf{A}(z)$ takes the explicit form \cite{SasamotoW97}
 \begin{equation} \label{eq:SSEP_vector_A}
  \mathbf{A}(z) = \begin{pmatrix}
                   -z+E \\ z+D
                  \end{pmatrix}.
 \end{equation}
\end{example}
We will see in section \ref{sec:examples_MA} more examples of such vectors $\mathbf{A}(z)$ with a richer expansion and a bigger number of components.

\subsubsection{Zamolodchikov-Faddeev relation} \label{subsec:ZF}

The question that we address now is the one of encoding the commutation relations of the generators entering the definition of the vector
$\mathbf{A}(z)$. It was first noticed in \cite{SasamotoW97} and then investigated in \cite{CrampeRV14} that the Zamolodchikov-Faddeev relation
play a central role in the matrix product construction. In the same spirit as the FRT formalism (which permits to present the commutation relations 
of the generators of the quantum groups out of the associated $R$-matrix), this relation is a way to efficiently encode the algebraic
relations required by the matrix ansatz, with the help of the $R$-matrix associated to the integrable model.
\begin{definition}
 A Zamolodchikov-Faddeev algebra is an algebra generated by the algebraic elements entering the definition of the vector $\mathbf{A}(z)$ subject
 to the Zamolodchikov-Faddeev (ZF) relation\footnote{Note that the ZF algebra is in general defined for a vector $\mathbf{A}(z)$ which is a Laurent
 series with respect to $z$.
 We are interested in this manuscript in a vector $\mathbf{A}(z)$ with a finite expansion
 with respect to the spectral parameter $z$, which thus corresponds to a subclass of the general ZF framework.}
 \begin{equation}\label{eq:ZF}
 \check{R}\left(\frac{z_1}{z_2}\right) \mathbf{A}(z_1) \otimes \mathbf{A}(z_2) = \mathbf{A}(z_2) \otimes \mathbf{A}(z_1).
 \end{equation}
\end{definition}
In words, the action of the matrix $\check{R}$ induces an exchange of the spectral parameters $z_1$ and $z_2$. The same kind of idea will 
appear again in chapter \ref{chap:four}, when dealing with the quantum Knizhnik-Zamolodchikov equation.
This ZF relation is well-known in the context of integrable quantum field theory. 
It was first introduced in \cite{ZamolodchikovZ79}. 
It appears as a generalization of the commutation of creation operators of bosons (which corresponds to $\check{R}(z) = 1$) 
or the anti-commutation of creation operators of fermions (which corresponds to $\check{R}(z)=-1$).
\begin{proposition}
 The Zamolodchikov-Faddeev relation \eqref{eq:ZF} defines an associative algebra if the matrix $\check{R}$ satisfies the braided Yang-Baxter
 equation.
\end{proposition}
\proof
The order of the spectral parameters $z_1$, $z_2$ and $z_3$ in the product $\mathbf{A}(z_3) \otimes \mathbf{A}(z_2) \otimes \mathbf{A}(z_1)$
can be reversed in two different ways. We can first exchange $z_1$ and $z_2$, then exchange $z_1$ and $z_3$ and finally exchange $z_2$ and $z_3$
like this
\begin{eqnarray*}
 \mathbf{A}(z_3) \otimes \mathbf{A}(z_2) \otimes \mathbf{A}(z_1) & = & 
 \check{R}_2\left(\frac{z_1}{z_2}\right) \mathbf{A}(z_3) \otimes \mathbf{A}(z_1) \otimes \mathbf{A}(z_2) \\
 & = & \check{R}_2\left(\frac{z_1}{z_2}\right)\check{R}_1\left(\frac{z_1}{z_3}\right) \mathbf{A}(z_1) \otimes \mathbf{A}(z_3) \otimes \mathbf{A}(z_2) \\
 & = & \check{R}_2\left(\frac{z_1}{z_2}\right)\check{R}_1\left(\frac{z_1}{z_3}\right)\check{R}_2\left(\frac{z_2}{z_3}\right) 
 \mathbf{A}(z_1) \otimes \mathbf{A}(z_2) \otimes \mathbf{A}(z_3),
\end{eqnarray*}
where we recall that $\check{R}_1=\check{R} \otimes \id$ and $\check{R}_2=\id \otimes \check{R}$.

The other way is to first exchange $z_2$ and $z_3$, then exchange $z_1$ and $z_3$ and finally exchange $z_1$ and $z_2$:
\begin{eqnarray*}
 \mathbf{A}(z_3) \otimes \mathbf{A}(z_2) \otimes \mathbf{A}(z_1) & = & 
 \check{R}_1\left(\frac{z_2}{z_3}\right) \mathbf{A}(z_2) \otimes \mathbf{A}(z_3) \otimes \mathbf{A}(z_1) \\
 & = & \check{R}_1\left(\frac{z_2}{z_3}\right)\check{R}_2\left(\frac{z_1}{z_3}\right) \mathbf{A}(z_2) \otimes \mathbf{A}(z_1) \otimes \mathbf{A}(z_3) \\
 & = & \check{R}_1\left(\frac{z_2}{z_3}\right)\check{R}_2\left(\frac{z_1}{z_3}\right)\check{R}_1\left(\frac{z_1}{z_2}\right) 
 \mathbf{A}(z_1) \otimes \mathbf{A}(z_2) \otimes \mathbf{A}(z_3).
\end{eqnarray*}
The consistency between these two different ways of performing the computation ({\it i.e} the associativity of the algebra) is 
ensured by the fact that $\check{R}$ satisfies the braided Yang-Baxter equation.
\finproof

\begin{remark}
 Another consistency relation, which arises when applying the ZF relation twice, is ensured by the unitarity \eqref{eq:unitarity_2spectralparameters}
 of the matrix $\check{R}$:
 \begin{equation}
  \mathbf{A}(z_2) \otimes \mathbf{A}(z_1) = \check{R}\left(\frac{z_1}{z_2}\right) \mathbf{A}(z_1) \otimes \mathbf{A}(z_2)
  = \check{R}\left(\frac{z_1}{z_2}\right) \check{R}\left(\frac{z_2}{z_1}\right)\mathbf{A}(z_2) \otimes \mathbf{A}(z_1).
 \end{equation}
\end{remark}

\begin{remark}
 If there exist a vector $v(z)$ (with scalar entries) satisfying $\check{R}\left(\frac{z_1}{z_2}\right)v(z_1) \otimes v(z_2)= v(z_2) \otimes v(z_1)$,
 (i.e which is a scalar representation of the ZF algebra), then it is possible to construct a sub-algebra of the quantum group associated with 
 $\check{R}$ that fulfills the requirement of the ZF algebra \eqref{eq:ZF}. More precisely, if $\mathbf{T}(z)$ is a matrix (with algebraic entries)
 that satisfies the RTT relation \eqref{eq:RTT_relation}, then the vector defined by $\mathbf{A}(z)=\mathbf{T}(z)v(z)$ satisfies \eqref{eq:ZF}.
 More generally if we have a representation $\mathbf{A}(z)^{(q)}$ of the ZF algebra on a vector space $\cV_q$
 and a representation $\mathbf{T}(z)^{(q')}$ of the RTT algebra on a vector space $\cV_{q'}$
 then we can build a new representation $\mathbf{T}(z)^{(q')}\mathbf{A}(z)^{(q)}$ of the ZF algebra on the vector space $\cV_{q'} \otimes \cV_q$.
 See chapter \ref{chap:four} and \cite{CrampeRV14} for more details.
\end{remark}

We now turn to a more physical interpretation of the ZF relation and shed some light on its connection with the telescopic relation
\eqref{eq:telescopic_relation_bulk}.
\begin{proposition} \label{prop:ZF_to_telescopic}
 Let $\mathbf{A}(z)$ satisfies the ZF relation \eqref{eq:ZF}.
 Define the vectors $\mathbf{X}=\mathbf{A}(1)$ and $\mathbf{\overline{X}}=\theta \mathbf{A}'(1)$ (where $\theta$ is the proportionality 
 coefficient involved in \eqref{eq:derivative_bulk_local_jump} and $\cdot '$ denotes the derivation with respect to the spectral parameter). 
 Then we have the telescopic relation
 \begin{equation}
 m \mathbf{X} \otimes \mathbf{X} = \mathbf{X} \otimes \mathbf{\overline{X}} - \mathbf{\overline{X}} \otimes \mathbf{X}.
\end{equation}
\end{proposition}
\proof
The telescopic relation is obtained by taking the derivative with respect to $z_1$ in the ZF relation \eqref{eq:ZF} and setting $z_1=z_2=1$.
We then need to apply the definition of the vectors $\mathbf{X}$ and $\mathbf{\overline{X}}$ and to use the fact that $m=\theta \check{R}'(1)$.
\finproof

This precise connection between the ZF relation and the bulk telescopic relation of the matrix ansatz was first stated in \cite{SasamotoW97}
for the SSEP and in \cite{CrampeRV14} for the general case.  
We now present how this framework can be applied to the TASEP, ASEP and SSEP.
\begin{example}
 We come back to the TASEP case. We write the ZF relation with the matrix $\check{R}$, defined in \eqref{eq:TASEP_R}, 
 and the vector $\mathbf{A}(z)$, defined in \eqref{eq:TASEP_vector_A},
 to determine the algebraic constraints satisfied by the generators $e$ and $d$.
 A direct computation yields
 \begin{equation}
  \check{R}\left(\frac{z_1}{z_2}\right) \mathbf{A}(z_1) \otimes \mathbf{A}(z_2) = \mathbf{A}(z_2) \otimes \mathbf{A}(z_1) 
  \quad \Leftrightarrow \quad \begin{pmatrix}
                   0 \\ \frac{z_2-z_1}{z_2}(d e-1) \\ \frac{z_1-z_2}{z_2}(d e-1) \\ 0
                  \end{pmatrix} = 0 \quad \Leftrightarrow \quad d e=1.
 \end{equation}
 The matrices $E$ and $D$ are thus given by
 \begin{equation}
  \begin{pmatrix}
   E \\ D 
  \end{pmatrix} = \mathbf{X} = \mathbf{A}(1) = \begin{pmatrix}
                                                1+e \\ 1+d
                                               \end{pmatrix}.
 \end{equation}
 We can then recover easily from the relation $d e=1$ that 
 \begin{equation}
 DE=(1+d)(1+e)=1+e+d+d e=1+e+1+d=E+D,
 \end{equation}
 in agreement with the statement \eqref{eq:TASEP_algebraic_relations}.
 This framework allows us also to recover the expression of the hat vector 
 \begin{equation}
  \mathbf{\overline{X}}= \theta \mathbf{A}'(1)=-\begin{pmatrix}
                                                 1 \\ -1
                                                \end{pmatrix}=
                                                \begin{pmatrix}
                                                 -1 \\ 1
                                                \end{pmatrix},
 \end{equation}
 where we recall that the value of $\theta$ associated to the TASEP is $-1$. 
\end{example}

\begin{example}
 In the ASEP case the matrix $\check{R}$ is defined in \eqref{eq:ASEP_R} and the vector $\mathbf{A}(z)$ in \eqref{eq:TASEP_vector_A}. A straightforward
 computation allows us to establish that the ZF relation is equivalent to the relation $pd e-qed=p-q$.
 The expressions of the matrices $E$ and $D$ are again derived through
  \begin{equation}
  \begin{pmatrix}
   E \\ D 
  \end{pmatrix} = \mathbf{X} = \mathbf{A}(1) = \begin{pmatrix}
                                                1+e \\ 1+d
                                               \end{pmatrix}.
 \end{equation}
 The relation $pd e-qed=p-q$ implies immediately that
 \begin{equation}
  pDE-qED=(1+d)(1+e)-q(1+e)(1+d)=(p-q)(1+e+d)+pd e-qed=(p-q)(E+D),
 \end{equation}
 in agreement with \ref{ex:ASEP_MA}.
 We can also derive the expression of the hat vector 
 \begin{equation}
  \mathbf{\overline{X}}= \theta \mathbf{A}'(1)=(q-p)\begin{pmatrix}
                                                 1 \\ -1
                                                \end{pmatrix} = 
                                                \begin{pmatrix}
                                                 q-p \\ p-q
                                                \end{pmatrix},
 \end{equation}
 where we recall that the value of $\theta$ associated to the ASEP is $q-p$.
\end{example}

\begin{example}
 In the SSEP case, the matrix $\check{R}$ has been introduced in \eqref{eq:SSEP_R} and the vector $\mathbf{A}(z)$ is given in \eqref{eq:SSEP_vector_A}.
 We write the ZF relation (with additive spectral parameters, in agreement with the Yang-Baxter equation \eqref{eq:Yang_Baxter_additif}), 
 to determine the commutation relation satisfied by the generators $E$ and $D$. By direct computation we show that
 \begin{equation}
  \check{R}\left(z_1-z_2\right) \mathbf{A}(z_1) \otimes \mathbf{A}(z_2) = \mathbf{A}(z_2) \otimes \mathbf{A}(z_1) 
  \quad \Leftrightarrow \quad DE-ED=E+D,
 \end{equation}
 which is in agreement with \ref{ex:SSEP_MA}.
 We can also derive the expression of the hat vector 
 \begin{equation}
  \mathbf{\overline{X}}= \theta \mathbf{A}'(0)=\begin{pmatrix}
                                                 -1 \\ 1
                                                \end{pmatrix},
 \end{equation}
 where we recall that the value of $\theta$ associated to the SSEP is $1$.
\end{example}

The examples presented above should not mislead the reader to think that the ZF relation is equivalent to the telescopic relation 
\eqref{eq:telescopic_relation_bulk}. This is the case only when the hat vector $\mathbf{\overline{X}}$ has scalar entries (as in the three
examples presented above). 
But let us stress that when $\mathbf{\overline{X}}$ is not scalar, the ZF relation often contains far more information that the telescopic relation.
In the next section \ref{sec:examples_MA} we will encounter an example in which the telescopic relation does not give enough information 
to compute efficiently the stationary weights of the model (using only the algebraic relations, as it has been done in \eqref{eq:TASEP_particular_weight} 
for instance, and not using the explicit representation of the matrices $X_i$). In such model 
some additional information given by the ZF relations are needed to simplify the computations (for instance information telling us how to commute
an operator $X_i$ with a hat operator $\overline{X}_j$). Moreover, we will observe in the particular examples of section \ref{sec:examples_MA}, 
that the ZF relations often provide a change of generator basis that is very convenient for the computation of physical quantities. 
The reader may refer to section \ref{sec:examples_MA} for more details.

\subsubsection{Ghoshal-Zamolodchikov relations}

The previous sub-section dealt with the encoding of the commutation relations that must be verified by the matrices $X_0,\dots,X_N$ 
involved in the matrix ansatz. We address here the question of the algebraic relations between the boundary vectors
$\llangle W|$ and $|V\rrangle$ and the matrices $X_i$. Similarly to the bulk case, where (for integrable models) 
the bulk telescopic relation can be upgraded to a
spectral parameter dependent relation (the ZF relation) containing more information, the boundary telescopic relations derive also 
from more general relations: the Ghoshal-Zamolodchikov (GZ) relations. They are intuitively interpreted as the boundary counterparts 
of the ZF relation. They are expressed using the $K$ matrices associated to the integrable model under consideration.

\begin{definition}
 For matrices $K$ and $\overline K$ satisfying the reflection equations \eqref{eq:reflection_equation} and \eqref{eq:reflection_equation_reversed}, 
 the Ghoshal-Zamolodchikov relations read
 \begin{equation} \label{eq:GZ}
  \llangle W| \left( K(z)\mathbf{A}\left(\frac{1}{z}\right)-\mathbf{A}(z) \right)=0, \qquad 
  \left(\overline{K}(z)\mathbf{A}\left(\frac{1}{z}\right)-\mathbf{A}(z) \right)|V\rrangle = 0.
 \end{equation}
\end{definition}
Note that in these relations the vector $|V\rrangle$ (respectively $\llangle W|$) is a (possibly infinite dimensional) column vector
(row vector respectively). The entries of $\mathbf{A}(z)$ are (possibly infinite dimensional) matrices acting on the vectors 
$\llangle W|$ and $|V\rrangle$ with the usual matrix product.

These relations appeared first in the context of quantum field theory to deal with integrable boundaries \cite{GhoshalZ94}. 
In this context $|V\rrangle$ is interpreted as the vacuum state of the theory. The relevance of these relations for integrable stochastic 
processes was first noticed in \cite{SasamotoW97} and then investigated in \cite{CrampeRV14}.

\begin{proposition}
 The two Ghoshal-Zamolodchikov relations \eqref{eq:GZ} are consistent with the Zamolodchikov-Faddeev relation \eqref{eq:ZF}. 
\end{proposition}
\proof
Starting from the quantity $\llangle W|\mathbf{A}(z_2) \otimes \mathbf{A}(z_1)$, there are indeed two different ways to change $z_1 \longrightarrow 1/z_1$
and $z_2 \longrightarrow 1/z_2$. The first one is to change $z_2$ to $1/z_2$ using the GZ relation, then exchange $1/z_2$ and $z_1$ using
the ZF relation, and then change $z_1$ to $1/z_1$ and finally exchange $1/z_1$ and $1/z_2$:
\begin{eqnarray*}
 \llangle W|\mathbf{A}(z_2) \otimes \mathbf{A}(z_1) & = & K_1(z_2) \llangle W|\mathbf{A}\left(\frac{1}{z_2}\right) \otimes \mathbf{A}(z_1) \\
 & = &  K_1(z_2) \check{R}(z_1 z_2) \llangle W|\mathbf{A}(z_1) \otimes \mathbf{A}\left(\frac{1}{z_2}\right) \\
 & = &  K_1(z_2) \check{R}(z_1 z_2) K_1(z_1) \llangle W|\mathbf{A}\left(\frac{1}{z_1}\right) \otimes \mathbf{A}\left(\frac{1}{z_2}\right) \\
 & = &  K_1(z_2) \check{R}(z_1 z_2) K_1(z_1) \check{R}\left(\frac{z_1}{z_2}\right)
 \llangle W|\mathbf{A}\left(\frac{1}{z_2}\right) \otimes \mathbf{A}\left(\frac{1}{z_1}\right),
\end{eqnarray*}
where we recall that $K_1(z)=K(z)\otimes \id$.

The other way to perform the transformation is to first exchange $z_1$ and $z_2$, then change $z_1$ to $1/z_1$, then exchange $1/z_1$ and $z_2$
and finally change $z_2$ to $1/z_2$:
\begin{eqnarray*}
 \llangle W|\mathbf{A}(z_2) \otimes \mathbf{A}(z_1) & = & \check{R}\left(\frac{z_1}{z_2}\right) \llangle W|\mathbf{A}(z_1) \otimes \mathbf{A}(z_2) \\
& = & \check{R}\left(\frac{z_1}{z_2}\right) K_1(z_1) \llangle W|\mathbf{A}\left(\frac{1}{z_1}\right) \otimes \mathbf{A}(z_2) \\
& = & \check{R}\left(\frac{z_1}{z_2}\right) K_1(z_1) \check{R}(z_1 z_2) \llangle W|\mathbf{A}(z_2) \otimes \mathbf{A}\left(\frac{1}{z_1}\right) \\
& = & \check{R}\left(\frac{z_1}{z_2}\right) K_1(z_1) \check{R}(z_1 z_2) K_1(z_2)
\llangle W|\mathbf{A}\left(\frac{1}{z_2}\right) \otimes \mathbf{A}\left(\frac{1}{z_1}\right).
\end{eqnarray*}
The consistency between these two different ways of doing the transformation is ensured by the fact that the matrix $K$ satisfies the 
reflection equation \eqref{eq:reflection_equation}. Similarly there are also two ways to transform $z_1$ to $1/z_1$ and $z_2$ to $1/z_2$ in the expression
$\mathbf{A}(z_2) \otimes \mathbf{A}(z_1) |V\rrangle$. The consistency is again ensured by the fact that the matrix $\overline{K}$ 
solves the reversed reflection equation \eqref{eq:reflection_equation_reversed}.
\finproof

\begin{remark}
 Another consistency relation, which arises when applying a GZ relation twice, is ensured by the unitarity \eqref{eq:Kmatrix_unitarity}
 of the matrices $K$ and $\overline{K}$:
 \begin{equation}
 \llangle W|\mathbf{A}(z) = K(z) \llangle W|\mathbf{A}\left(\frac{1}{z}\right) = K(z) K\left(\frac{1}{z}\right)\llangle W|\mathbf{A}(z),
 \end{equation}
 and
  \begin{equation}
 \mathbf{A}(z)|V\rrangle = \overline{K}(z)\mathbf{A}\left(\frac{1}{z}\right)|V\rrangle 
 = \overline{K}(z) \overline{K}\left(\frac{1}{z}\right)\mathbf{A}(z)|V\rrangle.
 \end{equation}
\end{remark}

We now turn to a more physical interpretation of the GZ relations and shed some light on their connection with the boundary telescopic relations
\eqref{eq:telescopic_relation_boundaries}.
\begin{proposition} \label{prop:GZ_to_telescopic}
 Let $\mathbf{A}(z)$, $\llangle W|$ and $|V\rrangle$ satisfying the GZ relations \eqref{eq:GZ}.
 Define the vectors $\mathbf{X}=\mathbf{A}(1)$ and $\mathbf{\overline{X}}=\theta \mathbf{A}'(1)$ (where $\theta$ is the proportionality 
 coefficient involved in \eqref{eq:derivative_bulk_local_jump}, \eqref{eq:Kmatrix_localjump} and \eqref{eq:Kbmatrix_localjump}). 
 Then we have the telescopic relations \eqref{eq:telescopic_relation_boundaries}:
 \begin{equation}
 \llangle W|B \mathbf{X} = \llangle W|\mathbf{\overline{X}}, \qquad  \overline{B}\mathbf{X}|V\rrangle = -\mathbf{\overline{X}}|V\rrangle.
\end{equation}
\end{proposition}
\proof
The telescopic relations are obtained by taking the derivative with respect to $z$ in the GZ relations \eqref{eq:GZ} and setting $z=1$.
We then need to apply the definition of the vectors $\mathbf{X}$ and $\mathbf{\overline{X}}$ and to use the fact that $B=\frac{\theta}{2} K'(1)$
and $\overline{B}=-\frac{\theta}{2} \overline{K}'(1)$.
\finproof

We know illustrate these GZ relations on the familiar examples that are the TASEP, the ASEP and the SSEP.
\begin{example}
 Coming back again to the TASEP case, we write the GZ relations with the boundary matrices $K$ and $\overline{K}$ defined in \eqref{eq:TASEP_K} 
 and \eqref{eq:TASEP_Kb} and the vector $\mathbf{A}(z)$ defined in
 \eqref{eq:TASEP_vector_A} to determine the algebraic constraints between by the generators $e$ and $d$ and
 the boundary vectors $\llangle W|$ and $|V\rrangle$.
 A direct computation yields
\begin{equation}
 \llangle W|K(z)\mathbf{A}\left(\frac{1}{z}\right)=\llangle W|\mathbf{A}(z) \quad \Leftrightarrow \quad 
 \llangle W| \begin{pmatrix}
  \frac{(1-z^2)(\alpha e+\alpha-1)}{\alpha z-\alpha-z} \\
  -\frac{(1-z^2)(\alpha e+\alpha-1)}{\alpha z-\alpha-z}
 \end{pmatrix}=0 \quad \Leftrightarrow \quad \llangle W|(\alpha e +\alpha-1)=0.
\end{equation}
In the same way 
\begin{equation}
 \overline{K}(z)\mathbf{A}\left(\frac{1}{z}\right)|V\rrangle= \mathbf{A}(z)|V\rrangle \quad \Leftrightarrow \quad 
 \begin{pmatrix}
  \frac{(z^2-1)(\beta d+\beta-1)}{z(\beta z-\beta+1)} \\
  -\frac{(z^2-1)(\beta d+\beta-1)}{z(\beta z-\beta+1)}
 \end{pmatrix}|V\rrangle=0 \quad \Leftrightarrow \quad (\beta d+\beta-1)|V\rrangle=0.
\end{equation}
 We can then recover easily, from the relation $\llangle W|(\alpha e+\alpha-1)=0$ and having in mind the identity $E=1+e$, that 
 \begin{equation}
 \llangle W|E = \llangle W|(1+e)=\frac{1}{\alpha}\llangle W|
 \end{equation}
 in agreement with the statement \eqref{eq:TASEP_algebraic_relations}.
 Similarly using relation $(\beta d+\beta-1)|V\rrangle=0$ we can show
 \begin{equation}
  D|V\rrangle = (1+d)|V\rrangle = \frac{1}{\beta}|V\rrangle
 \end{equation}
which is also in agreement with \eqref{eq:TASEP_algebraic_relations}. 
\end{example}

\begin{example}
 In the ASEP case the matrices $K$ and $\overline{K}$ are defined in \eqref{eq:ASEP_K} and \eqref{eq:ASEP_Kb}
 and the vector $\mathbf{A}(z)$ in \eqref{eq:TASEP_vector_A}. A straightforward
 computation allows us to establish that the GZ relations are equivalent to the relations
 \begin{equation}
  \llangle W|(\alpha e-\gamma d)=(\gamma-\alpha+p-q)\llangle W|
 \end{equation} 
 and 
 \begin{equation}
  (\beta d-\delta e)|V\rrangle=(\delta-\beta+p-q)|V\rrangle.
 \end{equation}
Using the identities $E=1+e$ and $D=1+d$ these relations imply immediately that
 \begin{equation}
  \llangle W|\left(\alpha E-\gamma D\right)=(p-q)\llangle W|
 \end{equation}
 and 
 \begin{equation}
  \left(\beta D-\delta E\right)|V\rrangle=(p-q)|V\rrangle 
 \end{equation}
 which are in agreement with \eqref{eq:ASEP_relation_boundaries}.
\end{example}

\begin{example}
 In the SSEP case, the matrices $K$ and $\overline{K}$ have been introduced in \eqref{eq:SSEP_K} and \eqref{eq:SSEP_Kb}
 and the vector $\mathbf{A}(z)$ is given in \eqref{eq:SSEP_vector_A}.
 We write the GZ relations (with additive spectral parameters, in agreement with the reflection equation \eqref{eq:reflection_equation_additif}), 
 to determine the algebraic relations 
 satisfied by the generators $E$ and $D$ and the boundary vectors $\llangle W|$ and $|V\rrangle$. By direct computation we show that
 \begin{equation}
  \llangle W|K(z)\mathbf{A}(-z)=\llangle W|\mathbf{A}(z) \quad \Leftrightarrow \quad  \llangle W|(\alpha E-\gamma D)=\llangle W|
 \end{equation}
 and
 \begin{equation}
  \overline{K}(z)\mathbf{A}(-z)|V\rrangle= \mathbf{A}(z)|V\rrangle \quad \Leftrightarrow \quad (\beta D-\delta E)|V\rrangle = |V\rrangle,
 \end{equation}
 in agreement with \ref{eq:SSEP_algebraic_relations}.
\end{example}

\subsection{Ground state of the transfer matrix} \label{subsec:inhomogeneous_ground_state}

\subsubsection{Inhomogeneous ground state}

We stressed in the previous sections the direct relation between the ZF and GZ relations on one hand, and the steady state of the Markov matrix 
of integrable models on the other hand. We explained in chapter \ref{chap:two} that the Markov matrix associated to an integrable process
belongs to a family of commuting operators generated by the transfer matrix. Because of this commutation property, the whole family of operators
shares the same eigenvectors. In particular the steady state of the Markov matrix should be an eigenvector of the transfer matrix. In order 
to address this problem in full generality, we will be interested in the transfer matrix with inhomogeneity parameters 
(introduced in chapter \ref{chap:two}). 
This motivates the following definition.
\begin{definition} \label{def:inhomogeneous_ground_state}
For $\mathbf{A}(z)$, $\llangle W|$ and $|V\rrangle$ satisfying the ZF relation \eqref{eq:ZF} and the GZ relations \eqref{eq:GZ},
we define the inhomogeneous ground state\footnote{the notation $\ket{\cS(z_1,z_2,\dots,z_L)}$ should not be confused with the components 
$\cS(\tau_1,\tau_2,\dots,\tau_L)$ of the undeformed steady state. The distinction between these two objects should be clear from the context.}
\begin{equation} \label{eq:inhomogeneous_ground_state}
 \ket{\cS(z_1,z_2,\dots,z_L)}=\frac{1}{Z_L(z_1,z_2,\dots,z_L)}
 \llangle W| \mathbf{A}(z_1) \otimes \mathbf{A}(z_2) \otimes \dots \otimes \mathbf{A}(z_L) |V\rrangle,
\end{equation}
where $Z_L(z_1,z_2,\dots,z_L) = \llangle W|C(z_1)C(z_2) \dots C(z_L)|V\rrangle$ with
\begin{equation}
 C(z)=A_0(z)+A_1(z)+\dots+A_N(z).
\end{equation}
\end{definition}

$\ket{\cS(z_1,z_2,\dots,z_L)}$ is a inhomogeneous deformation of the steady state $\steady$ (of the Markov matrix obtained by 
taking the derivative of the transfer matrix with respect to the spectral parameter). The following proposition clarifies this statement.

\begin{proposition} \label{pro:deformed_steady_state}
 $\steady = \ket{\cS(1,1,\dots,1)}$ is the steady state of the model, i.e $M\steady=0$, 
 where $M$ is given by \eqref{eq:transfer_matrix_to_Markov_matrix_open}.
\end{proposition}
\proof
 This derives directly from the propositions \ref{prop:ZF_to_telescopic}, \ref{prop:GZ_to_telescopic} and \ref{prop:steady_state}. 
\finproof

\begin{proposition} \label{pro:inhomogeneous_normalization}
 The normalization $Z_L(z_1,z_2,\dots,z_L)$ is symmetric under permutation of variables $z_i \leftrightarrow z_j$ and 
 inversion of variable $z_i \rightarrow 1/z_i$.
\end{proposition}
This property shed some light on the possible connection of $Z_L(z_1,z_2,\dots,z_L)$ with the theory of symmetric functions. This relation
will be clarified and explored in chapter \ref{chap:four}.

\begin{proposition} \label{pro:scattering_matrices}
 For all $i=1,\dots,L$ we have
  \begin{equation}
   t(z_i |z_1,\dots,z_L)\ket{\cS(z_1,z_2,\dots,z_L)}=\ket{\cS(z_1,z_2,\dots,z_L)}
  \end{equation}
 and
 \begin{equation}
   t(1/z_i |z_1,\dots,z_L)\ket{\cS(z_1,z_2,\dots,z_L)}=\ket{\cS(z_1,z_2,\dots,z_L)}.
  \end{equation}
 The specialized transfer matrices
 \begin{eqnarray*}
  t(z_i |z_1,\dots,z_L) & = & \check R_{i-1,i}\left(\frac{z_i}{z_{i-1}}\right) \dots \check R_{1,2}\left(\frac{z_i}{z_1}\right) K_1(z_i) \\
  & \times & \check R_{1,2}(z_iz_1)\dots \check R_{i-1,i}(z_iz_{i-1})\check R_{i,i+1}(z_iz_{i+1}) \dots \check R_{L-1,L}(z_iz_L) \\
  & \times & \overline{K}_L\left(\frac{1}{z_i}\right) \check R_{L-1,L}\left(\frac{z_i}{z_L}\right) \dots \check R_{i,i+1}\left(\frac{z_i}{z_{i+1}}\right)
 \end{eqnarray*}
 and
 \begin{eqnarray*}
  t(1/z_i |z_1,\dots,z_L) & = & \check R_{i,i+1}\left(\frac{z_{i+1}}{z_i}\right) \dots \check R_{L,1,L}\left(\frac{z_L}{z_i}\right)\overline{K}_L(z_i)\\
  & \times & \check R_{L-1,L}\left(\frac{1}{z_iz_L}\right) \dots \check R_{i,i+1}\left(\frac{1}{z_iz_{i+1}}\right)
  \check R_{i-1,i}\left(\frac{1}{z_iz_{i-1}}\right) \dots \check R_{1,2}\left(\frac{1}{z_iz_1}\right) \\
  & \times & K_1\left(\frac{1}{z_i}\right)\check R_{1,2}\left(\frac{z_1}{z_i}\right) \dots \check R_{i-1,i}\left(\frac{z_{i-1}}{z_i}\right)
 \end{eqnarray*}
 are sometimes called the scattering matrices. Those matrices will appear again in chapter \ref{chap:four} in the context of 
 quantum Knizhnik-Zamolodchikov equations.
\end{proposition}
\proof
 Analyzing the explicit form of $t(z_i |z_1,\dots,z_L)$ we see that when acting on $\ket{\cS(z_1,z_2,\dots,z_L)}$ it takes the inhomogeneity parameter
 $z_i$ and pushes it to the right through all the $z_j$ (by successive permutation because of the ZF relation satisfied by $\mathbf{A}(z)$). 
 It then inverts $z_i$ into $1/z_i$ at the right boundary (because of the GZ relation).
 $1/z_i$ is then pushed to the left and again inverted at the left boundary, to finally being placed in its initial position, completing in this 
 manner the whole circle. More precisely
 \begin{eqnarray*}
   t(z_i |z_1,\dots,z_L)\ket{\cS(z_1,\dots,z_L)} & = & \check R_{i-1,i}\left(\frac{z_i}{z_{i-1}}\right) \dots \check R_{1,2}\left(\frac{z_i}{z_1}\right) K_1(z_i) \\
  & & \times \check R_{1,2}(z_iz_1)\dots \check R_{i-1,i}(z_iz_{i-1})\check R_{i,i+1}(z_iz_{i+1}) \dots \check R_{L-1,L}(z_iz_L) \\
  & & \times \overline{K}_L\left(\frac{1}{z_i}\right) \ket{\cS(z_1,\dots,z_{i-1},z_{i+1},\dots,z_L,z_i)} \\
  & = & \check R_{i-1,i}\left(\frac{z_i}{z_{i-1}}\right) \dots \check R_{1,2}\left(\frac{z_i}{z_1}\right) K_1(z_i) \\
  & & \times \ket{\cS(\frac{1}{z_i},z_1,\dots,z_{i-1},z_{i+1},\dots,z_L)} \\
  & = & \ket{\cS(z_1,\dots,z_L)}.
 \end{eqnarray*}
The proof is exactly the same for the scattering matrix $t(1/z_i |z_1,\dots,z_L)$ but the circle is made the other way around.
\finproof

\begin{remark}
Note that the definition \ref{def:inhomogeneous_ground_state} and the propositions \ref{pro:deformed_steady_state}, 
\ref{pro:inhomogeneous_normalization} and \ref{pro:scattering_matrices} can be easily adapted to the periodic case (we chose not to 
give them explicitly here to avoid repetitions). The reader may refer to subsection \ref{subsec:inhomogeneous_periodic_TASEP}
where an example of model defined on the periodic lattice is studied in details.
\end{remark}

Now, a question arises naturally: is the inhomogeneous ground state $\ket{\cS(z_1,z_2,\dots,z_L)}$ an eigenvector of the 
transfer matrix $t(z |z_1,\dots,z_L)$ for a generic $z$?
In all the models that we have encountered so far (ASEP, TASEP, SSEP) the answer is yes, but we are still lacking for a general proof of this fact.
Indeed in all the models that we have studied, a specific (model dependent) analysis of the degree and symmetries of the transfer matrix was needed
to prove this fact. The common way to tackle the problem is to remark that the transfer matrix often has rational entries in the spectral parameter
$z$. Knowing that $\ket{\cS(z_1,z_2,\dots,z_L)}$ is an eigenvector of $t(z |z_1,\dots,z_L)$ for $z=z_i^{\pm 1}$ and studying some symmetry of the 
transfer matrix is often enough to prove by degree considerations that the inhomogeneous ground state is an eigenvector of the transfer matrix for all 
$z$. We give below several examples.

\subsubsection{Some examples}

\begin{example}
 We begin to present the case of the ASEP defined on a lattice with open boundaries. We have for this model
 \begin{equation}
  t(z|z_1,\dots,z_L)\ket{\cS(z_1,\dots,z_L)} = \lambda(z|z_1,\dots,z_L)\ket{\cS(z_1,\dots,z_L)}
 \end{equation}
 with
 \begin{eqnarray*}
  & & \lambda(z|z_1,\dots,z_L) = 1+(pq)^{L-1} \left(\prod_{i=1}^L \frac{(1-z z_i)(z_i-z)}{(p-qz z_i)(pz_i-qz)}\right) \\
 & \times & \frac{(1-z^2)(\delta p^2+(\beta-\delta+q-p)pqz-\beta q^2z^2)(\gamma p^2+(\alpha-\gamma+q-p)pqz-\alpha q^2z^2)}
  {(p^2-q^2z^2)(\delta z^2+(\beta-\delta+q-p)z-\beta)(\gamma z^2+(\alpha-\gamma+q-p)z-\alpha)}.
 \end{eqnarray*}
 This relation is proved using the symmetry 
 \begin{equation}
  t(z|z_1,\dots,z_L)=\left(\lambda(z|z_1,\dots,z_L)-1\right)t(\frac{p}{qz}|z_1,\dots,z_L).
 \end{equation}
 More details about this symmetry relation can be found in \cite{CrampeRV14}.
\end{example}

\begin{example}
 Concerning the SSEP defined on a lattice with open boundaries, the situation is very similar. We have indeed for this model
 \begin{equation}
  t(z|z_1,\dots,z_L)\ket{\cS(z_1,\dots,z_L)} = \lambda(z|z_1,\dots,z_L)\ket{\cS(z_1,\dots,z_L)}
 \end{equation}
 with
 \begin{equation*}
 \lambda(z|z_1,\dots,z_L) = 1 + \left( \prod_{i=1}^L \frac{(z-z_i)(z+z_i)}{(z+1-z_i)(z+1+z_i)} \right)
  \frac{z((\alpha+\gamma)(z+1)-1)((\beta+\delta)(z+1)-1)}{(z+1)((\alpha+\gamma)z+1)((\beta+\delta)z+1)}.
 \end{equation*}
  This relation is proved using the symmetry 
 \begin{equation}
  t(z|z_1,\dots,z_L)=\left(\lambda(z|z_1,\dots,z_L)-1\right)t(-1-z|z_1,\dots,z_L).
 \end{equation}
 More details about this symmetry relation can be found in \cite{CrampeRV14}.
\end{example}

\subsubsection{Inhomogeneous periodic TASEP} \label{subsec:inhomogeneous_periodic_TASEP}

 We study, in this subsection, the ground state of the periodic inhomogeneous transfer matrix. Instead of addressing the problem in full generality
 (which would have been a bit redundant with the open case), we prefer to present the method through a specific simple example:
 the periodic TASEP. The inhomogeneous transfer matrix associated to the single species periodic TASEP 
 was introduced in chapter \ref{chap:two}. It has been proven to be itself the Markov matrix encoding a discrete time process.
 We are now interested in computing the inhomogeneous ground state of this transfer matrix (note that in this particular case the inhomogeneous
 ground state can be called stationary state because the associated eigenvalue is equal to $1$ thanks to the Markovian property of the 
 transfer matrix). The results presented here are mostly extracted from \cite{CrampeMRV15inhomogeneous} and the reader is invited to refer to this
 paper for the details.
 
The building block of the stationary state is the following vector
\begin{equation} \label{eq:TASEP_vector_v}
 v(z)=\left(
 \begin{array}{c}
 z\\1
 \end{array}
\right).
\end{equation}
It satisfies the Zamolodchikov-Faddeev relation 
\begin{equation} \label{eq:TASEP_ZF_scalar}
 R_{12}(z_1/z_2)v_1(z_1)v_2(z_2)=v_1(z_1)v_2(z_2) \, . 
\end{equation}
The stationary state of the process $M(z|\mathbf{z})$ (defined in \eqref{eq:TASEP_discrete_time_Markov}) is constructed from 
\begin{equation} \label{eq:TASEP_steady_state_periodic}
 \steady = v_1(z_1)v_2(z_2)\dots v_L(z_L).
\end{equation}
Recall that the subscripts denote which component of the tensor space the vector $v$ belongs to.
To be more precise, since the process conserves the number of particles, there are $L$ independent sectors
(as mentioned previously we do not consider the empty sector), each corresponding to a given number $m$
of particles  in the system. The stationary state is hence degenerate: there is one stationary state for each sector. 
Moreover, the exact normalization $Z^{(m)}$ of the stationary state depends on the sector we are considering.  Then, each stationary state is given
by the components of $\steady$  corresponding to the sector and correctly normalized. 

The following calculation justifies that $\steady$ given in \eqref{eq:TASEP_steady_state_periodic}
 is the stationary state  of the system. Indeed, for $i=1,2,...,L$ we have
\begin{eqnarray*}
 t(z_i|\mathbf{z}) \steady & = & 
 R_{i,i-1}(\frac{z_i}{z_{i-1}})\dots R_{i,1}(\frac{z_i}{z_{1}})R_{i,L}(\frac{z_i}{z_L})\dots R_{i,i+1}(\frac{z_i}{z_{i+1}})v_1(z_1)\dots v_L(z_L)\\
 & = & v_1(z_1)v_2(z_2)\dots v_L(z_L) \\
 & = & \steady,
\end{eqnarray*}
where $t(z|\mathbf{z})$ has been introduced in \eqref{eq:inhomogeneous_transfer_matrix} .
The first equality is obtained using the regularity property of the R-matrix whereas the second equality is obtained using $L-1$
times the property \eqref{eq:TASEP_ZF_scalar}. 
Clearly, $t(z|\mathbf{z})$ is a polynomial of degree less or equal to $L$ in $z$. It is possible to show (using  the graphical interpretation  for instance)
that in the sectors where there is at least one particle, {the degree of $t(z|\mathbf{z})$ is in fact at most equal to $L-1$.} 
Hence we can deduce through 
interpolation arguments that in these sectors, the stationary state is given by the corresponding components of $\steady$. 
Note  that when we take the homogeneous limit $z_i \rightarrow 1$ we recover the uniform stationary distribution of the continuous time TASEP.

From the knowledge of the stationary distribution, we can calculate various physical quantities. 
We shall see that some observables can be expressed as symmetric polynomials in the inhomogeneity
parameters $z_1,\ldots,z_L$.
  
The weight of the configuration $(\tau_1,\dots,\tau_L)$ is readily obtained  using \eqref{eq:TASEP_steady_state_periodic}: 
\begin{equation}
 \mathcal{S}(\tau_1,\dots,\tau_L)=\prod_{i=1}^{L}\left(\delta_{1,\tau_i}+z_i\delta_{0,\tau_i}\right).
\end{equation}

In the sector with $m$ particles, the normalization factor of the stationary state is obtained by summing the weights of all the configurations
with $m$ particles
\begin{equation}\label{eq:TASEP_Z_periodic}
 Z^{(m)}=\sum\limits_{\substack{ I \subset  \{1,\dots,L\} \\ |I|=L-m}}^{}\ {\prod_{i \in I} z_i \ =\ e_{L-m}(z_1,\dots,z_L)} \, ,
\end{equation}
 where $e_{L-m}$ is the elementary symmetric homogeneous polynomial of degree $L-m$. The normalization factor can be written as a  Schur polynomial:
\begin{equation}
 Z^{(m)}=s^A_{1^{L-m}}(z_1,\dots,z_L),\quad \mbox{where}\quad 1^{L-m}=(\underbrace{1,...,1}_{L-m},\underbrace{0,...,0}_{m}).
\end{equation}
We remind the definition of the Schur polynomial (of type A) associated with 
 a partition  $\lambda=(\lambda_1,\dots,\lambda_L)$ with $\lambda_1\geq\dots\geq\lambda_L\geq0$:
\begin{equation}
s^A_{\lambda}(z_1,\dots,z_L)=
\frac{\det \left((z_j)^{L-i+\lambda_i} \right)_{i,j}}{\det \left(z_j^{L-i} \right)_{i,j}}.
\end{equation}
This expression of   the partition function in terms of the Schur polynomial allows us to relate the value of $Z^{(m)}$ 
in the homogeneous limit ($z_i\rightarrow 1$) 
with the dimension of the representation $\pi^A(\lambda)$ of $sl(L)$ labeled by the Young tableau $[\lambda]$.
 Indeed, using the Weyl character formula (see e.g. \cite{Varadarajan13,FrappatSS00} for a review), we obtain 
\begin{equation}\label{eq:TASEP_rsl1}
s^A_{\lambda}(1,\dots,1)=\prod_{1\leq j\leq i\leq L} \frac{\lambda_j-\lambda_i+i-j}{i-j}=\text{dim}(\pi^A(\lambda))\;.
\end{equation}
In particular, we have 
\begin{equation}\label{eq:TASEP_rsl2}
 Z^{(m)}\Big|_{z_1=\dots=z_L=1}=\text{dim}(\pi^A(1^{L-m}))=\left(\begin{array}{c} L \\m\end{array}\right)
\end{equation}
in accordance with a direct computation starting from \eqref{eq:TASEP_Z_periodic}. 

In the sector with $m$ particles, the particle density at site $i$ is obtained by summing the weights of all the configurations with $m$ particles,
one of them being at site $i$
\begin{equation} \label{eq:TASEP_density_periodic}
 \langle \tau_i \rangle =\frac{e_{L-m}(z_1,\dots,z_{i-1},z_{i+1},\dots,z_L)}{e_{L-m}(z_1,\dots,z_L)}\;.
\end{equation}
We can show that $\displaystyle \sum_{i=1}^L \langle \tau_i \rangle=m$ as expected.

  The higher correlation functions take also a very simple form. The correlations between the sites $i_1<i_2<\dots<i_\ell$ is given by
\begin{equation} \label{eq:TASEP_corr_periodic}
 \langle \tau_{i_1}\tau_{i_2}\dots \tau_{i_\ell} \rangle =
 \frac{e_{L-m}(z_1,\dots,z_{i_1-1},z_{i_1+1},\dots,z_{i_2-1},z_{i_2+1},\dots,z_{i_\ell-1},z_{i_\ell+1},\dots,z_L)}{e_{L-m}(z_1,\dots,z_L)}\;.
\end{equation}
For $\ell>m$, the correlation functions vanish as expected since the number of particles is $m$ and the correlation
functions for more than $m$ particles has no meaning.

\subsubsection{Inhomogeneous open TASEP}

 We now move to the case of the single species open TASEP, whose inhomogeneous transfer matrix had been introduced in chapter \ref{chap:two}.
 It has been shown that this transfer matrix defines a discrete time Markov process. The results displayed are taken from 
 \cite{CrampeMRV15inhomogeneous}.
 
The stationary state is expressed as a matrix product. Indeed, the probability of the configuration 
$\mathcal{C}=(\tau_1,\dots,\tau_L)$ can be written as
\begin{equation}
 \mathcal{S}(\tau_1,\dots,\tau_L)=\frac{1}{Z_L}\llangle W| \prod_{i=1}^{L} ((1-\tau_i)E(z_i)+\tau_iD(z_i)) |V \rrangle \;,
\end{equation}
where the normalization factor is 
$Z_L=\llangle W|C(z_1)\dots C(z_L) |V\rrangle$, with $C(z)=E(z)+D(z)$.
 We recall that the algebraic elements $E(z)$ and $D(z)$ are the entries of the vector 
\begin{equation}
 \mathbf{A}(z)=\left(
 \begin{array}{c}
 E(z)\\ D(z)
 \end{array}
\right). 
\end{equation}
Using this vector, the stationary state can be written in a more compact way \eqref{eq:inhomogeneous_ground_state}

The vector $A(z)$ allows us to write easily the exchange relations between $E(z)$ and $D(z)$ through the ZF relation:
\begin{equation} \label{eq:TASEP_ZF_open}
 \check R(z_1/z_2)A(z_1) \otimes A(z_2) = A(z_2) \otimes A(z_1).
\end{equation}
Written explicitly it gives four relations
\begin{equation} \label{eq:TASEP_ZF_components}
 \left\{
 \begin{array}{l}
 \left[E(z_1),E(z_2)\right]=0\\[1.ex] 
 \left[D(z_1),D(z_2)\right]=0\\[1.ex]
 \frac{z_1}{z_2}D(z_1)E(z_2)=D(z_2)E(z_1)\\[1.ex]
 D(z_1)E(z_2)=\frac{z_2}{z_2-z_1}\left[ E(z_2)D(z_1)-E(z_1)D(z_2) \right]
 \end{array}
 \right.
\end{equation}
Let us remark that the third relation is implied by the fourth.

The relations between the boundary vectors $\llangle W|$, $|V\rrangle$ and the algebraic elements $E(z)$, $D(z)$ 
are given by the GZ relations
\begin{equation} \label{eq:TASEP_GZ_open}
 \llangle W | K(z)A(1/z)=\llangle W| A(z), \qquad \bar K(z)A(1/z) |V \rrangle = A(z) |V \rrangle  \;.
\end{equation}

Written explicitly we obtain four relations
\begin{equation} \label{eq:TASEP_GZ_components}
 \left\{
 \begin{array}{l}
 \llangle W|E(z)=\llangle W|\frac{(a+z)z}{za+1}E(1/z) \\[1.ex]
 \llangle W|E(z)=\frac{z(a+z)}{z^2-1}\llangle W|(D(1/z)-D(z) )
 \end{array}
 \right.
 \text{ and  }
  \left\{
 \begin{array}{l}
 D(z)|V\rrangle= \frac{zb+1}{z(b+z)}D(1/z)|V\rrangle \\[1.ex]
 D(z)|V\rrangle=\frac{bz+1}{z^2-1}(E(z)-E(1/z) )|V\rrangle
 \end{array}
 \right.
\end{equation}
Let us remark that for each boundary, the first relation is implied by the second one.

We now prove that the vector $\steady$ is the stationary state of the transfer matrix. From proposition \ref{pro:scattering_matrices} we know that
\begin{equation}
 t(z_i|\mathbf{z})\steady = \steady \;,
\end{equation}
and also
\begin{equation}
 t(1/z_i|\mathbf{z})\steady = \steady \;.
\end{equation}
We show below, using the graphical representation, that the numerators of the entries of $t(z|\mathbf{z})$ are polynomials of degree
less than $L+2$. Since the equation $t(z|\mathbf{z})\steady = \steady$ is satisfied for $2L$ different values of $z$ ($z=z_1,1/z_1,\dots,z_L,1/z_L$),
we get through interpolation arguments that for $L\geq 3$, $t(z|\mathbf{z})\steady = \steady$ for all $z$. For $L<5$ we checked by computer that 
this equation is satisfied.

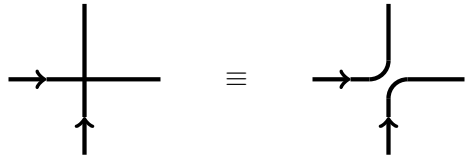
\begin{figure}[ht]
\begin{center}
 \begin{tikzpicture}[scale=1]
\draw[->,ultra thick] (0,0) -- (0.5,0); \draw[ultra thick] (0.5,0) -- (2,0);
\draw[->,ultra thick] (1,-1) -- (1,-0.5); \draw[ultra thick] (1,-0.5) -- (1,1);
\draw[->,ultra thick] (4,0) -- (4.5,0); \draw[ultra thick] (4.5,0) -- (4.75,0); \draw[ultra thick] (5.25,0) -- (6,0);
\draw[->,ultra thick] (5,-1) -- (5,-0.5); \draw[ultra thick] (5,-0.5) -- (5,-0.25); \draw[ultra thick] (5,0.25) -- (5,1);
\draw[ultra thick] (4.75,0) arc (-90:0:0.25); \draw[ultra thick] (5.25,0) arc (90:180:0.25);
\node at (3,0) [] {$\equiv$} ;
 \end{tikzpicture}
 \end{center}
 \caption{Vertex with non-intersecting thick lines.\label{fig:TASEP_identification_vertex}}
\end{figure}

We use the graphical representation given in figures \ref{fig:TASEP_matrix_elements}, 
\ref{fig:TASEP_matrix_elements_K} and \ref{fig:TASEP_matrix_elements_Ktilde}, interpreting the last vertex of
fig. \ref{fig:TASEP_matrix_elements} as in fig. \ref{fig:TASEP_identification_vertex}.
We can see that any matrix element $\langle \cC'|t(z|\mathbf{z})|\cC\rangle$ can be decomposed into continuous thick lines that do not intersect, 
as illustrated in the figure \ref{fig:TASEP_thicklines}.
\begin{figure}[ht]
\begin{center}
 \begin{tikzpicture}[scale=0.7]
\draw (0,-2) -- (0,2) ;
\draw (12,-2) -- (12,2) ;
\foreach \i in {-1.6,-1.2,...,2.1}
{\draw (-0.4,\i-0.4) -- (0,\i) ; \draw (12+0.4,\i-0.4) -- (12,\i) ;}
\draw[->,dashed] (2,-2) -- (1,-1); \draw[dashed] (1,-1) -- (0,0); \draw[->,dashed] (0,0) -- (1,1); \draw[dashed] (1,1) -- (2,2);
\draw[->,ultra thick] (10,2) -- (11,1); \draw[ultra thick] (11,1) -- (12,0); \draw[->,dashed] (12,0) -- (11,-1); \draw[dashed] (11,-1) -- (10,-2);
\draw[dashed] (2,2) -- (3,2); \draw[ultra thick] (3,2) -- (4.75,2); \draw[ultra thick] (5.25,2) -- (8.75,2); \draw[ultra thick] (9.25,2) -- (10,2);
\draw[dashed] (2,-2) -- (3,-2); \draw[dashed] (7,-2) -- (10,-2); \draw[ultra thick] (3,-2) -- (4.75,-2); \draw[ultra thick] (5.25,-2) -- (7,-2);
\draw[->,dashed] (3,-3) -- (3,-2.5); \draw[dashed] (3,-2.5) -- (3,-2); \draw[->,dashed] (3,2) -- (3,2.5); \draw[dashed] (3,2.5) -- (3,3); \draw[ultra thick] (3,-2) -- (3,2);
\draw[->,ultra thick] (5,-3) -- (5,-2.5); \draw[ultra thick] (5,-2.5) -- (5,-2.25); \draw[ultra thick] (5,-1.75) -- (5,1.75); \draw[->,ultra thick] (5,2.25) -- (5,2.5); \draw[ultra thick] (5,2.5) -- (5,3);
\draw[->,dashed] (7,-2) -- (7,2.5); \draw[dashed] (7,2.5) -- (7,3); \draw[->,ultra thick] (7,-3) -- (7,-2.5); \draw[ultra thick] (7,-2.5) -- (7,-2);
\draw[->,ultra thick] (9,-3) -- (9,-2.5); \draw[ultra thick] (9,-2.5) -- (9,1.75); \draw[->,ultra thick] (9,2.25) -- (9,2.5); \draw[ultra thick] (9,2.5) -- (9,3);
\draw[ultra thick] (5,-2.25) arc (0:90:0.25); \draw[ultra thick] (5,-1.75) arc (180:270:0.25);
\draw[ultra thick] (4.75,2) arc (270:360:0.25); \draw[ultra thick] (5.25,2) arc (90:180:0.25);
\draw[ultra thick] (8.75,2) arc (270:360:0.25); \draw[ultra thick] (9.25,2) arc (90:180:0.25);
 \end{tikzpicture}
 \caption{Decomposition of a matrix element into thick continuous lines.\label{fig:TASEP_thicklines}}
 \end{center}
\end{figure}

The vertices that do not enter a thick line have degree 0 in $z$. Thus, the total degree of the matrix element is the sum of the degree of each
thick line, plus the degree coming from the boundaries that we look at separately. We first consider lines that are in the 'bulk' 
(that do not include a boundary). Looking at the weights of the vertices in fig. \ref{fig:TASEP_matrix_elements}, one can see that the incoming part of 
a line always carry a degree 1. Moreover, a line corresponding to a degree $n$ must cross at least $n$ exit lines, 
the first on the left being always of degree 0. This implies that the bulk part is at most of degree $L$. 
Figure  \ref{fig:TASEP_matrix_elements_Ktilde} shows that the right boundary does not change this counting. 
The left boundary can add at most a degree 2, as it can be seen on figure  \ref{fig:TASEP_matrix_elements_K}. Altogether, this
leads to a total degree $L+2$.

We now construct an explicit representation of the algebraic elements $E(z)$ and $D(z)$. 
 We  introduce the shift operators $e$ and $d$ such that $de=1$, which are well known in the context of the continuous time open
  TASEP since they enter the construction of the original matrix ansatz \cite{DerridaEHP93}.
We define  
\begin{equation}
 \widetilde{E}(z)=z+e \,\,\,\, \hbox{ and } \,\,\,\,
  \widetilde{D}(z)=1/z+d \, ,
\label{eq:TASEP_ReprExpl}
\end{equation}
and the boundary vectors such that 
\begin{equation}\label{eq:bound-ed}
\llangle\widetilde W|\,e = a\,\llangle\widetilde W|  \,\,\,\, \hbox{ and } \,\,\,\,
d |\widetilde V\rrangle=b\,|\widetilde V\rrangle \, .
\end{equation}
It is  readily verified that   the 
Zamolodchikov-Faddeev relation \eqref{eq:TASEP_ZF_open} and 
the Ghoshal-Zamo\-lo\-dchikov relations \eqref{eq:TASEP_GZ_open} are satisfied. 
From the results of \cite{DerridaEHP93} giving explicit forms for $e$, $d$, $\llangle\widetilde W|$ and $|\widetilde V\rrangle$, we deduce 
that $\llangle\widetilde W|$ and $|\widetilde V\rrangle$ can be chosen such that $\llangle\widetilde W|\widetilde V\rrangle\neq 0$
which guarantees the non-vanishing of $\steady$.

The matrix ansatz allows us to calculate the stationary probability of any given configuration. 
Let $1\leq j_1<\dots<j_r\leq L$ be integers and $\mathcal{C}(j_1,\dots,j_r)$ be the configuration
$(\tau_1,\dots,\tau_L)$ with $\tau_i=1$ if $i=j_k$ and $\tau_i=0$ otherwise.
We define the (non-normalized)
 weight of the word with $D(j_1),\dots, D(j_r)$ at positions $j_1,\dots,j_r$ as 
\begin{eqnarray}
 W_L(j_1,\dots,j_r) & = & Z_L \times \mathcal{S}(\mathcal{C}(j_1,\dots,j_r)) \\
 & = & \llangle W | \dots D(z_{j_1}) \dots D(z_{j_2}) \dots D(z_{j_r}) \dots |V\rrangle,
\end{eqnarray}
where the dots stand for $E(z_i)$ operators.

The same quantity can be calculated using  the explicit representation \eqref{eq:TASEP_ReprExpl}:
\begin{equation}
 \widetilde{W}_L(j_1,\dots,j_r)= \llangle\widetilde W| \dots \left(d +\frac{1}{z_{j_1}}\right) \dots \left(d +\frac{1}{z_{j_2}}\right)
 \dots \left(d +\frac{1}{z_{j_r}}\right) \dots |\widetilde V\rrangle.
\end{equation}
 The weight  $W_L(j_1,\dots,j_r)$ computed from  the relations
  \eqref{eq:TASEP_ZF_components} and  \eqref{eq:TASEP_GZ_components} is proportional to the weight 
 $\widetilde{W}_L(j_1,\dots,j_r)$ computed using the explicit representation 
 (thanks to the uniqueness of the steady-state guaranteed by the Perron-Frobenius theorem). Thus, we have 
\begin{equation} \label{eq:TASEP_equivalence_rep_explicite}
 W_L(j_1,\dots,j_r)=f(z_1,\dots,z_L)\widetilde{W}_L(j_1,\dots,j_r).
\end{equation}
 The multiplicative coefficient is obtained by comparing the weights of the empty
 configuration:
\begin{equation}
 f(z_1,\dots,z_L)=
 \frac{\llangle W| E(z_1)\dots E(z_L) |V\rrangle}
{ (a+z_1)\dots (a+z_L)\llangle W|V\rrangle} \, ; 
\end{equation}
$f(z_1,\dots,z_L)$ is symmetric under the permutation of the $z_i$ thanks to \eqref{eq:TASEP_ZF_components} but also under the transformation
$z_i \mapsto 1/z_i$ thanks to \eqref{eq:TASEP_GZ_components}.
{The group generated by these transformations is denoted by $BC_L$ in reference to the Weyl group
of the root system of the Lie algebra $sp(2L)$.}
 
The expression of  $\widetilde{W}_L(j_1,\dots,j_r)$ is given by
\begin{eqnarray*}
 \widetilde{W}_L(j_1,\dots,j_r) & = &  \frac{z_1\dots z_L}{z_{j_1} \dots z_{j_r}}
 \sum_{p=0}^{r}\,b^{p}\sum_{q_{r-p}=j_{r-p}}^{L}\frac1{z_{q_{r-p}}}\,\sum_{q_{r-p-1}=j_{r-p-1}}^{q_{r-p}-1}\frac1{z_{q_{r-p-1}}} \dots 
 \sum_{q_1=j_1}^{q_2-1}\frac{1}{z_{q_1}} \nonumber \\
 & & \times \prod_{l_0=1}^{j_1-1}\left(1+\frac{a}{z_{l_0}}\right)\prod_{l_1=q_1+1}^{j_2-1}\left(1+\frac{a}{z_{l_1}}\right)
 \dots \prod_{l_{r-p}=q_{r-p}+1}^{j_{r-p+1}-1}\left(1+\frac{a}{z_{l_{r-p}}}\right)
\end{eqnarray*}
By convention when $p=r$ in the first sum, there is no summation over the $q_i$ and the formula  reduces to
$b^r\prod_{l_0=1}^{j_1-1}\left(1+\frac{a}{z_{l_0}}\right)$. We also set $j_{r+1}=L+1$ in the last product when $p=0$. 
The proof of this formula is obtained by induction on the size $L$, using the identity
\begin{eqnarray*}
 \widetilde D(z_{j_r}) \widetilde E(z_{j_r+1})&=& \left( \frac{1}{z_{j_r}}+d \right)\left(z_{j_r+1}+e \right)=
 \frac{1}{z_{j_r}}\left(z_{j_r+1}+e \right)+z_{j_r+1}\left(\frac{1}{z_{j_r+1}}+d \right)\\
& =& \frac1{z_{j_r}} \widetilde E(z_{j_r+1}) + z_{j_r+1} \widetilde D(z_{j_r+1}) \, . 
\end{eqnarray*}

Another important step is to compute the normalization factor of the probability distribution.
 We define  $C(z)=E(z)+D(z)$ and $\widetilde{C}(z)=\widetilde{E}(z)+\widetilde{D}(z)$.
The normalization factors are thus given by $Z_L(z_1,\dots,z_L)=\llangle W|C(z_1)\dots C(z_L)|V\rrangle$, and
$\widetilde{Z}_L(z_1,\dots,z_L)=\llangle \widetilde W|\widetilde{C}(z_1)\dots \widetilde{C}(z_L)|\widetilde V\rrangle$ for the explicit representation.
Thanks to the property \eqref{eq:TASEP_equivalence_rep_explicite}, we have 
\begin{equation}
 Z_L(z_1,\dots,z_L)=f(z_1,\dots,z_L)\widetilde{Z}_L(z_1,\dots,z_L).
\end{equation}
$Z_L$ and $\widetilde{Z}_L$ are symmetric under $BC_L$.

Our goal is to get an analytic expression for the normalization factor.
{For this purpose, for any sequence of complex numbers  $\mathbf{u}=(u_1,u_2,...)$, we define the shifted product by
\begin{equation}
(z|\mathbf{u})^{k}=\begin{cases} (z-u_1)(z-u_2)\cdots(z-u_k)\,,\quad k>0, \\ 1 \mbox{ if } k=0,\end{cases}
\end{equation}}
We have the result (the proof can be found in \cite{CrampeMRV15inhomogeneous}):
\begin{equation}\label{eq:TASEP_schurCshifte}
 \widetilde{Z}_L(z_1,\dots,z_L)= 
 \frac{\det \left((z_j|\mathbf{v})^{L+2-i}-(1/z_j|\mathbf{v})^{L+2-i} \right)_{i,j}}{\det \left(z_j^{L+1-i}-(1/z_j)^{L+1-i} \right)_{i,j}}
\quad \mbox{with}\quad \mathbf{v}=(-a,-b,0,\dots,0).
\end{equation}
When $a,b=0$, we recognize {in the L.H.S. of  equation \eqref{eq:TASEP_schurCshifte}, the expression of the Schur polynomial of type C 
associated to the partition $1^L=(1,....,1)$. We remind that}
the Schur polynomial of type C associated with the partition $\lambda=(\lambda_1,\dots,\lambda_L)$ 
with $\lambda_1\geq \dots \lambda_L\geq 0$ is defined by:
\begin{equation}\label{eq:TASEP_sC}
\widetilde{Z}_L(z_1,\dots,z_L)\Big|_{a=b=0}=s^C_{\lambda}(z_1,\dots,z_L)= \frac{\det \left((z_j)^{L+1-i+\lambda_i}-(1/z_j)^{L+1-i+\lambda_i} \right)_{i,j}}{\det \left(z_j^{L+1-i}-(1/z_j)^{L+1-i} \right)_{i,j}}\;.
\end{equation}
As in the periodic case where the normalization factor is linked to the Schur polynomial 
of type $A$ and to the  representation
of the Lie algebra $sl(L)$ (see \eqref{eq:TASEP_rsl1} and \eqref{eq:TASEP_rsl2}), the normalization factor \eqref{eq:TASEP_sC} is given in terms of
the Schur polynomial of type $C$ and is associated to representation of the Lie algebra $sp(2L)$. These observations will 
 allow us to use results of the Lie algebra theory to take the homogeneous limit $z_i\rightarrow 1$.   

When $a,b$ are arbitrary, we need to use some generalizations of the Schur polynomial,  called {\it shifted (or factorial) Schur polynomials}.
A shifted  Schur polynomial of type A, is defined, for any sequence of complex numbers $\mathbf{u}=(u_1,u_2,...)$ \cite{KoikeT87}, as follows:
\begin{eqnarray}\label{eq:TASEP_sA2}
s^A_{\lambda}(z_1,\dots,z_L|\mathbf{u}) &=& \frac{\det \left((z_j|\mathbf{u})^{L-i+\lambda_i} \right)_{i,j}}{\det \left(z_j^{L-i} \right)_{i,j}}\;.
\end{eqnarray}
Similarly, we define the shifted (or factorial) Schur polynomial of type C as\footnote{{It is easy to see that 
$s^C_{\lambda}(z_1,\dots,z_L|\mathbf{u})$ is indeed a polynomial in the variables $z_1,\dots,z_L$.}}
\begin{equation}\label{eq:TASEP_sC2}
s^C_{\lambda}(z_1,\dots,z_L|\mathbf{u})= \frac{\det \left((z_j|\mathbf{u})^{L+1-i+\lambda_i}-(1/z_j|\mathbf{u})^{L+1-i+\lambda_i} \right)_{i,j}}
{\det \left(z_j^{L+1-i}-(1/z_j)^{L+1-i} \right)_{i,j}}\;.
\end{equation}
Thus,  the normalization factor \eqref{eq:TASEP_schurCshifte} is the shifted Schur polynomial
 of type C associated to the partition $1^L$ and to the sequence $\mathbf{v}=(-a,-b,0,0,\dots)$.
For this particular partition, the shifted Schur polynomial of type C can be expanded on the usual type C Schur polynomials as 
(the proof can be found in \cite{CrampeMRV15inhomogeneous})
\begin{equation}\label{eq:TASEP_ex}
\widetilde{Z}_L(z_1,\dots,z_L)=s^C_{1^L}(z_1,\dots,z_L|\mathbf{v})=\sum_{n=1}^{L+1} \frac{a^n-b^n}{a-b}\, s^C_{1^{L+1-n}}(z_1,\dots,z_L).
\end{equation}

Using the explicit representation, we can also determine the particle density at site $i$ from the stationary measure:
\begin{eqnarray*}
 \langle \tau_i(z_1,\dots,z_L)\rangle & = & 
 \left(b+\frac{1}{z_i}\right) \frac{\widetilde{Z}_{L-1}(z_1,\dots,z_{i-1},z_{i+1},\dots,z_L)}{\widetilde{Z}_L(z_1,\dots,z_L)} + \\
 (1-ab) & \times & \sum_{k=0}^{L-i-1} 
 \frac{\left. \widetilde{Z}_{i-1+k}(z_1,\dots,z_{i-1},z_{i+1},\dots,z_{i+k})\right|_{b=0}
 \left. \widetilde{Z}_{L-i-1-k}(z_{i+2+k},\dots,z_L)\right|_{a=0}}{\widetilde{Z}_L(z_1,\dots,z_L)}.
\end{eqnarray*}

In the homogeneous limit $z_i \rightarrow 1$,  $ s^C_{1^{L+1-n}}(1,\dots,1)$ is equal to the dimension of the $sp(2L)$ representation   
associated to the partition $1^{L+1-n}$:
{
\begin{equation}\label{eq:TASEP_dimC}
s^C_{1^{L+1-n}}(1,\dots,1)=\frac{n}{L+1}\,
\left( \begin{array}{c}  2L+2\\  L+1-n  \end{array} \right)
=dim\left(\pi^C(1^{L+1-n})\right).
\end{equation}
This expression is obtained through the general formula (see \cite{FrappatSS00} for instance):
\begin{equation}
dim\left(\pi^C(\lambda)\right)= \prod_{1\leq i<j\leq L} 
\left(\frac{\lambda_i-\lambda_j+j-i}{j-i}\ \frac{\lambda_i+\lambda_j+2L+2-j-i}{2L+2-j-i}\right)\ \prod_{i=1}^{ L} \frac{\lambda_i+L+1-i}{L+1-i}.
\end{equation}
} 
Then from \eqref{eq:TASEP_ex}, we get 
\begin{equation}\label{eq:TASEP_ZL.C}
\widetilde{Z}_L(1,1,...,1) = \frac{g_L(a)-g_L(b)}{a-b}
\mbox{ with } g_L(x)=\sum_{n=0}^{L} \frac{n+1}{L+1}\,
\left( \begin{array}{c}  2L+2\\  L-n  \end{array} \right)\,x^{n+1}
\end{equation}
in accordance with the results known for continuous time open TASEP \cite{DerridaEHP93}
\begin{equation} \label{eq:TASEP_ZL.Derrida}
Z_L =  \frac{h_L(\frac1{\alpha})-h_L(\frac1{\beta})}{\frac{1}{\alpha}-\frac{1}{\beta}}
\mbox{ with } h_L(x)=\sum_{p=1}^{L} \frac{p}{2L-p}
\left( \begin{array}{c}  2L-p\\  L  \end{array} \right)\,x^{p+1}.
\end{equation}
{ The equality between these two expressions is ensured by 
the identity
\begin{equation} \label{eq:TASEP_egaliteZL}
g_L(x-1)=h_L(x)-\frac1L\left(\begin{array}{c} 2L\\L-1\end{array}\right)
\end{equation}
} and recalling that $a=\frac1{\alpha}-1$ and $b=\frac1{\beta}-1$. 

In particular, when $a=b=0$ (i.e. $\alpha=\beta=1$),  $Z_L$
 is equal to the  dimension of the representation $1^L$,  which is  the Catalan number
$ C_{L+1}=\frac{1}{L+2}\left( \begin{array}{c}
                              2L+2\\
                              L+1
                             \end{array} \right).
$

The model considered  here has non-diagonal boundary matrices (see \eqref{eq:TASEP_K}) and cannot be completely diagonalized by
the usual integrability methods (like CBA or ABA). In the last few years, various specific techniques have been developed for solving such problems, 
based on the functional Bethe ansatz \cite{Nepomechie03,MurganN05,FrahmGSW10}, the coordinate Bethe ansatz \cite{CrampeR12,CrampeRS10,CrampeRS11,Simon09}, 
the separation of variables \cite{FrahmSW08,Niccoli12,FaldellaKN14}, the q-Onsager approach \cite{BaseilhacK07,BaseilhacB13} and 
the algebraic Bethe ansatz \cite{BelliardCR13,BelliardC13}. Recently, a generalization \cite{CaoYSW13,WenYCCY15,Nepomechie13} of the TQ relations expresses 
the eigenvalues of problems in terms of solutions of a new type of Bethe equations, called \textup{inhomogeneous Bethe equations.}

In what follows, we shall find an unexpected relation between the `partition function' $Z_L$ and the Baxter $Q$ operator 
appearing in the $TQ$-relations.

In \cite{Crampe15}, the Bethe equations corresponding to the eigenvalues and the eigenvectors of the transfer matrix \eqref{eq:inhomogeneous_transfer_matrix_open}
associated to the open TASEP have been computed using the modified algebraic Bethe ansatz.
In this context, the eigenvalue of the transfer matrix \eqref{eq:inhomogeneous_transfer_matrix_open} corresponding to the stationary state
$\Lambda(z)$ is given by
\begin{equation}\label{eq:TASEP_L}
 \Lambda(z)=z^{L+1}\frac{b+z}{bz+1} \prod_{k=1}^L \frac{zu_k-1}{u_k-z}
 -\frac{(z^2-1)}{(bz+1)}\prod_{j=1}^L\left[(z-z_j)(z-\frac{1}{z_j})\right] \prod_{k=1}^L \frac{u_k}{u_k-z}\;,
\end{equation}
where $u_k$ are called Bethe roots and are solutions of the following Bethe equations
\begin{equation}\label{eq:TASEP_beinh}
\prod_{p=1}^L\frac{(u_j-z_p)(u_jz_p-1)}{u_jz_p}=  (u_j+b)\  \prod_{\genfrac{}{}{0pt}{}{k=1}{k \neq j}}^{L}
 \left(u_j-\frac{1}{u_k}\right)\ ,\qquad\text{for $j=1,2,\dots,L$}\;.
\end{equation}
Let us stress that here the Bethe roots are the ones corresponding to the steady state. The corresponding Bethe equations \eqref{eq:TASEP_beinh}
were found in \cite{Crampe15} not to depend on the boundary parameter $a$.
The Bethe roots corresponding to other states obey 
different Bethe equations (see \cite{Crampe15}, or \cite{deGierE05} for the homogeneous case) which involve parameter $a$.
We  now introduce the following function:
\begin{equation}\label{eq:TASEP_Q}
 Q(z)=\prod_{k=1}^L\left(\frac{1}{u_k}-\frac{1}{z}\right)\;.
\end{equation}
This function (up to a coefficient $z^L$) is linked to the polynomial $Q$ of Baxter \cite{Baxter82}: its zeros are the Bethe roots.
The eigenvalue \eqref{eq:TASEP_L} can be written in terms of the Q function as follows
\begin{equation}\label{eq:TASEP_LQ}
 \Lambda(z)=\frac{z(b+z)Q(1/z)}{(bz+1)Q(z)}
 -\frac{(z^2-1)}{(bz+1)Q(z)}\prod_{j=1}^L\left[(z-z_j)(\frac{1}{zz_j}-1)\right]\;.
\end{equation}
For the steady-state  eigenvector, we know  that $\Lambda(z)=1$. Hence $Q(z)$ satisfies
\begin{equation}\label{eq:TASEP_TQ}
z(z+b)Q(1/z)-(1+bz)Q(z)=(z^2-1) \prod_{j=1}^L\left[(z-z_j)(\frac{1}{zz_j}-1)\right]\;.
\end{equation}
This equation, called TQ relation,  allows us to compute explicitly the function Q:
 for a given $L$, it can be shown that equation \eqref{eq:TASEP_TQ} has a unique solution of the form \eqref{eq:TASEP_Q}.

For the model studied here, the function Q and the normalization factor $\widetilde Z(z_1,\dots,z_L)$ are closely related. Namely, we get
\begin{eqnarray}
 Q(z)= \widetilde Z(z_1,\dots,z_L)\big|_{a\rightarrow -1/z}&=&\sum_{n=1}^{L+1} \frac{(-1/z)^n-b^n}{-1/z-b}\, s^C_{1^{L+1-n}}(z_1,\dots,z_L)
 \label{eq:TASEP_QZ1}\\
 &=&\sum_{p=0}^L \left(-\frac{1}{z}\right)^p\  \sum_{n=0}^{L-p} b^n\ \  s^C_{1^{L-n-p}}(z_1,\dots,z_L)\;.\label{eq:TASEP_QZ2}
\end{eqnarray}
where we have used the explicit form of $\widetilde Z$ given in \eqref{eq:TASEP_ex}.
To prove this result, we remark that $Q(x)$ given by \eqref{eq:TASEP_QZ2} has the  form \eqref{eq:TASEP_Q}
and we show that it satisfies the TQ relation \eqref{eq:TASEP_TQ}. We  have 
\begin{eqnarray}
 z(z+b)Q(1/z)-(1+bz)Q(z)&=&-z\sum_{n=1}^{L+1}\left((-z)^n-(-1/z)^n\right)s^C_{1^{L+1-n}}(z_1,\dots,z_L) \nonumber \\
 &=&(z^2-1)s^C_{1^{L}}(z_1,\dots,z_L|z,1/z,0,\dots,0)\label{eq:TASEP_TQp}\;.
\end{eqnarray}
Finally, we readily check  that $s^C_{1^{L}}(z_1,\dots,z_L|z,1/z,0,\dots,0)$ vanishes at the points $z=z_j$ and $z=1/z_j$
which allows us to conclude that \eqref{eq:TASEP_TQp} is equal to the L.H.S. of \eqref{eq:TASEP_TQ}.

 Relation \eqref{eq:TASEP_QZ1} means that the Bethe roots (zeros of the function $Q$) are linked to the zeros
of the steady-state normalization factor in the complex plane of the transition rate $a$. These zeros 
appeared previously (for the homogeneous case) in \cite{BlytheE02,BlytheE03,BenaDL05} as Lee-Yang zeros and allows the generalization of the Lee-Yang theory 
for the phase transition of non equilibrium system. Therefore, relation \eqref{eq:TASEP_QZ1} expresses
 an unexpected relation between two objects arising from very different contexts. 
 
\section{Application to integrable models: examples} \label{sec:examples_MA}

The goal of this section is to provide several examples of physical systems that can be investigated using the framework presented in this chapter.
These systems, comprising reaction diffusion processes or involving several species of particles, will hopefully by their diversity illustrate
all the different notions introduced.
The study conducted below will mainly concern the matrix product construction of the steady state, taking advantage of the integrability
of the models as stressed previously. A particular attention will also be put on the exact computation of physical quantities, in order to stress 
the efficiency and usefulness of the matrix ansatz for physical applications.

\subsection{A diffusive model with evaporation and condensation} \label{subsec:DiSSEP}

We consider in this subsection a one parameter generalization of the SSEP. The model involves a single species of particle that can diffuse
on a one dimensional lattice coupled with two reservoirs at its extremities. In the bulk the particles can jump with equal probability rate 
to the left or right neighboring site provided that it is empty (exclusion constraint). 
The injection and extraction rates at the boundaries are asymmetric to model the coupling with the particle reservoirs.
In addition to the usual dynamics of the SSEP, we allow the annihilation and creation of particle pairs on two adjacent sites in the bulk,
with equal probability rates
(provided again that the two target sites are empty when performing a pair creation, to respect the exclusion constraint). 
This pair creation and annihilation is said to be {\it dissipative} because it does not conserve the number of particles in the bulk.
This explains the name of the model: Dissipative Symmetric Simple Exclusion Process (DiSSEP).
This model has been exactly solved on the periodic lattice in \cite{GrynbergNS94}. The model with open boundaries has been defined in \cite{CrampeRV14} and studied 
in details in \cite{CrampeRRV16}. We present here the main results, following the lines of \cite{CrampeRRV16}.

\subsubsection{Presentation of the model}

\begin{figure}[htb]
\begin{center}
 \begin{tikzpicture}[scale=0.7]
\draw (-2,0) -- (12,0) ;
\foreach \i in {-2,-1,...,12}
{\draw (\i,0) -- (\i,0.4) ;}
\draw[->,thick] (-2.4,0.9) arc (180:0:0.4) ; \node at (-2.,1.8) [] {$\alpha$};
\draw[->,thick] (-1.6,-0.1) arc (0:-180:0.4) ; \node at (-2.,-0.8) [] {$\gamma$};
\draw  (1.5,0.5) circle (0.3) [fill,circle] {};
\draw  (4.5,0.5) circle (0.3) [fill,circle] {};
\draw  (5.5,0.5) circle (0.3) [fill,circle] {};
\draw  (8.5,3.1) circle (0.3) [fill,circle] {};
\draw  (9.5,3.1) circle (0.3) [fill,circle] {};
\draw[->,thick] (1.4,1) arc (0:180:0.4); \node at (1.,1.8) [] {$1$};
\draw[->,thick] (1.6,1) arc (180:0:0.4); \node at (2.,1.8) [] {$1$};
\node at (5,1.1) [rotate=-90] {$\Big{\{}$};
\draw[->,thick] (5,1.3) -- (5,2.8); \node at (5.4,2.2) [] {$\lambda^2$};
\node at (9,2.5) [rotate=90] {$\Big{\{}$};
\draw[->,thick] (9,2.3) -- (9,0.8); \node at (9.4,1.6) [] {$\lambda^2$};
\draw[->,thick] (11.6,1) arc (180:0:0.4) ; \node at (12.,1.8) [] {$\beta$};
\draw[->,thick] (12.4,-0.1) arc (0:-180:0.4) ; \node at (12.,-0.8) [] {$\delta$};
 \end{tikzpicture}
 \end{center}
 \caption{Dynamical rules of the DiSSEP.}
 \label{fig:DiSSEP}
\end{figure}
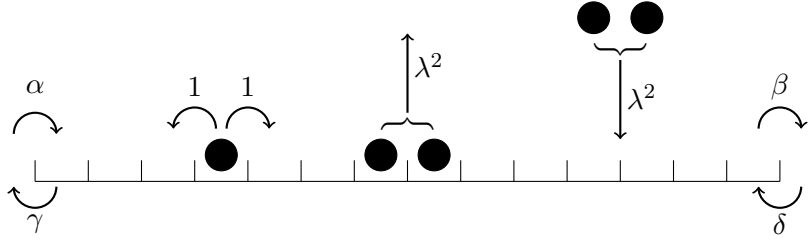

The precise stochastic dynamics is given as follows.
 During an infinitesimal time $dt$, a particle in the bulk can jump to the left or to the right neighboring site with probability $dt$ 
 if it is unoccupied. A pair of neighbor particles can also be annihilated with probability $\lambda^2 \times dt$ and be created on unoccupied 
 neighbor sites with probability $\lambda^2 \times dt$ (see figure \ref{fig:DiSSEP}).
 At the two extremities of the lattice the dynamics is modified to take into account 
 the interaction with the reservoirs: at the first site (connected with the left reservoir), during time $dt$, a particle is injected with 
 probability $\alpha \times dt$ if the site is empty and extracted with probability $\gamma \times dt$ if it is occupied. 
 The dynamics is similar at last site (connected with the right reservoir) with injection rate $\delta$ and extraction rate $\beta$.
 The dynamical rules can be summarized in the following table where $0$ stands for vacancy and $1$ stands for a particle.
 The transition rates between the configurations are written above the arrows.

 \begin{equation} \label{eq:DiSSEP_rules}
 \begin{array}{|c |c| c| }
 \hline \text{Left} & \text{Bulk} & \text{Right} \\
 \hline
 0\, \xrightarrow{\ \alpha \ }\, 1&  01\, \overset{\ 1\ }{\longleftrightarrow} \,10&1\, \xrightarrow{\ \beta \ } \,0\\
 1\, \xrightarrow{\ \gamma \ }\, 0&00\, \overset{\ \lambda^2 \ }{\longleftrightarrow}\, 11&0\, \xrightarrow{\ \delta\ }\, 1\\ \hline
 \end{array}
 \end{equation}

 The Markov matrix encoding the stochastic dynamics of the process is given as usual as the sum of operators acting
 locally on the lattice:
\begin{equation} \label{eq:DiSSEP_Markov_matrix_decomposition}
 M=B_1+\sum_{k=1}^{L-1}m_{k,k+1}+\overline{B}_L,
\end{equation}
where the local jump operators $B$, $\overline{B}$ and $m$ are given by
\begin{equation} \label{eq:DiSSEP_local_jump_operators}
B =\left( \begin {array}{cc} 
-\alpha&\gamma\\ 
\alpha&-\gamma
\end {array} \right)\quad ,\qquad  m=\left( \begin {array}{cccc} 
-\lambda^2&0&0&\lambda^2\\ 
0&-1&1&0\\
0&1&-1&0\\
\lambda^2&0&0&-\lambda^2
\end {array} \right)
\quad\text{,}\qquad
\overline{B} =\left( \begin {array}{cc} 
-\delta&\beta\\ 
\delta&-\beta
\end {array} \right)\;.
\end{equation}
We recall that we already encountered these local operators in the examples \eqref{eq:DiSSEP_m} and \eqref{eq:ASEP_B_Bb} in chapter \ref{chap:two}.

We choose the coefficient  of condensation and evaporation to be $\lambda^2$ and not $\lambda$ for later convenience. 
Let us remark that the SSEP is recovered when the creation/annihilation rate $\lambda^2$ vanishes. 
The limit $\lambda^2\rightarrow\infty$ provides a model with only condensation and evaporation.

The system is driven out of equilibrium 
by the boundaries. As shown in \eqref{eq:DiSSEP_mean_lattice_current} and \eqref{eq:DiSSEP_mean_condensation_current}, 
there are particle currents in the stationary state for generic 
boundary rates $\alpha$, $\beta$, $\gamma$ and $\delta$. We will see below \eqref{eq:DiSSEP_density_thermo} that these choices of rates 
describe particle reservoirs with densities
\begin{equation} \label{densities}
 \rho_{l}=\frac{\alpha}{\alpha+\gamma}, \quad \text{and} \quad \rho_r=\frac{\delta}{\beta+\delta}.
\end{equation}

\begin{remark}
 The system will converge in the long time limit to a thermodynamic equilibrium if and only if the densities of the particle reservoirs at the 
 boundaries are both equal to $1/2$. This translates to $\alpha=\gamma$ and $\beta=\delta$.
 The detailed balance condition is indeed only satisfied in this case. This can be interpreted by the fact that the system is in some way coupled
 to a particle reservoir in the bulk because of the evaporation and condensation. This reservoir has a fixed density $1/2$ because the evaporation
 and condensation processes happen with the same probability rate. The thermodynamic equilibrium can thus occurs if and only if the two
 other particle reservoirs at the boundaries have the same density $1/2$.
\end{remark}

\begin{remark}
The model displays several symmetries that are listed below:
\begin{itemize}
\item The evaporation/condensation probability rate is chosen to be $\lambda^2$, hence all the results should be 
invariant under the transformation $\lambda\rightarrow -\lambda$. 
\item The left/right symmetry of the chain is given by the transformations $\alpha\leftrightarrow\delta$, $\gamma\leftrightarrow\beta$ 
and a change of numbering of the sites $i\to L+1-i$.
\item The vacancy-particle symmetry translates into $\alpha\leftrightarrow\gamma$ and $\delta\leftrightarrow\beta$.
\end{itemize}
\end{remark}

The DiSSEP defines an integrable model. We indeed saw in chapter \ref{chap:two} that there exists an $R$ matrix \eqref{eq:DiSSEP_R} satisfying 
the Yang Baxter equation and such that $m=2\lambda \check R'(1)$.
There exist also boundary matrices $K$ and $\overline{K}$, introduced in \eqref{eq:DiSSEP_K} and \eqref{eq:DiSSEP_Kb} respectively,
satisfying the reflection equation and such that $B=\lambda K'(1)$ and $\overline{B}=-\lambda \overline{K}'(1)$. 
Following the lines of chapter \ref{chap:two}, we can construct a transfer matrix
\begin{equation}
 t(z)= tr_0(\widetilde K_0(z) R_{0,L}(z)\dots R_{0,1} K_0(z) R_{1,0}(z) \dots R_{L,0}(z))
\end{equation}
where
\begin{equation}
 \widetilde K(z)= tr_0\left(\overline{K}_0\left(\frac{1}{z}\right)\left(\left(R_{0,1}(z^2)^{t_1}\right)^{-1}\right)^{t_1}P_{0,1}\right).
\end{equation}
This transfer matrix satisfies the key relation $[t(z),t(z')]=0$ and is directly related to the Markov matrix $M$ through
\begin{equation}
 M=\lambda t'(1).
\end{equation}

\subsubsection{Matrix ansatz}

This subsection is devoted to the construction of the steady state of the model in a matrix product form. Following the general procedure 
developed in this chapter for integrable stochastic processes, we start by introducing a vector $\mathbf{A}(z)$ with algebraic entries and 
a finite expansion with respect to the spectral parameter $z$:
\begin{equation}
 \mathbf{A}(z)=\begin{pmatrix}
                G_1 z+G_2+\frac{G_3}{z} \\
                -G_1 z+G_2-\frac{G_3}{z}
               \end{pmatrix}.
\end{equation}

To determine the commutation relations of the generators $G_i$'s, we write the ZF relation
\begin{eqnarray*}
& & \check R\left(\frac{z_1}{z_2}\right) \mathbf{A}(z_1) \otimes \mathbf{A}(z_2) = \mathbf{A}(z_2) \otimes \mathbf{A}(z_1) \quad \Leftrightarrow \quad \\
& & \frac{z_1^2-z_2^2}{z_1z_2}
\begin{pmatrix}
\frac{\Big(z_1z_2+1\Big)\Big((\lambda+1)G_2G_1+(\lambda-1)G_1G_2\Big)+\Big(z_2(\lambda-1)-z_1(\lambda+1)\Big)\Big(G_1G_3-G_3G_1\Big)}
{z_2(\lambda-1)-z_1(\lambda+1)} \\
\frac{\Big(z_1z_2-1\Big)\Big((\lambda+1)G_2G_1+(\lambda-1)G_1G_2\Big)-\Big(z_1(\lambda+1)+z_2(\lambda-1)\Big)\Big(G_1G_3-G_3G_1\Big)}
{z_1(\lambda+1)+z_2(\lambda-1)} \\
\frac{\Big(z_1z_2-1\Big)\Big((\lambda+1)G_2G_1+(\lambda-1)G_1G_2\Big)+\Big(z_1(\lambda+1)+z_2(\lambda-1)\Big)\Big(G_1G_3-G_3G_1\Big)}
{-z_1(\lambda+1)-z_2(\lambda-1)} \\
\frac{\Big(z_1z_2+1\Big)\Big((\lambda+1)G_2G_1+(\lambda-1)G_1G_2\Big)+\Big(z_1(\lambda+1)-z_2(\lambda-1)\Big)\Big(G_1G_3-G_3G_1\Big)}
{z_1(\lambda+1)-z_2(\lambda-1)}
\end{pmatrix} = 0.
\end{eqnarray*}
This can be concisely and equivalently rewritten with the three following relations
\begin{equation} \label{eq:DiSSEP_commutation relations_G}
 [G_1,G_3]=0, \quad G_2G_1=\phi \ G_1G_2, \quad \text{and} \quad G_3G_2=\phi \ G_2G_3, \quad  \text{with} \quad \phi=\frac{1-\lambda}{1+\lambda}.
\end{equation}
The relations on the boundaries are obtained by writing the GZ relations. On the left boundary vector we have
\begin{eqnarray*}
& & \llangle W|K(z)\mathbf{A}\left(\frac{1}{z}\right)=\llangle W| \mathbf{A}(z) \quad \Leftrightarrow \quad \\
& &  \llangle W| \begin{pmatrix}
  \frac{\Big(1-z^4\Big) \Big((\alpha+\gamma+2\lambda)G_1+(\alpha+\gamma-2\lambda)G_3+(\alpha-\gamma)G_2 \Big)}
  {z\Big((\alpha+\gamma+2\lambda)z^2-\alpha-\gamma+2\lambda \Big)} \\
   \frac{\Big(z^4-1\Big) \Big((\alpha+\gamma+2\lambda)G_1+(\alpha+\gamma-2\lambda)G_3+(\alpha-\gamma)G_2 \Big)}
  {z\Big((\alpha+\gamma+2\lambda)z^2-\alpha-\gamma+2\lambda \Big)}
 \end{pmatrix} = 0.
\end{eqnarray*}
On the right boundary the GZ relation reads
\begin{eqnarray*}
& & \overline{K}(z)\mathbf{A}\left(\frac{1}{z}\right)|V\rrangle = \mathbf{A}(z)|V\rrangle \quad \Leftrightarrow \quad \\
& & \begin{pmatrix}
  \frac{\Big(1-z^4\Big) \Big((\beta+\delta+2\lambda)G_3+(\beta+\delta-2\lambda)G_1+(\delta-\beta)G_2 \Big)}
  {z\Big((\beta+\delta-2\lambda)z^2-\beta-\delta-2\lambda \Big)} \\
   \frac{\Big(z^4-1\Big) \Big((\beta+\delta+2\lambda)G_3+(\beta+\delta-2\lambda)G_1+(\delta-\beta)G_2 \Big)}
  {z\Big((\beta+\delta-2\lambda)z^2-\beta-\delta-2\lambda \Big)}
 \end{pmatrix} |V\rrangle = 0.
\end{eqnarray*}
These two relations are equivalent to
\begin{equation} \label{eq:DiSSEP_boundaries_relations_G}
 \left\{ \begin{aligned}
          & \llangle W |\big( G_1-c\;G_2-a\;G_3 \big)=0\,, \\
          &\big( G_3-b\;G_1-d\;G_2 \big) | V \rrangle =0
         \end{aligned}
 \right.
 \mbox{ with }
 \left\{ \begin{aligned}
& a=\frac{2\lambda-\alpha-\gamma}{2\lambda+\alpha+\gamma}\,,\quad 
& c=\frac{\gamma-\alpha}{2\lambda+\alpha+\gamma}\,,
\\
& b=\frac{2\lambda-\delta-\beta}{2\lambda+\delta+\beta}\,, 
& d=\frac{\beta-\delta}{2\lambda+\delta+\beta}\,.
 \end{aligned}
 \right.
\end{equation}
All this construction allows us to express the steady state as the matrix product state
\begin{equation}
 \steady=\frac{1}{Z_L}\llangle W|\mathbf{X}\otimes \dots \otimes \mathbf{X}|V\rrangle
\end{equation}
where
\begin{equation}
 \mathbf{X}=\begin{pmatrix}
              E \\ D
             \end{pmatrix}:= \mathbf{A}(1).
\end{equation}
We define also the auxiliary vector
\begin{equation}
 \overline{\mathbf{X}}=\begin{pmatrix}
              -H \\ H
             \end{pmatrix}:= 2\lambda \mathbf{A}'(1).
\end{equation}
We thus have the explicit relations 
\begin{equation} \label{eq:DiSSEP_MA_change_generators_basis}
 \left\{ \begin{aligned}
          & E=G_1+G_2+G_3, \\
          & D=G_2-G_1-G_3, \\
          & H=2\lambda(G_3-G_1).
         \end{aligned}
 \right.
\end{equation}
The commutation relations on the $G_i$'s \eqref{eq:DiSSEP_commutation relations_G} translates into commutation relations on the 
$E$, $D$, $H$ generators:
\begin{equation} \label{eq:DiSSEP_commutation_relations_DE}
 [D,E]=EH+HD, \quad \text{and} \quad [H,E]=[H,D]=\lambda^2(D^2-E^2).
\end{equation}
These relations are equivalent to the very useful telescopic relation
\begin{equation} \label{eq:DiSSEP_telescopic_bulk_DE}
w \left( \begin{array}{c}
         E \\
         D
        \end{array} \right) \otimes
        \left( \begin{array}{c}
         E \\
         D
               \end{array} \right) = 
        \left( \begin{array}{c}
         E \\
         D
               \end{array} \right) \otimes 
        \left( \begin{array}{c}
         -H \\
         H
        \end{array} \right)-
        \left( \begin{array}{c}
         -H \\
         H
               \end{array} \right) \otimes 
        \left( \begin{array}{c}
         E \\
         D
        \end{array} \right)\;.
\end{equation}
Notice here that, in contrast with the SSEP case (see \cite{Derrida07}) where $H$ is a scalar, the commutation relations
between $H$ and $E, \ D$ are not trivial.
The action of the generators $E$, $D$ and $H$ on the boundary vectors $\llangle W|$ and $|V\rrangle$ are obtained
by direct translation of the action of the generators $G_i$'s \eqref{eq:DiSSEP_boundaries_relations_G}:
\begin{equation} \label{eq:DiSSEP_boundaries_relations_DE}
 \llangle W|\left(\alpha E-\gamma D \right)=\llangle W|H, \quad \text{and} \quad \left(\delta E-\beta D\right)|V\rrangle = -H|V\rrangle.
\end{equation}
It is equivalent to
\begin{equation} \label{eq:DiSSEP_telescopic_boundaries_DE}
 \llangle W|B \left( \begin{array}{c}
         E \\
         D
               \end{array} \right) = 
 \llangle W|\left( \begin{array}{c}
         -H \\
         H
        \end{array} \right), \quad \text{and} \quad 
\overline{B} \left( \begin{array}{c}
              E \\
              D
                    \end{array} \right)|V\rrangle = 
 -\left( \begin{array}{c}
         -H \\
         H
        \end{array} \right)|V\rrangle.       
\end{equation}
Note that the relations satisfied by the generators $E$, $D$ and $H$ are strictly equivalent to the relations verified by the $G_i$'s,
they just correspond to the change of basis \eqref{eq:DiSSEP_MA_change_generators_basis}.
We gave them in order to make the connection with the usual presentation of the matrix ansatz and to make the comparison with the SSEP 
easier. Nevertheless we will not use the $E$, $D$, $H$ basis in the following and stick with the $G_i$'s basis because the computations
will be more efficient.

The explicit representation of the boundary vectors $\llangle W|$ and $|V\rrangle$ and of the generators $G_i$'s (or equivalently of 
the generators $E$, $D$ and $H$) can be found in \cite{CrampeRV14}. We do not present it here because it does not simplify the computation
of the physical quantities (that can be efficiently done using the algebraic relations \eqref{eq:DiSSEP_commutation relations_G} and 
\eqref{eq:DiSSEP_boundaries_relations_G}, see next subsection) and it does not shed any new light on the stationary distribution. 
Nevertheless the existence of such explicit representation is necessary to prove the existence of the generators and boundary vectors.

\begin{remark}
 As previously seen in this chapter (see subsection \ref{subsec:inhomogeneous_ground_state} for details), we can define the inhomogeneous ground state
 \begin{equation}
  \ket{\cS(z_1,\dots,z_L)}=\frac{\llangle W|\mathbf{A}(z_1)\otimes \dots \otimes \mathbf{A}(z_L)|V\rrangle}{Z_L(z_1,\dots,z_L)},
 \end{equation}
 where the normalization 
 \begin{equation}
  Z_L(z_1,\dots,z_L)=\llangle W|C(z_1)\dots C(z_2)|V\rrangle, \quad \mbox{with} \quad C(z)=A_0(z)+A_1(z)+A_2(z).
 \end{equation}
This inhomogeneous deformation of the steady state (we recall that $\ket{\cS(1,\dots,1)}$ is the steady state) is an eigenvector of the 
inhomogeneous transfer matrix associated to the model
\begin{equation}
 t(z|z_1,\dots,z_L)\ket{\cS(z_1,\dots,z_L)}=\lambda(z|z_1,\dots,z_L)\ket{\cS(z_1,\dots,z_L)},
\end{equation}
where the eigenvalue is equal to
\begin{equation}
 \lambda(z|z_1,\dots,z_L)=1+\phi^{2L}\frac{(z^4-1)(az^2+\phi^2)(bz^2+\phi^2)}{(z^4-\phi^4)(z^2+a)(z^2+b)}
 \prod_{i=1}^L \frac{((zz_i)^2-1)(z^2-z_i^2)}{((zz_i)^2-\phi^2)(z^2-(\phi z_i)^2)}.
\end{equation}
This can be proven (see subsection \ref{subsec:inhomogeneous_ground_state} for details) using degree consideration and the symmetry property
on the transfer matrix
\begin{equation}
 t(z|z_1,\dots,z_L)=\big(\lambda(z|z_1,\dots,z_L)-1\big)t(\frac{\phi}{z}|z_1,\dots,z_L).
\end{equation}
\end{remark}

\subsubsection{Computation of physical quantities}

We now focus on the computation of physical quantities in the stationary state. We will see that the matrix ansatz algebra, introduced
to express the steady state, reveals to be very efficient in computing correlation functions and mean values of the currents. 
Before studying the general model, we focus on the case $\lambda=1$ ($\phi=0$), 
where the calculations simplify drastically: it corresponds to the free fermion point of the model we introduced.

For $\lambda=1$, all the eigenvalues and the eigenvectors can be computed easily as shown in the following proposition.
\begin{proposition}
For $L\geq3$, the $2^L$ eigenvectors are characterized by the 
set $\boldsymbol{\epsilon}=(\epsilon_1,\epsilon_2,\dots,\epsilon_L)$ with $\epsilon_i=\pm 1$ and are given by 
\begin{equation}
 \Omega(\boldsymbol{\epsilon})=v(\epsilon_1,\epsilon_2,\alpha,\gamma)\otimes\begin{pmatrix}
                                                   1\\ \epsilon_2
                                                  \end{pmatrix}\otimes\begin{pmatrix}
                                                   1\\ \epsilon_3
                                                  \end{pmatrix}\otimes\cdots\otimes\begin{pmatrix}
                                                   1\\ \epsilon_{L-1}
                                                  \end{pmatrix}\otimes v(\epsilon_L,\epsilon_{L-1},\delta,\beta)
\end{equation}
where $v(\epsilon,\epsilon',\mu,\nu)=\begin{pmatrix}  \epsilon'+\nu\\  1+\mu+f(\epsilon,\epsilon',\mu+\nu) \end{pmatrix}$
and $f(\epsilon,\epsilon',\tau)=\epsilon\epsilon'-1-\frac{\tau}{2}(1-\epsilon)$. The corresponding eigenvalues are
\begin{equation}
 \Lambda(\boldsymbol{\epsilon})=f(\epsilon_1,\epsilon_2,\alpha+\gamma)
 + \sum_{j=2}^{L-2}( \epsilon_j\epsilon_{j+1}-1)+f(\epsilon_L,\epsilon_{L-1},\delta+\beta)\;.
\end{equation}
\end{proposition}
Let us remark that the ASEP on a ring with Langmuir kinetics has been treated similarly in \cite{SatoN16}. 

\proof
This can be checked by a direct computation.
\finproof

\begin{corollary}
From the previous results, we deduce that the stationary state is 
\begin{equation}
 \Omega(+,+,\dots,+)= \frac{1}{Z_L}\ \   \begin{pmatrix}  1+\gamma\\ 1+\alpha
                                                  \end{pmatrix}\otimes\begin{pmatrix}
                                                   1\\ 1
                                                  \end{pmatrix}^{\otimes L-2}\otimes\begin{pmatrix}
                                                   1+\beta\\ 1+\delta
                                                  \end{pmatrix} \end{equation}
where $Z_L=2^{L-2}(2+\alpha+\gamma)(2+\beta+\delta)$ is the normalisation such that the entries be probabilities.
\end{corollary}
\proof
We have indeed $\Lambda(+,+,\dots,+)=0.$
\finproof

From this stationary state, we can compute the mean value of the injected current by the left reservoir (resp. by the right reservoir)
\begin{equation}\label{eq:DiSSEP_mean_injected_current}
 \langle j_{\text{left}}\rangle=\frac{\alpha-\gamma}{2+\alpha+\gamma} \qquad(\text{resp. }  
 \langle j_{\text{right}}\rangle =\frac{\delta-\beta}{2+\delta+\beta})\;.
\end{equation}
We see that the current 
has the sign of $\alpha-\gamma$ (resp. $\delta-\beta$). As expected, it goes to the left when extraction is promoted, and to the right when injection is proeminent. It vanishes for $\alpha=\gamma$. The lattice current in the bulk vanishes.

We can also compute easily the first excited state whose eigenvalue provides the relaxation rate. Indeed, the greatest non vanishing eigenvalue is
\begin{equation} \label{eq:DiSSEP_gap_lambda1}
 \begin{cases}
 G=-4 & \mbox{if } \alpha+\gamma>2 \mbox{ and } \beta+\delta>2\\
 G=-2-\beta-\delta& \mbox{if } \alpha+\gamma>2\mbox{ and } \beta+\delta<2\\
 G=-2-\alpha-\gamma& \mbox{if } \alpha+\gamma<2 \mbox{ and } \beta+\delta>2\\
 G=-\alpha-\gamma-\beta-\delta& \mbox{if } \alpha+\gamma<2\mbox{ and }\beta+\delta<2
 \end{cases}
\end{equation}
These results shall be generalized in the thermodynamic limit in chapter \ref{chap:five} for any $\lambda$ using the Bethe equations. 
The general result displayed on figure \ref{fig:g3} matches the above values of the gap for $\lambda=1$ ($\phi=0$).

For this particular choice of $\lambda$, it is also possible to get the generating function of the cumulants of the current 
entering in the system from the left reservoir (the same result is also obtained by symmetry for the right reservoir). 
For general $\lambda$, one obtains the variance in \eqref{eq:DiSSEP_variance_lattice_current}. 
It is well established (see chapter \ref{chap:one} for instance) that this generating function is the greatest eigenvalue
of the following deformed Markov matrix
\begin{equation} \label{eq:DiSSEP_deformed_Markov_matrix_lambda1}
 M=B_1(\mu)+\sum_{k=1}^{L-1}m_{k,k+1}+\overline{B}_L,
\end{equation}
where the local jump operator $B(\mu)$ is deformed as follows
\begin{equation} \label{eq:DiSSEP_localBs}
B(\mu) =\left( \begin {array}{cc} 
-\alpha&\gamma e^{-\mu}\\ 
\alpha e^{\mu}&-\gamma
\end {array} \right)\;.
\end{equation}
One can show by direct computation that the greatest eigenvalue is given by\footnote{We recover that $E(0)=0$ in agreement with
the fact that at $\mu=0$ the greatest eigenvalue is vanishing because of the Markovian property. Note that $E(\mu)$ depends only on 
the boundary rates of the left reservoir $\alpha$ and $\gamma$, and not on the boundary rates of the right reservoir $\beta$ and $\delta$.}
\begin{equation}\label{eq:DiSSEP_generating_function}
 E(\mu)=-\frac{2+\alpha+\gamma}{2}+\frac{1}{2}\sqrt{4+4\alpha e^{\mu}+4\gamma e^{-\mu}+(\alpha+\gamma)^2}
\end{equation}
with the eigenvector 
\begin{equation}
 \Omega(\mu)=     \begin{pmatrix}  1+\gamma e^{-\mu}\\ 1+\alpha +E(\mu)
                                                  \end{pmatrix}\otimes\begin{pmatrix}
                                                   1\\ 1
                                                  \end{pmatrix}^{\otimes L-2}\otimes\begin{pmatrix}
                                                   1+\beta\\ 1+\delta
                                                  \end{pmatrix}\;. \end{equation}

\begin{figure}[htbp]
\begin{center}
\includegraphics[width=80mm,height=80mm]{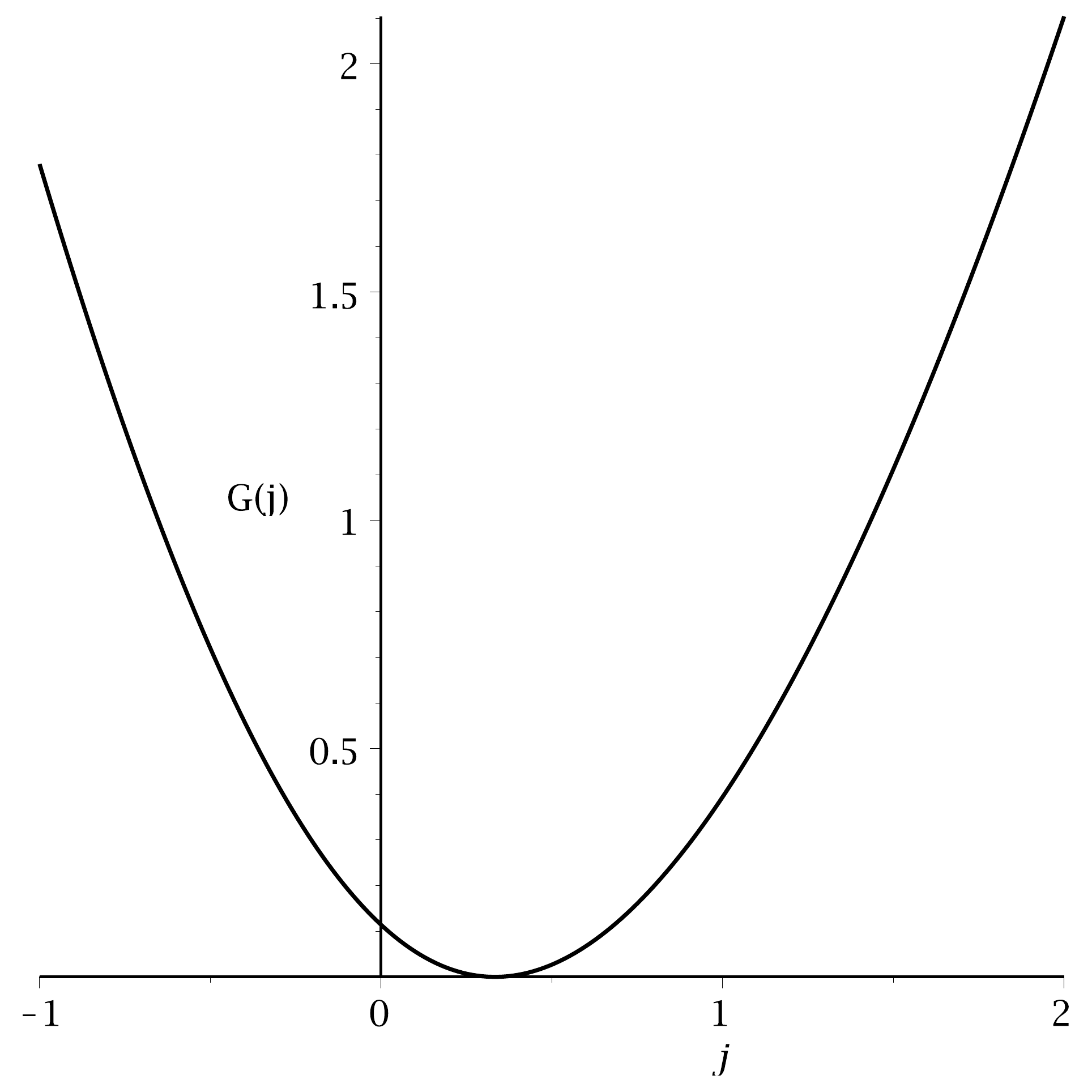}
\end{center}
\caption{Example of large deviation function $G(j)$ (on the plot $\alpha=2,\gamma=0.5$). \label{fig:DiSSEP_large_deviation}}
\end{figure}

The rate-function $G(j)$ associated to the current is the Legendre transformation of this generating function of the cumulants:
 \begin{equation} 
G(j)=\mu^* j-E(\mu^*)\mbox{ , }\frac{d}{d\mu}E(\mu)\Big|_{\mu=\mu^*}=j 
\end{equation}
Its explicit form can be extracted from the expression of $E(\mu)$ and is given in the following proposition
\begin{proposition}
\begin{equation}\label{eq:DiSSEP_large_deviation}
 G(j)=1+\frac{\alpha+\gamma}{2}-\sqrt{1+\Delta(j)+\left(\frac{\alpha+\gamma}{2}\right)^2}
 +j\ln\left(\frac{\Delta(j)}{2\alpha}+\frac{j}{\alpha}\sqrt{1+\Delta(j)+\left(\frac{\alpha+\gamma}{2}\right)^2} \right)
\end{equation}
where 
\begin{equation}
 \Delta(j)=2j^2+\sqrt{4(\alpha\gamma+j^2+j^4)+j^2(\alpha+\gamma)^2}\;.
\end{equation}
\end{proposition}
Let us stress that \eqref{eq:DiSSEP_large_deviation} represents an exact result on the large deviation function of the current on the left boundary.
The function $G(j)$ is convex and vanishes when $j$ is equal to the mean value of the current on the left boundary given by 
\eqref{eq:DiSSEP_mean_injected_current} as expected, 
see figure \ref{fig:DiSSEP_large_deviation}. Note that it is not Gaussian.

We now come back to the general case $\lambda \neq 1$ ($\phi \neq 0$). We start by stating a formula, giving the value of a general 
word in the generators $G_i$'s, that will be of prime utility to compute correlation functions and mean values of particle currents.
It thus appears as one of the main analytical result obtained concerning the DiSSEP.
\begin{proposition}
For all integers $p,q,r \geq 0$ we have the equality
\begin{equation}\label{eq:DiSSEP_formula_general_words}
 \frac{\llangle W|G_1^pG_2^qG_3^r|V\rrangle}{\llangle W|G_2^{p+q+r}|V\rrangle}=
 \frac{\displaystyle \prod_{\ell=0}^{p-1}(c\ \phi^{p-1-\ell}+ad\ \phi^{q+r+\ell})\ \prod_{n=0}^{r-1}(d\ \phi^{r-1-n}+bc\ \phi^{q+p+n})}
{\displaystyle\prod_{k=q}^{p+q+r-1}(1-ab\ \phi^{2k})}\ ,
\end{equation}
\end{proposition}
Note that the value of any word in the generators $G_i$'s (not necessarily ordered as above) can be easily obtained from this formula because
the commutation relations among the $G_i$'s are very simple \eqref{eq:DiSSEP_commutation relations_G}. The reordering of the $G_i$'s 
makes appear only a power of $\phi$. 
\proof
In order to compute $\llangle W|G_1^pG_2^qG_3^r|V\rrangle$, we use a change of generators defined as follows
\begin{equation} \label{eq:GtoLR}
 L_i=\frac{G_1}{\phi^i}-aG_3\phi^i, \quad \text{and} \quad R_k=\frac{G_3}{\phi^k}-bG_1\phi^k.
\end{equation}
This is built so that $L_i$ and $R_k$ fulfill the following relations (derived straightforwardly from \eqref{eq:DiSSEP_commutation relations_G}
and \eqref{eq:DiSSEP_boundaries_relations_G})
\begin{eqnarray} \label{eq:relLR}
&&G_2^iL_i=L_0G_2^i, \quad R_kG_2^k=G_2^kR_0, \quad [L_i,G_1]=[L_i,G_3]=[R_k,G_1]=[R_k,G_3]=0\qquad \nonumber\\
&\text{and}&  \llangle W|L_0=c\llangle W|G_2, \quad R_0|V\rrangle = dG_2|V\rrangle.\label{eq:relLR2}
\end{eqnarray}
The change of generators \eqref{eq:GtoLR} can be inverted to get
\begin{equation} \label{eq:LRtoG}
 G_1=\frac{\phi^{k+i}}{1-ab\phi^{2(k+i)}}\left(\frac{L_i}{\phi^k}+a\phi^iR_k\right), \quad \text{and} \quad 
 G_3=\frac{\phi^{k+i}}{1-ab\phi^{2(k+i)}}\left(b\phi^kL_i+\frac{R_k}{\phi^i}\right).
\end{equation}
We can now begin the computation
\begin{eqnarray*}
 \llangle W|G_1^pG_2^qG_3^r|V\rrangle & = & \frac{\phi^q}{1-ab\phi^{2q}}\llangle W|\left(\frac{L_0}{\phi^q}+aR_q\right)G_1^{p-1}G_2^qG_3^r|V\rrangle \\
 & = &  \frac{1}{1-ab\phi^{2q}}\llangle W|cG_2G_1^{p-1}G_2^qG_3^r|V\rrangle+\frac{1}{1-ab\phi^{2q}}\llangle W|G_1^{p-1}G_2^qG_3^rad\phi^qG_2|V\rrangle \\
 & = & \frac{c\phi^{p-1}+ad\phi^{q+r}}{1-ab\phi^{2q}}\llangle W|G_1^{p-1}G_2^{q+1}G_3^r|V\rrangle.
\end{eqnarray*}
The first equality is obtained using \eqref{eq:LRtoG} with $i=0$ and $k=q$ to transform the leftmost $G_1$. The second equality relies on the 
relations \eqref{eq:relLR}. We get the last one using \eqref{eq:DiSSEP_commutation relations_G}.
This relation provides a recursion for $\llangle W|G_1^pG_2^qG_3^r|V\rrangle$ that we can iterate to obtain
\begin{equation} \label{eq:recG1}
 \llangle W|G_1^pG_2^qG_3^r|V\rrangle = \left(\prod_{l=0}^{p-1} \frac{c\phi^{p-1-l}+ad\phi^{r+q+l}}{1-ab\phi^{2(q+l)}}\right)\llangle W|G_2^{q+p}G_3^r|V\rrangle.
\end{equation}
Performing similar computations with $G_3$  we obtain the following recursive relation
\begin{eqnarray*}
 \llangle W|G_2^{q+p}G_3^r|V\rrangle & = & \frac{\phi^{q+p}}{1-ab\phi^{2(q+p)}}\llangle W|G_2^{q+p}G_3^{r-1}\left(bL_{q+p}+\frac{R_0}{\phi^{q+p}}\right)|V\rrangle \\
& = & \frac{d\phi^{r-1}+bc\phi^{q+p}}{1-ab\phi^{2(q+p)}}\llangle W|G_2^{q+p+1}G_3^{r-1}|V\rrangle\;,
\end{eqnarray*}
to get
\begin{equation} \label{eq:recG3}
 \llangle W|G_2^{q+p}G_3^r|V\rrangle = \left(\prod_{n=0}^{r-1}\frac{d\phi^{r-1-n}+bc\phi^{q+p+n}}{1-ab\phi^{2(q+p+n)}}\right)\llangle W|G_2^{p+q+r}|V\rrangle.
\end{equation}
Recombining \eqref{eq:recG1} and \eqref{eq:recG3} together, the desired result \eqref{eq:DiSSEP_formula_general_words} is proved.
\finproof

Let us stress that since $G_1,G_2,G_3$ form a basis, the knowledge of all words built on them allows us to reconstruct all words 
built on $E$ and $D$ using the two first relations of \eqref{eq:DiSSEP_MA_change_generators_basis}. 
In particular, we are able to compute exactly physical observables, as it is illustrated below.

\begin{proposition}
The one-point function (i.e the mean particle density) reads for $1\leq i \leq L$
\begin{eqnarray} \label{eq:DiSSEP_one_point_function}
 \langle \tau_i \rangle & = & \frac{1}{2}\frac{\llangle W| G_2^{i-1}(-G_1+G_2-G_3)G_2^{L-i} |V\rrangle}{\llangle W| G_2^L |V\rrangle} \\
& = & \frac{1}{2}-\frac{c\phi^{i-1}+ad\phi^{L+i-2}+d\phi^{L-i}+bc\phi^{2L-i-1}}{2(1-ab\phi^{2L-2})}\:,
\label{eq:DiSSEP_one_point_function_bis}
\end{eqnarray}
the connected two-point correlation function, for $1\leq i<j\leq L$, is given by
\begin{eqnarray}\label{eq:DiSSEP_two_point_function}
 \langle \tau_i \tau_j \rangle_c & = & \langle \tau_i \tau_j \rangle - \langle \tau_i \rangle \langle \tau_j \rangle \\
 & = & \frac{\llangle W| C^{i-1}DC^{j-i-1}DC^{L-j} |V\rrangle}{\llangle W| C^L |V\rrangle}
 -\frac{\llangle W| C^{i-1}DC^{L-i} |V\rrangle}{\llangle W| C^L |V\rrangle}\frac{\llangle W| C^{j-1}DC^{L-j} |V\rrangle}{\llangle W| C^L |V\rrangle}\qquad \nonumber \\
 & = & \frac{\phi^{L+j-i-3}(1-\phi^2)(1+b\phi^{2(L-j)})(1+a\phi^{2(i-1)})(d+bc\phi^{L-1})(c+ad\phi^{L-1})}{4(1-ab\phi^{2(L-1)})^2(1-ab\phi^{2(L-2)})},
 \label{eq:DiSSEP_two_point_function_bis}
\end{eqnarray}
and the connected three-point correlation function, for $1\leq i<j<k\leq L$, is equal to
\begin{eqnarray} \label{eq:DiSSEP_three_point_function}
 \langle \tau_i \tau_j \tau_k\rangle_c & = & \langle \tau_i \tau_j \tau_k\rangle 
 - \langle \tau_i \rangle \langle \tau_j \tau_k \rangle- \langle \tau_j \rangle \langle \tau_i \tau_k \rangle- \langle \tau_k \rangle \langle \tau_i \tau_j \rangle 
 +2 \langle \tau_i\rangle\langle \tau_j \rangle\langle \tau_k\rangle \\
 &=&-\frac{\phi^{L+k-i-5}(1-\phi^2)^2(1+b\phi^{2(L-k)}) (1+a\phi^{2(i-1)}) (d+bc\phi^{L-1}) (c+ad\phi^{L-1}) }
         {8(1-ab\phi^{2(L-1)})^3(1-ab\phi^{2(L-2)})(1-ab\phi^{2(L-3)})}\nonumber\\
 &\times&\big[ \phi^{L-j}(d+bc\phi^{L-3})(1+2a\phi^{2(j-1)}+ab\phi^{2(L-1)})
 \label{eq:DiSSEP_three_point_function_bis}\\
 &&\quad+\phi^{j-1}(c+ad\phi^{L-3})(1+2b\phi^{2(L-j)}+ab\phi^{2(L-1)})  \big].   \nonumber     
\end{eqnarray}
\end{proposition}
Remark that for generic $i,j,k$, the two- and three-point correlation functions satisfy both a set of closed linear relations:
\begin{eqnarray} \label{eq:DiSSEP_reccurence-three_point}
&&(1-\lambda^2) \Big(\langle \tau_{i-1} \tau_{j} \rangle_c+\langle \tau_{i+1} \tau_{j} \rangle_c+\langle \tau_{i} \tau_{j-1} \rangle_c+\langle \tau_{i} \tau_{j+1} \rangle_c\Big)
=4(1+\lambda^2) \langle \tau_{i} \tau_{j} \rangle_c\,,\\
&&(1-\lambda^2) \Big(\langle \tau_{i-1} \tau_{j} \tau_{k} \rangle_c+\langle \tau_{i+1} \tau_{j} \tau_{k}\rangle_c+\langle \tau_{i} \tau_{j-1} \tau_{k}\rangle_c+\langle \tau_{i} \tau_{j+1} \tau_{k}\rangle_c
+\langle \tau_{i} \tau_{j} \tau_{k-1}\rangle_c\nonumber\\
&&\qquad\qquad+\langle \tau_{i} \tau_{j} \tau_{k+1}\rangle_c\Big)
=6(1+\lambda^2) \langle \tau_{i} \tau_{j} \tau_k\rangle_c
\end{eqnarray}

\begin{proposition}
The mean particle lattice current between sites $i$ and $i+1$ (i.e the mean number of particles, counted algebraically, that jump from site $i$ to
site $i+1$ per unit of time) is given by the exact expression 
\begin{eqnarray}
 \langle j_{lat}^{i\rightarrow i+1} \rangle & = & \frac{\llangle W|C^{i-1}\left(DE-ED\right)C^{L-i-1}|V\rrangle}{\llangle W|C^L|V\rrangle} \\
 & = & \frac{1-\phi}{2} \frac{bc\phi^{2L-i-2}+d\phi^{L-i-1}-ad\phi^{L+i-2}-c\phi^{i-1}}{1-ab\phi^{2L-2}}.
 \label{eq:DiSSEP_mean_lattice_current}
\end{eqnarray}
Counting positively the pairs of particles which condensate on the lattice and negatively the pairs which evaporate, we get for
the mean evaporation-condensation current on sites $i$ and $i+1$ (i.e the mean number of particles, counted algebraically, that condensate
on sites $i$ and $i+1$ per unit of time)
\begin{eqnarray}
 \langle j_{cond}^{i, i+1} \rangle & = & 2\frac{\llangle W|C^{i-1}\lambda^2\left(E^2-D^2\right)C^{L-i-1}|V\rrangle}{\llangle W|C^L|V\rrangle} \\
 & = & \frac{(1-\phi)^2}{1+\phi} \frac{bc\phi^{2L-i-2}+d\phi^{L-i-1}+ad\phi^{L+i-2}+c\phi^{i-1}}{1-ab\phi^{2L-2}}.
 \label{eq:DiSSEP_mean_condensation_current}
\end{eqnarray}
\end{proposition}

Note that the above expressions behave as expected under the three symmetries:
\begin{enumerate}
\item The symmetry $\lambda\to-\lambda$, that translates into $\phi\to1/\phi$, $a\to1/a$, $c\to-c/a$, $b\to1/b$ and $d\to-d/b$, leaves them invariant.
\item The left/right symmetry, that becomes $a\leftrightarrow b$, $c\leftrightarrow d$ and $i\to L+1-i$, changes the sign of the lattice current, keeps the condensation current and the density invariant.
\item The particle-hole symmetry, which reads $a\to a$, $b\to b$, $c\to-c$ and $d\to-d$, changes the sign of both currents and transforms $ \langle n_i \rangle$ into $1- \langle n_i \rangle$.
\end{enumerate}

The physical quantities computed above are not all independent. 
The particle conservation law at site $i$ reads
\begin{equation}\label{eq:DiSSEP_particle_conservation_discret}
 \langle j_{lat}^{i-1\rightarrow i} \rangle-\langle j_{lat}^{i\rightarrow i+1} \rangle+
 \frac{1}{2}\langle j_{cond}^{i-1, i} \rangle+\frac{1}{2}\langle j_{cond}^{i, i+1} \rangle=0,
\end{equation}
which can be seen on the matrix product ansatz using relations \eqref{eq:DiSSEP_commutation_relations_DE}.
From the identity $[D,E]=[D,C]$ one then deduces 
\begin{equation} \label{eq:DiSSEP_drho_discret}
\langle j_{lat}^{i\rightarrow i+1} \rangle = \langle \tau_{i}\rangle - \langle \tau_{i+1}\rangle ,
\end{equation}
and using $E^2 -D^2 = C^2-CD-DC$ one  gets
\begin{equation}\label{eq:DiSSEP_jcond_discret}
\langle j_{cond}^{i, i+1} \rangle=2\lambda^2\,\Big(1-\langle \tau_{i}\rangle - \langle \tau_{i+1}\rangle\Big).
\end{equation}
From these three relations, one obtains
\begin{eqnarray} \label{eq:DiSSEP_relation_currents}
&&\langle j_{cond}^{i-1,i} \rangle-\langle j_{cond}^{i, i+1} \rangle 
+2\lambda^2 \Big(\langle j_{lat}^{i-1\rightarrow i} \rangle+\langle j_{lat}^{i\rightarrow i+1} \rangle\Big) = 0\\
&&\label{eq:DiSSEP_relation_density}
\langle \tau_{i-1} \rangle + \langle \tau_{i+1} \rangle -2\langle \tau_i \rangle
 +\lambda^2\Big(2-\langle \tau_{i-1} \rangle - \langle \tau_{i+1} \rangle -2 \langle \tau_i \rangle \Big)=0.
\end{eqnarray}

\begin{remark}
The one point correlation function verifies a closed set of equation as for the SSEP, in contrast with the ASEP case 
where the equations couple the one point function and the two points function. This property remains valid for the higher order
correlation functions \eqref{eq:DiSSEP_reccurence-three_point}, which allows in principle to compute them. However, for the multi-points correlation functions, 
solving this set of equation can be very hard. This points out the usefulness of the matrix product ansatz which makes the calculations much easier.
\end{remark}
 
\paragraph*{Fluctuations of the currents}

As mentioned previously, there are closed linear relations between the two- and three-point correlation functions which allow one to compute
the cumulant of the currents. In this section, we present the computations of the second cumulant of the lattice current between sites $i$ and $i+1$.
Let us note that it depends on the site, because of the evaporation-condensation process. As usual for such a purpose,
we use the deformed Markovian matrix defined as follows:
\begin{equation}
 M^{\mu}=B_1+\sum_{k=1}^{i-1}m_{k,k+1}+m_{i,i+1}^{\mu}+\sum_{k=i+1}^{L-1}w_{k,k+1}+\overline{B}_L
 \mbox{ with }
 m^{\mu} = \begin{pmatrix} -\lambda^2 & 0 & 0& \lambda^2 \\ 0 & -1 & e^{\mu} & 0 \\ 0 & e^{-\mu} & -1 & 0 \\  \lambda^2 & 0 & 0& -\lambda^2
 \end{pmatrix}.
\end{equation}
Let $ |\Psi^{\mu}\rangle$ be the eigenstate of $M^{\mu}$ with highest  eigenvalue
\begin{equation}
M^{\mu} \, |\Psi^{\mu}\rangle = E(\mu)\,  |\Psi^{\mu}\rangle\,.\label{eq:DiSSEP_eigenvalue_s}
\end{equation}
$E(\mu)$ is the generating function for the cumulants of the lattice current between sites $i$ and $i+1$.
We introduce {the following notation for vectors}
\begin{eqnarray}
\langle\{j\}|&=&\Big(\langle0|+\langle1|\Big)^{\otimes (j-1)} \otimes \langle1|\otimes \Big(\langle0|+\langle1|\Big)^{\otimes (L-j)}\\
\langle\{j,k\}|&=&\Big(\langle0|+\langle1|\Big)^{\otimes (j-1)} \otimes \langle1|\otimes \Big(\langle0|+\langle1|\Big)^{\otimes (k-j-1)} \otimes 
\langle1|\otimes \Big(\langle0|+\langle1|\Big)^{\otimes (L-k)}\\
\vdots && \nonumber
\end{eqnarray}
In words, $\langle\{j_1,j_2,...,j_M\}|$ represents configurations with one particle at site $j_1$, $j_2$, ..., $j_M$, and anything else 
on the other sites. Remark that this definition applies whatever the order on $j_1,...,j_M$, and thus extends the one given in the above equations.  
By extension, we note $\langle\emptyset|=\big(\langle0|+\langle1|\big)^{\otimes L}$. Then, we define the components:
\begin{equation}
T_j(\mu)\equiv T_j = \frac{\langle\{j\} |\Psi^{\mu}\rangle}{\langle\emptyset|\Psi^{\mu}\rangle}
\mbox{ ; }
U_{jk}(\mu)\equiv U_{jk} =  \frac{\langle\{j,k\}|\Psi^{\mu}\rangle}{\langle\emptyset|\Psi^{\mu}\rangle}
\mbox{ and }
V_{jkl}(\mu)\equiv V_{jkl} =  \frac{ \langle\{j,k,l\}|\Psi^{\mu}\rangle}{\langle\emptyset|\Psi^{\mu}\rangle}.
\end{equation}
{Note that by construction, $U$ and $V$ are symmetric, e.g. $U_{jk} =U_{kj} $.}
Now, projecting equation \eqref{eq:DiSSEP_eigenvalue_s} on $\langle\emptyset|$, we get
\begin{equation}
E(\mu) = (e^{-\mu}-1)(T_{i+1}-U_{i,i+1})+(e^{\mu}-1)(T_{i}-U_{i,i+1}).
\label{eq:DiSSEP_orders_lattice_current}
\end{equation}
We also project equation \eqref{eq:DiSSEP_eigenvalue_s} on $\langle\{j\} |$ for $j=1$, $1<j<i$ and $i+1<j<L$, $j=i$, $j=i+1$,  and $j=L$.
We get respectively:
\begin{eqnarray}
E(\mu)\,T_1 &=&\alpha (1-T_1)-\gamma T_1 +\lambda^2(1-T_1-T_2)+T_2-T_1
\nonumber \\
&&+ (e^{-\mu}-1)(U_{1,i+1}-V_{1,i,i+1})
+(e^{\mu}-1)(U_{1,i}-V_{1,i,i+1})\,,
\label{eq:DiSSEP_T1s}\\
E(\mu)\,T_j &=& \lambda^2(2-2T_j-T_{j+1}-T_{j-1}) +T_{j+1}-2T_j+T_{j-1} 
\nonumber \\
&&+ (e^{-\mu}-1)(U_{j,i+1}-V_{j,i,i+1})+(e^{\mu}-1)(U_{j,i}-V_{j,i,i+1})\,,
\label{eq:DiSSEP_Tjs}\\
E(\mu)\,T_i &=& \lambda^2(2-2T_i-T_{i+1}-T_{i-1}) +T_{i+1}-2T_i+T_{i-1} 
\nonumber \\
&&+ (e^{-\mu}-1)(T_{i+1}-U_{i,i+1})\,,
\label{eq:DiSSEP_Tis}\\
E(\mu)\,T_{i+1} &=& \lambda^2(2-2T_{i+1}-T_{i+2}-T_{i}) +T_{i+2}-2T_{i+1}+T_{i} 
\nonumber \\
&&+(e^{\mu}-1)(T_{i}-U_{i,i+1})\,,
\label{eq:DiSSEP_TIs}\\
E(\mu)\,T_L &=&\delta (1-T_L)-\beta T_L +\lambda^2(1-T_L-T_{L-1})+T_{L-1}-T_L
\nonumber \\
&&+ (e^{-\mu}-1)(U_{i+1,L}-V_{i,i+1,L})
+(e^{\mu}-1)(U_{i,L}-V_{i,i+1,L})\,.
\label{eq:DiSSEP_TLs}
\end{eqnarray}
These equations are solved iteratively, expanding all quantities as series in $\mu$.
We set 
\begin{eqnarray*}
 E(\mu) & = & E^{(0)}+\mu\,E^{(1)}+\frac{\mu^2}{2}E^{(2)}+o(\mu^2)\,,\\
 T_j(\mu) & = & T_j^{(0)}+\mu\,T_j^{(1)}+o(\mu)\,, \\
 U_{j,k}(\mu) & = & U_{j,k}^{(0)}+\mu\,U_{j,k}^{(1)}+o(\mu)\,.
\end{eqnarray*}
In the above expansions, $E^{(0)}=0$ is the greatest eigenvalue of the undeformed Markov matrix and 
$E^{(1)}=\langle j_{lat}^{i\rightarrow i+1} \rangle$ is the mean value of the lattice current measured between the site $i$ and $i+1$, 
where the deformation occurs.
We recall that $\langle j_{lat}^{i\rightarrow i+1} \rangle$ has been computed in \eqref{eq:DiSSEP_mean_lattice_current}. The value of 
$T_j^{(0)}= \langle \tau_j \rangle$ has also been already calculated, see \eqref{eq:DiSSEP_one_point_function_bis}. 
Similarly, $U_{j,k}^{(0)}$ is linked to 
the two-points correlation function, see \eqref{eq:DiSSEP_two_point_function_bis}.

We wish to compute $E(\mu)$ up to order 2, which corresponds to the variance of the lattice current. 
We get it through the expansion of \eqref{eq:DiSSEP_orders_lattice_current} up to order 2:
\begin{eqnarray}
E^{(1)} &=& T_i^{(0)}-T_{i+1}^{(0)}\,,\label{eq:DiSSEP_order1_lattice_current}
\\
E^{(2)} &=& 2\Big( T_i^{(1)}-T_{i+1}^{(1)} \Big) +T_i^{(0)}+T_{i+1}^{(0)} -2\,U_{i,i+1}^{(0)}\,.
\label{eq:DiSSEP_order2_lattice_current}
\end{eqnarray}
Equation \eqref{eq:DiSSEP_order1_lattice_current} just reproduces the relation \eqref{eq:DiSSEP_drho_discret} between the mean values
of the lattice current and of the density. 

To get $T_j^{(1)}$, one considers equations \eqref{eq:DiSSEP_T1s}-\eqref{eq:DiSSEP_TLs} at first order in $\mu$. They only involve  
$T_j^{(1)}$, $T_j^{(0)}$ and $U_{jk}^{(0)}$, and can be solved recursively in $T_j^{(1)}$. We get
\begin{equation*}
\left(\begin{array}{c} T_{i+1}^{(1)} \\[1.2ex] T_{i}^{(1)} \end{array}\right)=\frac{\phi^L}{1-ab\phi^{2L-2}}
\left(\begin{array}{cc} \displaystyle\frac{b\phi^{L-i-1}+\phi^{i+1-L}}{\phi^2-1}&\displaystyle \frac{a\phi^{i}+\phi^{-i}}{\phi^2-1} \\[2.1ex]
\displaystyle\frac{ b\phi^{L-i}+\phi^{i-L}}{\phi^2-1}&\displaystyle \frac{a\phi^{i-1}+\phi^{1-i}}{\phi^2-1}\end{array}\right)
 \left(\begin{array}{c} \displaystyle{\sum_{l=0}^{i-1}\Big(a\phi^l+\phi^{-l}\Big)I_{l+1}} \\[2ex]
 \displaystyle{\sum_{l=0}^{L-i-1}\Big(b\phi^l+\phi^{-l}\Big)I_{L-l}} \end{array}\right) 
\end{equation*}
with
\begin{eqnarray}
I_{j} &=&\frac{(\phi+1)^2}{4\phi}\times \begin{cases}
E^{(1)}\, T_i^{(0)}+T_{i+1}^{(0)}-U_{i,i+1}^{(0)}\,,\quad &\mbox{for } j=i
\\[1ex]
E^{(1)}\, T_{i+1}^{(0)}-T_{i}^{(0)}+U_{i,i+1}^{(0)}\,,\quad &\mbox{for } j=i+1
\\[1ex]
E^{(1)}\, T_j^{(0)}+U_{j,i+1}^{(0)}-U_{j,i}^{(0)}\,,\quad &\mbox{otherwise}
\end{cases}
\end{eqnarray}
Plugging these values into \eqref{eq:DiSSEP_order2_lattice_current}, we get the analytical expression of the variance of the lattice current:
\begin{eqnarray}
&&E^{(2)}\ =\  T_i^{(0)}+T_{i+1}^{(0)}-2\,U_{i,i+1}^{(0)}\label{eq:DiSSEP_variance_lattice_current} \\
&&\hspace{-5mm}+ \frac2{1+\phi}\,\left\{ \frac{\phi^i(b\phi^{2L-2i-1}-1)}{1-ab\phi^{2L-2}}\, \sum_{\ell=0}^{i-1} (a\phi^\ell+\phi^{-\ell})I_{\ell+1}
-\frac{\phi^{L-i}(a\phi^{2i-1}-1)}{1-ab\phi^{2L-2}}\, \sum_{\ell=0}^{L-i-1} (b\phi^\ell+\phi^{-\ell})I_{L-\ell}\right\}.
\nonumber
\end{eqnarray}
Using the explicit form of $I_j$, one can compute the sums in \eqref{eq:DiSSEP_variance_lattice_current} to perform the thermodynamic limit for
$E^{(2)}$, see chapter \ref{chap:five}.
Let us conclude this subsection by mentioning that the higher cumulants may be computed in principle by similar methods. However, 
the computations become much harder and are 
beyond the scope of this manuscript.

\paragraph*{Comparison with SSEP}

As mentioned previously the {DiSSEP} is a deformation of the SSEP which can be easily recovered when taking $\lambda=0$.
This limit already reveals at the level of the matrix product ansatz algebra: the commutation relations between $E$, $D$ and $H$ 
\eqref{eq:DiSSEP_commutation_relations_DE}
become simpler when $\lambda=0$. We have indeed $[H,E]=[H,D]=0$. Hence $H$ can be chosen equal to the identity. In this case we recover the well known
relation $[D,E]=D+E$, relevant in the construction of the steady state of the SSEP. Remark that the generators $G_1$, $G_2$ and $G_3$ in contrast
``diverge'' when taking the limit $\lambda \rightarrow 0$ (this can be seen by inverting the change of basis \eqref{eq:DiSSEP_MA_change_generators_basis}).

We can also take the limit $\lambda \rightarrow 0$ at the level of the physical observables.
For the one and two points correlation function we get
 \begin{equation*}
 \lim\limits_{\lambda \rightarrow 0} \ \langle \tau_i \rangle = \frac{\rho_l\left(L+B-i\right)+\rho_r\left(i-1+A\right)}{L+A+B-1},
 \mbox{ with } A=\frac{1}{\alpha+\gamma}\,,\ B=\frac{1}{\beta+\delta}
 \end{equation*}
and 
\begin{equation*}
 \lim\limits_{\lambda \rightarrow 0} \ \langle \tau_i \tau_j \rangle_c = -\frac{\left(i-1+A\right)\left(B+L-j\right)(\rho_l-\rho_r)^2}
 {\left(L+A+B-1\right)^2\left(L+A+B-2\right)},
\end{equation*}
which are in agreement with the known expressions for the SSEP \cite{DerridaDR04,Derrida07}.
For the lattice and evaporation-condensation currents we get
\begin{equation*}
 \lim\limits_{\lambda \rightarrow 0} \ \langle j_{lat}^{i\rightarrow i+1} \rangle = \frac{\rho_l-\rho_r}{L+A+B-1}
 \quad \text{and} \quad \lim\limits_{\lambda \rightarrow 0} \ \langle j_{cond}^{i, i+1} \rangle = 0,
\end{equation*}
also in agreement with the SSEP results.

We can also take carefully the limit of the expression \eqref{eq:DiSSEP_variance_lattice_current} to recover the variance of
the lattice current for the SSEP model \cite{DerridaDR04}
\begin{align}
&E^{(2)}_{SSEP} = \frac{\rho_l+\rho_r}{L+A+B-1}+\frac{(A-3A^2+2A^3+B-3B^2+2B^3)(\rho_l-\rho_r)^2}{3(L+A+B-1)^3(L+A+B-2)}
\nonumber \\
&\!\!
-\frac{(\rho_l-\rho_r)^2}{3(L+A+B-1)^2(L+A+B-2)}+\frac{\rho_l^2+\rho_r^2}{(L+A+B-1)(L+A+B-2)}-\frac{2(\rho_l^2+\rho_l\rho_r+\rho_r^2)}{3(L+A+B-2)}.
\nonumber
\end{align}

To summarize, the DiSSEP is an integrable one parameter generalization of the SSEP, where particles pairs are allowed to condensate and evaporate 
with the same probability rates. From a physical point of view, it provides an example of exclusion process in which the lattice current 
is not conserved along the lattice but depends on the place it is measured. It appears as a toy model for which a lot of physical quantities 
can be exactly computed but it displays also an interesting phenomenology, with non-trivial density profiles. From a more mathematical point of 
view, it provides a simple example of matrix product expression for the stationary state, which involves non-scalar 'hat operators'. It also points
out the usefulness of the ZF and GZ relations to select an efficient generators basis and a convenient change of parameters. 
It also stresses the fact that the computations of physical observables can be achieved using only the algebraic relations between the generators
and not necessarily using an explicit representation. We will come back to this model in chapter \ref{chap:five} to study its properties in 
the thermodynamic limit.

\subsection{An open two species TASEP} \label{subsec:2TASEP}

We now turn to a detailed study of another example of stochastic model that we already encountered several times in this manuscript, through its 
local jump operators and its $R$ and $K$ matrices. This model is called the two-species totally asymmetric exclusion process (2-TASEP) with open
boundaries and was first introduced and studied in \cite{CrampeMRV15}. The integrable boundaries associated to the two species
ASEP were classified there and the matrix product structure of the stationary state was investigated on a particular example. 
An explicit matrix product representation of the steady state of the models was given in \cite{CrampeEMRV16} together with the rigorous derivation 
a the phase diagrams and the exact computation of physical quantities. Following the lines of \cite{CrampeMRV15} and \cite{CrampeEMRV16}, we 
present here the main results.
     
\subsubsection{Presentation of the model}

The model is a Markov process that enters the framework of exclusion process introduced in chapter \ref{chap:two}. 
It describes two species of particles in 
stochastic evolution on a one-dimensional lattice comprising $L$ sites and coupled with two reservoirs at the boundaries. Each site
$i=1,\ldots,L$ can be in one of three states $\tau_i=0$, $1$ or $2$.
As usual, state $0$ may be considered as an empty site or hole. State $1$
is interpreted as a species of heavy, slow particles (sometimes called second class particles). State $2$ corresponds to a
species of light, fast particles (sometimes called first class particles).  
At each pair of nearest neighbor sites in the bulk, the exchange rates read
\begin{equation}
 1\ 0 \xrightarrow{1} 0\ 1 \quad,\qquad 2\ 0  \xrightarrow{1}
 0\ 2\quad,\qquad 2\ 1  \xrightarrow{1} 1\ 2\;.
\end{equation}
(Note that several other labelling conventions for the three particle species have been employed in the literature e.g. \cite{ProlhacEM09}.)
The sites $1$ and $L$ are in contact with boundary reservoirs and particles are
exchanged at  different rates at the boundaries. For generic values of
these  boundary rates, the system is not integrable (in contrast with
the 1-species TASEP, which is integrable for arbitrary  boundary
rates).  Finding in this general case an exact solution looks hopeless. However,  in
\cite{CrampeMRV15},   using a systematic procedure (the resolution of the reflection equation), all  possible  boundary
rates for  the 2-species TASEP that preserve   integrability were classified. Amongst  such  models, some  had  been studied  earlier:
the first open two-species matrix product solutions were derived in \cite{EvansFGM95}; in \cite{DerridaE97} the boundary conditions for which
the stationary state may be expressed using the matrices $D$, $E$, $A$ of \cite{DerridaEHP93} and \cite{DerridaJLS93} were deduced;
in \cite{AyyerLS09,AyyerLS12} the restricted class of {\it semi-permeable} boundaries, 
in which second class particles can neither enter nor leave the system was studied.  
In all of these cases a matrix product representation of the stationary state was found 
involving the quadratic algebra used by Derrida, Evans, Hakim and
Pasquier \cite{DerridaEHP93} in their exact solution of the 1-species exclusion process with open boundaries. 

In the present example, we construct a matrix ansatz for integrable 2-TASEP with open boundaries that allow all species of
particles to  enter and  leave the system. The algebraic structures required will be much more involved than the fundamental quadratic
algebra of \cite{DerridaEHP93}.  

 We shall study  two classes of 2-species TASEP models with the following
 boundary rates
\begin{eqnarray}
\begin{array}{c c c}
&\text{left boundary}\hspace{1cm}&\text{right boundary}\\
& 2 \xrightarrow{\hspace{2mm}\alpha\hspace{2mm}} 1\hspace{1cm} & 2 \xrightarrow{\hspace{2mm}\beta\hspace{2mm}} 0\\
(M_1):& 0 \xrightarrow{\hspace{2mm}\alpha\hspace{2mm}} 1\hspace{1cm}&  1 \xrightarrow{\hspace{2mm}\beta\hspace{2mm}} 0\\
& 0 \xrightarrow{1-\alpha} 2\hspace{1cm}&   1 \xrightarrow{1-\beta} 2
 \end{array}\label{P1}
\end{eqnarray}
or 
\begin{eqnarray}
\begin{array}{c  c c}
& \text{left boundary}\hspace{1cm}&\text{right boundary}\\
 &2\xrightarrow{\hspace{2mm}\alpha\hspace{2mm}} 1\hspace{1cm} & 2 \xrightarrow{\hspace{2mm}\beta\hspace{2mm}} 0\\
(M_2): &0 \xrightarrow{\hspace{2mm}\alpha\hspace{2mm}} 1\hspace{1cm}&  1 \xrightarrow{\hspace{2mm}\beta\hspace{2mm}} 0\\
 &0 \xrightarrow{1-\alpha} 2\hspace{1cm}& 
 \end{array}\label{P2}
\end{eqnarray}
 Hereafter, the  two  different models will be denoted by $(M_1)$ and
 $(M_2)$.  
Note that in  the classification of \cite{CrampeMRV15} the left
 boundary conditions were referred to as $L_2$ and the right hand
 boundary conditions for  $(M_1)$ or $(M_2)$ were referred to as $R_2$ and
 $R_3$ respectively. It is a simple matter to translate our results for $(M_2)$ to the case
of right boundary $R_2$ and left boundary $L_3$. The final case of
right boundary $R_3$ and left boundary $L_3$ leaves the stationary state devoid of holes and thus reduces to a one-species TASEP.

The physical interpretation of the boundary conditions is as follows.
In both models $(M_1)$, $(M_2)$ the left-hand boundary conditions correspond to a boundary reservoir containing only first and second class particle
with densities $\alpha$ and $1-\alpha$ respectively,  with no holes.
In model ($M_1$) the right-hand boundary conditions correspond to a reservoir containing second-class particles and holes
with densities $1-\beta$ and $\beta$ respectively, with no first-class particles.
In model ($M_2$) the right-hand boundary conditions correspond  to a reservoir containing first-class particles and holes
with densities $1-\beta$ and $\beta$ respectively, with no second-class particles.

\begin{figure}[htb]
\begin{center}
 \begin{tikzpicture}[scale=0.7]
\draw (-4,0) -- (14,0) ;
\foreach \i in {-4,-3,...,14}
{\draw (\i,0) -- (\i,0.4) ;}
\draw[->,thick] (-4.4,0.9) arc (180:0:0.4) ; \node at (-4.,1.8) [] {$1-\alpha$};
\draw[<-,dashed,thick] (-3.6,-0.1) arc (0:-180:0.4) ; \node at (-4.,-1.) [] {$\alpha$};
\draw  (-1.5,0.5) circle (0.3) [circle] {};
\draw  (1.5,0.5) circle (0.3) [fill,circle] {};
\draw  (4.5,0.5) circle (0.3) [fill,circle] {};
\draw  (5.5,0.5) circle (0.3) [fill,circle] {};
\draw  (6.5,0.5) circle (0.3) [circle] {};
\draw  (9.5,0.5) circle (0.3) [circle] {};
\draw  (10.5,0.5) circle (0.3) [fill,circle] {};
\draw  (13.5,0.5) circle (0.3) [fill,circle] {};
\draw[->,thick] (-1.6,1) arc (0:180:0.4); \draw[thick] (-1.8,1.2) -- (-2.2,1.6) ; \draw[thick] (-1.8,1.6) -- (-2.2,1.2) ;
\draw[->,thick] (-1.4,1) arc (180:0:0.4); \node at (-1.,1.8) [] {$1$};
\draw[->,thick] (1.4,1) arc (0:180:0.4); \draw[thick] (1.2,1.2) -- (0.8,1.6) ; \draw[thick] (1.2,1.6) -- (0.8,1.2) ;
\draw[->,thick] (1.6,1) arc (180:0:0.4); \node at (2.,1.8) [] {$1$};
\draw[<->,thick] (5.6,1) arc (180:0:0.4); \node at (6.,1.8) [] {$1$};
\draw[->,thick] (6.6,1) arc (180:0:0.4); \node at (7.,1.8) [] {$1$};
\draw[<->,thick] (9.6,1) arc (180:0:0.4); \draw[thick] (9.8,1.2) -- (10.2,1.6) ; \draw[thick] (9.8,1.6) -- (10.2,1.2) ;
\draw[->,thick] (10.6,1) arc (180:0:0.4); \node at (11.,1.8) [] {$1$};
\draw[->,thick] (13.6,1) arc (180:0:0.4) ; \node at (14.,1.8) [] {$\beta$};
 \end{tikzpicture}
 \end{center}
 \caption{Dynamical rules of the $2$-species TASEP. The empty sites stands for species $0$,
  circles for species $1$ and  bullets for species $2$. 
  On the left boundary the continuous line means injection of bullets whereas the dashed line means injection of circles. \label{fig:mASEP}}
\end{figure}
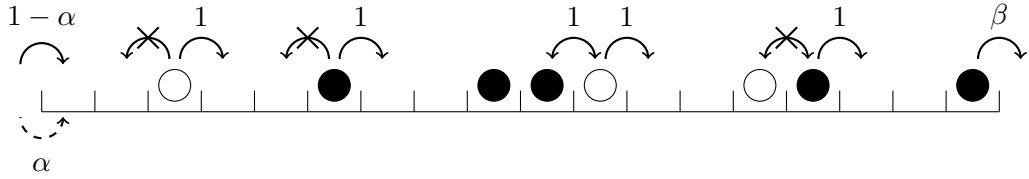

The 2-TASEP is a finite Markov process that reaches a unique steady-state in the long time limit, in which each configuration
 has the  stationary probability  (or weight)  $\cS(\tau_1,\tau_2,\dots,\tau_L)$.  The  column-vector $\steady$ of
 length $3^L$, whose components are the probabilities $\cS(\tau_1,\tau_2,\dots,\tau_L)$, satisfies the stationary master equation
\begin{equation}\label{eq:2TASEP_master_equation}
 M^{(3)}\steady=0,
\end{equation}
where $M^{(3)}$ is the $3^L \times 3^L$ Markov matrix for the 2-TASEP system. It can be decomposed as 
\begin{equation}
\label{eq:2TASEP_decomM3}
M^{(3)} =  B_1^{(3)} + \sum_{\ell=1}^{L-1} m^{(3)}_{\ell,\ell+1} + \overline B_L^{(3)}\,,
\end{equation}
with the local bulk update operator acting on nearest neighbor sites  
\begin{eqnarray}
m^{(3)}=\begin{pmatrix}
   .&.&.&.&.&.&.&.&.\\
   .&.&.&1&.&.&.&.&.\\
   .&.&.&.&.&.&1&.&.\\
   .&.&.&-1&.&.&.&.&.\\
   .&.&.&.&.&.&.&.&.\\
   .&.&.&.&.&.&.&1&.\\
   .&.&.&.&.&.&-1&.&.\\
   .&.&.&.&.&.&.&-1&.\\
   .&.&.&.&.&.&.&.&.
  \end{pmatrix}
\end{eqnarray}
where the points in the matrix stand for vanishing entries. The boundary operators read
\begin{equation}\label{eq:2TASEP_defB}
 B^{(3)}=\begin{pmatrix}
 -1   &0 &0\\
 1-\alpha & -\alpha &0\\
 \alpha& \alpha&0
    \end{pmatrix}\,,
 \qquad   \widehat B^{(3)}=\begin{pmatrix}
 0  &\beta &\beta\\
 0 & -\beta &1-\beta\\
 0& 0&-1
    \end{pmatrix}\,,\qquad    
 \check B^{(3)}=\begin{pmatrix}
 0  &\beta &\beta\\
 0 & -\beta &0\\
 0& 0&-\beta
    \end{pmatrix}\,.
\end{equation}
These operators are written as usual in the local state basis $(0,1,2)$ which is the  natural choice corresponding to increasing order of priority 
in the update rules. We recall that in equation \eqref{eq:2TASEP_decomM3}, the subscripts indicate on which sites of the lattice
the local operators act non-trivially, and the right boundary  matrix $\overline B^{(3)}$ corresponds to
$\widehat B^{(3)}$ for the processes $(M_1)$  and  $\check B^{(3)}$ for $(M_2)$. 
As a rule, the superscripts in \eqref{eq:2TASEP_decomM3} (and later in \eqref{eq:2TASEP_markov2}) indicate
the number of possible states at a site, i.e. the number of species plus one.
  
The Markov matrix $M^{(3)}$ defines an integrable stochastic process
As usual, the main objects needed to deal with integrability are the R-matrix encoding the bulk dynamics and the K-matrices encoding 
the boundaries rates.

For the 2-species TASEP, we recall that the braided R-matrix reads 
\begin{equation}
\check R^{(3)}(z)=1+(1-z)\, m^{(3)}, 
\end{equation}
with the property $-\check R^{(3)'}(1)=m^{(3)}$. The reader may refer to example \eqref{eq:2TASEP_R} for the detailed properties of this $R$-matrix.
The K-matrix for the left boundary is
\begin{eqnarray}
  K^{(3)}(z)=\begin {pmatrix} {z}^{2}&0&0\\ \noalign{\medskip}-{\frac {a z \left( {z}^{2}-1 \right) }{za+1}}&{\frac {z \left( a+z \right) }{za+
1}}&0\\ \noalign{\medskip}-{\frac {{z}^{2}-1}{za+1}}&-{\frac {{z}^{2}-1}{za+1}}&1\end {pmatrix} 
\end{eqnarray}
and the ones for the two choices of right boundary are
\begin{eqnarray}
\widehat K^{(3)}(z)=\begin {pmatrix}  1&{\frac {{z}^{2}-1}{ \left( b+z \right) z}}&{\frac {{z}^{2}-1}{ \left( b+z \right) z}}\\ 
0&{\frac {bz+1}{ \left( b+z \right) z}}&{\frac {b \left( {z}^{2}-1\right) }{ \left( b+z \right) {z}^{2}}}\\ 
 0&0&{z}^{-2}\end {pmatrix} 
\mbox{ and }
\check K^{(3)}(z)=\begin {pmatrix} 1&{\frac {{z}^{2}-1}{ \left( b+z \right) z}}&{\frac {{z}^{2}-1}{ \left( b+z \right) z}}\\ 
0&{\frac {bz+1}{ \left( b+z \right) z}}&0\\ 
0&0&{\frac{bz+1}{ \left( b+z \right) z}}
\end {pmatrix}  .
\end{eqnarray}
One obtains
\begin{equation}
 -\frac{1}{2}K^{(3)'}(1)=B\mbox{ , }\frac{1}{2}\widehat K^{(3)'}(1)=\widehat B\mbox{ and }\frac{1}{2}\check{K}^{(3)'}(1)=\check{B}^{(3)}.
\end{equation}
The reader may refer to example \eqref{eq:2TASEP_K_example} for the detailed properties of these $K$ matrices.

As explained in chapter \ref{chap:two}, we can construct a transfer matrix from these building blocks
 \begin{equation}
  t(z)=tr_0 \left(\widetilde{K}_0(z) R_{0,L}(z) \dots R_{0,1}(z)
  K_0(z) R_{1,0}(z) \dots R_{L,0}(z) \right),
 \end{equation}
 where $\widetilde{K}_0(z)$ satisfies\footnote{Note that $\widetilde{K}_0(z)$ cannot be obtained directly from \eqref{eq:Ktilde_from_Kb} here 
 because $R_{0,1}(z)^{t_1}$ is not invertible.} \eqref{eq:Kb_from_Ktilde}.
 
 It generates a family of commuting operators and is linked to the Markov matrix through
 \begin{equation}
 -\frac{1}{2}t'(1)=M^{(3)}.
 \end{equation}

\subsubsection{Matrix ansatz}

We are now interested in the stationary state of the model.
Finding the steady state of the Markov matrix \eqref{eq:2TASEP_M3} amounts to solving a linear system that grows exponentially with the size 
of the lattice. Following the general method developed in this chapter for integrable models, we will construct the stationary weights
in a matrix product from {\it i.e.},
the weight of the configuration $\mathcal{C}=(\tau_1,\dots,\tau_L)$ in the stationary state will be written as 
\begin{equation} \label{eq:2TASEP_steady_state}
 \cS(\tau_1,\dots,\tau_L)=\frac{1}{Z_L}\llangle W| X_{\tau_1}X_{\tau_2}\cdots X_{\tau_L} |V\rrangle,
\end{equation}
where $Z_L=\llangle W|(X_0+X_1+X_2)^L|V\rrangle$ is the normalization factor. 
This will basically allow us to reduce the {\it a priori} exponential complexity of the steady state (see the general construction 
with rooted trees in chapter \ref{chap:one}) to a polynomial computation.
 Following the general procedure exposed in section \ref{sec:Matrix_ansatz}, we know that the operators $X_0,X_1$ and $X_2$
 can be constructed from
\begin{equation} \label{eq:2TASEP_XfromA}
 \mathbf{A}(1) = \mathbf{X} = \begin{pmatrix}
 X_0\\
 X_1\\
 X_2
 \end{pmatrix}\;, 
\end{equation}
 where the vector $\mathbf{A}(z)$  satisfies the ZF and GZ relations \eqref{eq:ZF} and \eqref{eq:GZ} that we recall here
\begin{eqnarray}
  && \check R^{(3)} (z_1/z_2) \mathbf{A}(z_1)\otimes \mathbf{A}(z_2)=\mathbf{A}(z_2)\otimes \mathbf{A}(z_1),\label{eq:2TASEP_ZF}\\
  && \llangle W| K^{(3)}(z)\,\mathbf{A}(1/z)= \llangle W|\mathbf{A}(z)
\mbox{ and } \overline K^{(3)}(z) \mathbf{A}(1/z)|V\rrangle = \mathbf{A}(z) |V\rrangle\;,
\qquad\label{eq:2TASEP_GZ}
\end{eqnarray}
where $\overline K^{(3)}(z)$ is $\widehat K^{(3)}(z)$ or $\check K^{(3)}(z)$ depending on the right boundary considered.
For more details about the use of these relations in the context of Markov chains see \cite{CrampeRV14} or section \ref{sec:Matrix_ansatz}.
We recall that, taking the derivative of these relations w.r.t. $z_1$ and setting $z_1=z_2$, we recover the telescopic relations \eqref{eq:2TASEP_mXX}
and \eqref{eq:2TASEP_BXX} used to prove the matrix ansatz:
\begin{equation} \label{eq:2TASEP_telescopic_relation_bulk}
 m^{(3)} \mathbf{X} \otimes \mathbf{X} = \mathbf{X} \otimes \overline{\mathbf{X}} - \overline{\mathbf{X}} \otimes \mathbf{X}
\end{equation}
for the bulk and
\begin{equation} \label{eq:2TASEP_telescopic_relation_boundaries}
 \llangle W| B^{(3)} \mathbf{X} = \llangle W| \overline{\mathbf{X}}, \quad \mbox{and} \quad 
 \overline{B}^{(3)} \mathbf{X}|V\rrangle = -\overline{\mathbf{X}} |V\rrangle 
\end{equation}
for the boundaries, where 
\begin{equation} \label{eq:2TASEP_XbfromA}
 \begin{pmatrix}
  \overline{X}_0 \\
  \overline{X}_1 \\
  \overline{X}_2
 \end{pmatrix} =
\overline{\mathbf{X}} = -\mathbf{A}'(1).
\end{equation} 

\begin{remark}
The exchange relations among the $X_i$'s are obtained from these telescopic relations \eqref{eq:2TASEP_telescopic_relation_bulk}.
However, as already pointed out in section \ref{sec:Matrix_ansatz}, the generators $\overline{X}_i$'s are not necessarily scalars 
(they will {\it not} be  scalars in the present case, see below). 
Therefore, more relations are required to close the algebra generated by the $X_i$'s (the telescopic relation \eqref{eq:2TASEP_telescopic_relation_bulk}
does not tell us how to exchange $X_i$ with $\overline{X}_j$ for instance). 
A systematic way to deal with this question \cite{CrampeRV14} is to use the relations \eqref{eq:2TASEP_ZF} and \eqref{eq:2TASEP_GZ}. 
\end{remark}
 
 We thus need to build explicitly the vector $\mathbf{A}(z)$ and the boundary vectors $\llangle W|$ and $|V\rrangle$
 satisfying \eqref{eq:2TASEP_ZF} and \eqref{eq:2TASEP_GZ}. 
 For such a purpose, we assume the following expansion for the vector $\mathbf{A}(z)$:
\begin{equation} \label{eq:2TASEP_expansionA}
 \mathbf{A}(z)=\begin{pmatrix}
 z^2+G_9z+G_8+G_7/z\\
 G_6z+G_5+G_4/z\\
 G_3z+G_2+G_1/z+1/z^2
 \end{pmatrix}\;, 
\end{equation}
where the $G_i$'s belong to a non-commuting algebra. 

\begin{remark}
This expansion appears as arbitrary at first sight but it is empirically chosen as follows. For a given expansion, the ZF relation \eqref{eq:2TASEP_ZF}
provides the commutation relations among the $G_i$'s, see for instance the algebraic relations \eqref{eq:2TASEP_rel_exchange} corresponding
to the specific expansion above \eqref{eq:2TASEP_expansionA}. The relations on the boundary vectors $\llangle W|$ and $|V\rrangle$ are obtained
through the GZ relations \eqref{eq:2TASEP_GZ}, see for instance \eqref{eq:2TASEP_rel_boundaries} and \eqref{eq:2TASEP_rel_boundaries_bis}. 
All these relations are in principle sufficient to 
fix the value of any word $\llangle W|G_{i_1}G_{i_2} \dots G_{i_k}|V\rrangle$ up to a global normalization (in the same way as we were able to 
compute the stationary weights of the single species open TASEP using only the algebraic relations). If the algebra generated by the $G_i$'s
is not rich enough ({\it i.e} if there is not enough generators in the expansion of $\mathbf{A}(z)$), we noticed that all the words of length
3 built from the $G_i$'s vanish\footnote{The ZF and GZ relations implies the telescopic relations and thus the relation 
$M^{[3)}\llangle W|\mathbf{X}\otimes \mathbf{X}|V\rrangle=0$. We are hence left with two possibilities: either the matrix product state gives 
the correct stationary state of the model or it is identically vanishing.}.
We have observed that 9 generators \eqref{eq:2TASEP_expansionA} is the minimal choice that ensures
that not all words of length 3 identically vanish. 
\end{remark} 
 
Now that the expansion of the vector $\mathbf{A}(z)$ is guessed and fixed, 
 the algebra satisfied by the nine generators $G_i$ is found by writing each 
component of the ZF relation and identifying the coefficients of the polynomials in $z_1$ and $z_2$.
The  generators $G_i$ satisfy a quadratic algebra, given by the following {\it exchange} relations:
 \begin{align}
&\left[G_1,G_2\right]=0, \nonumber\\
&\left[G_1,G_3\right]=0, &&\left[G_2,G_3\right]=0,\nonumber\\
&\ G_1G_4=G_5, &&\ G_2G_4=G_6, &&\ G_3G_4=0,\nonumber \\
&\left[G_1,G_5\right]= G_6-G_4G_2,&&\  G_2G_5=G_1G_6, &&\ G_3G_5=0, && \left[G_4,G_5\right]=0,\nonumber\\
&\left[G_1,G_6\right]= -G_4\,G_3, && \left[G_2,G_6\right]= -G_5G_3, &&\ G_3G_6=0, && \left[G_4,G_6\right]=0,\nonumber\\
&\  G_1G_7=G_8, &&\ G_2G_7=G_9, &&\ G_3G_7=1, &&\ G_4G_7=0,\nonumber \\
&\left[G_1,G_8\right]=G_9-G_7G_2, &&\ G_2G_8=G_1G_9, &&\ G_3G_8=G_1, &&\left[G_4,G_8\right]=-G_7G_5,\nonumber\\
&\left[G_1,G_9\right]=1-G_7G_3, &&\left[G_2,G_9\right]=G_1-G_8G_3, &&\ G_3G_9=G_2, &&\left[G_4,G_9\right]=-G_7G_6,\nonumber\\
&\left[G_5,G_6\right]=0,\label{eq:2TASEP_rel_exchange} \\
&\ G_5G_7=0, &&G_6G_7=0, \nonumber\\
& \ G_5G_8=G_4G_9,&&G_6G_8=G_4,&&\left[G_7,G_8\right]=0,\nonumber  \\
& \left[G_5,G_9\right]=G_4-G_8G_6,&&G_6G_9=G_5,&&\left[G_7,G_9\right]=0,&&\left[G_8,G_9\right]=0.\nonumber
 \end{align} 

\begin{remark}
From the knowledge of the  exchange relations for the $G_i$'s, the
 algebra generated by the $X_i$'s can  be obtained  using \eqref{eq:2TASEP_XfromA}, i.e
 \begin{equation}
\begin{aligned}
 X_0 &= 1+G_9+G_8+G_7 \\
 X_1 &= G_6+G_5+G_4 \\
 X_2 &= 1+G_3+G_2+G_1
\end{aligned}\label{eq:2TASEP_XenG}
\end{equation}
 and  \eqref{eq:2TASEP_rel_exchange}. As already mentioned, to write the relations defining this algebra we need to introduce six more generators
$\bar X_i$ and $\bar{\bar{X_i}}$ corresponding to 
\begin{equation}
\begin{pmatrix}
 \bar X_1\\
\bar X_2\\
\bar X_3
\end{pmatrix} = -\mathbf{A}'(1) =
\begin{pmatrix}
 -2-G_9+G_7 \\
 -G_6+G_4 \\
 -G_3+G_1+2
\end{pmatrix}
\,,\quad \begin{pmatrix}
\bar{\bar{X_1}}\\
\bar{\bar{X_2}}\\
\bar{\bar{X_3}}
 \end{pmatrix} = \mathbf{A}''(1)=
 \begin{pmatrix}
 2+2G_7 \\
 2G_4 \\
 2G_1+6
\end{pmatrix}
 \end{equation}
 Note that these generators $\bar X_i$ are not scalar. 
 As an example, we give some of the relations satisfied by these generators
 \begin{eqnarray*}
 & & X_2 X_1 = X_1 \bar{X_2}-\bar{X_1}X_2 = \bar{X_2} X_1-X_2\bar{X_1} \\
 & & X_3 X_1 = X_1 \bar{X_3}-\bar{X_1}X_3 = \bar{X_3} X_1-X_3\bar{X_1} \\
 & & \dots \\
 & & 2\bar{X_2} X_1 = X_1 \bar{\bar{X_2}}-\bar{\bar{X_1}}X_2 = \bar{\bar{X_2}} X_1-X_2\bar{\bar{X_1}} \\
 & & 2\bar{X_3} X_1 = X_1 \bar{\bar{X_3}}-\bar{\bar{X_1}}X_3 = \bar{\bar{X_3}} X_1-X_3\bar{\bar{X_1}} \\
 & & \dots 
 \end{eqnarray*}
 The  algebra generated by $X_i$, $\bar X_i$ and $\bar{\bar{X_i}}$ is the same as the one generated by the $G_i$'s,
 written in a different  basis. We  chose to present 
 the commutation relations in the $G_i$ basis because they are simpler.
\end{remark}

The action of the  $G_i$'s on the boundary vectors
 is derived  from the GZ relations \eqref{eq:GZ}. For the left vector we have
\begin{equation} \label{eq:2TASEP_rel_boundaries}
  \begin{array}{l}
 \llangle W|\left(G_4-a\right)=0, \\
 \llangle W|\,G_7=0, \\
 \llangle W|\left(G_8-1\right)=0, \\
 \llangle W|\left(a(G_1-G_3)-G_5\right)=0, \\
 \llangle W|\left(G_3+G_6+G_9-G_1-a\right)=0.
  \end{array}
 \end{equation}
 For the right vector, depending on the model under consideration we have
 \begin{equation} \label{eq:2TASEP_rel_boundaries_bis}
  \begin{array}{ll}
  \mbox{For} \quad (M_1): &  \mbox{For} \quad (M_2): \\
   & \\
 G_3\,|V\rrangle =0, \hspace{1cm}& \left(G_3-b\right)\,|V\rrangle =0,\\
 \left(G_2-1\right)\,|V\rrangle =0, \hspace{1cm}& \left(G_5-bG_4\right)\,|V\rrangle =0,\\
 \left(G_6-b\right)\,|V\rrangle =0, \hspace{1cm}& G_6\,|V\rrangle =0, \\
 \left(b(G_7-G_9)+G_5\right)\,|V\rrangle =0, \hspace{1cm}& \left(b(b-G1)+G_2-1\right)|V\rrangle =0,\\
 \left(G_1+G_4+G_7-G_9-b\right)\,|V\rrangle =0, \hspace{1cm}& \left(G_1+G_4+G_7-G_9-b\right)\,|V\rrangle =0.
  \end{array}
 \end{equation}
 Note that the relations and generators $G_i$ are not all independent. 
For instance, $G_3G_5=0$ and $\left(G_2-1\right)|V\rrangle =0$ can be deduced from 
the other exchange and boundary relations.

We present below an explicit representation of the generators $G_i$'s and of the boundary vectors $\llangle W|$ and $|V\rrangle$, 
that was found in \cite{CrampeEMRV16}. This representation is build as tensor product of several copies of the DEHP algebra.
It proves rigorously the existence of the algebraic structure 
\eqref{eq:2TASEP_rel_exchange}-\eqref{eq:2TASEP_rel_boundaries_bis} and hence of the matrix product expression of the stationary state.
This representation can also be useful to perform explicit computation of physical quantities. This fact will be illustrated with the computation 
of the normalization \eqref{eq:2TASEP_ZL}. 
 
We now stress that using only the algebraic relations \eqref{eq:2TASEP_rel_exchange}-\eqref{eq:2TASEP_rel_boundaries_bis} (and not the explicit 
representation of the $G_i$'s and boundary vectors), 
any expression containing $X_0$, $X_1$ and $X_2$ and placed  between $\llangle W|$ and $|V\rrangle$
can be reduced to a multiple of $\llangle W|V\rrangle$.
Therefore, the weights of any configuration and the partition function of the model
can be in principle calculated from these relations. A computer program allowed us to obtain this way all the weights for systems of small sizes. 
We present below example of weights of configurations of arbitrary length that can be exactly computed using only these algebraic relations. 
These examples are given for the particular case $\alpha=1/2$ and $\beta=1$ ({\it i.e} for $a=1$ and $b=0$) for which the models 
$(M_1)$ and $(M_2)$ coincide. Indeed the computations simplify in this case  
 
 $\bullet$ A  word containing only  the letters $X_1$ and $X_2$ is evaluated as 
\begin{equation}
 \llangle W|Y^{(k)}(X_1,X_2)X_2^p|V\rrangle = \frac{k+2}{p+k+2}
 \left(
 \begin{array}{c}
 2p+k+1\\
 p
 \end{array}
\right) \llangle W|V\rrangle,\qquad \text{for } p,\,k=0,1,2,\dots,
\end{equation}
where $Y^{(k)}(X_1,X_2)$ is the empty word for $k=0$ and is a word of length $k$ containing only 
 the letters $X_1$ and $X_2$ and ending with $X_1$.

$\bullet$ A word starting with a power of $X_0$ on the left and followed by a combination of $X_1$ and $X_2$ is calculated as
\begin{equation}
 \llangle W|X_0^p\,Y^{(k)}(X_1,X_2)|V\rrangle = \frac{2p+1}{p+1} \left( \begin{array}{c}
 2p \\p \end{array} \right)\,
 \llangle W|V\rrangle,\qquad \text{for } p,\,k=0,1,2,\dots,
\end{equation}
where $Y^{(k)}(X_1,X_2)$ is as above.

 $\bullet$    The partition function is given by 
\begin{equation}
 Z_L=\llangle W|C^L |V\rrangle=(2L+1)A_L A_{L+1} \llangle W|V\rrangle,
\end{equation}
where $C=X_1+X_2+X_3$ and $A_L=\frac{1}{L+1}\left( \begin{array}{c}
 2L \\L
 \end{array} \right)$ is the Catalan number.

Note that in principle it should be possible to obtain similar formulas for general values of the parameters $\alpha$ and $\beta$.
Nevertheless the computations and recursive relations given by the algebraic relations
\eqref{eq:2TASEP_rel_exchange}-\eqref{eq:2TASEP_rel_boundaries_bis} appear as much more involved and their study lies beyond the scope 
of this manuscript.

\subsubsection{Identification}

Interesting physical quantities, such as the mean particles currents and densities of each species, can be exactly computed using
an identification, or coloring argument, formalized in \cite{AyyerLS12}. 
The idea is to identify two species of particles, or a species of particles with holes,
to simplify the model. The model obtained comprises a single species of particle and can thus be identified as a single species open TASEP.
All possible identifications are not consistent with the dynamics of the 2-TASEP. For instance it is not possible to identify particles of 
species $2$ (the fastest particles) with holes, because the particles of species $1$ behave differently with respect to holes (that they overtake)
and to particles of species $2$ (that overtake them). 

The 2-TASEP models $(M_1)$ and $(M_2)$ can both be mapped to the one-species TASEP model using two possible identifications:
\begin{enumerate}
\item One can identify holes and species 1 to get a one-species TASEP model for which the phase diagram is given in table
 \eqref{eq:2TASEP_phase_diag_1-TASEP}. The boundary conditions read
\begin{eqnarray} \label{2TASEP_identification_bis}
\begin{array}{c c c}
&\text{left boundary}\hspace{1cm}&\text{right boundary}\\[1ex]
(M_1): & (0,1) \xrightarrow{\hspace{2mm}\alpha\hspace{2mm}} 2\hspace{1cm} & 2 \xrightarrow{\hspace{2mm}1\hspace{2mm}} (0,1)\\[1ex]
(M_2): & (0,1) \xrightarrow{\hspace{2mm}\alpha\hspace{2mm}} 2\hspace{1cm} & 2 \xrightarrow{\hspace{2mm}\beta\hspace{2mm}} (0,1)
 \end{array}
\end{eqnarray}
\item One can identify species 1 and 2 to get another version of the one-species TASEP model. In that case the two models $(M_1)$
  and $(M_2)$ produce the same boundary conditions:
\begin{eqnarray}\label{2TASEP_identification}
\begin{array}{c c c}
&\text{left boundary}\hspace{1cm}&\text{right boundary}\\[1ex]
(M_1)\,\&\,
(M_2): & 0 \xrightarrow{\hspace{2mm}1\hspace{2mm}} (1,2)\hspace{1cm} & (1,2) \xrightarrow{\hspace{2mm}\beta\hspace{2mm}} 0
 \end{array}
\end{eqnarray}
\end{enumerate}

 For the two-species TASEP, we denote by $\langle j_1 \rangle$, $\langle j_2 \rangle$
the mean particle currents  in the stationary state for the particles of
species 1 and 2 respectively ($\langle j_0 \rangle$ denotes the mean current of holes).  The currents are counted positively
when particles flow from the left to the right.  In the same way, $\langle\rho_1^{(i)}\rangle$ and $\langle\rho_2^{(i)}\rangle$ 
denote the mean densities of particles of species 1 and 2 respectively on site $i$ ($\langle\rho_0^{(i)}\rangle$ denotes the density of holes). 

The exclusion property ensures that 
\begin{equation}
 \langle j_0 \rangle+\langle j_1 \rangle+\langle j_2 \rangle=0, \quad \mbox{and} \quad 
 \langle\rho_0^{(i)}\rangle+\langle\rho_1^{(i)}\rangle+\langle\rho_2^{(i)}\rangle=1.
\end{equation}

Identification 1. allows us to compute the current $\langle j_2 \rangle$ and the
density $\langle\rho_2^{(i)}\rangle$, while identification 2. yields the current
$\langle j_1 \rangle + \langle j_2 \rangle=-\langle j_0 \rangle$ and the density $\langle\rho_1^{(i)}\rangle+\langle\rho_2^{(i)}\rangle$. 

The identification procedure can be understood at the level of Markov matrices in the following way. We present here the case of the 
identification of species $1$ and species $2$ (but the case of identification of holes with species $1$ can be treated similarly). When 
doing the aforementioned identification, we end up with the single species open TASEP defined in \ref{2TASEP_identification} for
both the models $(M_1)$ and $(M_2)$. The stochastic evolution rules of this single species model can be, as usual, encoded in a Markov matrix
written as the sum of local operators
\begin{equation}
M^{(2)} =  B_1^{(2)} + \sum_{\ell=1}^{L-1} m^{(2)}_{\ell,\ell+1} + \overline B_L^{(2)}\,,
\label{eq:2TASEP_Markov1species}
\end{equation}
 where the local bulk Markov matrix between site $\ell$ and $\ell+1$
 and the boundary matrices are given by
\begin{equation}\label{eq:2TASEP_markov2}
 m^{(2)}=\begin{pmatrix}
          0&0&0&0\\
          0&0&1&0\\
          0&0&-1&0\\
          0&0&0&0
         \end{pmatrix}\ ;\quad  B^{(2)}=\begin{pmatrix} -1& 0 \\ 1& 0 \end{pmatrix}\ ;\quad 
         \overline B^{(2)}=\begin{pmatrix} 0& \beta \\ 0& -\beta \end{pmatrix}\;.
\end{equation}
We now define the rectangular matrix
\begin{equation}
 u = \begin{pmatrix}
      1 & 0 & 0 \\
      0 & 1 & 1
     \end{pmatrix}.
\end{equation}
The role of this matrix can be intuitively understood as summing the two last entries of a vector of size $3$. We have indeed
\begin{equation}
 \begin{pmatrix}
  1 & 0 & 0 \\
  0 & 1 & 1
 \end{pmatrix}
 \begin{pmatrix}
  x \\ y \\ z
 \end{pmatrix} =
 \begin{pmatrix}
  x \\ y+z
 \end{pmatrix}.
\end{equation}
The validity of the identification $2$ lies in the following intertwining relations for the local jump operators
\begin{equation} \label{eq:2TASEP_intertwining_local_operators}
  u \ B^{(3)}=B^{(2)} u, \qquad u \otimes u \ m^{(3)}=m^{(2)} u \otimes u, \qquad u \ \overline{B}^{(3)}=\overline{B}^{(2)}. 
\end{equation}
This is a rigorous way of saying that the process by the identification of species $1$ and species $2$ (which corresponds 
to summing the two last entries) is given by \ref{2TASEP_identification}.
This can be immediately upgraded to an intertwining relation on the Markov matrices. If we define 
$U=u \otimes \dots \otimes u$ 
then we have
\begin{equation}
 U \ M^{(3)} = M^{(2)} U.
\end{equation}
Such intertwining relation have already been used for the multi-species ASEP on the ring \cite{AritaAMP11,AritaAMP12,CantiniDGW15}.

\begin{remark}
Note that the relations \eqref{eq:2TASEP_intertwining_local_operators} on the local jump operators can be upgraded at the level 
of $R$ and $K$ matrices
\begin{equation} 
  u \ K^{(3)}(z)=K^{(2)}(z) u, \qquad u \otimes u \ R^{(3)}(z)=R^{(2)}(z) u \otimes u, \qquad u \ \overline{K}^{(3)}(z)=\overline{K}^{(2)}(z). 
\end{equation}
where we have introduced the  $K$-matrices for the single-species open TASEP 
\begin{eqnarray}
  K^{(2)}(z)=\begin {pmatrix}
 {z}^{2}&0\\1-{z}^{2}&1
\end{pmatrix}\mbox{ and }
\overline K^{(2)}(z)=\begin {pmatrix}
          1&{\frac {{z}^{2}-1}{ \left( b+z \right) z}}\\0&{\frac {bz+1}{ \left( b+z \right) z}}
         \end {pmatrix}.
\end{eqnarray}
These reflection matrices are related to the boundary matrices through
\begin{equation}
 -\frac{1}{2}\left.\frac{d}{dz}K^{(2)}(z)\right|_{z=1}=B^{(2)}\mbox{ , }
 \frac{1}{2}\left.\frac{d}{dz}\overline K^{(2)}(z)\right|_{z=1}=\overline B^{(2)}.
\end{equation}
\end{remark}
These intertwining relations have strong implications on connection between steady states of the two species and single species models.
Define the vector $\mathbf{X}^{(2)}= u \ \mathbf{X}$, where $\mathbf{X}$ introduced in \eqref{eq:2TASEP_XfromA} statisfies the telescopic
relations in the bulk \eqref{eq:2TASEP_telescopic_relation_bulk} and on the boundaries \eqref{eq:2TASEP_telescopic_relation_boundaries}.
Then a direct computation, using the intertwining property \eqref{eq:2TASEP_intertwining_local_operators}, yields
\begin{equation}
 m^{(2)} \mathbf{X}^{(2)}\otimes\mathbf{X}^{(2)}=\mathbf{X}^{(2)}\otimes\overline{\mathbf{X}}^{(2)}-\overline{\mathbf{X}}^{(2)}\otimes\mathbf{X}^{(2)}
\end{equation}
and
\begin{equation}
 \llangle W|B^{(2)} \mathbf{X}^{(2)}=\llangle W|\overline{\mathbf{X}}^{(2)}, \qquad 
 \overline{B}^{(2)} \mathbf{X}^{(2)} |V\rrangle =-\overline{\mathbf{X}}^{(2)}|V\rrangle,
\end{equation}
where $\overline{\mathbf{X}}^{(2)}= u \ \overline{\mathbf{X}}$ with $\overline{\mathbf{X}}$ defined in \eqref{eq:2TASEP_XbfromA}.

\begin{remark}
 Once again this can be promoted to spectral parameter dependent relations. If we define $\mathbf{A}^{(2)}(z)=u \ \mathbf{A}(z)$ where the size three
 vector $\mathbf{A}(z)$ satisfies the ZF and GZ relation associated to the two species model, then the size two vector $\mathbf{A}^{(2)}(z)$
 satisfies the ZF and GZ relation associated to the single species model. Note that going the other direction is much harder: finding a size three vector,
 that fulfill ZF and GZ relations, from a size two vector, that satisfies these relations for the single species model, requires a more involved 
 construction which will be exposed when dealing with the explicit representation.
\end{remark}

This immediately implies that the matrix product state 
\begin{equation} \label{eq:2TASEP_reduced_MA}
 \llangle W| \mathbf{X}^{(2)} \otimes \dots \otimes \mathbf{X}^{(2)} |V\rrangle =
 \llangle W| \begin{pmatrix}
              X_0 \\ X_1 + X_2
             \end{pmatrix} \otimes \dots \otimes 
             \begin{pmatrix}
              X_0 \\ X_1 + X_2
             \end{pmatrix} |V\rrangle
\end{equation}
is the (unnormalized) steady state of $M^{(2)}$. The Perron-Frobenius theorem tells us that the steady state of $M^{(2)}$ is unique: 
the previous matrix product state \eqref{eq:2TASEP_reduced_MA} is thus equal to the usual matrix product steady state of the single species 
open TASEP (constructed from the matrices $E$ and $D$, see \cite{DerridaEHP93}) up to a global normalization (that is easily evaluated on 
a particular configuration)
\begin{equation} \label{eq:2TASEP_proportionality_relation}
 \llangle W| \mathbf{X}^{(2)} \otimes \dots \otimes \mathbf{X}^{(2)} |V\rrangle=
 \frac{\llangle W| X_0^L |V\rrangle}{\llangle W|E^L|V\rrangle}
 \llangle W| \begin{pmatrix}
              E \\ D
             \end{pmatrix} \otimes \dots \otimes
             \begin{pmatrix}
              E \\ D
             \end{pmatrix} |V\rrangle,
\end{equation}
where we recall that the matrices $E$ and $D$ satisfy
\begin{equation}
 DE=D+E, \quad \llangle W|E=\llangle W|, \quad D|V\rrangle=\frac{1}{\beta}|V\rrangle,
\end{equation}
because we are in the particular case $\alpha=1$. Note that we made a slight abuse of notation, writing always $\llangle W|$ and $|V\rrangle$ for 
both the boundary vectors for the well-known single species solution \cite{DerridaEHP93} and also for 
the two species model, but the reader has to be careful that they are note equal. Nevertheless there should be hopefully no ambiguity in 
the relation \eqref{eq:2TASEP_proportionality_relation}, the boundary vectors can indeed be distinguished by the matrices they are acting on.

From \eqref{eq:2TASEP_proportionality_relation} we deduce immediately that
\begin{eqnarray*}
 \langle j_1\rangle + \langle j_2\rangle & = & \frac{\llangle W|(X_0+X_1+X_2)^{i-1}(X_1+X_2)X_0(X_0+X_1+X_2)^{L-i-1}|V\rrangle}
 {\llangle W|(X_0+X_1+X_2)^L|V\rrangle} \\
 & = & \frac{\llangle W|(E+D)^{i-1}DE(E+D)^{L-i-1}|V\rrangle}{\llangle W|(E+D)^L|V\rrangle} \\
 & = & \frac{\cZ_{L-1}(1,\beta)}{\cZ_L(1,\beta)},
\end{eqnarray*}
where
\begin{equation}
 \cZ_L(\alpha,\beta)=\frac{\alpha\beta}{\alpha+\beta-1}\sum_{p=1}^L B_{L,k} 
 \frac{\left(\frac{1}{\alpha}\right)^{p+1}-\left(\frac{1}{\beta}\right)^{p+1}}{\frac{1}{\alpha}-\frac{1}{\beta}}
\end{equation}
is the normalization of the single species open TASEP computed in \eqref{eq:TASEP_normalization}. The combinatorial coefficients 
\begin{equation}
 B_{n,k} = \frac{k(2n-1-k)!}{n!(n-k)!}
\end{equation}
were defined in \eqref{eq:TASEP_ballot_number}. Note that the formula of the normalization have a slight difference with the one given in 
\eqref{eq:TASEP_normalization} because we evaluate explicitly here the scalar product $\llangle W|V\rrangle = \alpha\beta/(\alpha+\beta-1)$ 
using the representation \eqref{eq:TASEP_representation_vectors}. There is also a slight change of notation with the calligraphic $\cZ_L$ 
to distinguish it from $Z_L$ the normalization of the two species model. We precise also in the notation the injection rate $\alpha$ and 
the extraction rate $\beta$ because these parameters may vary when doing the different identification.

Thanks to \eqref{eq:2TASEP_proportionality_relation} we can also compute
\begin{eqnarray}
 \langle \rho_1^{(i)}+\rho_2^{(i)} \rangle & = & \frac{\llangle W| (X_0+X_1+X_2)^{i-1}(X_1+X_2)(X_0+X_1+X_2)^{L-i}|V\rrangle}
 {\llangle W|(X_0+X_1+X_2)^L|V\rrangle} \\
 & = & \frac{\llangle W|(E+D)^{i-1} D (E+D)^{L-i}|V\rrangle}{\llangle W|(E+D)^L|V\rrangle} \\
 & = & \sum_{k=1}^{L-i} B_{k,1}\frac{\cZ_{L-k}(1,\beta)}{\cZ_L(1,\beta)}
 +\frac{\cZ_{i-1}(1,\beta)}{\cZ_L(1,\beta)}\sum_{k=1}^{L-i} B_{L-i,k}\frac{1}{\beta^{k+1}}.
\end{eqnarray}
The last equality is obtained using the result \eqref{eq:TASEP_density}.

We can repeat exactly the same procedure with the identification 1 (defined in \ref{2TASEP_identification}), that we recall give
access to the mean particle current $\langle j_2 \rangle$ and the mean particle density $\langle \rho_2^{(i)} \rangle$.
The mean particle current of species $2$ is thus given by 
\begin{equation}
 \langle j_2 \rangle=\frac{\cZ_{L-1}(\alpha,1)}{\cZ_L(\alpha,1)}, \quad \mbox{ for } \quad (M_1) \quad \mbox{ and } \quad 
 \langle j_2 \rangle=\frac{\cZ_{L-1}(\alpha,\beta)}{\cZ_L(\alpha,\beta)}, \quad \mbox{ for } \quad (M_2).
\end{equation}
The mean particle density of species $2$ at site $i$ is equal to 
\begin{equation}
 \langle \rho_2^{(i)} \rangle = \sum_{k=1}^{L-i+1} B_{k,1}\frac{\cZ_{L-k}(\alpha,1)}{\cZ_L(\alpha,1)}
\end{equation}
for model $(M_1)$ and 
\begin{equation}
 \langle \rho_2^{(i)} \rangle = \sum_{k=1}^{L-i} B_{k,1}\frac{\cZ_{L-k}(\alpha,\beta)}{\cZ_L(\alpha,\beta)}
 +\frac{\cZ_{i-1}(\alpha,\beta)}{\cZ_L(\alpha,\beta)}\sum_{k=1}^{L-i} B_{L-i,k}\frac{1}{\beta^{k+1}}
\end{equation}
for model $(M_2)$.

To summarize, gathering the results obtained through both identifications, we obtain that 

$\bullet$  The mean particle currents are given by 
\begin{equation}
 \langle j_0 \rangle = -\frac{\cZ_{L-1}(1,\beta)}{\cZ_L(1,\beta)}, \quad 
 \langle j_1 \rangle = \frac{\cZ_{L-1}(1,\beta)}{\cZ_L(1,\beta)}-\frac{\cZ_{L-1}(\alpha,1)}{\cZ_L(\alpha,1)}
 \ \ \text{ and } \ \ \langle j_2 \rangle =\frac{\cZ_{L-1}(\alpha,1)}{\cZ_L(\alpha,1)}.
\end{equation}
for the model $(M_1)$ and 
\begin{equation}
 \langle j_0 \rangle = -\frac{\cZ_{L-1}(1,\beta)}{\cZ_L(1,\beta)}, \quad 
 \langle j_1 \rangle = \frac{\cZ_{L-1}(1,\beta)}{\cZ_L(1,\beta)}-\frac{\cZ_{L-1}(\alpha,\beta)}{\cZ_L(\alpha,\beta)}
 \ \ \text{ and } \ \ \langle j_2 \rangle=\frac{\cZ_{L-1}(\alpha,\beta)}{\cZ_L(\alpha,\beta)}.
\end{equation}
for the model $(M_2)$.

$\bullet$ The average densities of particles at site $i$ are given by
\begin{eqnarray}
 \langle \rho_0^{(i)}\rangle &=& 1-\sum_{k=1}^{L-i} B_{k,1}\frac{\cZ_{L-k}(1,\beta)}{\cZ_L(1,\beta)}
 -\frac{\cZ_{i-1}(1,\beta)}{\cZ_L(1,\beta)}\sum_{k=1}^{L-i} B_{L-i,k}\frac{1}{\beta^{k+1}}, \\ 
 \langle \rho_1^{(i)}\rangle &=& \sum_{k=1}^{L-i} B_{k,1}\frac{\cZ_{L-k}(1,\beta)}{\cZ_L(1,\beta)}
 +\frac{\cZ_{i-1}(1,\beta)}{\cZ_L(1,\beta)}\sum_{k=1}^{L-i} B_{L-i,k}\frac{1}{\beta^{k+1}} \\
 & & \quad -\sum_{k=1}^{L-i+1} B_{k,1}\frac{\cZ_{L-k}(\alpha,1)}{\cZ_L(\alpha,1)}, \\
 \langle \rho_2^{(i)}\rangle &=& \sum_{k=1}^{L-i+1} B_{k,1}\frac{\cZ_{L-k}(\alpha,1)}{\cZ_L(\alpha,1)}\;.
\end{eqnarray}
 for the model $(M_1)$ and 
\begin{eqnarray}
 \langle \rho_0^{(i)}\rangle &=& 1-\sum_{k=1}^{L-i} B_{k,1}\frac{\cZ_{L-k}(1,\beta)}{\cZ_L(1,\beta)}
 -\frac{\cZ_{i-1}(1,\beta)}{\cZ_L(1,\beta)}\sum_{k=1}^{L-i} B_{L-i,k}\frac{1}{\beta^{k+1}}, \\ 
 \langle \rho_1^{(i)}\rangle &=& \sum_{k=1}^{L-i} B_{k,1}\frac{\cZ_{L-k}(1,\beta)}{\cZ_L(1,\beta)}
 +\frac{\cZ_{i-1}(1,\beta)}{\cZ_L(1,\beta)}\sum_{k=1}^{L-i} B_{L-i,k}\frac{1}{\beta^{k+1}} \\
 & & \quad -\sum_{k=1}^{L-i} B_{k,1}\frac{\cZ_{L-k}(\alpha,\beta)}{\cZ_L(\alpha,\beta)}
 -\frac{\cZ_{i-1}(\alpha,\beta)}{\cZ_L(\alpha,\beta)}\sum_{k=1}^{L-i} B_{L-i,k}\frac{1}{\beta^{k+1}}, \\
 \langle \rho_2^{(i)}\rangle &=& \sum_{k=1}^{L-i} B_{k,1}\frac{\cZ_{L-k}(\alpha,\beta)}{\cZ_L(\alpha,\beta)}
 +\frac{\cZ_{i-1}(\alpha,\beta)}{\cZ_L(\alpha,\beta)}\sum_{k=1}^{L-i} B_{L-i,k}\frac{1}{\beta^{k+1}}\;.
\end{eqnarray} 
for the model $(M_2)$. In chapter \ref{chap:five} we will study the large size limit of these results which will allow us to compute 
the phase diagrams of both models.

In the particular case $\alpha=1/2$ and $\beta=1$ that we studied previously, the results are much simpler
(we recall that for these specific values of $\alpha$ and $\beta$ the models $(M_1)$ and $(M_2)$ coincide)

$\bullet$  The mean particle currents are given by 
\begin{equation}
 \langle j_0 \rangle = -\frac{L+2}{2(2L+1)}, \quad \langle j_1\rangle = \frac{1}{2(2L+1)} \ \ \text{ and } \ \ \langle j_2\rangle = \frac{L+1}{2(2L+1)}.
\end{equation}

$\bullet$ The average densities of particles at site $i$ are given by
\begin{eqnarray}
 \langle \rho_0^{(i)}\rangle &=& \frac{1}{A_{L+1}}\sum_{k=0}^{i-1}A_kA_{L-k}, \\ 
 \langle \rho_1^{(i)}\rangle &=& \frac{1}{A_{L+1}}\sum_{k=i}^{L}\frac{L-k+1}{L+2}A_kA_{L-k}, \\
 \langle \rho_2^{(i)}\rangle &=& \frac{1}{A_{L+1}}\sum_{k=i}^{L}\frac{k+1}{L+2}A_kA_{L-k}\;.
\end{eqnarray}

 We just saw that densities and particle currents can be computed using the identification method. 
 However, we stress that the algebraic relations \eqref{eq:2TASEP_rel_exchange}-\eqref{eq:2TASEP_rel_boundaries_bis} are necessary to compute all
 the individual weights and correlations between different type of
 particles: this cannot be obtained from  the identification procedure.  
 
 We mention finally that similar  algebras, with at most
 nine generators, can be defined to compute the weights of the stationary state of the other integrable 2-TASEP models found in this manuscript
 (see discussion in \cite{CrampeMRV15}).
 
 We now address the problem of finding an explicit representation of the nine generators and boundary vectors.

\subsubsection{Explicit representation of the matrix ansatz algebra}

This section is devoted to the construction of an explicit representation for the boundary vectors $\llangle W|$ and $|V\rrangle$ and 
for the generators $G_i$'s (which will thus give a representation for the matrices $X_0$, $X_1$ and $X_2$).

 This representation for the 2-TASEP will be constructed  in terms of
 tensor products of the fundamental operators $A$, $d$ and
 $e$ that appear in the solution of the one-species
 exclusion process \cite{DerridaEHP93}.  These operators  $A$, $d$ and
 $e$ define a quadratic algebra and satisfy  
 \begin{equation}
 d\,e = 1\,,\quad A=1-e\,d\,,\quad
 d\,A = 0\,,\quad  A\,e = 0\,.\label{eq:2TASEP_ed} 
 \end{equation} 
 The
 relation with  the operators $D$ and $E$ of  \cite{DerridaEHP93}  is  $d
 = D -1, \,  e = E -1 $ and $ A = DE - ED$. 

\hfill\break
 We also define  the following parameters 
\begin{equation}
a=\frac{1-\alpha}{\alpha}\quad \mbox{and}\quad  b=\frac{1-\beta}{\beta}\;.
\end{equation}

Finally, we shall need four commuting copies of the algebra \eqref{eq:2TASEP_ed}, $(e_n,d_n,A_n)$, $n=1,2,3,4$.
A simple way to achieve this is to make four-fold
tensor products:
\begin{equation}
\begin{aligned}
e_1 = e \otimes 1\otimes 1 \otimes 1,  \\ 
e_2 = 1 \otimes e\otimes 1 \otimes 1 ,  \\ 
e_3 = 1\otimes 1 \otimes e \otimes  1,  \\ 
e_4 = 1\otimes 1\otimes 1 \otimes e ,
\label{eq:2TASEP_tensor}
\end{aligned}
\end{equation}
and similarly for $d_n$ and $A_n$.

We are now in a position to present explicit matrices for the 2-TASEP with open boundaries

\begin{proposition}
\begin{eqnarray}
 X_0 &=&\Big(1+a A_1 A_2+e_2 d_3 \Big)(1+e_4) + \Big( e_2+e_3+a A_1A_2e_3+e_1A_3\Big)(1+d_4),
\label{eq:2TASEP_X0} \qquad 
\\
X_1 &=&a d_1A_2\,(1+e_4)+ \Big(  a d_1A_2e_3+a A_2A_3\Big)(1+d_4),
\\
X_2 &=&\Big( d_2+d_3 \Big)(1+e_4)+ \Big( 1+d_2e_3+e_1d_2A_3 \Big)(1+d_4).
\label{eq:2TASEP_X2}
\end{eqnarray}
We get the following realization for the 9 generators $G$
\begin{eqnarray} 
 &&G_1=d_3e_4+e_1d_2A_3+d_2e_3+d_4\,;\quad G_2=d_3+d_2e_4+d_2e_3d_4+e_1d_2A_3d_4\,;\quad G_3=d_2\qquad \nonumber \\
 &&G_4=\lambda(d_1A_2e_3+A_2A_3)\ ;\quad G_5=\lambda(d_1A_2e_4+d_1A_2e_3d_4+A_2A_3d_4) \ ;\quad G_6=\lambda d_1A_2 \nonumber \\
 &&G_7=\lambda A_1A_2e_3+e_2\quad;\quad G_8=\lambda A_1A_2e_4+e_2d_3e_4+\lambda A_1A_2e_3d_4+e_2d_4+e_1A_3+e_3 \nonumber \\
 && G_9=\lambda A_1A_2+e_2d_3+e_4+e_1A_3d_4+e_3d_4 \label{eq:2TASEP_representation_G}
\end{eqnarray}
\end{proposition}
\proof
It can be easily checked using a symbolic calculation program \cite{Vermaseren00} that the representation presented here indeed obeys the commutation 
relations \eqref{eq:2TASEP_rel_exchange}. However, we give below a more elegant proof of this fact, taking advantage of a factorization property of 
the vector $\mathbf{A}(z)$, that is obtained using the identification procedure. 
\finproof

\begin{remark}
It is important to realize that the integer suffices on
the left and right hand sides of  \eqref{eq:2TASEP_X0}--\eqref{eq:2TASEP_X2}  are unrelated:
on the left $\tau =0, 1, 2$  corresponds to particle species
whereas on the right $n=1, 2, 3, 4$  labels the tensor product as in
the example (\ref{eq:2TASEP_tensor}).
\end{remark}

To construct the vectors $\llangle W|$ and $|V\rrangle$, we first define the elementary vectors
$\llangle x|$ and $|x\rrangle$ that obey
\begin{equation}
\llangle x|\,e = x\,\llangle x| \mbox{ and } d\,|x\rrangle = x\,|x\rrangle\,.
\label{eq:2TASEP_vectx}
\end{equation}
It is known \cite{DerridaEHP93} that explicit representations of such elementary vectors exit. Here we use the representation exposed previously
in this chapter \eqref{eq:TASEP_representation_vectors} where 
\begin{equation}
\llangle x|y\rrangle = \frac1{1-xy}\,.
\end{equation}

\begin{proposition}
The left boundary  vector reads:
\begin{equation}
  \llangle W|_{1234} =\llangle 1|_1\llangle 0|_2\llangle 0 |_3\llangle 0|_4\;,
\end{equation}
where the indices indicate again which copy of the $(A,d,e)$
algebra acts on the vector.  To make the notation less cluttered, we
shall simply write  $\llangle W|$  instead of  $\llangle
W|_{1234}$. Note that the left  vector is the same  for  the models
$(M_1)$ and  $(M_2)$.

The right boundary  vector  depends on the choice of the dynamics at
the  right boundary  (i.e. on the choice of the model $(M_1)$ or
$(M_2)$). We have 
\begin{eqnarray}
|V (M_1) \rrangle_{1234} &=& |\frac{b}{a}\rrangle_1\, |0\rrangle_2\,
|1\rrangle_3\, |b\rrangle_4 \,\, \,\mbox{ for } (M_1) \\ 
\mbox{ and } |V (M_2)\rrangle_{1234} &=& \, |0\rrangle_1\, |b\rrangle_2 \,|1\rrangle_3\,
|b\rrangle_4  \,\,\,\,  \mbox{ for } (M_2).
\end{eqnarray}
We shall simply   write  $|V\rrangle$ for the right vector, without
specifying  the  indices and  which model we consider. This should
be unambiguous from the context. 
\end{proposition}
\proof
It can be checked by direct computation that the relations \eqref{eq:2TASEP_rel_boundaries} and \eqref{eq:2TASEP_rel_boundaries_bis} 
are indeed satisfied. However, we provide below a simpler proof of this fact using a factorization of the vector $\mathbf{A}(z)$. 
\finproof

\subsubsection{Factorized form for the representation}

The expressions \eqref{eq:2TASEP_X0}--\eqref{eq:2TASEP_X2} for the $X_\tau$'s can be written in a factorized form
which will be useful to compute the normalization but also to prove the validity of the representation.

\begin{definition}
We define the Lax operators by $\LL^{(3)}(z)=L^{(3)}(z)\widetilde L^{(3)}(z)$ where
\begin{eqnarray}\label{eq:2TASEP_lax3}
  L^{(3)}(z)=\begin{pmatrix} z+\lambda A_1 A_2&z e_1&z e_2\\ 
  \lambda d_1 A_2&\lambda  A_2&0\\ 
  d_2& e_1 d_2&1
\end{pmatrix}\mbox{ and }
\widetilde L^{(3)}(z)= \begin{pmatrix}1& e_3\\
0& A_3\\ 
d_3/z& 1/z
\end{pmatrix},
 \end{eqnarray}
 and $\LL^{(2)}(z)=L^{(2)}(z)\widetilde L^{(2)}(z)$ where
 \begin{eqnarray}
  L^{(2)}(z)=\begin{pmatrix} z & ze_4\\ 
  d_4& 1
\end{pmatrix}\mbox{ and }
\widetilde L^{(2)}(z)= \begin{pmatrix}1\\
1/z
\end{pmatrix}.
\end{eqnarray}
\end{definition}

\begin{proposition}
The vector $\mathbf{A}(z)$ can be factorized as
\begin{equation}\label{eq:2TASEP_XLL}
 \mathbf{A}(z)=\LL^{(3)}(z)\LL^{(2)}(z)=L^{(3)}(z)\widetilde L^{(3)}(z)L^{(2)}(z)\widetilde L^{(2)}(z).
\end{equation}
\end{proposition}
\proof
This property can be directly check using the explicit expression of the vector $\mathbf{A}(z)$ given by the expansion \eqref{eq:2TASEP_expansionA}
and the explicit representation of the $G_i$'s \eqref{eq:2TASEP_representation_G}.
\finproof

As mentioned previously, we can prove the ZF and GZ relations \eqref{eq:2TASEP_ZF} and \eqref{eq:2TASEP_GZ} by direct computations. 
However, using the factorization
\eqref{eq:2TASEP_XLL}, we can split the proof of these relations into simpler ones, involving only one or two copies of the DEHP algebra each, instead
of dealing simultaneously with the four copies.

\begin{proposition}
One can show that the following relations hold
\begin{eqnarray}
&& \check  R^{(3)}(z_1/z_2) L^{(3)}(z_1) \otimes L^{(3)}(z_2)=L^{(3)}(z_2) \otimes L^{(3)}(z_1) \check R^{(3)}(z_1/z_2),\label{eq:2TASEP_RLL1}\\
&& \check  R^{(2)}(z_1/z_2) L^{(2)}(z_1) \otimes L^{(2)}(z_2)=L^{(2)}(z_2) \otimes L^{(2)}(z_1)\check R^{(2)}(z_1/z_2),\\
&& \check R^{(3)}(z_1/z_2) \widetilde L^{(3)}(z_1) \otimes \widetilde L^{(3)}(z_2)=\widetilde L^{(3)}(z_2) \otimes  \widetilde L^{(3)}(z_1)\check R^{(2)}(z_1/z_2),\\
&& \check  R^{(2)}(z_1/z_2) \widetilde L^{(2)}(z_1) \otimes \widetilde L^{(2)}(z_2)=\widetilde L^{(2)}(z_2) \otimes \widetilde L^{(2)}(z_1),
\end{eqnarray}
where we used the braided R-matrix for the single-species TASEP, built on the local operator $m^{(2)}$  (see \eqref{eq:2TASEP_markov2}): 
\begin{equation}
\check R^{(2)}(x)=1+(1-x)\, m^{(2)}.
\end{equation}
\end{proposition}
\proof
This is directly checked by a straightforward computation.
\finproof

Note that these ``generalized'' RLL relations not only allow us to factorize the difficulty to deal with the four-fold tensor space but also 
permit to reduce step by step the size of the $L$ matrices through intertwining relations involving the $R$ matrix of the 2 species model and 
of the single species model. These kind of relations are deeply inspired by the identification procedure exposed previously.

\begin{corollary}
These identities imply
\begin{eqnarray}
&& \check  R^{(3)}(z_1/z_2) \LL^{(3)}(z_1) \otimes \LL^{(3)}(z_2)=\LL^{(3)}(z_2) \otimes \LL^{(3)}(z_1)\check  R^{(2)}(z_1/z_2), \label{eq:2TASEP_RLL2}\\
&& \check  R^{(2)}(z_1/z_2) \LL^{(2)}(z_1) \otimes \LL^{(2)}(z_2)=\LL^{(2)}(z_2) \otimes \LL^{(2)}(z_1).\label{eq:2TASEP_RLL3}
\end{eqnarray}
\end{corollary}

\begin{corollary}
 The vector $\mathbf{A}(z)$ satisfies the ZF relation.
\end{corollary}
\proof
Using the factorization \eqref{eq:2TASEP_XLL} and applying the relations \eqref{eq:2TASEP_RLL2} and \eqref{eq:2TASEP_RLL3} successively, we 
prove the ZF relation \eqref{eq:2TASEP_ZF}.
\finproof

\begin{proposition}
The following relations hold
\begin{equation}\label{eq:2TASEP_k1}
\begin{aligned}
& \llangle W|_{123} \, K(z) \LL^{(3)}(1/z) = \llangle W|_{123}\,\LL^{(3)}(z)  K^{(2)}(z) 
\mbox{ ; } \llangle W|_{4} \, K^{(2)}(z) \LL^{(2)}(1/z) = \llangle W|_{4}\,\LL^{(2)}(z)
\qquad\\
&
\overline K(z) \LL^{(3)}(1/z)\,|V\rrangle_{123} = \LL^{(3)}(z) \overline K^{(2)}(z)\, |V\rrangle_{123} \mbox{ ; }
\overline K^{(2)}(z) \LL^{(2)}(1/z)\,|V\rrangle_{4} = \LL^{(2)}(z) \, |V\rrangle_{4} \;,
\end{aligned}
\end{equation}
\end{proposition}

Finally, relations \eqref{eq:2TASEP_k1} imply equations \eqref{eq:2TASEP_GZ}.

\begin{remark}
Note that $\widetilde L(z)$ can be obtained from $L(z)$. 
We define the transposition in the space of generators as follows
\begin{equation}
 e^t=d\mbox{ , }d^t=e\mbox{ , } A^t=A \mbox{ and } \llangle x|^t=|x\rrangle
\end{equation}
Let us remark that starting from an $L^{(3)}(z)$ solution to the relation \eqref{eq:2TASEP_RLL1}, the matrix
\begin{equation}
\overline{L}^{(3)}(z)= U L^{(3)}(1/z)^t U\mbox{ where } U=\begin{pmatrix}
                                                         0&0&1\\0&1&0\\1&0&0
                                                        \end{pmatrix}
\end{equation}
is also a solution of \eqref{eq:2TASEP_RLL1}. We have used the following property of the matrix $R^{(3)}(z)$
\begin{equation}
 U_1U_2 R_{21}^{(3)}(z) U_1U_2=R^{(3)}(z)\;.
\end{equation}
This symmetry was already pointed out in \eqref{eq:mASEP_symmetry_R}.
Starting from the realization \eqref{eq:2TASEP_lax3} for $L^{(3)}(z)$, one gets
\begin{equation}
 \overline{L}^{(3)}(z)= \begin{pmatrix} 1&d_1e_2&e_2\\
 0&\lambda A_2& \lambda e_1A_2\\
 d_2/z& d_1/z&1/z+\lambda A_1A_2
\end{pmatrix}\;.
\end{equation}
The trivial representation for the $e,d,A$ algebra is defined as $e=d=1$ and $A=0$. 
These values are consistent with the relation \eqref{eq:2TASEP_ed} and the definition of $A$.
In the $ \overline{L}^{(3)}(z)$ matrix, we may choose the trivial representation for the generators in the space 1 
(\textit{i.e.} $e_1=d_1=1$ and $A_1=0$). Changing the name of space 2 to space 3 
and putting $\lambda=1$, one establishes a link with the matrix $\widetilde L^{(3)}(z)$:
\begin{equation}
 \overline{L}^{(3)}(z)\Big|_{e_1=d_1=1,A_1=0,\lambda=1}= \begin{pmatrix} 1&e_3&e_3\\
 0&A_3&  A_3\\
 d_3/z& 1/z&1/z
\end{pmatrix} 
= \widetilde L^{(3)}(z)\begin{pmatrix}
 1&0&0\\
 0&1&1
\end{pmatrix}\;.
\end{equation}
The procedure to choose the trivial representation to get a simpler matrix has been used previously for the periodic case in \cite{CantiniDGW15}.
\end{remark}

Let us remark that taking the derivative of the RLL relations w.r.t. $z_1$ and setting $z_1=z_2=1$, we obtain generalized telescopic relations.
For instance, \eqref{eq:2TASEP_RLL2} implies that 
\begin{eqnarray}\label{eq:2TASEP_tel3}
 m^{(3)} \LL^{(3)}\otimes \LL^{(3)}-\LL^{(3)}\otimes \LL^{(3)} m^{(2)}=  \LL^{(3)'}\otimes \LL^{(3)}- \LL^{(3)}\otimes \LL^{(3)'}\;.
\end{eqnarray}
and \eqref{eq:2TASEP_RLL3} implies 
\begin{eqnarray}\label{eq:2TASEP_tel2}
 m^{(2)} \LL^{(2)}\otimes \LL^{(2)}=  \LL^{(2)'}\otimes \LL^{(2)}- \LL^{(2)}\otimes \LL^{(2)'}\;,
\end{eqnarray}
where 
\begin{equation}
  \LL^{(3)}:= \LL^{(3)}(1), \quad  \LL^{(3)'}:= \LL^{(3)'}(1), \quad \LL^{(2)}:=\LL^{(2)}(1), \quad \mbox{and} \quad \LL^{(2)'}:=\LL^{(2)'}(1).
\end{equation}

Remark that this generalized ``hat relation" already appeared in the solution of the periodic multi-species ASEP \cite{AritaAMP12}. 
The present matrix $\LL^{(3)}$ provides a new solution to this relation, adapted for the open boundaries case (see below). 

Combining equations \eqref{eq:2TASEP_tel3} and \eqref{eq:2TASEP_tel2} yields the usual bulk relation
\begin{equation}\label{eq:2TASEP_mXX}
m^{(3)}\,\mathbf{X} \otimes \mathbf{X} = 
\mathbf{X} \otimes \overline{\mathbf{X}}-\overline{\mathbf{X}} \otimes \mathbf{X},
\end{equation}
where
\begin{equation}
 \mathbf{X}=\LL^{(3)}\LL^{(2)}, \quad \mbox{and} \quad \overline{\mathbf{X}}=-\LL^{(3)}\LL^{(2)'}-\LL^{(3)'}\LL^{(2)}
\end{equation}
which consistent with $\mathbf{X}=\mathbf{A}(1)$ and $\overline{\mathbf{X}}=-\mathbf{A}'(1)$.

In the same way taking the derivative of generalized GZ relations \eqref{eq:2TASEP_k1} w.r.t. $z$ and setting $z=1$, we obtain generalized
telescopic relations on the boundaries.
 The following identities hold for $\LL^{(3)}$
\begin{eqnarray}\label{eq:2TASEP_b1}
&& \llangle W|_{123} \Big( B \LL^{(3)}-\LL^{(3)}  B^{(2)} \Big) =  -\llangle W|_{123}\LL^{(3)'}  \\
&& \label{eq:2TASEP_b1bis}
\Big( \overline B \LL^{(3)}- \LL^{(3)} \overline B^{(2)}\Big) |V\rrangle_{123}=  \LL^{(3)'}|V\rrangle_{123} \;.
\end{eqnarray}
Similar relations also exist for $\LL^{(2)}$
\begin{eqnarray}
 \llangle 0|_{4}\, B^{(2)} \LL^{(2)}=  -\llangle 0|_{4}\,\LL^{(2)'}  \\
 \overline B^{(2)} \LL^{(2)}|b\rrangle_{4}=  \LL^{(2)'} |b\rrangle_{4} \;.\label{eq:2TASEP_b2}
\end{eqnarray}
 Combining equations \eqref{eq:2TASEP_b1}-\eqref{eq:2TASEP_b2} leads us  to 
 the boundary relations
 \begin{equation} \label{eq:2TASEP_BXX}
\llangle W| B\,\mathbf{X}= \llangle W|\overline{\mathbf{X}}
\mbox{ and } \overline B \mathbf{X}|V\rrangle = -\overline{\mathbf{X}}|V\rrangle.
\end{equation}

 In \cite{Arita12} a multi-species ASEP with reflective open boundaries has been studied, and another solution (with vanishing hat operators)
of the generalized ``hat relation" \eqref{eq:2TASEP_tel3}
was provided to construct the steady state in a matrix product form. 
In this model the boundary counterparts of the ``hat relation" \eqref{eq:2TASEP_b1}-\eqref{eq:2TASEP_b1bis} 
were automatically satisfied because the boundary matrices $B$ and $\overline B$ vanish.

We recall that matrix product expression for the stationary probability vector reads
\begin{equation}
 \steady = \frac{1}{Z_L} 
 \llangle W|\boldsymbol{X} \otimes \boldsymbol{X}\otimes 
 \cdots\otimes \boldsymbol{X}\,|V\rrangle\,.
 \label{eq:2TASEP_M3}
\end{equation}
 The factorization $\mathbf{X}=\LL^{(3)}\LL^{(2)}$ leads to 
\begin{equation} \label{eq:2TASEP_Pfac}
\steady = \frac1{Z_L} \PP^{(3)}\    \PP^{(2)} \,,
\end{equation}
where
\begin{equation}
 \PP^{(3)}=
 \llangle W|_{123}\ \LL^{(3)}\otimes \cdots \otimes \LL^{(3)}\,|V\rrangle_{123}\mbox{ and }
 \PP^{(2)}=
 \llangle W|_{4}\ \LL^{(2)}\otimes\cdots\otimes \LL^{(2)}\,|V\rrangle_{4}\;.
\end{equation}
 Here,  $\PP^{(3)}$ is a $3^L \times 2^L$ matrix and $\PP^{(2)}$
  is a $2^L$-component vector so that $\cS$ is a $3^L$-component
  vector as expected. We also remark that $\PP^{(2)}$ (up to a normalisation)
 is identical to the steady-state vector of the one species TASEP
  with open boundaries. Therefore,  we have 
\begin{equation} \label{eq:2TASEP_Stat1TASEP}
  M^{(2)} \PP^{(2)}=0 \;,
\end{equation}
where $M^{(2)}$ is the Markov matrix of the one-species TASEP.
We also have the intertwining relation
\begin{equation}
 M^{(3)} \PP^{(3)}=\PP^{(3)} M^{(2)}.
\end{equation}

Finally, thanks to the factorization property \eqref{eq:2TASEP_XLL} of $\mathbf{A}(z)$ 
we know that
\begin{equation} \label{eq:2TASEP_facto_X}
\mathbf{X}=\mathbf{A}(1)=L^{(3)}\widetilde L^{(3)}L^{(2)}\widetilde L^{(2)},
\end{equation}
with
$L^{(3)}:=L^{(3)}(1)$, $\widetilde L^{(3)}:=\widetilde L^{(3)}(1)$, $L^{(2)}:=L^{(2)}(1)$ and $\widetilde L^{(2)}:=\widetilde L^{(2)}(1)$.
 The stationary state can thus be further decomposed as 
\begin{eqnarray}
\steady &=&  \frac1{Z_L} \PP^{(3)}\    \PP^{(2)}=\frac1{Z_L}  P^{(3)}\ \widetilde P^{(3)}\ P^{(2)}\ \widetilde P^{(2)}\;,
\end{eqnarray}
with 
\begin{eqnarray}
 &&P^{(3)}=\llangle W|_{12}\ L^{(3)}\otimes L^{(3)} \otimes\cdots \otimes L^{(3)}\,|V\rrangle_{12} \\
 &&\widetilde P^{(3)}=\llangle W|_{3}\ \widetilde L^{(3)}\otimes \widetilde L^{(3)}\otimes\cdots \otimes \widetilde L^{(3)}\,|V\rrangle_{3}\\
 &&P^{(2)}=\llangle W|_{4}\ L^{(2)}\otimes L^{(2)}\otimes\cdots \otimes L^{(2)}\,|V\rrangle_{4} \\
 &&\widetilde P^{(2)}=\begin{pmatrix}
            1 \\
             1
           \end{pmatrix}\otimes \begin{pmatrix}
            1 \\
             1
           \end{pmatrix}\otimes\dots \otimes \begin{pmatrix}
            1 \\
             1
           \end{pmatrix} \;.
\end{eqnarray}
Let us note that $P^{(3)}$ is a $3^L \times 3^L$ matrix, $\widetilde P^{(3)}$ is a $3^L \times 2^L$ matrix, $P^{(2)}$ is a $2^L\times 2^L$ matrix and
$\widetilde P^{(2)}$ is a $2^L$-component vector with constant components. 

\subsubsection{Computation of the normalization from the explicit representation}

We may now use the factorisation 
properties  of the previous subsection to calculate the normalisation $Z_L$ of the stationary
probabilities.  
The results we obtain are
\begin{equation} \label{eq:2TASEP_ZL}
 \begin{aligned}
 &Z_L = \frac{a}{a-b}\,\cZ_L(\alpha,1)\,\cZ_L(1,\beta)  \,\,\,\,  \mbox{ for } (M_1)
 \\[1ex]
 &Z_L = (1-ab)\,\cZ_L(\alpha,\beta)\,\cZ_L(1,\beta) \mbox{ for } (M_2)
 \end{aligned}
 \end{equation} 
 where $\cZ_L(\alpha,\beta)$ is the partition function of the
 open one-species TASEP with injection rate $\alpha$ and extraction
 rate $\beta$. Its exact expression \cite{DerridaEHP93} is given by
 \begin{equation}
  \label{eq:2TASEP_Norme1TASEP}
 \cZ_L(\alpha,\beta) =  \llangle
 a|(2+e+d)^L|b\rrangle =  \llangle a|(D +
 E)^L|b\rrangle =  \sum_{p=0}^L \frac{ p \, (2 L - p -1)!}{L! (L-p)!}
 \frac{ \left( \frac{1}{\alpha} \right)^{p+1} -  \left(
   \frac{1}{\beta} \right)^{p+1} } { \frac{1}{\alpha}  -
   \frac{1}{\beta}  } \,\llangle a|b\rrangle \,  .  
  \end{equation}
 From the matrix ansatz, we know that   
 \begin{equation}
 Z_L=\llangle
 W|\,(X_0+X_1+X_2)^L\,|V\rrangle.  
 \end{equation}
 Using  the factorization \eqref{eq:2TASEP_facto_X}, we obtain  
 \begin{equation}
 X_0+X_1+X_2=
 (1,1,1)\cdot \begin{pmatrix} X_{0} \\ X_{1} \\ X_{2}\end{pmatrix} \,
  = (1,1,1) L^{(3)}\,\widetilde L^{(3)} \LL^{(2)} . 
  \end{equation}
  We  first  compute
 \begin{equation}
 (1,1,1)\cdot L^{(3)} =
 \Big(1+a(A_1+d_1)A_2+d_2\,,\ e_1(1+d_2)
 +a A_2,\ 1+e_2\Big).
 \end{equation}
 Then, from  the relations $\llangle 1|(A+d)=\llangle 1|$
 and $\llangle 1|e=\llangle 1|$,  we deduce
 \begin{equation}
 \llangle  1|_1 (1,1,1)\cdot L^{(3)} =
 \Big(1+aA_2+d_2\,,\ 1+aA_2+d_2,\ 1+e_2\Big)\llangle
 1|_1
 \end{equation}
 This implies  that the space 1 drops out (because neither
 $\widetilde L^{(3)}$ nor $ \LL^{(2)}$ act on it).  Remarking that  
 \begin{eqnarray*}
 \Big(1+aA_2+d_2\,,\ 1+aA_2+d_2,\ 1+e_2\Big)\widetilde
 L^{(3)} &=&
  \Big(1+aA_2+d_2\,,\ 1+e_2\Big)\, 
\begin{pmatrix} 1 &1 & 0 \\ 0&0&1\end{pmatrix} 
 \widetilde L^{(3)} \qquad
 \\
 &=& \Big(1+aA_2+d_2\,,\ 1+e_2\Big)\, 
 \begin{pmatrix} 1 &A_3+e_3 \\ d_3&1\end{pmatrix}
\end{eqnarray*}
  and using $(A+e)|1\rrangle=|1\rrangle$
  and $d|1\rrangle=|1\rrangle$, we have 
\begin{equation}
 \begin{pmatrix} 1 &A_3+e_3 \\ d_3&1\end{pmatrix}|1\rrangle_3 
 = |1\rrangle_3 \begin{pmatrix} 1 &1 \\ 1&1\end{pmatrix}
 =|1\rrangle_3 \begin{pmatrix} 1  \\ 1\end{pmatrix} ( 1 ,1 )
 \end{equation}
 so that space 3 also drops out. Gathering the different results, we obtain 
 \begin{equation}
 \begin{cases}
 Z_L = \llangle 1|b/a\rrangle_1\  \llangle 0|(2+aA_2+d_2+e_2)^L
 |0\rrangle_2 \ \llangle 0 |1\rrangle_3\
 \llangle 0|(2+e_4+d_4)^L|b\rrangle_4
  \quad \mbox{ for } (M_1)
 \\[1ex]
 Z_L = \llangle 1|0\rrangle_1\  \llangle 0|(2+aA_2+d_2+e_2)^L|b\rrangle_2 \ \llangle 0 |1\rrangle_3\
 \llangle 0|(2+e_4+d_4)^L|b\rrangle_4
   \quad \quad \, 
  \mbox{ for }  (M_2). \end{cases}
 \end{equation}
  We conclude the derivation  of \eqref{eq:2TASEP_ZL} by using  \eqref{eq:2TASEP_Norme1TASEP} and by observing that 
\begin{equation}\label{eq:2TASEP_norm.ZL}
 \llangle 0|(2+aA+d+e)^L|b\rrangle=
 \frac{\llangle 0|b\rrangle}{\llangle a|b\rrangle}\cZ_L(\alpha,\beta) 
\end{equation}
because  the operators $\widetilde e=aA+e$ and $d$ obeys 
the same algebraic rules as $e$ and $d$, but
now  $\llangle 0|$ is a left eigenvector of   $\widetilde e$ with eigenvalue $a$. 

To summarize, the $2$-TASEP models studied here provide the first example of integrable multi-species open TASEP in which none the particles currents
of the species vanishes in the stationary state. The stationary state is analytically expressed in a matrix product form, which is written 
using an algebra involving $9$ generators. This translates into the presence of non-scalar 'hat operators'. The telescopic relations are not sufficient
to encode all the algebraic relations satisfied by these $9$ generators. These algebraic relations are provided by the ZF and GZ relations together 
with a convenient change of generators basis. It allows us to compute some physical quantities or some particular stationary weights using 
only these algebraic relations. 
The integrable structure of the model gives a guideline to construct an explicit representation of the matrix product algebra. A lot remains 
to be understood to construct such explicit representations for general integrable multi-species open (T)ASEP.
We will come back to these $2$-TASEP models in chapter \ref{chap:five} to study their properties in the thermodynamic limit. We will be in particular 
interested in their phase diagrams.

\subsection{An open multi-species SSEP}

This section is devoted to the study of diffusive multi-species open model that was introduced recently in \cite{Vanicat17}.
The interest of this model lies in the fact that it describes a system driven out-of-equilibrium by two reservoirs, and also in the fact
that there is an arbitrary fixed number of species. We will see that the algebraic structure of the non-equilibrium steady state provides
very convenient tools for physical quantities computations.
The discussion and results exposed here are deeply based on the content of \cite{Vanicat17}, with minor modifications. The reader is thus 
encouraged to refer to this paper for details.

\subsubsection{Presentation of the model} \label{subsubsec:mSSEP_presentation}

We consider a system involving $N$ species of particles 
which diffuse on a one dimensional lattice comprising $L$ sites. Each site $i$ can be in $N+1$ different states $\tau_i=0,\dots,N$ 
depending on its occupancy. More precisely, we set $\tau_i=s$
if the site at position $i$ carries a particle of species $s$, with $1\leq s \leq N$, and we set $\tau_i=0$ if the site is empty.
Hence a configuration of the lattice will be denoted by a $L$-uplet $\bm\tau=(\tau_1,\dots, \tau_L) \in \left\{0,\dots,N\right\}^L$.

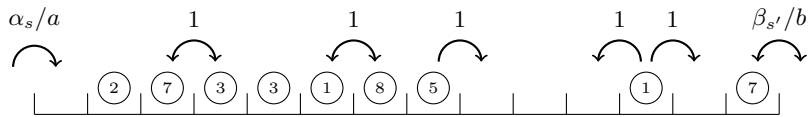
\begin{figure}[htb]
\begin{center}
 \begin{tikzpicture}[scale=0.7]
\draw (-2,0) -- (12,0) ;
\foreach \i in {-2,-1,...,12}
{\draw (\i,0) -- (\i,0.4) ;}
\draw[->,thick] (-2.4,0.9) arc (180:0:0.4) ; \node at (-2.,1.8) [] {\begin{footnotesize}$\alpha_s/a$\end{footnotesize}}; 
\draw  (-0.5,0.5) circle (0.3) [circle] {}; \node at (-0.5,0.5) [] {\begin{tiny}$2$\end{tiny}};
\draw  (0.5,0.5) circle (0.3) [circle] {}; \node at (0.5,0.5) [] {\begin{tiny}$7$\end{tiny}};
\draw  (1.5,0.5) circle (0.3) [circle] {}; \node at (1.5,0.5) [] {\begin{tiny}$3$\end{tiny}};
\draw  (2.5,0.5) circle (0.3) [circle] {}; \node at (2.5,0.5) [] {\begin{tiny}$3$\end{tiny}};
\draw  (3.5,0.5) circle (0.3) [circle] {}; \node at (3.5,0.5) [] {\begin{tiny}$1$\end{tiny}};
\draw  (4.5,0.5) circle (0.3) [circle] {}; \node at (4.5,0.5) [] {\begin{tiny}$8$\end{tiny}};
\draw  (5.5,0.5) circle (0.3) [circle] {}; \node at (5.5,0.5) [] {\begin{tiny}$5$\end{tiny}};
\draw  (9.5,0.5) circle (0.3) [circle] {}; \node at (9.5,0.5) [] {\begin{tiny}$1$\end{tiny}};
\draw  (11.5,0.5) circle (0.3) [circle] {}; \node at (11.5,0.5) [] {\begin{tiny}$7$\end{tiny}};
\draw[<->,thick] (0.6,1) arc (180:0:0.4); \node at (1.,1.8) [] {\begin{footnotesize}$1$\end{footnotesize}};
\draw[<->,thick] (3.6,1) arc (180:0:0.4); \node at (4.,1.8) [] {\begin{footnotesize}$1$\end{footnotesize}};
\draw[->,thick] (5.6,1) arc (180:0:0.4); \node at (6.,1.8) [] {\begin{footnotesize}$1$\end{footnotesize}};
\draw[<-,thick] (8.6,1) arc (180:0:0.4); \node at (9.,1.8) [] {\begin{footnotesize}$1$\end{footnotesize}};
\draw[->,thick] (9.6,1) arc (180:0:0.4); \node at (10.,1.8) [] {\begin{footnotesize}$1$\end{footnotesize}};
\draw[<->,thick] (11.6,1) arc (180:0:0.4) ; \node at (12.,1.8) [] {\begin{footnotesize}$\beta_{s'}/b$\end{footnotesize}}; 
 \end{tikzpicture}
 \end{center}
 \caption{Dynamical rules of the open multi-species SSEP.}
 \label{fig:mSSEP}
\end{figure}

The dynamics is stochastic. During an infinitesimal time $dt$, in the bulk, there is a probability $dt$ that a particle of a given species
jumps to its left or right neighbor site, providing that it is empty. 
There is also a probability $dt$ that two particles
of different species, located on two adjacent sites, exchange their positions.
At the left boundary, a particle of species $s'$ (or a hole) located on the first site can be replaced by a particle of species $s$ with 
probability $dt \times \alpha_s/a$. A particle of species $s'$ can also be absorbed in the left reservoir ({\it i.e} replaced by a hole) 
with probability $dt \times \alpha_0/a$
In the same way, at the right boundary,  a particle of species $s'$ (or a hole) located on the last site
can be replaced by a particle of species $s$ with probability $dt \times \beta_s/b$. A particle of species $s'$ can also be absorbed in the right
reservoir ({\it i.e} replaced by a hole) 
with probability $dt \times \beta_0/b$. See figure \ref{fig:mSSEP} for a graphical illustration.

 Later on, the parameters $\alpha_1,\dots,\alpha_N$ (respectively $\beta_1,\dots,\beta_N$) will be interpreted as the particles densities 
 at the left (respectively right) reservoir. The parameter $\alpha_0$ (respectively $\beta_0$) is interpreted as the density of holes at the 
 left (respectively right) reservoir. We have thus the constraints
 \begin{equation} \label{eq:mSSEP_sum_densities}
  \sum_{\tau=0}^N \alpha_{\tau} =1, \quad \mbox{and} \quad \sum_{\tau=0}^N \beta_{\tau} =1.
 \end{equation} 
 The number $a$ (respectively $b$) will be seen as the distance between the left reservoir 
 and the first site (respectively the distance between the right reservoir and the last site), the lattice spacing being one in the bulk.  

The update rules of the stochastic process described above are summarized in the following table where the rates
of the allowed transitions are depicted above the arrows:
 \begin{equation}
 \begin{array}{|c |c| c| }
 \hline \text{Left} & \text{Bulk} & \text{Right} \\
 \hline
\rule{0pt}{4ex} \tau'\, \xrightarrow{\ \alpha_{\tau}/a\ }\, \tau&  \tau'\tau\, \xrightarrow{\ 1\ } \,\tau\tau'&\tau'\, \xrightarrow{\ \beta_{\tau}/b\ } \,\tau\\ [1ex]
 0\leq \tau,\tau'\leq N&0\leq \tau,\tau' \leq N&0\leq \tau,\tau' \leq N\\ \hline
 \end{array}
 \end{equation}
Let us remark that the dynamics in the bulk is symmetric between the different species of particle and the holes: we cannot distinguish the different
species of particles or even particles and holes from the bulk dynamics. It is thus possible to relabel the species or even to interpret species $s$
as holes and holes as species $s$ (modifying accordingly the label of injection/extraction rates). This observation led us to present the expression
of physical quantities (computed below) in the most symmetric way. Note that we already stick with this symmetry willing in the presentation 
of the injection/extraction rates on the boundaries (we did note single out the notation of ''injection/extraction`` rates for the holes: we introduced
the reservoirs densities of holes $\alpha_0$ and $\beta_0$).
 
Let us stress that the injection and extraction rate of each species at the boundaries are not the most general. The particular model
presented here is motivated by the fact that it is integrable (see discussion below). It turns out that it has a nice physical interpretation.
Taking into account the constraints \eqref{eq:mSSEP_sum_densities}, we are left with $2\times N$ free parameters. 
For a generic choice of these parameters, the system will be driven out of equilibrium by the two reservoirs.

\begin{remark}
 The system will reach, in the long time limit, a thermodynamic equilibrium if and only if the reservoir densities of 
 each species of particle are the same on the left and on the right, namely: $\alpha_s=\beta_s$, for all $1\leq s \leq N$
 (that implies also, because of \eqref{eq:mSSEP_sum_densities}, that $\alpha_0=\beta_0$).  The detailed
 balance condition is indeed satisfied only in this case.
 \end{remark}

\subsubsection{Markov matrix and integrability.}

In this subsection we recall briefly the mathematical formalism (exposed precisely in chapter \ref{chap:two})
needed to write the probability density function of the model and its time
evolution (master equation) in a concise vector form. This will be also of great help to compute and express in a simple form the 
stationary probability density function.

Following the discussion of chapter \ref{chap:two}, we first attach to each site of the lattice a vector space $\mathbb{C}^{N+1}$
with basis $|0\rangle,|1\rangle,\dots,|N\rangle$, where
$|\tau\rangle=(\underbrace{0,\dots,0}_{\tau},1,\underbrace{0,\dots,0}_{N-\tau})^t$. The set of all configurations of the lattice is thus 
embedded in $\underbrace{\mathbb{C}^{N+1}\otimes\cdots\otimes\mathbb{C}^{N+1}}_{L}$ with natural basis $|\tau_1\rangle\otimes \cdots \otimes |\tau_L\rangle$,
where $\tau_i=0,1,...,N$.
We denote by $P_t(\tau_1,\dots,\tau_L)$ the probability for the system to be in configuration $(\tau_1,\dots,\tau_L)$ at time $t$. 
These probabilities can be encompassed in a single vector
\begin{equation}
 |P_t\rangle = \begin{pmatrix}
                P_t(0,\dots,0,0) \\
                P_t(0,\dots,0,1) \\
                \vdots \\
                P_t(N,\dots,N,N)
               \end{pmatrix}
             = \sum_{0\leq \tau_1,\dots,\tau_L \leq N} P_t(\tau_1,\dots,\tau_L)  \, |\tau_1\rangle\otimes \cdots \otimes |\tau_L\rangle.
\end{equation}
This allows us to write in a compact form the master equation, governing the time evolution of the probability density
\begin{equation} \label{eq:mSSEP_master_equation}
 \frac{d |P_t\rangle}{dt}=M|P_t\rangle,
\end{equation}
where the Markov matrix $M$ is given by
\begin{equation} \label{eq:mSSEP_Markov_matrix_decomposition}
 M=B_1+\sum_{i=1}^{L-1}m_{i,i+1} +\overline{B}_L.
\end{equation}
The matrices $B$, $\overline{B}$ and $m$ are the local jump operators. The indices denote the sites,
or equivalently the copies of $\mathbb{C}^{N+1}$, on which
the operators act non trivially (they act as the identity in the other copies). The matrix $B$ encodes the dynamics at the left boundary 
and acts on the first site as 
\begin{equation}
 B|\tau'\rangle =-\frac{1}{a}|\tau'\rangle+ \sum_{0\leq \tau\leq N} \frac{\alpha_s}{a}|\tau\rangle, \qquad 0\leq \tau'\leq N,
\end{equation}
which leads to the explicit expression 
\begin{equation} \label{eq:mSSEP_mat_B}
 B=\frac{1}{a}\begin{pmatrix}
     \alpha_0-1 & \alpha_0 & \alpha_0 & \hdots & \hdots & \alpha_0 \\
     \alpha_1 & \alpha_1-1 & \alpha_1 & \hdots & \hdots & \alpha_1 \\
     \alpha_2 & \alpha_2 & \alpha_2-1 & \hdots & \hdots & \alpha_2 \\
     \vdots &  \vdots &   & \ddots &  & \vdots \\
     \alpha_{N-1} &  \alpha_{N-1} & \hdots & \hdots &  \alpha_{N-1} -1 &  \alpha_{N-1} \\
     \alpha_N  & \alpha_N & \hdots & \hdots &  \alpha_N & \alpha_N-1
   \end{pmatrix}.
\end{equation}
In the same way, the matrix $\overline B$ encodes the dynamics at the right boundary 
and acts on the last site as 
\begin{equation}
 \overline B|\tau'\rangle = -\frac{1}{b}|\tau'\rangle+\sum_{0\leq \tau\leq N} \frac{\beta_s}{b}|\tau\rangle,  \qquad 0\leq \tau'\leq N,
\end{equation}
which leads to the explicit expression 
\begin{equation} \label{eq:mSSEP_mat_Bb}
 \overline B=\frac{1}{b}\begin{pmatrix}
     \beta_0-1 & \beta_0 & \beta_0 & \hdots & \hdots & \beta_0 \\
     \beta_1 & \beta_1-1 & \beta_1 & \hdots & \hdots & \beta_1 \\
     \beta_2 & \beta_2 & \beta_2-1 & \hdots & \hdots & \beta_2 \\
     \vdots &  \vdots &   & \ddots &  & \vdots \\
     \beta_{N-1} &  \beta_{N-1} & \hdots & \hdots &  \beta_{N-1} -1 &  \beta_{N-1} \\
     \beta_N  & \beta_N & \hdots & \hdots &  \beta_N & \beta_N-1
   \end{pmatrix}.
\end{equation}
Finally the matrix $m$ acts on two adjacent sites and encodes the dynamics in the bulk as
\begin{equation} \label{eq:bulk_jump_operator}
 m |\tau'\rangle \otimes |\tau\rangle = |\tau\rangle \otimes |\tau'\rangle - |\tau'\rangle \otimes |\tau\rangle.
\end{equation}
It can be expressed as $m=P-1$, where $P$ is the permutation operator, namely $P|v\rangle \otimes |w\rangle = |w\rangle \otimes |v\rangle$ if 
$|v\rangle,|w\rangle \in \mathbb{C}^{N+1}$.

\begin{remark} \label{rem:mSSEP_onespecies}
 The well-known SSEP model with one species of particles plus holes is recovered from this framework for $N=1$ 
 (one has then to identify species $1$ with holes). The present parameters are in this case related to the usual one $\alpha$, $\beta$, 
 $\gamma$ and $\delta$ by
 $\alpha_0=\gamma/(\alpha+\gamma)$, $\alpha_1=\alpha/(\alpha+\gamma)$, $\beta_0=\beta/(\beta+\delta)$, $\beta_1=\delta/(\beta+\delta)$,
 $a=1/(\alpha+\gamma)$ and $b=1/(\beta+\delta)$. Note that this corresponds to the change of variable already used to study the one species SSEP,
 see for instance \cite{Derrida07}.
 \end{remark}

 This model is integrable. We can indeed construct an $R$-matrix satisfying the Yang-Baxter equation (with additive spectral parameter)
 \eqref{eq:Yang_Baxter_additif}
 \begin{equation}
  R(z)=\frac{z+P}{z+1}.
 \end{equation}
It relates to the bulk local jump operator through $P.R'(0)=m$. It fulfills also the regularity, unitarity and Markov conditions.
We can also define two reflection matrices $K$ and $\overline{K}$ 
(that we already encountered in \eqref{eq:mSSEP_K_Kb_general}) satisfying the reflection equations 
\eqref{eq:reflection_equation_additif} and \eqref{eq:reflection_equation_reversed} (with additive spectral parameters) respectively 
\begin{equation}
 K(z)=1+\frac{2za}{z+a}B \quad \mbox{and} \quad \overline{K}(z)=1+\frac{2zb}{z-b}\overline{B}.
\end{equation}
They satisfy also the regularity, unitarity and Markov conditions. 
The boundary local jump operators can be recovered by taking the derivative $K'(0)/2=B$ and $-\overline{K}'(0)/2=\overline{B}$.

As presented in chapter \ref{chap:two}, these objects are the building blocks of the transfer matrix $t(z)$
(see \eqref{eq:transfer_matrix_open_additive}), which generates a family of commuting operators (including the Markov matrix).

\subsubsection{Matrix ansatz}

This section  is devoted to the construction of the stationary state of the model. More precisely we want to compute the vector $\steady$
which satisfies the stationary version of the master equation \eqref{eq:mSSEP_master_equation}, that is $M\steady =0$.

Following the Leitmotiv of this chapter, the integrability of the model can be exploited to express this vector
in a matrix product form, {\it i.e} the probability to observe a configuration 
$(\tau_1,\dots,\tau_L)$ in the steady state can be written as
\begin{equation} \label{eq:mSSEP_matrix_product}
 \mathcal{S}(\tau_1,\dots,\tau_L)=\frac{1}{Z_L}\llangle W|X_{\tau_1}X_{\tau_2}\dots X_{\tau_L} |V\rrangle, 
\end{equation}
where we recall that $Z_L=\llangle W|C^L|V\rrangle$ is a normalization, so that the entries of $\steady$ sum to $1$. We recall also the notation
\begin{equation} \label{eq:mSSEP_def_C}
C=X_1+\dots+X_N.
\end{equation}

For the matrix product state \eqref{eq:mSSEP_matrix_product} to compute the stationary distribution correctly, the operators $X_0,\dots,X_N$
and the boundary vectors $\llangle W|$ and $|V\rrangle$ have to satisfy precise algebraic relations.
As presented in this chapter, these relations take their root in the ZF and GZ relations.
More precisely the matrices $X_0,\dots,X_N$ are obtained through the relation
\begin{equation}
 \mathbf{X}=\begin{pmatrix}
             X_0 \\ X_1 \\ \vdots \\ X_N
            \end{pmatrix} = \mathbf{A}(0) 
\end{equation}
where $\mathbf{A}(z)$ satisfies the ZF relation $\check R(z_1-z_2) \mathbf{A}(z_1)\otimes\mathbf{A}(z_2)=\mathbf{A}(z_2)\otimes\mathbf{A}(z_1)$.
The boundary 
vectors $\llangle W|$ and $|V\rrangle$ are chosen to satisfy the GZ relations $\llangle W|K(z)\mathbf{A}(-z)=\llangle W|\mathbf{A}(z)$ 
and $\overline{K}(z)\mathbf{A}(-z)|V\rrangle=\mathbf{A}(z)|V\rrangle$.
The ZF relation implies the bulk telescopic relation 
\begin{equation} \label{eq:mSSEP_tel}
 m \mathbf X \otimes \mathbf X
= \mathbf X \otimes \overline{ \mathbf X}-\overline{ \mathbf X} \otimes \mathbf X,
\end{equation}
with $\overline{\mathbf{X}}=\mathbf{A}'(0)$. The GZ relations provide the boundary telescopic relations
\begin{equation} \label{eq:mSSEP_tel_W}
 \llangle W| B \mathbf X =  \llangle W| \overline{ \mathbf X},
\end{equation}
and 
\begin{equation} \label{eq:mSSEP_tel_V}
 \overline B \mathbf X |V \rrangle = -\overline{ \mathbf X}|V \rrangle.
\end{equation}
Altogether this ensures that the vector 
\begin{equation} \label{eq:mSSEP_steady}
 \steady=\frac{1}{Z_L}\llangle W| \mathbf X \otimes \mathbf X \otimes \dots \otimes \mathbf X |V\rrangle
\end{equation}
is the steady state of the model.

The first step is to guess an ''optimal`` (in the sense of the smallest number of generators) expansion for the vector $\mathbf{A}(z)$.
\begin{definition}
 We define the vector 
\begin{equation}
 \mathbf{A}(z)=\begin{pmatrix}
                z\lambda_0 +X_0 \\ z\lambda_1 +X_1 \\ \vdots \\ z\lambda_N +X_N
               \end{pmatrix} = z\overline{\mathbf{X}}+\mathbf{X},
\end{equation}
with
\begin{equation}
 \overline{\mathbf{X}}=\begin{pmatrix}
                        \lambda_0 \\ \lambda_1 \\ \vdots \\ \lambda_N
                       \end{pmatrix} = \mathbf{A}'(0).
\end{equation}
\end{definition}

Now we are equipped to write explicitly the ZF relation and identify the coefficients of the different powers of $z$ in each components. 
We obtain that the following property.
\begin{proposition}
The ZF relation is equivalent to the fact that
the operators $X_1,\dots,X_N$ belong to a Lie algebra\footnote{The Lie algebra \eqref{eq:mSSEP_lie_algebra} is not semi-simple since there is an abelian ideal of rank $N$ generated by the elements 
$\lambda_0 X_{\tau}-\lambda_{\tau} X_0$ for $1\leq \tau \leq N$. Hence it does not belongs to the well known classification of semi-simple Lie algebras.
It could be interesting to study the decomposition into solvable and semi-simple parts of this algebra but this is beyond the scope of this manuscript.}.
They satisfy the commutation relations
\begin{equation} \label{eq:mSSEP_lie_algebra}
 [X_{\tau},X_{\tau'}]=\lambda_{\tau} X_{\tau'}-\lambda_{\tau'} X_{\tau}, \qquad 0\leq \tau,\tau'\leq N,
\end{equation}
\end{proposition}

The last step is to write explicitly the GZ relations to derive the algebraic relations between the matrices $X_0,\dots,X_N$ and the boundary vectors
$\llangle W|$ and $|V\rrangle$. By again collecting the coefficients of the powers of $z$ in each components of the GZ relations we obtain the 
following properties.
\begin{proposition}
The action of the operators $X_{\tau}$ on the left boundary vector $\llangle W|$ is given by
\begin{equation}\label{eq:mSSEP_rel_W}
 \llangle W| \big( \alpha_{\tau} C-X_{\tau} \big) = a\lambda_{\tau} \llangle W|, \qquad 0\leq \tau\leq N,
\end{equation}
where $C$ is defined in \eqref{eq:mSSEP_def_C}.
Note that these $N$ relations are not all independent (the sum of these equations is trivial),
only $N-1$ are necessary.
In the same way the action of the operators $X_s$ on the right boundary vector $|V\rrangle$ read
\begin{equation} \label{eq:mSSEP_rel_V}
 \big( \beta_{\tau} C-X_{\tau} \big)|V \rrangle = -b\lambda_{\tau} |V \rrangle, \qquad 0\leq \tau\leq N.
\end{equation}
Again, only $N-1$ of these equations are independent.
\end{proposition}

The computation of length-$1$ words $\llangle W|X_{\tau}|V\rrangle$, for $0\leq \tau\leq N$, from the relations \eqref{eq:mSSEP_rel_W} and 
\eqref{eq:mSSEP_rel_V} yields the equations
\begin{equation}
 (\alpha_{\tau}-\beta_{\tau})\llangle W|C|V\rrangle=(a+b)\lambda_{\tau} \llangle W|V\rrangle, \qquad 0\leq \tau\leq N.
\end{equation}
These equations fix the values of the numbers $\lambda_{\tau}$ (up to a common factor) as
\begin{equation} \label{eq:mSSEP_lambda}
 \lambda_{\tau}=\alpha_{\tau}-\beta_{\tau}, \qquad 0\leq \tau \leq N.
\end{equation}
These particular values of the $\lambda_{\tau}$'s solve the previous set of equations and ensure that the length-$1,2,3$ words
$\llangle W|X_{\tau}|V\rrangle$, $\llangle W|X_{\tau}X_{\tau'}|V\rrangle$, $\llangle W|X_{\tau}X_{\tau'}X_{\tau''}|V\rrangle$
are completely fixed by the algebraic relations \eqref{eq:mSSEP_lie_algebra}, \eqref{eq:mSSEP_rel_W} and \eqref{eq:mSSEP_rel_V} (up to 
a global factor $\llangle W|V\rrangle$), are non vanishing and provide the correct stationary weights (this was check using formal computation software). 
 
Unfortunately, we were not able to find an explicit representation for the operators $X_{\tau}$ and the boundary vectors $\llangle W|$ and $|V\rrangle$.
However, we will show that the commutation relations \eqref{eq:mSSEP_lie_algebra} and the relations on the boundary vectors
\eqref{eq:mSSEP_rel_W} and \eqref{eq:mSSEP_rel_V} allow us to compute the currents and correlation functions, 
see subsection \ref{subsec:mSSEP_physical_quantities}, and to prove an additivity principle, see chapter \ref{chap:five}.

\begin{remark}
Once again the matrix ansatz solution of the usual SSEP with one species of particles and holes can be obtained
 for $N=1$, by doing the same change of parameters as mentioned in the remark \ref{rem:mSSEP_onespecies}, and setting
 $D= X_1/\lambda_1$ and $E=-X_0/\lambda_0=X_0/\lambda_1$. They satisfy $DE-ED=D+E$ and $\llangle W|(\alpha E-\gamma D)=\llangle W|$,
 $(\delta E-\beta D)|V\rrangle = -|V\rrangle$. 
\end{remark}

\begin{remark}
We already mentioned that the system reaches a thermodynamic equilibrium if and only if $\alpha_{\tau}=\beta_{\tau}$ for all $0\leq \tau\leq N$.
In this case we have $\lambda_{\tau}=0$ for all $0\leq \tau\leq N$, which implies that the operators $X_{\tau}$ commute one with each other and can be
chosen proportional to the identity operator. We hence set $X_{\tau}:=r_{\tau}$, with $r_0,\dots,r_N$ real numbers. It is straightforward to check
that $r_{\tau}=\alpha_{\tau}=\beta_{\tau}$ satisfy the boundary relations \eqref{eq:mSSEP_rel_W} and \eqref{eq:mSSEP_rel_V}. 

The steady state is given by
\begin{equation} \label{eq:mSSEP_steady_thermo}
 \steady=\begin{pmatrix}
          r_0 \\ \vdots \\ r_N
         \end{pmatrix} \otimes 
         \begin{pmatrix}
          r_0 \\ \vdots \\ r_N
         \end{pmatrix} \otimes \dots \otimes 
         \begin{pmatrix}
          r_0 \\ \vdots \\ r_N
         \end{pmatrix}.
\end{equation}
This shows that in the thermodynamic equilibrium, the occupation numbers $\tau_1,\dots,\tau_L$ are independent and 
identically distributed random variables.
\end{remark}

\subsubsection{Computation of physical quantities} \label{subsec:mSSEP_physical_quantities}

The algebraic structure of the stationary state described in \eqref{eq:mSSEP_lie_algebra}, \eqref{eq:mSSEP_rel_W} and \eqref{eq:mSSEP_rel_V} 
proves very powerful in the computation of physical 
quantities such as the correlation functions and the particle currents. The first step is to evaluate the normalization $Z_L$.

\begin{proposition}
Assuming that the scalar product of the boundary vectors $\llangle W|V\rrangle =1$, the normalization of the steady state defined
by $Z_L=\llangle W|C^L|V\rrangle$ is equal to
\begin{equation} \label{eq:mSSEP_normalisation}
 Z_L=\frac{\Gamma(a+b+L)}{\Gamma(a+b)},
\end{equation}
where the gamma function satisfies the functional relation $\Gamma(x+1)=x\Gamma(x)$.
\end{proposition}

\proof
We first remark that because of constraints \eqref{eq:mSSEP_sum_densities}, we have
\begin{equation}
 \sum_{\tau=0}^N\lambda_{\tau}=\sum_{\tau=0}^N \alpha_{\tau}-\sum_{\tau=0}^N\beta_{\tau}=1-1=0.
\end{equation}
It allows us to compute
\begin{equation}
 [X_{\tau},C]=\sum_{\tau'=0}^N [X_{\tau},X_{\tau'}]=\lambda_{\tau} \sum_{\tau'=0}^N X_{\tau'} -X_{\tau} \sum_{\tau'=0}^N\lambda_{\tau'},
\end{equation}
and leads to the very useful relation
\begin{equation} \label{eq:mSSEP_com_C}
 [X_{\tau},C]=\lambda_{\tau} C, \quad \mbox{or equivalently} \quad X_{\tau} C=C(X_{\tau}+\lambda_{\tau}).
\end{equation}
Using this equality $n$ times we obtain 
\begin{equation} \label{eq:mSSEP_com_Ci}
 X_{\tau} C^n=C^n (X_{\tau} +n\lambda_{\tau}).
\end{equation}
We are now equipped to compute the normalization 
\begin{eqnarray}
 Z_L & = & \llangle W| C^L |V \rrangle = \frac{a\lambda_0}{\alpha_0}Z_{L-1}+\frac{1}{\alpha_0} \llangle W|X_0 C^{L-1} |V \rrangle \nonumber \\
 & = & \frac{\lambda_0}{\alpha_0}(a+L-1)Z_{L-1}+\frac{1}{\alpha_0} \llangle W|C^{L-1}X_0 |V \rrangle \nonumber \\
 & = & \frac{\lambda_0}{\alpha_0}(a+b+L-1)Z_{L-1}+\frac{\beta_0}{\alpha_0}Z_L. \label{eq:mSSEP_calcul_ZL}
\end{eqnarray}
The first line is obtained thanks to relation \eqref{eq:mSSEP_rel_W} for $\tau=0$.  
We get the second line through relation \eqref{eq:mSSEP_com_Ci} for $\tau=0$ and $n=L-1$. 
The last equality is established using \eqref{eq:mSSEP_rel_V} for $\tau=0$. 
Finally \eqref{eq:mSSEP_calcul_ZL} can be rearranged and leads to the recursive relation
\begin{equation}
 Z_L=(a+b+L-1)Z_{L-1}.
\end{equation}
Keeping in mind that $Z_0=\llangle W|V\rrangle =1$, we can solve the previous relation and we obtain \eqref{eq:mSSEP_normalisation}.
\finproof 

We now turn to the study of the mean stationary current of the particles of species $s$ between site $i$ and $i+1$. It is defined by
the average algebraic number of particles of species $s$ crossing the bound between sites $i$ and $i+1$ per unit of time:
\begin{equation}
  \langle j_s \rangle =  \frac{\llangle W| C^{i-1}X_s(C-X_s)C^{L-i-1} |V \rrangle}{Z_L}-\frac{\llangle W| C^{i-1}(C-X_s)X_sC^{L-i-1} |V \rrangle}{Z_L}.
\end{equation}
Once again the algebraic structure of the steady state allows us to derive an exact expression for this quantity.
\begin{proposition}
The analytical expression of the mean current of particle of species $s$ between site $i$ and $i+1$ is given by
\begin{equation} \label{eq:mSSEP_current}
 \langle j_s \rangle=\frac{\lambda_s}{L-1+a+b},
\end{equation}
which is independent of the site $i$, as expected from the conservation of the particles number in the bulk.
Note that this expression remains also valid for the mean current of holes in the system
\begin{equation}
 \langle j_0 \rangle=\frac{\lambda_0}{L-1+a+b}.
\end{equation}
We recover immediately by summing these exact expression the property given by the exclusion constraint
\begin{equation}
 \langle j_0\rangle+\langle j_1\rangle+\dots+\langle j_N \rangle=0.
\end{equation}
\end{proposition}

\proof 
We present the proof for $0\leq \tau\leq N$, which includes the current of all the particle species and of the holes. We have
\begin{eqnarray*}
 \langle j_{\tau} \rangle & = &  \frac{\llangle W| C^{i-1}[X_{\tau},C-X_{\tau}]C^{L-i-1} |V \rrangle}{Z_L} 
  =   \frac{\llangle W| C^{i-1}[X_{\tau},C]C^{L-i-1} |V \rrangle}{Z_L} 
  =  \lambda_{\tau}\frac{Z_{L-1}}{Z_L},
\end{eqnarray*}
where the last equality is obtained thanks to \eqref{eq:mSSEP_com_C}.
Hence using \eqref{eq:mSSEP_normalisation} we get the desired expression \eqref{eq:mSSEP_current}.
\finproof 

\begin{remark}
In the thermodynamic equilibrium case, that is when $\lambda_{\tau}=0$ for all $\tau$, all the particles (and holes)
currents vanish, as expected.
\end{remark}

Other physical quantities of prime interest are the equal time correlation functions in the stationary state. We recall that
for a given configuration, we set $\rho_{\tau}^{(i)}=1$ if the site $i$ is in local configuration $\tau$ and $\rho_{\tau}^{(i)}=0$ else. 
The algebraic structure of the steady state, revealed by the matrix product formulation, offers a very efficient framework to 
compute the equal time multi-points correlation functions in the stationary state 
$\langle \rho_{s_1}^{(i_1)}\rho_{s_2}^{(i_2)}\dots \rho_{s_k}^{(i_k)}\rangle $, where $\langle \cdot \rangle$ stands for the expectation 
with respect to the stationary measure. We will compute below only the one and 
two points correlation functions, which are of particular interest for a physical point of view. In principle closed expressions for 
the higher order correlation functions can also be derived using the computational techniques presented below.

The one point function $\langle \rho_s^{(i)} \rangle$ (respectively $\langle \rho_0^{(i)} \rangle$) represents the mean density 
of particles of a given species $s$ (respectively of holes) at a given site $i$. 
It can be expressed through the matrix product formalism as
\begin{equation}
 \langle \rho_{\tau}^{(i)} \rangle =  \frac{\llangle W| C^{i-1}X_{\tau}C^{L-i} |V \rrangle}{Z_L}, \qquad 0\leq \tau\leq N.
\end{equation}
\begin{proposition}
Using the algebraic structure (see the proof below), it can be reduced to the closed expression
\begin{equation} \label{eq:mSSEP_density}
 \langle \rho_{\tau}^{(i)} \rangle =\frac{(b+L-i)\alpha_{\tau}+(a+i-1)\beta_{\tau}}{a+b+L-1}.
\end{equation}
Note that the density profile is the linear interpolation between the left reservoir with density $\alpha_{\tau}$ located at distance $a$ from 
the first site and the right reservoir with density $\beta_{\tau}$ located at distance $b$ from the last site. We recover the Fourier law.
\end{proposition}
\proof
The mean particle density of species $s$ (or holes) at site $i$ can be computed using the algebraic structure given by the matrix product form
\begin{eqnarray}
 \langle \rho_{\tau}^{(i)} \rangle & = & \frac{\llangle W| C^{i-1}X_{\tau}C^{L-i} |V \rrangle}{Z_L} 
  =  (L-i)\lambda_{\tau} \frac{Z_{L-1}}{Z_L}+\frac{\llangle W| C^{L-1}X_{\tau} |V \rrangle}{Z_L} \\
 & = & (b+L-i)\lambda_{\tau} \frac{Z_{L-1}}{Z_L}+\beta_{\tau} 
  =  \frac{(b+L-i)\alpha_{\tau}+(a+i-1)\beta_{\tau}}{a+b+L-1}.
\end{eqnarray}
The second equality is obtained using relation \eqref{eq:mSSEP_com_Ci}. We use then \eqref{eq:mSSEP_rel_V} to get the second line of the equation
and the last equality is established thanks to expression \eqref{eq:mSSEP_normalisation}.
\finproof

The two-point correlation function can also be written in a matrix product form 
\begin{equation} 
 \langle \rho_{\tau}^{(i)}\rho_{\tau'}^{(j)} \rangle=\frac{\llangle W| C^{i-1}X_{\tau}C^{j-i-1}X_{\tau'}C^{L-j} |V \rrangle}{Z_L}.
\end{equation}
It leads to the following proposition.
\begin{proposition}
We have a factorized expression for the connected two-point function
\begin{eqnarray}
\langle \rho_{\tau}^{(i)}\rho_{\tau'}^{(j)} \rangle_c & := & 
\langle \rho_{\tau}^{(i)}\rho_{\tau'}^{(j)} \rangle-\langle \rho_{\tau}^{(i)} \rangle \langle \rho_{\tau'}^{(j)} \rangle \nonumber \\
& = & -\lambda_{\tau}\lambda_{\tau'}\frac{(a+i-1)(b+L-j)}{(a+b+L-1)^2(a+b+L-2)}. \label{eq:mSSEP_2pts_function}
\end{eqnarray}
\end{proposition}
The formulas \eqref{eq:mSSEP_density} and \eqref{eq:mSSEP_2pts_function} are very similar to the ones derived for the usual one-species SSEP 
\cite{DerridaDR04,Derrida07} and appear as direct generalisation for the multi-species case.

\proof
For the two-point function, using again \eqref{eq:mSSEP_com_Ci} and \eqref{eq:mSSEP_rel_V}, we have 
\begin{eqnarray}
 \langle \rho_{\tau}^{(i)}\rho_{\tau'}^{(j)} \rangle & = & \frac{\llangle W| C^{i-1}X_{\tau}C^{j-i-1}X_{\tau'}C^{L-j} |V \rrangle}{Z_L} \\
 & = & \lambda_{\tau'}(L-j+b) \frac{\llangle W| C^{i-1}X_{\tau}C^{L-i-1}|V \rrangle}{Z_L}+\beta_{\tau'} \langle \rho_{\tau}^{(i)} \rangle
 \label{eq:mSSEP_two_points_function_intermediate}
\end{eqnarray}
Replacing $L$ by $L-1$ in the expression \eqref{eq:mSSEP_density} we obtain
\begin{eqnarray*}
 \frac{\llangle W| C^{i-1}X_{\tau}C^{L-i-1}|V \rrangle}{Z_L} & = & \frac{Z_{L-1}}{Z_L}\frac{(b+L-1-i)\alpha_{\tau}+(a+i-1)\beta_{\tau}}{a+b+L-2} \\
 & = & \frac{Z_{L-1}}{Z_L}\left(  \langle \rho_{\tau}^{(i)} \rangle -\lambda_{\tau}\frac{i-1+a}{(L-1+a+b)(L-2+a+b)} \right)
\end{eqnarray*}
Substituting back in \eqref{eq:mSSEP_two_points_function_intermediate} leads to
\begin{equation}
 \langle \rho_{\tau}^{(i)}\rho_{\tau'}^{(j)} \rangle = \langle \rho_{\tau}^{(i)} \rangle \langle \rho_{\tau'}^{(j)} \rangle
 -\lambda_{\tau}\lambda_{\tau'}\frac{(a+i-1)(b+L-j)}{(a+b+L-1)^2(a+b+L-2)},
\end{equation}
which concludes the proof.
\finproof 

To summarize, we introduced here an integrable multi-species open SSEP. The dynamics on the boundaries has a simple physical interpretation:
it models the interaction with a particles reservoirs with fixed densities of each species. The stationary state of the model is 
given analytically in terms of a simple matrix product expression. The generators indeed belong to a Lie algebra, which allows us to compute 
exactly physical quantities using only these algebraic relations (and without any explicit representation). We will encounter again this 
model in chapter \ref{chap:five} while studying its properties in the hydrodynamic limit.

In the whole present chapter, we dealt with matrix product expressions of the steady states of one dimensional exclusion processes. We encountered
several matrix ansatz algebras and we explained their connections to the integrable structure of the models. The matrix product 
expressions allowed us to compute exactly physical quantities in the stationary state. But these quantities were only related to 'static' observables
(such as particles densities) or to the average value of 'dynamic' observables (such as particles currents). To get access to the full
statistics of these 'dynamic' observables, we need (as explained in chapter \ref{chap:one}) to study a current counting deformation 
of the Markov matrices. We are going to see in the next chapter, how the matrix product states can be used to obtain the ground state of such deformed
Markov matrices. It will be achieved through the study of the specific example of the single species open ASEP.
As a byproduct, we will derive a matrix product expression of some symmetric polynomials: the Koornwinder polynomials (in a particular case).
This stresses, once more, the wide range of applications of the matrix product states.

\chapter{q-KZ equation and fluctuations of the current} \label{chap:four}

We saw in the previous chapter the efficiency of the matrix product states to encode exactly the 
stationary state of a wide range of stochastic processes. But we already stressed that the potential applications of such matrix product
states goes far beyond the computation of steady states associated to Markov matrices. We now present another application of 
the matrix product formalism in out-of-equilibrium statistical physics. The goal is to extend the previous matrix ansatz construction of
steady states to the computation of the fluctuations of the particles current. It was argued in chapter \ref{chap:one} that the generating 
function of the cumulants of the current can be computed through the ground state of a current counting deformation of the Markov matrix. 
We construct in the present chapter a further deformation of this ground state in a matrix product form that is conjectured to converge
to the ground state in some limit. 
The interest of such construction lies both on the physical and mathematical side. 

The physical motivation arises from the possible interpretation of the large deviation function of the current (that is obtained from the 
generating function through a Legendre transformation, see chapter \ref{chap:one}) as a dynamical generalization of the 
thermodynamic potential \cite{Touchette09} (this large deviation function is expected to exhibit singularities at the dynamical phase transitions). 
It is thus important to derive the simplest expression as possible of the current generating function.

The mathematical motivation arises from the use of matrix product states to construct solutions to quantum deformation of the 
Knizhnik-Zamolodchikov equations ($q$KZ equations) that were introduced in the context of representation theory of quantum affine algebra 
\cite{FrenkelR92}. These are a set of difference equations depending on the parameters
$t$ and $s$ for periodic systems and also on the parameters $a$, $b$, $c$ and $d$ for open boundaries systems.
These equations involve a vector, whose components depends on several variables, and relate precisely the permutation of variables in the vector
with the mixing of components of the same vector. Once again the integrability plays a central role in the setting of the qKZ equations and
ensures their consistency. It has been established that when the parameters satisfy the constraint
\begin{equation}
 t^{k+1}s^{r-1}abcd=1
\end{equation}
for some integers $k$ and $r$, their exist solutions to the $q$KZ equations that are polynomial in the variables
\cite{FeiginJMM02,FeiginJMM03,KasataniP07,ShigechiU07}. 
The parameter $k$ is called the rank and the parameter $r$ is called the level of the equation. 
These polynomial solutions are often related \cite{KasataniT07} to representations of the double quantum affine algebra \cite{Kasatani08,Noumi95,Sahi99}
and expressed in terms of Macdonald or Koornwinder polynomials \cite{FeiginJMM03,Cantini15,CantiniDGW15} 
The latter polynomials play a central role in combinatorics and representation theory \cite{Macdonald98}.

It is now well established that, in the context of stochastic processes, these $q$KZ equations for special case $s=1$
play a central role in the computation of the stationary states \cite{Cantini15,CantiniDGW15,Cantini16,CantiniGDGW16}.
This fact is deeply related (at least for stochastic interacting particles systems) to the construction exposed 
in chapter \ref{chap:three} with the ZF and GZ relations \cite{CrampeRV14,SasamotoW97}. The main novelties presented in this chapter are on one hand
an interplay between the parameter $s$ involved in the qKZ equations and a current counting deformation parameter $\xi$ of the Markov matrix and on
the other hand the matrix product construction of the associated polynomial solutions to the qKZ equations.
On the mathematical side it leads to a matrix product construction of some Koornwinder polynomials. On the physical side it allows 
to construct the ground state of the deformed Markov matrix and gives access to the generating function of the cumulants of the current. 
The qKZ equations point out once again a bridge between the theory of symmetric polynomials and physical observables in integrable stochastic processes
\cite{GarbaliDW16,BorodinP16}.
The results presented here are mainly extracted from the work \cite{FinnV16}.

\section{Current counting deformation and q-KZ equation}

We focus our study on the ASEP \cite{Spitzer70,Liggett85} for two reasons.
On the physical side, it has become over the last decades a paradigmatic model in
non-equilibrium statistical mechanics \cite{Derrida98,ChouMZ11}. It is an example of a physical system exhibiting a
macroscopic current in a stationary regime.  Such systems, which cannot be described by the usual thermal
equilibrium formalism, can be seen as the simplest out-of-equilibrium situation one can imagine
\cite{KatzLS84,KrapivskyRB10,SchmittmannZ95}. On the mathematical side, as seen in chapter \ref{chap:two}, 
the ASEP enjoys the property of being integrable.

Several exact results have been obtained for the current large deviation function of the ASEP using a deformed current-counting transition matrix
\cite{DerridaEM95,deGierE11,LazarescuM11,GorissenLMV12,Lazarescu13jphysA,LazarescuP14}.  The works
\cite{DerridaEM95,LazarescuM11,GorissenLMV12,Lazarescu13jphysA,LazarescuP14} build upon the matrix product
method \cite{DerridaEHP93}, used to compute the stationary state of the undeformed ASEP.  In this chapter we
continue in this vein, relying particularly on the integrability of the  ASEP \cite{Baxter82,Sklyanin88} and
the connection between integrability and the matrix product method \cite{SasamotoW97,CrampeRV14}.  We note
also the approach of \cite{deGierE11}, in which the Bethe ansatz was used to obtain the cumulant
generating function in the thermodynamic limit.

More precisely, the integrable structure of the ASEP gives rise to a connection with Hecke algebras, that are at 
the heart of the definition of Macdonald and
Koornwinder polynomials \cite{CantiniDGW15,Cantini15,CantiniGDGW16}.  The Macdonald polynomials are associated with the
periodic system, and in \cite{CantiniDGW15} this connection is exploited to derive a matrix product formula
for the symmetric Macdonald polynomials.  The connection between the open system and
Koornwinder polynomials was first identified in \cite{Cantini15}, then fully established in
\cite{CantiniGDGW16}.  However a matrix product formula, and the link to the general form of
the Koornwinder polynomials is still lacking.

Our aim is to exploit the integrable structure of the ASEP with deformed current-counting matrix,
to make a connection to the general form of the Koornwinder polynomials.  This in turns leads to a connection
between the symmetric Koornwinder polynomials and the generating function of the cumulants of the current.  We
do this by introducing scattering relations and $q$KZ equations with a further deformation, through which we
define a twice deformed ground state vector.  We give a matrix product construction of this ground state
vector and of the symmetric Koornwinder polynomial associated with it.  This leads us to
conjecture a beautiful relation between the generating function of the cumulants of the current, and a
certain limit of symmetric Koornwinder polynomials.

We consider here the ASEP with only {\it partial} asymmetry, but it would be interesting also to consider
the {\it totally} asymmetric simple exclusion process (TASEP)\footnote{That is with $q = 0$ in the model, as
defined below.}. The TASEP exhibits broadly similar behaviour to the general ASEP physically, but often the involved mathematical expressions are much
simpler.  The stationary state of the TASEP can be expressed in matrix product form \cite{DerridaEHP93}, but
was also given by directly solving certain recursion relations \cite{SchutzD93}.  The results related to
current fluctuations in the TASEP \cite{LazarescuM11} are also much simpler than those for the general ASEP.
In our notation, the TASEP relates to the $t \to \infty$ limit of the Koornwinder polynomials, which has been
previously studied \cite{OrrS13}. Thus it would
be interesting to study this limit in the TASEP context, and to see if any simplifications occur.

In the following subsections we recall briefly the main tools that are needed in this work: 
(i) the ASEP, the current-counting deformation of the associated Markov matrix, and the link with 
the generating function of cumulants of the current, (ii) the Hecke algebra and Koornwinder polynomials, (iii) the
integrable structure of the ASEP.
 
\subsection{Deformed Markov matrix}

\subsubsection{Definition of the model and Gallavotti-Cohen symmetry}

\begin{figure}[htb]
    \centering
    \includegraphics{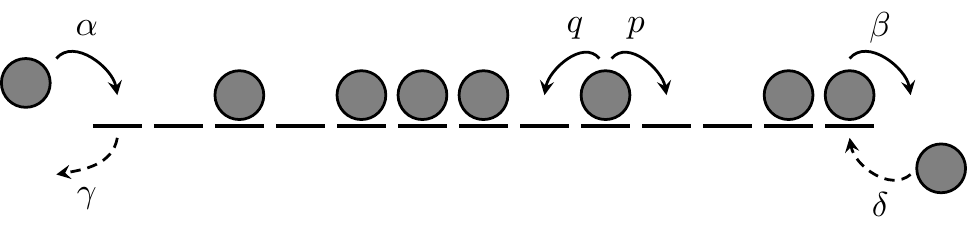}
    \caption{Transition rates for the ASEP with open boundaries.}
    \label{fig:aseprules}
\end{figure}

The open boundary ASEP is a stochastic model that we introduced previously through the local jump operators \eqref{eq:ASEP_m} and 
\eqref{eq:ASEP_B_Bb}.
 We recall that, in the bulk of the lattice particles hop right one site with rate
$p$, and left with rate $q$, so long as the target site is empty (the exclusion rule).  With open boundaries,
particles may enter and exit at the first and last sites.  If site $1$ is empty (occupied), a particle is
injected (extracted) with rate $\alpha$ ($\gamma$).  At site $L$, particles are extracted with rate $\beta$ and
injected with rate $\delta$. These rules are summarised in figure~\ref{fig:aseprules}.

We recall that we attach to each site $i$ a boolean variable $\tau_i \in \{0,1\}$ indicating if the site is empty ($\tau_i =0$) 
or occupied ($\tau_i = 1$).  The state of a single site is represented by a vector $\ket{\tau_i} \in
\mathbb{C}^2$, where
\begin{equation*}
    \ket{0} = \begin{pmatrix}
        1
        \\
        0
    \end{pmatrix},
    \qquad
    \ket{1} = \begin{pmatrix}
        0
        \\
        1
    \end{pmatrix}.
\end{equation*}
The state of the lattice is given by a vector
$\ket{\bm\tau} = \ket{\tau_1, \ldots, \tau_L} \in \left(\mathbb{C}^2\right)^{\otimes L}$ with
\begin{equation*}
    \ket{\tau_1, \ldots \tau_L} = \ket{\tau_1} \otimes \ldots \otimes \ket{\tau_L}.
\end{equation*}
The ASEP transition rates are then encoded in the transition matrix $M(\xi = 1)$, where\footnote{The unusual
normalisation is to ease the notation in later sections.}
\begin{equation}
    M(\xi)
    =
    \sqrt{\alpha \gamma} B_1(\xi) + \sum_{i=1}^{L-1} \sqrt{p q} w_{i,i+1} + \sqrt{\beta \delta} \overline{B}_L,
\end{equation}
and
\begin{equation}
    \sqrt{\alpha \gamma} B(\xi) = \begin{pmatrix}
        -\alpha & \xi^{-1} \gamma \\
        \xi \alpha & -\gamma
    \end{pmatrix},
    \qquad
    \sqrt{\beta \delta} \overline{B} = \begin{pmatrix}
        -\delta & \beta \\
        \delta & -\beta
    \end{pmatrix},
    \qquad
    \sqrt{p q} w = \begin{pmatrix}
        0 & 0 & 0 & 0
        \\
        0 & -q & p & 0
        \\
        0 & q & -p & 0
        \\
        0 & 0 & 0 & 0
    \end{pmatrix}.
\end{equation}
The indices on the matrices indicate the sites on which they act non trivially. The matrix $M(\xi)$ is stochastic only
for $\xi = 1$, but the introduction of this fugacity allows the study of the current generating function, as will be
discussed below.  The stochastic matrix $M(\xi = 1)$ has a unique eigenvector with eigenvalue $0$, that is
\begin{equation*}
    M(1) \ket{\Psi} = 0,
    \qquad
    \ket{\Psi} = \sum_{\bm\tau} \psi_{\bm\tau} \ket{\bm\tau}.
\end{equation*}
Normalizing this vector gives the stationary distribution of the system: letting
\begin{equation*}
    \mathcal{Z} = \langle 1 | \Psi \rangle,
    \qquad
    \bra{1} = (1, 1)^{\otimes L},
\end{equation*}
the stationary probability of a configuration $\bm\tau$ is
\begin{equation*}
   \cS(\bm{\tau}) = \frac{1}{\mathcal{Z}} \psi_{\bm\tau}.
\end{equation*}

If we now consider the deformed transition matrix $M(\xi)$, then the deformed ground state vector
satisfies
\begin{equation*}
    M(\xi) \ket{\Psi(\xi)} = \Lambda_0(\xi) \ket{\Psi(\xi)},
\end{equation*}
with $\Lambda_0(\xi) \to 0$ as $\xi \to 1$. As already stressed in chapter \ref{chap:one}, the eigenvalue $\Lambda_0(\xi)$ 
for general $\xi$ is an object of prime interest in the context of out-of-equilibrium statistical physics, because of its connection to the
generating function of the cumulants of the current, $E(\mu)=\Lambda_0(e^{\mu})$. It has been shown recently
\cite{LazarescuM11,GorissenLMV12,Lazarescu13jphysA,LazarescuP14} that the cumulants of the current for finite systems can be extracted
analytically at any order at the price of solving non-linear implicit equations.
 We recall that the Legendre transformation
of $E(\mu)$ provides the large deviation function of the particle current in the stationary state,
\begin{equation*}
G(j)=\min\limits_{\mu}\big(\mu j-E(\mu)\big),
\end{equation*}
which is expected to be a possible generalisation of thermodynamic potential to non-equilibrium systems
\cite{Touchette09}.
In words, $G(j)$ describes the non-typical fluctuations of the
mean particle flux\footnote{More precisely, we recall that, if we denote by $Q_T$ the algebraic number of particles exchanged between
the system and the left reservoir during the time interval $[0,T]$, then $G(j)$ is characterised by
$P(Q_T/T=j)\sim \exp(-TG(j))$ for large $T$.}.
   The reader can refer to chapter \ref{chap:one} or to \cite{GorissenLMV12} for more details.

The eigenvalue $\Lambda_0(\xi)$ is invariant under the Gallavotti--Cohen symmetry \cite{GallavottiC95,LebowitzS99}
\begin{equation}  \label{eq:Koor_GC_sym}
    \xi \to \xi' = \frac{\gamma \delta}{\alpha \beta} \left(\frac{q}{p}\right)^{L-1} \xi^{-1}.
\end{equation}
This translates immediately into a symmetry on the large deviation function of the particle current.
\begin{equation}
 G(j)-G(-j)=j\ln \left(\frac{\gamma \delta}{\alpha \beta} \left(\frac{q}{p}\right)^{L-1}\right).
\end{equation}
This symmetry arises from the relation between the transition matrix and its transpose:
\begin{equation}  \label{eq:Koor_GC_Transpose}
    M(\xi') = U_{\text{GC}} M(\xi)^T U_{\text{GC}}^{-1},
\end{equation}
where $\xi'$ is as defined in \eqref{eq:Koor_GC_sym} and
\begin{equation}  \label{eq:Koor_U_GC}
    U_{\text{GC}}
    =
    \begin{pmatrix}
        1 & 0
        \\
        0 & \frac{\delta}{\beta}\left(\frac{q}{p}\right)^{L-1}
    \end{pmatrix}
    \otimes
    \begin{pmatrix}
        1 & 0
        \\
        0 & \frac{\delta}{\beta}\left(\frac{q}{p}\right)^{L-2}
    \end{pmatrix}
    \otimes \dots \otimes
    \begin{pmatrix}
        1 & 0
        \\
        0 & \frac{\delta}{\beta}
    \end{pmatrix}.
\end{equation}
The relation \eqref{eq:Koor_GC_Transpose} implies that $M(\xi)$ and $M(\xi')$ have the same spectrum and thus the largest
eigenvalue is the same:
\begin{equation*}
    \Lambda_0(\xi') = \Lambda_0(\xi).
\end{equation*}
As a further consequence of this symmetry, given a solution of the left eigenvalue problem
\begin{equation*}
    \bra{\Phi({\xi})} M(\xi) = \Lambda(\xi) \bra{\Phi(\xi)}.
\end{equation*}
there is a corresponding solution of the right eigenvalue problem
\begin{equation*}
    M(\xi') \ket{\Psi({\xi'})} = \Lambda(\xi) \ket{\Psi(\xi')},
\end{equation*}
with
\begin{equation*}
    \ket{\Psi(\xi')} = U_{\text{GC}} \ket{\Phi(\xi)},
\end{equation*}
and $\xi'$ as defined in \eqref{eq:Koor_GC_sym}.
Note that here and in the following we use the convention $\bra{\cdot}^T=\ket{\cdot}$, where $^T$ denotes the usual transposition.
We will explain in the following sections the connection that can be made between the ground state $\ket{\Psi(\xi)}$ and the 
theory of Koornwinder polynomials that we present now.

\subsubsection{Integrability of the deformed Markov matrix}

We recall that the ASEP is an integrable model -- the deformed transition matrix $M(\xi)$ belongs to an infinite family of
commuting matrices \cite{Sklyanin88} (see also \cite{CrampeRV14,CantiniGDGW16}). 
The generating function of these commuting matrices is called the transfer matrix. The key ingredients to construct this transfer matrix are
matrices $\check{R}(z)$, $K(z)$, and $\overline{K}(z)$ satisfying the braided Yang--Baxter relation \eqref{eq:braided_ybe},
the reflection equation \eqref{eq:reflection_equation} and the reversed reflection equation \eqref{eq:reflection_equation_reversed} respectively.
They also satisfy the unitarity and regularity properties.

For the open boundary ASEP the matrices are given by
\begin{align}
    \check{R}(z) & = 1 + r(z) m,
    \label{eq:Koor_Rcheck}
    \\
    K(x; \xi) & = 1 + k(z; t_0^{1/2},u_0^{1/2}) B(\xi),
    \label{eq:Koor_K}
    \\
    \overline{K}(z) & = 1 + k(z^{-1}; t_L^{1/2}, u_L^{1/2}) \overline{B},
    \label{eq:Koor_Kbar}
\end{align}
where
\begin{equation}
    r(z) = \frac{z - 1}{t^{-1/2} z - t^{1/2}},
    \qquad
    k(z; t_i^{1/2}, u_i^{1/2}) = \frac{z^2 - 1}{t_i^{-1/2} z^2 - (u_i^{1/2} - u_i^{-1/2}) z - t_i^{1/2}},
\end{equation}
and $w$, $B(\xi)$, $\overline{B}$ are written in terms of the Hecke parameters as
\begin{equation}
B(\xi) = \begin{pmatrix}
    -t_0^{1/2} & \xi^{-1} t_0^{-1/2} \\
    \xi t_0^{1/2} & -t_0^{-1/2}
\end{pmatrix},
\qquad
\overline{B} = \begin{pmatrix}
    -t_L^{-1/2} & t_L^{1/2} \\
    t_L^{-1/2} & -t_L^{1/2}
\end{pmatrix},
\qquad
m = \begin{pmatrix}
        0 & 0 & 0 & 0
        \\
        0 & -t^{-1/2} & t^{1/2} & 0
        \\
        0 & t^{-1/2} & -t^{1/2} & 0
        \\
        0 & 0 & 0 & 0
\end{pmatrix}.
\label{eq:Koor_BBBarwHecke}
\end{equation}
We will write $K(z)$ for $K(z; \xi)$, except when it is necessary to distinguish between values of $\xi$.
Note that the local (physical) transition matrices are obtained from \eqref{eq:Koor_Rcheck} -- \eqref{eq:Koor_Kbar} through
\begin{equation*}
    \sqrt{p q} w = (q-p)\check{R}'(1),
    \qquad
    \sqrt{\alpha \gamma} B(\xi) = \frac{1}{2}(q - p) K'(1; \xi),
    \qquad
    \sqrt{\beta \delta} \overline{B} = -\frac{1}{2}(q - p) \overline{K}'(1),
\end{equation*}
and the Gallavotti--Cohen symmetry on the transition matrices now reads
\begin{equation*}
\begin{aligned}
    \check{R}_i(z) & = U_{\text{GC}} \check{R}_i(z)^T U_{\text{GC}}^{-1},
    \\
    K_1(z; \xi') & = U_{\text{GC}} K_1(z; \xi)^T U_{\text{GC}}^{-1},
    \\
    \overline{K}_L(z) & = U_{\text{GC}} \overline{K}_L(z)^T U_{\text{GC}}^{-1},
\end{aligned}
\end{equation*}
with $\xi'$ and $U_{\text{GC}}$ defined in \eqref{eq:Koor_GC_sym} and \eqref{eq:Koor_U_GC} respectively.

\subsection{Second deformation}

\subsubsection{Scattering matrices}

Instead of the transfer matrix approach, one can define scattering matrices \cite{ZinnJustin07,Cantini15},
although the two methods are closely related.  We first define a modified left boundary matrix
\begin{equation}
    \widetilde{K}(z) = K(s^{-1/2}z),
    \label{eq:Koor_Ktilde}
\end{equation}
in order to introduce the Hecke parameter $s$.  The matrix $\widetilde{K}(z)$ satisfies
deformed unitary and reflection relations
\begin{align*}
    \widetilde{K}(s z) \widetilde{K}\left(z^{-1}\right) & = 1,
    \\
    \check{R}_1(z_2/z_1) \widetilde{K}_1(z_2) \check{R}_1(s^{-1} z_1 z_2) \widetilde{K}_1(z_1)
    & =
    \widetilde{K}_1(z_1) \check{R}_1(s^{-1} z_1 z_2) \widetilde{K}_1(z_2) \check{R}_1(z_2/z_1),
\end{align*}
and has the Gallavotti--Cohen symmetry
\begin{equation*}
    \widetilde{K}_1(z; \xi') = U_{\text{GC}} \widetilde{K}_1(z; \xi)^T U_{\text{GC}}^{-1}.
\end{equation*}

For $1 \le i \le L$, define the scattering matrices
\begin{equation} \label{eq:Koor_scatteringmatrix}
\begin{aligned}
    \mathcal{S}_i(\mathbf{z}) =
        & \check{R}_{i-1}\left(\frac{z_{i-1}}{sz_{i}}\right) \ldots \check{R}_1\left(\frac{z_1}{sz_i}\right)
        \\
        & \cdot \widetilde{K}_1\left(\frac{1}{z_i}\right)
        \\
        & \cdot \check{R}_1\left(\frac{1}{z_i z_1}\right) \ldots \check{R}_{i-1}\left(\frac{1}{z_i z_{i-1}}\right)
          \cdot \check{R}_i\left(\frac{1}{z_i z_{i+1}}\right) \ldots \check{R}_{L-1}\left(\frac{1}{z_i z_L}\right)
        \\
        & \cdot \overline{K}_L(z_i)
        \\
        & \cdot \check{R}_{L-1}\left(\frac{z_L}{z_i}\right) \ldots \check{R}_i\left(\frac{z_{i+1}}{z_{i}}\right).
\end{aligned}
\end{equation}
Using the Yang--Baxter, reflection relations, and unitarity, we see that the scattering matrices satisfy a
deformed commutation relation
\begin{equation*}
    \mathcal{S}_i(\ldots, s z_j, \ldots) \mathcal{S}_j(z_1, \ldots, z_L)
    =
    \mathcal{S}_j(\ldots, s z_i, \ldots) \mathcal{S}_i(z_1, \ldots, z_L).
\end{equation*}
When $s = 1$, $[\mathcal{S}_i(\mathbf{z}), \mathcal{S}_j(\mathbf{z})] = 0$ for all $i$, $j$, and in fact in that case there
is a direct relation to the transfer matrix approach \cite{Sklyanin88,CrampeRV14,CantiniGDGW16}\footnote{Note that for $s \neq1$ there is
no obvious link between the scattering matrices and the usual transfer matrix, as far as we know.}: 
\begin{equation}
 \mathcal{S}_i(\mathbf{z})=t(z_i | \mathbf{z}),
\end{equation}
where $t(z | \mathbf{z})$ is the usual transfer matrix with spectral parameter $z$ and inhomogeneity parameters $ \mathbf{z}=z_1,\dots,z_L$
(the reader may refer to chapter \ref{chap:two} for a precise definition).
 
At $s=1$, we also have
the important relations
\begin{equation} \label{eq:Koor_properties_scatteringmatrices}
    \mathcal{S}_i(\mathbf{z})|_{s = z_1 = \ldots = z_L = 1} = 1,
    \qquad
    \frac{\partial}{\partial z_i} \mathcal{S}_i(\mathbf{z})|_{s = z_1 = \ldots = z_L = 1}
    = \frac{2}{p-q} M(\xi).
\end{equation}

Considering $s$ general again, we would like to find solutions of the scattering relation
\begin{equation} \label{eq:Koor_scattering}
    \mathcal{S}_i(\mathbf{z}) \ket{\Psi(\ldots, z_i, \ldots)} = \ket{\Psi(\ldots, s z_i, \ldots)},
\end{equation}
where
\begin{equation}
    \ket{\Psi(\mathbf{z})} = \sum_{\bm\tau} \psi_{\bm\tau}(\mathbf{z}) \ket{\bm\tau}.
\end{equation}
Taking the derivative of \eqref{eq:Koor_scattering} with respect to $z_i$ and specialising with $z_1 = \ldots = z_L
= s = 1$, this would imply
\begin{equation}  \label{eq:Koor_MxiOnPsi}
    M(\xi) \ket{\Psi(\mathbf{1})} = 0.
\end{equation}
For $\xi = 1$ this is the unnormalized stationary vector of the ASEP with eigenvalue 0.  For $\xi \ne 1$, the
ground state eigenvalue is non-zero, and so \eqref{eq:Koor_MxiOnPsi} should not have a solution at this point (that
is $s = 1$, $\xi \ne 1$).  However, in section \ref{sec:Koor_currentKoornwinder}, we will discuss how there could be
a solution of \eqref{eq:Koor_scattering} for $s \to 1$, $\xi \ne 1$, and how it relates to the current-counting
eigenvalue.

\subsubsection{q-KZ equation} 

It can be checked directly that sufficient conditions for a solution of the scattering relation
\eqref{eq:Koor_scattering} are
\begin{align}
    & \check{R}_i(z_{i+1}/z_{i}) \ket{\Psi(\ldots, z_i, z_{i+1}, \ldots)} = \ket{\Psi(\ldots, z_{i+1}, z_i, \ldots)},
      \qquad 1 \le i \le L - 1,
      \label{eq:Koor_qKZi}
    \\
    & \widetilde{K}_1(z_1^{-1}) \ket{\Psi(z_1^{-1}, z_2, \ldots)} = \ket{\Psi(s z_1, z_2, \ldots) },
      \label{eq:Koor_qKZ1}
    \\
    & \overline{K}_L(z_L) \ket{\Psi(\ldots, z_{L-1}, z_L)} = \ket{\Psi(\ldots, z_{L-1}, 1/z_L) }.
      \label{eq:Koor_qKZL}
\end{align}
Note that the Yang--Baxter, reflection, and unitary conditions ensure the consistency of this definition.  We
will refer to \eqref{eq:Koor_qKZi} -- \eqref{eq:Koor_qKZL}  as the $q$KZ equations, although in our notation the $q$ has
been replaced by the parameter $s$.  These $q$-difference equations were first introduced in \cite{FrenkelR92}
and appear as $q$-deformation of the KZ equations \cite{KnizhnikZ84}.

Motivated by the connection to the ASEP stationary state, we make the following definition:
\begin{definition}
We call a solution
\begin{equation*}
    \ket{\Psi(\mathbf{z}; s, \xi)} = \sum_{\bm\tau} \psi_{\bm\tau}(\mathbf{z}; s, \xi) \ket{\bm\tau}
\end{equation*}
of equations \eqref{eq:Koor_qKZi} -- \eqref{eq:Koor_qKZL} a twice deformed inhomogeneous ground state vector, with deformation
parameters $s$ and $\xi$.
\end{definition}
As indicated at the end of the previous section, such a vector with $s = \xi = 1$ is the inhomogeneous ground
state vector of the open boundary ASEP, and can be constructed in matrix product form \cite{CrampeRV14,CrampeMRV15inhomogeneous} or
from specialised non-symmetric Koornwinder polynomials \cite{CantiniGDGW16}.  We will show that more general
solutions exist when $s$ and $\xi$ obey some relations.

We define left $q$KZ equations
\begin{align}
    & \bra{\Phi(\ldots, z_i, z_{i+1}, \ldots)} \check{R}_i(z_{i+1}/z_{i}) = \bra{\Phi(\ldots, z_{i+1}, z_i, \ldots)},
      \qquad 1 \le i \le L - 1,
      \label{eq:Koor_leftqKZi}
    \\
    & \bra{\Phi(z_1^{-1}, z_2, \ldots)} \widetilde{K}_1(z_1^{-1}) = \bra{\Phi(s z_1, z_2, \ldots) },
      \label{eq:Koor_leftqKZ1}
    \\
    & \bra{\Phi(\ldots, z_{L-1}, z_L)} \overline{K}_L(z_L) = \bra{\Phi(\ldots, z_{L-1}, 1/z_L) },
      \label{eq:Koor_leftqKZL}
\end{align}
with
\begin{equation}
    \bra{\Phi(\mathbf{z})} = \sum_{\bm\tau} \phi_{\bm\tau}(\mathbf{x}) \bra{\bm\tau}.
    \label{eq:Koor_vec_Phi}
\end{equation}
These would imply a solution of a left scattering equation (analogous to \eqref{eq:Koor_scattering}) with a
scattering matrix, defined by reversing the order of matrices in the definition \eqref{eq:Koor_scatteringmatrix}.

The Gallavotti--Cohen symmetry allows us to relate solutions of the left and right $q$KZ equations.
\begin{lemma}
For any vector $\bra{\Phi(\mathbf{z}; s, \xi)}$ satisfying the left $q$KZ equations \eqref{eq:Koor_leftqKZi} --
\eqref{eq:Koor_leftqKZL}, the vector
\begin{equation}
    \ket{\Psi(\mathbf{z}; s, \xi')} = U_{\text{GC}} \ket{\Phi(\mathbf{z}; s, \xi)},
\end{equation}
with
\begin{equation}
    \xi' = t_0^{-1} t_L^{-1} t^{-(L-1)} \xi^{-1},
\end{equation}
is a solution to the right $q$KZ equations \eqref{eq:Koor_qKZi} -- \eqref{eq:Koor_qKZL}.
\label{lemma:Koor_leftrightsol}
\end{lemma}
\proof
This is checked by transposing the left $q$KZ equations and using the Gallavotti--Cohen symmetry on the
$\check{R}$, $\widetilde{K}$ and $\overline{K}$ matrices.
\finproof
         
\section{Koornwinder polynomials and link with q-KZ equation}

We now introduce the symmetric and non-symmetric Koornwinder polynomials, which form the other main theme of
this chapter. The symmetric Koornwinder polynomials \cite{Koornwinder92, vanDiejen96} are a family of
multivariate orthogonal polynomials generalising the Askey--Wilson polynomials.  The symmetric
Koornwinder polynomials can be constructed from their non-symmetric counterparts, which arise from the
polynomial representation of the affine Hecke algebra of type $C_L$ \cite{Sahi99, Stokman00}.

\subsection{Non-symmetric Koornwinder polynomials}

\subsubsection{Hecke algebra and Noumi representation} 

The affine Hecke algebra of type $C_L$ is generated by elements $T_0, T_1, \ldots, T_L$, with parameters
$t^{1/2}$, $t_0^{1/2}$ and $t_L^{1/2}$.  The generators satisfy the quadratic relations,
\begin{align*}
    & \left(T_0 - t_0^{1/2}\right)\left(T_0 + t_0^{-1/2}\right) = 0,
    \\
    & \left(T_i - t^{1/2}\right)\left(T_i + t^{-1/2}\right) = 0, \qquad 1 \le i \le L-1,
    \\
    & \left(T_L - t_L^{1/2}\right)\left(T_L + t_L^{-1/2}\right) = 0,
\end{align*}
the braid relations
\begin{align*}
    T_1 T_0 T_1 T_0 & = T_0 T_1 T_0 T_1
    \\
    T_i T_{i+1} T_i & = T_{i+1} T_i T_{i+1}, \qquad 1 \le i \le L - 2,
    \\
    T_L T_{L-1} T_L T_{L-1} & = T_{L-1} T_L T_{L-1} T_L,
\end{align*}
and otherwise commute.  That is
\begin{equation*}
    T_i T_j = T_j T_i, \qquad |i - j| \ge 2.
\end{equation*}
The algebra contains a family of mutually commuting elements \cite{Sahi99,Lusztig89}
\begin{equation}
    Y_i = T_i \ldots T_{L-1} T_L \ldots T_0 T_1^{-1} \ldots T_{i-1}^{-1},
    \qquad
    1 \le i \le L.
    \label{eq:Koor_Yi}
\end{equation}

We are interested in the representation of this algebra due to Noumi \cite{Noumi95} (see also \cite{Sahi99}),
acting on Laurent polynomials in $z_1, \ldots, z_L$.  The Noumi representation contains three additional
parameters, $u_0^{1/2}$, $u_L^{1/2}$, and $s^{1/2}$, and is defined in terms of operators $s_i$ acting on the $x_i$
as
\begin{equation}
    s_0: z_1 \to s z_1^{-1}, \qquad s_L: z_L \to z_L^{-1},
        \qquad s_i: z_i \leftrightarrow z_{i+1}, \quad 1 \le i \le L - 1.
\label{eq:Koor_si_def}
\end{equation}
The elements $s_0, s_1, \ldots s_L$ generate the affine Weyl $W$ group of type $C_L$.  The {\it finite} Weyl
group $W_0$ is the subgroup generated by $s_1, \ldots, s_L$.

Then in the Noumi representation, the generators of the affine Hecke algebra are given by
\begin{equation}
\begin{aligned}
    & T_0^{\pm 1} = t_0^{\pm 1/2} - t_0^{-1/2}\frac{(z_1 - a)(z_1 - b)}{z_1(z_1 - s z_1^{-1})}(1 - s_0),
        \\
    & T_i^{\pm 1} = t^{\pm 1/2} - \frac{t^{1/2} z_i - t^{-1/2} z_{i+1}}{(z_i - z_{i+1})}(1 - s_i),
       \qquad 1 \le i \le L - 1,
        \\
    & T_L^{\pm 1} = t_L^{\pm 1/2} + t_L^{-1/2}\frac{(c z_L - 1)(d z_L - 1)}{z_L(z_L - z_L^{-1})}(1 - s_L),
\end{aligned}
\label{eq:Koor_Ti_reps}
\end{equation}
with
\begin{equation*}
    a = s^{1/2} t_0^{1/2}u_0^{1/2}, \quad b = -s^{1/2} t_0^{1/2} u_0^{-1/2},
    \quad
    c = t_L^{1/2}u_L^{1/2}, \quad d = -t_L^{1/2} u_L^{-1/2}.
\end{equation*}
One can check directly that the definitions \eqref{eq:Koor_Ti_reps} satisfy the relations of the Hecke algebra.
Formally, we define the field $\mathbb{F} = \mathbb{C}(s^{1/2}, t^{1/2}, t_0^{1/2}, u_0^{1/2}, t_L^{1/2}, u_L^{1/2})$, and
let $\mathcal{R} = \mathbb{F}[z_1, \ldots, z_L]$ be the ring of Laurent polynomials in $L$ variables over
$\mathbb{F}$.  The map sending the generators of the Hecke algebra to the operators defined in
\eqref{eq:Koor_Ti_reps} gives a representation of the algebra on $\mathcal{R}$ \cite{Sahi99}.

Later we will see that to relate the ASEP to the Noumi representation of the Hecke algebra we should take
\begin{equation*}
    t^{1/2} = \sqrt{\frac{p}{q}},
    \qquad
    t_0^{1/2} = \sqrt{\frac{\alpha}{\gamma}},
    \qquad
    t_L^{1/2} = \sqrt{\frac{\beta}{\delta}},
\end{equation*}
and
\begin{equation*}
    u_0^{1/2} - u_0^{-1/2} = \frac{p - q + \gamma - \alpha}{\sqrt{\alpha \gamma}},
    \qquad
    u_L^{1/2} - u_L^{-1/2} = \frac{p - q + \delta - \beta}{\sqrt{\beta \delta}}.
\end{equation*}
For the remainder of this paper we will use this parameterisation in preference to the physical parameters of
the ASEP, or the combinations $a$, $b$, $c$, $d$ appearing in \eqref{eq:Koor_Ti_reps}.

\subsubsection{Non-symmetric Koornwinder polynomials}

Before defining the non-symmetric Koornwinder polynomials, we will introduce some notation and definitions
concerning integer vectors, $\lambda \in \mathbb{Z}^L$, with
\begin{equation*}
    \lambda = (\lambda_1, \ldots, \lambda_L).
\end{equation*}
We call such a vector a {\it composition}.  For a given composition, $\lambda$, we write monomials
\begin{equation*}
    \mathbf{z}^\lambda = z_1^{\lambda_1} \ldots z_L^{\lambda_L}.
\end{equation*}
A {\it partition} is a composition satisfying
\begin{equation*}
    \lambda_1 \ge \lambda_2 \ge \ldots \ge \lambda_L \ge 0.
\end{equation*}
We denote by $\lambda^+$ the unique partition obtained from a composition $\lambda$ by reordering and changing
signs so that the entries are non-negative and in decreasing order.

There are two partial orderings on compositions that will be relevant \cite{Stokman00}.  First define the
\textit{dominance order}:  for $\mu, \lambda \in \mathbb{Z}^L$
\begin{equation*}
    \mu \le \lambda \iff \sum_{i=1}^j(\mu_i - \lambda_i) \le 0, \qquad \forall j, \, 1 \le j \le L.
\end{equation*}
Then $\mu < \lambda$ if $\mu \le \lambda$ and $\mu \ne \lambda$.
The second partial ordering `$\preceq$` is defined as
\begin{equation*}
    \mu \preceq \lambda \iff \mu^+ < \lambda^+ \text{ or }
        \left(\mu^+ = \lambda^+ \text{ and } \mu \le \lambda\right).
\end{equation*}
Then, $\mu \prec \lambda$ if $\mu \preceq \lambda$ and $\mu \ne \lambda$.

\begin{definition}
The non-symmetric Koornwinder polynomial $E_\lambda(\mathbf{z})$, indexed by composition $\lambda$, is the
unique Laurent polynomial satisfying
\begin{equation*}
\begin{aligned}
    Y_i E_\lambda(\mathbf{z}) & = y(\lambda)_i E_\lambda(\mathbf{z}), \quad 1 \le i \le L, \\
    E_\lambda(\mathbf{z})
    & =
    \mathbf{z}^\lambda + \sum_{\lambda' \prec \lambda} c_{\lambda \lambda'} \mathbf{z}^{\lambda'},
\end{aligned}
\end{equation*}
where $Y_i$ is defined in \eqref{eq:Koor_Yi} in the Noumi representation, $y(\lambda)_i$ is the eigenvalue, and $c_{\lambda
\lambda'}$ are coefficients.
\end{definition}
The composition $\lambda$ determines the eigenvalue $y(\lambda)_i$ \cite{Sahi99}.  The
two following cases will appear directly in this work:
\begin{itemize}
    \item
For $m > 0$, $\lambda = \left((-m)^L\right) = (-m, \ldots, -m)$,
\begin{equation} 
    y(\lambda)_i = t_0^{-1/2} t_L^{-1/2} s^{-m} t^{-(i - 1)}.
    \label{eq:Koor_ynegm}
\end{equation}

    \item
For $m \ge 0$, $\lambda = \left(m^L\right) = (m, \ldots, m)$,
\begin{equation}
    y(\lambda)_i = t_0^{1/2} t_L^{1/2} s^{m} t^{L - i}.
    \label{eq:Koor_yposm}
\end{equation}
\end{itemize}
However, other non-symmetric Koornwinder polynomials will appear implicitly, and we define the following
space:
\begin{definition}
For a partition $\lambda$ of length $L$, define $\mathcal{R}^\lambda$ as the space spanned by
$\{E_\mu | \mu \in \mathbb{Z}^L, \mu^+ = \lambda\}$.
\end{definition}

\subsection{Symmetric Koornwinder polynomials}

\subsubsection{Finite difference operator and symmetric Koornwinder polynomials}

The symmetric Koornwinder polynomials were introduced in \cite{Koornwinder92},
 as eigenfunctions of the $s$-difference operator
\begin{equation} \label{eq:Koor_operatorD}
 D=\sum_{i=1}^{L} g_i(\mathbf{z})(T_{s,i}-1)+\sum_{i=1}^{L} g_i(\mathbf{z}^{-1})(T_{s,i}^{-1}-1),
\end{equation}
where $g_i(\mathbf{z})$ is defined by
\begin{equation} \label{eq:Koor_function_g_i}
 g_i(\mathbf{z})= \frac{(1-az_i)(1-bz_i)(1-cz_i)(1-dz_i)}{(1-z_i^2)(1-sz_i^2)}
 \prod_{\genfrac{}{}{0pt}{}{j=1}{j \neq i}}^{L} \frac{(1-tz_iz_j^{-1})(1-tz_iz_j)}{(1-z_iz_j^{-1})(1-z_iz_j)},
\end{equation}
and $T_{s,i}$ is the $i^{th}$ $s$-shift operator
\begin{equation}
 T_{s,i}f(z_1,\dots,z_i,\dots,z_L)=f(z_1,\dots,sz_i,\dots,z_L).
\end{equation}

\begin{definition}
For a partition $\lambda$, the symmetric Koornwinder polynomial $P_\lambda(\mathbf{z})$ is characterised by
the eigenvalue equation
\begin{equation} \label{eq:Koor_eigenvalue_eq_symmetric_Koornwinder}
 D P_\lambda = d_\lambda P_\lambda,
\end{equation}
with eigenvalue
\begin{equation}
 d_\lambda = \sum_{i=1}^{L} \left[t_0 t_L t^{2L-i-1}(s^{\lambda_i}-1)+t^{i-1}(s^{-\lambda_i}-1)\right],
\end{equation}
and where the coefficient of $\mathbf{x}^\lambda$ in $P_\lambda$ is equal to $1$.
\end{definition}

\subsubsection{Link with the non-symmetric polynomials}

The symmetric Koornwinder polynomials are $W_0$-invariant (that is, invariant under the action of $s_1,
\ldots, s_L$, defined in \eqref{eq:Koor_si_def}), and their relation to the non-symmetric Koornwinder polynomials
was given in \cite{Sahi99}.
\begin{theorem}[Corollary 6.5 of \cite{Sahi99}]
The symmetric Koornwinder polynomial $P_\lambda$ can be characterised as the unique $W_0$-invariant polynomial in
$\mathcal{R}^\lambda$ which has the coefficient of $\mathbf{z}^\lambda$ equal to 1.
\label{theorem:Koor_SahiSymmetric}
\end{theorem}

\subsection{Link with the q-KZ equation}

\subsubsection{Reformulation of the q-KZ equation}

We use the Noumi representation of the Hecke algebra to write the $q$KZ equations in component form.  To
specify a lattice configuration $\bm\tau$ we use `$\circ$' for an empty site ($\tau_i = 0$), `$\bullet$'
for a filled site ($\tau_i = 1$) and '$\dots$' for unspecified values. Then, for example, we write $\psi_{\circ\ldots}$ to indicate the weight for
any configuration with the first site empty ($\tau_1 = 0$)
\begin{lemma}
The $q$KZ equations \eqref{eq:Koor_qKZi} -- \eqref{eq:Koor_qKZL} for the deformed ground state vector are equivalent to
the following exchange relations on the components:
\begin{align}
    T_0 \psi_{\circ \ldots} & = \xi^{-1} t_0^{-1/2} \psi_{\bullet \ldots},
    \label{eq:Koor_qkzExchangeLeft}
    \\
    T_L \psi_{\ldots \bullet} & = t_L^{-1/2} \psi_{\ldots \circ},
    \label{eq:Koor_qkzExchangeRight}
\end{align}
and for $1 \le i \le L - 1$,
\begin{align}
    T_i \psi_{\ldots \circ \circ \ldots} & = t^{1/2} \psi_{\ldots \circ \circ \ldots},
    \label{eq:Koor_qkzExchangeBulk00}
    \\
    T_i \psi_{\ldots \bullet \bullet \ldots} & = t^{1/2} \psi_{\ldots \bullet \bullet \ldots},
    \label{eq:Koor_qkzExchangeBulk11}
    \\
    T_i \psi_{\ldots \bullet \circ \ldots} & = t^{-1/2} \psi_{\ldots \circ \bullet \ldots},
    \label{eq:Koor_qkzExchangeBulk01}
\end{align}
where the marked sites are in positions $i$, $i+1$.
\label{lemma:Koor_rightExchange}
\end{lemma}
\proof
This can be checked directly.
\finproof

Note that the parameters $s$ and $\xi$ both enter through \eqref{eq:Koor_qkzExchangeLeft}, with $s$ contained
within the $T_0$ operator.

\begin{lemma}
The left $q$KZ equations \eqref{eq:Koor_leftqKZi} -- \eqref{eq:Koor_leftqKZL} for a vector of form \eqref{eq:Koor_vec_Phi}
are equivalent to the following exchange relations on the components:
\begin{align}
    T_0 \phi_{\circ \ldots} & = \xi t_0^{1/2} \phi_{\bullet \ldots},
    \label{eq:Koor_leftqkzExchangeLeft}
    \\
    T_L \phi_{\ldots \bullet} & = t_L^{1/2} \phi_{\ldots \circ},
    \label{eq:Koor_leftqkzExchangeRight}
\end{align}
and for $1 \le i \le L - 1$,
\begin{align}
    T_i \phi_{\ldots \circ \circ \ldots} & = t^{1/2} \phi_{\ldots \circ \circ \ldots},
    \label{eq:Koor_leftqkzExchangeBulk00}
    \\
    T_i \phi_{\ldots \bullet \bullet \ldots} & = t^{1/2} \phi_{\ldots \bullet \bullet \ldots},
    \label{eq:Koor_leftqkzExchangeBulk11}
    \\
    T_i \phi_{\ldots \bullet \circ \ldots} & = t^{1/2} \phi_{\ldots \circ \bullet \ldots},
    \label{eq:Koor_leftqkzExchangeBulk01}
\end{align}
where the marked sites are in positions $i$, $i+1$.
\label{lemma:Koor_leftqkzExchange}
\end{lemma}

\subsubsection{Reference state}

\begin{lemma}
For any vector $\ket{\Psi(\mathbf{z}; s, \xi)}$ satisfying the $q$KZ equations \eqref{eq:Koor_qKZi} --
\eqref{eq:Koor_qKZL}, the empty lattice weight $\psi_{\circ \ldots \circ}$ (where the '$\dots$' here stands for empty sites)
is an eigenfunction of the $Y_i$ operators \eqref{eq:Koor_Yi}, satisfying
\begin{equation}
    Y_i \psi_{\circ \ldots \circ}
        = \xi^{-1} t_0^{-1/2} t_L^{-1/2} t^{-(i-1)} \psi_{\circ \ldots \circ}.
    \label{eq:Koor_asepEigenref}
\end{equation}
\label{lemma:Koor_asepEigenref}
\end{lemma}
\proof
This follows by direct computation with the exchange relations in Lemma~\ref{lemma:Koor_rightExchange}.
\finproof

Lemma~\ref{lemma:Koor_asepEigenref} immediately suggests the following connection to the non-symmetric Koornwinder polynomials:
\begin{enumerate}
    \item
Taking $\xi = s^m$, $m > 0$, the eigenvalue in \eqref{eq:Koor_asepEigenref} is given by \eqref{eq:Koor_ynegm},
corresponding to the non-symmetric Koornwinder polynomial labelled by the composition $\left((-m)^L \right)$.

    \item
Taking $\xi = t_0^{-1} t_L^{-1} t^{-(L-1)} s^{-m}$, $m \ge 0$, the eigenvalue instead corresponds to
\eqref{eq:Koor_yposm}, for the composition $\left(m^L\right)$.
\end{enumerate}
Note that case $2$ is obtained from case $1$ by sending
\begin{equation*}
    \xi \to t_0^{-1} t_L^{-1} t^{-(L-1)} \xi^{-1},
\end{equation*}
which is exactly the Gallavotti--Cohen symmetry \eqref{eq:Koor_GC_sym}.
  In section~\ref{sec:Koor_matrixproduct} we will
give a direct matrix product construction of the inhomogeneous ground state vector for case 1, that is $\xi =
s^m$.  To solve case 2, we will use the Gallavotti--Cohen symmetry on solutions of {\it left} $q$KZ
equations, which we will present next.
We note that an alternative approach, as followed in \cite{Kasatani10,CantiniGDGW16}, would be to take $\psi_{\circ\ldots\circ}$
as the non-symmetric Koornwinder polynomial given in case 1 or case 2, then show that a solution of the
exchange relations \eqref{eq:Koor_qkzExchangeLeft} -- \eqref{eq:Koor_qkzExchangeBulk01} can be constructed from this
reference state.

\begin{lemma}
For any vector $\bra{\Phi(\mathbf{z}; s, \xi)}$ satisfying the left $q$KZ equations \eqref{eq:Koor_leftqKZi} --
\eqref{eq:Koor_leftqKZL}, the empty lattice weight $\phi_{\circ \ldots \circ}$ is an eigenfunction of the $Y_i$
operators \eqref{eq:Koor_Yi}, satisfying
\begin{equation}
    Y_i \phi_{\circ \ldots \circ}
        = \xi t_0^{1/2} t_L^{1/2} t^{L-i} \phi_{\circ \ldots \circ}.
    \label{eq:Koor_asepleftEigenref}
\end{equation}
\label{lemma:Koor_asepleftEigenref}
\end{lemma}

Again, the same two constraints on $\xi$ and $s$ appear, but with the correspondence to the non-symmetric
Koornwinder polynomials reversed: taking $\xi = s^m$, $m \ge 0$, would correspond to the composition
$\left(m^L\right)$; taking $\xi = t_0^{-1} t_L^{-1} t^{-(L-1)} s^{-m}$, $m > 0$, would correspond to the
composition $\left((-m)^L)\right)$.
 
\section{Matrix product solution to the q-KZ equation} \label{sec:Koor_matrixproduct}

\subsection{Construction of the solution}

\subsubsection{General construction}

The matrix product ansatz for the twice deformed inhomogeneous ground state vectors is written
\begin{equation}
 \ket{\Psi(\mathbf{z}; s, \xi)} = \bbra{W} \mathbb{S} \mathbb{A}(z_1) \otimes \ldots \otimes \mathbb{A}(z_L) \kket{V},
 \label{eq:Koor_mpaForm}
\end{equation}
with
\begin{equation*}
    \mathbb{A}(z) = \begin{pmatrix}
        A_0(z)
        \\
        A_1(z)
    \end{pmatrix}.
\end{equation*}
The entries $A_0(z)$, $A_1(z)$ as well as $\mathbb{S}$ are operators in some auxiliary algebraic space,
and the left and right vectors $\bbra{W}$ and $\kket{V}$ contract this space to give a scalar value.

Writing out \eqref{eq:Koor_mpaForm} gives the $2^L$ component vector
\begin{equation*}
    \ket{\Psi(\mathbf{z}; s, \xi)}
    = \begin{pmatrix}
        \bbra{W} \mathbb{S} A_0(z_1) \ldots A_0(z_{L-1}) A_0(z_L) \kket{V}
        \\
        \bbra{W} \mathbb{S} A_0(z_1) \ldots A_0(z_{L-1}) A_1(z_L) \kket{V}
        \\
        \bbra{W} \mathbb{S} A_0(z_1) \ldots A_1(z_{L-1}) A_0(z_L) \kket{V}
        \\
        \vdots
        \\
        \bbra{W} \mathbb{S} A_1(z_1) \ldots A_1(z_{L-1}) A_1(z_L) \kket{V}
    \end{pmatrix},
\end{equation*}
with entries
\begin{equation*}
\psi_{\bm\tau}(\mathbf{z}; s, \xi)=   \bbra{W} \mathbb{S} A_{\tau_1}(z_1) \ldots A_{\tau_{L-1}}(z_{L-1}) A_{\tau_L}(z_L) \kket{V}.
\end{equation*}

\begin{lemma}
Sufficient conditions for a vector of form \eqref{eq:Koor_mpaForm} to satisfy the $q$KZ equations \eqref{eq:Koor_qKZi} --
\eqref{eq:Koor_qKZL} are the following:
\begin{align}
    \check{R}\left(\frac{z_{i+1}}{z_i}\right) \mathbb{A}(z_i) \otimes \mathbb{A}(z_{i+1})
    & =
    \mathbb{A}(z_{i+1}) \otimes \mathbb{A}(z_i),
    \label{eq:Koor_ZF}
    \\
    \widetilde{K}\left(z_1^{-1}\right) \bbra{W} \mathbb{S} \mathbb{A}\left(z_1^{-1}\right)
    & =
    \bbra{W} \mathbb{S} \mathbb{A}\left(s z_1\right),
    \label{eq:Koor_deformedGZ1}
    \\
    \overline{K}(z_L) \mathbb{A}(z_L) \kket{V}
    & =
    \mathbb{A}\left(z_L^{-1}\right) \kket{V}.
    \label{eq:Koor_deformedGZL}
\end{align}
\end{lemma}
Equation \eqref{eq:Koor_ZF} is the Zamolodchikov--Faddeev (ZF) algebra \cite{ZamolodchikovZ79,Faddeev80}.  Equations
\eqref{eq:Koor_deformedGZ1}, \eqref{eq:Koor_deformedGZL} are a deformation of the Ghoshal--Zamolodchikov (GZ) relations
\cite{GhoshalZ94}.  The undeformed GZ relations are obtained by setting $\mathbb{S}$ to the identity and $s = 1$.
The matrix product ansatz for the open boundary ASEP can be expressed as a solution of the undeformed
relations, and solutions for related models have also been found and studied \cite{CrampeRV14,CrampeRRV16}.

\subsubsection{Definition of the algebra} \label{subsec:Koor_representation}

We now give an explicit construction of the $q$KZ solution when $\xi = s^m$, $m \ge 1$.  We first define certain algebraic objects
through the relations they satisfy.
\begin{definition}
We define algebraic objects\footnote{These are $s$-deformations of the usual $q$-bosons that we recover for $q=t$ and $s=t^2$.} 
satisfying the following relations: operators $a$, $a^\dagger$ and $S$:
\begin{equation}
\begin{aligned}
    & a a^\dagger - t a^\dagger a = 1 - t,
    \\
    & a S = \sqrt{s} S a,
    \\
    & S a^\dagger = \sqrt{s} a^\dagger S.
\end{aligned}
\label{eq:Koor_oscAlgBulk}
\end{equation}
And paired boundary vectors $\bbra{w}$ and $\kket{v}$:
\begin{equation}
\begin{aligned}
    \bbra{w} \left( t_0^{1/2}a-t_0^{-1/2}a^\dagger \right)
        & = \bbra{w} \left( u_0^{1/2}-u_0^{-1/2} \right),
  \\
    \left( t_L^{1/2}a^\dagger-t_L^{-1/2}a \right) \kket{v}
        & =  \left( u_L^{1/2}-u_L^{-1/2} \right) \kket{v},
\label{eq:Koor_oscAlgBoundary}
\end{aligned}
\end{equation}
and $\bbra{\widetilde{w}}$ and $\kket{\widetilde{v}}$:
\begin{equation}
\begin{aligned}
    \bbra{\widetilde w} \left( t_0^{1/2}a-t_0^{-1/2}a^\dagger \right)
        & = \bbra{\widetilde w} \left( t_0^{1/2}-t_0^{-1/2} \right),
        \\
    \left( t_L^{1/2}a^\dagger-t_L^{-1/2} a \right) \kket{\widetilde v}
        & =  \left( t_L^{1/2}-t_L^{-1/2} \right)\kket{\widetilde v}.
\end{aligned}
\label{eq:Koor_oscAlgBoundaryTwid}
\end{equation}
\label{def:oscAlg}
\end{definition}
Elements of this algebra have appeared in many places in the context of the ASEP. 
The first algebraic relation of \eqref{eq:Koor_oscAlgBulk} and the relations \eqref{eq:Koor_oscAlgBoundary} were first stated in \cite{Sandow94} 
to study the stationary state of the open ASEP. This work shed new light on the DEHP algebra introduced in \cite{DerridaEHP93} by 
showing that it can be recast in a form of a $q$-deformed oscillator algebra by an appropriate shift and normalisation of the generators.
The representation of the algebraic elements involved in the first relation of \eqref{eq:Koor_oscAlgBulk} and in the
relations \eqref{eq:Koor_oscAlgBoundary} were found in \cite{Sandow94}, and permitted explicit computations. In
particular the author of that work
pointed out the relevance of the parametrisation used here. More precisely the parameters $\kappa_+(\alpha,\gamma)$ and $\kappa_+(\beta,\delta)$ 
defined in \cite{Sandow94} by $\kappa_+(x,y)=\frac{1}{2x}\left(y-x+p-q+\sqrt{(y-x+p-q)^2+4xy}\right)$ play a central role in the representation of the 
algebra, and are relevant in describing the phase transitions of the system. The precise relations with the parameters used here are
$\kappa_+(\alpha,\gamma)=u_0^{1/2}t_0^{-1/2}$ and $\kappa_+(\beta,\delta)=u_L^{1/2}t_L^{-1/2}$.

The other relations
\eqref{eq:Koor_oscAlgBulk}, \eqref{eq:Koor_oscAlgBoundary} and \eqref{eq:Koor_oscAlgBoundaryTwid} appear previously in
\cite{GorissenLMV12,Lazarescu13jphysA,Lazarescu13,LazarescuP14} to compute the fluctuations of the current. 
We recall an infinite dimensional representation of this algebra: $\kket{v},\kket{\widetilde v},\dots$ are vectors of a Fock space endowed 
with the usual scalar product. In this manuscript the scalar product of two vectors $\kket{x}$ and $\kket{y}$ of this Fock space is denoted
by $\bbra{x} \cdot \kket{y}$.
The operators $a$ and $a^\dagger$ are linear operators on this Fock space.
Let us stress here that the creation operator $a^\dagger$ is not the Hermitian conjugate of the annihilation operator $a$ (it is 
a standard notation which appears often in the literature, see for instance \cite{Sandow94}).  

The representation of the algebra is defined on the Fock space $\text{Span}\{ \kket{k} \}_{k=0}^\infty$.  The
bulk matrices are given by
\begin{align*}
    & a = \sum_{k=1}^\infty (1 - t^k) \kket{k - 1} \bbra{k},
    \qquad
    a^\dagger = \sum_{k=0}^\infty \kket{k + 1} \bbra{k},
    \\
    & S  = \sum_{k=0}^\infty s^{k/2} \kket{k} \bbra{k}.
\end{align*}
The boundary vectors are written
\begin{equation*}
    \bbra{w} = \sum_{k=0}^\infty w_k \bbra{k},
    \qquad
    \kket{v} = \sum_{k=0}^\infty v_k \kket{k},
\end{equation*}
and
\begin{equation*}
    \bbra{\widetilde{w}} = \sum_{k=0}^\infty \widetilde{w}_k \bbra{k},
    \qquad
    \kket{\widetilde{v}} = \sum_{k=0}^\infty \widetilde{v}_k \kket{k}.
\end{equation*}
As a consequence of the boundary relations, the coefficients appearing in $\bbra{w}$, $\kket{v}$ satisfy the
recursion relations
\begin{align*}
    & w_{k+1} + t_0^{1/2}(u_0^{1/2} - u_0^{-1/2}) w_k - t_0(1-t^k)w_{k-1} = 0,
    \\
    & (t;t)_{k+1} v_{k+1} + t_L^{1/2}(u_L^{1/2} - u_L^{-1/2}) (t;t)_k v_k - t_L (1-t^k) (t;t)_{k-1} v_{k-1} = 0,
\end{align*}
with $w_{-1} = v_{-1} = 0$.  We have used the $t$-Pochhammer symbol
\begin{equation*}
    (x;t)_n = \prod_{k=0}^{n-1} (1 - t^k x).
\end{equation*}
The $t$-Pochhammer symbol can be defined for $n = \infty$ if $t < 1$ :
\begin{equation*}
    (x;t)_\infty = \prod_{k=0}^\infty (1 - t^k x),
\end{equation*}
and we use the notation
\begin{equation*}
    (x, y, z, \ldots ;t)_\infty = (x;t)_\infty (y;t)_\infty (z;t)_\infty \ldots
\end{equation*}
The $t$-Hermite polynomials are given by
\begin{equation*}
    H_n(x, y) = \sum_{k=0}^n \frac{(t;t)_n}{(t;t)_k (t;t)_{n-k}} x^k y^{n-k},
\end{equation*}
and satisfy the recursion relation
\begin{equation*}
    H_{n+1}(x, y) - (x + y) H_n(x, y) + x y (1 - t^n) H_{n-1}(x, y) = 0.
\end{equation*}
Thus we find
\begin{equation}
\begin{aligned}
    w_n & = H_n(t_0^{1/2} u_0^{-1/2}, -t_0^{1/2} u_0^{1/2}),
    \\
    v_n & = \frac{H_n(t_L^{1/2} u_L^{-1/2}, -t_L^{1/2} u_L^{1/2})}{(t;t)_n}.
\end{aligned}
\label{eq:Koor_wnvn}
\end{equation}
The coefficients $\widetilde{w}_n$, $\widetilde{v}_n$ are obtained by setting $u_i = t_i$ in the equation above.

To compute the normalisation factors in the $q$KZ solution, we will make use of the $t$-Mehler formula
\begin{equation*}
    \sum_{n=0}^\infty H_n(x, y) H_n(w, z) \frac{\lambda^n}{(t;t)_n}
    =
    \frac{(x y w z \lambda^2 ;t)_\infty}{(x w \lambda, x z \lambda, y w \lambda, y z \lambda ;t)_\infty}.
\end{equation*}
For the normalisation, we need to compute
\begin{equation*}
    \bbra{w} S^a \kket{v} = \sum_{n=0}^\infty w_n \left(s^{n /2}\right)^a v_n.
\end{equation*}
Using the coefficients \eqref{eq:Koor_wnvn} and the $t$-Mehler formula gives
\begin{equation}
\begin{aligned}
   &  \bbra{w} S^a \kket{v} = 
   \\
   & \frac{\left(t_0^{1/2} t_L^{1/2} s^a ;t\right)_\infty}
    {\left(
        t_0^{1/2} u_0^{-1/2} t_L^{1/2} u_L^{-1/2} s^{a/2},
        -t_0^{1/2} u_0^{-1/2} t_L^{1/2} u_L^{1/2} s^{a/2},
        -t_0^{1/2} u_0^{1/2} t_L^{1/2} u_L^{-1/2} s^{a/2},
        t_0^{1/2} u_0^{1/2} t_L^{1/2} u_L^{1/2} s^{a/2} ;t
    \right)_\infty}.
\end{aligned}
\label{eq:Koor_infNormFactor}
\end{equation}
We also need to compute $\bbra{\widetilde{w}} S^a \kket{\widetilde{v}}$, but this is obtained from
\eqref{eq:Koor_infNormFactor} by setting $u_i = t_i$, $i = 0, L$.

\subsubsection{Construction of the solutions}

In the matrix product construction we will need several commuting copies $(a_n,a^\dagger_n,S_n)$ of the algebra $(a,a^\dagger,S)$
defined in \eqref{eq:Koor_oscAlgBulk}. Similarly to the construction \eqref{eq:2TASEP_tensor}, a simple way to build these commuting copies is 
to define
\begin{equation}
\begin{aligned}
a_n = \underbrace{1 \otimes \dots \otimes 1}_{n-1}\otimes a \otimes 1 \otimes 1 \otimes \dots ,  \\ 
a^\dagger_n = \underbrace{1 \otimes \dots \otimes 1}_{n-1}\otimes a^\dagger \otimes 1 \otimes 1 \otimes \dots ,  \\ 
S_n = \underbrace{1 \otimes \dots \otimes 1}_{n-1}\otimes S \otimes 1 \otimes 1 \otimes \dots,
\label{eq:Koor_commuting_copies}
\end{aligned}
\end{equation}

We introduce also the corresponding boundary vectors
\begin{equation}
\begin{aligned}
\bbra{w}_n = \underbrace{1 \otimes \dots \otimes 1}_{n-1}\otimes \bbra{w} \otimes 1 \otimes 1 \otimes \dots ,  \\ 
\bbra{\widetilde w}_n = \underbrace{1 \otimes \dots \otimes 1}_{n-1}\otimes \bbra{\widetilde w} \otimes 1 \otimes 1 \otimes \dots ,  \\ 
\kket{v}_n = \underbrace{1 \otimes \dots \otimes 1}_{n-1}\otimes \kket{v} \otimes 1 \otimes 1 \otimes \dots, \\
\kket{\widetilde v}_n = \underbrace{1 \otimes \dots \otimes 1}_{n-1}\otimes \kket{\widetilde v} \otimes 1 \otimes 1 \otimes \dots,
\label{eq:Koor_commuting_copies_vectors}
\end{aligned}
\end{equation}

The subscript indices on $a,a^\dagger,S$ and $\bbra{w},\bbra{\widetilde w},\kket{v},\kket{\widetilde v}$ in \eqref{eq:Koor_commuting_copies}
and \eqref{eq:Koor_commuting_copies_vectors} thus denote the position of the operator or vector in the tensor 
product.

In the following, as a matter of convention, we will reserve the symbol `$\otimes$' for objects belonging to
the space of lattice configurations (see \eqref{eq:Koor_mpaForm} for example), 
and use the index notation to denote the tensor product in the auxiliary algebraic space (see \eqref{eq:Koor_Sm},\eqref{eq:Koor_Am} for example).
The aim is to provide a notational distinction between these two spaces and two tensor products.

Building on these commuting copies of the algebra, we define
\begin{align}
    \mathbb{S}^{(m)} & = (S_{2m-1})^{2m-1} \times (S_{2m-2})^{2m-2} \times \ldots \times (S_3)^3 \times (S_2)^2 \times S_1,
    \label{eq:Koor_Sm}
    \\
    \mathbb{A}^{(m)}(z)
        & = L_{2m-1,2m-2}(z) \cdot \ \ldots \ \cdot L_{5,4}(z) \cdot L_{3,2}(z) \cdot b(z),
    \label{eq:Koor_Am}
\end{align}
with
\begin{equation}
    L_{k+1,k}(z) = \begin{pmatrix}
        1 & a_{k+1}
        \\
        z a^\dagger_{k+1} & z
    \end{pmatrix}\cdot \begin{pmatrix}
        1/z & a_k/z
        \\
         a^\dagger_k & 1
    \end{pmatrix},
    \qquad
    b(z) = \begin{pmatrix}
        1/z+a_1
        \\
        z+a^\dagger_1
    \end{pmatrix}.
    \label{eq:Koor_Lbdef}
\end{equation}
The symbol $\cdot$ indicates the usual matrix product in the physical space.
For example, expanding the definition of $L_{k+1,k}(z)$ gives
\begin{equation*}
    L_{j,k}(z) = \begin{pmatrix}
        1/z + a_{j} a^\dagger_k
        &
        a_k/z + a_{j} 
        \\
        a^\dagger_{j} + z a^\dagger_k
        &
        a^\dagger_{j} a_k + z
    \end{pmatrix}.
\end{equation*}
To lighten the notation, we will sometimes write $L(z)$ instead of $L_{k+1,k}(z)$ when there is no ambiguity.

We also define boundary vectors
\begin{align}
    \bbra{W^{(m)}}
    & =
        \bbra{w}_{2m-1}\bbra{\widetilde w}_{2m-2} \ \ldots \ \bbra{w}_3 \bbra{\widetilde w}_2 \bbra{w}_1
    \label{eq:Koor_Wm}
    \\
    \kket{V^{(m)}}
    &=
        |v\rrangle_{2m-1} \kket{\widetilde v}_{2m-2} \ \ldots \ \kket{v}_3 \kket{\widetilde v}_2 \kket{v}_1.
    \label{eq:Koor_Vm}
\end{align}

\begin{proposition}
For integer $m > 0$ and $\xi = s^m$, 
\begin{equation}
    \ket{\Psi^{(m)}(\mathbf{z};s)}
    =
    \frac{
        1
    }
    {
        \Omega^{(m)}
    }
    \bbra{W^{(m)}}
        \mathbb{S}^{(m)} \mathbb{A}^{(m)}(z_1) \otimes \ldots \otimes \mathbb{A}^{(m)}_L(z_L) 
    \kket{V^{(m)}},
    \label{eq:Koor_Psim}
\end{equation}
with normalization factor
\begin{equation}
    \Omega^{(m)} = \bbra{W^{(m)}} \mathbb{S}^{(m)} \kket{V^{(m)}},
    \label{eq:Koor_Omegam}
\end{equation}
is a solution of the $q$KZ equations \eqref{eq:Koor_qKZi} -- \eqref{eq:Koor_qKZL}.
\end{proposition}
Note that the dependence on $\xi$ has disappeared in the vector $\ket{\Psi^{(m)}(\mathbf{z};s)}$ because of the constraint $\xi = s^m$. 
\proof
The normalization factor $\Omega^{(m)}$ can be chosen freely, but we must show that the choice
\eqref{eq:Koor_Omegam} is non-zero.  To do so, we compute $\Omega^{(m)}$ using the infinite dimensional
representation of the algebra defined in \eqref{eq:Koor_oscAlgBulk} -- \eqref{eq:Koor_oscAlgBoundaryTwid}, 
see subsection \ref{subsec:Koor_representation}.  Then to prove that $\ket{\Psi^{(m)}(\mathbf{z};s)}$ is a $q$KZ
solution, it is sufficient to show that \eqref{eq:Koor_ZF}, \eqref{eq:Koor_deformedGZ1}, \eqref{eq:Koor_deformedGZL} are
satisfied.

By a direct computation, using the algebraic relations \eqref{eq:Koor_oscAlgBulk}, it can be checked that the
vector $b(z)$ and the matrix $L_{k+1,k}(z)$ satisfy the relations
\begin{align*}
 \check{R}(z_{i+1}/z_{i})b(z_{i})\otimes b(z_{i+1}) & = b(z_{i+1}) \otimes b(z_{i}),
 \\
 \check{R}(z_{i+1}/z_{i}) L(z_{i}) \otimes L(z_{i+1}) & = L(z_{i+1}) \otimes L(z_{i})\check{R}(z_{i+1}/z_{i}).
\end{align*}
These elementary exchange relations can be used successively several times to give \eqref{eq:Koor_ZF}.
On the right boundary, using relations \eqref{eq:Koor_oscAlgBoundary},
\eqref{eq:Koor_oscAlgBoundaryTwid} gives
\begin{align*}
 \overline{K}(z_L) b(z_L)|v\rrangle_1  & = b(1/z_L)|v\rrangle_1,
  \\
 \overline{K}(z_L) L_{k+1,k}(z_L)|v\rrangle_{k+1} | \widetilde v \rrangle_k
    & = L_{k+1,k}(1/z_L)\overline{K}(z_L)|v\rrangle_{k+1} | \widetilde v \rrangle_k.
\end{align*}
Using these properties several times, it is straightforward to prove \eqref{eq:Koor_deformedGZL}.
Finally, on the left boundary, the vector $b(z)$ satisfies
\begin{equation*}
 \llangle w|_1 S_1 \left.\widetilde{K}(z_1^{-1})\right|_{\xi=s} b(z_1^{-1}) = \llangle w|_1 S_1 b(s z_1),
\end{equation*}
and the matrix $L_{k+1,k}(z)$ satisfies
\begin{align*}
 \llangle w|_{k+1} \llangle \widetilde w|_k & (S_{k+1})^{2a+1}  (S_k)^{2a} \left.\widetilde{K}(z_1^{-1})\right|_{\xi=s^{a+1}} L_{k+1,k}(z_1^{-1})
 \\
 & = 
 \llangle w|_{k+1} \llangle \widetilde w|_k (S_{k+1})^{2a+1} (S_k)^{2a} L_{k+1,k}(s z_1) \left.\widetilde{K}(z_1^{-1})\right|_{\xi=s^a}.
\end{align*}
In words, the last equation means that the parameter $\xi$ is multiplied by a factor $s$ when the matrix $L$
passes through the matrix $\widetilde{K}$.  Thus by imposing the constraint $\xi = s^m$ and applying these
relations successively, relation \eqref{eq:Koor_deformedGZ1} follows.
\finproof

\subsubsection{Construction of the solutions to the left qKZ equation}
We now present briefly the construction of row vector solutions of the {\it left} $q$KZ equations \eqref{eq:Koor_leftqKZi} --
\eqref{eq:Koor_leftqKZL} in the matrix product form
\begin{equation}
    \bra{\Phi(\mathbf{z}; s, \xi)} = \bbra{W} \mathbb{S} \mathbb{A}(z_1) \otimes \ldots \otimes \mathbb{A}(z_L) \kket{V},
    \label{eq:Koor_leftmpaForm}
\end{equation}
with
\begin{equation}
    \mathbb{A}(z) = \Big( \ A_0(z) \ , \   A_1(z) \ \Big).
\end{equation}

\begin{lemma}
Sufficient conditions for a vector of form \eqref{eq:Koor_leftmpaForm} to satisfy the left $q$KZ equations
\eqref{eq:Koor_leftqKZi} -- \eqref{eq:Koor_leftqKZL} are the following:
\begin{align}
    \mathbb{A}(z_i) \otimes \mathbb{A}(z_{i+1}) \check{R}\left(\frac{z_{i+1}}{z_i}\right) 
    & =
    \mathbb{A}(z_{i+1}) \otimes \mathbb{A}(z_i),
    \label{eq:Koor_leftZF}
    \\
    \bbra{W} \mathbb{S} \mathbb{A}\left(z_1^{-1}\right)\widetilde{K}\left(z_1^{-1}\right) 
    & =
    \bbra{W} \mathbb{S} \mathbb{A}\left(s z_1\right),
    \label{eq:Koor_leftdeformedGZ1}
    \\
    \mathbb{A}(z_L) \kket{V} \overline{K}(z_L)
    & =
    \mathbb{A}\left(z_L^{-1}\right) \kket{V}.
    \label{eq:Koor_leftdeformedGZL}
\end{align}
\end{lemma}

We will construct a solution at $\xi = s^m$, with $m \ge 0$ an integer.  We define the following objects:
\begin{equation}
\begin{aligned}
    b_{\text{left}}(z) & = (1, 1),
    \\
    \mathbb{A}^{(m)}_{\text{left}}(z)
        & = b_{\text{left}}(z) \cdot L_{1,2}(1/z) \cdot \ \ldots \ \cdot L_{2m-1,2m}(1/z),
    \\
    \mathbb{S}^{(m)}_{\text{left}} & = S_1 \times (S_2)^2 \times \ldots \times (S_{2m-1})^{2m-1} \times (S_{2m})^{2m},
    \\
    \bbra{W^{(m)}_{\text{left}}}
    & = \bbra{w}_1\bbra{\widetilde w}_2\bbra{w}_3 \ \ldots \ \bbra{w}_{2m-1}\bbra{\widetilde w}_{2m},
    \\
    \kket{V^{(m)}_{\text{left}}}
    & = \kket{v}_1\kket{\widetilde v}_2\kket{v}_3 \ \ldots \ \kket{v}_{2m-1}\kket{\widetilde v}_{2m}.
\end{aligned}
\end{equation}
The algebraic objects are as given in Definition~\ref{def:oscAlg}, and $L(z)$ is defined in \eqref{eq:Koor_Lbdef}.

\begin{proposition}
For integer $m \ge 0$ and $\xi = s^m$, 
\begin{equation}
    \bra{\Phi^{(m)}(\mathbf{z}; s)}
    =
    \frac{1}{\Omega^{(m)}_{\text{left}}}
    \bbra{W^{(m)}_{\text{left}}}
        \mathbb{S}^{(m)}_{\text{left}} \mathbb{A}^{(m)}_{\text{left}}(z_1) \otimes
        \ldots \otimes
        \mathbb{A}^{(m)}_{\text{left}}(z_L) 
    \kket{V^{(m)}_{\text{left}}},
    \label{eq:Phim}
\end{equation}
with normalization factor
\begin{equation}
    \Omega^{(m)}_{\text{left}}
        = \bbra{W^{(m)}_{\text{left}}} \mathbb{S}^{(m)}_{\text{left}} \kket{V^{(m)}_{\text{left}}},
\end{equation}
is a solution of the left $q$KZ equations \eqref{eq:Koor_leftqKZi} -- \eqref{eq:Koor_leftqKZL}.
\end{proposition}
\proof
We give the elementary exchange relations, which imply \eqref{eq:Koor_leftZF} -- \eqref{eq:Koor_leftdeformedGZL},
and thus a solution of the left $q$KZ equations.

In the bulk,
\begin{align*}
    b_{\text{left}}(z_i)\otimes b_{\text{left}}(z_{i+1})\check{R}(z_{i+1}/z_{i})
    & =
    b_{\text{left}}(z_{i+1})\otimes b_{\text{left}}(z_i),
    \\
    L(1/z_{i}) \otimes L(1/z_{i+1})\check{R}(z_{i+1}/z_{i})  & = \check{R}(z_{i+1}/z_{i}) L(1/z_{i+1})\otimes L(1/z_{i}),
\end{align*}
from which \eqref{eq:Koor_leftZF} follows.  On the right boundary
\begin{align*}
    b_{\text{left}}(z_L) \overline{K}(z_L)  & =  b_{\text{left}}(1/z_L),
    \\
L_{k,k+1}(1/z_L)\overline{K}(z_L)|v\rrangle_{k}|\widetilde v\rrangle_{k+1} & =\overline{K}(z_L)L_{k,k+1}(z_L)|v\rrangle_{k}|\widetilde v\rrangle_{k+1},
\end{align*}
from which \eqref{eq:Koor_leftdeformedGZL} follows.  On the left boundary
\begin{equation*}
 b_{\text{left}}(1/z_1) \left.\widetilde{K}(z_1^{-1})\right|_{\xi=1}
 =
 b_{\text{left}}(s z_1),
\end{equation*}
and 
\begin{align*}
 & \llangle w|_{k}\llangle \widetilde w|_{k+1} (S_k)^{2a-1} (S_{k+1})^{2a} L_{k,k+1}(z_1) \left.\widetilde{K}(z_1^{-1})\right|_{\xi=s^{a}}
 \\
 & = 
 \llangle w|_{k}\llangle \widetilde w|_{k+1} (S_k)^{2a-1} (S_{k+1})^{2a} \left.\widetilde{K}(z_1^{-1})\right|_{\xi=s^{a-1}} L(1/(s z_1)).
\end{align*}
With the constraint $\xi = s^m$, these imply \eqref{eq:Koor_leftdeformedGZ1}.
\finproof

\subsection{Computation of Koornwinder polynomials}

We still need to show that the construction gives a non-zero vector.  Before doing so, we introduce some
notation, then look at some examples.

\begin{definition}
For a lattice configuration $\bm\tau = (\tau_1, \ldots, \tau_L)$, define the composition
$\lambda^{(m)}(\bm\tau)$, with
\begin{equation*}
    \lambda^{(m)}(\bm\tau)_i = \begin{cases}
        -m, & \tau_i = 0,
        \\
        m,  & \tau_i = 1.
    \end{cases}
\end{equation*}
The corresponding partition is $\lambda^{(m)+}(\bm\tau) = \left(m^L\right)$.
\end{definition}

\begin{example}
With $m=1$, $\mathbb{A}^{(1)}(z) = b(z)$.  For $L=1$,
\begin{equation*}
\begin{aligned}
    \Omega^{(1)} \ket{\Psi^{(1)}(z_1; s)}
    & =
    \bbra{w}_1 S_1 b(z_1) \kket{v}_1
    \\
    & =
    \begin{pmatrix}
        \bbra{w}_1 S_1 \left(\frac{1}{z_1} + a_1 \right)\kket{v}_1
        \\
        \bbra{w}_1 S_1 \left(z_1 + a^\dagger_1 \right)\kket{v}_1
    \end{pmatrix},
\end{aligned}
\end{equation*}
which is a real valued vector of size $2$.
For $L=2$,
\begin{equation*}
\begin{aligned}
    \Omega^{(1)} \ket{\Psi^{(1)}(z_1, z_2; s)}
    & =
    \bbra{w}_1 S_1 b(z_1) \otimes b(z_2) \kket{v}_1
    \\
    & =
    \begin{pmatrix}
        \bbra{w}_1 S_1 \left(\frac{1}{z_1} + a_1\right) \left(\frac{1}{z_2} + a_1 \right) \kket{v}_1
        \\
        \bbra{w}_1 S_1 \left(\frac{1}{z_1} + a_1\right) \left(z_2 + a^\dagger_1 \right) \kket{v}_1
        \\
        \bbra{w}_1 S_1 \left(z_1 + a^\dagger_1\right) \left(\frac{1}{z_2} + a_1 \right) \kket{v}_1
        \\
        \bbra{w}_1 S_1 \left(z_1 + a^\dagger_1\right) \left(z_2 + a^\dagger_1 \right) \kket{v}_1
    \end{pmatrix},
\end{aligned}
\end{equation*}
which is a real valued vector of size $4$.
In general,
\begin{equation*}
    \Omega^{(1)} \ket{\Psi^{(1)}(\mathbf{z}; s)}
    =
    \bbra{w}_1 S_1 b(z_1) \otimes  \ldots b(z_L) \kket{v}_1.
\end{equation*}
Note that the normalization  $\Omega^{(1)} = \bbra{w}_1 S_1 \kket{v}_1$ ensures that each component $\psi^{(1)}_{\bm\tau}$
has leading term $\mathbf{z}^{\lambda^{(1)}(\bm\tau)}$ with coefficient $1$, and all other terms correspond
to compositions $\mu$ with
\begin{equation*}
    \mu^+ < \lambda^{(1)+}(\bm\tau) = \left(1^L\right).
\end{equation*}
\label{ex:Koor_Psim1}
\end{example}

\begin{example}
With $m=2$,
\begin{equation*}
    \mathbb{A}^{(2)}(z) = L_{3,2}(z) \cdot b(z).
\end{equation*}
Then for $L=1$,
\begin{align*}
    \Omega^{(2)}\ket{\Psi^{(2)}(z_1; s)}
    & =
    \bbra{w}_3\bbra{\widetilde{w}}_2\bbra{w}_1 \
    (S_3)^3 (S_2)^2 S_1 \
    \Big(L_{3,2}(z_1) \cdot b(z_1)\Big) \
    \kket{v}_3\kket{\widetilde{v}}_2\kket{v}_1
    \\
    & =
    \bbra{w}_3\bbra{\widetilde w}_2 \ (S_3)^3 (S_2)^2 \ L_{3,2}(z_1) \ \kket{v}_3\kket{\widetilde v}_2 \ . \ \bbra{w}_1 S_1 b(z_1) \kket{v}_1
    \\
    & =
    \bbra{w}_3\bbra{\widetilde w}_2 \ (S_3)^3 (S_2)^2 \ L_{3,2}(z_1) \ \kket{v}_3\kket{\widetilde v}_2 \ . \ \Omega^{(1)} \ket{\Psi^{(1)}(z_1; s)},
\end{align*}
with
\begin{eqnarray*}
   & & \bbra{w}_3\bbra{\widetilde w}_2 \ (S_3)^3 (S_2)^2 \ L_{3,2}(z_1) \ \kket{v}_3\kket{\widetilde v}_2 \\
   & = &
    \bbra{w}_3\bbra{\widetilde{w}}_2(S_3)^3 (S_2)^2
    \begin{pmatrix}
        z_1^{-1} + a_3 a^\dagger_2
        &
        z_1^{-1} a_2 + a_3
        \\
        a^\dagger_3 + z_1  a^\dagger_2
        &
        a^\dagger_3  a_2 + z_1
    \end{pmatrix}_1
    \kket{v}_3\ket{\widetilde{v}}_2.
\end{eqnarray*}
For $L=2$,
\begin{align*}
    \Omega^{(2)}\ket{\Psi^{(2)}(z_1, z_2; s)}
    =
    \bbra{w}_3\bbra{\widetilde{w}}_2\bbra{w}_1 \ 
    (S_3)^3 (S_2)^2 S_1 \
    & \Big(L_{3,2}(z_1) \cdot b(z_1)\Big)
    \\
    & \otimes \Big(L_{3,2}(z_2) \cdot b(z_2)\Big) \
    \kket{v}_3\kket{\widetilde{v}}_2\kket{v}_1.
\end{align*}
The matrix $L_{3,2}(z_2)$ can be brought past $b(z_1)$ as they are in different physical spaces, and
their entries are in different auxiliary algebraic spaces.  Thus we obtain
\begin{equation*}
    \Omega^{(2)}\ket{\Psi^{(2)}(z_1, z_2; s)}
    =
    \bbra{w}_3\bbra{\widetilde w}_2 \ (S_3)^3 (S_2)^2 \ L_{3,2}(z_1) \otimes L_{3,2}(z_2) \ \kket{v}_3\kket{\widetilde v}_2 \
    . \ \left(\Omega^{(1)} \ket{\Psi^{(1)}(z_1, z_2; s)} \right),
\end{equation*}
where
\begin{eqnarray*}
  & &  \bbra{w}_3\bbra{\widetilde w}_2 \ (S_3)^3 (S_2)^2 L_{3,2}(z_1) \otimes L_{3,2}(z_2) \ \kket{v}_3\kket{\widetilde v}_2 \\
  & = &
    \bbra{w}_3\bbra{\widetilde w}_2 \ (S_3)^3 (S_2)^2 \
     \begin{pmatrix}
        z_1^{-1} + a_3 a^\dagger_2
        &
        z_1^{-1} a_2 + a_3
        \\
        a^\dagger_3 + z_1  a^\dagger_2
        &
        a^\dagger_3 a_2 + z_1
    \end{pmatrix}
\otimes \begin{pmatrix}
        z_2^{-1} + a_3 a^\dagger_2
        &
        z_2^{-1} a_2 + a_3
        \\
        a^\dagger_3 + z_2 a^\dagger_2
        &
        a^\dagger_3 a_2 + z_2
    \end{pmatrix} \
    \kket{v}_3\kket{\widetilde{v}}_2,
\end{eqnarray*}
is a real valued $4\times 4$ matrix
The normalization factor is
\begin{equation*}
    \Omega^{(2)}
    = \bbra{w}_3\bbra{\widetilde{w}}_2\bbra{w}_1 \
        (S_3)^3 (S_2)^2 S_1 \ \kket{v}_3\kket{\widetilde{v}}_2\kket{v}_1,
\end{equation*}
and it can be checked directly for $L=1,2$ that
each component $\psi^{(2)}_{\bm\tau}$
has leading term $\mathbf{z}^{\lambda^{(2)}(\bm\tau)}$ with coefficient $1$, and all other terms correspond
to compositions $\mu$ with
\begin{equation*}
    \mu^+ < \lambda^{(2)+}(\bm\tau) = \left(2^L\right).
\end{equation*}
\label{ex:Koor_Psim2}
\end{example}
We now give the general form.
\begin{theorem}
\label{theorem:Koor_Psisol}
The $q$KZ equations have a solution when $\xi = s^m$, written recursively on $m$:
For $m > 1$
\begin{eqnarray}\label{eq:Koor_Psimrec}
 & & \hspace{-1.5cm} \ket{\Psi^{(m)}(\mathbf{z};s)} =
   \frac{1}{\bbra{w}S^{2m-1}\kket{v} \bbra{\tilde{w}}S^{2m-2}\kket{\tilde{v}}} 
 \bbra{w}_{2m-1}\bbra{\tilde{w}}_{2m-2} \ (S_{2m-1})^{2m-1}(S_{2m-2})^{2m-2} \\
 & & \hspace{2cm} L_{2m-1,2m-2}(z_1) \otimes \ldots \otimes L_{2m-1,2m-2}(z_L) \ \kket{v}_{2m-1}\kket{\tilde{v}}_{2m-2} 
 \ \cdot \ \ket{\Psi^{(m-1)}(\mathbf{z};s)}, \nonumber 
\end{eqnarray}
with
\begin{equation}
    \ket{\Psi^{(1)}(\mathbf{z};s)}
    =
    \frac{1}{\bbra{w} S \kket{v}}\bbra{w}_1 \ S_1 \ b(z_1) \otimes \ldots \otimes b(z_L) \ \kket{v}_1.
    \label{eq:Koor_Psi1rec}
\end{equation}
The components of the solution, $\psi^{(m)}_{\bm\tau}(\mathbf{x};s)$, have leading term
$\mathbf{z}^{\lambda^{(m)}(\bm\tau)}$, and all other terms correspond to compositions $\mu$ with
\begin{equation*}
    \mu^+ < \lambda^{(m)+}(\bm\tau) = \left(m^L\right).
\end{equation*}
\end{theorem}
\proof
The recursive form \eqref{eq:Koor_Psimrec}, \eqref{eq:Koor_Psi1rec} is obtained by a reordering of the matrix product
form \eqref{eq:Koor_Psim}, as in Example~\ref{ex:Koor_Psim2}.

The second part of the claim, on the degree and normalization of components of the solution, can be proven
inductively.  We assume the property holds at $m-1$ and use \eqref{eq:Koor_Psimrec} to obtain the solution at $m$.
That is, we multiply by the `increment' matrix
\begin{eqnarray*}
 & & \frac{1}{\bbra{w}S^{2m-1}\kket{v} \bbra{\tilde{w}}S^{2m-2}\kket{\tilde{v}}} 
  \bbra{w}_{2m-1}\bbra{\tilde{w}}_{2m-2} \ (S_{2m-1})^{2m-1}(S_{2m-2})^{2m-2}  \\
 & & \hspace{5cm} L_{2m-1,2m-2}(z_1) \otimes \ldots \otimes L_{2m-1,2m-2}(z_L) \ \kket{v}_{2m-1}\kket{\tilde{v}}_{2m-2},
\end{eqnarray*}
which is a real valued $2^L \times 2^L$ matrix.
The following points can be deduced by writing \eqref{eq:Koor_Psimrec} and the increment matrix in component form:
\begin{itemize}
    \item
A term $\mathbf{z}^\mu$ with $\mu^+ = \left(m^L\right)$ can only be produced from the leading order terms
of the $m-1$ solution, which correspond to the partition $\left((m-1)^L\right)$, and thus we can ignore
sub-leading terms.

    \item
The diagonal entries of the increment matrix produce the term $\mathbf{z}^{\lambda^{(m)}(\bm\tau)}$ with
coefficient $1$ (plus lower order terms) in $\psi^{(m)}_{\bm\tau}$, from the corresponding
component of the $m-1$ solution.

    \item
The off-diagonal entries of the increment matrix, acting on the leading order term of a component of the $m-1$ solution,
either reduces the degree or leaves it unchanged.
\end{itemize}
These points are sufficient to show that the degree and normalization properties hold at $m$, assuming they hold
at $m-1$.  As the $m=1$ case was checked in Example~\ref{ex:Koor_Psim1}, the properties hold for all $m$.
\finproof

Similar results hold for the solution of the left $q$KZ equations at $s = \xi^m$ as stated in the following theorem.

\begin{theorem}
For integer $m \ge 0$ and $\xi = s^m$, solutions of the left $q$KZ equations can be constructed in matrix
product form, and can be defined recursively.  For $m > 0$
\begin{eqnarray}
  & & \hspace{-1cm} \bra{\Phi^{(m)}(\mathbf{z};s)} =
    \frac{1}{\bbra{w} S^{2m-1} \kket{v} \bbra{\tilde{w}} S^{2m} \kket{\tilde{v}}}
    \bra{\Phi^{(m-1)}(\mathbf{z};s)} \ \cdot \
   \bbra{w}_{2m-1}\bbra{\widetilde w}_{2m} \ (S_{2m-1})^{2m-1} (S_{2m})^{2m} \nonumber \\
  & & \hspace{4cm} L_{2m-1,2m}\left(\frac{1}{z_1}\right) \otimes \ldots \otimes  L_{2m-1,2m}\left(\frac{1}{z_L}\right) \kket{v}_{2m-1} \kket{\widetilde v}_{2m}
    \label{eq:Koor_Phimrec}
\end{eqnarray}
with
\begin{equation}
    \bra{\Phi^{(0)}(\mathbf{z};s)} = \bra{1} = (1, 1)^{\otimes L}
\end{equation}
The solution is non-zero: the component of the solution, $\phi^{(m)}_{\bm\tau}(\mathbf{z};s)$, contains the term
$\mathbf{z}^{-\lambda^{(m)}(\bm\tau)}$ with coefficient $1$, and all terms correspond to compositions $\mu$ with
\begin{equation*}
    \mu^+ \le \lambda^{(m)+}(\bm\tau) = \left(m^L\right).
\end{equation*}

\label{theorem:Koor_Phisol}
\end{theorem}
\proof
The proof of this theorem is very similar to that for
Theorem~\ref{theorem:Koor_Psisol}.  We will, however, comment briefly on the terms appearing in each component
$\phi^{(m)}_{\bm\tau}(\mathbf{x})$, and the normalization.  To do so, we look at the normalized `increment
matrix' taking the $m-1$ solution to the $m$ solution:
\begin{eqnarray*}
  & & \hspace{-1cm} \frac{1}{\bbra{w} S^{2m-1} \kket{v} \bbra{\tilde{w}} S^{2m} \kket{\tilde{v}}}
   \bbra{w}_{2m-1}\bbra{\widetilde w}_{2m} \ (S_{2m-1})^{2m-1} (S_{2m})^{2m} \nonumber \\
  & & \hspace{5cm} L_{2m-1,2m}\left(\frac{1}{z_1}\right) \otimes \ldots \otimes  L_{2m-1,2m}\left(\frac{1}{z_L}\right) \kket{v}_{2m-1} \kket{\widetilde v}_{2m}
\end{eqnarray*}
The diagonal entries of this matrix contain the term $\mathbf{z}^{-\lambda^{(1)}(\bm\tau)}$ with coefficient
$1$, which produce the term $\mathbf{z}^{-\lambda^{(m)}(\bm\tau)}$ in $\phi^{(m)}_{\bm\tau}$.  In the top row of
the increment matrix, the entry in column $\bm\tau'$ has leading term
\begin{equation}
    \frac{\bbra{\widetilde w}S^{2m}{a^{\sum_i \tau'_i}} \kket{\widetilde v}}
         {\bbra{\widetilde{w}} S^{2m} \kket{\widetilde{v}}} z_1 \ldots z_N,
    \label{eq:xaaacoeff}
\end{equation}
and as a consequence, each component $\phi^{(m)}_{\bm\tau}$ contains the term $\mathbf{z}^{(m^N)}$ with the same
coefficient as in \eqref{eq:xaaacoeff}.
\finproof

\begin{corollary}
For $m > 0$ and $\xi = t_0^{-1} t_L^{-1} t^{-(L-1)} s^{-m}$, the right $q$KZ equations
\eqref{eq:Koor_qKZi} -- \eqref{eq:Koor_qKZL} have solution
\begin{equation}
    \ket{\Psi(\mathbf{z}; s, \xi = t_0^{-1} t_L^{-1} t^{-(L-1)} s^{-m})}
    =
    U_{\text{GC}} \ket{\Phi^{(m)}(\mathbf{z}; s)}.
\end{equation}
\label{corr:Koor_PsiPhisol}
\end{corollary}
\proof
This follows from Lemma~\ref{lemma:Koor_leftrightsol}.
\finproof

The $m=0$ case is a bit special:
$\ket{\Psi(\mathbf{z}; s, \xi = t_0^{-1} t_L^{-1} t^{-(L-1)})} = U_{\text{GC}} \ket{1}$. The solution does not depend on 
$\mathbf{z}$ and $s$. Imposing in addition that $\xi=1$, i.e. $t_0t_Lt^{L-1}=1$, gives
a very simple ASEP stationary state. Indeed the system is at thermal equilibrium in this case: written in the usual ASEP parameters
the constraint is $\frac{\alpha\beta}{\gamma\delta}\left(\frac{p}{q}\right)^{L-1}=1$.

\subsubsection{Normalization and symmetric Koornwinder polynomials}

We can now make the connection between solutions of the $q$KZ equations, and the symmetric and non-symmetric
Koornwinder polynomials.

\begin{lemma}
The component $\psi^{(m)}_{\circ \ldots \circ}$ of the vector $\ket{\Psi^{(m)}(\mathbf{z},s)}$ is the
non-symmetric Koornwinder polynomial $E_{((-m)^L)}$.  All other components can be constructed through the
relations
\begin{equation}
\begin{aligned}
    \psi^{(m)}_{\circ \ldots \circ \bullet} & = t_L^{-1/2} T_L^{-1} \psi^{(m)}_{\circ \ldots \circ \circ},
    \\
    \psi^{(m)}_{\ldots \bullet \circ \ldots}
    &
    = t^{-1/2} T_i^{-1} \psi^{(m)}_{\ldots \circ \bullet \ldots},
    \qquad 1 \le i \le L-1.
\end{aligned}
\label{eq:Koor_constructpsis}
\end{equation}
The set of all components $\{\psi^{(m)}_{\bm\tau}\}$ forms a basis for $\mathcal{R}^{(m^L)}$, the space
spanned by non-symmetric Koornwinder polynomials $\{E_\mu | \mu \in \mathbb{Z}^L, \mu^+ = (m^L)\}$.
\label{lemma:Koor_psibasis}
\end{lemma}
\proof
By Theorem~\ref{theorem:Koor_Psisol}, and Lemma~\ref{lemma:Koor_asepEigenref} with $\xi = s^m$, $\psi^{(m)}_{\circ
\ldots \circ}$ is an eigenfunction of the $Y_i$ operators, and is a Laurent polynomial with the required
degree and normalisation.  Thus by uniqueness, we can identify $\psi^{(m)}_{\circ \ldots \circ} =
E_{((-m)^L)}.$  The relations \eqref{eq:Koor_constructpsis} come from the exchange relations
\eqref{eq:Koor_qkzExchangeBulk01} and \eqref{eq:Koor_qkzExchangeRight}.

The preceding parts of this lemma give the preconditions for
Proposition 1 and Corollary 1 of \cite{CantiniGDGW16}, from which it follows that
$\{\psi^{(m)}_{\bm\tau}\}$ forms a basis for $\mathcal{R}^{(m^L)}$.
\finproof

\begin{lemma}
The component $\phi^{(m)}_{\circ \ldots \circ}$ of the vector $\bra{\Phi^{(m)}(\mathbf{z},s)}$ is the
non-symmetric Koornwinder polynomial $E_{(m^L)}$.  All other components can be constructed through the
relations
\begin{align*}
    \phi^{(m)}_{\circ \ldots \circ \bullet} & = t_L^{1/2} T_L^{-1} \phi^{(m)}_{\circ \ldots \circ \circ},
    \\
    \phi^{(m)}_{\ldots \bullet \circ \ldots}
    &
    = t^{1/2} T_i^{-1} \phi^{(m)}_{\ldots \circ \bullet \ldots},
    \qquad 1 \le i \le L-1.
\end{align*}
The set of all components $\{\phi^{(m)}_{\bm\tau}\}$ forms a basis for $\mathcal{R}^{(m^L)}$.
\end{lemma}
\proof
This follows in the same way, with reference to Theorem~\ref{theorem:Koor_Phisol}, and Lemmas
\ref{lemma:Koor_asepleftEigenref} and \ref{lemma:Koor_leftqkzExchange}.
\finproof

\begin{lemma}
Given a solution $\ket{\Psi(\mathbf{z}; s, \xi)}$ of the $q$KZ equations \eqref{eq:Koor_qKZi} --
\eqref{eq:Koor_qKZL}, the sum of components
\begin{equation}
    \mathcal{Z}(\mathbf{z}; s, \xi) = \langle 1 | \Psi(\mathbf{z}; s, \xi) \rangle
\end{equation}
is $W_0$ invariant.
\label{lemma:Koor_ZW0}
\end{lemma}
\proof
We first note that $\bra{1}$ is a left eigenvector of $\check{R}_i$, $\overline{K}_L$ with eigenvalue 1 (see
\eqref{eq:Koor_Rcheck}, \eqref{eq:Koor_Kbar}).  Then applying $\bra{1}$ to the bulk and right boundary $q$KZ equations
\eqref{eq:Koor_qKZi}, \eqref{eq:Koor_qKZL} we see that $\mathcal{Z}(\mathbf{z}; s, \xi)$ is invariant under $s_i$, $1
\le i \le L$, and hence is $W_0$ invariant.
\finproof

\begin{theorem} \label{thm:Koor_symmetricKoornAsNormalisation}
The sum of components of $\ket{\Psi^{(m)}(\mathbf{z}; s)}$ is the
symmetric Koornwinder polynomial $P_{\left(m^L\right)}$.  That is
\begin{equation}
    P_{\left(m^L\right)}(\mathbf{z}) = \mathcal{Z}^{(m)}(\mathbf{z}; s),
\end{equation}
where
\begin{equation}
    \mathcal{Z}^{(m)}(\mathbf{z}; s)= \langle 1 | \Psi^{(m)}(\mathbf{z}; s) \rangle,
\end{equation}
and $\ket{\Psi^{(m)}(\mathbf{z}; s)}$ is the $q$KZ solution with $\xi = s^m$, constructed as in
Theorem~\ref{theorem:Koor_Psisol}.
\end{theorem}
\proof
By Lemmas \ref{lemma:Koor_psibasis} and \ref{lemma:Koor_ZW0}, $\mathcal{Z}^{(m)}(\mathbf{z},; s)$ is $W_0$ invariant,
and belongs to the space $\mathcal{R}^{(m^L)}$, and from Theorem~\ref{theorem:Koor_Psisol}, we see that it contains
$\mathbf{z}^{(m^L)}$ with coefficient 1.  The result then follows from the characterisation of symmetric
Koornwinder polynomials in \cite{Sahi99}, quoted in Theorem~\ref{theorem:Koor_SahiSymmetric}.
\finproof

We note that Theorem~\ref{thm:Koor_symmetricKoornAsNormalisation} implies a matrix product construction for the
symmetric Koornwinder polynomial $P_{(m^L)}$.  Direct computations from this form would be difficult, but the
structure leads to certain conjectures that we discuss in Section~\ref{sec:Koor_currentKoornwinder}.  We also note
that an integral form for the polynomial $P_{(m^L)}$ is already known \cite{Mimachi01}.  The solution of the
left $q$KZ equation is also related to the same symmetric Koornwinder polynomial.

\begin{theorem} \label{thm:Koor_symmetricKoornAsNormalisation2}
The sum of components of $U_{\text{GC}} \ket{\Phi^{(m)}(\mathbf{z}; s)}$ is proportional
to the symmetric Koornwinder polynomial $P_{\left(m^L\right)}$.  That is
\begin{equation}
    P_{\left(m^L\right)}(\mathbf{z}) \propto \mathcal{Z}^{(m)}(\mathbf{z}; s, \xi'),
\end{equation}
where
\begin{equation}
    \mathcal{Z}^{(m)}(\mathbf{z}; s, \xi')
        = \langle 1 | U_{\text{GC}} \ket{\Phi^{(m)}(\mathbf{z}; s)},
\end{equation}
with $\xi' = t_0^{-1} t_L^{-1} t^{-(L-1)} s^{-m}$.
\end{theorem}
\proof
Note that $U_{\text{GC}} \ket{\Phi^{(m)}(\mathbf{z}; s)}$ is the solution of the right $q$KZ equations
at $\xi'$, and the proof follows as in Theorem~\ref{thm:Koor_symmetricKoornAsNormalisation}.  However, because of
the structure of the components $\phi^{(m)}_{\bm\tau}$ (see Theorem~\ref{theorem:Koor_Phisol} and
\eqref{eq:Phim}), and the multiplication by matrix $U_{\text{GC}}$, the coefficient of
$\mathbf{z}^{(m^L)}$ in $\mathcal{Z}^{(m)}(\mathbf{z}; s, \xi')$ is not 1.  Thus the identification with
$P_{(m^L)}$ can only be made up to normalization.
\finproof
 
\section{Current fluctuations conjecture} \label{sec:Koor_currentKoornwinder}

\subsection{Quasi-classical limit}

The aim of this section is to make contact between the machinery developed previously, and the generating
function of the cumulants of the current. The idea is quite simple and arises from the following observation:
the constraint $\xi=s^m$ that was imposed in order to solve the $q$KZ equations, can be satisfied by setting
$s=\xi^{1/m}$, leaving $\xi$ free instead of $s$, which then implies $s \to 1$ as $m \to \infty$.  It appears
then natural to think that the scattering relation \eqref{eq:Koor_scattering} may degenerate, in this $s \to 1$
limit, to an eigenvector equation.  Then as $m \to \infty$, the vector
$\ket{\Psi^{(m)}(\mathbf{z};s=\xi^{1/m})}$ should thus converge in some sense to an eigenvector of the
scattering matrix.  To move towards this direction, we make the following conjectures.
 
\begin{conjecture}
 It is conjectured that 
 \begin{eqnarray}
  & & \lim\limits_{m\rightarrow \infty}\frac{\ket{\Psi^{(m)}(\mathbf{z};s=\xi^{1/m})}}{\mathcal{Z}^{(m)}(\mathbf{z};s=\xi^{1/m})}
  = \ket{\Psi_0(\mathbf{z};\xi)}, \\
  & & \lim\limits_{m\rightarrow \infty} \frac{\ln(\xi)}{m}\ln \left(\mathcal{Z}^{(m)}(\mathbf{z};s=\xi^{1/m})\right) 
  = F_0(\mathbf{z};\xi),
 \end{eqnarray}
 with $\ket{\Psi_0}$ and $F_0$  regular functions of $\mathbf{z}$.
\end{conjecture}
These conjectures are supported by strong numerical evidences (up to 3 sites) and by the fact that the matrix
product construction of $\ket{\Psi^{(m)}(\mathbf{z};s)}$ is similar to the one presented in
\cite{GorissenLMV12,Lazarescu13jphysA}. In those works, the authors developed a method called the
``perturbative matrix ansatz'', which allowed them to approximate the ground state of $M(\xi)$, at any order
in the current counting parameter $\xi$. Let us also mention that these kind of results have already been
observed in the context of $q$KZ equations of different models, and are known as the ``quasi-classical'' limit
\cite{Reshetikhin1992,ReshetikhinV94}.
 
Note that the second part of the conjecture can be immediately rewritten in terms of symmetric Koornwinder
polynomials as
\begin{equation}
  \lim\limits_{m\rightarrow \infty} \frac{\ln(\xi)}{m}\ln \left(P_{(m^L)}(\mathbf{z};s=\xi^{1/m})\right) 
  = F_0(\mathbf{z};\xi).
\end{equation}

In the following these conjectures will be considered as facts and properties will be deduced from them. But
one has to keep in mind that the validity of the deduced results relies obviously on the validity of these
conjectures.
 
\subsection{Generating function of the cumulants of the current}
 
 \begin{proposition}
  The function $F_0(\mathbf{z};\xi)$ is $W_0$ invariant and its derivative, with respect to any of the $x_i$,
  is invariant under the Gallavotti-Cohen symmetry
  $\xi \to \xi' = t_0^{-1}t_L^{-1}t^{-(L-1)} \xi^{-1}$.
 \end{proposition}
 \proof
  The $W_0$ invariance directly follows from the $W_0$ invariance of the symmetric Koornwinder polynomials. 
  The Gallavotti-Cohen symmetry follows from theorems \ref{thm:Koor_symmetricKoornAsNormalisation} and \ref{thm:Koor_symmetricKoornAsNormalisation2}
  which give that 
\begin{align*}
    P_{(m^L)}(\mathbf{z};s=\zeta^{1/m}) & = \mathcal{Z}^{(m)}(\mathbf{z}; s = \zeta^{1/m}, \xi = \zeta)
    \\
    & \propto \mathcal{Z}^{(m)}(\mathbf{z}; s = \zeta^{1/m}, \xi = t_0^{-1} t_L^{-1} t^{-(L-1)} \zeta^{-1})
\end{align*}
  with a proportionality coefficient independent 
  of $\mathbf{z}$. Taking the large $m$ limit it translates into the fact that $F_0(\mathbf{z};\xi')=F_0(\mathbf{z};\xi)+c$ with $c$ a constant 
  term independent of $\mathbf{z}$, which concludes the proof.
 \finproof

 In the following we will specify when needed the dependence on $s$ and $\xi$ of the scattering matrix $\mathcal{S}_i(\mathbf{z};s,\xi)$ defined
 in \eqref{eq:Koor_scatteringmatrix}.
 
 \begin{proposition}
 The vector $\ket{\Psi_0(\mathbf{z};\xi)}$ is an eigenvector of the scattering matrices evaluated at $s=1$,
 with
 \begin{equation} \label{eq:Koor_eigenvector_scatteringmatrices}
  \mathcal{S}_i(\mathbf{z};1,\xi)\ket{\Psi_0(\mathbf{z};\xi)}=
  \exp\left( z_i \frac{\partial F_0}{\partial z_i}(\mathbf{z};\xi)\right) \ket{\Psi_0(\mathbf{z};\xi)}.
 \end{equation}
 \end{proposition}
 \proof
  Our starting point is the scattering relation \eqref{eq:Koor_scattering} applied with $s=\xi^{1/m}$.  We divide by 
  $\mathcal{Z}^{(m)}(\mathbf{z};s=\xi^{1/m})$ to obtain
  \begin{eqnarray*}
  & &  \mathcal{S}_i(\mathbf{z};s=\xi^{1/m},\xi) 
  \frac{\ket{\Psi^{(m)}(\ldots, z_i, \ldots;s=\xi^{1/m})}}{\mathcal{Z}^{(m)}(\ldots, z_i, \ldots;s=\xi^{1/m})} \\
  & & \hspace{1cm} = \frac{\mathcal{Z}^{(m)}(\ldots, \xi^{1/m}z_i, \ldots;s=\xi^{1/m})}{\mathcal{Z}^{(m)}(\ldots, z_i, \ldots;s=\xi^{1/m})}
  \frac{\ket{\Psi^{(m)}(\ldots, \xi^{1/m}z_i, \ldots;s=\xi^{1/m})}}{\mathcal{Z}^{(m)}(\ldots, \xi^{1/m}z_i, \ldots;s=\xi^{1/m})}.
\end{eqnarray*}
We then have the limits 
 \begin{eqnarray*}
  & & \lim\limits_{m\rightarrow \infty}\mathcal{S}_i(\mathbf{z};s=\xi^{1/m},\xi)
  =\mathcal{S}_i(\mathbf{z};1,\xi), \\
  & & \lim\limits_{m\rightarrow \infty}\frac{\ket{\Psi^{(m)}(\ldots, z_i, \ldots;s=\xi^{1/m})}}{\mathcal{Z}^{(m)}(\ldots, z_i, \ldots;s=\xi^{1/m})}
  = \ket{\Psi_0(\ldots, z_i, \ldots;\xi)}, \\
  & & \lim\limits_{m\rightarrow \infty}\frac{\ket{\Psi^{(m)}(\ldots, \xi^{1/m}z_i, \ldots;s=\xi^{1/m})}}{\mathcal{Z}^{(m)}(\ldots, \xi^{1/m}z_i, \ldots;s=\xi^{1/m})}
  = \ket{\Psi_0(\ldots, z_i, \ldots;\xi)}, \\
  & & \lim\limits_{m\rightarrow \infty} \frac{\mathcal{Z}^{(m)}(\ldots, \xi^{1/m}z_i, \ldots;s=\xi^{1/m})}{\mathcal{Z}^{(m)}(\ldots, z_i, \ldots;s=\xi^{1/m})}
  = \exp\left( z_i \frac{\partial F_0}{\partial z_i}(\mathbf{z};\xi)\right),
 \end{eqnarray*}
 which yield the desired result.
 \finproof

 \begin{proposition}
 The vector $\ket{\Psi_0(\mathbf{1};\xi)}$ is an eigenvector of the deformed Markov matrix, with
 \begin{equation}
  M(\xi)\ket{\Psi_0(\mathbf{1};\xi)}=\frac{p-q}{2}\frac{\partial^2 F_0}{\partial
  z_i^2}(\mathbf{1};\xi)\ket{\Psi_0(\mathbf{1};\xi)}.
 \end{equation}
This implies immediately the following expression for the generating function of the cumulants of the
 current:
 \begin{equation}
  E(\mu)=\Lambda_0(e^{\mu})=\frac{p-q}{2}\frac{\partial^2 F_0}{\partial z_i^2}(\mathbf{1};e^{\mu}).
 \end{equation}
 \end{proposition}
\proof
This is proven by taking the derivative of \eqref{eq:Koor_eigenvector_scatteringmatrices} with respect to $z_i$ and
then setting $z_1=\dots=z_L=1$. One has to make basic use of the properties given in
\eqref{eq:Koor_properties_scatteringmatrices}, and notice the fact that $\frac{\partial F_0}{\partial
z_i}(\mathbf{1};\xi)=0$ because $F_0$ is $W_0$ invariant.
\finproof

Note that despite revealing a beautiful connection between the symmetric functions and the fluctuations of the
current, in practice, the last expression does not help to compute the cumulants of the current because a
closed expression for $F_0$ is missing.
 
However a step can be made toward an exact expression of $F_0$, using the characterization of the symmetric Koornwinder polynomials
 as eigenfunctions of the finite difference operator $D$ defined in \eqref{eq:Koor_operatorD}. 
 The eigenvalue $d_{(m^L)}$ of this operator associated to the symmetric Koornwinder polynomial $P_{(m^L)}$ is given for $s=\xi^{1/m}$ by
 \begin{equation}
 d_0(\xi)=\frac{1-t^L}{1-t}(\xi-1)(t_0t_Lt^{L-1}-1/\xi).
 \end{equation}
It is straightforward to check that $d_0(\xi)$ is invariant under the Gallavotti-Cohen symmetry, that is $d_0(\xi)=d_0(\xi')$.

In the following we will explicitly write the dependence on $s$ of the functions $g_i(\mathbf{z};s)$ defined in \eqref{eq:Koor_function_g_i}.
\begin{proposition}
We have the following characterization of the function $F_0$:
\begin{eqnarray*}
 & &\sum_{i=1}^{L} g_i(\mathbf{z};1)\left[\exp\left( z_i \frac{\partial F_0}{\partial z_i}(\mathbf{z};\xi)\right)-1\right]
 +\sum_{i=1}^{L} g_i(\mathbf{z}^{-1};1)\left[\exp\left(- z_i \frac{\partial F_0}{\partial z_i}(\mathbf{z};\xi)\right)-1\right] \\
 & & =d_0(\xi)
\end{eqnarray*}
\end{proposition}
\proof
 This follows directly from the relation \eqref{eq:Koor_eigenvalue_eq_symmetric_Koornwinder}
 applied for the symmetric Koornwinder polynomial $P_{(m^L)}$ and $s=\xi^{1/m}$. Dividing the 
 latter relation by $P_{(m^L)}(\mathbf{z};s=\xi^{1/m})$ and taking the large $m$ limit yield the desired result.
\finproof

It would be interesting to understand if this characterization of the function $F_0$ can be related to a Baxter $t-Q$ relation \cite{Baxter82}.
It has been shown in \cite{CrampeMRV15inhomogeneous} that the normalisation of the stationary state (corresponding to the case $s=\xi=1$)
satisfies a $t-Q$ difference equation.
We can mention also in this context the work \cite{LazarescuP14} where the authors constructed a Baxter Q operator for the open ASEP
(with current-counting deformation) and derived the corresponding $t-Q$ relations.

\chapter{Hydrodynamic limit} \label{chap:five}

In the previous chapters we focused essentially on the study of systems defined on a finite size lattice. We computed exactly physical observables
in these models. A lot of efforts were put on the comprehension of the mathematical structures related to the stochastic processes,
in particular the algebraic and combinatorial aspects of the Markov matrices and of the stationary states. The emphasis was mainly
put on the mathematical methods involved in the exact computations rather than on the physical interpretation of the results. 

This chapter focuses mainly on the physical description of the models 
and tries to follow the usual purpose of statistical physics.
The goal is to try to bridge the gap between the microscopic scale and the macroscopic scale. 
At the microscopic scale, we know precisely the interactions between each elementary components of the system
(typically the Coulomb interactions between atoms forming a plasma). 
At the macroscopic scale, we would like to identify, for a system composed of a huge number of components, 
a few number of relevant physical macroscopic observables (typically the temperature or the pressure).
We would like to infer physical laws or principles satisfied by these observables, 
in order to describe efficiently the behavior of the system at this scale 
(typically an equation of state such as the law of ideal gas).

In our precise case the system is composed by a big number of particles evolving on a lattice and 
the microscopic dynamics is defined by the Markov matrix associated to the model. From this stochastic dynamics defined at the particle scale,
we would like to infer the macroscopic behavior of the system. In other words, we would like to construct a simple formalism to answer easily 
questions like: what is the mean particles flow (and also its fluctuations) through the system? or is it likely to observe a 
high concentration of particles at some place? We can easily see the potential wide range of applications (in population dynamics,
traffic flow, biology...) of such theory.
The essential feature of such description lies in the fact the system comprises a huge number of individual components, so that the 
details of the microscopic dynamics can be averaged out at the macroscopic scale to give rise to a universal macroscopic description of the system.

The first step, that we need to focus on, is thus to take what is commonly called the thermodynamic limit, or the hydrodynamic limit. 
It consists in letting the size $L$ of the lattice go to infinity. We will in particular determine the behavior of the physical observables,
computed exactly for a finite size lattice in chapter \ref{chap:three}, in this limit. We will see that they give precious hints 
concerning the macroscopic behavior of the system, such as transitions between different regimes, called phase transitions 
(for instance a transition from fluid traffic flow to jammed traffic flow) 
or such as the relaxation time toward the stationary state.

We will see in a second time that, while taking this large size $L \rightarrow \infty$ limit, we can define macroscopic variables 
(that can be heuristically interpreted as local particle density or local particle current) whose stochastic evolution 
follows simple laws as $L  \rightarrow \infty$. We will typically observe large deviation principles for these macroscopic variables 
that are expected to be an efficient generalization of thermodynamic potentials to non-equilibrium systems. This could be though as a prelude
toward a general theory to describe out-of-equilibrium systems.

\section{Hydrodynamic limit and density profile}

\subsection{Hydrodynamic limit}

\subsubsection{Continuous limit of the lattice}

We want to let the lattice size $L$ go to infinity, {\it i.e} to have an infinite number of sites in the system. We precise here the 
mathematical meaning of this limit. The lattice is seen as embedded in the segment $[0,1]$ of the real line. The site $i$ is located 
at coordinate 
\begin{equation}
 x=\frac{i}{L}.
\end{equation}
When the number of sites $L$ increases and tends to infinity, the distance between two adjacent sites, which is equal to $1/L$,
decreases and converges toward $0$. For a given $x\in [0,1]$, it is straightforward to check that the coordinate of the site $\lfloor Lx \rfloor$
(where $\lfloor \cdot \rfloor$ denotes the floor function), which is equal to $\lfloor Lx \rfloor /L$, converges toward $x$ when $L$ goes
to infinity. In this thermodynamic limit we can thus define physical quantities depending on a continuous variable $x\in [0,1]$.

For instance we can define a function 
\begin{equation}
 \overline{\rho}_{\tau}(x)=\lim\limits_{L\rightarrow \infty} \langle \rho_{\tau}^{(\lfloor Lx \rfloor)} \rangle,
\end{equation}
which stands for the mean particle density of species $\tau$ at position $x \in [0,1]$ in the thermodynamic limit.

We define also the limit of the two-point function
\begin{equation}
 \overline{\rho}_{\tau,\tau'}(x,y)=\lim\limits_{L\rightarrow \infty} \langle \rho_{\tau}^{(\lfloor Lx \rfloor)}\rho_{\tau'}^{(\lfloor Ly \rfloor)} \rangle,
\end{equation}
and more generally the multi-point correlation function
\begin{equation}
 \overline{\rho}_{\tau^{(1)},\dots,\tau^{(k)}}(x_1,\dots,x_k)=
 \lim\limits_{L\rightarrow \infty} \langle \rho_{\tau^{(1)}}^{(\lfloor Lx_1 \rfloor)} \dots \rho_{\tau^{(k)}}^{(\lfloor Lx_k \rfloor)} \rangle.
\end{equation}

The limits of the connected correlation functions are more subtle because we might need to rescale the function with a $L$-dependent factor
to obtain a non-vanishing (and non-diverging) real value when performing the large $L$ limit.
For instance for the connected two-point function we may have
\begin{equation}
 \overline{\rho}_{\tau,\tau'}^{c}(x,y)  = 
 \lim\limits_{L\rightarrow \infty} f(L) \langle \rho_{\tau}^{(\lfloor Lx \rfloor)}\rho_{\tau'}^{(\lfloor Ly \rfloor)} \rangle_c, 
\end{equation}
where the function $f$ depends on the model under consideration (see examples below).

Similar issues appear when defining the thermodynamic limit of the particle current on the lattice, where the rescaling of the 
current may also depend on the model under consideration. We will encounter two different classes of model in the examples below: models
in which the lattice current does not need to be rescaled and that are called ballistic, and models in which the lattice current has 
to be multiplied by a factor $L$ and that are called diffusive.

Note that in the case of a single species model, {\it i.e} for $N=1$, we will lighten the notation by writing  $\overline{\rho}(x)$,
$\overline{\rho}(x,y)$, $\overline{\rho}^{c}(x,y)$,... instead of $\overline{\rho}_{1}(x)$,
$\overline{\rho}_{1,1}(x,y)$, $\overline{\rho}_{1,1}^{c}(x,y)$,...

In what follows, we provide a detailed study of the thermodynamic limit of the models that we encountered in chapter \ref{chap:three}: the DiSSEP,
the 2-TASEP and the multi-species SSEP.

\subsubsection{Limit of observables in the DiSSEP} \label{subsubsec:DiSSEP_thermo}

In this subsection, we study the thermodynamic limit of the DiSSEP. Once again this work was realized in details in \cite{CrampeRRV16} and 
we expose hereafter the results obtained there. The computations will be based on the 
exact expressions derived for a finite size lattice in chapter \ref{chap:three}.
We are interested in the case where there exists a competition between the diffusion of particles and 
the evaporation/condensation of pairs. It is indeed possible that, when performing the thermodynamic limit in a naive way, the effect of one process 
completely overcomes the other process.

In order to maintain the competition in the continuous limit, we have to scale properly the parameters of the model.
In other words, the mean time for a particle to go through the lattice by diffusion must be comparable to the time 
for it to be evaporated.

In order to evaluate quantitatively this competition,
let us write the time evolution of the one-point correlation function for $1<i<L$
\begin{eqnarray}
 \frac{d \langle \tau_i \rangle_t}{dt} & = & \langle \tau_{i-1}(1-\tau_i)\rangle_t + \langle (1-\tau_i)\tau_{i+1} \rangle_t -\langle (1-\tau_{i-1})\tau_i\rangle_t - 
 \langle \tau_i(1-\tau_{i+1})\rangle_t \label{eq:DiSSEP_time_densitybis}\\
 & & +\lambda^2\Big( \langle (1-\tau_{i-1})(1-\tau_i)\rangle_t + \langle (1-\tau_i)(1-\tau_{i+1})\rangle_t - \langle \tau_{i-1}\tau_i\rangle_t - 
 \langle \tau_i\tau_{i+1}\rangle_t \Big)\nonumber \\
 & = & \langle \tau_{i-1} \rangle_t + \langle \tau_{i+1} \rangle_t -2\langle \tau_i \rangle_t
 +\lambda^2\Big(2-\langle \tau_{i-1} \rangle_t - \langle \tau_{i+1} \rangle_t -2 \langle \tau_i \rangle_t \Big),\label{eq:DiSSEP_time_density}
\end{eqnarray}
where $\langle \cdot \rangle_t$ stands for the expectation with respect to the probability density $P_t(\cC)$ whose time evolution obeys the 
master equation.
Note that although the two-point correlation functions cancel when going from \eqref{eq:DiSSEP_time_densitybis} to \eqref{eq:DiSSEP_time_density},   
the mean field approximation is not exact in the sense that the connected two-point function does not vanish, 
see \eqref{eq:DiSSEP_two_point_function_bis}.

We want to take the large $L$ limit in equation \eqref{eq:DiSSEP_time_density}. 
For $x\in [0\,,\,1]$, we define 
\begin{equation} \label{eq:DiSSEP_definition_time_density}
\overline\rho_t(x)=\lim\limits_{L\rightarrow \infty}\langle \tau_{\lfloor Lx \rfloor} \rangle_{L^2t}.
\end{equation}
Note that the time has been speeded up by a factor $L^2$ which is necessary in diffusive systems to observe non-trivial behavior.
We obtain
\begin{equation}
 \frac{\partial\overline\rho_t}{\partial t}(x)=\frac{\partial^2 \overline\rho_t}{\partial x^2}(x)+2\Big(\lim\limits_{L\rightarrow \infty}L^2\lambda^2\Big)(1-2\overline\rho_t(x))
\end{equation}
We see on the previous equation that we have to take $\lambda=\lambda_0/L$ in order to have a balance between diffusion and creation-annihilation.
In this case we obtain
\begin{equation} \label{eq:DiSSEP_time_density_thermo}
 \frac{\partial \overline\rho_t}{\partial t}= \frac{\partial^2 \overline\rho_t}{\partial x^2}+2\lambda_0^2(1-2\overline\rho_t)
\end{equation}
with the boundary conditions $\overline\rho_t(0)=\rho_l$ and $\overline\rho_t(1)=\rho_r$. 
This equation shows that the correlation length for this scaling
is finite. Indeed, the stationary density shown in equation \ref{eq:DiSSEP_density_thermo} decays as
$\overline{\rho}(x)-1/2\sim \exp(-2\lambda_0x)$ for $x$ far from the boundaries. The correlation length can thus be defined as $1/(2\lambda_0)$.

Without rescaling $\lambda$  w.r.t $L$ and without speeding up the time in \eqref{eq:DiSSEP_definition_time_density}
(or for $\lambda=\lambda_0/L^\mu$ with $\mu<1$ and speeding up the time with a factor $L^{2\mu}$
in \eqref{eq:DiSSEP_definition_time_density}), 
the diffusive term drops out and the density satisfies
\begin{equation} \label{eq:DiSSEP_time_density_thermo1}
 \frac{\partial \overline\rho_t}{\partial t}= 2\lambda^2(1-2\overline\rho_t)\;.
\end{equation}
In the case where $\lambda=\lambda_0/L^\mu$ for  $\mu>1$ (and keeping the definition \eqref{eq:DiSSEP_definition_time_density} as it is), 
the system becomes a pure diffusive model for large $L$  and one gets for the density
\begin{equation} \label{eq:DiSSEP_time_density_thermo2}
 \frac{\partial \overline\rho_t}{\partial t}= \frac{\partial^2 \overline\rho_t}{\partial x^2}\;.
\end{equation}

\paragraph*{Thermodynamic limit of the observables}
We are now interested in evaluating the thermodynamic limit of the physical quantities computed for a finite size lattice in chapter \ref{chap:three}.
We will perform this limit in the case where there is a competition between diffusion and evaporation/condensation, {\it i.e} when 
we take $\lambda=\lambda_0/L$.

We are first interested in the one point correlation function in the continuous limit.
\begin{proposition}
The mean particle density in the thermodynamic limit is given for $x\in [0,1]$ by the exact expression
\begin{equation} \label{eq:DiSSEP_density_thermo}
\overline{\rho}(x)= \lim\limits_{L\rightarrow \infty} \langle \tau_{\lfloor Lx \rfloor} \rangle 
=\frac{1}{2}+\frac{1}{2\sinh 2\lambda_0} \left(q_1 e^{-2\lambda_0(x-1/2)}+q_2 e^{2\lambda_0(x-1/2)} \right)\;,
\end{equation}
where
\begin{equation}\label{def:DiSSEP_q1}
q_1=\left(\rho_l+\rho_r-1\right)\sinh(\lambda_0)-\left(\rho_r-\rho_l\right)\cosh(\lambda_0),
\end{equation}
and 
\begin{equation}\label{def:DiSSEP_q2}
q_2=\left(\rho_l+\rho_r-1\right)\sinh(\lambda_0)+\left(\rho_r-\rho_l\right)\cosh(\lambda_0)\;.
\end{equation}
It is easy to check that it satisfies the stationary version of \eqref{eq:DiSSEP_time_density_thermo}. 
\end{proposition}
\proof
It can be directly computed from the expression \eqref{eq:DiSSEP_one_point_function} for a finite size lattice.
\finproof

We can also compute the connected two-point correlation function in this limit. We can see that it scales as $\frac1L$,
{\it i.e } it has weak correlations.
\begin{proposition}
For $x,y\in [0,1]$, $x<y$, we have the following analytical expression of the connected two-point correlation function in the thermodynamic limit
\begin{eqnarray}
& & \overline\rho^c(x,y) = \lim\limits_{L\rightarrow \infty} L \times \langle \tau_{\lfloor Lx \rfloor}\tau_{\lfloor Ly \rfloor} \rangle_c =  
\frac{2\lambda_0  \,q_1q_2}{\left(\sinh 2\lambda_0 \right)^3}\,\sinh 2\lambda_0(1-y) \sinh 2\lambda_0 x \,.\qquad
\end{eqnarray}
\end{proposition}
For $\lambda_0<<1$, this connected two-point correlation function behaves algebraically w.r.t. $x$ and $y$ whereas it behaves exponentially and 
is short range for $\lambda_0>>1$. 
\proof
Once again this is directly evaluated from the finite size lattice expression \eqref{eq:DiSSEP_two_point_function}.
\finproof

We can also study the particle lattice current and condensation current.
\begin{proposition}
The thermodynamic limit of the mean particle currents are given for $x\in [0,1]$ by
\begin{equation}
 \overline j_{lat}(x):= \lim\limits_{L\rightarrow \infty} L \times \langle j_{lat}^{\lfloor Lx \rfloor \rightarrow \lfloor Lx \rfloor +1} \rangle =
 \frac{\lambda_0}{\sinh 2\lambda_0} \left(q_1 e^{-2\lambda_0(x-1/2)}-q_2 e^{2\lambda_0(x-1/2)} \right)\;,
\end{equation}
and 
\begin{equation}
 \overline j_{cond}(x):= \lim\limits_{L\rightarrow \infty} L^2 \times \langle j_{cond}^{\lfloor Lx \rfloor, \lfloor Lx \rfloor +1} \rangle =
  \frac{-2\lambda_0^2}{\sinh 2\lambda_0} \left(q_1 e^{-2\lambda_0(x-1/2)}+q_2 e^{2\lambda_0(x-1/2)} \right)\;.
\end{equation}
\end{proposition}
\proof
These formulas are derived from the explicit expression of the currents for a finite size lattice 
\eqref{eq:DiSSEP_mean_lattice_current} and \eqref{eq:DiSSEP_mean_condensation_current}
\finproof

Note that these expressions are consistent with the fact that when the system reaches a thermodynamic equilibrium,
that is for $\rho_l=\rho_r=1/2$ (or equivalently $q_1=q_2=0$), both currents vanish.

The particle conservation law \eqref{eq:DiSSEP_particle_conservation_discret} becomes in the thermodynamic limit
\begin{equation*}
 -\frac{d\overline j_{lat}}{dx}(x)+\overline j_{cond}(x)=0,
\end{equation*}
which is satisfied by the expressions above.
In the same way, relations \eqref{eq:DiSSEP_drho_discret} and \eqref{eq:DiSSEP_jcond_discret} become in the thermodynamic limit:
\begin{equation}\label{eq:DiSSEP_rho_cond_thermo}
\frac{d\overline\rho}{dx}(x)+\overline j_{lat}(x)=0\,,\qquad \overline j_{cond}(x)=2\lambda_0^2\Big(1-2\overline\rho(x)\Big).
\end{equation}

\begin{figure}[pht]
  \begin{center}
    \subfloat[Density]{
      \includegraphics[width=0.4\textwidth,height=0.3\textheight]{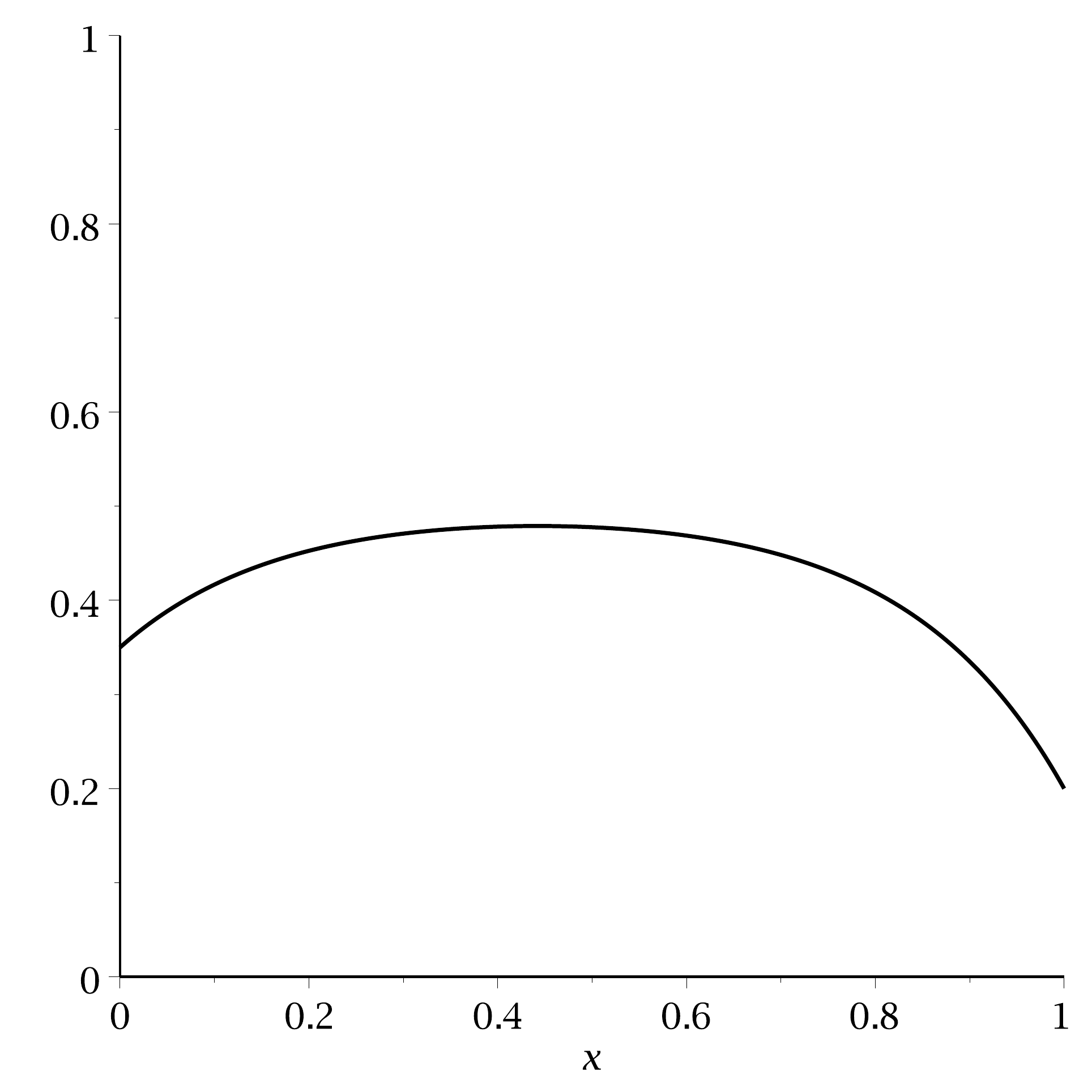}
      \label{sub:densite_concave}
                         }\hfill
    \subfloat[Mean value (---) and variance ($\cdots$) of the lattice current]{
      \includegraphics[width=0.4\textwidth,height=0.3\textheight]{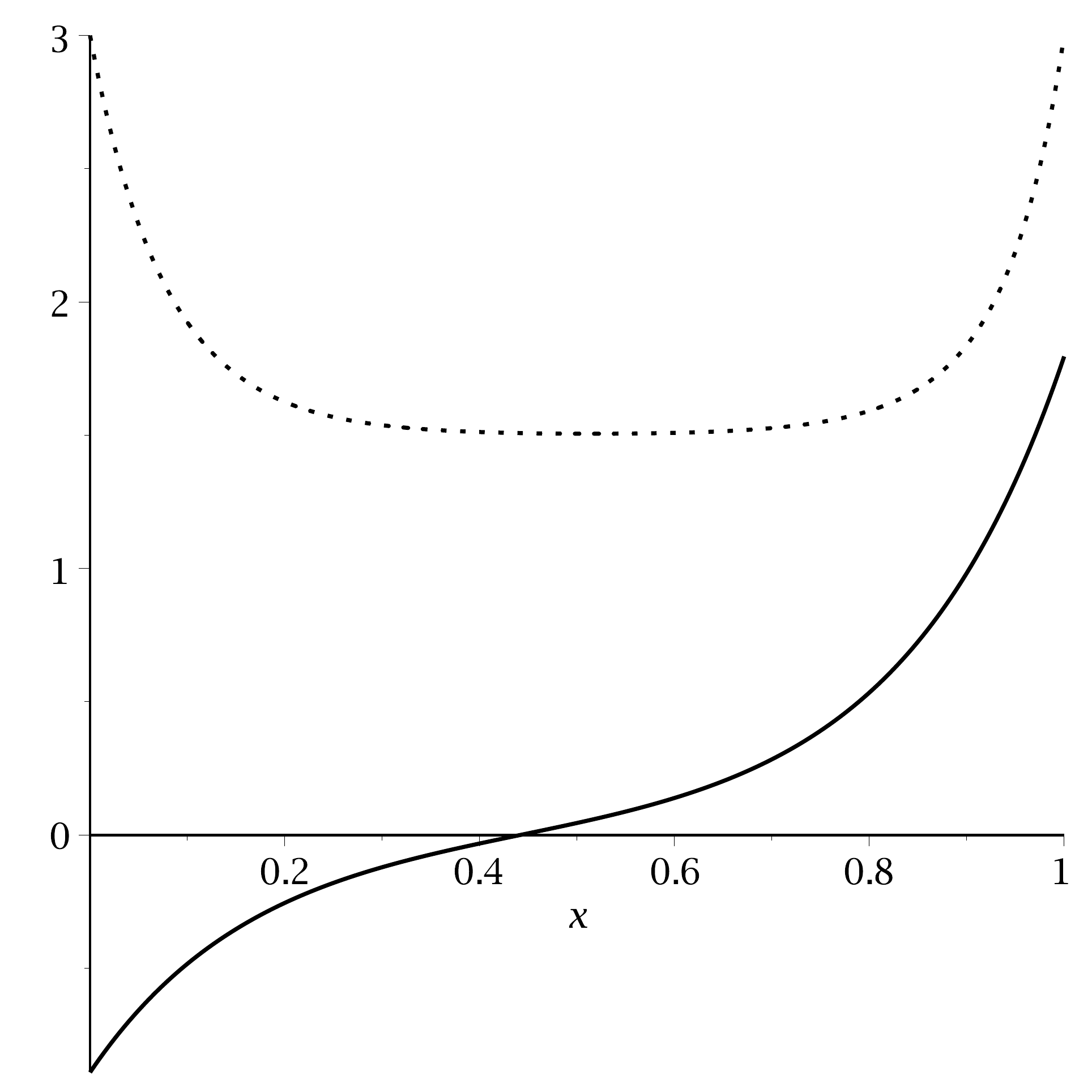}
      \label{sub:courant_concave}
                         }
    \caption{Plot of the density and of the lattice current for $\rho_l=0.35$, $\rho_r=0.2$ and $\lambda_0=3$.}
    \label{fig:concave}
  \end{center}
\end{figure}

\begin{figure}[pht]
  \begin{center}
    \subfloat[Density]{
      \includegraphics[width=0.4\textwidth,height=0.3\textheight]{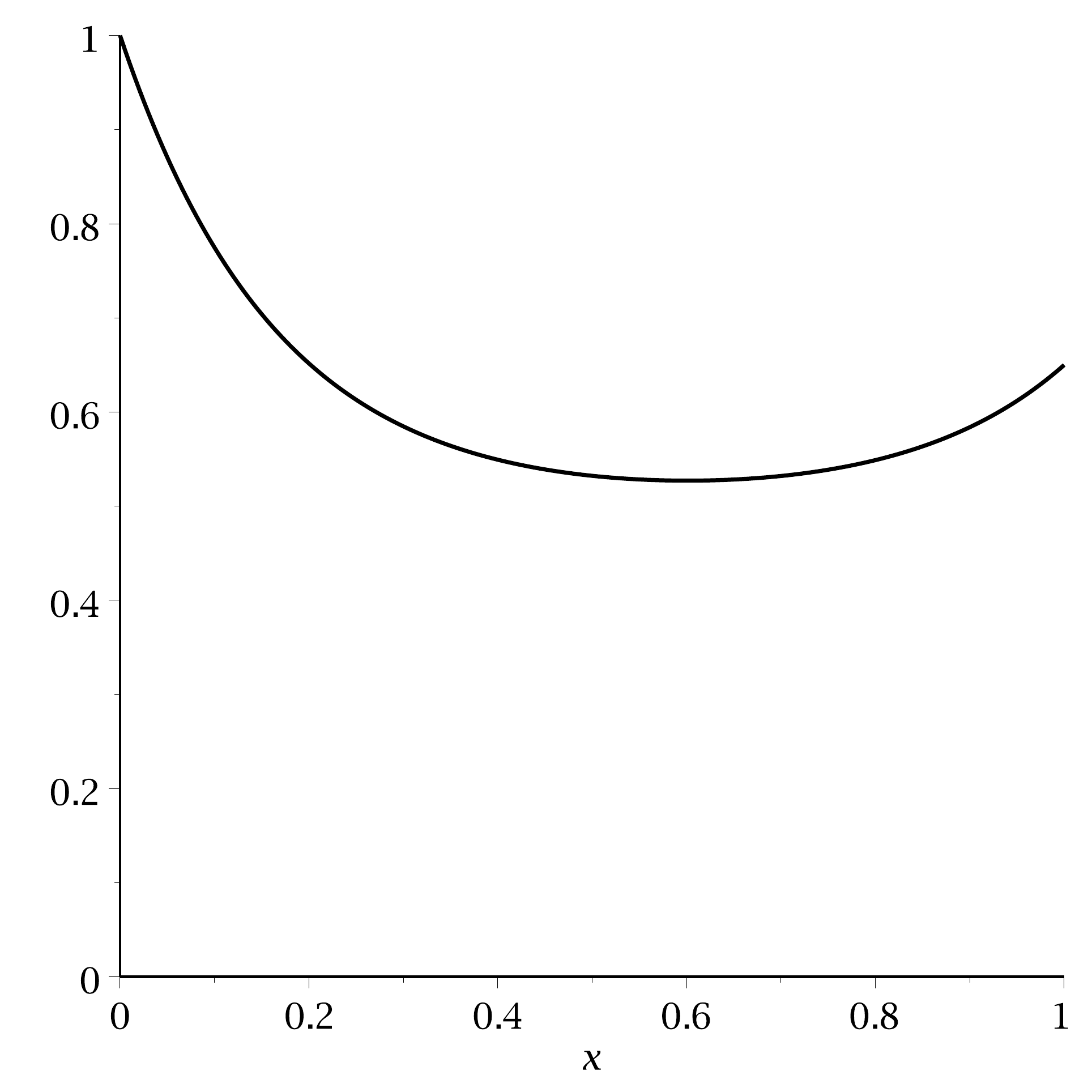}
      \label{sub:densite_convexe}
                         }\hfill
    \subfloat[Mean value (---) and variance ($\cdots$) of the lattice current]{
      \includegraphics[width=0.4\textwidth,height=0.3\textheight]{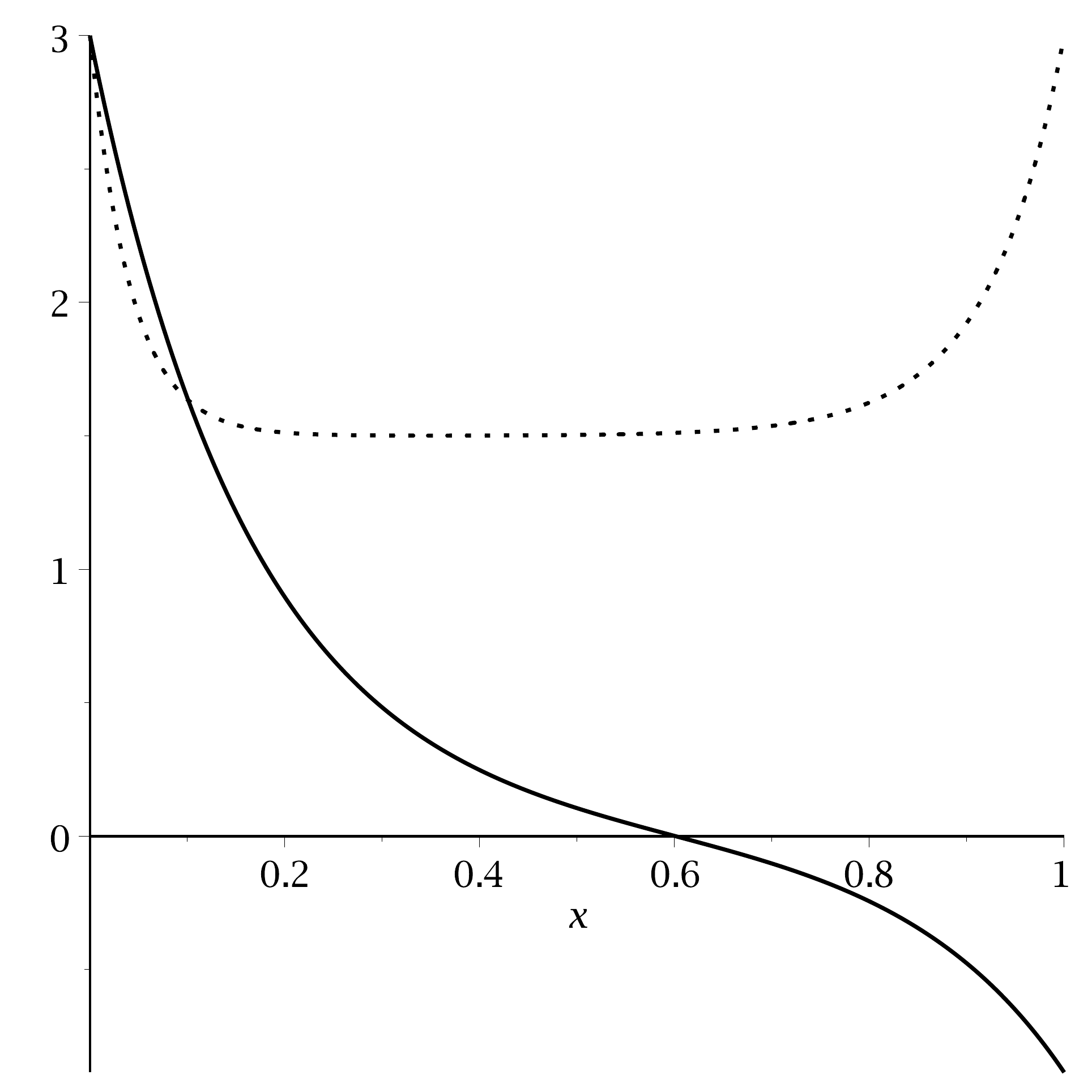}
      \label{sub:courant_convexe}
                         }
    \caption{Plot of the density and of the lattice current for $\rho_l=1$, $\rho_r=0.65$ and $\lambda_0=3$.}
    \label{fig:convexe}
  \end{center}
\end{figure}

\begin{figure}[pht]
  \begin{center}
    \subfloat[Density]{
      \includegraphics[width=0.4\textwidth,height=0.3\textheight]{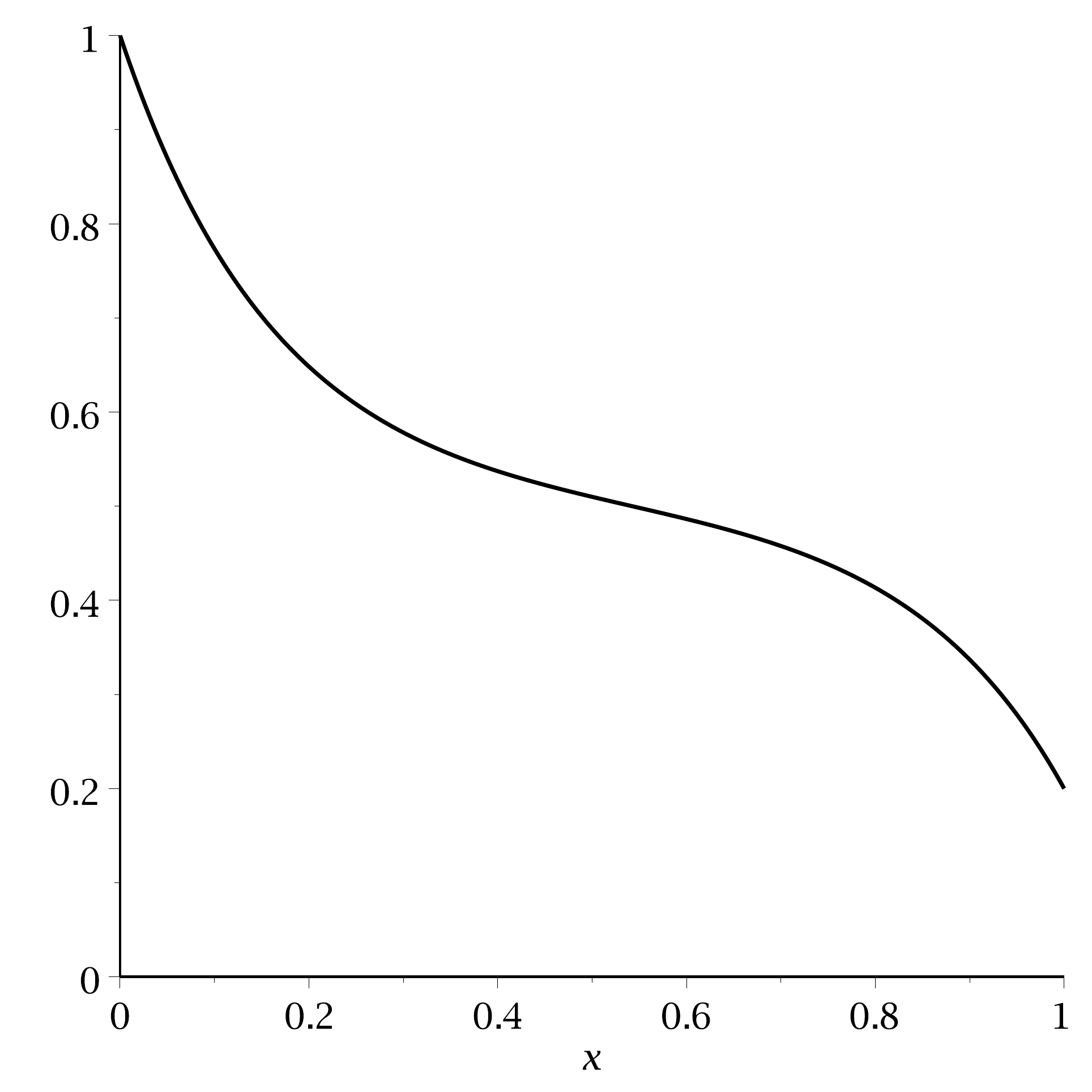}
      \label{sub:densite_monotone}
                         }\hfill
    \subfloat[Mean value (---) and variance ($\cdots$) of the lattice current]{
      \includegraphics[width=0.4\textwidth,height=0.3\textheight]{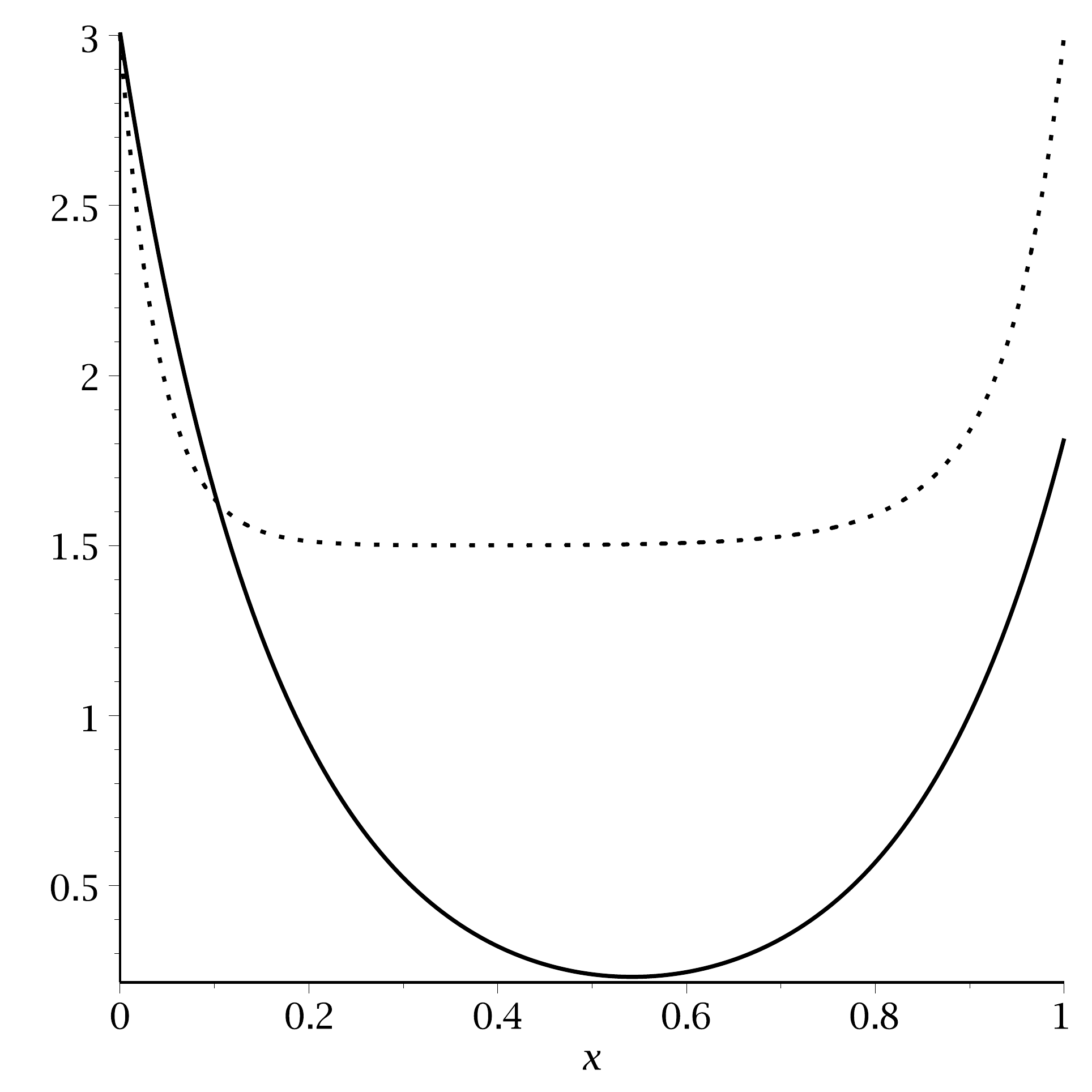}
      \label{sub:courant_monotone}
                         }
    \caption{Plot of the density and of the lattice current for $\rho_l=1$, $\rho_r=0.2$ and $\lambda_0=3$.}
    \label{fig:monotone}
  \end{center}
\end{figure}

\paragraph*{Behavior of the density and the currents}
Depending on the values of $q_1$ and $q_2$ defined in \eqref{def:DiSSEP_q1} and \eqref{def:DiSSEP_q2}, the behavior of the density may change: 
 \begin{itemize}
 \item the density is not monotonic when $e^{-2\lambda_0}< \frac{q_1}{q_2}< e^{2\lambda_0}$, which implies that $q_1$ and $q_2$ have the same sign. 
 In that case, it possesses an extremum
 at $\overline{x}$ 
 satisfying $e^{4\lambda_0(\overline x -1/2)}=\frac{q_1}{q_2}$. The lattice current vanishes at this point.  
 \begin{itemize}
 \item the density presents a maximum
 \begin{equation}
  \overline{\rho}(\overline x)=\frac12-\frac{\sqrt{q_1q_2}}{\sinh(2\lambda_0)}
 \end{equation}
 when $q_1,q_2<0$. Let us remark that in this case, the density is everywhere smaller than $1/2$.
 Example of such behavior can be seen on figure \ref{fig:concave}.
 
The lattice current changes direction at the point $\overline x$, as expected since the lattice current 
 goes from high density to low density.
 At this point, the condensation current is minimal but positive, since the density is smaller than $1/2$, so that condensation is promoted.
 \item It presents a minimum
 \begin{equation}
  \overline{\rho}(\overline x)=\frac12+\frac{\sqrt{q_1q_2}}{\sinh(2\lambda_0)}
 \end{equation}
 when $q_1,q_2>0$. In this case, the density is everywhere greater than $1/2$. 
 
 The condensation current is negative but maximal, so that the evaporation is minimal. As previously, the lattice current changes sign at $\overline x$, 
 still going from high density to low density. 
  Example of such behavior can be seen on figure \ref{fig:convexe}.
 \end{itemize}
 \item The density is monotonic from $\rho_l$ to $\rho_r$ when $\frac{q_1}{q_2}<e^{-2\lambda_0} $ or $\frac{q_1}{q_2} > e^{2\lambda_0}$.
 In this case, the lattice current never vanishes.  Example of such behavior can be seen on figure \ref{fig:monotone}.

\end{itemize}
The condensation current follows the same pattern, due to the relation \eqref{eq:DiSSEP_rho_cond_thermo}.
The lattice current behaves as follows:
 \begin{itemize}
 \item it is not monotonic when $e^{-2\lambda_0}< -\frac{q_1}{q_2}< e^{2\lambda_0}$, which implies that $q_1$ and $q_2$ have opposite sign. 
There is an extremum at $\overline{x}$ 
 satisfying $e^{4\lambda_0(\overline x -1/2)}=-\frac{q_1}{q_2}$. The condensation current vanishes at this point.  
 \begin{itemize}
 \item  When $q_1<0$, the lattice current presents a maximum
 \begin{equation}
  \overline j_{lat}(\overline x)=-\frac{2\lambda_0}{\sinh(2\lambda_0)}\sqrt{-q_1q_2}.
 \end{equation} 
 \item When $q_1>0$, it presents a minimum (see figure \ref{fig:monotone})
 \begin{equation}
 \overline j_{lat}(\overline x)=\frac{2\lambda_0}{\sinh(2\lambda_0)}\sqrt{-q_1q_2}.
 \end{equation}
 \end{itemize}
 \item The lattice current is monotonic  when $-\frac{q_1}{q_2}<e^{-2\lambda_0} $ or $-\frac{q_1}{q_2} > e^{2\lambda_0}$, 
 see figures \ref{fig:concave} and \ref{fig:convexe}.
\end{itemize}

We now turn to the study of the behavior of the lattice current in the large system size limit.
\begin{proposition}
The thermodynamic limit of the variance of the lattice current, computed exactly in \eqref{eq:DiSSEP_variance_lattice_current} for any size, 
takes the form:
\begin{eqnarray}\label{eq:DiSSEP_variance_lattice_current_thermo}
E^{(2)}(x) &=& 2q_1\,q_2\,\lambda_0^{2}\left\{(2x-1)\,\frac {\sinh\big( 2\,\lambda_0\,(2\,x-1)\big) }{ \left( \sinh( 2\,\lambda_0)  \right) ^{3}}
- \frac { \cosh( 2\,\lambda_0) \,\cosh\big( 2\,\lambda_0\,(2x-1)\big)
 +1}{ \left( \sinh( 2\,{\lambda_0})  \right) ^{4}}\right\}
 \nonumber \\
 -& & \hspace{-5mm}
q_2^{2}\lambda_0\,{\frac{{e^{4\,\lambda_0\,x}}+
{e^{-4\,\lambda_0\, ( 1-x ) }} -{e^{4\,\lambda_0\, \left( 2\,x-1 \right) }}+3}{4 \left( \sinh( 2\,\lambda_0) 
 \right) ^{3}}}
 -q_1^{2}\lambda_0\,{\frac{e^{4\,\lambda_0\, ( 1-x ) }+e^{-4\,\lambda_0\,x}-e^{4\,\lambda_0\, ( 1-2\,x ) } +3
 }{4 \left( \sinh( 2\,\lambda_0)  \right) ^{3}}}
 \nonumber \\
+& & \hspace{-5mm} \frac{\lambda_0\,\cosh
 \left( 2\,\lambda_0\,x \right) \cosh \big( 2\,\lambda_0\,( 1-x )  \big) }{\sinh( 2\,\lambda_0) }.
\end{eqnarray}
\end{proposition}
As all physical quantities of the model, the variance is invariant under the transformation $q_1\,\leftrightarrow\,q_2$ and $x\to 1-x$, 
which is the left-right symmetry. The particle-hole symmetry amounts to change $q_1\to-q_1$ and $q_2\to-q_2$: it leaves $E^{(2)}$ invariant, 
transforms $\overline\rho(x)$ into $1-\overline\rho(x)$ and changes the sign of the currents. 
The symmetry $\lambda\to-\lambda$ reads $\lambda_0\to-\lambda_0$ and $q_1\leftrightarrow(-q_2)$ and leaves all quantities invariant.

\begin{remark}
By taking the limit $\lambda_0 \rightarrow 0$ in the previous quantities, i.e in the limit where the evaporation/condensation is negligible, 
we recover the well-known SSEP expressions \cite{DerridaDR04}:
\begin{eqnarray*}
&&\lim\limits_{\lambda_0 \rightarrow 0} \overline\rho(x)= \rho_l(1-x)+\rho_r x,\qquad
\lim\limits_{\lambda_0 \rightarrow 0} \overline\rho^c(x,y) =  -x(1-y)(\rho_l-\rho_r)^2,\\
&&\lim\limits_{\lambda_0 \rightarrow 0} \overline j_{lat}(x) = \rho_l-\rho_r,\qquad\qquad\quad\
\lim\limits_{\lambda_0 \rightarrow 0} E^{(2)}(x) = \rho_l+\rho_r -\frac23(\rho_l^2+\rho_l\rho_r+\rho_r^2).
\end{eqnarray*}
\end{remark}

We are finally interested in the study of the dynamical properties of the model: using the Bethe equations 
\eqref{eq:DiSSEP_bethe_equations1} and \eqref{eq:DiSSEP_bethe_equations2}, 
we study the approach to the stationary state at large times for a large system. We must compute the eigenvalue, denoted by $G$, 
for the first excited state (\textit{i.e.} the one with the 
greatest non-vanishing eigenvalue). 

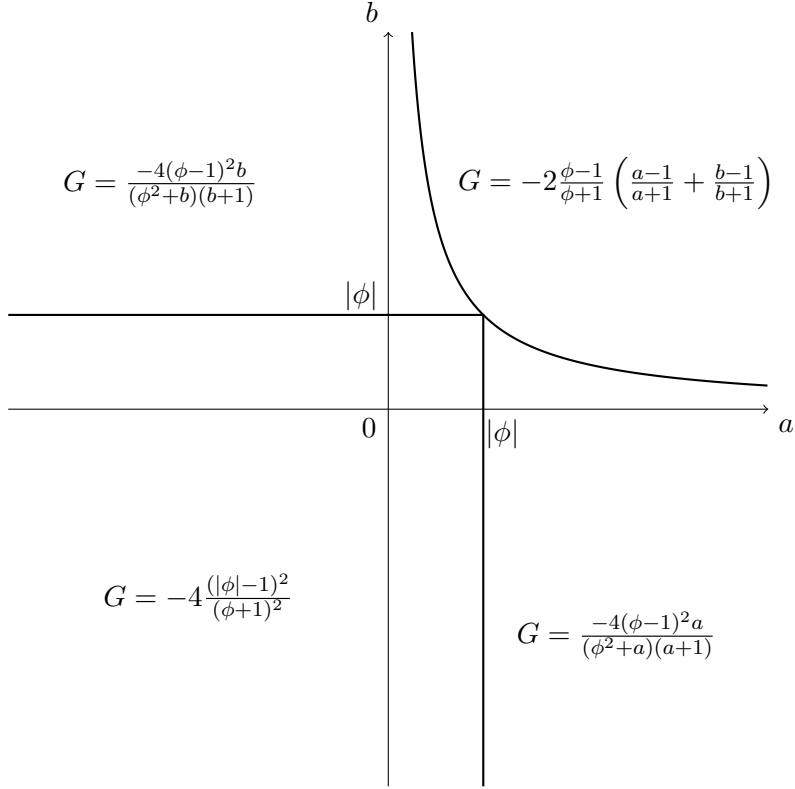
\begin{figure}[htb]
\begin{center}
 \begin{tikzpicture}[scale=5]
\draw [->] (-1,0) -- (1,0)  ;
\draw [->] (0,-1) -- (0,1)  ;
\node [below right] at (1,0) {$a$};
\node [above left] at (0,1) {$b$};
 \node at (-0.05,-0.05) {0};
\draw[thick, domain=0.0625:1,samples=300] plot (\x, { 0.25^2/\x   });
\draw[thick] (-1,0.25) -- (0.25,0.25) -- (0.25,-1) ;
\node at (0.6,-0.6) {$G=\frac{-4(\phi-1)^2a}{(\phi^2+a)(a+1)}$};
\node at (0.6,0.6) {$G=-2\frac{\phi-1}{\phi+1}\left(\frac{a-1}{a+1}+\frac{b-1}{b+1}\right)$};
\node at (-0.6,0.6) {$G=\frac{-4(\phi-1)^2b}{(\phi^2+b)(b+1)}$};
\node at (-0.5,-0.5) {$G=-4\frac{(|\phi|-1)^2}{(\phi+1)^2}$};
\draw[dotted] (0.25,0) --(0.25,0.25) --(0,0.25)  ;
\node [below] at (0.3,0) {$|\phi|$};
\node [left] at (0,0.3) {$|\phi|$};
 \end{tikzpicture}
 \end{center}
\caption{Value of the gap G depending on the parameters $a$, $b$ and $\phi$. The equation of the curve is $b=\phi^2/a$.
(This particular figure is drawn for $\phi=1/4$ ($\lambda=0.6$) even though similar behavior is valid for any $\phi$).}
 \label{fig:g3}
\end{figure}

\begin{figure}[htb]
\begin{center}
\begin{minipage}{0.6\textwidth}
 \includegraphics[width=\textwidth,height=0.4\textheight]{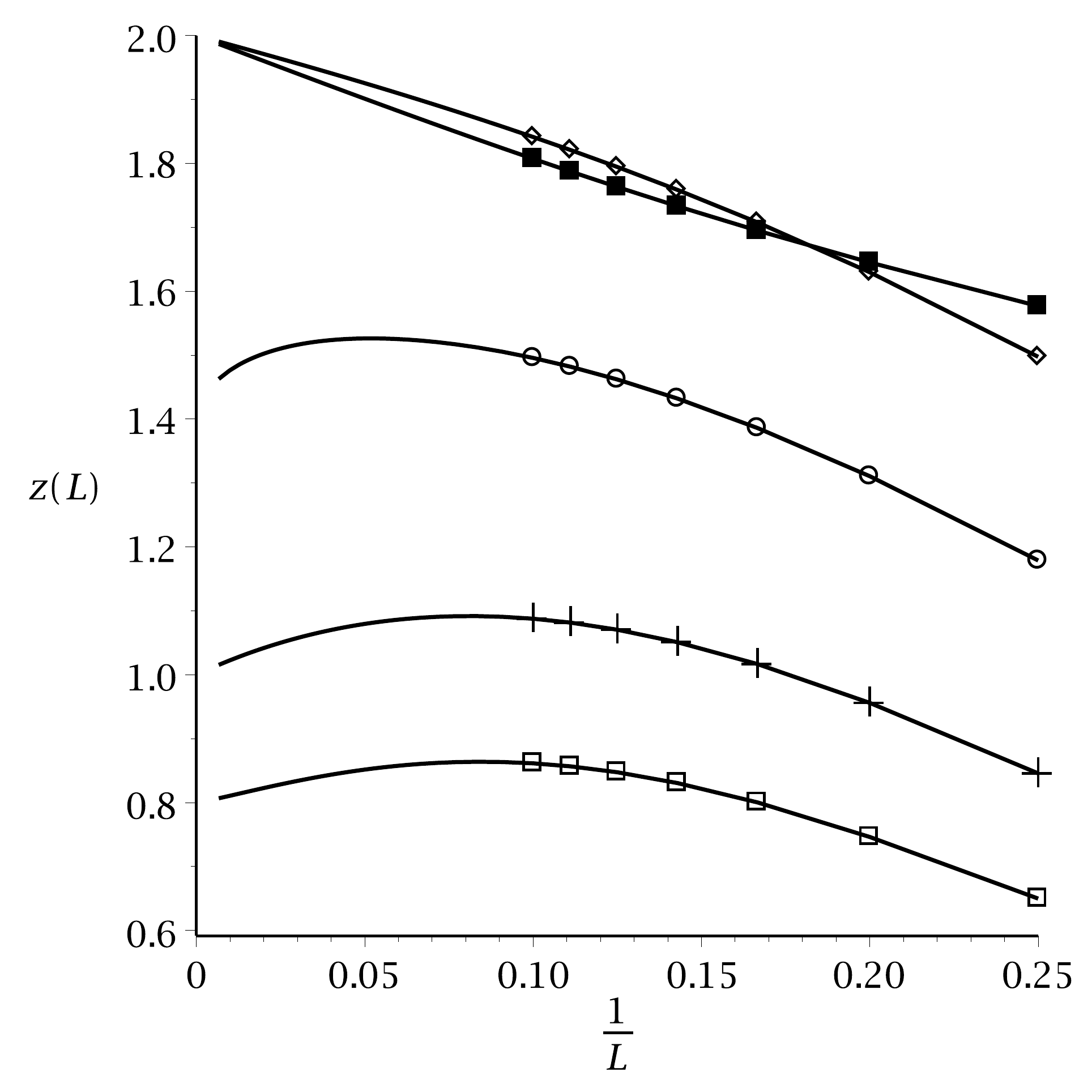}
\end{minipage}
\begin{minipage}{0.2\textwidth}
 \begin{tikzpicture}[scale=2]
\node at (.2,.7) {$\rule{2mm}{2mm}: \ \mu=2.$};
\node at (.2,.2) {$\Diamond :\ \mu=1.$};
\node at (.2,-.5) {$\circ :\ \mu=0.7$};
\node at (.2,-1.2) {$+: \ \mu=0.5$};
\node at (.2,-1.7) {$\Box: \ \mu=0.4$};
 \end{tikzpicture}
\end{minipage}
\end{center}
\caption{Behavior of the gap in the thermodynamic limit when $\lambda$ behaves as $\frac{1}{L^{\mu}}$: plot 
of $z(L)=\frac{\ln(G_L)-\ln(G_{L-1})}{\ln(L-1)-\ln({L})}$ as a function of $\frac1L$.
The lines represent the values obtained from Bethe ansatz (for $4\leq L\leq  150$), while the dots correspond to direct diagonalisation of $H$
(for $4\leq L\leq10$).}
 \label{fig:mu}
\end{figure}
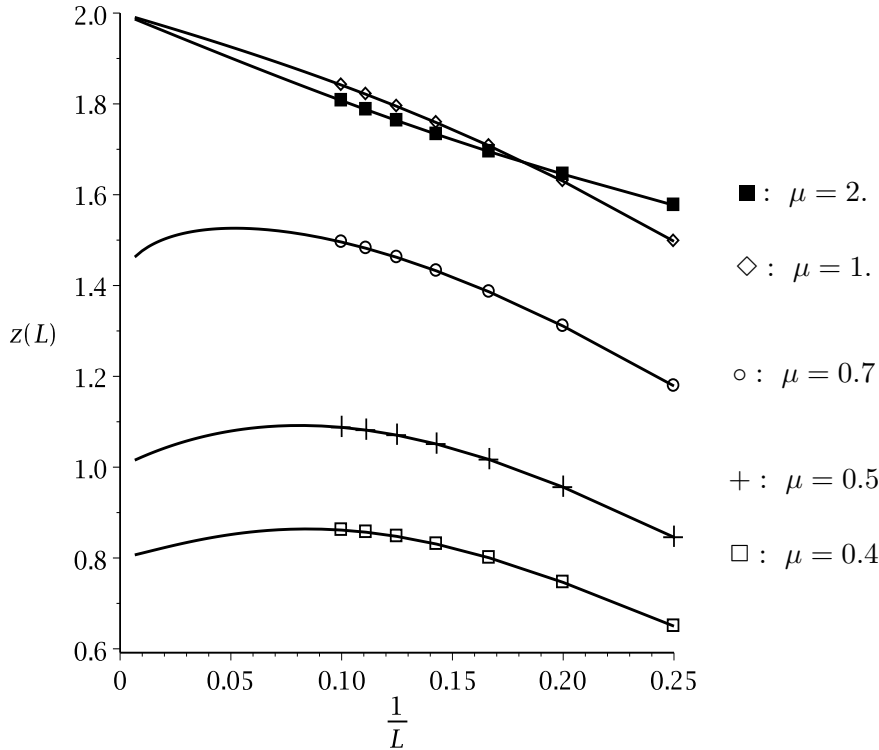

We start by presenting the main results for the gap then we give the sketch of the numerical evidences for them.
\begin{itemize}
\item In the case when evaporation rate $\lambda$ is independent of the size of the system $L$, there is a non-vanishing gap. 
The values of this gap depends on
the boundaries parameters and on $\lambda$. We present these different values of the gaps on Figure \ref{fig:g3}. 
They are consistent with the analytical result obtained for $\lambda=1$, see \eqref{eq:DiSSEP_gap_lambda1}.
\item
If the rate $\lambda$ behaves as $\frac{1}{L^\mu}$ for large system, the model is gapless and we get
\begin{equation}\label{eq:DiSSEP_gap}
 G\sim \frac{1}{L^{2\mu}} \mbox{ for } 0<\mu<1 \mbox{ and }  G\sim \frac{1}{L^{2}} \mbox{ for } \mu\geq 1\;.
\end{equation}
We show in figure \ref{fig:mu} numerical evidence for such a behavior. We plot 
\begin{equation}
z(L)=\frac{\ln(G_L)-\ln(G_{L-1})}{\ln(L-1)-\ln({L})} 
\end{equation}
as a function of $\frac1L$: as $\frac1L$ tends to 0, it tends to $2\mu$ (resp. 2) for $\mu<1$ (resp. $\mu\geq1$).
The $\frac{1}{L^{2}}$ behavior of the gap for $\mu> 1$ is expected since 
the system becomes in this case a diffusive model in the thermodynamic limit as discussed around equation \eqref{eq:DiSSEP_time_density_thermo2}.
\end{itemize}

To prove these result, we must study in detail the Bethe equations \eqref{eq:DiSSEP_bethe_equations1} and \eqref{eq:DiSSEP_bethe_equations2}.
The comparison of the eigenvalues obtained by the exact diagonalisation of M or by the numerical resolutions of the Bethe equations 
for small system (up to $10$ sites),
 show that the gap is obtained for $N=1$ in \eqref{eq:DiSSEP_eigenvalue2} and \eqref{eq:DiSSEP_bethe_equations2} or is equal to 
 $G=-\alpha-\beta-\gamma-\delta=-2\frac{\phi-1}{\phi+1}\left(\frac{a-1}{a+1}+\frac{b-1}{b+1}\right)$ 
(which corresponds to $N=0$ in \eqref{eq:DiSSEP_eigenvalue1} and \eqref{eq:DiSSEP_bethe_equations1}).
We assume that this behavior holds for any $L$ then we must solve only \eqref{eq:DiSSEP_bethe_equations1} for $N=1$.
This Bethe equation can be written as the vanishing of a polynomial of degree $2L+2$ w.r.t. $v_1$.
This polynomial has two obvious roots $\phi$ and $-\phi$ which are not physical since they corresponds to a vanishing ``eigenvector''. 
The remaining factor
is a polynomial of degree $2L$ w.r.t. $v_1$ which can be transformed, thanks to \eqref{eq:DiSSEP_eigenvalue2} (and up to a normalization), 
to a polynomial of degree $L$ w.r.t. $E$.
Then, the Bethe equation \eqref{eq:DiSSEP_bethe_equations2} for $N=1$ becomes
\begin{eqnarray}\label{eq:DiSSEP_spectral_gap_polynomial}
 \sum_{p=0}^L \frac{(1+\phi)^{2p}\ E^p}{4^p}&&\sum_{q=0}^{L-p}\phi^{2q}\left[
 ab\ \begin{pmatrix}p+q\\q\end{pmatrix}\begin{pmatrix} L-q-2\\p-2\end{pmatrix}+
 (a+b)\ \begin{pmatrix}p+q-1\\q\end{pmatrix}\begin{pmatrix} L-q-1\\p-1\end{pmatrix}\right. \nonumber\\
 &&\left.\qquad\qquad +\begin{pmatrix}p+q-2\\q\end{pmatrix}\begin{pmatrix} L-q\\p\end{pmatrix}
 \right]=0
\end{eqnarray}
The L.H.S. of the previous equation is a factor of the characteristic polynomial of the Hamiltonian $H$ \eqref{eq:DiSSEP_H_XXZ} or of the 
Markov matrix $M$ \eqref{eq:DiSSEP_Markov_matrix_decomposition}.
It is possible now to find numerically the roots of the polynomial \eqref{eq:DiSSEP_spectral_gap_polynomial}
for large system (up to $150$ sites) and pick up the largest ones.
Performing this computation for different values of $\lambda$ and of the boundary parameters, we obtain the results for the gap summarized previously,
see figure \ref{fig:g3}. 

\subsubsection{Limit of observables in 2-TASEP}
 
 We are now interested in another example that we encountered in chapter \ref{chap:three}: the open two-species TASEP. We expressed 
 the stationary state in a matrix product form and computed exactly the mean particle currents and the particle densities for a finite 
 size lattice. The goal is now to study this model in the thermodynamic limit. The results presented here are mainly extracted from the 
 work \cite{CrampeEMRV16}.

 The stationary state of the exclusion process can exhibit different
 qualitative features and different analytical expressions 
 for macroscopic quantities in the infinite size limit, $L \to \infty$.
 The system  is said to exhibit various phases, that depend on the
 values of the boundary exchange rates.  These different phases can
 be discriminated by the values of the currents and by the shapes of
 the density profiles. More refined features, such as correlations
 length  or even dynamical behavior,  can  even lead us to define
 subphases (see \cite{BlytheE07} for details and references). We stressed in chapter \ref{chap:three} that the 
 value of the mean particle currents and the mean density profiles can be exactly obtained, from 
 the the study of single species open TASEP, through the identification procedure. The first step 
 to compute analytically the asymptotic behavior of the particle currents and densities in the two species TASEP
 is thus to determine the asymptotic behavior of these quantities in the single species TASEP. 
 
The phase diagram of the one-species TASEP has been well-known for a long time; first determined using a mean-field approximation
\cite{Krug91,DerridaDM92}, it was rigorously established  and precisely investigated after the finding of the exact solution 
\cite{DerridaEHP93,SchutzD93,BlytheE07}.  
The phase diagram is determined by the behavior of the stationary current $\langle j\rangle$ 
and bulk density of particles in the limit $L\to\infty$ \cite{BlytheE07}. 
The different phases are summarized in table \eqref{eq:2TASEP_phase_diag_1-TASEP}.

\begin{eqnarray} \label{eq:2TASEP_phase_diag_1-TASEP}
\begin{array}{c|c|c|c} 
\mbox{Region}& \mbox{Phase} & \mbox{Current }  \overline j & \mbox{Bulk density}  \\ \hline
\rule{0pt}{3.54ex}\alpha<\beta,\,\alpha<\frac{1}{2} & \mbox{Low-density (LD)} & \alpha(1-\alpha) & \alpha \\[1ex]
\beta<\alpha,\,\beta<\frac{1}{2} & \mbox{High-density (HD)} & \beta(1-\beta) & 1-\beta \\[1ex]
\alpha>\frac{1}{2},\,\beta>\frac{1}{2} & \mbox{Maximal current (MC)} & \frac{1}{4} & \frac{1}{2}
 \end{array}
\vspace*{1ex}
\end{eqnarray}

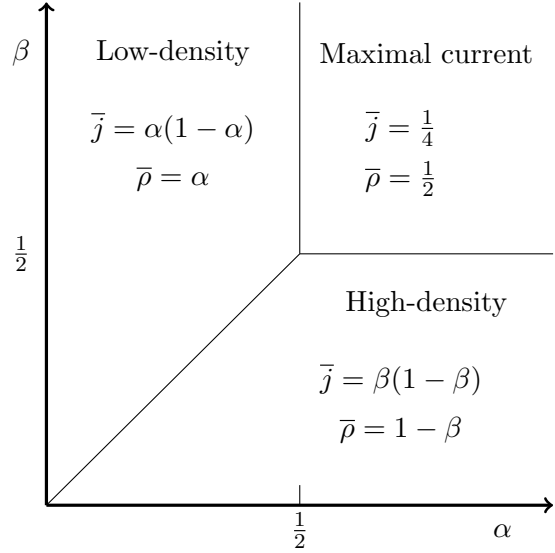
\begin{figure}[htb]
\begin{center}
 \begin{tikzpicture}[scale=2/3]
\draw[->,very thick] (0,0)--(10,0);
\draw[->,very thick] (0,0)--(0,10);
\draw (5,5)--(5,10);
\draw (5,5)--(10,5);
\draw (0,0)--(5,5);
\draw (5,0)--(5,0.4);
\node at (9,-0.5) [] {$\alpha$} ;
\node at (-0.5,9) [] {$\beta$} ;
\node at (5,-0.5) [] {$\frac{1}{2}$} ;
\node at (-0.5,5) [] {$\frac{1}{2}$} ;
\node at (2.5,9) [] {Low-density};
\node at (7.5,9) [] {Maximal current};
\node at (7.5,4) [] {High-density};
\node at (2.5,7.5) [] {$\overline{j}=\alpha(1-\alpha)$};
\node at (2.5,6.5) [] {$\overline\rho=\alpha$};
\node at (7,7.5) [] {$\overline{j}=\frac14$};
\node at (7,6.5) [] {$\overline\rho=\frac{1}{2}$};
\node at (7,2.5) [] {$\overline{j}=\beta(1-\beta)$};
\node at (7,1.5) [] {$\overline\rho=1-\beta$};
 \end{tikzpicture}
 \end{center}
 \caption{Phase diagram of the single species open TASEP.}
 \label{fig:TASEP_phase_diagram}
\end{figure}

We recall briefly the main steps of the asymptotic study of the particle current and density. The goal is to take 
the large size limit $L \to \infty$ in the exact expression \eqref{eq:TASEP_current} and \eqref{eq:TASEP_density} derived for a finite size lattice.
These quantities are expressed in terms of the normalization $Z_n$ \eqref{eq:TASEP_normalization}. The key point is thus to evaluate 
the asymptotic behavior of this normalization. Following the lines of \cite{DerridaEHP93} we first define
\begin{equation} \label{eq:TASEP_Rn}
 R_n(x)=\sum_{p=2}^{n+1} \frac{(p-1)(2n-p)!}{n!(n+1-p)!} x^p,
\end{equation}
which is of particular interest because
\begin{equation}
 Z_n=\llangle W|C^n|V\rrangle= \frac{R_n(1/\beta)-R_n(1/\alpha)}{1/\beta-1/\alpha}\llangle W|V\rrangle.
\end{equation}
An asymptotic expression of $R_n(1/\beta)$, respectively of $R_n(1/\alpha)$, can be obtained by determining the value of $p$
which gives the dominant contribution to the sum \eqref{eq:TASEP_Rn}:
\begin{equation} \label{eq:TASEP_Rn_asymptotic}
 R_n\left(\frac{1}{\beta}\right) \simeq \begin{cases} \vspace{2mm}
                                         \frac{1}{\sqrt{\pi}(2\beta-1)^2}\frac{4^n}{n^{3/2}}, \quad \mbox{for } \beta >\frac{1}{2} \\ \vspace{2mm}
                                         \frac{2}{\sqrt{\pi}}\frac{4^n}{n^{1/2}}, \quad \mbox{for } \beta=\frac{1}{2} \\ 
                                         (1-2\beta)\frac{1}{\beta^{n+1}(1-\beta)^{n+1}}, \quad \mbox{for } \beta <\frac{1}{2}.
                                        \end{cases}
\end{equation}
This provides an asymptotic expression of the normalization $Z_n$ depending on the value of $\alpha$ and $\beta$. We present here the case 
where $\alpha \leq \beta$. By symmetry, the case $\beta< \alpha$ can then be obtained by interchanging $\alpha$ and $\beta$ in the expressions below.
We have
\begin{equation}
 Z_n \simeq \begin{cases} \vspace{2mm}
             \frac{\alpha\beta}{\sqrt{\pi}(\beta-\alpha)}\left(\frac{1}{(2\alpha-1)^2}-\frac{1}{(2\beta-1)^2}\right)\frac{4^n}{n^{3/2}}, 
             \quad \mbox{for } \frac{1}{2}<\alpha<\beta, \\ \vspace{2mm}
             \frac{\alpha^2}{\sqrt{\pi}(2\alpha-1)^3}\frac{4^n}{n^{3/2}}, 
             \quad \mbox{for } \frac{1}{2}<\alpha=\beta, \\ \vspace{2mm}
             \frac{2\beta}{\sqrt{\pi}(2\beta-1)}\frac{4^n}{n^{1/2}},
             \quad \mbox{for } \alpha=\frac{1}{2}<\beta, \\ \vspace{2mm}
             4^n,
             \quad \mbox{for } \alpha=\beta=\frac{1}{2}, \\ \vspace{2mm}
             \frac{\beta(1-2\alpha)}{(\beta-\alpha)(1-\alpha)}\frac{1}{\alpha^n(1-\alpha^n)}, 
             \quad \mbox{for } \alpha<\frac{1}{2} \mbox{ and }\alpha<\beta, \\ \vspace{2mm}
             \frac{(1-2\alpha)^2}{(1-\alpha)^2}\frac{n}{\alpha^n(1-\alpha^n)},
             \quad \mbox{for } \alpha=\beta<\frac{1}{2}.
            \end{cases}
\end{equation}
We are now in position to give the asymptotic behavior of the particle current.
\begin{proposition}
 In the limit $L \to \infty$, the expression of the mean particle current \eqref{eq:TASEP_current} tends to 
 \begin{equation} \label{eq:TASEP_current_thermo}
 \overline{j}:= \lim\limits_{L\rightarrow \infty} \langle j \rangle = \begin{cases} \vspace{2mm}
      \frac{1}{4}, \quad \mbox{for } \alpha \geq \frac{1}{2}, \mbox{ and } \beta \geq \frac{1}{2}, \\ \vspace{2mm}
      \alpha(1-\alpha), \quad \mbox{for } \alpha < \frac{1}{2}, \mbox{ and } \beta>\alpha, \\ \vspace{2mm}
      \beta(1-\beta), \quad \mbox{for } \beta < \frac{1}{2}, \mbox{ and } \alpha>\beta.
     \end{cases}
 \end{equation}
\end{proposition}
The previous proposition show that the system displays three different macroscopic behavior (called phases) depending on the value 
of the injection/extraction rates.
These phases are represented in figure \ref{fig:TASEP_phase_diagram}.

The region where $\overline{j}=1/4$ is called the maximal-current phase and the region where $\overline{j}=\alpha(1-\alpha)$, 
respectively $\overline{j}=\beta(1-\beta)$, is 
called the low-density phase, respectively high-density phase. This denomination will make sense below through the computation of the
asymptotic behavior of the particle density.
A careful study of the exact expression of the particle density for a finite size lattice \eqref{eq:TASEP_density} combined with the asymptotic
behavior of the quantity $R_n(1/\beta)$ \eqref{eq:TASEP_Rn_asymptotic} yields \cite{DerridaEHP93} the shape of the particle density
near the right boundary
\begin{equation}
\lim\limits_{L\rightarrow \infty} \langle \tau_{L-j} \rangle = \begin{cases} \vspace{2mm}
                               \frac{1}{2}-\frac{1}{2\sqrt{\pi j}}+\cO\left(\frac1{j^{3/2}}\right), 
                               \quad \mbox{for } \alpha,\beta > \frac{1}{2} \quad \mbox{(MC)} \\ \vspace{2mm}
                               \alpha + \cO\left(\exp\left(-\frac{j}{\xi}\right)\right)
                                \quad \mbox{for } \alpha<\beta<\frac{1}{2} \quad \mbox{(LD)} \\ \vspace{2mm}
                               \alpha + \cO\left(\frac1{j^{3/2}}\,\exp\left(-\frac{j}{\xi}\right)\right)
                                \quad \mbox{for } \alpha<\frac{1}{2}<\beta \quad \mbox{(LD)} \\ \vspace{2mm}
                               1-\beta
                               \quad \mbox{for } \beta<\frac{1}{2}, \ \beta<\alpha \quad \mbox{(HD)}
                              \end{cases}
\end{equation}
Note that the shape of the particle density near the left boundary can be obtained from the previous results using the symmetry
$\langle \tau_{j} \rangle=1-\left.\langle \tau_{L+1-j} \rangle \right|_{\alpha \leftrightarrow \beta}$.

We stress also that on the particular line defined by $\alpha=\beta$ and $\alpha,\beta<1/2$, called the shock line, 
the shape of the particle density is 
different from the one of the low and high density phases, it is a linear interpolation between the reservoir densities $\alpha$ and 
$1-\beta=1-\alpha$. More precisely we have
\begin{equation}
 \lim\limits_{L\rightarrow \infty} \langle \tau_{\lfloor Lx \rfloor} \rangle = \alpha + (1-2\alpha)x.
\end{equation}

The  phase diagrams of the  $(M_1)$ and $(M_2)$ models can be
determined rigorously  without having to compute exactly the steady-state
probabilities.  Indeed,  the various phases  of these two
species models can extracted from the knowledge of the one-species
TASEP phase diagram, by using the two possible identifications described in \eqref{2TASEP_identification_bis} and \eqref{2TASEP_identification}
and called identification 1. and 2. respectively.

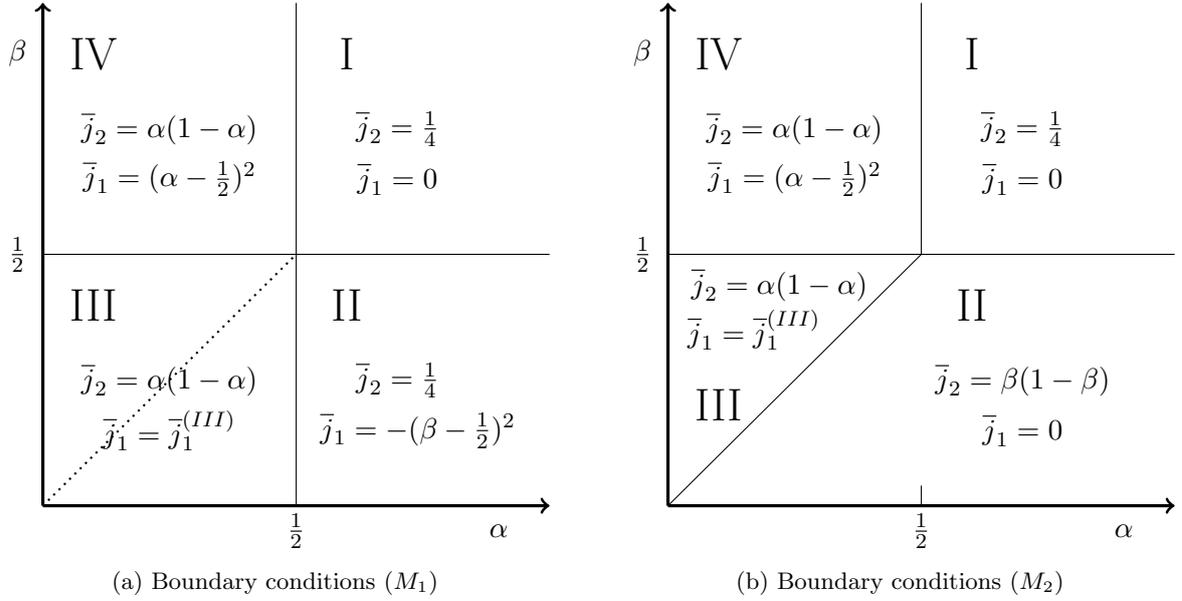
\begin{figure}[ht]
\begin{center}
\subfloat[Boundary conditions  $(M_1)$]{
 \begin{tikzpicture}[scale=2/3]
\draw[->,very thick] (0,0)--(10,0);
\draw[->,very thick] (0,0)--(0,10);
\draw (5,0)--(5,10);
\draw (0,5)--(10,5);
\draw[ dotted, thick] (0,0)--(5,5);
\node at (9,-0.5) [] {$\alpha$} ;
\node at (-0.5,9) [] {$\beta$} ;
\node at (5,-0.5) [] {$\frac{1}{2}$} ;
\node at (-0.5,5) [] {$\frac{1}{2}$} ;
\node at (1,9) [] {\LARGE{IV}};
\node at (6,9) [] {\LARGE{I}};
\node at (1,4) [] {\LARGE{III}};
\node at (6,4) [] {\LARGE{II}};
\node at (2.5,7.5) [] {$\overline j_2=\alpha(1-\alpha)$};
\node at (2.5,6.5) [] {$\overline j_1=(\alpha-\frac12)^2$};
\node at (7,7.5) [] {$\overline j_2=\frac14$};
\node at (7,6.5) [] {$\overline j_1=0$};
\node at (2.5,2.5) [] {$\overline j_2=\alpha(1-\alpha)$};
\node at (2.5,1.5) [] {$\overline j_1=\overline j_1^{(III)}$};
\node at (7,2.5) [] {$\overline j_2=\frac14$};
\node at (7.4,1.5) [] {$\overline j_1=-(\beta-\frac12)^2$};
 \end{tikzpicture}
}\hfill
\subfloat[Boundary conditions $(M_2)$]{
 \begin{tikzpicture}[scale=2/3]
\draw[->,very thick] (0,0)--(10,0);
\draw[->,very thick] (0,0)--(0,10);
\draw (5,5)--(5,10);
\draw (0,5)--(10,5);
\draw (0,0)--(5,5);
\draw (5,0)--(5,0.4);
\node at (9,-0.5) [] {$\alpha$} ;
\node at (-0.5,9) [] {$\beta$} ;
\node at (5,-0.5) [] {$\frac{1}{2}$} ;
\node at (-0.5,5) [] {$\frac{1}{2}$} ;
\node at (1,9) [] {\LARGE{IV}};
\node at (6,9) [] {\LARGE{I}};
\node at (1,2) [] {\LARGE{III}};
\node at (6,4) [] {\LARGE{II}};
\node at (2.5,7.5) [] {$\overline j_2=\alpha(1-\alpha)$};
\node at (2.5,6.5) [] {$\overline j_1=(\alpha-\frac12)^2$};
\node at (7,7.5) [] {$\overline j_2=\frac14$};
\node at (7,6.5) [] {$\overline j_1=0$};
\node at (2.2,4.4) [] {$\overline j_2=\alpha(1-\alpha)$};
\node at (1.7,3.5) [] {$\overline j_1=\overline j_1^{(III)}$};
\node at (7,2.5) [] {$\overline j_2=\beta(1-\beta)$};
\node at (7,1.5) [] {$\overline j_1=0$};
 \end{tikzpicture}
}
\caption{Phase diagrams of the 2-TASEP with open boundaries for the boundary conditions
 $(M_1)$  and  $(M_2)$ (we have  used the notation $\overline j_1^{(III)}=(\beta-\alpha)(1-\alpha-\beta)$).\label{fig:2TASEP_phase}}
 \end{center}
\end{figure}

We recall that identification 1. allows us to compute the current $\langle j_2 \rangle$ and the
density $\langle\rho_2^{(i)}\rangle$, whereas identification 2. yields the current
$\langle j_1\rangle+\langle j_2\rangle=-\langle j_0\rangle$ and the density $\langle\rho_1^{(i)}\rangle+\langle\rho_2^{(i)}\rangle$. 
Gathering these results, we obtain the phase diagrams depicted in figure \ref{fig:2TASEP_phase}.

\paragraph*{Phase diagram of the  $(M_1)$ model}

 The phase diagram of the  $(M_1)$ model is displayed  in
 Figure~\ref{fig:2TASEP_phase}(a).  It comprises   four phases.   Using the
 identification procedure, we observe that  $\overline\rho_2$ behaves as the
 density for the one-species TASEP with boundary rates $(\alpha,1)$,
 while $\overline \rho_1+\overline \rho_2$ behaves as the density for the one-species
 TASEP with boundary rates $(1,\beta)$.  The values of the currents and particle densities
 in each phase (see fig~\ref{fig:2TASEP_phase}(a)) are readily  found by this
 identification. 
\begin{description}
\item[Phase I: ]  For $\alpha>\frac{1}{2}$ and $\beta>\frac{1}{2}$, 
  particles of species $2$ exhibit a maximal current,
  whereas the current of the particles of species $1$ vanishes. The bulk
 density of particles of species $2$ and of holes is equal to 1/2, while
 the number of particles of species $2$ in the bulk is vanishingly small.
The density profiles of particles of species $1$ and $2$ are characterised by
power law decays to the bulk values:
\begin{equation*}
\lim\limits_{L\rightarrow \infty} \langle \rho_2^{(L-j)} \rangle =\frac12-\frac{1}{2\sqrt{\pi j}} +\cO\left(\frac1{j^{3/2}}\right)
\mbox{ and } \lim\limits_{L\rightarrow \infty} \langle \rho_1^{(L-j)} \rangle =\cO\left(\frac1{j^{3/2}}\right).
\end{equation*}
The system is similar to the one-species TASEP in its maximal current phase.
\item[Phase II:]  For $\alpha>\frac{1}{2}$ and $\beta<\frac{1}{2}$, 
  none of the currents $\overline j_1,\overline j_2$ and $\overline j_0$ vanishes. This is a genuine
 2-TASEP phase with  boundaries  permeable to all the species. 
 The  two species  and the holes coexist in the bulk
 with non-zero bulk densities and density profiles characterised by power-law decays:
\begin{equation*}
\lim\limits_{L\rightarrow \infty} \langle \rho_2^{(L-j)} \rangle=\frac12-\frac{1}{2\sqrt{\pi j}} +\cO\left(\frac1{j^{3/2}}\right) \mbox{ and }
\lim\limits_{L\rightarrow \infty} \langle \rho_1^{(L-j)} \rangle=\frac12-\beta+\frac{1}{2\sqrt{\pi j}}+\cO\left(\frac1{j^{3/2}}\right).
\end{equation*}
 First-class  particles are in their maximal
 current phase.
 Boundary effects are long-range for species 1 and 2.
\item[Phase III:] For $\alpha<\frac{1}{2}$ and $\beta<\frac{1}{2}$, 
 we obtain a `massive' phase in which  boundary effects are localized: 
 after a finite correlation length,   the system reaches  its bulk behaviour,
 \begin{equation*}
\lim\limits_{L\rightarrow \infty} \langle \rho_2^{(L-j)} \rangle=\alpha+\cO\left(\frac1{j^{3/2}}\,\exp\left(-\frac{j}{\xi}\right)\right)  \mbox{ and } 
\lim\limits_{L\rightarrow \infty} \langle \rho_1^{(L-j)} \rangle=1-\alpha-\beta+\cO\left(\frac1{j^{3/2}}\,\exp\left(-\frac{j}{\xi}\right)\right).
\end{equation*}
 The current of second-class particles $\overline j_2$ vanishes along the line
 $\alpha = \beta <\frac{1}{2}$  and changes its sign across this line.
\item[Phase IV:] This phase, obtained for  $\alpha<\frac{1}{2}$
  and $\beta>\frac{1}{2}$,  is massive  for first-class  particles
 but  `massless' (exhibiting long-range correlations characterised by power laws) for  second-class particles and holes. Here again, the
  two species  and the holes  coexist in the bulk:
 \begin{equation*}
\lim\limits_{L\rightarrow \infty} \langle \rho_2^{(L-j)} \rangle=\alpha+\cO\left(\frac1{j^{3/2}}\,\exp\left(-\frac{j}{\xi}\right)\right)  \mbox{ and } 
\lim\limits_{L\rightarrow \infty} \langle \rho_1^{(L-j)} \rangle=\frac12-\alpha-\frac{1}{2\sqrt{\pi j}}+\cO\left(\frac1{j^{3/2}}\right).
\end{equation*}
Holes are in their  maximal current phase $\overline j_0 = -1/4$.

\end{description}

\paragraph*{Phase diagram of the  $(M_2)$ model}

The phase diagram of the  $(M_2)$ model also comprises four phases,
displayed in Figure~\ref{fig:2TASEP_phase}(b). The diagram is qualitatively
different from that of the $(M_1)$ model. Here, $\overline \rho_2$ behaves as
the density for the one-species TASEP with boundary rates
$(\alpha,\beta)$, while $\overline\rho_1+\overline\rho_2$ behaves as the density for the
one-species TASEP with boundary rates $(1,\beta)$.
\begin{description}
\item[Phase I:]  For $\alpha>\frac{1}{2}$ and $\beta>\frac{1}{2}$, 
 first-class particles exhibit a maximal current. This phase is
 similar  to Phase I of model $(M_1)$.
\item[Phase II:] This phase is obtained for  $\beta <\alpha <\frac{1}{2}$. 
Particles of species $2$ are in their high density phase. The density of
particles of species $1$ in the bulk vanishes.
\begin{equation*}
\lim\limits_{L\rightarrow \infty} \langle \rho_2^{(L-j)} \rangle=1-\beta  \mbox{ and } 
\lim\limits_{L\rightarrow \infty} \langle \rho_1^{(L-j)} \rangle=0 .
\end{equation*}
\item[Phase III:]  For  $\alpha < \beta <\frac{1}{2}$, the two species and the holes
 are simultaneously present with non-vanishing currents. 
 The current of particles of species one $\overline j_1$ is strictly positive.
 This phase is  massive for the two  classes of particles and the holes: 
\begin{equation*}
\lim\limits_{L\rightarrow \infty} \langle \rho_2^{(L-j)} \rangle=\alpha+\cO\left(\exp\left(-\frac{j}{\xi}\right)\right)  \mbox{ and } 
\lim\limits_{L\rightarrow \infty} \langle \rho_1^{(L-j)} \rangle=1-\alpha-\beta+\cO\left(\,\exp\left(-\frac{j}{\xi}\right)\right) .
\end{equation*}
\item[Shock Line:] This line corresponds to  $\alpha = \beta <\frac{1}{2}$.
The density profiles $\overline\rho_1$ and $\overline\rho_2$ display a linear behaviour
that reflect a coexistence between a low density and a high density
regions: 
 \begin{equation}
\lim\limits_{L\rightarrow \infty} \langle \rho_2^{(\lfloor Lx \rfloor)} \rangle= \alpha+x(1-2\alpha)  \mbox{ and } 
\lim\limits_{L\rightarrow \infty} \langle \rho_1^{(\lfloor Lx \rfloor)} \rangle=(1-2\alpha)(1-x).
\end{equation}
The density profile of particles of species $2$ takes the values
 $\alpha$ and  $1-\alpha$ with a discontinuous shock between the two
 regions. The particles of species $1$ have a plateau density of
 $1-2\alpha$ to the left of the shock  and zero  density  to the right shock. 
This means effectively
 that in the stationary state in the infinite system limit only the left reservoir is active
 as far as particles of species $1$  are concerned.
\item[Phase IV] This phase, obtained for  $\alpha<\frac{1}{2}$
  and $\beta>\frac{1}{2}$, is similar to Phase IV
 of the  $(M_1)$  model.
\end{description}

\subsubsection{Limit of observables in the multi-species SSEP} 

We now turn to the study of the multi-species SSEP in the thermodynamic limit. We compute in particular the large size limit $L\rightarrow \infty$ 
of the mean particle currents, of the mean particle densities and of the two-point functions.

\begin{proposition}
The thermodynamic limit of the mean particle current of species $\tau$ ($0\leq \tau \leq N$) is given by the exact expression
 \begin{equation}
  \overline{j}_{\tau}:= \lim\limits_{L\rightarrow \infty} L \times \langle j_{\tau} \rangle = \lambda_{\tau}.
 \end{equation}
\end{proposition}

\proof
This formula is derived by a straightforward computation, starting from the expression of the particle current on the finite size lattice
\eqref{eq:mSSEP_current}.
\finproof

Note that the formula above is valid for $\tau=0$ and gives the mean current of holes on the lattice.
We observed in chapter \ref{chap:three}, with the exact computations on the finite size lattice, that the exclusion constraint implied
$\langle j_{0} \rangle+\dots+\langle j_{N} \rangle=0$. This result is directly recovered, in the thermodynamic limit, from the formula above 
\begin{equation}
 \sum_{\tau=0}^N \overline{j}_{\tau} = \sum_{\tau=0}^N \lambda_{\tau} = 0,
\end{equation}
where the last equality is obtained using the definition \eqref{eq:mSSEP_lambda} and the constraints \eqref{eq:mSSEP_sum_densities}.

Note also that we multiplied the mean particle current $\langle j_{\tau} \rangle$ with a factor $L$ when performing the large size limit
to obtain a finite and non vanishing value. This is characteristic of diffusive system in contrast with bulk driven systems (that are called
ballistic) where the scaling is not needed, see for instance \eqref{eq:TASEP_current_thermo}.

\begin{proposition}
The thermodynamic limit of the mean particle density of species $\tau$ ($0\leq \tau \leq N$) at position $x\in [0,1]$ on the lattice is given by 
the exact expression
 \begin{equation} \label{eq:mSSEP_density_thermo}
 \overline{\rho}_{\tau}(x):= \lim\limits_{L\rightarrow \infty} \langle \rho_{\tau}^{(\lfloor Lx \rfloor)} \rangle
 = \alpha_{\tau}(1-x)+\beta_{\tau}x.
 \end{equation}
\end{proposition}

\proof
This is obtained through a direct computation, using the expression of the particle density for a finite size lattice \eqref{eq:mSSEP_density}.
\finproof

Note that the proposition above provides for $\tau=0$ the density of holes on the lattice. 
The exclusion constraint that we had on the finite size lattice $\langle \rho_{0}^{(i)} \rangle + \dots + \langle \rho_{N}^{(i)} \rangle = 1$ is
easily observed in the thermodynamic limit
\begin{eqnarray}
 \sum_{\tau=0}^N \overline{\rho}_{\tau}(x)=(1-x)\sum_{\tau=0}^N \alpha_{\tau} +x\sum_{\tau=0}^N \beta_{\tau} = (1-x)+x=1,
\end{eqnarray}
where we made use of the constraints \eqref{eq:mSSEP_sum_densities}.

We observe that the mean particle density of species $\tau$ is the linear interpolation between the density of species $\tau$ in the left reservoir 
$\alpha_{\tau}$ and the density of species $\tau$ in the right reservoir $\beta_{\tau}$. We have indeed 
$\overline{\rho}_{\tau}(0)=\alpha_{\tau}$ and $\overline{\rho}_{\tau}(1)=\beta_{\tau}$.

\begin{proposition}
 The large size limit of the connected two-point function between a particle of species $\tau$ and a particle of species $\tau'$ located
 respectively at $x$ and $y$ on the lattice (with $0 \leq x \leq y \leq 1$) is given by the exact expression
 \begin{equation}
  \overline{\rho}_{\tau,\tau'}(x,y):=
  \lim\limits_{L\rightarrow \infty} L \times \langle \rho_{\tau}^{(\lfloor Lx \rfloor)}\rho_{\tau'}^{(\lfloor Ly \rfloor)} \rangle_c
  =-\lambda_{\tau}\lambda_{\tau'}x(1-y).
 \end{equation}
\end{proposition}

\proof
This is again obtained by direct computation starting from the formula \eqref{eq:mSSEP_2pts_function} for the finite size lattice.
\finproof

Note that we multiplied the connected two-point function with a factor $L$ when performing the large size limit. It shows that the correlations in 
the system decrease algebraically as $1/L$.

\subsection{Large deviation functional of the density profile}

The previous subsection was devoted to the evaluation of the physical quantities 
(mean particle currents, mean particle densities, two-point function,...), that have been computed exactly for a finite size lattice,
in the thermodynamical limit. That was performed on a series of examples studied in details on a finite size lattice in chapter \ref{chap:three}.
We observed that in this large size limit, the expressions of the physical quantities became simpler. 
Now we would like to push further this study of the thermodynamic limit with a coarse-grained description of the physical models.
The idea is to define local average of the occupation variables $\rho_0^{(i)},\dots,\rho_N^{(i)}$ (we recall that these variables define 
uniquely the configuration of the system). These averages will be computed over several sites $i$ giving rise to a coarse-grained variable.
In the spirit of the law of large number (where the average of independent identically distributed random variables
converges to the expectation with probability $1$), 
we expect these coarse-grained variables to behave nicely in the large size limit, typically to display a deterministic behavior. 
It means intuitively that, in the thermodynamic limit, we are almost sure to observe a density profile that is equal to a typical density profile.
We will see that (on particular example) we can even get a refined result and evaluate the probability to deviate from this typical profile. 
This probability is, on the example treated below, exponentially small in the system size and will lead to a large deviation principle.
As explained in chapter \ref{chap:one}, the large deviation functional could be a generic and powerful formalism to describe the macroscopic behavior
of out-of-equilibrium statistical physics systems.

\subsubsection{Definition}

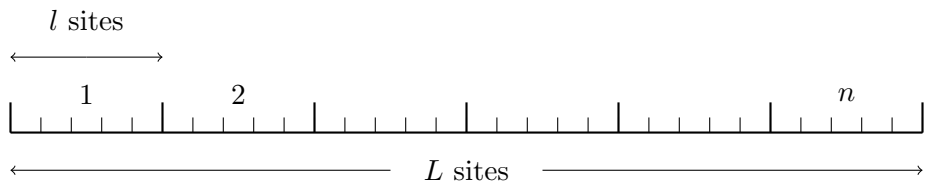
\begin{figure}[htb]
\begin{center}
 \begin{tikzpicture}[scale=1.0]
\draw[thick] (0,0) -- (12,0) ;
\foreach \i in {0,2,...,12}
{\draw[thick] (\i,0) -- (\i,0.4) ;}
\foreach \i in {0,0.4,...,12}
{\draw (\i,0) -- (\i,0.2) ;}
\draw[->] (5,-0.5)--(0,-0.5);
\draw[->] (7,-0.5)--(12,-0.5);
\node at (6,-0.5) [] {$L$ sites};
\draw[->] (1,1)--(0,1);
\draw[->] (1,1)--(2,1);
\node at (1,1.5) [] {$l$ sites};
\node at (1,0.5) [] {$1$};
\node at (3,0.5) [] {$2$};
\node at (11,0.5) [] {$n$};
 \end{tikzpicture}
 \end{center}
 \caption{The system of length $L$ is divided into $n$ boxes of length $l$.}
 \label{fig:subdivision}
\end{figure}

The first step toward a coarse-grained description of the stationary state of the model in the thermodynamic limit is to define precisely
coarse-grained, or average, variables.
In order to formalize the problem, we split the full system which contains $L=nl$ sites into $n$ subsystems (called ``boxes'' below)
containing $l$ sites each, see figure \ref{fig:subdivision}. This leads to the following definition.

\begin{definition}
For all particle species (and holes) $\tau \in \{0,\dots,N\}$ and all box number
$1\leq k \leq n$, we define the average variable
\begin{equation}
 \rho_{\tau}^{\{k\}}=\frac{1}{l}\sum_{i=kl+1}^{(k+1)l}\rho_{\tau}^{(i)},
\end{equation}
which corresponds to the average number of particles of species $\tau$ in the box $k$ for a given configuration. 
\end{definition}

We expect these variables to display some deterministic behavior in the limit $l \rightarrow \infty$ (that is when the number of sites 
involved in the average goes to infinity).

\begin{definition}
It will be useful to introduce the row vector
$\boldsymbol{\rho}^{\{k\}}$ encompassing the average variables of each species (and holes) in the box $k$:
\begin{equation}
 \boldsymbol{\rho}^{\{k\}}=(\rho_0^{\{k\}},\dots,\rho_N^{\{k\}}).
\end{equation}
\end{definition}

Note that the average variables fulfill the exclusion constraint $\rho_0^{\{k\}}+\dots+\rho_N^{\{k\}}=1$, for all $1\leq k \leq n$ 
(this is directly derived from the exclusion constraint $\rho_0^{(i)}+\dots+\rho_N^{(i)}=1$ which holds on every site $i$).

The second step is to determine precisely which observable of these coarse-grained variables is both physically relevant and simple enough to be 
exactly computed in the thermodynamic limit.
We are interested in evaluating the probability of observing in the stationary state a given density profile in the limit of large system size $L$.
A density profile is defined by a fixed number of particles of each species (and holes) $\tau=0,\dots,N$ in each boxes $k=1,\dots,n$.
We thus want to study the joint probability law of the average variables. This motivates the following definition.

\begin{definition}
We denote by 
\begin{equation}
P_L(\{\boldsymbol{\rho}^{\{1\}},\boldsymbol{\rho}^{\{2\}},\dots,\boldsymbol{\rho}^{\{n\}}\}\ |\ \boldsymbol{\alpha},\boldsymbol{\beta})
\end{equation}
the probability to find simultaneously $\rho_{\tau}^{\{k\}}\times l$ particles
of species $\tau$ in the box $k$ for $\tau=0,\dots,N$ and $k=1,\dots,n$. 
 We have introduced in this notation the two row vectors of size $N+1$, encompassing
the particle (and hole) densities at the two reservoirs
\begin{equation}
 \boldsymbol{\alpha} = (\alpha_0,\dots,\alpha_N) \quad \mbox{and} \quad \boldsymbol{\beta} = (\beta_0,\dots,\beta_N).
\end{equation}
\end{definition}

 For $l$ large (and thus $L$ large), that is when the size of the boxes used to define the coarse-grained variables goes to infinity,
 we expect the joint probability 
 $P_L(\{\boldsymbol{\rho}^{\{1\}},\boldsymbol{\rho}^{\{2\}},\dots,\boldsymbol{\rho}^{\{n\}}\}\ |\ \boldsymbol{\alpha},\boldsymbol{\beta}) $ to follow 
 a large deviation principle
\begin{equation}
 P_L(\{\boldsymbol{\rho}^{\{1\}},\boldsymbol{\rho}^{\{2\}},\dots,\boldsymbol{\rho}^{\{n\}}\}\ |\ \boldsymbol{\alpha},\boldsymbol{\beta})\sim
 \exp\left(-L\mathcal{F}_n(\{\boldsymbol{\rho}^{\{1\}},\boldsymbol{\rho}^{\{2\}},
 \dots,\boldsymbol{\rho}^{\{n\}}\}\ |\ \boldsymbol{\alpha},\boldsymbol{\beta})\right).
\end{equation}
In the limit where we have in addition the number of boxes $n$ going also to infinity, we can define a continuous coordinate $x$ such that 
$k=xL$ and a vector $\boldsymbol{\rho}(x)=\boldsymbol{\rho}^{\{k\}}$. We obtain in this case a large deviation functional 
$\mathcal{F}(\{ \boldsymbol{\rho}(x)\}\ |\ \boldsymbol{\alpha},\boldsymbol{\beta})$
\begin{equation}
 P_L(\{\boldsymbol{\rho}(x)\}\ |\ \boldsymbol{\alpha},\boldsymbol{\beta})\sim
 \exp\left(-L\mathcal{F}(\{\boldsymbol{\rho}(x)\}\ |\ \boldsymbol{\alpha},\boldsymbol{\beta})\right).
\end{equation}

We will now study this joint probability law and show these large deviation principles on the particular example of the multi-species SSEP 
that we encountered many times in this manuscript. The results presented below are mainly taken out of the work \cite{Vanicat17}.

\subsubsection{Multi-species SSEP case}

We now turn to the computation of the joint probability law 
$P_L(\{\boldsymbol{\rho}^{\{1\}},\boldsymbol{\rho}^{\{2\}},\dots,\boldsymbol{\rho}^{\{n\}}\}\ |\ \boldsymbol{\alpha},\boldsymbol{\beta})$
for the multi-species SSEP. This computation can be drastically simplified by using a powerful tool: the additivity principle.
This was first introduced in \cite{DerridaLS02} (see also \cite{Derrida07} for the connection with matrix ansatz) 
to compute the large deviation functional of the single species open SSEP and has then been successfully
applied to compute the one of the single species open ASEP \cite{DerridaLS02bis,DerridaLS03}. 
The general idea is to relate the probability of observing a given configuration of 
the lattice in the stationary state with the probabilities of the subsystems that are obtained by cutting the original system into two pieces.
The difficult part lies in the correct tuning of the injection/extraction rates that have to be introduced at the level of the cutting.
We stress that this approach is completely rigorous in the sense that the formula obtained to relate the full system to the two 
subsystems is exact. The interest of this additivity principle is to reduce, or factorize, the complexity of the computations on the full 
system into computations on the subsystems that we expect to be simpler. The result that we present below holds for a finite size lattice of any
length. Note that similar ideas of additivity have also been used directly in the thermodynamic limit to compute large deviation functionals 
\cite{BodineauD04}
(which by analogy with the free energy should behave as an extensive quantity). More details about that will be presented altogether with
the macroscopic fluctuation theory.

\paragraph*{Additivity principle from matrix ansatz.}

In order to write an additivity principle, we first need the following definition.

\begin{definition}
We introduce the vector 
\begin{equation}
\overline{\boldsymbol{\rho}}(u)=(1-u)\boldsymbol{\alpha}+u\boldsymbol{\beta}
\end{equation}
which achieves the linear interpolation between the particle densities at the left (respectively right) reservoir $\boldsymbol{\alpha}$ 
(respectively $\boldsymbol{\beta}$).
\end{definition}

This vector thus contains the mean particle densities at position $u\in [0,1]$ on the lattice 
in the stationary state $\overline{\rho}_{\tau}(u)$, that have been computed in \eqref{eq:mSSEP_density_thermo}.

\begin{definition}
We denote by $\mathcal{S}_L(\tau_1,\dots,\tau_L | \boldsymbol{\alpha},\boldsymbol{\beta},a,b)$ the probability of the configuration
$(\tau_1,\dots,\tau_L)$ in the stationary state for a system of size $L$ with boundary parameters
$\boldsymbol{\alpha}$, $\boldsymbol{\beta}$, $a$ and $b$.
\end{definition}

We recall that the definition of the boundary parameters and of the dynamics of the model can be found in subsection \ref{subsubsec:mSSEP_presentation}.
Up to now the length of the lattice and the boundary parameters were omitted in the notation $\mathcal{S}(\tau_1,\dots,\tau_L)$
because there was no ambiguity, but this precision will make sense when formulating the additivity principle.
As already mentioned, the idea of the additivity principle is to express the stationary weights of a system of size $L$ 
in terms of the stationary weights of the two subsystems of size $L'$ and $L''$ (with $L=L'+L''$) obtained by cutting the original system in two parts. 
The results presented in this subsection are heavily inspired by what was done in \cite{Derrida07,Derrida11} for the usual SSEP 
(with one species of particles plus holes).  

\begin{proposition}
For the present model, the additivity principle reads\footnote{The integration contour is chosen to contain the pole at $u=1$ but not the pole at $u=0$.}
\begin{eqnarray} \label{eq:mSSEP_additivity_principle}
 \mathcal{S}_L(\tau_1,\dots,\tau_L | \boldsymbol{\alpha},\boldsymbol{\beta},a,b) & = & -\frac{\Gamma(a+b+L')\Gamma(L''+1)}{\Gamma(a+b+L)} \
 \oint_{u=1} \frac{du}{2i\pi}\frac{1}{u^{a+b+L'}(1-u)^{L''+1}} \\
 & \times &  \mathcal{S}_{L'}(\tau_1,\dots,\tau_{L'} | \boldsymbol{\alpha},\overline{\boldsymbol{\rho}}(u),a,b) \ 
 \mathcal{S}_{L''}(\tau_{L'+1},\dots,\tau_{L} | \overline{\boldsymbol{\rho}}(u),\boldsymbol{\beta},1-b,b) \nonumber 
\end{eqnarray}
\end{proposition}

This additivity property can be rewritten using the matrix product formalism. Since the algebraic relations \eqref{eq:mSSEP_lie_algebra},
 \eqref{eq:mSSEP_rel_W} and \eqref{eq:mSSEP_rel_V} involving the operators $X_{\tau}$ and 
the boundary vectors $\llangle W|$ and $|V\rrangle$ depend explicitly on the boundary parameters,
we need to introduce some more notations.
 We denote by $\widetilde X_{\tau}(u)$, $\llangle \widetilde W(u)|$ and $|\widetilde V(u)\rrangle$ the operators and boundary vectors 
 associated to the system with parameters 
 $\boldsymbol{\alpha}$ and $a$ for the left reservoir and $\overline{\boldsymbol{\rho}}(u)$ and $b$ for the right reservoir.
 Namely they satisfy \eqref{eq:mSSEP_lie_algebra}, \eqref{eq:mSSEP_rel_W} and \eqref{eq:mSSEP_rel_V} where $\beta_{\tau}$ has been replaced by
 $(1-u)\alpha_{\tau}+u\beta_{\tau}$ for all $\tau$:
 \begin{equation} 
 [\widetilde X_{\tau}(u),\widetilde X_{\tau'}(u)]
 =\widetilde \lambda_{\tau}(u) \widetilde X_{\tau'}(u)-\widetilde \lambda_{\tau'}(u)\widetilde X_{\tau}(u)
 =u\big(\lambda_{\tau} \widetilde X_{\tau'}(u)- \lambda_{\tau'}\widetilde X_{\tau}(u)\big), 
\end{equation}
where
\begin{equation}
 \widetilde \lambda_{\tau}(u)=\alpha_s-[(1-u)\alpha_{\tau}+u\beta_{\tau}]=u\lambda_{\tau}, 
\end{equation}
and for the boundaries
\begin{equation}
 \llangle \widetilde W(u)| \big( \alpha_{\tau} \widetilde C(u)-\widetilde X_{\tau}(u) \big) = au\lambda_{\tau} \llangle \widetilde W(u)|,
\end{equation}
and
\begin{equation}
 \big( [(1-u)\alpha_{\tau}+u\beta_{\tau}] \widetilde C(u)-\widetilde X_{\tau}(u) \big) | \widetilde V(u)\rrangle
 = -bu\lambda_{\tau} | \widetilde V(u)\rrangle, 
\end{equation}
where
\begin{equation}
 \widetilde C(u)=\widetilde X_0(u)+\dots+\widetilde X_N(u).
\end{equation}

 In the same way we denote by $\widehat X_{\tau}(u)$, $\llangle \widehat W(u)|$ and $|\widehat V(u)\rrangle$ the operators and boundary vectors 
 associated to the system with parameters 
 $\overline{\boldsymbol{\rho}}(u)$ and $1-b$ for the left reservoir and $\boldsymbol{\beta}$ and $b$ for the right reservoir.
 Namely they satisfy \eqref{eq:mSSEP_lie_algebra}, \eqref{eq:mSSEP_rel_W} and \eqref{eq:mSSEP_rel_V} where $\alpha_{\tau}$ has been replaced by
 $(1-u)\alpha_{\tau}+u\beta_{\tau}$ for all $\tau$ and $a$ has been replaced by $1-b$:
  \begin{equation} 
 [\widehat X_{\tau}(u),\widehat X_{\tau'}(u)]
 =\widehat \lambda_{\tau}(u) \widehat X_{\tau'}(u)-\widehat \lambda_{\tau'}(u)\widehat X_{\tau}(u)
 =(1-u)\big(\lambda_{\tau} \widehat X_{\tau'}(u)- \lambda_{\tau'}\widehat X_{\tau}(u)\big), 
\end{equation}
where
\begin{equation}
 \widehat \lambda_{\tau}(u)=[(1-u)\alpha_{\tau}+u\beta_{\tau}]-\beta_{\tau}=(1-u)\lambda_{\tau}, 
\end{equation}
and for the boundaries
\begin{equation}
 \llangle \widehat W(u)| \big( [(1-u)\alpha_{\tau}+u\beta_{\tau}] \widehat C(u)-\widehat X_{\tau}(u) \big) = (1-b)(1-u)\lambda_{\tau} \llangle \widehat W(u)|,
\end{equation}
and
\begin{equation}
 \big( \beta_{\tau} \widehat C(u)-\widehat X_{\tau}(u) \big) | \widehat V(u)\rrangle
 = -b(1-u)\lambda_{\tau} | \widehat V(u)\rrangle, 
\end{equation}
where
\begin{equation}
 \widehat C(u)=\widehat X_0(u)+\dots+\widehat X_N(u).
\end{equation}
 
 \begin{proposition}
 We have the formula:
\begin{eqnarray} \label{eq:mSSEP_additivity_principle_MA}
 \llangle W|X_{\tau_1}\dots X_{\tau_L}|V\rrangle & = &  -\oint_{u=1} \frac{du}{2i\pi}\frac{1}{u^{a+b+L'}(1-u)^{L''+1}} \\
 & \times &  \llangle \widetilde W(u)|\widetilde X_{\tau_1}(u)\dots \widetilde X_{\tau_{L'}}(u)|\widetilde V(u)\rrangle \ 
 \llangle \widehat W(u)|\widehat X_{\tau_{L'+1}}(u)\dots \widehat X_{\tau_{L}}(u)|\widehat V(u)\rrangle. \nonumber 
\end{eqnarray}
\end{proposition}

\proof
For $i=1,\dots,L'$ we perform the change of variables
\begin{equation}
X_{\tau_i}=\alpha_{\tau_i}C-L_{\tau_i} \quad \mbox{and} \quad 
\widetilde X_{\tau_i}(u)=\alpha_{\tau_i}\widetilde C(u)-\widetilde L_{\tau_i}(u).
\end{equation}
The new operators $L_{\tau}$ and $\widetilde L_{\tau}(u)$ behave conveniently on the left boundary
\begin{equation}
 \llangle W|L_{\tau}=a\lambda_{\tau} \llangle W| \quad \mbox{and} \quad
\llangle \widetilde W(u)| \widetilde L_{\tau}(u)=ua\lambda_{\tau} \llangle \widetilde W(u)|.
\end{equation}
When we expand the product $X_{\tau_1}\dots X_{\tau_{L'}}$ (respectively the product $\widetilde X_{\tau_1}(u)\dots \widetilde X_{\tau_{L'}}(u)$),
we can push the $L_{\tau}$ (respectively the $\widetilde L_{\tau}(u)$) to the left through the $C$'s (respectively the $\widetilde C(u)$'s) using  
the relation $[L_{\tau},C]=-\lambda_{\tau} C$ (respectively the relation $[\widetilde L_{\tau}(u),\widetilde C(u)]=-u\lambda_{\tau} \widetilde C(u)$).
At the end the expansion of $X_{\tau_1}\dots X_{\tau_{L'}}$ involve monomials of the form 
$\lambda_{s_1}\dots \lambda_{s_k}L_{s_{k+1}}\dots L_{s_{n'}}C^{L'-n'}$. The expansion of the product 
$\widetilde X_{\tau_1}(u)\dots \widetilde X_{\tau_{L'}}(u)$ is exactly the same but with the previous monomial replaced by 
$u^k\lambda_{s_1}\dots \lambda_{s_k}\widetilde L_{s_{k+1}}(u)\dots \widetilde L_{s_{n'}}(u)\widetilde C(u)^{L'-n'}$.

In the same way for $i=L'+1,\dots,L'+L''$ we perform the change of variables 
\begin{equation}
X_{\tau_i}=\beta_{\tau_i}C-R_{\tau_i} \quad \mbox{and} \quad  
\widehat X_{\tau_i}(u)=\beta_{\tau_i}\widehat C(u)-\widehat R_{\tau_i}(u).
\end{equation}
The new operators $R_{\tau}$ and $\widehat R_{\tau}(u)$ behave conveniently on the right boundary
\begin{equation} 
R_{\tau}|V\rrangle=-b\lambda_{\tau} |V\rrangle \quad \mbox{and} \quad
\widehat R_{\tau}(u)|\widehat V(u)\rrangle=-(1-u)b\lambda_{\tau} |\widehat V(u)\rrangle.
\end{equation}
Following the same idea as previously, the expansion of $X_{\tau_{L'+1}}\dots X_{\tau_{L}}$ involve monomials of the form 
$\lambda_{s_1}\dots \lambda_{s_k}C^{L''-n''}R_{s_{k+1}}\dots R_{s_{n''}}$. The expansion of the product 
$\widetilde X_{\tau_{L'+1}}(u)\dots \widetilde X_{\tau_{L}}(u)$ is exactly the same but with the previous monomial replaced by \newline 
$(1-u)^k\lambda_{s_1}\dots \lambda_{s_k}\widehat C(u)^{L''-n''}\widehat R_{s_{k+1}}(u)\dots \widehat R_{s_{n''}}(u)$.

Putting all these expansions together, we see that finally it remains to prove
\begin{eqnarray}
 \llangle W|C^{L'+L''-n'-n''}|V\rrangle & = & -\oint_{u=1} \frac{du}{2i\pi}\frac{1}{u^{a+b+L'-n'}(1-u)^{1+L''-n''}} \\
 & \times & \llangle \widetilde W(u)|\widetilde C(u)^{L'-n'}|\widetilde V(u)\rrangle
 \llangle \widehat W(u)|\widehat C(u)^{L''-n''}|\widehat V(u)\rrangle. \nonumber 
\end{eqnarray}
This is established using result \eqref{eq:mSSEP_normalisation} and the fact that
\begin{eqnarray}
 \oint_{u=1} \frac{du}{2i\pi}\frac{1}{u^{a+b+L'-n'}(1-u)^{1+L''-n''}} & = & 
 -\frac{(-1)^{L''-n''}}{(L''-n'')!}\left. \frac{d^{L''-n''}}{du^{L''-n''}} \frac{1}{u^{a+b+L'-n'}} \right|_{u=1} \\
 & = & -\frac{\Gamma(a+b+L'+L''-n'-n'')}{\Gamma(a+b+L'-n')\Gamma(1+L"-n")}.
\end{eqnarray}

\finproof

\paragraph*{Large deviation functional of the density profile.}
We are now equipped to study the large deviation functional of the density profile. As a warm-up, we start with  
the particular case of the thermodynamic equilibrium, i.e. when $\boldsymbol{\alpha}=\boldsymbol{\beta}:=\boldsymbol{r}=(r_0,\dots,r_N)$, 
see \eqref{eq:mSSEP_steady_thermo}, which is much easier than the non-equilibrium case.

\begin{proposition}
In the thermodynamic equilibrium case, the large deviation functional is given by
\begin{equation} \label{eq:mSSEP_large_dev_thermo}
 \mathcal{F}(\{ \boldsymbol{\rho}(x)\}\ |\ \boldsymbol{r},\boldsymbol{r})= \int_0^1 dx \ B(\boldsymbol{\rho}(x),\boldsymbol{r}),
\end{equation}
where
\begin{equation}
 B(\boldsymbol{\rho},\boldsymbol{r})= \sum_{\tau=0}^{N} \rho_{\tau}\ln \left(\frac{\rho_{\tau}}{r_{\tau}}\right)
\end{equation}
\end{proposition}

We recall that $r_0+\dots+r_N=1$ and $\rho_0(x)+\dots+\rho_N(x)=1$ for all $x$. Note that $ B(\boldsymbol{\rho}(x),\boldsymbol{r})$ is 
nothing else but the Kullback-Leibler divergence between the two discrete measure $\boldsymbol{\rho}(x)$ and $\boldsymbol{r}$.

\proof
In the thermodynamic equilibrium case the stationary distribution is given by \eqref{eq:mSSEP_steady_thermo}. Hence we can 
easily evaluate
\begin{equation}
 P_L(\{\boldsymbol{\rho}^{\{1\}},\boldsymbol{\rho}^{\{2\}},\dots,\boldsymbol{\rho}^{\{n\}}\}\ |\ \boldsymbol{r},\boldsymbol{r})=
 \prod_{k=1}^n \frac{l!}{(l\rho_0^{\{k\}})!\dots (l\rho_N^{\{k\}})!} r_0^{l\rho_0^{\{k\}}}\dots r_N^{l\rho_N^{\{k\}}}.
\end{equation}
Then using the Stirling formula we obtain
\begin{equation}
 \lim\limits_{l\rightarrow \infty} -\frac{1}{L}
\ln P_L(\{\boldsymbol{\rho}^{\{1\}},\boldsymbol{\rho}^{\{2\}},\dots,\boldsymbol{\rho}^{\{n\}}\}\ |\ \boldsymbol{r},\boldsymbol{r})=
 \frac{1}{n}\sum_{k=1}^n \sum_{\tau=0}^N \rho_{\tau}^{\{k\}}\ln \left(\frac{\rho_{\tau}^{\{k\}}}{r_{\tau}}\right).
\end{equation}
The limit of large $n$ thus gives
\begin{equation}
 \lim\limits_{n\rightarrow \infty} \lim\limits_{l\rightarrow \infty} -\frac{1}{L}
 \ln P_L(\{\boldsymbol{\rho}^{\{1\}},\boldsymbol{\rho}^{\{2\}},\dots,\boldsymbol{\rho}^{\{n\}}\}\ |\ \boldsymbol{r},\boldsymbol{r})=
 \int_0^1 dx \ \sum_{\tau=0}^{N} \rho_{\tau}(x)\ln \left(\frac{\rho_{\tau}(x)}{r_{\tau}}\right),
\end{equation}
which yields the desired result.
\finproof 

The non-equilibrium case $\boldsymbol{\alpha} \neq \boldsymbol{\beta}$ is more involved.
\begin{proposition}
The large deviation functional of the density profile is given by
\begin{equation} \label{eq:mSSEP_large_dev}
 \mathcal{F}(\{ \boldsymbol{\rho}(x)\}\ |\ \boldsymbol{\alpha},\boldsymbol{\beta})= 
 \int_0^1 dx \ \Big[B(\boldsymbol{\rho}(x),\overline{\boldsymbol{\rho}}(u(x)))+ \ln u'(x)\Big],
\end{equation}
where
$u$ is the monotonic solution of the differential equation 
\begin{equation} \label{eq:mSSEP_diff_eq_F}
 \frac{u''(x)}{(u'(x))^2}+\sum_{\tau=0}^N \lambda_{\tau} \frac{\rho_{\tau}(x)}{\overline{\rho}_{\tau}(u(x))}=0
\end{equation}
satisfying boundary conditions $u(0)=0$ and $u(1)=1$.
\end{proposition}

Before proving this expression, we formulate two remarks.

\begin{remark}
We can deduce from this expression that the most probable density profile is given by $\boldsymbol{\rho}(x)=\overline{\boldsymbol{\rho}}(x)$.
The differential equation is indeed solved by the function $u(x)=x$ in this case because $\lambda_0+\dots+\lambda_N=0$.
Injecting in \eqref{eq:mSSEP_large_dev} makes the large deviation function vanish.
\end{remark}

\begin{remark}
The thermodynamic equilibrium case can be of course recovered from the general case. Indeed we have 
$\overline{\boldsymbol{\rho}}(u)=\boldsymbol{\alpha}=\boldsymbol{\beta}=\boldsymbol{r}$ for all $u$. 
Moreover the differential equation \eqref{eq:mSSEP_diff_eq_F} reduces to $u''(x)=0$ because $\lambda_s=0$ for all $s$ in this case.
It is solved by the function $u(x)=x$. Injecting in \eqref{eq:mSSEP_large_dev} leads to \eqref{eq:mSSEP_large_dev_thermo} as expected.
\end{remark}

We now present the proof of \eqref{eq:mSSEP_large_dev}

\proof
The proof presented here follows heavily the lines of the proof written in \cite{Derrida07,Derrida11} for the one species SSEP.
For the sake of simplicity, we will present the proof for the case where $a+b=1$, but the generalization to the other cases is straightforward.

We want to evaluate the probability 
$P_L(\{\boldsymbol{\rho}^{\{1\}},\boldsymbol{\rho}^{\{2\}},\dots,\boldsymbol{\rho}^{\{n\}}\}\ |\ \boldsymbol{\alpha},\boldsymbol{\beta})$
to find $\rho_{\tau}^{\{k\}}\times l$ particles (or holes) of species $\tau$ in the box $k$ for $\tau=0,\dots,N$ and $k=1,\dots,n$. This is done by summing 
the probabilities of all the configurations satisfying these constraints. For each of these configurations, we use the additivity 
principle \eqref{eq:mSSEP_additivity_principle} to divide the system into two part of size $L'=kl$ (containing $k$ boxes) and $L''=(n-k)l$
(containing $n-k$ boxes), for a fixed $1\leq k\leq n$. We thus obtain
\begin{equation}
\begin{aligned}
& P_{nl}\left(\{\boldsymbol{\rho}^{\{1\}},\dots,\boldsymbol{\rho}^{\{n\}}\}\, |\, \boldsymbol{\alpha},\boldsymbol{\beta}\right)
  =  -\frac{(kl)!((n-k)l)!}{(nl)!} \
 \oint \frac{du}{2i\pi}\frac{1}{u^{kl+1}(1-u)^{(n-k)l+1}} \\
& \hspace{3cm} \times  P_{kl}\left(\{\boldsymbol{\rho}^{\{1\}},\dots,\boldsymbol{\rho}^{\{k\}}\}\, |\, \boldsymbol{\alpha},\overline{\boldsymbol{\rho}}(u)\right) \, 
P_{(n-k)l}\left(\{\boldsymbol{\rho}^{\{k+1\}},\dots,\boldsymbol{\rho}^{\{n\}}\}\, |\, \overline{\boldsymbol{\rho}}(u),\boldsymbol{\beta}\right) 
\end{aligned}
\end{equation}
In the large $l$ limit, evaluating the previous expression at the saddle point,
we obtain the following equation for the large deviation function 
\begin{equation}
\begin{aligned}
 & \mathcal{F}_n\left(\{\boldsymbol{\rho}^{\{1\}},\dots,\boldsymbol{\rho}^{\{n\}}\}\, |\, \boldsymbol{\alpha},\boldsymbol{\beta}\right)
 = \max\limits_{0<u<1} \quad \frac{k}{n}\ln\left(\frac{nu}{k}\right)+\frac{n-k}{n}\ln\left(\frac{n(1-u)}{n-k}\right)\\
& + \frac{k}{n} \mathcal{F}_k\left(\{\boldsymbol{\rho}^{\{1\}},\dots,\boldsymbol{\rho}^{\{k\}}\}\, |\, \boldsymbol{\alpha},\overline{\boldsymbol{\rho}}(u)\right)
+ \frac{n-k}{n}\mathcal{F}_{n-k}\left(\{\boldsymbol{\rho}^{\{k+1\}},\dots,\boldsymbol{\rho}^{\{n\}}\}\, |\, \overline{\boldsymbol{\rho}}(u),\boldsymbol{\beta}\right)
\end{aligned}
\end{equation}
We repeat $n$ times the same procedure to obtain 
\begin{equation}
 \begin{aligned}
  & \mathcal{F}_n\left(\{\boldsymbol{\rho}^{\{1\}},\dots,\boldsymbol{\rho}^{\{n\}}\}\, |\, \boldsymbol{\alpha},\boldsymbol{\beta}\right)
 = \max\limits_{0=u_0<u_1<\dots<u_n=1} \frac{1}{n} \sum_{k=1}^n 
 \mathcal{F}_1\left(\boldsymbol{\rho}^{\{k\}}\, |\, \overline{\boldsymbol{\rho}}(u_{k-1}),\overline{\boldsymbol{\rho}}(u_k)\right) \\
 & \hspace{10cm} +\ln\left((u_{k}-u_{k-1})n\right)
 \end{aligned}
\end{equation}
In the large $n$ limit, we can define the continuous variable $x=k/n$ and a function $u$ such that $u(x)=u_k$. The sequence $u_k$ being monotone,
the difference $u_k-u_{k-1}$ is small in this limit. Hence we have that $\overline{\boldsymbol{\rho}}(u_{k-1})\simeq \overline{\boldsymbol{\rho}}(u_k)$
and we can replace $\mathcal{F}_1\left(\boldsymbol{\rho}^{\{k\}}\, |\, \overline{\boldsymbol{\rho}}(u_{k-1}),\overline{\boldsymbol{\rho}}(u_k)\right)$
by the equilibrium value 
$\mathcal{F}_1\left(\boldsymbol{\rho}^{\{k\}}\, |\, \overline{\boldsymbol{\rho}}(u_k),\overline{\boldsymbol{\rho}}(u_k)\right)=
B\left(\boldsymbol{\rho}^{\{k\}}\, |\, \overline{\boldsymbol{\rho}}(u_k)\right)$. We thus obtain
\begin{equation}
 \mathcal{F}(\{ \boldsymbol{\rho}(x)\}\ |\ \boldsymbol{\alpha},\boldsymbol{\beta})= 
\max\limits_{u(x)} \int_0^1 dx \ \Big[B(\boldsymbol{\rho}(x),\overline{\boldsymbol{\rho}}(u(x)))+ \ln u'(x)\Big],
\end{equation}
where the maximum is evaluated over the increasing functions $u$ satisfying $u(0)=0$ and $u(1)=1$.
The Euler-Lagrange equation associated with the maximization over $u$ of this functional gives the differential equation \eqref{eq:mSSEP_diff_eq_F}.
\finproof

Let us stress that exact computation, from 
finite size lattice, of the large deviation functional of the density profile has only be achieved on a few out-of-equilibrium models,
including the SSEP \cite{DerridaLS01,DerridaLS02} and the ASEP \cite{DerridaLS02bis,DerridaLS03}.
 
\section{Macroscopic fluctuation theory}

The computations done in the previous subsection were all about the stationary state of the models in the large size limit. 
We are now interested in the study of the full dynamics of the models in the thermodynamic limit. The idea is quite the same as for the 
study of the density profile in the stationary state. We would like to introduce coarse-grained variables (such as macroscopic particle density
or current) that will depend on space and time. We expect that all the details of the microscopic dynamics of the model will be averaged out and
that these coarse grained variables will satisfy deterministic equations depending only on a small number of physically relevant parameters in the 
large size limit.

For instance it has been proved rigorously for the ASEP with weak asymmetry, {\it i.e} $p-q=\nu/L$, that the macroscopic density $\rho(x,t)$
satisfies the Burger equation with viscosity\footnote{This equation can be heuristically derived by writing the time evolution of the mean 
particle density and performing a mean-field approximation, {\it i.e} simplifying the two-point correlation functions that appear into the 
product of two one-point functions.}
\begin{equation}
 \frac{\partial \rho}{\partial t} = \frac{\partial^2 \rho}{\partial x^2}-2\nu \frac{\partial \rho(1-\rho)}{\partial x}.
\end{equation}
In the particular case of the SSEP, {\it i.e} when $\nu=0$, the macroscopic density satisfies the heat equation
\begin{equation}
 \frac{\partial \rho}{\partial t} = \frac{\partial^2 \rho}{\partial x^2}.
\end{equation}

These results on deterministic equations tell that the probability to observe a given time evolution, path history, of 
the coarse-grained variables tends to $1$ in the thermodynamic limit if this time evolution satisfies the deterministic equation and tends to $0$
otherwise.
We are now interested in a refined result. We would like to evaluate the probability of rare events, when
the temporal evolution of the macroscopic variables deviates from the typical evolution governed by the deterministic equation. These 
rare events show up with an exponentially weak probability in the system size $L$, giving rise to a large deviation principle. This approach is 
called the macroscopic fluctuation theory (MFT) and is the whole topic of the remaining part of this chapter. We will see that this framework gives
in principle access to the fluctuations of the particle current and density profile at the price of minimizing an action or equivalently 
at the price of solving a system of non linear coupled partial differential equations (the associated Euler-Lagrange equations). 
The MFT has been in particular successfully applied to characterize dynamical phase transitions, see for instance \cite{Lazarescu15,Lazarescu17}.
For the sake of simplicity, we start by presenting this theory on the case of single species diffusive models (the 
theory was first introduced to describe this class of systems) and we show its efficiency to derive the fluctuations of the current and density 
in the stationary state. The goal is to present the main tools and techniques that are provided by this theory. We then present extensions of 
this theory (to dissipative systems or to multi-species systems) that are suitable to describe the models studied all along this manuscript: the
DiSSEP and the multi-species SSEP. We will in particular derive results about current and density fluctuations in the stationary state 
through the MFT framework and check the consistency with exact computations that were done for a finite size lattice in chapter \ref{chap:three}.
 
\subsection{Single species diffusive systems}

\subsubsection{General idea}

The macroscopic fluctuation theory (MFT) is a general approach that aims to describe out of equilibrium diffusive particle gases in the 
thermodynamic limit. It was 
developed a few years ago by Bertini, De Sole, Gabrielli, Jona-Lasinio and Landim \cite{BertiniDGJL01,BertiniDGJL02}, and has proven to be an efficient way 
to compute fluctuations of the current and of the density profile. One strength of this theory is to describe the diffusive systems through only 
two key parameters, the diffusion constant $D(\rho)$ and the conductivity $\sigma(\rho)$ which depend on the local particle density $\rho$.
These parameters can be determined case by case from the microscopic dynamics of the model.
See \cite{BertiniDGJL15} for a detailed review. Some validations from a 
microscopic point of view were realised for exactly solvable models including the SSEP \cite{DerridaLS01,DerridaLS02,Derrida07,Derrida11}.

The first step toward this macroscopic, or hydrodynamic, description is to define coarse-grained variables. We recall that we study here the case
of systems with a single species of particle. The framework of the MFT presented here allows to describe models with possibly several
particles on the same site (this number can have an upper bound to describe exclusion or not) and encompasses in particular the example of
the single species SSEP.
We recall that the occupation number $\rho_1^{(i)}$ is equal to the number of particles lying on site $i$.
We introduce the random variables $\rho^{(t,i)}:=\rho_1^{(i)}(\cC_t)$.
We recall also that the random variable $C_t$ denote the configuration of the system at time $t$. Its probability law 
satisfies the master equation of the model. In other words $\rho^{(t,i)}$ is equal to the number of particle on the site $i$ 
and at time $t$.

\begin{definition}
We define, for $L$ large, the macroscopic density of particles $\rho(x,t)$ at time\footnote{Note that
in order to do this hydrodynamic limit, we will rescale in all this section the time
with a factor $L^2$, 
as usual in this context of diffusive systems.} $t$
and at position $x\in[0,1]$ 
on the lattice by
\begin{equation}
 \rho(x,t)= \frac{1}{2 \sqrt{L}}\sum_{\lvert i-Lx\rvert\leq \sqrt{L}} \rho^{(L^2 t,i)}.
\end{equation}
\end{definition}
Note that the macroscopic density $\rho(x,t)$ is a random variable which is intuitively understood as the average number of particles per site
in a box of size $2\sqrt{L}$ (which explains the denominator in the definition) around site $Lx$ at time $L^2t$.

We also need to define the macroscopic current of particles.
We denote by $Q^{(t,i\rightarrow i+1)}$ the algebraic number of particles
that have crossed the bound between sites $i$ and $i+1$ (from left to right) during the time interval $[0,t]$. It allows us to give the 
following definition.
\begin{definition}
We introduce, for $L$ large,
\begin{equation}
 Q(x,t)= \frac{1}{2 L\sqrt{L}}\sum_{\lvert i-Lx\rvert\leq \sqrt{L}} Q^{(L^2 t,i\rightarrow i+1)}.
\end{equation}
\end{definition}

\begin{definition}
The macroscopic particle current $j(x,t)$ at time $t$ and at position $x$ is then defined as 
\begin{equation}
 j(x,t)=\frac{\partial}{\partial t}Q(x,t).
\end{equation}
\end{definition}

We mentioned that the hydrodynamic description of the model relies on two key parameters. This motivates the following definition.
\begin{definition} \label{def:diffusion_conductivity}
 We introduce the diffusion constant $D(\rho)$ that satisfies for $\rho_l$ and $\rho_r$ both close to the value $\rho$
 \begin{equation}
  \lim\limits_{t\rightarrow \infty} \frac{\langle Q^{(t,i\rightarrow i+1)} \rangle}{t} = \frac{D(\rho)(\rho_l-\rho_r)}{L} 
 \end{equation}
 and the conductivity $\sigma(\rho)$ that satisfies for $\rho_l=\rho_r=\rho$
 \begin{equation}
  \lim\limits_{t\rightarrow \infty} \frac{\langle \left(Q^{(t,i\rightarrow i+1)}\right)^2 \rangle}{t} = \frac{\sigma(\rho)}{L} 
 \end{equation}
\end{definition}

These two parameters are model dependent and has to be computed case by case from the microscopic dynamics of the model. They 
have been computed exactly in particular for the SSEP and for free random walkers (also called Brownian particles). We recall that the 
model of free random walkers is defined as a particular case of the zero range process with $p=q=1$ and $w_n=n$ (see remark \ref{rem:ZRP}).

\begin{proposition}
 For the single species SSEP we have
 \begin{equation}
  D(\rho)=1, \qquad \sigma(\rho)=2\rho(1-\rho),
 \end{equation}
and for free random walkers we have
 \begin{equation}
  D(\rho)=1, \qquad \sigma(\rho)=2\rho.
 \end{equation}
\end{proposition}

We are now equipped to present the main result associated with MFT. 

\begin{proposition}
The probability to 
observe a given path history of the macroscopic density and current profiles $\{\rho(x,t),j(x,t)\}$ during a time interval $[t_1,t_2]$ 
satisfies the large deviation principle
\begin{equation} \label{eq:proba_MFT}
 P\left(\{\rho(x,t),j(x,t)\}\right) \sim
 \exp \left[ -L \int_{t_1}^{t_2} dt \int_0^1 dx \frac{(j(x,t)+D(\rho(x,t))\partial_x\rho(x,t))^2}{2\sigma(\rho(x,t))} \right],
\end{equation}
where the fields satisfy the usual conservation law 
\begin{equation}
 \frac{\partial }{\partial t}\rho(x,t)=-\frac{\partial }{\partial x}j(x,t),
\end{equation}
and the boundary conditions
\begin{equation} \label{eq:boundary_conditions}
\rho(0,t)=\rho_l, \qquad \rho(1,t)=\rho_r.
\end{equation}
\end{proposition}

\begin{remark}
 The large deviation functional in \eqref{eq:proba_MFT} vanishes when the path history of the current and density profiles 
 follows the typical time evolution (which can be seen as the Fick's law)
 \begin{equation}
  j(x,t)=-D(\rho(x,t))\partial_x \rho(x,t).
 \end{equation}
\end{remark}

\begin{remark}
 The large deviation functional can be heuristically interpreted as the description of current and density profiles satisfying a Langevin-like
 equation
 \begin{equation}
  \begin{cases}
   \partial_t \rho(x,t)=\partial_x j(x,t) \\
   j(x,t)=-D(\rho(x,t))\partial_x \rho(x,t) +\sqrt{\sigma(\rho(x,t))}\xi(x,t),
  \end{cases}
 \end{equation}
 where $\xi(x,t)$ is a Gaussian white noise with $\langle \xi(x,t) \rangle = 0$ and $\langle \xi(x,t)\xi(x',t') \rangle = \delta(x-x')\delta(t-t')/L$.
\end{remark}

\begin{remark}
 In the weakly asymmetric simple exclusion process ({\it i.e} the ASEP with the weak asymmetry scaling $p-q=\nu/L$), the large deviation functional is modified to
 \begin{equation}
  \int_{t_1}^{t_2} dt \int_0^1 dx \frac{(j(x,t)+D(\rho(x,t))\partial_x\rho(x,t)-\nu\sigma(\rho(x,t)))^2}{2\sigma(\rho(x,t))}.
 \end{equation}
 For the ASEP without the weak asymmetry scaling, the situation is more complicated. The coarse grained variable
 $\rho(x,t)$ converges in the large size $L$ limit (with a ballistic scaling, {\it i.e} time is accelerated with a factor $L$ instead of a factor
 $L^2$ in the diffusive scaling case) to the inviscid Burger equation 
 \begin{equation}
  \partial_t \rho(x,t)=-(p-q) \partial_x \Big(\rho(x,t)(1-\rho(x,t))\Big)
 \end{equation}
 which is a well-known example of shock wave formation in partial differential equations theory. To the best of our knowledge, a macroscopic 
 fluctuation theory description of such model has not yet been provided. 
\end{remark}

Note that from a physical point of view, the large deviation functional can be intuitively understood as an action (associated to a path
history of the density and current profiles, that are the reduced phase space variables of the system in this coarse-grained description).
The probability of the path history then reads as exponential of minus an action (which makes the physicists think to the path integral formalism
in quantum mechanics). More precisely we have the following statement.

\begin{proposition}
The probability to observe at time $t_2$ a density profile $\rho_{f}(x)$ and a current profile $j_{f}(x)$, 
knowing that the density and current profiles were equal to $\rho_{i}(x)$ and $j_{i}(x)$, can be written as the path integral
\begin{equation} \label{eq:MFT_path_integral}
 P\Big(\rho_{f}(x),j_{f}(x) \ | \ \rho_{i}(x),j_{i}(x) \Big)= 
 \int \cD \rho \cD j \exp \left(-L\int_{t_1}^{t_2} dt\int_{0}^{1} dx\frac{(j+D(\rho)\partial_x \rho)^2}{2\sigma(\rho)}\right)
\end{equation}
where the functional integral is performed over fields $\rho(x,t)$ and $j(x,t)$ that satisfy the particles conservation law 
\begin{equation} \label{eq:MFT_conservation_law}
 \partial_t \rho = -\partial_x j,
\end{equation}
the boundary conditions
\begin{equation} \label{eq:MFT_boundary_conditions}
 \rho(0,t)=\rho_l, \qquad \rho(1,t)=\rho_r,
\end{equation}
and the initial and final conditions
\begin{equation} \label{eq:MFT_initial_final_conditions}
 \rho(x,t_1)=\rho_{i}(x), \quad j(x,t_1)=j_{i}(x) \quad \mbox{and} \quad \rho(x,t_2)=\rho_{f}(x), \quad j(x,t_2)=j_{f}(x).
\end{equation}
\end{proposition}

In the large system size $L\rightarrow \infty$, the path integral \eqref{eq:MFT_path_integral} can be evaluated through a saddle point
(for density and current profiles that minimize the action). This yields the following result.

\begin{proposition}
We have the large deviation principle
\begin{equation} \label{eq:MFT_large_deviation_density_current}
 P\Big(\rho_{f}(x),j_{f}(x) \ | \ \rho_{i}(x),j_{i}(x) \Big) \sim 
 \exp \left(-L \min\limits_{\rho,j} \int_{t_1}^{t_2} dt \int_{0}^{1} dx \frac{(j+D(\rho)\partial_x \rho)^2}{2\sigma(\rho)}\right), 
\end{equation}
where the minimum is taken over fields $\rho(x,t)$ and $j(x,t)$ satisfying the constraints \eqref{eq:MFT_conservation_law}, 
\eqref{eq:MFT_boundary_conditions} and \eqref{eq:MFT_initial_final_conditions}.
\end{proposition}

We thus have to minimize an action over two fields that are coupled through the equation \eqref{eq:MFT_conservation_law}. This problem
of minimization under constraints can be tackled using the Lagrange multipliers. We introduce a third field $\pi(x,t)$ that will play the role 
of the Lagrange multipliers. The problem is reduced to the minimization of
\begin{equation} \label{eq:MFT_modified_action}
 \int_{t_1}^{t_2} dt \int_{0}^{1} dx \left[\frac{(j+D(\rho)\partial_x \rho)^2}{2\sigma(\rho)}
 +(\partial_t \rho+\partial_x j)\pi \right]:=  \int_{t_1}^{t_2} dt \int_{0}^{1} dx \ \cL(\rho,\partial_x \rho,\partial_t \rho,j,\partial_x j), 
\end{equation}
over the fields $\rho$ and $j$ that are now considered as independent. It is now possible to write down the Euler-Lagrange 
equations associated to this modified action. This yields the two following equations.
\begin{proposition}
The Euler-Lagrange equations satisfied by the optimal profiles $\hat \rho(x,t)$ and $\hat j(x,t)$ that
minimize the action \eqref{eq:MFT_modified_action} are 
\begin{equation}
 \hat j=\sigma(\hat \rho)\partial_x \pi - D(\hat \rho)\partial_x \hat \rho 
\end{equation}
and
\begin{equation}
 \partial_t \pi= -D(\hat \rho)\partial_{xx}\pi -\frac{1}{2}\sigma'(\hat \rho)(\partial_x \pi)^2.
\end{equation}
\end{proposition}

\proof
We have indeed
\begin{equation}
 \frac{d}{dx}\partial_{\partial_x j} \cL = \partial_{j} \cL,
\end{equation}
which can be simplified to
\begin{equation}
 \partial_x \pi = \frac{\hat j+D(\hat \rho)\partial_x \hat \rho}{\sigma(\hat \rho)}.
\end{equation}
This yields the first equation of the proposition. The second Euler-Lagrange equation is given by
\begin{equation}
 \frac{d}{dt}\partial_{\partial_t \rho} \cL + \frac{d}{dx}\partial_{\partial_x \rho} \cL = \partial_{\rho} \cL,
\end{equation}
which leads, through an explicit computation of the partial derivatives, to
\begin{equation}
 \partial_t \pi + \frac{d}{dx}\left[ D(\hat \rho)\frac{\hat j+D(\hat \rho)\partial_x \hat \rho}{\sigma(\hat \rho)} \right] =
 D'(\hat \rho)\partial_x \pi - \frac{1}{2}\sigma'(\hat \rho)(\partial_x \pi)^2.
\end{equation}
This can be simplified using the first Euler-Lagrange equation and yields the second equation of the proposition.
\finproof

\begin{proposition}
The optimal density and current profiles $\hat \rho$ and $\hat j$ that minimize the action \eqref{eq:MFT_large_deviation_density_current}
are obtained by solving the following system of coupled non-linear partial differential equations
\begin{equation}
 \partial_t \hat \rho = \partial_x \left(D(\hat \rho)\partial_x \hat \rho \right) - \partial_x \left(\sigma(\hat \rho)\partial_x \pi \right)
\end{equation}
and
\begin{equation}
 \partial_t \pi = -D(\hat \rho)\partial_{xx} \pi - \frac{1}{2}\sigma'(\hat \rho)\left(\partial_x \pi \right)^2,
\end{equation}
where $\pi$ is an auxiliary field and $\hat j$ is obtained through
\begin{equation}
 \hat j = \sigma(\hat \rho) \partial_x \pi - D(\hat \rho)\partial_x \hat \rho.
\end{equation}
\end{proposition}

\proof 
We just need to prove the first equation. It is obtained starting from
\begin{equation}
 \hat j = \sigma(\hat \rho) \partial_x \pi - D(\hat \rho)\partial_x \hat \rho,
\end{equation}
taking the derivative with respect to $x$ and using the fact that $\partial_t \hat \rho = -\partial_x \hat j$.
\finproof

These equations allow us in principle to compute large deviation functionals of the current and density profiles directly at the macroscopic scale
without having to study in details the microscopic dynamics of the model and to deal often with intractable combinatorial problems. 
The microscopic dynamics of the system of completely encapsulated in the diffusion $D$ and the conductivity $\sigma$. The formalism presented
here can in principle apply to a wide range of diffusive systems. But in practice these coupled non-linear partial differential equations are very
hard to solve. Even for the SSEP, for which the value of $D(\rho)$ and $\sigma(\rho)$ have been computed exactly and take a rather simple expression,
the exact solutions to these coupled differential equations are not known to the best of our knowledge.
Nevertheless, we will see below that several techniques have been developed within this MFT framework to compute explicitly current
and density fluctuations in the particular case of the stationary state.

\subsubsection{Stationary state}

We are now interested in what can be deduced from the large deviation principle \eqref{eq:proba_MFT} about the properties of the stationary state
of the model. It appears indeed plausible that knowing information about the full dynamics (through this estimation of a path history probability)
may give insight on the steady state. 
More precisely, two quantities are of prime interest in non-equilibrium stationary state, because they could be a generalization of 
the thermodynamic potentials far from equilibrium: the large deviation function of the particle current and the large deviation functional
of the density profile. It turns out that these quantities can be in principle obtained, for a wide range of diffusive models, using the MFT 
formalism.

It was shown in \cite{BertiniDGJL01} that the large deviation functional of the density profile in the stationary state is obtained by determining
the optimal density and current profile $\rho(x,t)$ and $j(x,t)$ which produces an atypical density profile starting from the stationary typical 
profile $\overline{\rho}(x)$. More precisely we have the statement
\begin{proposition}
The large deviation functional of the density profile can be expressed in the MFT formalism by
\begin{equation} \label{eq:MFT_large_deviation_density}
 \cF(\{\rho(x)\})= \min\limits_{\rho(x,t),j(x,t)} \int_{-\infty}^{T} dt \int_{0}^{1} dx
 \frac{(j(x,t)+D(\rho(x,t))\partial_x \rho(x,t))^2}{2\sigma(\rho(x,t))},
 \end{equation}
 where the minimum is taken over the density and current profiles $\rho(x,t)$ and $j(x,t)$ that satisfy the particle conservation law 
 \begin{equation}
  \partial_t \rho(x,t) = -\partial_x j(x,t),
 \end{equation}
 the boundary conditions
 \begin{equation}
  \rho(0,t)=\rho_l, \qquad \rho(1,t)=\rho_r,
 \end{equation}
and the limit conditions
\begin{equation}
 \rho(x,-\infty)= \overline{\rho}(x), \qquad \rho(x,T)=\rho(x). 
\end{equation}
The profile $\overline{\rho}(x)$ denotes the mean density profile of the model.
\end{proposition}

\begin{remark}
 Note that this correspond to large deviation principle we saw previously \eqref{eq:MFT_large_deviation_density_current} with
 $t_1=-\infty$, $t_2=T$, $\rho_{i}(x)=\overline{\rho}(x)$, $j_{i}(x)=\overline{j}$ and $\rho_{f}=\rho(x)$ (the value of $j_{f}(x)$ is not 
 fixed, we have also to minimize over this quantity). 
\end{remark}

\begin{definition}
We introduce the following function
 \begin{equation} \label{eq:MFT_definition_U}
  U(x)= \frac{\delta \cF(\{\rho(x)\})}{\delta \rho(x)}.
 \end{equation}
\end{definition}

\begin{proposition}
 The function $U'$ has to satisfy the Hamilton-Jacobi equation
 \begin{equation}
  \int_{0}^{1} dx \left[ \left(\frac{D(\rho)\rho'}{\sigma(\rho)}-U'\right)^2-\left(\frac{D(\rho)\rho'}{\sigma(\rho)}\right)^2\right]
  \frac{\sigma(\rho)}{2}=0
 \end{equation}
\end{proposition}

\proof
 Following the lines of \cite{Derrida07}, we observe that the large deviation functional \eqref{eq:MFT_large_deviation_density}
 should not depend on the time $T$. We thus have that 
 \begin{equation}
   \cF(\{\rho(x)\})= \min\limits_{\delta\rho(x),j(x)} 
   \left[ \cF(\{\rho(x)-\delta \rho(x)\})+\delta T \int_{0}^{1} dx \frac{(j(x)+D(\rho(x))\rho'(x))^2}{2\sigma(\rho(x))} \right].
 \end{equation}
 The particle conservation law imposes that $\delta \rho(x)=-\delta T \times j'(x)$. The equation right above can thus be rewritten
 \begin{equation} \label{eq:MFT_large_deviation_density_intermediate}
   \cF(\{\rho(x)\})= \min\limits_{j(x)} 
   \left[ \cF(\{\rho(x)\})+\delta T \int_{0}^{1} dx \ U(x)j'(x)+\delta T \int_{0}^{1} dx \frac{(j(x)+D(\rho(x))\rho'(x))^2}{2\sigma(\rho(x))} \right].
 \end{equation}
 The optimal current profile $\hat j(x)$ is obtained through the Euler-Lagrange equation
 \begin{equation}
  \frac{\hat j(x)+D(\rho(x))\rho'(x)}{\sigma(\rho(x))}= U'(x).
 \end{equation}
Plugging back in \eqref{eq:MFT_large_deviation_density_intermediate} yields
\begin{equation}
 \int_{0}^{1} dx \left[U(x)\hat j'(x)+\frac{\sigma(\rho(x))}{2}U'(x)^2 \right]=0.
\end{equation}
We can integrate by part $U(x)\hat j'(x)$ using the fact that $U(0)=U(1)=0$ because of the boundary conditions. 
\finproof

Solving this Hamilton-Jacobi equation gives in principle access to the large deviation functional of the density profile $\cF(\{\rho(x)\})$
through the relation \eqref{eq:MFT_definition_U}. In practice for general values of the diffusion $D(\rho)$ and of the conductivity $\sigma(\rho)$
we don't know how to solve this equation. It has only been solved for a very few models including the SSEP \cite{Derrida07,Derrida11}, 
the weakly asymmetric simple exclusion process \cite{DeMasiPS89,EnaudD04}
and the Kipnis-Marchioro-Presutti model \cite{KipnisMP82,BertiniGL05}.

We now turn to the study of the fluctuations of the current in the stationary state. 
We are interested in the large deviation of the integrated current 
\begin{equation}
 \int_{0}^{1}dx \ Q(x,T) = \int_{0}^{T}dt \int_{0}^{1} dx \ j(x,t)
\end{equation}
when the time $T \rightarrow \infty$. One way to address the problem is to first study the generating function of the cumulants of the
integrated current $G(\tau)$ (the large deviation function can be obtained from the generating function through a Legendre transform, 
see chapter \ref{chap:one}).
We recall that $G(\tau)$ is defined as
\begin{equation}
 G(\tau)= \lim\limits_{T\rightarrow \infty} \lim\limits_{L\rightarrow \infty}\frac{1}{LT}
 \ln \mathbb{E}_{[0,T]} \left( \exp \left( \tau \int_{0}^{T}dt \int_{0}^{1} dx \ j(x,t) \right) \right)
\end{equation}
where $\mathbb{E}_{[0,T]}$ denotes the expectation value with respect to the probability distribution on the path history of current and density
profiles on the time interval $[0,T]$. The large deviation principle given by the MFT \eqref{eq:proba_MFT} allows us to reduce this computation
to an optimization problem
\begin{equation}
G(\tau)= \lim\limits_{T\rightarrow \infty} \frac{1}{T} \sup\limits_{\rho(x,t),j(x,t)}
\left[\tau\int_{0}^{T}dt\int_{0}^{1}dx\ j(x,t)-\int_{0}^{T}dt\int_{0}^{1}dx\frac{(j(x,t)+D(\rho(x,t))\partial_x\rho(x,t))^2}{2\sigma(\rho(x,t))}\right].
\end{equation}
If some constraints are assumed on the dynamics of the model (the reader is invited to refer to \cite{BertiniDGJL15} for details), this is equivalent to 
a stationary optimization problem where the optimal profiles can be chosen independent of time. In this case, due to the particle
conservation law, we have $\partial_x j = -\partial_t \rho = 0$. Hence the optimal current profile is constant in space and we have
\begin{eqnarray}
 G(\tau) & = & \sup\limits_{\rho(x),j} \left(\tau j- \int_{0}^{1}dx\frac{(j+D(\rho(x))\rho'(x))^2}{2\sigma(\rho(x))}\right) \\
 & = & \sup\limits_{j} \left(\tau j- \inf\limits_{\rho(x)}\int_{0}^{1}dx\frac{(j+D(\rho(x))\rho'(x))^2}{2\sigma(\rho(x))}\right).
\end{eqnarray}
We recognize exactly a Legendre transformation with respect to $j$ (see chapter \ref{chap:one}), and hence we can identify 
the large deviation of the integrated current. This yields the following proposition.
\begin{proposition}
The large deviation of the integrated current can be expressed as
 \begin{equation}
 \cF(j) = \min\limits_{\rho(x)}\int_{0}^{1} dx \frac{(j+D(\rho(x))\rho'(x))^2}{2\sigma(\rho(x))}
\end{equation}
\end{proposition}

\begin{proposition}
The large deviation function of the current is given by
\begin{equation} \label{eq:MFT_large_deviation_current_parametric}
 \cF(j)=j\int_{\rho_r}^{\rho_l} d\rho \frac{D(\rho)}{\sigma(\rho)}\left[ \frac{1+K\sigma(\rho)}{\sqrt{1+2K\sigma(\rho)}}-1\right],
\end{equation} 
where the constant $K$ is fixed by the condition 
\begin{equation} \label{eq:MFT_parametric_constraint}
 \int_{\rho_r}^{\rho_l} d\rho \frac{D(\rho)}{\sqrt{1+2K\sigma(\rho)}}=j.
\end{equation}
The large deviation function of the particle current is thus obtained in parametric form.
\end{proposition}

\proof
We want to minimize the action
\begin{equation}
 \min\limits_{\rho(x)}\int_{0}^{1} dx \frac{(j+D(\rho(x))\rho'(x))^2}{2\sigma(\rho(x))}
\end{equation}
over the density profile $\rho(x)$ for a fixed current $j$ (independent of the position $x$).
Expanding the square gives
\begin{eqnarray}
 & & \min\limits_{\rho(x)} \int_{0}^{1} dx \left[ \frac{j^2}{2\sigma(\rho(x))}+\frac{D(\rho(x))^2\rho'(x)^2}{2\sigma(\rho(x))}
 +\frac{jD(\rho(x))\rho'(x)}{\sigma(\rho(x))}\right] \\
 & = & \min\limits_{\rho(x)} \int_{0}^{1} dx \left[ \frac{j^2}{2\sigma(\rho(x))}+\frac{D(\rho(x))^2\rho'(x)^2}{2\sigma(\rho(x))}\right] 
 - \int_{\rho_r}^{\rho_l} d\rho \frac{D(\rho)}{\sigma(\rho)}
\end{eqnarray}
We are thus only left with the minimization of the first term in the previous sum (the second one is 
now independent of the profile $\rho(x)$). The associated Euler-Lagrange equation reads
\begin{equation}
 j^2 \partial_{\rho} \left(\frac{1}{2\sigma(\hat \rho)}\right)+(\hat \rho')^2 \partial_{\rho}\left(\frac{D(\hat \rho)^2}{2\sigma(\hat \rho)}\right) =
 \partial_x \left(\frac{D(\hat \rho)^2 \hat \rho'}{\sigma(\hat \rho)}\right),
\end{equation}
where $\hat \rho(x)$ is the optimal density profile. 
Multiplying by $\hat \rho'(x)$ leads to
\begin{equation}
 \partial_x \left(\frac{j^2}{2\sigma(\hat \rho(x))}-\frac{D(\hat \rho(x))^2 \hat \rho'(x)^2}{2\sigma(\hat \rho(x))}\right)=0
\end{equation}
Hence there exists an integration constant $K$ such that
\begin{equation}
 \hat \rho'(x)^2=\frac{j^2(1+2K\sigma(\hat \rho(x)))}{D(\hat \rho(x))^2}
\end{equation}
The constant $K$ is fixed to satisfy the conditions $\hat \rho(0)=\rho_l$ and $\hat \rho(1)=\rho_r$ on the boundaries. If the value of the current $j$
does not deviate too much from the typical value $\overline{j}$, we expect the optimal profile to be monotone (more precisely decreasing 
if we assume that $\rho_l>\rho_r$).
We thus have
\begin{equation} \label{eq:MFT_dem_parametric_optimal_density}
 -\frac{\hat \rho'(x)D(\hat \rho(x))}{\sqrt{1+2K\sigma(\hat \rho(x))}}=j.
\end{equation}
Integrating this last equation for $x$ ranging from $0$ to $1$ leads to \eqref{eq:MFT_parametric_constraint}. Moreover, plugging the value 
of $\rho'(x)$ obtained in \eqref{eq:MFT_dem_parametric_optimal_density} in the action yields \eqref{eq:MFT_large_deviation_current_parametric}.
\finproof

\begin{remark}
 This parametric form of the current large deviation function was obtained in \cite{BodineauD04} with a different method 
 by assuming an additivity principle of the large deviation 
 function with respect to the system size. This principle reads
 \begin{equation}
  \cF_{L'+L''}(j,\rho_l,\rho_r)= \min\limits_{\rho}( \cF_{L'}(j,\rho_l,\rho)+ \cF_{L''}(j,\rho,\rho_r)),
 \end{equation}
 where $ \cF_{L}(j,\rho_l,\rho_r)$ denotes the large deviation functional of the current in a system of size $L$ with reservoir densities $\rho_l$
 and $\rho_r$.
 The equivalence between these two methods suggests that the additivity principle holds only when the optimal density profile 
 that produces a given current deviation is independent of time. 
\end{remark}

\subsection{Single species systems with dissipation}

\subsubsection{Large deviation functional}

The macroscopic fluctuation theory can be extended to a larger class of system with bulk dissipation \cite{BodineauL10,BodineauL12}. More precisely 
we are interested in models describing lattice gas with diffusive dynamics and
creation/annihilation of particles in the bulk, which are driven out of equilibrium by two reservoirs at different densities.
An example of such model is the DiSSEP that was presented and studied in details in chapter \ref{chap:three}.

We will start by presenting how the large deviation functional of the path history should be modified to include the creation/annihilation
dynamics in addition to the diffusive part that we already encountered in the previous subsection. This will be a short summary of the work
\cite{BodineauL10}, the reader is invited to study this reference for the details.
We will focus on the particular case of the DiSSEP. The MFT framework will be used to compute the variance of the current on the lattice that
will be checked against the exact value computed from the microscopic point of view in chapter \ref{chap:three} (and whose thermodynamic
limit was obtained in subsection \ref{subsubsec:DiSSEP_thermo}).

Let us start by briefly presenting the key ingredients of the MFT related to our model. A detailed presentation can be found in \cite{BodineauL10,BodineauL12}.
It has been shown that the microscopic behavior of the system can be averaged in the thermodynamic limit and can be described at the 
macroscopic level by a small number of relevant parameters: $D(\rho)$, $\sigma(\rho)$, $A(\rho)$ and $C(\rho)$.
These parameters depend on the microscopic dynamics of the model and have to be computed for each different model. 

The two first are related to the diffusive dynamics on the lattice: $D(\rho)$ is the diffusion coefficient and $\sigma(\rho)$ is
the conductivity already introduced in the previous subsection for purely diffusive systems.
For systems with creation and annihilation, these parameters are defined by considering the systems in which the creation/annihilation dynamics
has been switched off (leaving only the diffusive dynamics on), and applying on those reduced purely diffusive model the same 
definition \ref{def:diffusion_conductivity}.
\begin{proposition}
For the DiSSEP, the diffusive dynamics is the same as for the SSEP and hence the diffusion and conductivity take the 
values
\begin{equation}
D(\rho)=1 \quad  \mbox{and} \quad \sigma(\rho)=2\rho(1-\rho).
\end{equation}
\end{proposition}

The two other parameters $A(\rho)$ and $C(\rho)$ are related to the creation-annihilation dynamics. $A(\rho)$ can be understood 
intuitively as the mean number of particles annihilated per site and per unit of time when the density profile is identically flat
and equal to $\rho$ in the system whereas $C(\rho)$ stands for the mean number of particles created.
A rigorous definition of these parameters for general dissipative models can be found in \cite{BodineauL10}.
We give the precise definition in the case of the DiSSEP dynamics.
\begin{definition}
\begin{equation}
 \frac{A(\rho)}{L^2}=\langle \lambda^2\tau_{i-1}\tau_i \rangle_{\rho} + \langle \lambda^2\tau_i\tau_{i+1} \rangle_{\rho}
\end{equation}
and
\begin{equation}
 \frac{C(\rho)}{L^2}=\langle \lambda^2(1-\tau_{i-1})(1-\tau_i) \rangle_{\rho} + \langle \lambda^2(1-\tau_i)(1-\tau_{i+1}) \rangle_{\rho},
\end{equation}
where $\langle \cdot \rangle_{\rho}$ denotes the expectation value with respect to the Bernoulli probability density
\begin{equation}
 \begin{pmatrix}
  1-\rho \\ \rho
 \end{pmatrix} \otimes \dots \otimes
 \begin{pmatrix}
  1-\rho \\ \rho
 \end{pmatrix}.
\end{equation}
\end{definition}

Note that for general dissipative processes this definition should be adapted accordingly to the local creation/annihilation dynamics in the bulk.

\begin{proposition}
For the DiSSEP, annihilation rate $A(\rho)$ is given by
\begin{equation}
A(\rho)=2\lambda_0^2\rho^2
\end{equation}
and the creation rate $C(\rho)$ by
\begin{equation}
C(\rho)=2\lambda_0^2(1-\rho)^2.
\end{equation}
\end{proposition}

Before stating the large deviation principle associated to dissipative models, we first need to define the coarse-grained variable
corresponding to the creation annihilation current. We denote by $K^{(t,i)}$ the number of particles created minus the number of particles
annihilated at site $i$ during the time interval $[0,t]$. It allows us to give the 
following definition.
\begin{definition}
We introduce, for $L$ large,
\begin{equation}
 K(x,t)= \frac{1}{2 \sqrt{L}}\sum_{\lvert i-Lx\rvert\leq \sqrt{L}} K^{(L^2 t,i)}.
\end{equation}
\end{definition}

\begin{definition}
The macroscopic particle creation/annihilation current $k(x,t)$ at time $t$ and at position $x$ is then defined as 
\begin{equation}
 k(x,t)=\frac{\partial}{\partial t}K(x,t).
\end{equation}
\end{definition}

When the number of sites $L$ goes to infinity, the probability of observing a given history of the density profile $\rho$, of the lattice current
$j$ and of the creation-annihilation current $k$ during the time interval $[0,T]$, can by written as\footnote{The link with the quantities previously computed is given by the fact that in the stationary state the mean value of
$j(x)$ is $\overline{j}_{lat}(x)$ and the mean value of $k(x)$ is $\overline{j}_{cond}(x)$.}
\begin{proposition}
\begin{equation} \label{eq:DiSSEP_large_deviation_MFT}
 \mathbb{P}_{[0,T]}\left( \{ \rho,j,k \} \right) \sim \exp \left[-L\mathcal{I}_{[0,T]}(\rho,j,k) \right],
\end{equation}
with the large deviation functional
\begin{equation}
 \mathcal{I}_{[0,T]}(\rho,j,k)=\int_0^T dt \int_0^1 dx \left\{\frac{(j(x,t)+D(\rho(x,t))\partial_x\rho(x,t))^2}{2\sigma(\rho(x,t))}
 + \Phi\left(\rho(x,t),k(x,t)\right)\right\},
\end{equation}
where
\begin{equation}\label{eq:DiSSEP_Phi}
 \Phi(\rho,k)=\frac{1}{2}\left[ A(\rho)+C(\rho)-\sqrt{k^2+4A(\rho)C(\rho)}
 +k\ln\left(\frac{\sqrt{k^2+4A(\rho)C(\rho)}+k}{2C(\rho)}\right)\right].
\end{equation}
The quantities $\rho$, $j$ and $k$ are related through the conservation equation
\begin{equation}
\partial_t \rho(x,t)=-\partial_x j(x,t)+k(x,t),
\end{equation}
and the value of $\rho$ is fixed on the boundaries $\rho(0,t)=\rho_l$, $\rho(1,t)=\rho_r$.
\end{proposition}

Note that the factor $1/2$ in the definition of $\Phi$ is a slight modification in comparison to \cite{BodineauL10} due to the fact that we consider here
creation-annihilation of pairs of particles instead of creation-annihilation of single particles.

The minimum of the large deviation functional $\mathcal{I}_{[0,T]}$ is achieved when the particle currents take their typical values, that is
 $j(x,t)=-D(\rho(x,t))\partial_x\rho(x,t)$ and $k(x,t)=C(\rho(x,t))-A(\rho(x,t))$. The typical evolution of the density profile is
 hence given by
 \begin{equation}
 \partial_t \rho(x,t)=\partial_x \left( D(\rho(x,t))\partial_x\rho(x,t)\right)+C(\rho(x,t))-A(\rho(x,t))
 \end{equation}
 which matches \eqref{eq:DiSSEP_time_density_thermo} for the DiSSEP.

\subsubsection{Check with finite size lattice exact computations}

Using the previous formalism and following \cite{BodineauL10}, it is possible to compute the local variance of the lattice current $j$ in the stationary regime. 
Due to the fact that the dynamics of the model does not conserve the number of particle, this current and its fluctuations depend on the
position in the system. Hence, given a function $\tau(x)$, we want to compute the generating function of the cumulants of the integrated current
$\int_0^T dt \int_0^1 dx\; \tau(x)\; j(x,t)$ for $T$ going to infinity. This leads to the following definition.
\begin{definition}
We introduce the generating function
\begin{equation}
\mathcal{F}(\{ \tau(x) \} ) = \lim\limits_{T \rightarrow \infty}\lim\limits_{L \rightarrow \infty} \frac{1}{LT}
\ln \mathbb{E}_{[0,T]}\left( \exp \left( \int_0^T dt \int_0^1 dx \tau(x)j(x,t) \right) \right).
\end{equation}
\end{definition}
The previous expression can be simplified using \eqref{eq:DiSSEP_large_deviation_MFT} and a saddle point method. It reduces to maximize a functional
over the time dependent fields $\rho$, $j$ and $k$. Assuming that the extremum of this functional is achieved for time independent
profiles, we end up with the following expression (the reader is invited to refer to \cite{BodineauL10} for the details)
\begin{proposition}
\begin{equation}
\mathcal{F}(\{ \tau(x) \} ) = \sup\limits_{\rho,j} \left( \int_0^1 dx \tau(x)j(x)-\hat{\mathcal{I}}(\rho,j) \right),
\end{equation} 
with
\begin{equation}
\hat{\mathcal{I}}(\rho,j)= \int_0^1 dx \left( \frac{(j(x)+D(\rho(x))\partial_x\rho(x))^2}{2\sigma(\rho(x))}
 + \Phi\left(\rho(x),\partial_x j(x)\right) \right).
\end{equation} 
\end{proposition}

To compute the local variance of the lattice current $j$ at the point $y$, it is enough to take $\tau(x)=\delta(x-y)$ and expand 
$\mathcal{F}(\{ \epsilon \tau \} )$ up to order $\epsilon^2$. For a small perturbation $\epsilon$, the fields are expected to be close to
their typical value
\begin{equation}
\rho(x)=\overline{\rho}(x)+\epsilon\frac{f(x)}{D(\overline{\rho}(x))},
\qquad j(x)=-D(\overline{\rho}(x))\partial_x \overline{\rho}(x)+\epsilon q(x)
\end{equation}
with the constraint $f(0)=f(1)=0$ due to the boundaries.
We then obtain
\begin{equation}\label{eq:DiSSEP_developement_large_deviation}
\mathcal{F}(\{ \epsilon \tau \} )  =  -\epsilon D(\overline{\rho}(y))\partial_x \overline{\rho}(y) + \frac{\epsilon^2}{2}E^{(2)}(y) 
\end{equation}
with the variance of the lattice current at the point $y$ 
\begin{equation}
 E^{(2)}(y) =2\sup\limits_{q,f} \left\{ q(y)-\int_0^1 dx\left( \frac{(q(x)+f'(x))^2}{2\sigma(\overline{\rho}(x))}
 +\frac{(q'(x)+U(x)f(x))^2}{4(A(\overline{\rho}(x))+C(\overline{\rho}(x)))} \right) \right\}
\end{equation}
and $U(x)=\frac{A'(\overline{\rho}(x))-C'(\overline{\rho}(x))}{D(\overline{\rho}(x))}=4\lambda_0^2$\;.
We make the following change of variables to solve this optimization problem
\begin{equation}\label{eq:DiSSEP_definition_psi}
\varphi(x)=\frac{q(x)+f'(x)}{\sigma(\overline{\rho}(x))}, \qquad \psi(x)=\frac{q'(x)+U(x)f(x)}{2(A(\overline{\rho}(x))+C(\overline{\rho}(x)))},
\end{equation}
so that the Euler-Lagrange equations become for the DiSSEP:
\begin{equation}
 \left\{ \begin{aligned}
         & \psi'(x)=\varphi(x)-\delta(x-y) \\
         & \varphi'(x)=4\lambda_0^2 \psi(x).
         \end{aligned} \right. 
\end{equation}
Note that there are slight modifications in expressions \eqref{eq:DiSSEP_developement_large_deviation} 
and \eqref{eq:DiSSEP_definition_psi} with respect to \cite{BodineauL10}, 
in accordance with the modification of $\Phi$ 
(see discussion after \eqref{eq:DiSSEP_Phi}).

These equations can be solved analytically and we get
\begin{equation}
 \left\{ \begin{aligned}
         & \psi(x)=\frac{\theta(x\leq y)\sinh(2\lambda_0 x)\cosh(2\lambda_0(1-y))
         +\theta(x>y)\sinh(2\lambda_0(x-1))\cosh(2\lambda_0 y)}{\sinh 2\lambda_0}\\
         & \varphi(x)=\frac{2\lambda_0\left[\theta(x\leq y)\cosh(2\lambda_0 x)\cosh(2\lambda_0(1-y))
         +\theta(x>y)\cosh(2\lambda_0(x-1))\cosh(2\lambda_0 y) \right]}{\sinh 2\lambda_0}.
         \end{aligned} \right. 
         \label{eq:DiSSEP_phipsi}
\end{equation}
The function $q(x)$ can be also computed analytically by solving 
\begin{equation}
q''(x)-4\lambda_0^2 q(x)= \partial_x(2(A(\overline{\rho}(x))+C(\overline{\rho}(x)))\psi(x))-4\lambda_0^2 \sigma(\overline{\rho}(x))\varphi(x)\;.
\end{equation}
Note that it depends on $y$, see for instance the expressions \eqref{eq:DiSSEP_phipsi}. 
It allows us to deduce the expression of $q(x)$ at the special point $y$ (as needed in \eqref{eq:DiSSEP_developement_large_deviation})
\begin{equation}
 q(y)= \int_0^1 dx \left[ \sigma(\overline{\rho}(x))\varphi(x)^2+2(A(\overline{\rho}(x))+C(\overline{\rho}(x)))\psi(x)^2  \right], 
\end{equation}
with $\varphi$ and $\psi$ are given above. Hence for the DiSSEP, the variance of the current lattice computed from MFT is 
\begin{eqnarray*}
 E^{(2)}(y)
 & = & \int_0^1 dx \left[ \sigma(\overline{\rho}(x))\varphi(x)^2+2(A(\overline{\rho}(x))+C(\overline{\rho}(x)))\psi(x)^2  \right] \\
 & = & \frac{4\lambda_0^2}{\sinh^2 2\lambda_0}\left[ \cosh^2 2\lambda_0(1-y) \int_0^1 dx \left( \sigma(\overline{\rho}(x))+\sinh^2 2\lambda_0x \right)
 \right. \\
 & & \qquad \qquad \qquad \left.+\cosh^2 2\lambda_0y \int_0^1 dx \left( \sigma(\overline{\rho}(x))+\sinh^2 2\lambda_0(1-x) \right) \right]\;.
\end{eqnarray*}
Using the explicit form for $\sigma$, we show that this result obtained from MFT matches perfectly the previous result 
\eqref{eq:DiSSEP_variance_lattice_current_thermo} computed exactly from a microscopic description of the model.

The result obtained here points out the consistency of the MFT developed in \cite{BodineauL10}, for a system with diffusion and dissipation, with
exact computations performed for a finite size lattice.

\subsection{Multi-species diffusive systems: the multi-SSEP case}

\subsubsection{Large deviation functional}

In this section we propose, based on the exact microscopic computations, a hydrodynamic description of the multi-species SSEP
which extends the MFT to systems with several species of particles. We check the consistency with the exact results derived in the previous
subsection (about the large deviation functional of the density profile), 
the rigorous proof of the approach remains to be done.

The first step toward the hydrodynamic description of the multi-species SSEP is to defined coarse-grained, or average variables. 
In the same way as for the single species models, we introduce the random variables $\rho_{\tau}^{(t,i)}:=\rho_{\tau}^{(i)}(\cC_t)$.
We recall that the occupation number $\rho_{\tau}^{(i)}$ is equal to $1$ if the local configuration at site $i$ is equal to $\tau$ and 
$0$ otherwise. We recall also that the random variable $C_t$ denote the configuration of the system at time $t$. Its probability law 
satisfies the master equation of the model. In other words $\rho_{\tau}^{(t,i)}$ is equal to $1$ if there is a particle (or hole) of species 
$\tau$ at site $i$ and at time $t$ and $0$ otherwise.

\begin{definition}
We define, for $L$ large, the macroscopic density $\rho_{\tau}(x,t)$ of the species $\tau$ at time\footnote{Again,
in order to do this hydrodynamic limit, we will rescale in all this section the time
with a factor $L^2$, 
as usual in this context of diffusive systems.} 
$t$ and at position $x\in[0,1]$ 
on the lattice by
\begin{equation}
 \rho_{\tau}(x,t)=  \frac{1}{2 \sqrt{L}}\sum_{\lvert i-Lx\rvert\leq \sqrt{L}} \rho_{\tau}^{(L^2 t,i)}.
\end{equation}
We will also need the row vector
\begin{equation}
 \boldsymbol{\rho}(x,t)=(\rho_0(x,t),\dots,\rho_N(x,t)).
\end{equation}
\end{definition}
Note that the macroscopic density $\rho_{\tau}(x,t)$ is a random variable which is intuitively understood as the average number of particles of 
species $\tau$ in a box of size $2\sqrt{L}$ (which explains the denominator in the definition) around site $Lx$ at time $L^2t$.

We also need to define the macroscopic current of particles.
We denote by $Q_{\tau}^{(t,i\rightarrow i+1)}$ the algebraic number of particles of species $\tau$
that have crossed the bound between sites $i$ and $i+1$ (from left to right) during the time interval $[0,t]$. It allows us to give the 
following definition.
\begin{definition}
We introduce, for $L$ large,
\begin{equation}
 Q_{\tau}(x,t)= \frac{1}{2 L\sqrt{L}}\sum_{\lvert i-Lx\rvert\leq \sqrt{L}} Q_{\tau}^{(L^2 t,i\rightarrow i+1)}.
\end{equation}
\end{definition}

\begin{definition}
The macroscopic particle current $j_{\tau}(x,t)$ of species $\tau$ at time $t$ and at position $x$ is then defined as 
\begin{equation}
 j_{\tau}(x,t)=\frac{\partial}{\partial t}Q_{\tau}(x,t).
\end{equation}
We will also need the row vector
\begin{equation}
 \boldsymbol{j}(x,t)=(j_0(x,t),\dots,j_N(x,t)).
\end{equation}
\end{definition}

\paragraph*{Rate function for the multi-species SSEP.}

The idea of the MFT is to express the probability to observe certain density profiles $\boldsymbol{\rho}(x,t)$
and current profiles $\boldsymbol{j}(x,t)$ during the time interval $[t_1,t_2]$ as a large deviation principle.
We present now one of the main result of this paper, which gives a new perspective on the rate function of diffusive models with exclusion which
can be seen as that of a model of free particles but with an additional exclusion constraint.

\begin{proposition}
We have the large deviation principle
\begin{equation} \label{eq:mSSEP_proba_MFT}
 P\left(\{\boldsymbol{\rho}(x,t),\boldsymbol{j}(x,t)\}\right) \sim
 \exp \left[ -L \int_{t_1}^{t_2} dt \int_0^1 dx \sum_{\tau=0}^N \frac{(j_{\tau}(x,t)+\partial_x\rho_{\tau}(x,t))^2}{4\rho_{\tau}(x,t)} \right],
\end{equation}
where the fields satisfy the usual conservation law 
\begin{equation}
 \frac{\partial }{\partial t}\boldsymbol{\rho}(x,t)=-\frac{\partial }{\partial x}\boldsymbol{j}(x,t),
\end{equation}
the boundary conditions
\begin{equation} \label{eq:mSSEP_boundary_conditions}
\boldsymbol{\rho}(0,t)=\boldsymbol{\alpha}, \qquad \boldsymbol{\rho}(1,t)=\boldsymbol{\beta}
\end{equation}
and the additional exclusion constraints
\begin{equation} \label{eq:mSSEP_constraints}
\rho_0(x,t)+\dots+\rho_N(x,t)=1, \qquad j_0(x,t)+\dots+j_N(x,t)=0.
\end{equation}
\end{proposition}

The rate function \eqref{eq:mSSEP_proba_MFT} can be heuristically interpreted having in mind that, for Brownian particles,
the diffusion constant is $D(\rho)=1$ and the conductivity is $\sigma(\rho)=2\rho$. The functional \eqref{eq:mSSEP_proba_MFT} 
is exactly the one that describes
a model of independent Brownian particles of $N$ different species, but on top of that we impose the exclusion constraint \eqref{eq:mSSEP_constraints}
which translates the fact that there is at most one particle per site. We recall that in our notation the holes (empty sites) are interpreted
as a species of particles. This formula is supported by: (i) the consistency check with the large deviation
functional of the density profile in the stationary state done in the next subsection, (ii) the following remark. 

\begin{remark}
The well known case of the SSEP with a single species and holes can be recovered from \eqref{eq:mSSEP_proba_MFT} 
by setting $N=1$.
We recall that the holes are labelled by $0$ and the particles $1$.
We have in this case $j_1(x,t)=-j_0(x,t):=j(x,t)$ and $\rho_1(x,t)=1-\rho_0(x,t):=\rho(x,t)$ due to the constraints \eqref{eq:mSSEP_constraints}.
 Then the rate function in \eqref{eq:mSSEP_proba_MFT} becomes 
\begin{equation} 
\int_{t_1}^{t_2} dt \int_0^1 dx \frac{(j(x,t)+\partial_x\rho(x,t))^2}{4\rho(x,t)(1-\rho(x,t))},
\end{equation}
 which agrees with the known expression for the single species SSEP (recall that the diffusion constant is given by $D(\rho)=1$ 
 and the conductivity by $\sigma(\rho)=2\rho(1-\rho)$).
\end{remark}

\subsubsection{Check with finite size lattice exact computations}

Following what was done in \cite{Derrida07}, this framework allows us to express the probability to observe at time $T$ a density profile 
$\boldsymbol{\rho}(x)$ in the stationary state. We have to identify how this deviation is produced, i.e. we have to find 
the optimal path $\boldsymbol{\rho}(x,t)$ such that $\boldsymbol{\rho}(x,-\infty)=\overline{\boldsymbol{\rho}}(x)$ and 
$\boldsymbol{\rho}(x,T)=\boldsymbol{\rho}(x)$:
\begin{equation} \label{eq:mSSEP_large_devia_MFT}
 \mathcal{F}(\{ \boldsymbol{\rho}(x)\}\, |\, \boldsymbol{\alpha},\boldsymbol{\beta})=
 \min\limits_{\boldsymbol{\rho}(x,t),\boldsymbol{j}(x,t)}
 \int_{-\infty}^{T} dt \int_0^1 dx \sum_{\tau=0}^N \frac{(j_{\tau}(x,t)+\partial_x\rho_{\tau}(x,t))^2}{4\rho_{\tau}(x,t)}.
\end{equation}
Note that the probability to observe a deviation in the density profile $\boldsymbol{\rho}(x)$ does not depend on the time at which this deviation 
occurs. It means that \eqref{eq:mSSEP_large_devia_MFT} does not depend on $T$. 

Cutting the integration interval $(-\infty,T]$ in \eqref{eq:mSSEP_large_devia_MFT} into two pieces 
$(-\infty,T-\delta T]$ and $[T-\delta T,T]$ yields
\begin{equation} \label{eq:mSSEP_large_dev_diff}
\mathcal{F}(\{ \boldsymbol{\rho}(x)\}\, |\, \boldsymbol{\alpha},\boldsymbol{\beta})=
 \min\limits_{\delta\boldsymbol{\rho}(x),\boldsymbol{j}(x)} \left[
 \mathcal{F}(\{ \boldsymbol{\rho}(x)-\delta\boldsymbol{\rho}(x)\}\, |\, \boldsymbol{\alpha},\boldsymbol{\beta})+
 \delta T \int_0^1 dx \sum_{\tau=0}^N \frac{(j_{\tau}(x)+\rho_{\tau}'(x))^2}{4\rho_{\tau}(x)}\right],
\end{equation}
where we have used the definitions $\boldsymbol{\rho}(x)-\delta\boldsymbol{\rho}(x)=\boldsymbol{\rho}(x,T-\delta T)$ and
$\boldsymbol{j}(x)=\boldsymbol{j}(x,T)$. The conservation law reads $\delta\boldsymbol{\rho}(x)=-\boldsymbol{j}'(x)\times \delta T$.
The equation on the large deviation functional \eqref{eq:mSSEP_large_dev_diff} above suggests to introduce the following quantity.

\begin{definition}
We define 
\begin{equation}
 U_{\tau}(x)=\frac{\delta  \mathcal{F}(\{ \boldsymbol{\rho}(x)\}\, |\, \boldsymbol{\alpha},\boldsymbol{\beta})}{\delta \rho_{\tau}(x)}.
\end{equation}
\end{definition}
We can write using \eqref{eq:mSSEP_large_dev_diff} an equation satisfied by the $U_s(x)$'s. Indeed, maximising \eqref{eq:mSSEP_large_dev_diff} over
the current profile $\boldsymbol{j}(x)$ with the constraint \eqref{eq:mSSEP_constraints} yields
\begin{equation}
 j_{\tau}(x)=-\rho_{\tau}'(x)+2\rho_{\tau}(x)U_{\tau}'(x)-2\rho_{\tau}(x)\mu(x),
\end{equation}
with the Lagrange multiplier
\begin{equation}
 \mu(x)=\sum_{\tau=0}^N\rho_{\tau}(x)U_{\tau}'(x).
\end{equation}
Using the fact that $\sum_{\tau=0}^{N}j_{\tau}(0)U_{\tau}(0)=\sum_{\tau=0}^{N}j_{\tau}(1)U_{\tau}(1)=0$ (because of the boundary conditions \eqref{eq:mSSEP_boundary_conditions}),
we can perform an integration by part 
and derive an equation satisfied by the functions $U_{\tau}'(x)$.
\begin{proposition}
The functions $U_{\tau}'(x)$ should satisfy the Hamilton-Jacobi equation
\begin{equation} \label{eq:mSSEP_Hamilton}
 \int_0^1 dx \left[\sum_{\tau=0}^N \left(\rho_{\tau}'(x)U_{\tau}'(x)-\rho_{\tau}(x)U_{\tau}'(x)^2 \right)
 +\left(\sum_{\tau=0}^N\rho_{\tau}(x)U_{\tau}'(x)\right)^2\right] =0.
\end{equation}
\end{proposition}

We can check that the large deviation functional exactly computed in \eqref{eq:mSSEP_large_dev} indeed fulfills this equation.

\begin{proposition}
The function $U_{\tau}(x)$ obtained from the exact expression of the large deviation of the density profile \eqref{eq:mSSEP_large_dev} 
(derived from finite size lattice computations) that is given through direct computation by
\begin{equation}
 U_{\tau}(x)=\ln\left(\frac{\rho_{\tau}(x)}{\overline{\rho}_{\tau}(u(x))}\right)+1,
\end{equation}
where the function $u$ satisfies \eqref{eq:mSSEP_diff_eq_F}, is a solution to the Hamilton equation \eqref{eq:mSSEP_Hamilton}.
\end{proposition}

\proof
Using the constraints \eqref{eq:mSSEP_constraints} and the expression of $U_{\tau}(x)$, 
the differential equation \eqref{eq:mSSEP_diff_eq_F} can be rewritten 
\begin{equation}
 \frac{u''(x)}{u'(x)}=-\sum_{\tau=0}^N\rho_{\tau}(x)U_{\tau}'(x).
\end{equation}
This permits to show that 
\begin{equation}
 \left(\frac{u''}{u'}\right)'(x)=\sum_{\tau=0}^N \left(\rho_{\tau}'(x)U_{\tau}'(x)-\rho_{\tau}(x)U_{\tau}'(x)^2\right)+
 \left(\sum_{\tau=0}^N \rho_{\tau}(x)U_{\tau}'(x) \right)^2.
\end{equation}
Then we deduce that the left hand side of \eqref{eq:mSSEP_Hamilton} is equal to
\begin{equation}
 \int_0^1 dx\left(\frac{u''}{u'}\right)'(x)= \frac{u''(1)}{u'(1)}-\frac{u''(0)}{u'(0)}=0,
\end{equation}
because $u''(1)=u''(0)=0$ thanks to \eqref{eq:mSSEP_diff_eq_F}.
\finproof

\begin{remark}
 The computation presented above points out the consistency of the hydrodynamic description \eqref{eq:mSSEP_proba_MFT} with 
 the exact computations performed on the finite size lattice (through matrix ansatz). This is thus a strong hint on the validity of the 
 large deviation principle \eqref{eq:mSSEP_proba_MFT} in the multi-species case. 
\end{remark}

\renewcommand{\theequation}{R.\arabic{equation}}
\appendix

\chapter{R\'{e}sum\'{e} en fran\c cais}

\section*{Physique statistique des syst\`{e}mes hors d'\'{e}quilibre}

Un syst\`{e}me physique est dit \`{a} l'\'{e}quilibre thermodynamique s'il est \`{a} l'\'{e}quilibre par rapport \`{a} toutes les grandeurs physiques imaginables. 
Il doit par exemple \^{e}tre \`{a} l'\'{e}quilibre m\'{e}canique, thermique, \'{e}lectrodynamique, chimique. En d'autres termes c'est un syst\`{e}me pour lequel 
on n'observe aucun courant macroscopique d'aucune grandeur physique (comme un courant d'\'{e}nergie, de charge, de particules).

L'\'{e}tat de tels syst\`{e}mes est obtenu en maximisant l'entropie sous certaines contraintes, dict\'{e}es par l'interaction du syst\`{e}me avec son environnement
(l'\'{e}nergie moyenne peut \^{e}tre fix\'{e}e par exemple). Ce principe fondamental permet d'obtenir la c\'{e}l\`{e}bre distribution de Boltzmann
\begin{equation}
 \cS(\cC)=\frac{e^{-\beta E(\cC)}}{Z}, 
\end{equation}
o\`{u} $\cS(\cC)$ d\'{e}signe la probabilit\'{e} que le syst\`{e}me soit dans la configuration $\cC$ et $Z$ est la normalisation, appel\'{e}e fonction de partition.
Cela permet de d\'{e}finir un potentiel thermodynamique (\'{e}nergie libre) et de d\'{e}crire efficacement les propri\'{e}t\'{e}s macroscopiques du syst\`{e}me 
(transitions de phase par exemple).

A l'oppos\'{e}, un syst\`{e}me physique est dit hors d'\'{e}quilibre si il affiche des courants macroscopiques d'une ou plusieurs grandeurs physiques. 
L'arch\'{e}type de tel mod\`{e}le est donn\'{e} par deux r\'{e}servoirs de particules de densit\'{e}s diff\'{e}rentes reli\'{e}s par un tuyau. Le r\'{e}servoir de forte 
densit\'{e} se d\'{e}verse dans celui de faible densit\'{e} et on observe un courant de particules. Un syst\`{e}me peut \^{e}tre hors d'\'{e}quilibre parce qu'il est en 
phase de relaxation vers l'\'{e}quilibre ou alors parce qu'il est maintenu hors d'\'{e}quilibre par son environnement. On dit dans ce dernier cas que le syst\`{e}me 
est dans un \'{e}tat stationnaire hors d'\'{e}quilibre. Ce sont ces \'{e}tats stationnaires qui nous int\'{e}ressent dans ce manuscrit.

Il n'existe pas de cadre g\'{e}n\'{e}ral pour d\'{e}crire de tels syst\`{e}mes: on ne sait pas comment g\'{e}n\'{e}raliser le principe fondamental de maximisation de
l'entropie et la distribution de Boltzmann. L'objectif est donc d'\'{e}tudier des mod\`{e}les simples et de calculer exactement dans ces cas particuliers
la distribution de l'\'{e}tat stationnaire, afin d'apporter un \'{e}clairage nouveau sur la structure g\'{e}n\'{e}rale. 

Le cadre utilis\'{e} pour d\'{e}crire de tels syst\`{e}mes est celui des cha\^{i}nes de Markov. Le syst\`{e}me peut occuper un nombre fini de configurations. Pendant 
un intervalle de temps infinit\'{e}simal $dt$, le syst\`{e}me se trouvant dans la configuration $\cC$ a une probabilit\'{e} 
$m(\cC \rightarrow \cC')dt$ de sauter dans la configuration $\cC'$. 
L'\'{e}volution temporelle de la distribution de probabilit\'{e} du syst\`{e}me est dict\'{e}e par l'\'{e}quation ma\^{i}tresse
\begin{equation}
 \frac{d \cP_t(\cC)}{dt} = \sum_{\cC' \neq \cC} \cP_t(\cC') m(\cC' \rightarrow \cC) - \sum_{\cC' \neq \cC} \cP_t(\cC)m(\cC \rightarrow \cC'),
\end{equation}
o\`{u} $\cP_t(\cC)$ d\'{e}signe la probabilit\'{e} que le syst\`{e}me se trouve dans la configuration $\cC$ \`{a} l'instant $t$. Cette \'{e}quation lin\'{e}aire
peut simplement se r\'{e}\'{e}crire sous une forme matricielle en introduisant un vecteur $\ket{\cC}$ associ\'{e} \`{a} chaque configuration $\cC$,
un vecteur regroupant les probabilit\'{e}s de toutes les configurations
\begin{equation}
 \ket{\cP_t} = \sum_{\cC} \cP_t(\cC) \ket{\cC},
\end{equation}
une matrice de Markov
\begin{equation}
 M = \sum_{\cC',\cC} m(\cC \rightarrow \cC') \ket{\cC'} \bra{\cC},
\end{equation}
o\`{u} l'on d\'{e}finit $m(\cC \rightarrow \cC) = \sum_{\cC' \neq \cC} m(\cC \rightarrow \cC')$. L'\'{e}quation ma\^{i}tresse devient
\begin{equation}
 \frac{d\ket{\cP_t}}{dt} = M\ket{\cP_t}.
\end{equation}
On s'int\'{e}resse particuli\`{e}rement \`{a} l'\'{e}tat stationnaire associ\'{e}e \`{a} cette \'{e}quation (son existence est assur\'{e}e par le th\'{e}or\`{e}me de Perron-Frobenius),
on le notera $\steady$: il v\'{e}rifie $M\steady=0$.

Les syst\`{e}mes \`{a} l'\'{e}quilibre thermodynamique sont caract\'{e}ris\'{e}s par la relation de bilan d\'{e}taill\'{e}
\begin{equation}
 \cS(\cC)m(\cC \rightarrow \cC') = \cS(\cC')m(\cC' \rightarrow \cC),
\end{equation}
qui r\'{e}sout de mani\`{e}re \'{e}vidente la version stationnaire de l'\'{e}quation ma\^{i}tresse.
Elle permet notamment de retrouver la distribution de Boltzmann et elle assure la r\'{e}versibilit\'{e} temporelle du syst\`{e}me dans l'\'{e}tat stationnaire.

Le cas des syst\`{e}mes hors d'\'{e}quilibre est beaucoup plus complexe. La relation de bilan d\'{e}taill\'{e} peut \^{e}tre g\'{e}n\'{e}ralis\'{e}e mais elle ne conduit pas 
de mani\`{e}re \'{e}vidente \`{a} une expression simple de la distribution stationnaire. Elle donne cependant de fructueux r\'{e}sultats, 
avec le th\'{e}or\`{e}me de fluctuation (qui permet notamment de retrouver les relations de dissipation-fluctuation d'Einstein, les relations de r\'{e}ciprocit\'{e} 
d'Onsager). Elle sugg\`{e}re notamment l'\'{e}tude de la fonction g\'{e}n\'{e}ratrice du courant de particules dans l'\'{e}tat stationnaire qui semble \^{e}tre 
une g\'{e}n\'{e}ralisation naturelle des potentiels thermodynamiques au cas hors d'\'{e}quilibre.

On s'int\'{e}resse plus particuli\`{e}rement dans ce manuscrit \`{a} des mod\`{e}les d'exclusion. Ces mod\`{e}les sont d\'{e}finis sur des r\'{e}seaux unidimensionnels
avec un nombre fini de site $L$ sur lequel se d\'{e}placent des particules. Chaque site du r\'{e}seau est soit vide, soit occup\'{e} par au plus une particule 
(ce qui correspond \`{a} une r\`{e}gle d'exclusion de type Fermi).
Il y a $N$ esp\`{e}ces diff\'{e}rentes de particules. Le contenu du site $i$ est d\'{e}crit par une variable d'occupation locale $\tau_i\in \{0\dots N\}$. 
$\tau_i=0$ si le site est vide, et s'il est occup\'{e} $\tau_i=1,\dots,N$ suivant l'esp\`{e}ce de la particule qui l'occupe. 
Une configuration sur le r\'{e}seau est donc caract\'{e}ris\'{e}e par le $L$-uplet $\bm\tau=(\tau_1,\dots,\tau_L)$. Le r\'{e}seau peut \^{e}tre connect\'{e} \`{a} 
des r\'{e}servoirs de particules au niveau du site $1$ ainsi que du site $L$ ou alors avoir des conditions aux bords p\'{e}riodiques (g\'{e}om\'{e}trie 
d'anneau). On associe un vecteur $\ket{\tau}$ \`{a} chaque configuration locale $\tau=0,\dots,N$ et on note $V \simeq \CC^{N+1}$ l'espace vectoriel
engendr\'{e} par ces vecteurs.

L'\'{e}volution temporelle du syst\`{e}me est stochastique. Les particules peuvent sauter sur des sites voisins (s'ils sont libres), r\'{e}agir localement
avec d'autres esp\`{e}ces, ou alors \^{e}tre cr\'{e}\'{e}es ou d\'{e}truites, et tout cela avec des taux de probabilit\'{e} donn\'{e}s. La dynamique \'{e}tant locale, la matrice 
de Markov $M$ peut se d\'{e}composer comme une somme d'op\'{e}rateurs agissant localement sur le r\'{e}seau. Dans le cas p\'{e}riodique, cela donne
\begin{equation}
 M = \sum_{k=1}^L m_{k,k+1},
\end{equation}
o\`{u} $m$ est un op\'{e}rateur de saut local dans le bulk. C'est une matrice agissant sur l'espace tensoriel $V \otimes V$ et qui encode la dynamique 
du mod\`{e}le sur deux sites adjacents (une composante $V$ de l'espace tensoriel repr\'{e}sente en quelque sorte un site du r\'{e}seau).
La matrice de Markov agit elle sur l'espace tensoriel $V^{\otimes L}$ (qui repr\'{e}sente les $L$ sites du r\'{e}seau).
Les indices repr\'{e}sentent les composantes du produit tensoriel $V^{\otimes L}$ ({\it i.e} les sites) sur lesquels la matrice $m$ agit non trivialement.
Plus pr\'{e}cis\'{e}ment on a 
\begin{equation}
 m_{k,k+1} = \underbrace{1 \otimes \dots \otimes 1}_{k-1} \otimes m \otimes \underbrace{1 \otimes \dots \otimes 1}_{L-k-1}.
\end{equation}
Pour les syst\`{e}mes \`{a} bords ouverts, la matrice de Markov s'exprime comme
\begin{equation}
 M = B_1+\sum_{k=1}^{L-1} m_{k,k+1}+\overline{B}_L,
\end{equation}
o\`{u} $B$ et $\overline{B}$ sont des matrices agissant sur $V$ et qui encodent respectivement la dynamique avec le r\'{e}servoir de gauche et de droite.
La matrice
\begin{equation}
 B_1 = B \otimes \underbrace{1 \otimes \dots \otimes 1}_{L-1}
\end{equation}
agit sur le premier site du r\'{e}seau et de mani\`{e}re similaire $\overline{B}_L$ agit sur le dernier site du r\'{e}seau.

Nous nous focalisons dans ce manuscrit sur des mod\`{e}les d'exclusion qui sont exactement solubles. Le but est de calculer exactement la distribution 
stationnaire de tels mod\`{e}les. On s'int\'{e}ressera aussi aux fluctuations du courant de particules dans l'\'{e}tat stationnaire.

Les syst\`{e}mes int\'{e}grables donnent un cadre id\'{e}al pour construire de mani\`{e}re syst\'{e}matique de tels mod\`{e}les d'exclusion.
La proc\'{e}dure est d\'{e}taill\'{e}e dans la section suivante.

\section*{Int\'{e}grabilit\'{e}}

L'id\'{e}e essentielle des mod\`{e}les int\'{e}grables a pris naissance en m\'{e}canique classique avec l'observa\-tion que la pr\'{e}sence de grandeurs physiques 
conserv\'{e}es (typiquement l'\'{e}nergie) \'{e}tait tr\`{e}s utile \`{a} la r\'{e}solution exacte des \'{e}quations de Newton. Cette observation a \'{e}t\'{e} formalis\'{e}e 
pr\'{e}cis\'{e}ment (dans le cadre de la m\'{e}canique analytique) avec le th\'{e}or\`{e}me de Liouville. Celui-ci affirme que si un syst\`{e}me \'{e}voluant dans un 
espace de phase de dimension $2n$ poss\`{e}de $n$ grandeurs physiques ind\'{e}pendantes conserv\'{e}es, alors ses \'{e}quations du mouvement peuvent 
\^{e}tre r\'{e}solues par quadrature.

Cette id\'{e}e a \'{e}t\'{e} transpos\'{e}e aux syst\`{e}mes quantiques et aux cha\^{i}nes de Markov: dans ce cadre une grandeur conserv\'{e}e est un op\'{e}rateur qui commute 
avec l'Hamiltonien ou la matrice de Markov. Il n'existe pas d'\'{e}quivalent du th\'{e}or\`{e}me de Liouville dans ce cadre, assurant que l'\'{e}quation 
ma\^{i}tresse peut \^{e}tre exactement r\'{e}solue s'il y a assez de grandeurs conserv\'{e}es. N\'{e}anmoins leur pr\'{e}sence est un indice fort d'une possible 
r\'{e}solution analytique. 

Pour les mod\`{e}les d'exclusion, il existe un moyen syst\'{e}matique de g\'{e}n\'{e}rer simultan\'{e}ment une matrice de Markov avec des 
op\'{e}rateurs qui commutent avec elle (et qui commutent entre eux deux \`{a} deux). L'ensemble de ces op\'{e}rateurs et la matrice de Markov sont engendr\'{e}s
par une matrice de transfert $t(z)$, d\'{e}pendant d'un param\`{e}tre spectral $z$. La propri\'{e}t\'{e} essentielle de cette matrice de transfert est qu'elle 
commute pour des valeurs diff\'{e}rentes du param\`{e}tre spectral $[t(z),t(z')]=0$. Elle est reli\'{e}e tr\`{e}s simplement \`{a} la matrice de Markov $M$ par une 
relation du type $t'(1) \sim M$. La matrice de transfert est construite \`{a} partir d'un objet clef: la matrice $\check R(z)$, 
qui appara\^{i}t dans ce cadre comme la pierre angulaire de l'int\'{e}grabilit\'{e}. Elle agit dans l'espace tensoriel $V \otimes V$. 
Elle est solution de la c\'{e}l\`{e}bre \'{e}quation de Yang-Baxter (braid\'{e}e)
\begin{equation}
 \check R_{12}(z) \check R_{23}(z z')\check R_{12}(z')=\check R_{23}(z')\check R_{12}(z z')\check R_{23}(z)\;. 
\end{equation}
C'est une \'{e}quation portant sur des matrices agissant dans l'espace tensoriel $V \otimes V \otimes V$. Les indices d\'{e}signent les 
composantes de l'espace tensoriel dans lequel les matrices agissent non trivialement. Par exemple
\begin{equation}
 \check R_{12}(z) = \check R(z) \otimes 1 , \quad \check R_{23}(z) = 1 \otimes \check R(z).
\end{equation}
L'\'{e}quation de Yang-Baxter est essentielle pour prouver la propri\'{e}t\'{e} de commutation de la matrice de transfert. 
Dans le contexte des processus d'exclusion, la matrice $\check R(z)$ est directement connect\'{e}e \`{a} l'op\'{e}rateur de saut local $m$ par une 
relation du type $\check R'(1) \sim m$

Pour les syst\`{e}mes aux conditions aux bords p\'{e}riodiques la matrice de transfert s'exprime comme
\begin{equation}
 t(z) = tr_0(R_{0L}(z) \dots R_{01}(z)),
\end{equation}
o\`{u} $R(z)=P.\check R(z)$, avec $P$ l'op\'{e}rateur de permutation sur $V \otimes V$.

Pour les syst\`{e}mes \`{a} bords ouverts, il est n\'{e}cessaire d'introduire des matrices $K(z)$ et $\overline{K}(z)$, agissant sur $V$, 
qui assurent en quelque sorte l'int\'{e}grabilit\'{e} des conditions aux bords. Celles-ci doivent \^{e}tre solution de l'\'{e}quation de r\'{e}flexion 
 \begin{equation}
   \check R_{12}(z_1/z_2) K_1(z_1)\check R_{12}(z_1z_2)  K_1(z_2)\ = \ K_1(z_2)\check R_{12}(z_1z_2) K_1(z_1)\check R_{12}(z_1/z_2)\;,
 \end{equation}
 et d'une \'{e}quation similaire pour $\overline{K}(z)$.
C'est une \'{e}quation portant sur des matrices agissant sur $V \otimes V$. Les indices indiquent encore une fois sur quelle composante de 
l'espace tensoriel les matrices agissent non trivialement. Cette relation est essentielle pour prouver la propri\'{e}t\'{e} de commutation de la 
matrice de transfert dans le cas ouvert. Les matrices $K(z)$ et $\overline{K}(z)$ sont directement connect\'{e}es aux op\'{e}rateurs de saut
locaux $B$ et $\overline B$ par des relations du type $K'(1) \sim B$ et $\overline{K}'(1) \sim \overline{B}$.
Pour de tels syst\`{e}mes ouvert la matrice de transfert s'exprime comme
 \begin{equation}
  t(z)=tr_0(\widetilde{K}_0(z) R_{0L}(z) \dots R_{01}(z)K_0(z) R_{10}(z) \dots R_{L0}(z)),
 \end{equation}
o\`{u} $\widetilde K(z)$ s'exprime simplement en fonction de la matrice $\overline{K}(z)$.

Il appara\^{i}t donc tr\`{e}s important de d\'{e}terminer des solutions \`{a} l'\'{e}quation de Yang-Baxter et \`{a} l'\'{e}quation de r\'{e}flexion afin de d\'{e}couvrir de 
nouveaux mod\`{e}les hors d'\'{e}quilibre exactement solubles. Des progr\`{e}s ont \'{e}t\'{e} r\'{e}alis\'{e}s dans ce sens avec l'introduction de nouvelles 
structures alg\'{e}briques qui permettent de g\'{e}n\'{e}rer des solutions de ces deux \'{e}quations par une proc\'{e}dure de Baxt\'{e}risation. Cette proc\'{e}dure 
peut \^{e}tre r\'{e}sum\'{e}e comme \'{e}tant une m\'{e}thode pour cr\'{e}er une matrice $R(z)$ ou une matrice $K(z)$ \`{a} partir d'un op\'{e}rateur de saut local $m$
ou $B$ v\'{e}rifiant des relations alg\'{e}briques sp\'{e}cifiques, en ajoutant de mani\`{e}re judicieuse un param\`{e}tre spectral. Cela a permis par exemple 
de proposer une classe de conditions aux bords int\'{e}grables pour la g\'{e}n\'{e}ralisation multi-esp\`{e}ces de l'ASEP.

On s'int\'{e}resse dans la section suivante \`{a} une m\'{e}thode permettant de calculer exactement la distribution stationnaire des mod\`{e}les 
int\'{e}grables hors d'\'{e}quilibre.

\section*{Ansatz matriciel pour les \'{e}tats stationnaires hors d'\'{e}quilibre}

Depuis quelques d\'{e}cennies une technique s'est d\'{e}velopp\'{e}e pour calculer analytiquement la distribution stationnaire de syst\`{e}mes hors d'\'{e}quilibre,
appel\'{e}e ansatz matriciel. Cette m\'{e}thode a \'{e}t\'{e} introduite pour r\'{e}soudre exactement le processus d'exclusion simple totalement asym\'{e}trique (TASEP,
de l'acronyme anglais). Elle a depuis ce temps \'{e}t\'{e} largement utilis\'{e}e pour r\'{e}soudre d'autres mod\`{e}les d'exclusion.

L'id\'{e}e de l'ansatz matriciel est d'exprimer la distribution stationnaire comme un produit de matrices
\begin{equation}
 \cS(\tau_1,\dots,\tau_L) = \frac{1}{Z}\llangle W|X_{\tau_1}\dots X_{\tau_L}|V\rrangle, 
\end{equation}
avec $Z$ une normalisation assurant que les probabilit\'{e}s stationnaires se somment bien \`{a} $1$. $X_{\tau_i}$ est une matrice d\'{e}pendant du site $i$.
$\llangle W|$ est un vecteur ligne et $|V\rrangle$ est un vecteur colonne de sorte que la contraction de ce produit de matrices est un nombre r\'{e}el.
Cet ansatz peut se r\'{e}\'{e}crire de mani\`{e}re compacte sous forme vectorielle (qui nous est tr\`{e}s utile pour la suite), en introduisant le vecteur 
\begin{equation}
 \mathbf{X} = \begin{pmatrix}
             X_0 \\ X_1 \\ \vdots \\ X_N
            \end{pmatrix},
\end{equation}
contenant toutes les matrices. L'ansatz matriciel se r\'{e}\'{e}crit alors
\begin{equation}
 \steady = \llangle W| \mathbf{X} \otimes \dots \otimes \mathbf{X} |V\rrangle.
\end{equation}

Les matrices $X_0,\dots,X_N$ et les vecteurs $\llangle W|$ et $|V\rrangle$ doivent bien s\^{u}r satisfaire des relations alg\'{e}briques tr\`{e}s pr\'{e}cises
pour que ce produit de matrices calcule correctement les probabilit\'{e}s stationnaires.

Dans le cas des mod\`{e}les int\'{e}grables, il existe une mani\`{e}re syst\'{e}matique de d\'{e}terminer ces relations alg\'{e}briques, gr\^{a}ce \`{a} deux relations clefs:
la relation de Zamolodchikov-Faddeev
\begin{equation}
 \check R\left(\frac{z_1}{z_2}\right) \mathbf{A}(z_1) \otimes \mathbf{A}(z_2) = \mathbf{A}(z_2) \otimes \mathbf{A}(z_1),
\end{equation}
et les relations de Ghoshal-Zamolodchikov
\begin{equation}
 \llangle W| K(z)\mathbf{A}\left(\frac{1}{z}\right) = \llangle W| \mathbf{A}(z), \quad 
 \overline{K}(z) \mathbf{A}\left(\frac{1}{z}\right)|V\rrangle = \mathbf{A}(z)|V\rrangle.
\end{equation}
$\mathbf{A}(z)$ est un vecteur dont les entr\'{e}es sont des \'{e}l\'{e}ments d'une alg\`{e}bre non commutative.

Ces relations sont \`{a} la racine des relations t\'{e}lescopiques dans le bulk
\begin{equation}
 m \mathbf{X} \otimes \mathbf{X} = \mathbf{X} \otimes \overline{\mathbf{X}}-\overline{\mathbf{X}} \otimes \mathbf{X},
\end{equation}
et sur les bords
\begin{equation}
 \llangle W|B \mathbf{X}=\llangle W|\overline{\mathbf{X}}, \quad \overline{B}\mathbf{X}|V\rrangle= -\overline{\mathbf{X}},
\end{equation}
qui s'obtiennent respectivement en prenant la d\'{e}riv\'{e}e de la relation Zamolodchikov-Faddeev par rapport \`{a} $z_1$ et en imposant $z_1=z_2=1$, 
et en prenant la d\'{e}riv\'{e}e des relations de Ghoshal-Zamolodchikov par rapport \`{a} $z$ et en imposant $z=1$.
Le vecteur $\mathbf{X}$ est donn\'{e} par $\mathbf{X}=\mathbf{A}(1)$ et le vecteur $\overline{\mathbf{X}}$ est obtenu par une relation du type 
$\overline{\mathbf{X}} \sim \mathbf{A}'(1)$. Ces relations t\'{e}lescopiques 
permettent de prouver tr\`{e}s facilement que l'ansatz matriciel donne bien l'\'{e}tat stationnaire du mod\`{e}le (on obtient en effet une somme t\'{e}lescopique
en agissant avec la matrice de Markov sur l'\'{e}tat en produit de matrices).

Cette proc\'{e}dure peut \^{e}tre appliqu\'{e}e pour calculer l'\'{e}tat stationnaire de plusieurs nouveaux mod\`{e}les, qui ont \'{e}t\'{e} d\'{e}couvert en r\'{e}solvant l'\'{e}quation
de Yang-Baxter et l'\'{e}quation de r\'{e}flexion. L'un des mod\`{e}les est une g\'{e}n\'{e}ralisation du processus d'exclusion simple sym\'{e}trique (SSEP, de l'acronyme
anglais) o\`{u} des paires de particules peuvent condenser ou s'\'{e}vaporer. Un autre mod\`{e}le est un TASEP \`{a} deux esp\`{e}ces de particules avec des bords 
ouverts. Enfin le dernier mod\`{e}le est une g\'{e}n\'{e}ralisation multi-esp\`{e}ces (avec un nombre quelconque d'esp\`{e}ces) du SSEP avec bords ouverts.

En r\'{e}sum\'{e}, les relations de Zamolodchikov-Faddeev et de Ghoshal-Zamolodchikov ont permis d'introduire des ansatz matriciels avec une structure alg\'{e}brique 
riche (avec par exemple des op\'{e}rateurs ``chapeaux'' non scalaires). De plus des calculs exacts de grandeurs physiques (courants et densit\'{e}s de particules) 
ont pu \^{e}tre r\'{e}alis\'{e}s gr\^{a}ce \`{a} cet ansatz matriciel.

\section*{Equations qKZ et fluctuations du courant}

Nous venons de voir que l'ansatz matriciel est particuli\`{e}rement efficace pour calculer exactement l'\'{e}tat stationnaire de certains 
processus d'exclusion. Mais son champ d'application ne se limite pas \`{a} cela. Nous allons voir une autre de ses applications, toujours dans le 
contexte de la physique statistique hors d'\'{e}quilibre. Nous consid\`{e}rons le cas particulier de l'ASEP avec bords ouverts et nous nous 
int\'{e}ressons aux 
fluctuations du courant de particules dans l'\'{e}tat stationnaire. La fonction g\'{e}n\'{e}ratrice des cumulants de cette observable s'obtient en d\'{e}formant, 
\`{a} l'aide d'un param\`{e}tre $\xi$, la matrice de Markov du mod\`{e}le de fa\c con \`{a} pouvoir ``compter'' les particules qui sont inject\'{e}es
par le r\'{e}servoir de gauche dans le syst\`{e}me. La fonction g\'{e}n\'{e}ratrice est la plus grande valeur propre de cette matrice d\'{e}form\'{e}e.

Nous tentons de d\'{e}terminer cette plus grande valeur propre ainsi que le vecteur propre associ\'{e} (qu'on appelle le {\it ground state}) \`{a} l'aide 
des \'{e}quations de Knizhnik-Zamolodchikov $q$-d\'{e}form\'{e}es (\'{e}quations $q$KZ). Ces \'{e}quations peuvent \^{e}tre vues de mani\`{e}re simplifi\'{e}e comme des 
d\'{e}formations, par un param\`{e}tre $s$, des relations Zamolodchikov-Faddeev et Ghoshal-Zamolodchikov.

Le r\'{e}sultat essentiel est que ces \'{e}quations admettent une solution polynomiale, qui s'exprime sous forme de produit de matrices, lorsque la 
contrainte suivante est v\'{e}rifi\'{e}e:
\begin{equation}
 \xi = s^n, \quad \mbox{avec} \quad n \in \NN. 
\end{equation}
La normalisation du vecteur solution des \'{e}quations $q$KZ dans ce cas est identifi\'{e}e comme \'{e}tant un polyn\^{o}me de Koornwinder associ\'{e} \`{a} une 
partition particuli\`{e}re. Cela offre une expression en produit de matrices pour ce polyn\^{o}me.

De plus, nous avons conjectur\'{e} que dans la limite $n \rightarrow \infty$, le vecteur solution des \'{e}quations $q$KZ converge vers le ground state de la 
matrice de Markov d\'{e}form\'{e}e. Cela r\'{e}v\`{e}le une connexion in\'{e}dite entre les fluctuations du courant dans l'ASEP et la th\'{e}orie des polyn\^{o}mes 
sym\'{e}triques.

\section*{Limite hydrodynamique}

Un des objectifs premiers de la physique statistique est de d\'{e}crire de mani\`{e}re efficace le comportement macroscopique de syst\`{e}mes physiques dans 
la limite o\`{u} le nombre de leurs constituants tend vers l'infini. Le but est de d\'{e}terminer \`{a} partir des interactions entre les constituants \'{e}l\'{e}mentaires
au niveau microscopique, des variables macroscopiques (telles que la temp\'{e}rature ou la pression) ainsi que des lois physiques d\'{e}crivant 
l'\'{e}tat du syst\`{e}me (comme une \'{e}quation d'\'{e}tat par exemple).

Dans le cadre des processus d'exclusion, nous nous int\'{e}ressons donc \`{a} la limite thermodynamique, {\it i.e} lorsque le nombre de sites sur le r\'{e}seau 
tend vers l'infini. Nous calculons la limite thermodynamique des observables (courants et densit\'{e}s de particules notamment) pour les nouveaux mod\`{e}les 
introduits et \'{e}tudi\'{e}s sur le r\'{e}seau de taille finie. Nous calculons aussi de mani\`{e}re exacte la fonction de grande d\'{e}viation des profils de densit\'{e} pour 
le SSEP multi-esp\`{e}ces, en utilisant un principe d'additivit\'{e} prouv\'{e} gr\^{a}ce \`{a} l'ansatz matriciel.

Nous nous int\'{e}ressons ensuite \`{a} une th\'{e}orie d\'{e}velopp\'{e}e dans les derni\`{e}res ann\'{e}es, appel\'{e}e th\'{e}orie des fluctuations macroscopiques 
(MFT, de l'acronyme anglais), qui vise \`{a} donner un cadre g\'{e}n\'{e}ral pour d\'{e}crire les syst\`{e}mes diffusifs hors d'\'{e}quilibre dans la 
limite thermodynamique. L'id\'{e}e est d'introduire des variables macroscopiques de courants et de densit\'{e}s de particules (qui d\'{e}crivent 
la valeur moyenne du courant et de la densit\'{e} autour d'un point du r\'{e}seau). Par analogie avec la loi des grands nombres, ces variables sont suppos\'{e}es
avoir un comportement d\'{e}terministe dans la limite thermodynamique. Plus pr\'{e}cis\'{e}ment, il a \'{e}t\'{e} montr\'{e} que la probabilit\'{e} d'observer une trajectoire, 
une \'{e}volution temporelle, de ces variables macroscopiques v\'{e}rifie un principe de grande d\'{e}viation. D'un point de vue physique, la fonction 
de grande d\'{e}viation peut \^{e}tre interpr\'{e}t\'{e}e comme une action. Cela permet, au prix de r\'{e}soudre des \'{e}quations diff\'{e}rentielles non lin\'{e}aires 
(qui sont des \'{e}quations d'Euler-Lagrange obtenues par minimisation de l'action), de calculer les fluctuations du courant et du profil de densit\'{e}
dans l'\'{e}tat stationnaire.

Les r\'{e}sultats exacts obtenus pour le mod\`{e}le avec \'{e}vaporation et condensation de paires de particules sur le r\'{e}seau \`{a} taille finie ont \'{e}t\'{e} 
confront\'{e}s avec succ\`{e}s, dans la limite thermodynamique, avec les pr\'{e}dictions de la MFT. Cela constitue la premi\`{e}re v\'{e}rification de ce type 
pour les syst\`{e}mes avec cr\'{e}ation et annihilation de particules dans le bulk. Enfin les r\'{e}sultats obtenus pour la g\'{e}n\'{e}ralisation 
multi-esp\`{e}ces du SSEP \`{a} bords ouverts (notamment concernant la fonction de grande d\'{e}viation des profils de densit\'{e}s) ont permis de proposer 
une extension de la MFT \`{a} ce syst\`{e}me diffusif \`{a} plusieurs esp\`{e}ces de particules. Cela apporte un nouveau point de vue sur la transcription 
de la contrainte d'exclusion dans la fonctionnelle de grande d\'{e}viation (action) de la MFT.


\providecommand{\href}[2]{#2}\begingroup\raggedright\endgroup

\small{

\subsection*{Approche int\'{e}grabiliste des mod\`{e}les de physique statistique hors d'\'{e}quilibre.}

Malgr\'{e} son ind\'{e}niable succ\`{e}s pour d\'{e}crire les syst\`{e}mes physiques \`{a} l'\'{e}quilibre thermodynamique (gr\^{a}ce \`{a} la distribution 
de Boltzmann, refl\'{e}tant la maximisation de l'entropie, et permettant la construction syst\'{e}matique de potentiels thermodynamiques),
la physique statistique n'offre pas de cadre g\'{e}n\'{e}ral pour \'{e}tudier les ph\'{e}nom\`{e}nes hors d'\'{e}quilibre, i.e dans lesquels on 
observe un courant moyen non nul d'une grandeur physique (\'{e}nergie, charge, particules...).
L'objectif de la th\`{e}se est de d\'{e}crire de tels syst\`{e}mes \`{a} l'aide de mod\`{e}les tr\`{e}s simples mais qui retranscrivent n\'{e}anmoins 
les principales caract\'{e}ristiques physiques de ceux-ci. Ces mod\`{e}les sont constitu\'{e}s de particules se d\'{e}placant de mani\`{e}re al\'{e}atoire
sur un r\'{e}seau unidimensionnel connect\'{e} \`{a} des r\'{e}servoirs et soumises \`{a} un principe d'exclusion. L'enjeu est de calculer 
exactement l'\'{e}tat stationnaire du mod\`{e}le, notamment le courant de particules, ses fluctuations et plus particuli\`{e}rement sa fonction 
de grande d\'{e}viation (qui pourrait jouer le r\^{o}le d'un potentiel thermodynamique hors d'\'{e}quilibre).
Une premi\`{e}re partie de la th\`{e}se vise \`{a} construire des mod\`{e}les dits int\'{e}grables, dans lesquels il est possible de mener \`{a} bien des 
calculs exacts de quantit\'{e}s physiques. De nouveaux mod\`{e}les hors d'\'{e}quilibre sont propos\'{e}s gr\^{a}ce \`{a} la r\'{e}solution dans des cas particuliers
de l'\'{e}quation de Yang-Baxter et de l'\'{e}quation de r\'{e}flexion. De nouvelles structures alg\'{e}briques permettant la construction de ces solutions 
par une proc\'{e}dure de Baxt\'{e}risation sont introduites.
Une deuxi\`{e}me partie de la th\`{e}se consiste \`{a} calculer exactement l'\'{e}tat stationnaire de tels mod\`{e}les en utilisant l'ansatz matriciel. Les liens
entre cette technique et l'int\'{e}grabilit\'{e} du mod\`{e}le ont \'{e}t\'{e} mis en lumi\`{e}re au travers de deux relations clef: la relation de 
Zamolodchikov-Faddeev et la relation de Ghoshal-Zamolodchikov. L'int\'{e}grabilit\'{e} a aussi \'{e}t\'{e} exploit\'{e}e au travers des equations de 
Knizhnik-Zamolodchikov quantiques, afin de calculer les fluctuations du courant, mettant en lumi\`{e}re des connexions avec la th\'{e}orie des 
polyn\^{o}mes sym\'{e}triques (polyn\^{o}mes de Koornwinder en particulier).
Enfin une derni\`{e}re partie de la th\`{e}se porte sur la limite hydrodynamique des mod\`{e}les \'{e}tudi\'{e}s, i.e lorsque la maille du r\'{e}seau tend vers z\'{e}ro et 
que le nombre de constituants du syst\`{e}me tend vers l'infini. Les r\'{e}sultats exacts obtenus sur les mod\`{e}les \`{a} taille finie ont permis de v\'{e}rifier les 
pr\'{e}dictions de la th\'{e}orie des fluctuations macroscopiques (concernant les fluctuations du courant et du profil de densit\'{e} dans l'\'{e}tat stationnaire)
et de l'\'{e}tendre \`{a} des mod\`{e}les comprenant plusieurs esp\`{e}ces de particules.

\subsection*{Integrabilist approach of non-equilibrium statistical physics models}

Although statistical physics has been very successful to describe physical systems at thermal equilibrium (thanks to the Boltzmann distribution,
which reflects the maximization of the entropy, and allows one to construct in a systematic way thermodynamic potentials),
it remains elusive to provide an efficient framework to study phenomena that are out-of-equilibrium, i.e displaying non vanishing current of 
physical quantities (energy, charge, particles...).
The goal of the thesis is to describe such systems with very simple models which retain nevertheless their main physical features.
The models consist in particles evolving randomly on a one dimensional lattice connected to reservoirs and subject to hard-core repulsion.
The challenge lies in computing exactly the stationary state of the model, especially the particle current, its fluctuations and more
precisely its large deviation function (which is expected to play the role of an out-of-equilibrium thermodynamic potential).
In the first part of the thesis we construct models, called integrable, in which we can perform exact computations of physical quantities.
We introduce several new out-of-equilibrium models that are obtained by solving, in specific cases, the Yang-Baxter equation and the reflection 
equation. We provide new algebraic structures which allow us to construct the solutions through a Baxterisation procedure. 
In the second part of the thesis we compute exactly the stationary state of these models using a matrix ansatz. We shed light on the 
connection between this technique and the integrability of the model by pointing out two key relations: the Zamolodchikov-Faddeev 
relation and the Ghoshal-Zamolodchikov relation. The integrability is also exploited, through the quantum Knizhnik-Zamolodchikov equations,
to compute the fluctuations of the particles current, unrevealing connections with the theory of symmetric polynomials (the Koornwinder polynomials
in particular).
Finally the last part of the thesis deals with the hydrodynamic limit of the models, i.e when the lattice spacing tends to $0$ and the number
of particles tends to infinity. The exact results obtained for a finite size system allow us to check the validity of the predictions of 
the macroscopic fluctuations theory (concerning the fluctuations of the current and the density profile in the stationary state) and to extend
the theory to systems with several species of particles.}

\end{document}